\documentclass[b5paper,10pt,twoside,openright,english]{book}

\usepackage[sorting=none,citestyle=numeric-comp,backend=bibtex]{biblatex}
\usepackage{biblatex-style}
\usepackage{graphicx}
\usepackage{fullpage}
\usepackage{amsmath}
\usepackage{amsthm}
\usepackage{amsfonts}
\usepackage{amssymb}
\usepackage{enumerate}
\usepackage{graphics}
\usepackage{float}
\usepackage{placeins}
\usepackage{appendix}
\usepackage{subfig}
\usepackage{lineno}
\usepackage{multirow}
\usepackage{comment}
\usepackage{minitoc}
\usepackage{dirtytalk}
\usepackage{setspace}

\usepackage{hyperref}
\hypersetup{colorlinks, citecolor=black, filecolor=black, linkcolor=black, urlcolor=black}

\usepackage[margin=1in,headsep=0.2in]{geometry}
  
\usepackage{afterpage}
\newcommand\myemptypage{
    \null
    \thispagestyle{empty}
    \newpage
    }

\usepackage{fancyhdr}
\usepackage{titlesec}
\titleformat{\section} {\normalfont\fontsize{15}{15}\bfseries}{\thesection}{1em}{}
\titleformat{\subsection} {\normalfont\fontsize{15}{15}\bfseries}{\thesubsection}{1em}{}
\titleformat{\subsubsection} {\normalfont\fontsize{14}{15}\bfseries}{\thesubsubsection}{1em}{}

\usepackage{kpfonts}
\usepackage[T1]{fontenc}
\sffamily

\begin{document}

\pagenumbering{Roman}

\begin{titlepage}

\dominitoc

\begin{figure}
\centering \includegraphics[width=0.3\textwidth]{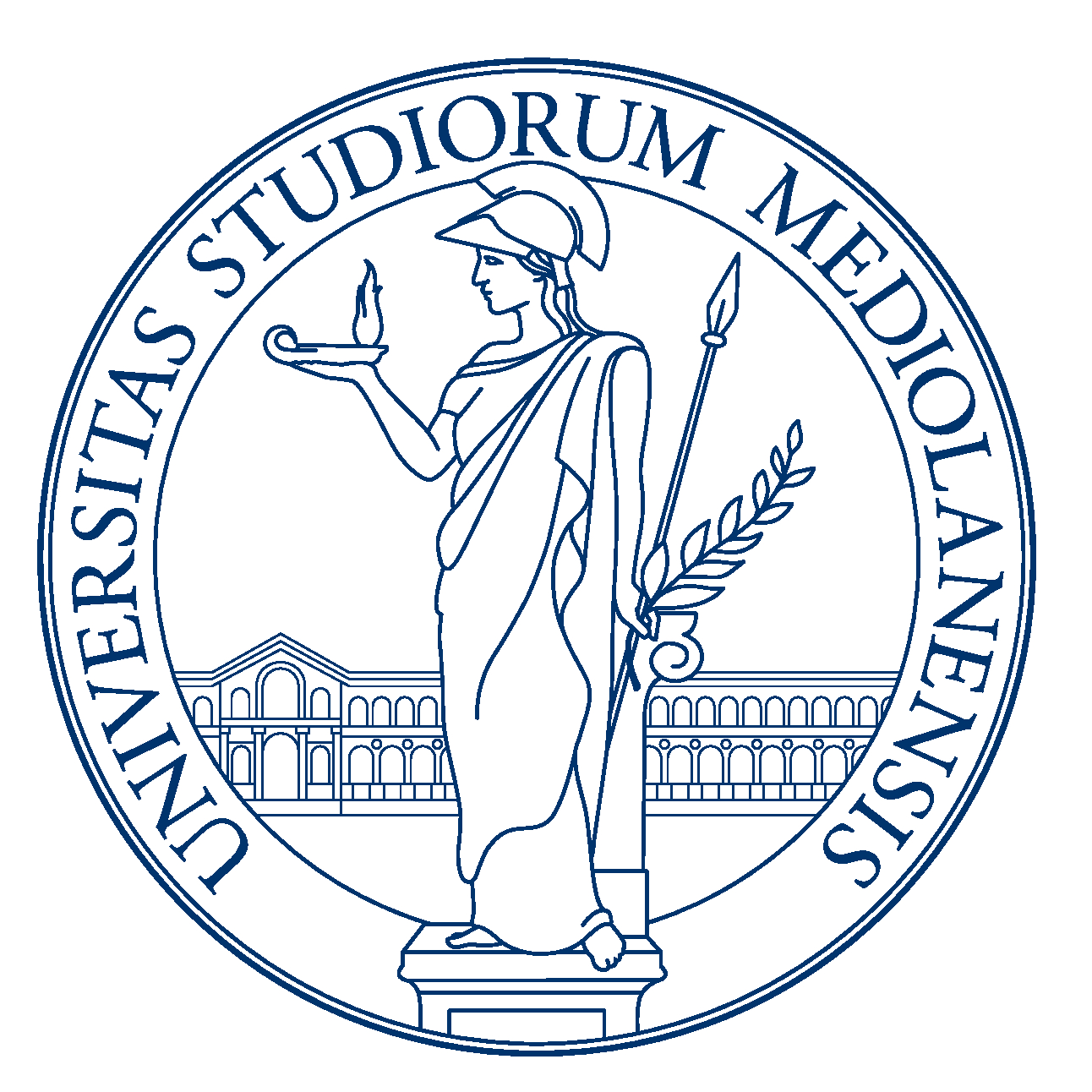} \vspace{0.4 cm}
\end{figure}

\begin{center} \huge \textsc{Universit\`a degli Studi di Milano} \\ \end{center}
\begin{center} \scalebox{0.7}{\huge Facolt\`a di Scienze Matematiche, Fisiche e Naturali} \\ \end{center}
\begin{center} \scalebox{0.6}{\huge Scuola di Dottorato in Fisica, Astrofisica e Fisica Applicata}\\ \end{center}
\begin{center} \scalebox{0.7}{\huge Ciclo XXXV}\\ \end{center}

\vspace{1.4 cm}
\begin{center}
\Huge \bf
Search for electroweak production of supersymmetric particles in compressed mass spectra with the ATLAS detector at the LHC
 \par \end{center}

\par \vspace{1 cm}
  
\begin{flushleft}
\begin{center} \huge \scalebox{1.1}{\textsc{Eric Ballabene}} \end{center}
\end{flushleft} 
 
\vfill \vspace{1 cm}
 
\begin{flushleft}
\Large \scalebox{1.1}{Thesis Supervisor: Tommaso Lari}\\
\Large \scalebox{1.1}{PhD Coordinator: Matteo Paris}\\
\Large \scalebox{1.1}{Academic Year: 2021/2022}
\end{flushleft}

\end{titlepage}

\myemptypage

\mbox{}
\vfill
\begin{flushleft}
{\bf Referees of the thesis}\\
Barbara Mele, \textit{Sapienza University of Rome}\\
Simone Gennai, \textit{University of Milano-Bicocca}\\
\vspace{1 cm}
{\bf Commission of the final examination}\\
Marco Paganoni, \textit{University of Milano-Bicocca}\\
Alan Barr, \textit{University of Oxford}\\
Laura Jeanty, \textit{University of Oregon}\\
\vspace{1 cm}
{\bf Final examination}\\
November 17, 2022\\
Universit\`a degli Studi di Milano

\end{flushleft}

\newpage

\tableofcontents

\fontsize{14}{14}\selectfont

\setcounter{page}{1}

\chapter{Introduction}
\pagenumbering{arabic}
\setcounter{page}{1}

In the previous century, it was observed from calculations that most galaxies would not move as they do if they did not contain a large amount of unseen matter, nowadays called \textit{Dark Matter} (DM), which composes almost 27\% of the entire universe. Since then, the quest for DM particles started with many indirect evidences of DM being reported, including the observation of gravitational lensing, the cosmic microwave background radiation, baryonic acoustic oscillations, and the formation and evolution of galaxies. Despite all of these evidences, DM has not been observed yet and its search is still ongoing. In addition to direct detection experiments, which search for the scattering of DM particles off atomic nuclei within a detector, there are indirect detection experiments, which look for the products of dark matter particle annihilations or decays. At the \textit{Large Hadron Collider} (LHC), the search for the indirect detection of DM particles is performed in larger and larger datasets of proton-proton collisions, which are used to both measure more and more accurately the properties of the known natural phenomena as well to try to unravel the mystery of the nature of DM. \\

From a theoretical point of view, the current framework of particle physics, the so-called \textit{Standard Model} (SM), does not provide a sufficient explanation of all the natural phenomena, e.g. the presence of dark matter in the universe, the mass of neutrinos, and the asymmetry between matter and antimatter in the universe.
Many alternative theories to the SM have been proposed, among which \textit{Supersymmetry} (SUSY) seems to be the most promising and well motivated, foreseeing the presence of a light neutral particle, the so-called neutralino $\tilde{\chi}_{0}$, which represents a valid candidate as DM constituent. SUSY predicts that universe is governed by a more fundamental symmetry than the one assumed by the SM, so that for every boson it exists a partner, called a superpartner, which has a fermionic nature, and, conversely, for every fermion there is a bosonic superparter. A new set of particles with different spins is so introduced, all to be discovered. If these particles were observed, we could solve the problems left behind by the SM. In the SM, for example, the mass of the Higgs boson should be of the same order as the maximum energy in the theory, however it has been measured to be just 125 GeV. SUSY particles are involved in loop corrections to the Higgs boson mass and they cancel out SM contribution in a natural way, leaving a mass that is compatible with the one measured. \\

In this thesis, two supersymmetric searches are presented, both assuming that the supersymmetric particles are produced through the electroweak interaction: the chargino analysis, targeting the pair production of charginos $\tilde{\chi}_{1}^{\pm}\tilde{\chi}_{1}^{\mp}$ decaying into SM $W$ bosons and neutralinos, and the displaced track analysis, searching for mildly displaced charged tracks arising from the decays of charginos $\tilde{\chi}_{1}^{\pm}$ and neutralinos $\tilde{\chi}_{2}^{0}$ into pions. These searches target a compressed phase space, where the mass difference between the next-to-lightest and lightest supersymmetric particle is relatively small. In the chargino search, the targeted difference in mass between charginos and neutralinos is close to the mass of the $W$ boson. In such phase space, the chargino pair production is kinematically similar to the $WW$ background, which makes it particularly interesting to be searched for, as the produced supersymmetric particles can be hiding behind a looking-alike SM process, but also experimentally challenging, as it is hard to discriminate the supersymmetric signal from the $WW$ background. In the displaced track search, the difference in mass between the produced supersymmetric particles and the lightest neutralinos goes down to 0.3 GeV and the tracks associated to pions from the decays of charginos and neutralinos are reconstructed as a few millimetres away from the primary vertex. \\

The ability to detect supersymmetric particles clearly depends on the detector performance, and in particular on the performance of the pixel detector, which is used for the measurement of the displacement of the tracks. The pixel detector is the closest one to the interaction point and so the most affected by the effects of radiation damage, which deteriorate the performance of the detector and its ability to correctly identify and reconstruct particles. The modelling of the effects of the radiation damage to the ATLAS pixel detector is presented. \\

The current ATLAS pixel detector has been exposed to a significant amount of radiation since the start of the LHC, and cannot be operated anymore for the next phase of LHC. Indeed, the whole inner detector will be replaced with a new one made of silicon pixels and strips. The future \textit{Inner Tracker} (ITk) of the ATLAS detector will also be able to cope with the higher luminosity of proton-proton collisions that will be recorded. In this context, a work carried out at CERN and involving the assembly and quality control of the pixel modules for ITk is presented. Several key activities are conducted to ensure sustained and reliable production rate of hybrid pixel modules. These start from the reception of the hardware components to the shipment to the loading sites and mainly focus on the quality control of the modules through visual inspection, metrology, flex to bare module assembly and electrical tests.

\chapter{Beyond the SM: Supersymmetry}
In this Chapter, a brief description of the main limitations and unsolved problems of the SM is reported. Supersymmetry is presented as a theoretical framework able to solve these problems and special attention is devoted to the minimal SUSY extension to the SM, the Minimal Supersymmetric Standard Model (MSSM) and its phenomenology.

\minitoc
\medskip

\section{The limitations of the SM}
\label{Section:SMLimitations}
The SM \cite{Halzen,Aitchison} is the theoretical framework that describes the elementary particles and their interactions. The SM is a quantum field theory based on the $\mathrm{SU(3)_{C} \times SU(2)_{L} \times U(1)_{Y}}$ symmetry group. Its success culminated with the discovery of the Higgs boson by the ATLAS \cite{ATLAS-HiggsBoson} and CMS \cite{CMS-HiggsBoson} experiments in 2012. More and more precise tests are confirming the SM predictions. \\
Despite all of these achievements, the SM has still different unsolved problems. It does not provide any suitable candidate for the DM \cite{DarkMatter-Evidence}, which is believed to compose almost 27\% of the universe \cite{DarkMatter-Planck}, and does not account for the Dark Energy (DE), which makes up almost 70\% of the universe. Other problems include the mass of the Higgs boson, the mass of the neutrinos \cite{NeutrinoProblem}, the matter/anti-matter asymmetry in the universe \cite{MatterAsymmetry}, the fermion mass hierarchy, the inclusion of gravity, and the unification of the other forces. \\
All these hints suggest that the SM is only an effective description of the natural phenomena at low energies, and it breaks if extended to higher energies. A limit of validity can be taken to be the scale where the couplings are closer to unification, the Grand Unified Theory (GUT) scale ($\Lambda_{\mathrm{GUT}} \approx 10^{16}$ GeV), or where the gravity effects can not be ignored anymore, the Planck scale ($\Lambda_{\mathrm{Planck}} \approx 10^{19}$ GeV). However, this introduces the \textit{hierarchy problem}: these scales are so different from the electroweak scale ($\Lambda_{\mathrm{EWK}} \approx 10^{2}$ GeV) where the SM has been tested.

\subsection{The presence of dark matter}
Precise cosmological measurements based on gravitational lensing, cosmic microwave background radiation, baryonic acoustic oscillations, and the formation and evolution of galaxies in the universe are compatible with the assumption that DM composes almost 27\% of the universe. It is called \textit{dark} because it is non-luminous and non-absorbing, while \textit{matter} is because its cosmological state equation is the same as for the ordinary matter ($\omega = P/\rho = 0$). The DE, instead, has the same cosmological state equation for energy ($\omega = P/\rho = -1$) and it accounts for cosmic inflation.\\
While the presence of DM is established, its nature is not known. A very plausible scenario is that it is a stable particle, or at least with a lifetime longer than the age of the universe, massive and neutral. There is no possible DM candidate in the SM, and the neutrinos, which are neutral stable and massive particles, would not be sufficient to account for all the DM presence. Neutrinos are always relativistic ($p/m \gg 1$) and this would make them a candidate for the so-called \textit{hot DM}, opposed to \textit{cold DM}, where $p/m \ll 1$. However, hot DM is generally considered not to be a good solution (at least as the only component of DM) since it does not fit models of galaxies formation. Therefore, an extension of the SM is needed in order to account for the DM.\\
Interesting candidates that could explain the current density of DM are the Weakly Interacting Massive Particles (WIMPs). It is possible to express the relative DM abundance ($\Omega_{\mathrm{DM}}$) in terms of the DM annihilation cross-section ($\sigma_{\mathrm{ann}}$),
\begin{equation}
\Omega_{\mathrm{DM}} \approx \frac{6 \cdot 10^{-27} \mathrm{cm^3 s^{-1}}}{<\sigma_{\mathrm{ann}}v>}
\end{equation}
where $<\sigma_{\mathrm{ann}}v>$ is the averaged cross-section with the thermal velocity. Considering $\Omega_{\mathrm{DM}} \approx 0.25$, a $\mathcal{O}(100)$ GeV DM candidate is needed with an electroweak cross-section.

Other DM candidates are the Axion-Like Particles (ALPs), pseudo-scalar particles which are generally very light, very weakly interacting and have a coupling to electromagnetism. They are abundant in string theory and are also predicted in many other Beyond SM (BSM) models.

\subsection{The naturalness problem}
The SM expected value of the Higgs boson mass $m_{h}$ is related to the vacuum expectation value ($v$) and the Higgs self-coupling ($\lambda$),
\begin{equation}
    m_{h}^2 =-2v^{2}\lambda.
\end{equation}
The ATLAS and CMS experiments have found that the Higgs boson has a mass of around 125 GeV, a value close to the electroweak scale. When considering one-loop corrections, the mass gains terms from each particle the Higgs boson couples with, directly or indirectly, and in particular the biggest contribution comes from the top quark, which has the highest mass.
The dominant corrections come from fermions and are
\begin{equation}
    \Delta m_{h}^2 = -\frac{|\lambda_{f}^2|}{8\pi^2}\Lambda_{\mathrm{UV}}^2 + \cdots
\end{equation}
where $\Lambda_{\mathrm{UV}}$ is the ultraviolet cut-off energy, the largest energy scale for which the SM is valid. The $\Delta m_{h}^2$ contribution diverges quadratically with the energy scale, therefore when considering higher scales (like the Plank Scale) all contributions must cancel out to avoid the correction exceeding the physical value of the Higgs mass observed at the electroweak scale. For this cancellation to happen, all the parameters must be finely tuned, to a level of around 1 part in many orders of magnitude according to the cut-off energy. The precise choice by nature of these particular values of the theory is called \textit{fine-tuning}. The fact that these parameters have these exact values is not a problem of internal consistency, but it is considered quite unnatural and could be a hint to the presence of an underlying structure, where the new logic imposes the observed values: in a natural theory no fine-tuning would be necessary. \textit{Naturalness} can be considered as a guiding principle of a theory, which ’t Hooft formulated as a condition for which a quantity in nature should be small only if the underlying theory becomes more symmetric as that quantity tends to zero \cite{NaturalnessHooft}. In other words, a small quantity going to zero increases the internal symmetry, thus protecting this parameter from getting increasingly large as the energy scale increases. Contrary to the fields of fermions and gauge bosons that benefit from the protection of the chiral and gauge symmetries, the Higgs boson is tied to the scale at which the symmetry of the electroweak theory is broken and receives radiative corrections that diverge quadratically with the energy cut-off \cite{NaturalnessDine}. The presence in the SM of the Higgs boson with a mass of the observed value poses a naturalness problem, representing a warning signal that the model might not be correct or have an underlying structure.

\subsection{Evolution of strength couplings}
In the SM, the three fundamental forces of the SM have different coupling constants. While at high energy the electromagnetic interaction and weak interaction unify into a single electroweak interaction, the unification does not happen with the strong force. This failure may suggest that the SM is only an effective description of the interactions at low energy while some new physics may happen at high energies, being able to match the experimental values at low energy. In many BSM theories, instead, the three fundamental forces of the SM appear after the breaking of some more general symmetry, e.g. a SU(5) symmetry, as proposed in the Georgi-Glashow model \cite{UnificationGlashow}. These theories predict that at high enough energies, e.g. the GUT energy, all the fundamental forces are merged into a single force with the same coupling constants. Unifying also gravity with the electronuclear interaction would provide a more comprehensive theory of everything rather than a GUT.

\subsection{The \texorpdfstring{$g-2$}{g-2} anomaly}
\begin{figure}[!htb]
\centering
\includegraphics[scale=0.1]{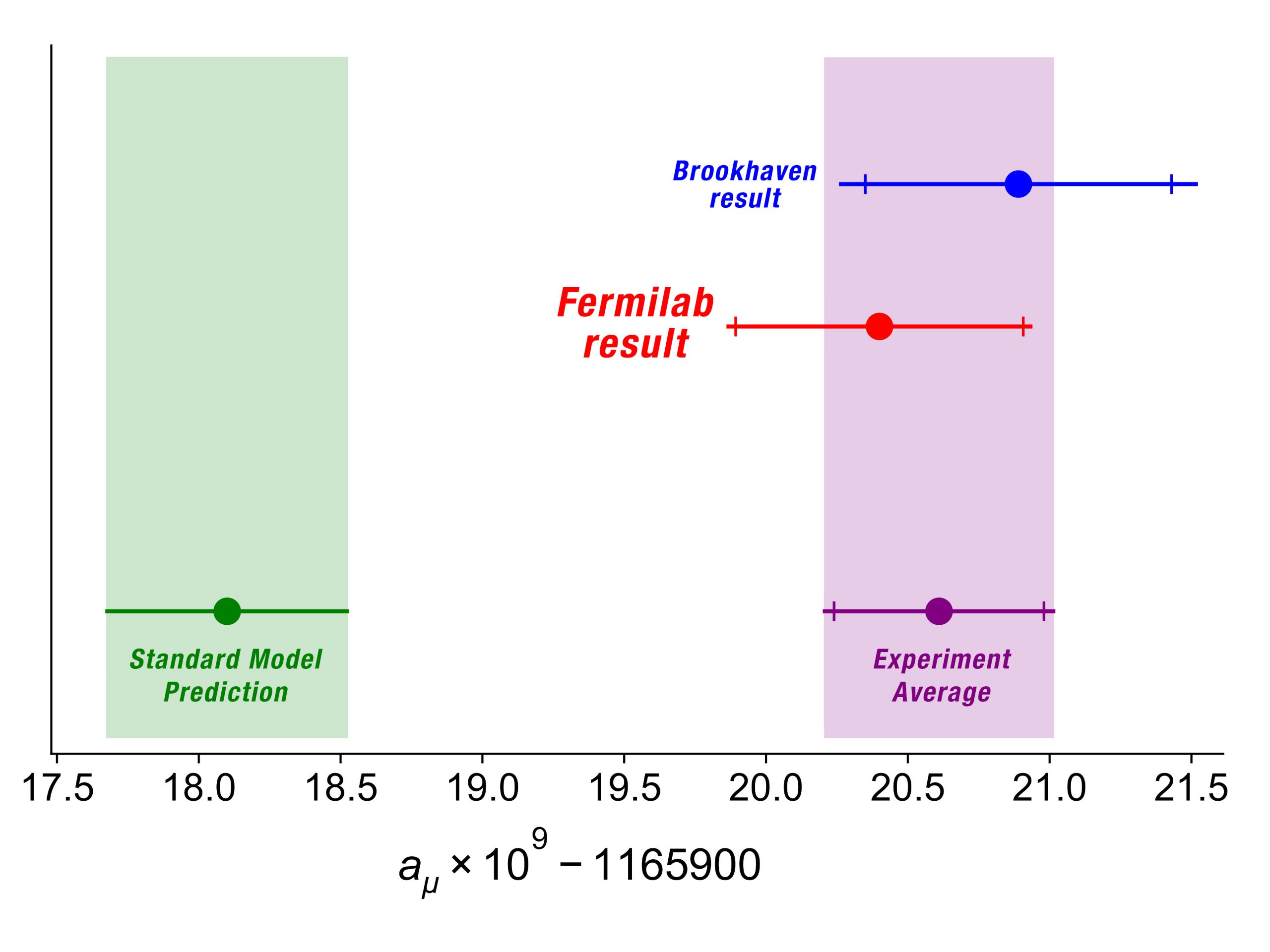}
\caption{Comparison between experimental values and theoretical prediction of the anomalous magnetic moment.
\label{fig:Valuesg-2}}
\end{figure}

In the SM, muons, like electrons, act as if they have a tiny internal magnet. In a strong magnetic field, the direction of the magnet of the muon precesses similarly to the axis of a spinning top or gyroscope. The strength of the internal magnet, the so-called $g$-factor, determines the rate that the muon precesses in an external magnetic field. This number has been calculated with ultra-high precision, and it is measured experimentally by making the muons to circulate in a circular orbit. Muons interact with the sea of subatomic particles continuously generated from the vacuum. Interactions with these short-lived particles affect the value of the $g$-factor, causing the precession of the muons to speed up or slow down very slightly. But if the vacuum sea of particles contains additional forces or particles not accounted for by the SM, that would tweak the muon $g$-factor further.
The SM predicts the $g$-factor to be very close to 2, and the difference $a_{\mu} = (g-2)/2$ is the so-called \textit{anomalous magnetic moment}. Fermilab and Brookhaven have provided very precise measurements of the $g$-factor. The predicted values and the experimental average from Fermilab and Brookhaven are (uncertainty in parenthesis) \cite{Fermilabg-2}:
\begin{eqnarray*}
&g^{\mathrm{theory}}= \scalebox{0.89}{\,\,\,2.00233183620(86)}\,\, &{a_{\mu}}^{\mathrm{theory}}= \scalebox{0.89}{\,\,\,0.00116591810(43)} \quad\,\,\\
&g^{\mathrm{exp-av.}}=  \scalebox{0.89}{2.00233184122(82)}\,\, &{a_{\mu}}^{\mathrm{exp-av.}}= \scalebox{0.89}{0.00116592061(41)}\quad\,\,
\end{eqnarray*}
and shown in Fig.~\ref{fig:Valuesg-2}.
The combined result from Fermilab and Brookhaven shows a 4.2$\sigma$ deviation from the SM prediction. The combination considers the Fermilab Run~1 dataset while a new analysis of the data from Run~2 and Run~3 (amounting to $\approx3$ times the Run~1 data) is expected to be completed within the next few years. The uncertainty from this new analysis is expected to be roughly half of the Run~1.

\subsection{Other problems in the SM}
There are other problems arising from the sole description of the SM. One is the observed asymmetry between matter and anti-matter in the universe, with matter being dominant over anti-matter. The baryon asymmetry of the universe $\eta^{B\bar{B}}$ can be defined as the difference between the number of baryons $n_{B}$ and antibaryons $n_{\bar{B}}$ divided by their sum,
\begin{equation}
\eta^{B\bar{B}} = \frac{n_{B}-n_{\bar{{B}}}}{n_{B}+n_{\bar{{B}}}}.
\end{equation}
While there was an initial balance between matter and anti-matter in the primordial universe just before antiprotons disappeared from the primordial plasma with $\eta^{B\bar{B}} \approx 10^{-10}$ \cite{MatterAntimatter}, matter largely dominates over antimatter in the universe today. Baryon number violation and a CP-violating term are needed to explain the observed asymmetry. In the SM, there is only one source of CP-violation, the CKM matrix, which does not account for all the observed asymmetry. New sources of CP violation are then needed to explain this phenomenon.\\
Furthermore, neutrinos are predicted to be massless in the SM but it has been observed from various measurements that the neutrinos oscillate implying a nonzero neutrino mass and neutrino flavour mixing~\cite{NeutrinoSNO,NeutrinoKamLAND,NeutrinoBorexino}. Differences in masses have been measured but absolute values are still not known, even if these masses must be smaller than 1 eV. This also implies that either the neutrino (anti-neutrino) has a Dirac right-handed (left-handed) partner or the neutrino is a Majorana spinor. No evidence has been found yet. Either way, the SM needs an extension to account for this.\\
Finally, the formulation of the gravitational theory is another problem. A renormalizable quantum field theory of gravity is not included in the SM, and gravity itself is not compatible with the theory of relativity. At the energies of TeV, however, gravity is negligible since its coupling constant is many orders of magnitude smaller, but its contribution becomes significant at $\Lambda_{\mathrm{Planck}}$. This energy corresponds to the mass of a particle with its Compton wavelength smaller than its Schwarzchild radius. Then, when approaching these energies, gravity has to be considered and every theory shall consider it. Different theorists have proposed many solutions, like String Theory or Quantum Loop Gravity. However, experiments allowing to prove these theories are still out of reach for now.

\section{Supersymmetry}
\label{Section:Supersymmetry}
At present, the most promising way to extend the SM and overcome the limitations of the SM described in Section~\ref{Section:SMLimitations} is to introduce supersymmetry. Supersymmetry can naturally explain the origin of the DM and solves the problem of quadratic divergences in the Higgs sector by introducing a new symmetry that connects different-spin particles. It also predicts a light Higgs boson (with the Higgs quartic coupling fixed by gauge couplings) and makes the measured couplings at the electroweak scale consistent with a GUT model. Additionally, it can explain the $g-2$ anomaly and the neutrinos being massive. To achieve this, supersymmetry doubles the particle spectrum of the SM and improves the properties of convergence of the corresponding field theory.\\

\subsection{Supermultiplets}
Supersymmetric particles are organized into \textit{supermultiplets} \cite{FayetHsnu, Fayet:1977yc}, each containing both a fermionic and a bosonic state (SM particles and their superpartners with spin differing by half unit).\\

The chiral supermultiplets contain the SM fermions (quarks and leptons) and their bosonic spin-0 superpartners (called {\it squarks} and {\it sleptons}), generically referred to as \textit{sfermions}. Squarks and sleptons are denoted similarly to the corresponding fermions, with the addition of a tilde to denote they are superpartners of SM particles. The chiral supermultiplets are listed in Table \ref{tab:chiral}. In particular, the fermionic superpartners of the Higgs boson are called \textit{higgsinos} and they are denoted by $\widetilde H_u$ and $\widetilde H_d$, with $\widetilde H_u^+$, $\widetilde H_u^0$ and $\widetilde H_d^0$, $\widetilde H_d^-$ weak isospin components. \\

\begin{table}[!htb]
\begin{center}
\begin{tabular}{c|c|c|c|c}
\noalign{\smallskip}\hline\noalign{\smallskip}
\multicolumn{2}{c}{Names} & spin 0 & spin 1/2 & $SU(3)_C ,\, SU(2)_L ,\, U(1)_Y$ \\ \noalign{\smallskip}\hline\noalign{\smallskip}
squarks, quarks & $Q$ & $({\widetilde u}_L\,\,\,\,\,\,{\widetilde d}_L)$ &
 $(u_L\,\,\,\,\,\,d_L)$ & $(\>{\bf 3},\>{\bf 2}\,\,,\,\,\frac{1}{6})$ \\
($\times 3$ families) & $\bar{u}$ & ${\widetilde u}^*_R$ & $u^\dagger_R$ & 
$(\,\,{\bf \overline 3},\,\, {\bf 1},\,\, -\frac{2}{3})$
\\ & $\bar{d}$ &${\bar{d}}^*_R$ & $d^\dagger_R$ & 
$(\,\,{\bf \overline 3},\,\, {\bf 1},\,\, \frac{1}{3})$
\\  \noalign{\smallskip}\hline\noalign{\smallskip}
sleptons, leptons & $L$ &$({\widetilde \nu}\,\,\,\,{\widetilde e}_L )$&
 $(\nu\,\,\,\,\,\,e_L)$ & $(\,\,{\bf 1},\,\,{\bf 2}\,\,,\,\,-\frac{1}{2})$
\\ 
($\times 3$ families) & $\bar{e}$
&${\widetilde e}^*_R$ & $e^\dagger_R$ & $(\,\,{\bf 1},\,\, {\bf 1},\,\,1)$
\\  \noalign{\smallskip}\hline\noalign{\smallskip}
Higgs, higgsinos &$H_u$ &$(H_u^+\,\,\,\,\,\,H_u^0 )$&
$(\widetilde H_u^+ \,\,\,\,\,\, \widetilde H_u^0)$& 
$(\,\,{\bf 1},\,\,{\bf 2}\,\,,\,\,+\frac{1}{2})$
\\ &$H_d$ & $(H_d^0 \,\,\,\,\,\, H_d^-)$ & $(\widetilde H_d^0 \,\,\,\,\,\, \widetilde H_d^-)$& 
$(\,\,{\bf 1},\>{\bf 2}\,\,,\,\,-\frac{1}{2})$
\\ \noalign{\smallskip}\hline\noalign{\smallskip}
\end{tabular}
\caption{Chiral supermultiplets in the Minimal Supersymmetric Standard Model. The superpartners of SM particles are denoted with a tilde. From Ref.~\cite{SupersymmetricPrimer}.}
\label{tab:chiral}
\vspace{-0.6cm}
\end{center}
\end{table}

The SM vector bosons reside in gauge supermultiplets with their fermionic superpartners, called \textit{gauginos}. The superpartner of the gluon $g$ is called gluino $\widetilde g$ while the superpartners of $W^+, W^0, W^-$ and $B^0$ gauge bosons are the fermionic $\widetilde W^+, \widetilde W^0, \widetilde W^-$ and $\widetilde B^0$, called {\it winos} and {\it bino}. The $\widetilde W^0$ and $\widetilde B^0$ gauge eigenstates mix to form the zino ($\widetilde Z$) and photino ($\widetilde \gamma$) mass eigenstates. The gauge supermultiplets are listed in Table \ref{tab:gauge}.

\begin{table}[!htb]
\begin{center}
\begin{tabular}{c|c|c|c}
\noalign{\smallskip}\hline\noalign{\smallskip}
Names & spin 1/2 & spin 1 & $SU(3)_C, \,\, SU(2)_L,\,\, U(1)_Y$\\
\noalign{\smallskip}\hline\noalign{\smallskip}
gluino, gluon &$ \widetilde g$& $g$ & $(\,{\bf 8},\,{\bf 1}\,\,,\,\, 0)$
\\ \noalign{\smallskip}\hline\noalign{\smallskip}
winos, $W$ bosons & $ \widetilde W^\pm\,\,\,\,\,\, \widetilde W^0 $&
 $W^\pm\,\,\,\,\,\, W^0$ & $(\>{\bf 1},\>{\bf 3}\>,\> 0)$
\\ \noalign{\smallskip}\hline\noalign{\smallskip}
bino, $B$ boson &$\widetilde B^0$&
 $B^0$ & $(\,\,{\bf 1},\,\,{\bf 1}\,\,,\,\, 0)$
\\ \noalign{\smallskip}\hline\noalign{\smallskip}
\end{tabular}
\caption{Gauge supermultiplets in the Minimal Supersymmetric Standard Model. The superpartners of SM gauge bosons are denoted with a tilde. From Ref.~\cite{SupersymmetricPrimer}.}
\label{tab:gauge}
\vspace{-0.45cm}
\end{center}
\end{table}

\subsection{R-parity conservation}
A new symmetry, called $R$-parity, is usually introduced to conserve baryon and lepton numbers~\cite{Rparity}. It is defined as 
\begin{eqnarray}
R = (-1)^{3({\rm B}-{\rm L}) + 2 s}
\label{defRparity}
\end{eqnarray}
where $B$, $L$, and $s$ are the baryon number, lepton number, and spin of the particle, respectively.

All particles in the SM have $R$ = 1, while all sparticles have $R$ = $-1$. If the $R$-parity is conserved, sparticles must be produced in pairs and decay into an odd number of sparticles. Moreover, the lightest supersymmetric particle (LSP) must be stable and, if neutral, represents a valid candidate as a dark matter constituent.

\subsection{The sparticles in the MSSM}
The sparticles in the MSSM amount to 32 distinct masses, not including the gravitino. The mass eigenstates of the MSSM are listed in Table \ref{tab:undiscovered}.
\begin{table}[!htb]
\begin{center}
\begin{tabular}{c|c|c|c|c}
\noalign{\smallskip}\hline\noalign{\smallskip}
Names & Spin & $R$ & Gauge Eigenstates & Mass Eigenstates \\ \noalign{\smallskip}\hline\noalign{\smallskip}
Higgs bosons & 0 & $+1$ & 
$H_u^0\,\,\,\, H_d^0\,\,\,\, H_u^+ \,\,\,\, H_d^-$ 
& 
$h^0\,\,\,\, H^0\,\,\,\, A^0 \,\,\,\, H^\pm$ \\
\noalign{\smallskip}\hline\noalign{\smallskip}
& & &${\widetilde u}_L\,\,\,\, {\widetilde u}_R\,\,\, \widetilde d_L\,\,\,\,\, \widetilde d_R$&(same) \\
squarks& 0&$-1$& ${\widetilde s}_L\,\,\,\, {\widetilde s}_R\,\,\,\, \widetilde c_L\,\,\,\,
\widetilde c_R$& (same) \\
& & &
$\widetilde t_L \,\,\,\,\widetilde t_R \,\,\,\,\widetilde b_L\,\,\,\, \widetilde b_R$ 
&
${\widetilde t}_1\,\,\,\, {\widetilde t}_2\,\,\,\, \widetilde b_1\,\,\,\, \widetilde b_2$ \\
\noalign{\smallskip}\hline\noalign{\smallskip}
& & &${\widetilde e}_L\,\,\,\, {\widetilde e}_R \,\,\,\,\widetilde \nu_e$&(same) \\
sleptons& 0&$-1$&${\widetilde \mu}_L\,\,\,\,{\widetilde \mu}_R\,\,\,\,\widetilde\nu_\mu$&(same) \\
& & &
$\widetilde \tau_L\,\,\,\, \widetilde \tau_R \,\,\,\,\widetilde \nu_\tau$ 
&
${\widetilde \tau}_1 \,\,\,\,{\widetilde \tau}_2 \,\,\,\,\widetilde \nu_\tau$ \\
\noalign{\smallskip}\hline\noalign{\smallskip}
neutralinos & $1/2$&$-1$ & 
$\widetilde B^0 \,\,\,\,\,\,\widetilde W^0\,\,\,\,\,\, \widetilde H_u^0\,\,\,\,\,\, \widetilde H_d^0$   
&
$\tilde{\chi}_{1}^{0}\,\,\,\, \tilde{\chi}_{2}^{0} \,\,\,\,\tilde{\chi}_{3}^{0} \,\,\,\, \tilde{\chi}_{4}^{0}$ \\
\noalign{\smallskip}\hline\noalign{\smallskip}
charginos & $1/2$&$-1$ & 
$\widetilde W^\pm\,\,\,\,\,\, \widetilde H_u^+ \,\,\,\,\,\,\widetilde H_d^-$ 
&
$\tilde{\chi}_{1}^\pm\,\,\,\,\,\,\tilde{\chi}_{2}^\pm $ \\
\noalign{\smallskip}\hline\noalign{\smallskip}
gluino & $1/2$ & $-1$ & $\widetilde g$ & (same) \\
\noalign{\smallskip}\hline\noalign{\smallskip}
$\genfrac{}{}{0pt}{}{\rm goldstino}{\rm (gravitino)}$ & $\genfrac{}{}{0pt}{}{1/2}{(3/2)} $&$-1$&$\widetilde G$ &(same) \\
\noalign{\smallskip}\hline\noalign{\smallskip}
\end{tabular}
\caption{The particles in the MSSM (with sfermion mixing for the first two families assumed to be negligible). From Ref.~\cite{SupersymmetricPrimer}.}
\label{tab:undiscovered}
\vspace{-0.4cm}
\end{center}
\end{table}%

\subsubsection{Neutralinos and charginos}
The higgsinos and electroweak gauginos mix with each other because of the effects of electroweak symmetry breaking. In particular:
\begin{itemize}
    \item {\it Neutralinos} are the four mass eigenstates formed from the combination of the neutral higgsinos ($\widetilde H_u^0$ and $\widetilde H_d^0$) and the neutral gauginos ($\widetilde B$, $\widetilde W^0$). They are denoted by $\tilde \chi_{i}^{0}$ ($i=1,2,3,4$), with the convention that $m_{\tilde{\chi}_{1}^{0}} < m_{\tilde \chi_{2}^{0}} <m_{\tilde{\chi}_{3}^{0}} <m_{\tilde{\chi}_{4}^{0}}$.
    \item {\it Charginos} are the four mass eigenstates formed from the combination of the charged higgsinos ($\widetilde H_u^+$ and $\widetilde H_d^-$) and winos ($\widetilde W^+$ and $\widetilde W^-$). They are denoted by $\tilde{\chi}^\pm_i$ ($i=1,2$), with the convention that $m_{\tilde{\chi}^\pm_1} < m_{\tilde{\chi}^\pm_2}$.
\end{itemize}

\subsubsection{Mass hierarchies}
\begin{figure}[!htb]
\centering
\includegraphics[scale=0.45]{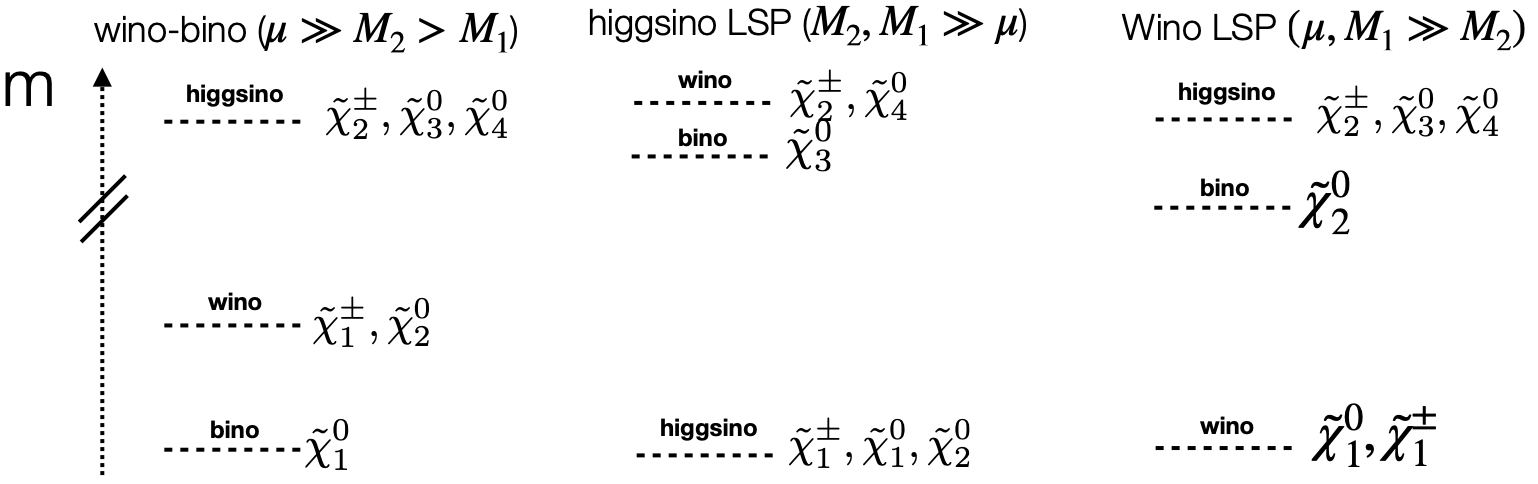}
\caption{The different mass hierarchies in the electroweak sector.} \label{fig:ewkhierarchy}
\end{figure}

The mass hierarchy in the electroweak sector is not unique and depends on the mass eigenstates of the particles. Three different mass hierarchies can be distinguished in the electroweak sector, as can be seen in Fig.~\ref{fig:ewkhierarchy}: the \textit{wino-bino}, the \textit{higgsino LSP} and the \textit{wino LSP}.

In the wino-bino mass hierarchy, the condition
\begin{eqnarray}
\mu \,\gg\,  M_{2}\, > M_{1},
\label{gauginolike}
\end{eqnarray}
is satisfied and the lightest neutralino is very nearly a \textit{bino-like} mass eigenstate $\tilde{\chi}_{1}^{0} \approx \widetilde B$, the next-to-lightest neutralino a \textit{wino-like} mass eigenstate $\tilde{\chi}_{2}^{0} \approx \widetilde W^0$, and $\tilde{\chi}_{3}^{0}, \tilde{\chi}_{4}^{0}$ are two \textit{higgsino-like} mass eigenstates $\tilde{\chi}_{3}^{0}, \tilde{\chi}_{4}^{0} \approx (\widetilde H_u^0 \pm \widetilde H_d^0)/\sqrt{2}$. The neutralino mass eigenvalues are given by~\cite{SupersymmetricPrimer}:
\begin{eqnarray}
m_{{\tilde{\chi}}_1^0}\!\! &=&\!\! M_1 - { \frac{m_Z^2 s^2_W (M_1 + \mu \sin 2 \beta )}{\mu^2 - M_1^2} } +\ldots \, ,\\
m_{{\tilde{\chi}}_2^0}\!\! &=&\!\! M_2 - { \frac{m_W^2 (M_2 + \mu \sin 2 \beta )}{\mu^2 - M_2^2}} +\ldots \, ,\\
m_{{\tilde{\chi}}_3^0}\!\! &=&\!\! |\mu| + \frac{m_Z^2  (I - \sin 2 \beta) (\mu + M_1 c^2_W +M_2 s^2_W)}{2 (\mu + M_1) (\mu + M_2) } + \ldots \, , \qquad\\
m_{{\tilde{\chi}}_4^0}\!\! &=&\!\!\! |\mu|  + \frac{m_Z^2  (I + \sin 2 \beta) (\mu - M_1 c^2_W - M_2 s^2_W)}{2 (\mu - M_1) (\mu - M_2) } +\ldots \, , \qquad
\end{eqnarray}
where $M_1$ and $M_2$ are assumed real and positive, and $\mu$ is real with sign$I = \pm 1$. In this limit, the chargino mass eigenstates consist of a wino-like $\tilde{\chi}_1^\pm$ and a higgsino-like $\tilde{\chi}_2^\pm$, with mass eigenvalues:
\begin{eqnarray}
m_{{\tilde{\chi}}_1^{\pm}} &=& M_2 - \frac{m_W^2 (M_2 + \mu \sin 2 \beta )}{\mu^2 - M_2^2 } +\ldots \\
m_{{\tilde{\chi}}_2^{\pm}} &=& |\mu | + \frac{I m_W^2 (\mu + M_2 \sin 2 \beta)}{\mu^2 - M^2_2 } +\ldots\,\,\, .
\end{eqnarray}
Interestingly, $\tilde{\chi}_1^{\pm}$ is degenerate with the neutralino $\tilde{\chi}_2^{0}$ and higgsino-like mass eigenstates $\tilde{\chi}_3^{0}$, $\tilde{\chi}_4^{0}$ and $\tilde{\chi}_2^{\pm}$ have masses of order $|\mu|$. \\

\subsection{Sparticle production}
At colliders, sparticles can be produced in pairs from parton collisions of electroweak strength or QCD strength.\\

The possible electroweak strength processes are:
\begin{itemize}
\item the production of a pair of charginos or neutralinos (see Fig. \ref{Fig:CC_NN}),
\begin{equation}
q \bar{q} \rightarrow \tilde{\chi}_i^+ \tilde{\chi}_j^-, \>\> \tilde{\chi}_{i}^{0} \tilde{\chi}_{j}^{0};
\label{Eq:CC_NN}
\end{equation}

\begin{figure}[!htb]
\centering
\includegraphics[scale=0.4]{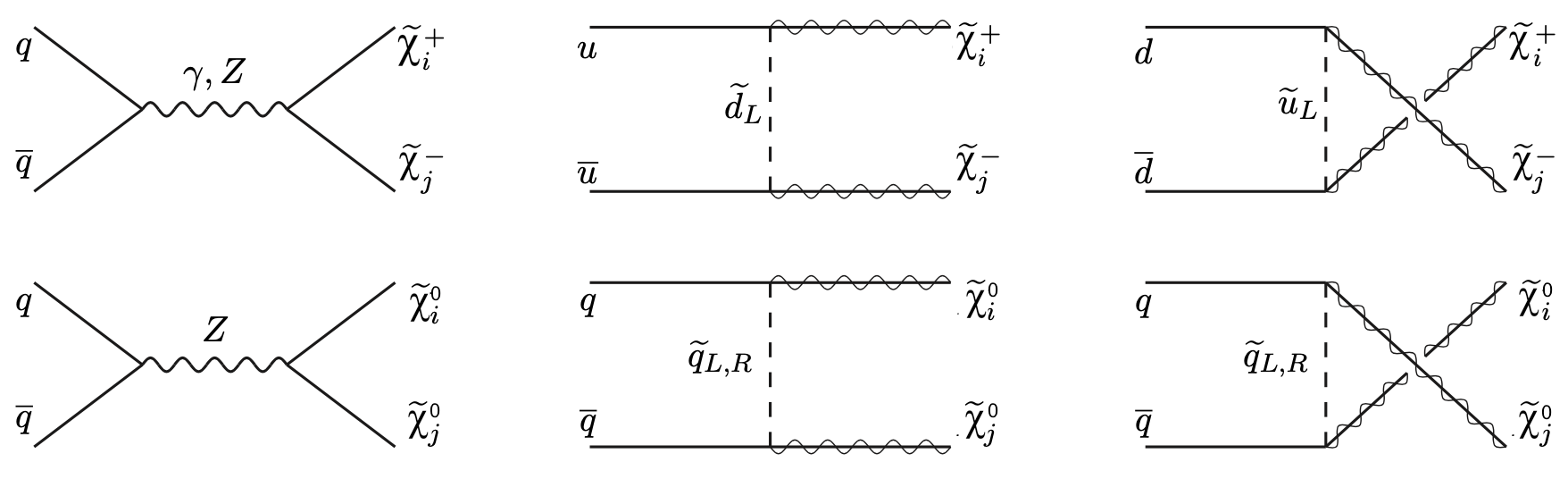}
\caption{Feynman diagrams for electroweak production of a pair of charginos or a pair of neutralinos at hadron colliders from quark-antiquark annihilation.}
\label{Fig:CC_NN}
\end{figure}

\item the production of a pair of a chargino and a neutralino (see Fig. \ref{Fig:CN_CN}),
\begin{equation}
u \bar{d} \rightarrow \tilde{\chi}_i^+ \tilde{\chi}_{j}^{0}, \qquad\quad d \bar{u} \rightarrow \tilde{\chi}_{i}^{-} \tilde{\chi}_{j}^{0};
\label{Eq:CN_CN}
\end{equation}

\begin{figure}[!htb]
\centering
\includegraphics[scale=0.4]{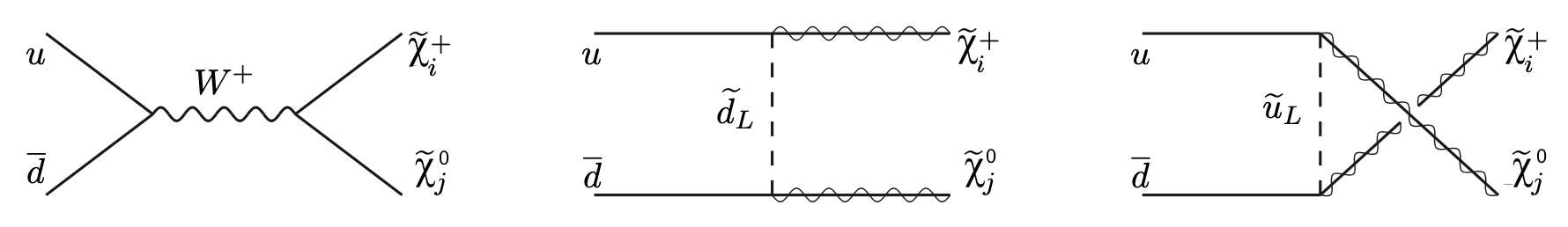}
\caption{Feynman diagrams for electroweak production of a pair of a chargino and a neutralino at hadron colliders from quark-antiquark annihilation.}
\label{Fig:CN_CN}
\end{figure}

\item the production of a pair of sleptons or a pair of neutralinos (see Fig. \ref{Fig:ll_nunu}),
\begin{equation}
q \bar{q} \rightarrow \tilde{\ell}^+_i \tilde{\ell}^-_j, \>\>\> \tilde{\nu}_\ell \tilde{\nu}^*_\ell;
\end{equation}

\begin{figure}[!htb]
\centering
\includegraphics[scale=0.4]{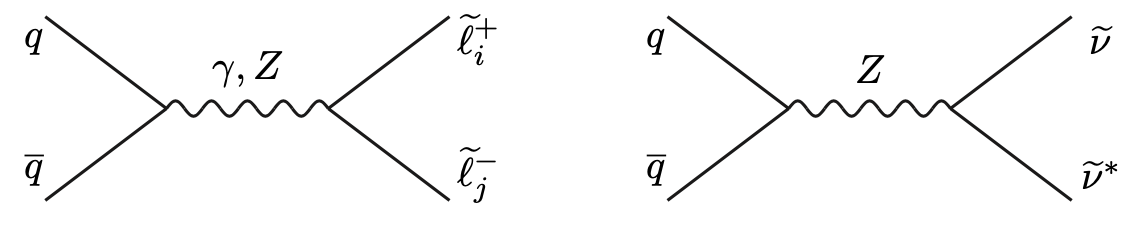}
\caption{Feynman diagrams for electroweak production of a pair of sleptons and a pair of sneutrinos at hadron colliders from quark-antiquark annihilation.}
\label{Fig:ll_nunu}
\end{figure}

\item the production of a pair of a slepton and a sneutrino (see Fig. \ref{Fig:lnu_lnu}),
\begin{equation}
u \bar{d} \rightarrow  \tilde{\ell}^+_L \tilde{\nu}_\ell \qquad\qquad\>
d \bar{u} \rightarrow  \tilde{\ell}^-_L \tilde{\nu}^*_\ell.
\label{Eq:lnu_lnu}
\end{equation}

\begin{figure}[!htb]
\centering
\includegraphics[scale=0.4]{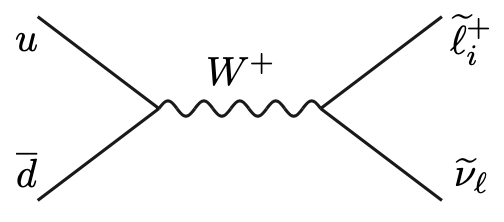}
\caption{Feynman diagrams for electroweak production of a pair of a slepton and a sneutrino at hadron colliders from quark-antiquark annihilation.}
\label{Fig:lnu_lnu}
\end{figure}
\end{itemize}

Due to the parton distribution functions, the most frequent parton collisions were quark-antiquark annihilation at the Tevatron and are gluon-gluon and gluon-quark fusion at the LHC. However, the supersymmetric signal is most likely produced by an inclusive combination of these parton collisions happening at the colliders. \\

\begin{figure}[!htb]
\centering
\includegraphics[scale=0.6]{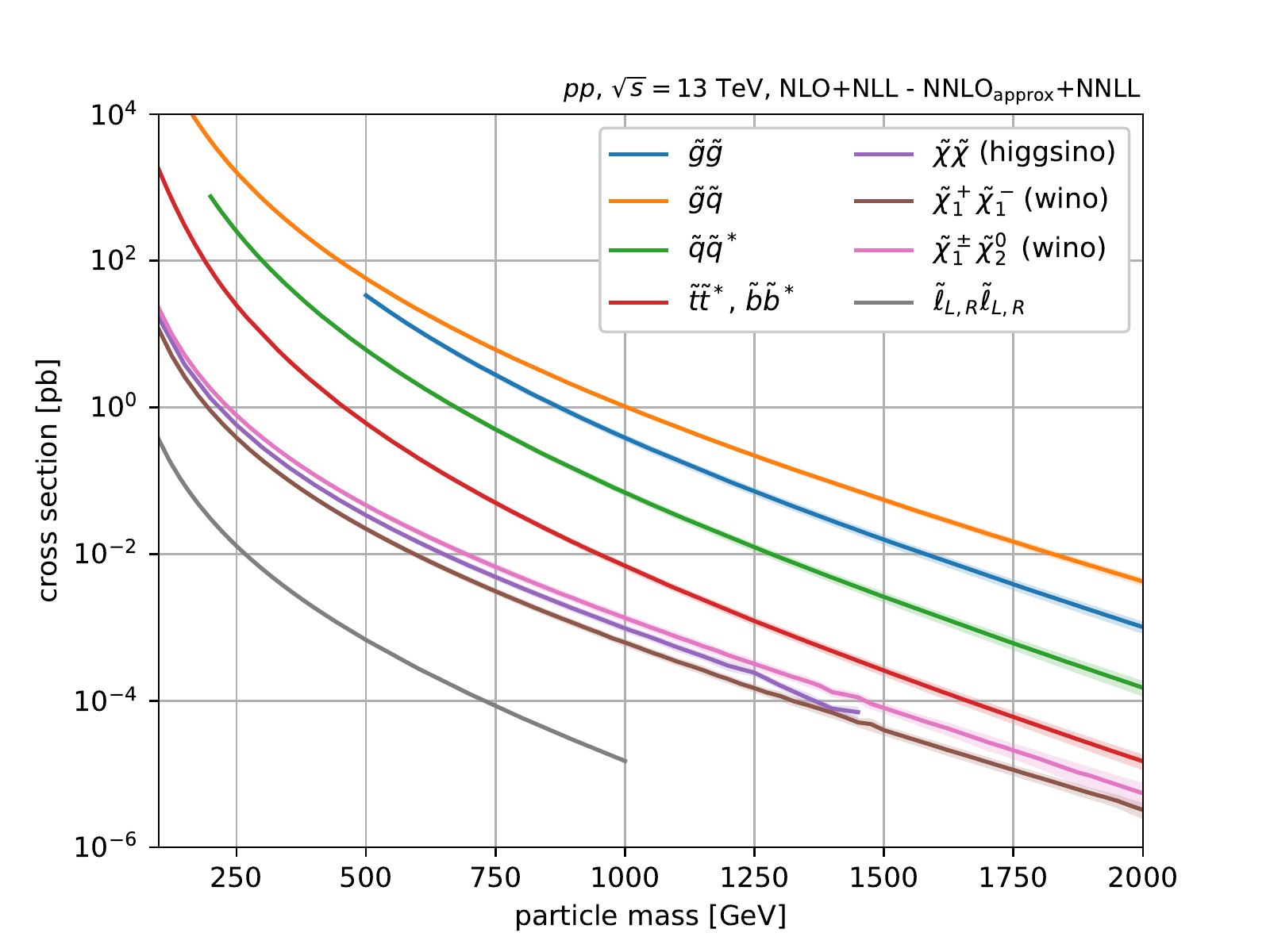}
\caption{Cross-section predictions for various supersymmetric processes in $pp$ collisions at $\sqrt{s} = 13$ TeV. Reported values are based on the agreement between the ATLAS and CMS collaborations, as well as with the LPCC SUSY cross-section working group. Calculations including the resummation of soft gluon emission at the next-to-leading logarithmic accuracy are used whenever available.}
\label{Fig:crosssec13TeV}
\end{figure}

Cross-sections for sparticle production at hadron colliders with $\sqrt{s}=13$ TeV are shown in Fig.~\ref{Fig:crosssec13TeV}. As can be seen, in all cases, the production cross-sections decrease as a function of the sparticle mass.

\subsection{Sparticle decay}
Different decay modes are available for charginos and neutralinos assuming that the $\tilde{\chi}_{1}^{0}$ is the LSP. Charginos and neutralinos can decay into lighter sparticles through a SM boson or through the lightest Higg boson, if the mass splitting is sufficiently large. Otherwise, they can decay into a pair with a slepton or a sneutrino, if the slepton and the sneutrino are sufficiently light, or into heavier Higgs boson and quark+squark pairs:
    \begin{equation}
    \tilde{\chi}_{i}^{0} \rightarrow Z \tilde{\chi}_{j}^{0}, \,\, W\tilde{\chi}_{j}^{\pm}, \,\, h^{0}\tilde{\chi}_{j}^{0}, \,\, \ell\tilde{\ell}, \,\, \nu\tilde{\nu}, \,\, A^{0}\tilde{\chi}_{j}^{0}, \,\, H^{0}\tilde{\chi}_{j}^{0}, \,\, H^{\pm}\tilde{\chi}_{j}^{\mp}, \,\, q\tilde{q}
    \end{equation}
    \begin{equation}
    \tilde{\chi}_{i}^{\pm} \rightarrow W \tilde{\chi}_{j}^{0}, \,\, Z\tilde{\chi}_{1}^{\pm}, \,\, h^{0}\tilde{\chi}_{1}^{\pm}, \,\, \ell\tilde{\nu}, \,\, \nu\tilde{\ell}, \,\, A^{0}\tilde{\chi}_{1}^{\pm}, \,\, H^{0}\tilde{\chi}_{1}^{\pm}, \,\, H^{\pm}\tilde{\chi}_{j}^{0}, \,\, q\tilde{q}'.
    \end{equation}
    
\begin{figure}[!ht]
    \centering
    \includegraphics[scale=0.4]{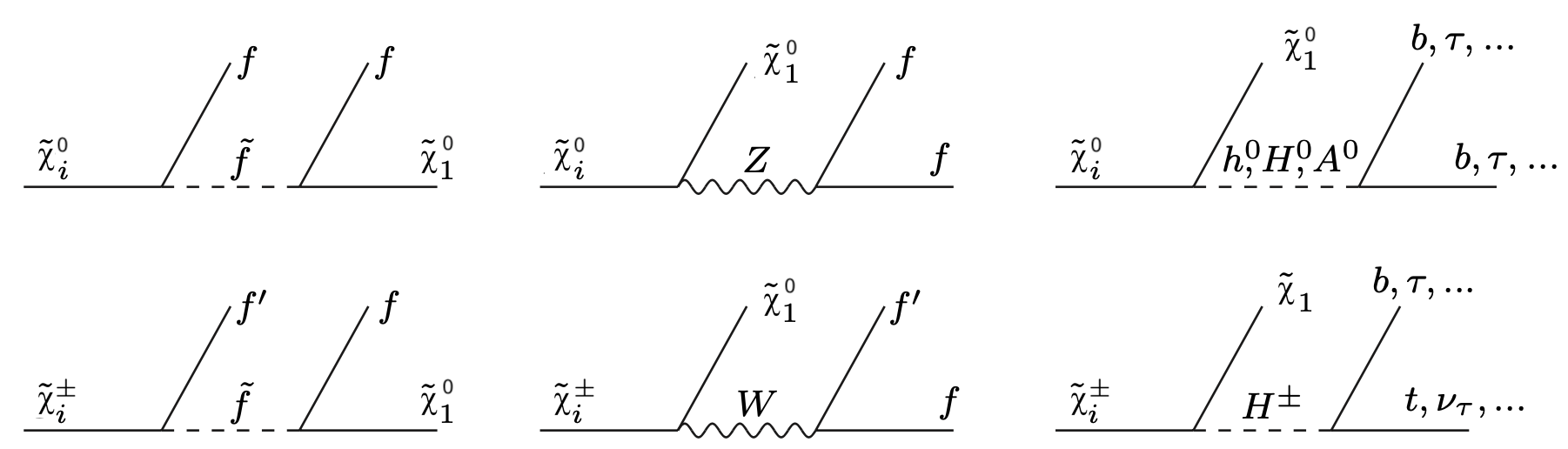}
    \caption{Feynman diagrams showing neutralino and chargino decays with $\tilde{\chi}_{1}^{0}$ in the final state.}
    \label{Fig:decayelectroweakinos}
    \end{figure}

    The Feynman diagrams for the neutralino and chargino decays with $\tilde{\chi}_{1}^{0}$ in the final state are shown in Fig.~\ref{Fig:decayelectroweakinos}. \\
    
\begin{figure}[!ht]
    \centering
    \includegraphics[scale=0.15]{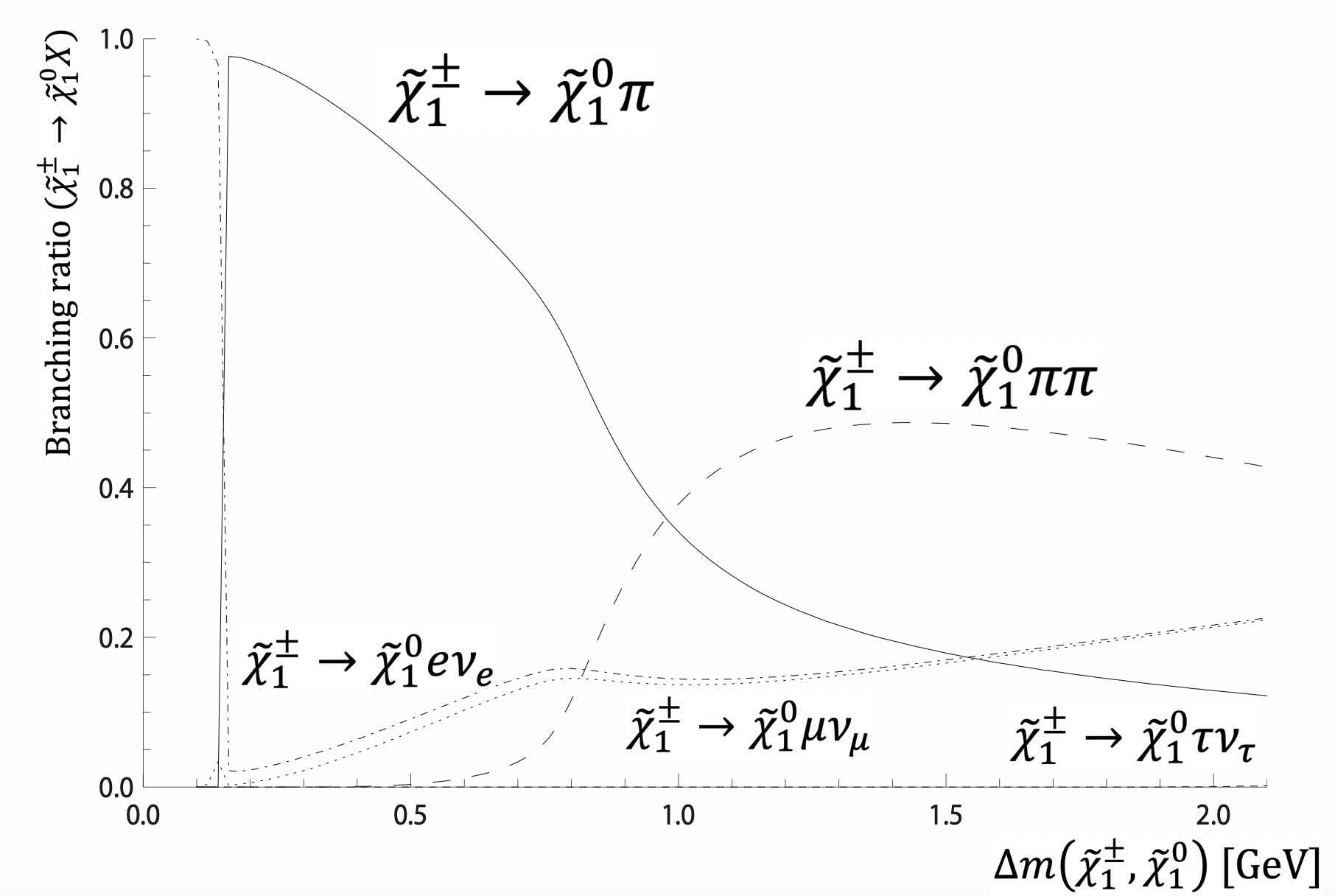}
    \includegraphics[scale=0.15]{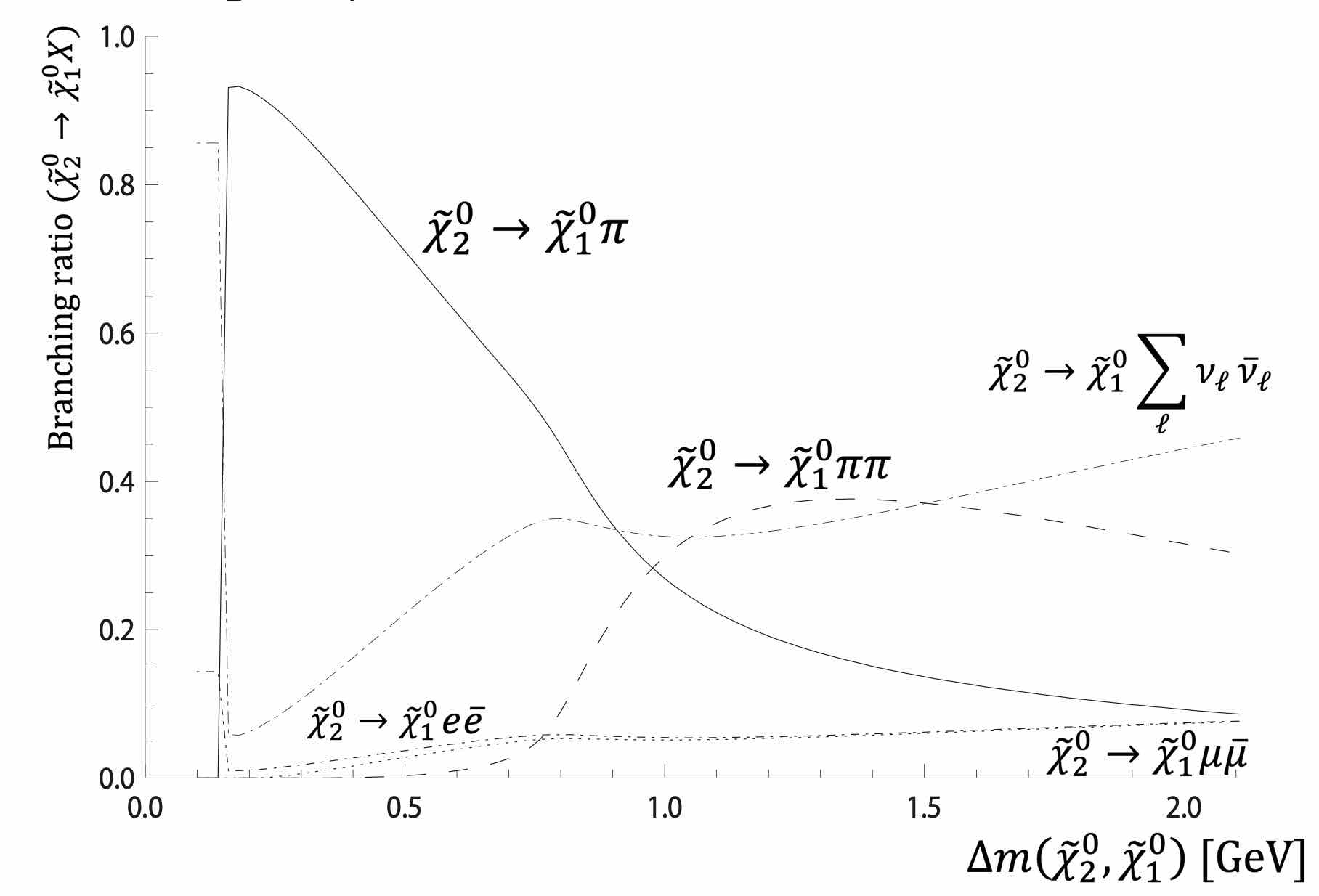}
    \caption{Branching ratios for chargino $\tilde{\chi}_{1}^{\pm}$ and neutralino $\tilde{\chi}_{2}^{0}$ decays as functions of the mass splittings $\Delta m(\tilde{\chi}_{1}^{\pm},\tilde{\chi}_{1}^{0})$ and $\Delta m(\tilde{\chi}_{2}^{0},\tilde{\chi}_{1}^{0})$. From Ref. \cite{HiggsinoBranchingRatio1,HiggsinoBranchingRatio2}.}
    \label{Fig:BR-Chargino1-Neutralino2}
    \end{figure}

    Higgsino-like charginos $\tilde{\chi}_{1}^{\pm}$ and next-to-lightest neutralinos $\tilde{\chi}_{2}^{0}$ are close in mass to the the $\tilde{\chi}_{1}^{0}$ they decay into. If the mass splitting is of the order of the GeV, they can decay producing pions or leptons:
    \begin{equation}
    \tilde{\chi}_{1}^{\pm} \rightarrow \tilde{\chi}_{1}^{0} \pi , \,\, \tilde{\chi}_{1}^{0}\pi\pi , \,\, \tilde{\chi}_{1}^{0}e\nu_{e} , \,\, \tilde{\chi}_{1}^{0}\mu\nu_{\mu} , \,\, \tilde{\chi}_{1}^{0}\tau\nu_{\tau} ,
    \end{equation}
    \begin{equation}
    \tilde{\chi}_{2}^{0} \rightarrow \tilde{\chi}_{1}^{0} \pi , \,\, \tilde{\chi}_{1}^{0}\pi\pi , \,\, \tilde{\chi}_{1}^{0}\sum_{\ell} \nu_{\ell} \nu_{\ell}, \,\, \tilde{\chi}_{1}^{0} ee , \,\, \tilde{\chi}_{1}^{0} \mu\mu.
    \end{equation}
    The branching ratios for the decays of charginos $\tilde{\chi}_{1}^{\pm}$ and neutralinos $\tilde{\chi}_{2}^{0}$ depend on the mass splittings $\Delta m(\tilde{\chi}_{1}^{\pm},\tilde{\chi}_{1}^{0})$ and $\Delta m(\tilde{\chi}_{2}^{0},\tilde{\chi}_{1}^{0})$ and are shown in Fig.~\ref{Fig:BR-Chargino1-Neutralino2}.

\chapter{LHC and the ATLAS Experiment}
\label{sec:lhcatlas}

In this Chapter, an overview of the experimental setup of the LHC, its accelerating facilities and the main experiments are presented. The data analysed in this thesis are taken from $\sqrt{s} = 13$ TeV proton-proton ($pp$) collisions produced by the LHC at CERN from 2015 to 2018, the so-called Run~2, and recorded by the ATLAS detector \cite{ATLAS}. A description of the ATLAS detector as composed of different sub-detectors is also presented.

\minitoc
\medskip

\section{The Large Hadron Collider}
\label{ssec:lhc}
The LHC is placed in the 27 km tunnel previously built for LEP at CERN and it is currently the most powerful accelerator in the world, able to accelerate protons with an energy up to 13.6 TeV in the centre of mass. The discovery potential at the TeV scale allows experimental physicists to study the fundamental processes of the Standard Model with great detail and to search for new physics never observed before. \\
The choice of a collider accelerating protons instead of electrons is motivated by the much higher energy achievable in the centre of mass. In fact, any charged particle radiates photons when accelerated. The energy loss of a particle in a circular motion, known as synchrotron radiation, depends on its mass $m$ as the following formula:
\begin{equation}
\frac{\mathrm{d}E}{\mathrm{d}t} \propto \frac{E^4}{m^4 R}
\end{equation}
where $E$ is the energy of the particle and $R$ is the radius of the orbit of the particle. As we can see, this implies that for fixed radius and fixed energy, as we would have at a collider, electrons lose $( m_p/m_e)^4 \approx 10^{12} $ times more energy than a proton. This sets a limit to the energy of accelerating electrons to around 100 GeV. To compensate for this we would have to build a much bigger ring for the electrons, causing the price of the project to increase. \\

Nonetheless, there are, at the current time, projects involving accelerators with electrons: CEPC \cite{CEPC} (in China) and the first stage of the FCC project, FCC-ee (at CERN) \cite{FCC1,FCC2,FCC3}. They both will have a circumference of around 100 km, and a cost significantly higher than LEP or LHC. Other projects involve linear accelerators using electrons, such as the CLIC \cite{CLIC} (at CERN) and ILC \cite{ILC-1,ILC-2,ILC-3,ILC-4} (in Japan). They both require a great length, e.g. CLIC is proposed to be 11 km long with an energy of 380 GeV in the first stage, while ILC will initially be 20 km long and operated with an energy of 250 GeV. A drawback of linear accelerators is represented by the fact that bunches can cross just one time and therefore the luminosity is smaller. 
While using protons is more advantageous for synchrotron radiation, it brings in new problems that are absent in an electron collider. This is due to the fact that protons have a complex structure. \\
Interesting collisions between protons can be of two different kinds: \textit{soft} or \textit{hard}.
\textit{Soft} collisions are the results of protons interacting as a whole, with low momentum transfer ($ \approx 500 $ MeV) and large cross-section. \textit{Hard} collision happens when the constituents of the protons, gluons and quarks, interact with each other. In this kind of events, we can produce new particles. \\

In an event of hard collision, the partonic centre of mass energy is unknown since quarks and gluons carry an arbitrary fraction of the proton momentum. For this reason, making kinematic calculations is more difficult. Moreover, the important events for physics are the ones from \textit{hard} collisions, but these have cross-sections orders of magnitude smaller than \textit{soft} collisions. For this reason, a high luminosity collider is needed. \\

The rate of events produced at LHC, $\mathrm{d}N/\mathrm{d}t$, is given by the product of the instantaneous luminosity $\mathcal{L}$ and the cross-section $\sigma$, 
\begin{equation}
\frac{\mathrm{d}N}{\mathrm{d}t} = \mathcal{L} \cdot {\sigma}.
\label{eq:istlumi}
\end{equation}

The instantaneous luminosity is a quantity related to the performance of the machine and is measured in cm$^{-1}$s$^{-1}$. It is related to the frequency $f$ of bunch crossing and the product of the number of particle in both bunches ($n_1$ and $n_2$) over the crossing area $A=4 \pi \sigma_x \sigma_y$,
\begin{equation}
\mathcal{L}=\frac{fn_1 n_2}{4 \pi \sigma_x \sigma_y}.
\label{eq:istlumi_long}
\end{equation}

The total number of events recorded during a given period of time can be obtained as the product of the integrated luminosity $\mathcal{L}_{\mathrm{int}}$ and the cross-section $\sigma$, where  $\mathcal{L}_{\mathrm{int}}$ is the instantaneous luminosity integrated over the period of time:
\begin{equation}
\mathcal{L}_{\mathrm{int}}= \int \mathcal{L} \, \mathrm{d}t.
\end{equation}

\begin{figure}[!htb]
\begin{center}
\includegraphics[scale=0.5]{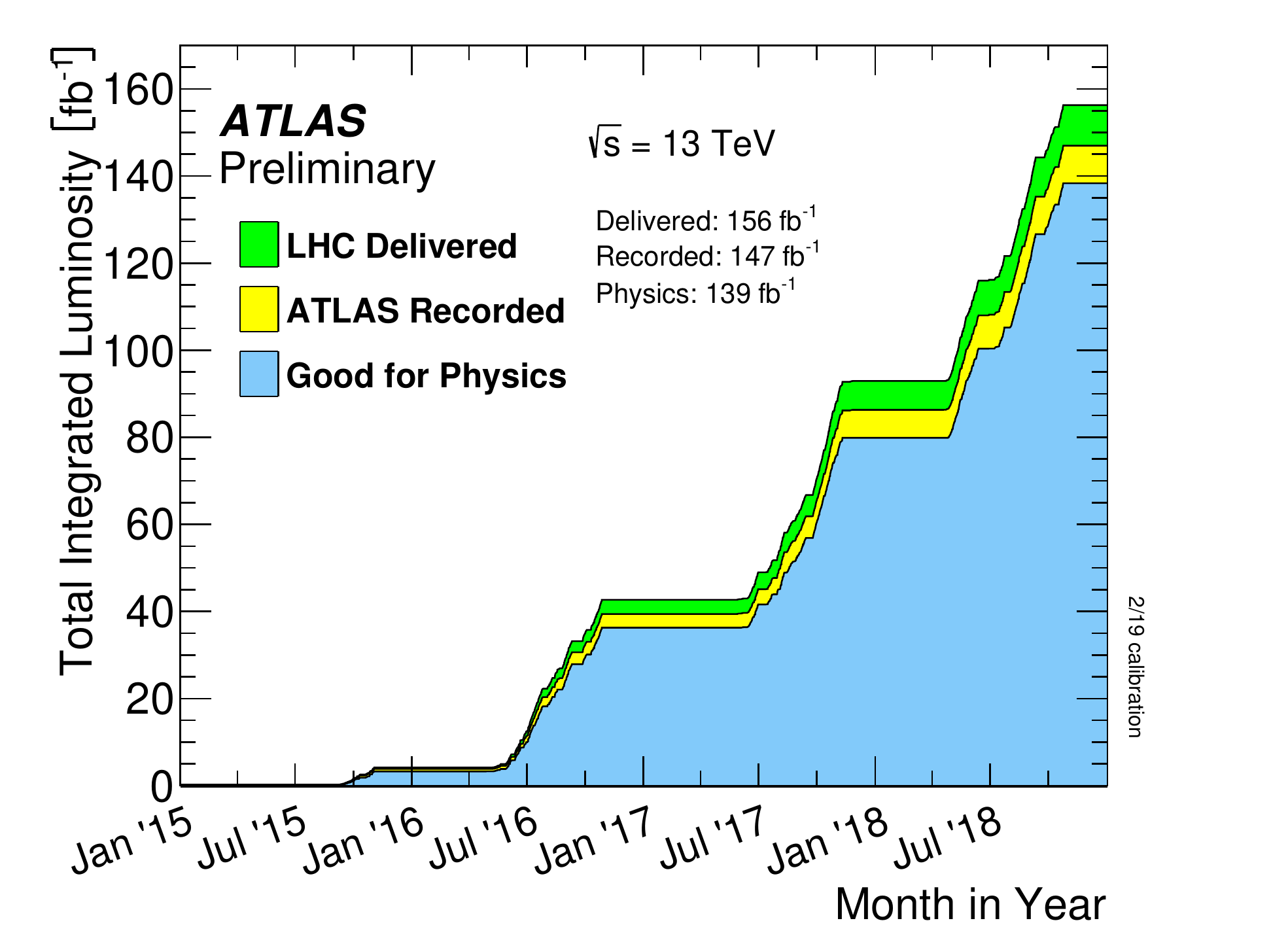}
\caption{Cumulative luminosity versus time delivered to ATLAS (green), recorded by ATLAS (yellow), and certified to be good quality data (blue) during stable beams for $pp$ collisions at 13 TeV centre-of-mass energy in 2015-2018.}
\label{fig:intlumivstimeRun2}
\end{center}
\end{figure}

LHC started colliding the first beams in 2009, used for tests and calibration. In March 2010 LHC reached the proton-proton collision energy of 7 TeV, starting a period of data-taking that lasted till 2011. During 2012 LHC reached an energy at center of mass of 8 TeV with a peak instantaneous luminosity of  8$\times 10^{33} $cm$^{-2}$s$^{-1}$. Together these periods are referred to as Run~1, and they amount to a total of 25 $\mathrm{fb}^{-1}$ data taken, for both the ATLAS and the CMS experiments. After a shutdown during 2013 and 2014 for technical improvements called \textit{Long Shutdown 1} (LS1), LHC started to operate again in 2015, reaching an energy in the centre-of-mass of 13 TeV and a peak luminosity of 5 $\times 10^{33} $cm$^{-2}$s$^{-1}$ with a separation of bunches of 25 ns. LHC delivered 4.2 $\mathrm{fb}^{-1}$ in 2015, and ATLAS recorded a total of 3.9 $\mathrm{fb}^{-1}$. From 2016 to 2018 LHC has steadily continued to deliver collisions at the same energy in the center-of-mass, delivering 38.5 $\mathrm{fb}^{-1}$ in 2016, 50.24 $\mathrm{fb}^{-1}$ in 2017, and 63.35 $\mathrm{fb}^{-1}$ in 2018. The luminosity recorded by ATLAS is instead: 35.56 $\mathrm{fb}^{-1}$ in 2016, 46.89 $\mathrm{fb}^{-1}$ in 2017, and 60.64 $\mathrm{fb}^{-1}$ in 2018, for a total in Run~2 of 146.9 $\mathrm{fb}^{-1}$. The dataset used in the analyses is a sub-sample where low quality events are removed. This sub-sample consists of 139 $\mathrm{fb}^{-1}$.\\
The integrated luminosity delivered by LHC during Run~2, recorded by ATLAS and certified to be good quality data is shown in Fig.~\ref{fig:intlumivstimeRun2}. \\

\begin{figure}[!htb]
\begin{center}
\includegraphics[width=1.\textwidth]{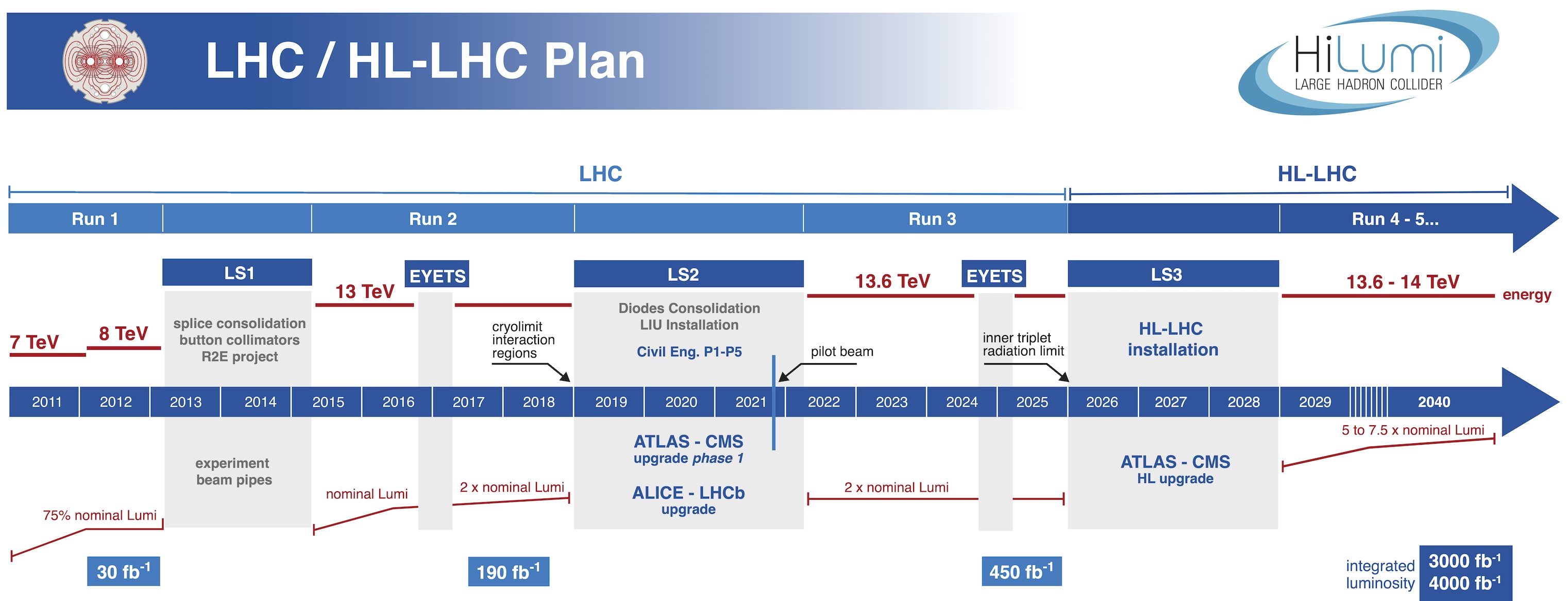}
\caption{The current schedule for the LHC and HL-LHC upgrade and run. Currently, the start of the HL-LHC run is foreseen in 2029. The long shutdowns, LS2 and LS3, will be used to upgrade both the accelerator and the detector hardware.}
\label{fig:LHC-schedule}
\end{center}
\end{figure}

Between 2019 and 2022 LHC was shut down to allow for additional improvement and maintenance of the acceleration facilities and the detectors. As can be seen in Fig~.\ref{fig:LHC-schedule}, this period is referred to as LS2 and was used by ATLAS to perform detector
upgrades that are described in Section~\ref{ssec:run3detector}. In 2022, the Run~3 has started and LHC is delivering $pp$ collisions with an energy in the centre of mass of 13.6 TeV. The expected nominal luminosity that is going to be delivered during Run~3 is about twice the one delivered in Run~2.\\

The schedule foresees a third long shutdown, LS3, to start in 2026 and to last for 3 years, finishing with the hardware and beam commissioning in early 2029. After LS3, a new phase of LHC, \textit{Phase-II}, will start and the \textit{High-Luminosity LHC} (HL-LHC) will deliver collisions with a record integrated luminosity, 4000 $\mathrm{fb}^{-1}$, corresponding to a factor of 10 beyond the design value of LHC. This new configuration relies on several key innovations that push accelerator technology beyond its present limits. These include cutting-edge superconducting magnets, compact superconducting cavities with ultra-precise phase control, new technology and physical processes for beam collimation, all of which required several years of dedicated R\&D effort on a global international level \cite{HL-LHC-TDR}. The HL-LHC era will allow physicists to study known mechanisms, such as the Higgs boson, in much greater detail and search for new phenomena that might be observed.

\subsection{The accelerator complex}
\begin{figure}[!htb]
\centering
\includegraphics[width=1.\textwidth]{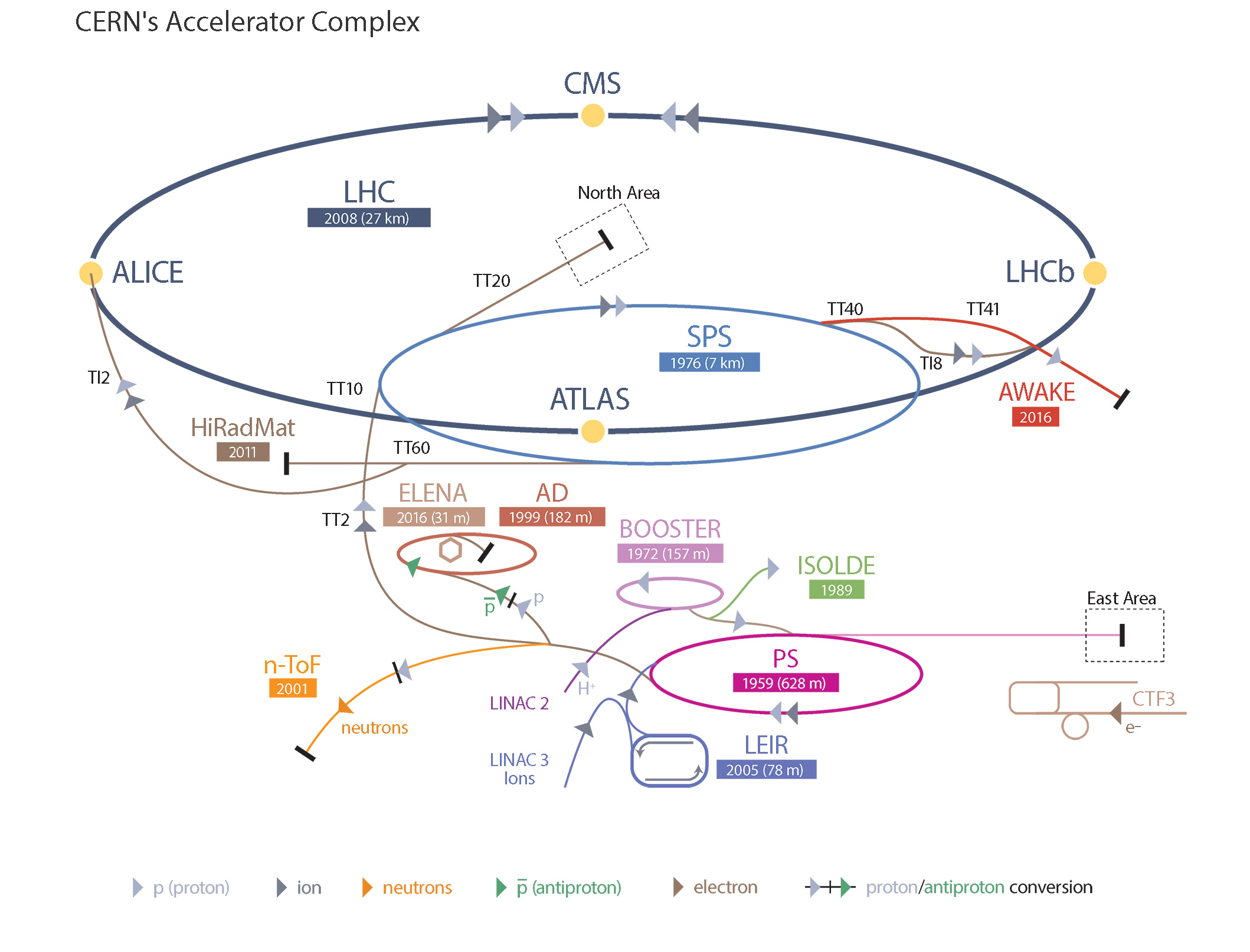}
\caption{Chain of accelerating facilities at CERN and most important experiments.}
\label{fig:LHCComplex}
\end{figure}

Before entering the ring of 27 km that is LHC, protons (or lead ions) are accelerated in different steps by a complex system of other machines. The accelerating chain (together with the experiments) is shown in Fig.~\ref{fig:LHCComplex}.

The first step is to produce protons, which is done by ionizing a hydrogen source. These protons are then accelerated up to 50 MeV by a linear accelerator: LINAC 2. Then they enter a circular machine: Proton Synchrotron Booster (PSB) where they are brought to an energy of 1.4 GeV, so that they can safely enter the Proton Synchrotron (PS). This is the first accelerator built at CERN in 1959 and it is still used to accelerate protons from 1.4 GeV to 26 GeV. The next step is passing protons to the Super Proton Synchrotron (SPS), the machine used in the '80s to discover the $W$ boson by the UA1 and UA2 collaborations, where protons are accelerated to 450 GeV. Once they have reached this energy, the protons are ready to be injected into the LHC.\\
For lead ions, the chain of machines is slightly different. They are accelerated first by the LINAC 3 (a linear accelerator) and then by the Low Energy Ion Ring (LEIR). At this point, they enter the PS and then the chain of process is the same as for the protons: SPS and then LHC.\\

\subsection{The experiments at LHC}
The are four main experiments at LHC, ALICE \cite{ALICE}, ATLAS \cite{ATLAS}, CMS \cite{CMS} and LHCb \cite{LHCb}. They are located in correspondence of the interaction points of the ring. However, the experiments have different purposes.

\begin{itemize}

    \item[\textbf{ALICE}] (A Large Ion Collider Experiment) is a detector designed to detect the products of heavy ion collisions. Its physical goal is to study quark-gluon plasma, which is the nature of strongly interacting matter at extreme energy densities. ALICE also studies proton-proton collisions both as a comparison to the Pb-Pb collisions and to do research in areas where it is competitive with the other experiments.

    \item[\textbf{ATLAS}] (A Toroidal LHC ApparatuS) and \textbf{CMS} (Compact Muon Solenoid) are two of the biggest detectors located at the LHC. They are multipurpose detectors, built for the search of the Higgs boson and to investigate new physics beyond the SM. Even if they use different technologies, their performances are similar, and the agreement between their results is a fundamental cross-check of the validity of their results.

    \item[\textbf{LHCb}] is dedicated to the study of the decay of states containing $b$ quarks, with a major interest in the symmetry between matter and anti-matter and the violation of the lepton flavour universality. Its design is quite different from the other detectors since it is not symmetrical along the interaction point. This is to enhance the detection of $b$ quarks. Moreover at LHCb rare decays of hadrons, such as  $B^0_s\rightarrow\mu^+\mu^-$, suppressed at leading order, but still possible are studied.

\end{itemize}

\section{The ATLAS experiment}
\label{ssec:atlas}
\begin{figure}[!htb]
\centering
\includegraphics[width=1.\textwidth]{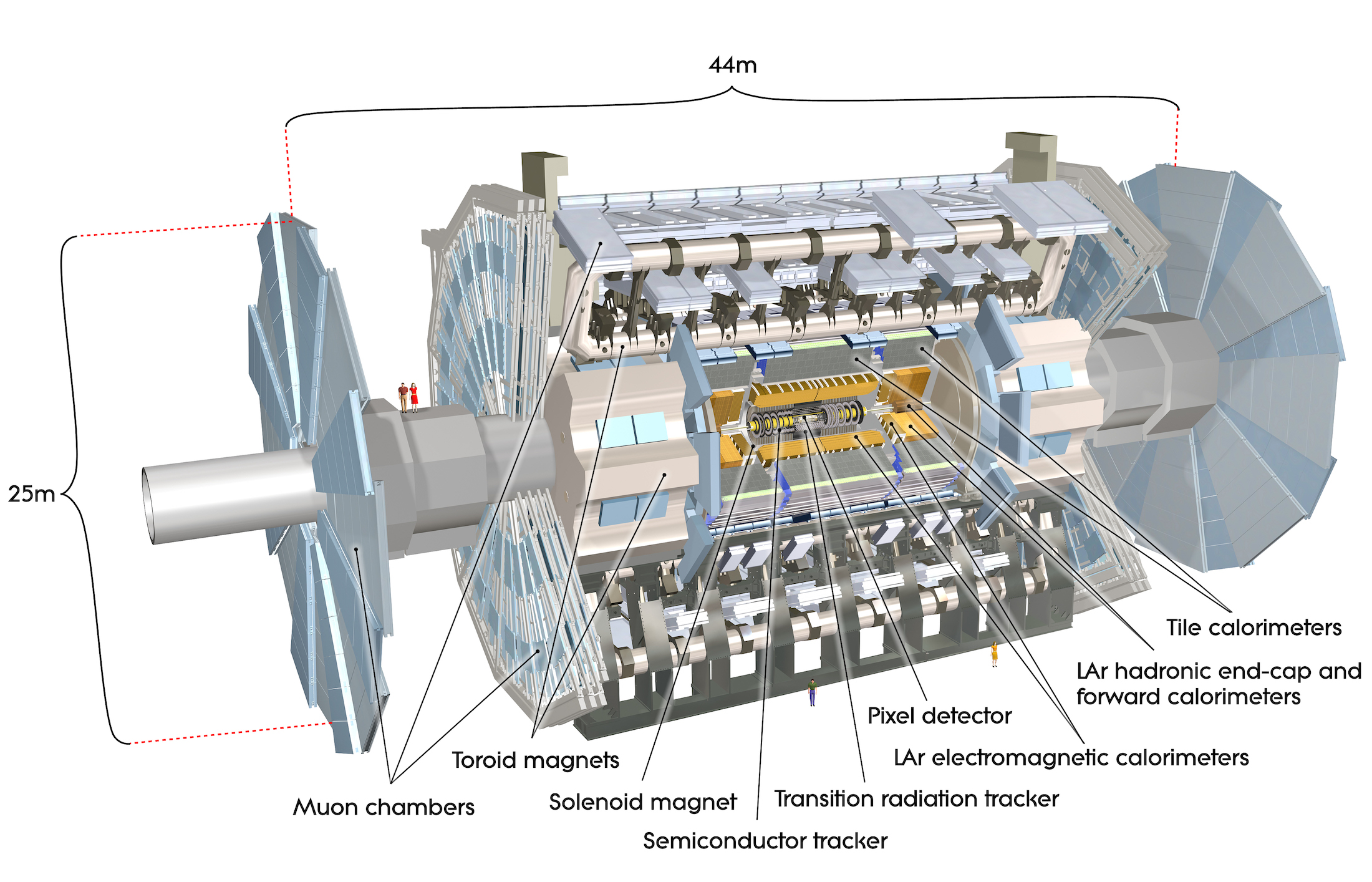}
\caption{Schematic view of the ATLAS detector}
\label{fig:ATLAS}
\end{figure}

The ATLAS detector is located at interaction Point 1 (see Fig.~\ref{fig:LHCComplex}) in a cavern almost 100 m underground. It has cylindrical symmetry along the beam axis, with a length of 44 m and a diameter of 25 m. It weighs almost 7000 tons. It was first proposed in 1994, but the construction didn't start until 2003, and it took until 2007 to be finished. It began operative in 2008 by looking at cosmic ray events, and since 2009 it records events from proton-proton and heavy ions collisions with a rate up to 400 Hz during Run~1. In Run~2 ATLAS reached a rate of recorded events of 1000 Hz.\\
ATLAS is a general purpose detector, which means it aims to reconstruct the nature and energy of all the particles generated in the collisions of protons (or ions) in a bunch crossing. To do so it must have some specific characteristics: first of all, it must be hermetic, this means that the detector must cover as much as possible of the solid angle around the interaction point so that the possibility of particles escaping through a non-instrumented zone is almost zero. To be able to identify all the particles, using only one detector is not a feasible way. For this reason, ATLAS (and most of the modern detectors) is divided into different sub-detectors, each of them specific for the study of one kind of particles. Moreover, the detector must be fast in response to the very high rate of events that occur. The rate of collisions is 40 MHz and it must be lowered to 1 kHz, in order to do that it is necessary to have a fast system of triggers (Section \ref{ssec:trigdata}). This constraint is due to available computational resources for the storage, reconstruction, and analysis of data.\\
A schematic representation of the ATLAS detector is shown in Fig.~\ref{fig:ATLAS}.
 
In the innermost part, just around the interaction point, it is located the Inner Detector (Section \ref{ssec:ID}), used for tracking particles, filled by a solenoidal magnetic field. The Inner Detector is divided into a barrel region in the middle and two end-caps. Over this there is the Calorimeter system (Section \ref{ssec:cal}), for the identification of hadronic and electromagnetic showering, and after that the Muon Spectrometer (Section \ref{ssec:muonspec}), built for the identification of muons. Also these parts are divided into a barrel region and two end-caps. A dedicated magnetic system is present for the Muon Spectrometer (Section \ref{ssec:magnet}). The last part is the Forward Detector which covers the most forward part near the beam (\ref{ssec:fwdet}).

\subsection{Coordinate System}
Before analysing the detector and its features it is useful to briefly describe the coordinate system that is used in ATLAS. In fact, ATLAS does not use a cartesian coordinate system ($x$,$y$,$z$) but, due to its geometry symmetry, it uses a right-handed cylindrical coordinate system ($\phi$,$\eta$,$z$). The origin of the system is located at the centre of the detector, the interaction point, and the $z$ axis is along the beam line. The azimuthal angle $\phi$ is defined by measuring around the $z$ axis perpendicularly to the beam. Instead of the polar angle $\theta$, it is more convenient to use the \textit{pseudorapidity} $\eta$, which is related to $\theta$ by
\begin{equation}
\eta = - \ln \tan \frac{\theta}{2}.
\end{equation}

Pseudorapidity is the limit for massless particles of the \textit{rapidity} $y$, which is given by
\begin{equation}
y = \frac{1}{2} \ln \left( \frac{E+p_z}{E-p_z}  \right).
\end{equation}

Rapidity and pseudorapidity are very useful variables in colliders because of their Lorentz transformation propriety. In fact, under a Lorentz boost $\beta$ along the $z$-axis, $y$ transforms by adding a constant in the following way
\begin{equation}
y'=y+\arctan(\beta)
\end{equation}
and the spectrum of high energy particles, $\mathrm{d}N/\mathrm{d}y$, is invariant for boost along the $z$ axis.\\

An important set of variables are the transverse one: momentum and energy. Those are defined as the momentum (or energy) perpendicular to the beam axis ($z$ axis). The transverse momentum $p_{\mathrm{T}}$ is defined as
\begin{equation}
p_{\mathrm{T}} = \sqrt{p_x^2 + p_y^2}.
\end{equation}

\subsection{Inner Detector}
\label{ssec:ID}
\begin{figure}[!htb]
\centering
\includegraphics[width=0.8\textwidth]{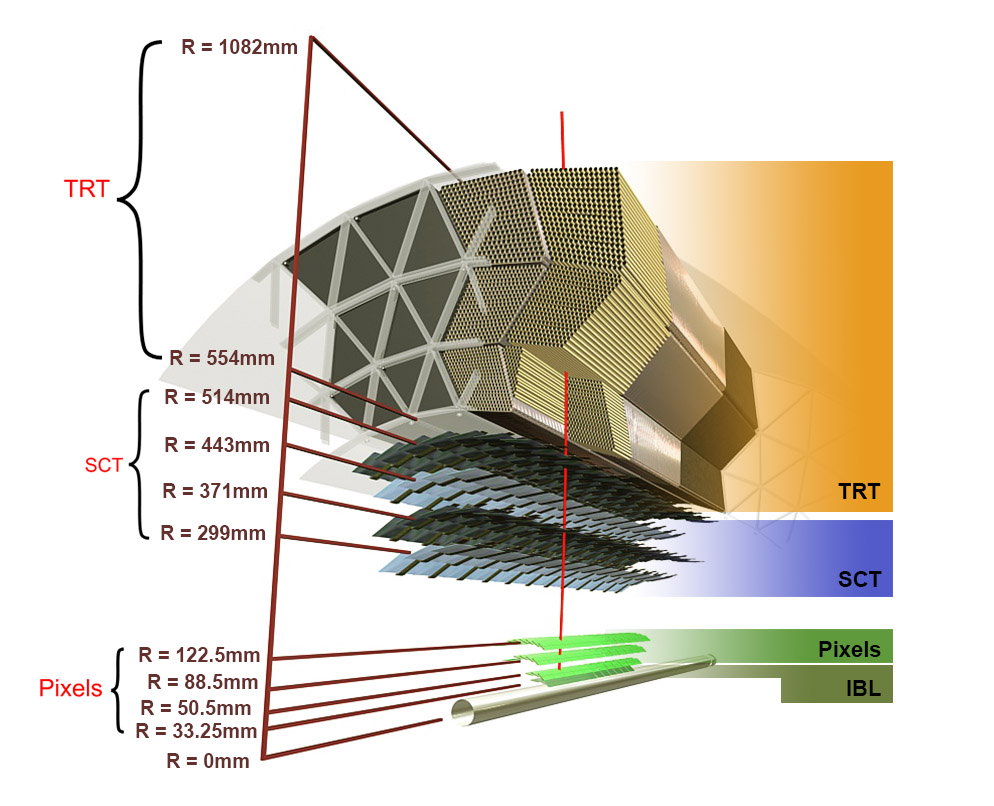}
\caption{Barrel part of ID of the ATLAS experiment with the Pixel, SCT and TRT sub-detectors.}
\label{fig:ATLASID-Barrel}
\end{figure}

\begin{figure}[!htb]
\centering
\includegraphics[width=0.8\textwidth]{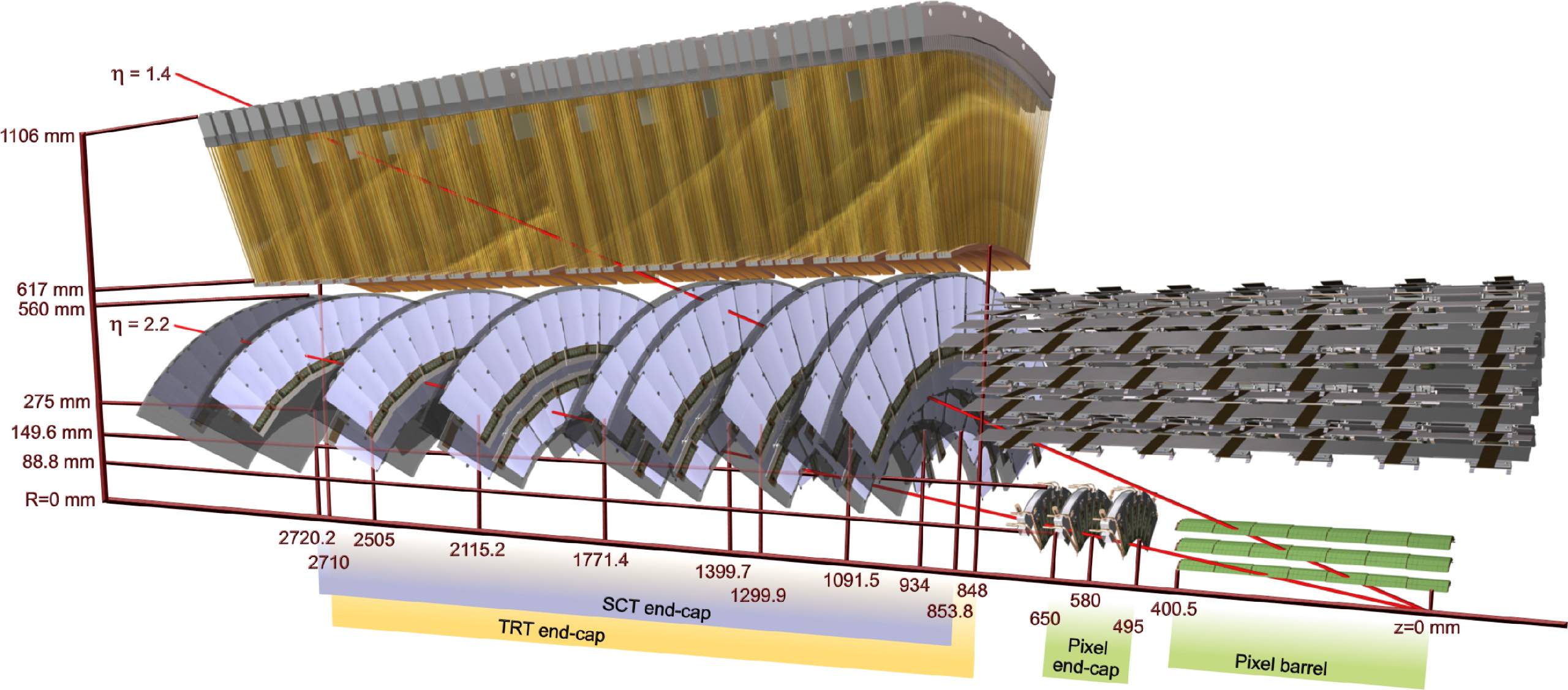}
\caption{End-cap part of ID of the ATLAS experiment with the Pixel, SCT and TRT sub-detectors.}
\label{fig:ATLASID-Endcap}
\end{figure}

The Inner Detector (ID) \cite{ATLAS-ID} system (shown in Figs. \ref{fig:ATLASID-Barrel}, \ref{fig:ATLASID-Endcap}) is the innermost part of the ATLAS detector. It is used for the reconstruction of the tracks and the interaction vertices of charged particles. Vertices can be both primary, from $pp$ collisions, and secondary, from long lived hadrons decays, such as hadrons containing $b$ quarks. For this purpose, the ID must have high spatial resolution and granularity, to discriminate all the tracks, and high resistance to radiation, since it is the closest system to the beam pipe. It consists of a silicon Pixel detector, the SemiConductor Tracker (SCT) made up of silicon microstrips and the Transition Radiation Tracker (TRT) made up of straw tubes. Its coverage is in the region $|{\eta}|<2.5$ and it is surrounded by a solenoid magnet that generates a 2~T field. The ID has been upgraded between Run~1 and Run~2 with a new layer of pixels: the Insertable B-Layer (IBL) \cite{ATLAS-IBL}. All of these components are described in this Section.

\subsubsection{Pixel Detector}
The ATLAS pixel detector \cite{ATLAS-ID} is the innermost part of the ID. It consists of four barrel layers and three disk layers per end-cap (six in total). The barrel layers are composed of $n^+$-in-$n$ planar oxygenated silicon detectors, and $n^+$-in-$p$ 3D silicon detectors. The innermost layer, the IBL, is located at just 3.3 cm from the beam axis and is made of pixels of 50 $\times$ 250 $\mu$m$^2$ in size, which are just 200 $\mu$m thick, while there are 3D pixels of 50 $\times$ 250 $\mu$m$^2$ in size that are instead 230 $\mu$m thick in the region with high $|z|$ of the IBL. The other layers are respectively at 5.05 cm (B-Layer), 8.85 cm (Layer1), and 12.255 cm (Layer2) from the beam axis and made of pixels of $50 \times 400$  $\mu$m$^2$ in size and 250 $\mu$m thick. They cover a range in pseudorapidity of $|{\eta}|<2.5 $. The end-caps (which close the barrel) are made of the same sensors as B-Layer/Layer1/Layer2 and have three disks each and 288 modules. The total number of readout channels is 86$\times 10^6$.\\
IBL was installed in ATLAS in May 2014 before the start of LHC Run~2, while the other three layers have been there since the beginning of Run~1. The motivation for this new layer arose from the high luminosity expected in Run~2 and beyond that would cause a much higher number of simultaneous collisions (pile-up) and a greater number of tracks, and therefore it becomes difficult to correctly assign the vertices to the right tracks.\\
A bias voltage of 150 to 600 V is required to completely deplete the semi-conductor. The current is constantly monitored and pixel calibrated. A signal is registered if it is over a certain threshold. When a particle deposits enough charge in a pixel to be over the threshold the front-end electronics stores the Time-over-Threshold (ToT), i.e. the time the signal from the pre-amplifier is over the threshold. This has a nearly linear dependence on the charge released in the pixel and therefore on the energy deposited by the particle. ToT is also useful for measuring the $\mathrm{d}E/\mathrm{d}x$ of the particles.\\

\subsubsection{Semi-Conductor Tracker}
The SCT \cite{ATLAS-SCT} is located between the pixel detector and the TRT. It is composed of a central barrel and two end-caps. The barrel is divided into four cylinders of radii from 299 mm to 514 mm that cover a range of $|{\eta|}< $1.1-1.4 for a length of 1492 mm. The end-caps disks are composed of 9 plates covering the remaining range $|{\eta}|< $ 2.5 with a radius of 56 cm. The SCT is made of silicon sensors as the pixel detector but strips are used instead of pixels to cover a bigger area. Strips in the barrel are $p$-on-$n$ semi-conductor, with a dimension of 64.0 $\times$ 63.6 mm and 80 $\mu$m strip pitch, while in the end-caps they have different geometry to maximise the coverage and the pitch change between 56.9 to 90.4 $\mu$m. The SCT covers 61 m$^2$ of silicon detectors with 6.3 million readout channels. SCT had a mean hit efficiency between 98\% and 98.5\% during 2018 data taking in the active part of the detector, while around 1.5\% of the subdetector is inactive. The precision on the azimuthal direction is 17 $\mu$m and 580 $\mu$m along the $z$ direction.

\subsubsection{Transition Radiation Tracker}
The TRT \cite{ATLAS-TRT} covers the last part of the ID. It is a straw-tube tracker with a diameter of 4 mm, made of Kapton and carbon fibres, filled with a gas mixture of Xenon, Argon, CO$_2$, and oxygen. In the middle of each tube, there is a gold-plated tungsten wire of 31 $\mu$m diameter. The TRT is composed of 52,544 tubes, each 1.5 m in length parallel to the beam axis. They cover a pseudorapidity range of $|{\eta}|<1.$ with a radius from 0.5 m to 1.1 m. The remaining range of pseudorapidity, $1<|\eta|<2$, and 0.8 m $<|z|<$ 2.7 m, is covered by the end-caps straws. These are perpendicular to the beam axis and are 0.4 m long. Each side of the end-caps contains 122 880 straws.\\
The edge of the wall is kept at -1.5 kV while the wire is at ground. In this way, every tube behaves as a proportional counter. The space between straws is filled with polymer fibres and foils, respectively in the barrel and in the end-caps, to enable high energy particles to emit radiation. This effect depends on the relativistic factor $\gamma = E/m $, so for electrons is stronger. This is helpful in the identification process. The TRT is complementary to the other silicon-based part of the ID, but its information is only on the R-$\phi$ plane and the resolution is about 120 $\mu$m. However the number of straws that a particle has to travel is 35, therefore even if the resolution on the single hit is low, the combination of all the hits gives a resolution on the momentum that is compatible with the one from the other silicon detector.

\subsection{Calorimeter}
\label{ssec:cal}
\begin{figure}[!htb]
\centering
\includegraphics[width=1.\textwidth]{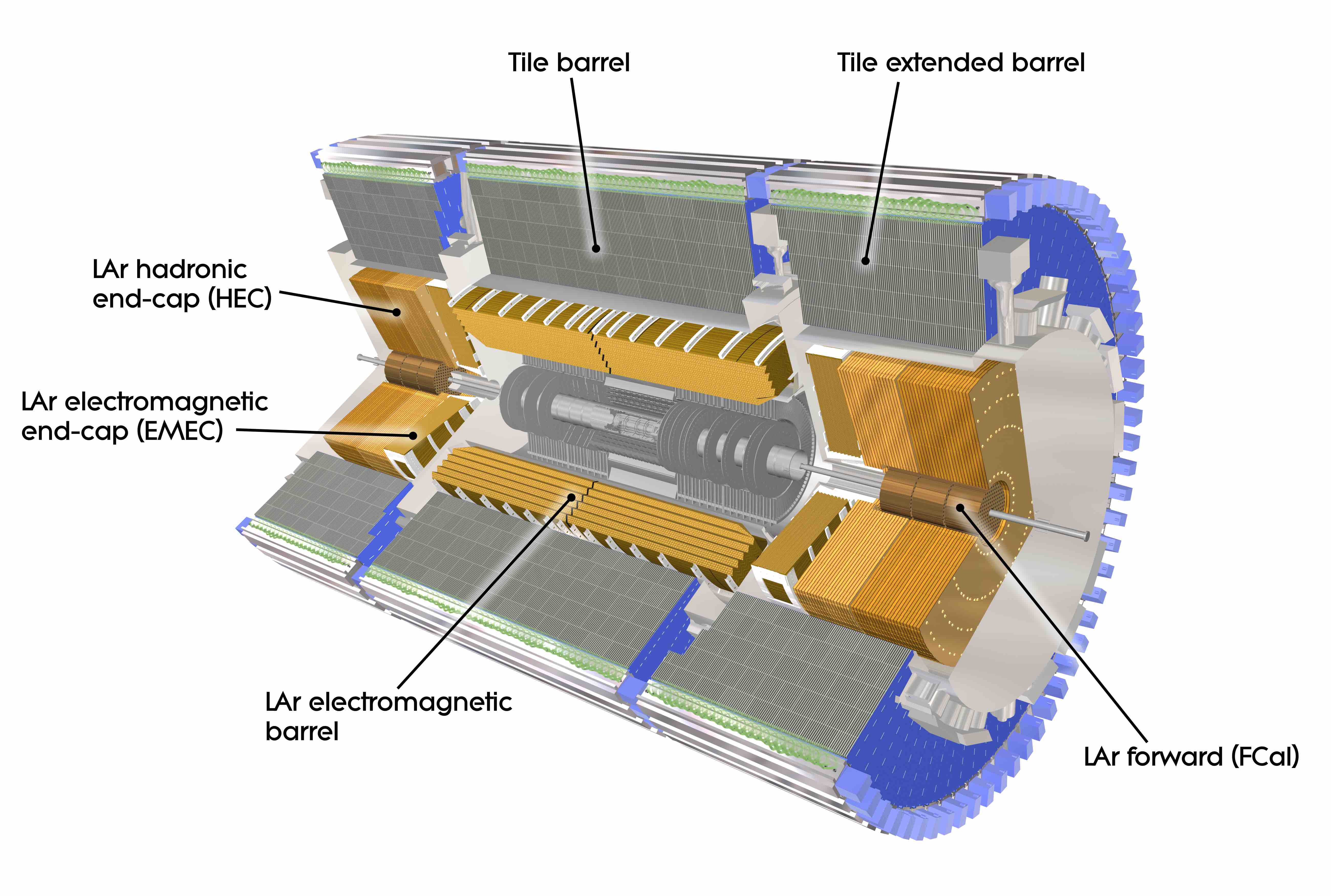}
\caption{Schematic view of the ATLAS calorimeter.}
\label{fig:ATLASCal}
\end{figure}

The next sub-system in the ATLAS detector is the calorimeter, shown in Fig. \ref{fig:ATLASCal}. 

Its purpose is to measure the energy of photons, electrons and hadrons. A particle should deposit all of its energy within the calorimeter to be correctly measured. Electromagnetic and hadronic showerings are the results of the interaction of electron/photon or hadrons in matter. These particles produce a cascade of other secondary particles less and less energetic. The sum of all this deposit is used to reconstruct the original particle energy. The shower produced by electrons and photons is different from the one produced by the hadrons, so different calorimeters are needed. \\
Three different calorimeters are installed in the ATLAS detector: the Electromagnetic CALorimenter (ECAL), just after the Solenoid, followed by the Hadronic CALorimeter (HCAL) and the Forward CALorimeter (FCAL). The first two are further divided into barrel and end-caps, and a letter, C or A, is used to distinguish the negative and positive pseudorapidity regions, respectively. \\
They are all sampling calorimeters without compensation (which is done offline) for the hadronic signal ($e/h>1$). The sampling is done using materials with high density where particles release most of their energy. The measure is done by sampling periodically the shape of the shower. This helps in better containing the particles inside the detector but as a drawback the energy resolution is lower. The segmentation of the calorimeters also allows the implementation of a position measurement. \\
Together they cover the range of $|\eta|<$ 4.9. It is important to have a large $\eta$ coverage because it helps reducing the momentum taken away by particles too forward to be detected, which would degrade the measurement of the missing transverse momentum.

\subsubsection{Electromagnetic Calorimeter}
\begin{figure}[!htb]
\centering
\includegraphics[width=0.8\textwidth]{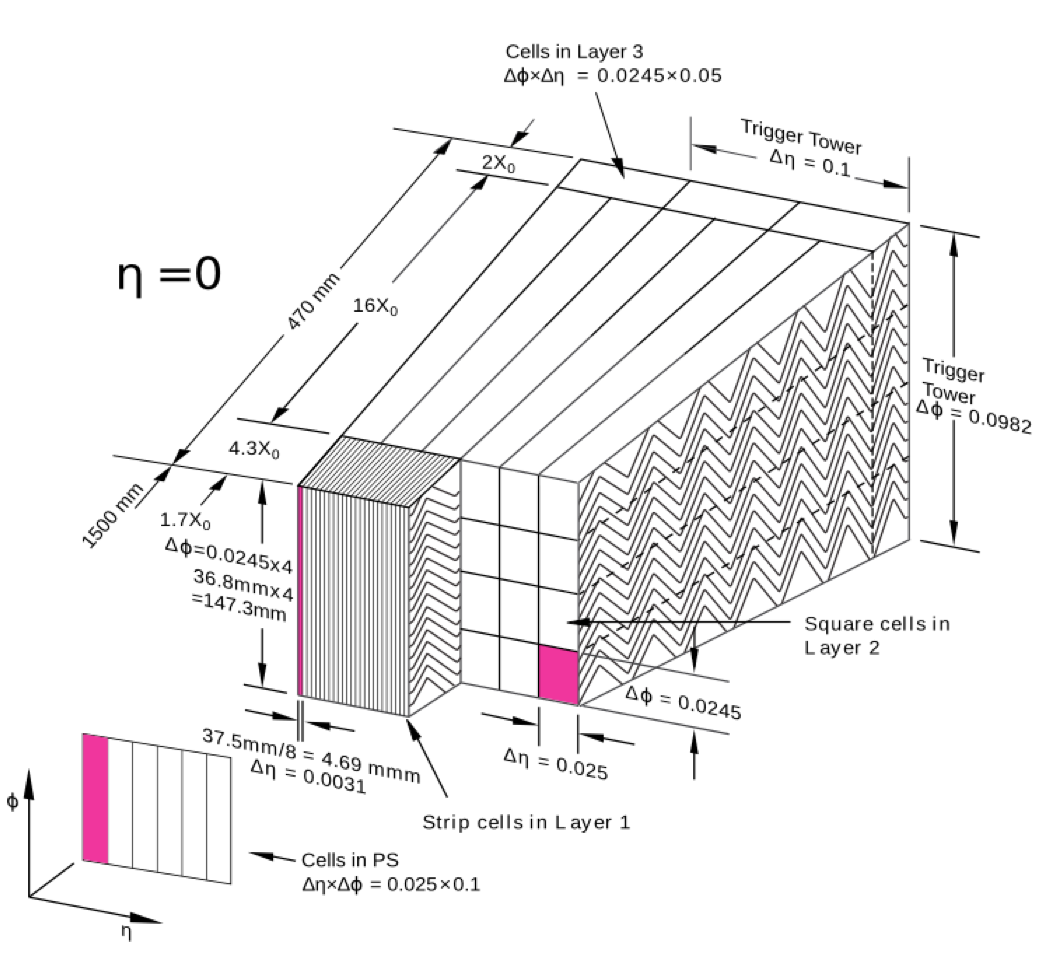}
\caption{View of the accordion geometry of the ECAL.}
\label{fig:ATLASECal}
\end{figure}

The ECAL \cite{ATLAS-ECAL} is composed of Liquid Argon (LAr) as an active medium and lead as passive material. The detectors are housed inside cryostats filled with LAr and kept at around 88 K. The ECAL has an \textit{accordion} geometry, as shown in Fig. \ref{fig:ATLASECal}, which allows to cover all the solid angle without an instrumentation gap and to collect a fast and azimuthally uniform response.

The ECAL is divided into a barrel region (EMB), covering a pseudorapidity range of $|\eta|<$ 1.475, and two the end-caps (EMEC), covering a pseudorapidity range of 1.375 $<|\eta|<$ 3.2. The ECAL extends over 24 electromagnetic radiation lengths ($X_0$), and it is segmented in three parts (front, middle and back) of increasing cell size and with a different radiation length (respectively 2, 20, 2 $X_0$) to efficiently discriminate between prompt photons and photons coming from $\pi^0$. Furthermore, a presampler is located in the region $|\eta|<$ 1.8 to provide information on the particle energy before the ECAL. Hadronic particles do not lose much energy in the ECAL since a hadronic interaction length is longer than the space between the interaction point and the calorimeters.\\
The energy resolution is given by

\begin{equation}
\frac{\sigma}{E}=\frac{10\%}{E}\oplus 0.7\%
\end{equation}
where $E$ is expressed in GeV, and $\oplus$ is the quadratic sum.

\subsubsection{Hadronic Calorimeter}
\begin{figure}[!htb]
\centering
\includegraphics[width=0.6\textwidth]{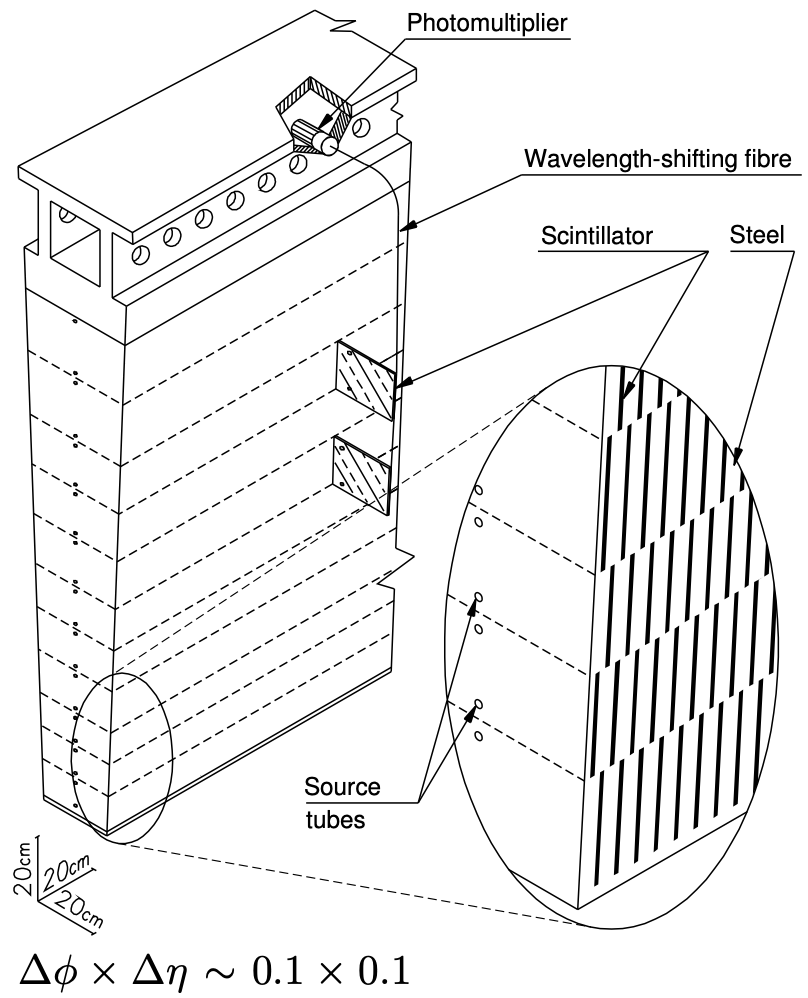}
\caption{View of the geometry of the HCAL.}
\label{fig:ATLASHCal}
\end{figure}

The HCAL is divided into a central part in the barrel, called \textit{Tile Calorimeter} (TileCal) \cite{ATLAS-TileCAL}, and two end-caps, called HEC. The two parts have a different composition, but are both sampling calorimeters.\\
The TileCal uses scintillators as sampling material and steel as absorber. The TileCal is composed of three sections: a Long Barrel (LB) in the range $|\eta|<$1.0 and two Extended Barrels (EBs) in the range $0.8<|\eta|<1.7$. Between the two there is a gap filled with scintillator to help recover energy otherwise lost in this transition region. The geometry of the hadronic calorimeter can be seen in Fig. \ref{fig:ATLASHCal}.

The HEC consists of two wheels divided into two longitudinal segments perpendicular to the beam covering the pseudorapidity range 1.5 $<|\eta|<$ 3.2. The wheels are made of copper as the passive material, and LAr as the active medium, due to the high radiation resistance requested in this area.\\
The energy resolution for the TileCal is
\begin{equation}
\frac{\sigma}{E}=\frac{50\%}{E}\oplus 3\% \, .
\end{equation}

\subsubsection{Forward Calorimeter}
The FCal covers the part the closest to the beam pipe, in the pseudorapidity range $3.1<|\eta|<4.9$. It measures both electromagnetic and hadronic particles. The geometry of the FCal is a cylindrical one, with three different modules. It is segmented using LAr as active medium while copper and tungsten are used as absorber. The choice of LAr in this region is due to the high dose of radiation in this region and LAr can be easily replaced without losses in performance. The FCal has a poor performance in particle reconstruction but is fundamental for the missing transverse energy and the reconstruction of very forward jets. The energy resolution for the FCal is 
\begin{equation}
\frac{\sigma}{E}=\frac{100\%}{E}\oplus 3.1\%
\end{equation}

\subsection{Muon spectrometers}
\label{ssec:muonspec}
\begin{figure}[!htb]
\centering
\includegraphics[width=0.7\textwidth]{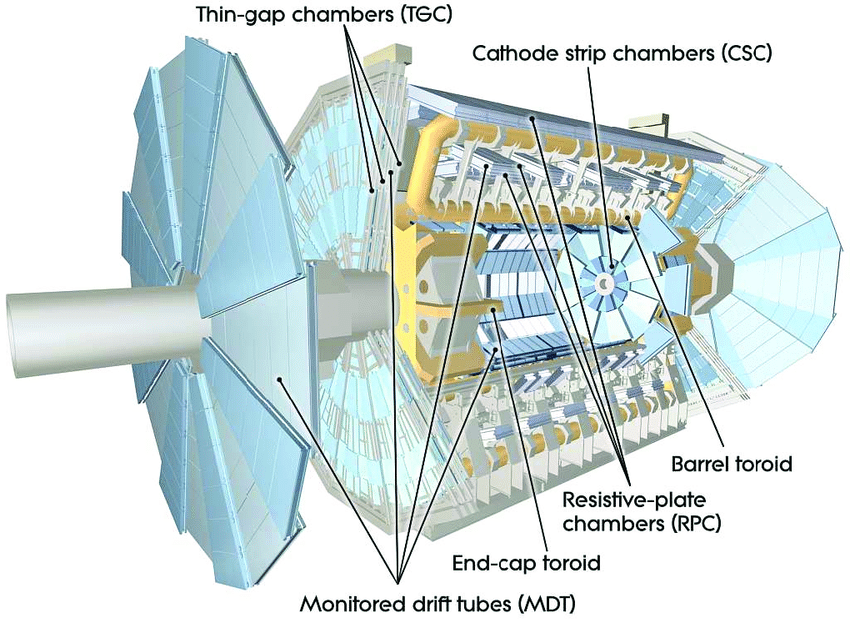}
\caption{Schematic view of the Muon Spectrometer.}
\label{fig:ATLASMuon}
\end{figure}

The Muon Spectrometer (MS) \cite{ATLAS-MUON} is shown in Fig. \ref{fig:ATLASMuon} and represents the last part of the ATLAS detector.

Muons can travel all the detector without being stopped as they lose energy almost always through ionization, and very little energy is spent while interacting in matter. Powerful magnetic fields are then needed to bend the trajectories and measure the momentum. The MS has also an important function as a trigger. Therefore it has to be fast and with high granularity.\\
The MS consists of one barrel that covers the pseudorapidity range $|\eta|<1.05$ and two end-caps that cover the pseudorapidity range 1.05 $<|\eta|<$ 2.7. Tracks are measured by three layers with a cylindrical geometry in the barrel region, while wheels perpendicular to the beam are present in the end-caps. The momentum measurement in the barrel region is done by the Monitored Drift Tubes (MDTs), while in the end-caps the measurement is done by the Cathode Strip Chambers (CSCs), multiwire proportional chambers with cathodes segmented into strips. Momentum measurement is done by measuring with high precision the coordinates of the curvature of the particles bent by the magnetic systems. Since $p=0.3BR$, where $B$ is the magnetic field expressed in Tesla, $R$ the curvature in meters, and $p$ the momentum of the particle in GeV, measuring $R$ and knowing exactly $B$ is possible to obtain $p$. The MDTs measure the $\phi$ coordinate while the CSC the $R$ one. They give a spatial resolution of 80 and 60 $\mu$m respectively for the MDT and CSC.\\
The trigger system is done by the Resistive Plate Chambers (RPCs) in the barrel and by the Thin Gap Chambers (TGCs) in the end-caps. Together they cover a pseudorapidity range $|\eta|<$ 2.4.\\
During the shutdown before Run~2, the MS has been upgraded (actually its original design) by adding some more chambers in the transition region between barrel and end-caps (1.0 $<|\eta|<$ 1.4). Moreover, other RPC and MDT chambers have been installed with tubes with a smaller radius to cope with the new higher rates.\\
The bending provided is about 2.5 Tm in the barrel and 6 Tm in the end-caps. This is provided by a system of three large superconducting air-core toroidal magnets. In the barrel, in the pseudorapidity range $|\eta|<$ 1.4 by the large barrel toroid, while for 1.6 $<|\eta|<$ 2.7 by the end-caps toroids. The region in between (1.4 $<|\eta|<$ 1.6) is deflected by a combination of the barrel and end-cap.

\subsection{The magnet system}
\label{ssec:magnet}
\begin{figure}[!htb]
\centering
\includegraphics[width=1.\textwidth]{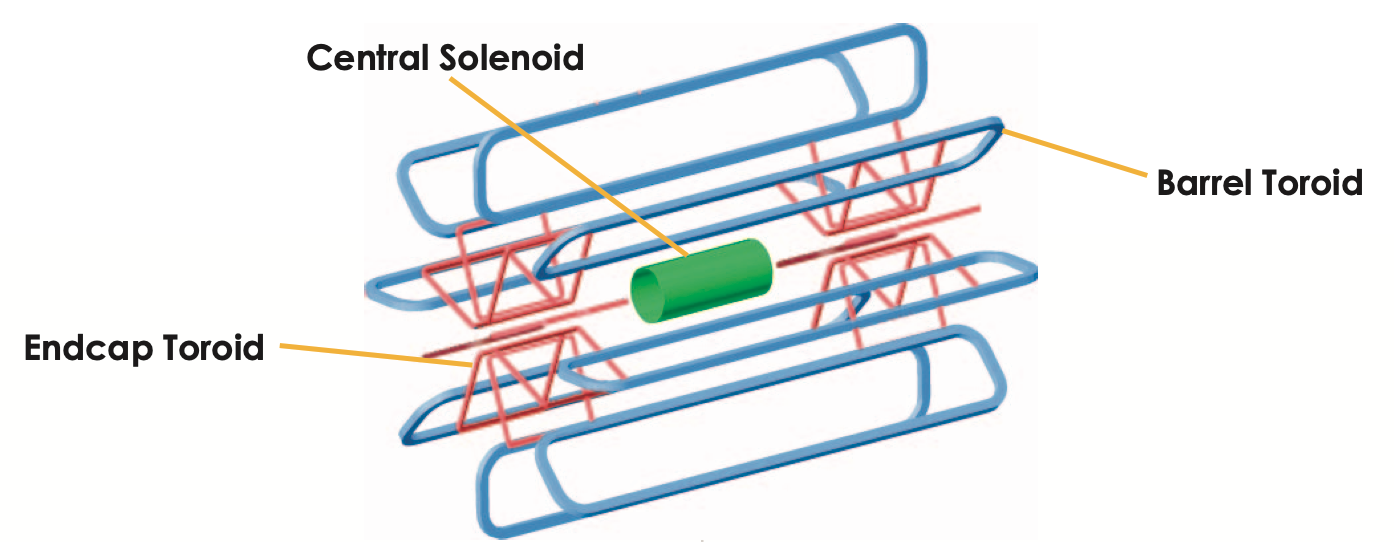}
\caption{Magnet system of the ATLAS detector.}
\label{fig:ATLASMag}
\end{figure}

The ATLAS magnetic system is 22 m in diameter and 26 m in length and it is composed of four different large superconducting magnets \cite{ATLAS-BarrelMAGNET,ATLAS-EndcapMAGNET,ATLAS-CentralMAGNET}, as shown in Fig.~\ref{fig:ATLASMag}.

The magnets are divided into one thin solenoid around the ID, and three large toroids, one in the barrel and two end-caps, arranged around the calorimeters. They provide bending power for the Muon Spectrometer. \\
\begin{itemize}
\item The solenoid is aligned to the beam axis and provides a 2 T magnetic field, it is as thin as possible (0.66 $X_0 $) to minimise the impact on the ECAL energy resolution. It is made of a single layer coil, wound with a high-strength Al stabilised NbTi conductor. The inner and outer diameters of the solenoid are respectively 2.46 m and 2.56 m. Its length is 5.8 m. The flux of the magnetic field is returned by the steel of the ECAL.
 
\item The toroids produce a magnetic field of 0.5 and 1 T for the barrel and end-caps regions, respectively, and they are sustained by a 25 kA current. Every toroid in the barrel has its own separated cryostat, while in the end-cap they all have a common cryogenic system. The toroids are made of a mixture of copper, niobium, aluminium and titanium.
\end{itemize}

\subsection{Forward detectors}
\label{ssec:fwdet}
\begin{figure}[!htb]
\centering
\includegraphics[width=1.\textwidth]{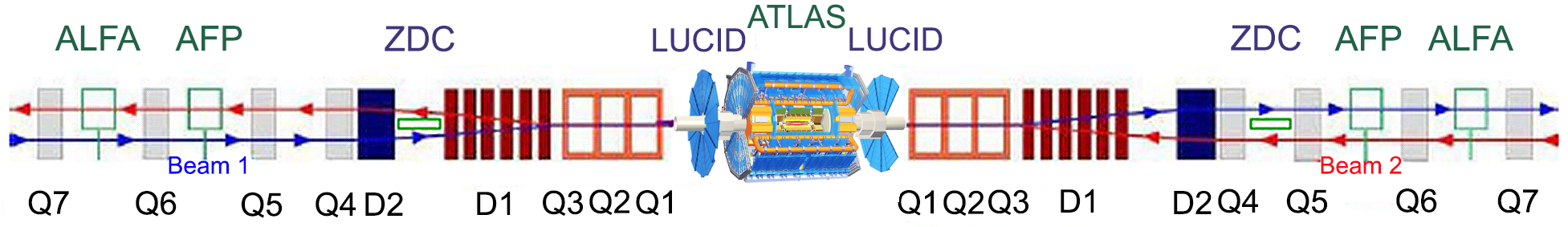}
\caption{Representation of the ATLAS forward detectors.}
\label{fig:ATLASForward}
\end{figure}

Four smaller detectors are present in the forward region, as shown in Fig.~\ref{fig:ATLASForward}, to help to measure the luminosity.

At 17 m from the interaction point in both directions there is LUCID (LUminosity measurement using Cerenkov Integrating Detector), a detector designed to detect the $pp$ scattering in the forward region (5.6 $ <|\eta|<$ 6.0). It is the main online relative luminosity monitor for ATLAS. Located at 140 m, there is the ZDC (Zero-Degree Calorimeter). It's made of layers of alternating tungsten plates and quartz rods. It covers a range of pseudorapidity of $|\eta|>$ 8.2. It is helpful for heavy-ions alignment. The third detector is AFP (ATLAS Forward Proton), which is used to identify protons that emerge intact from the $pp$ collisions. It is composed of tracking and timing silicon detectors 2 mm from the beam at 210 m from the ATLAS interaction point. Last, at 240 m lies the ALFA (Absolute Luminosity For ATLAS) detector. It is made of scintillating fibre trackers as close to the beam as 1 mm.

\subsection{Trigger and data acquisition}
\label{ssec:trigdata}

\begin{figure}[!htb]
\centering
\includegraphics[width=0.9\textwidth]{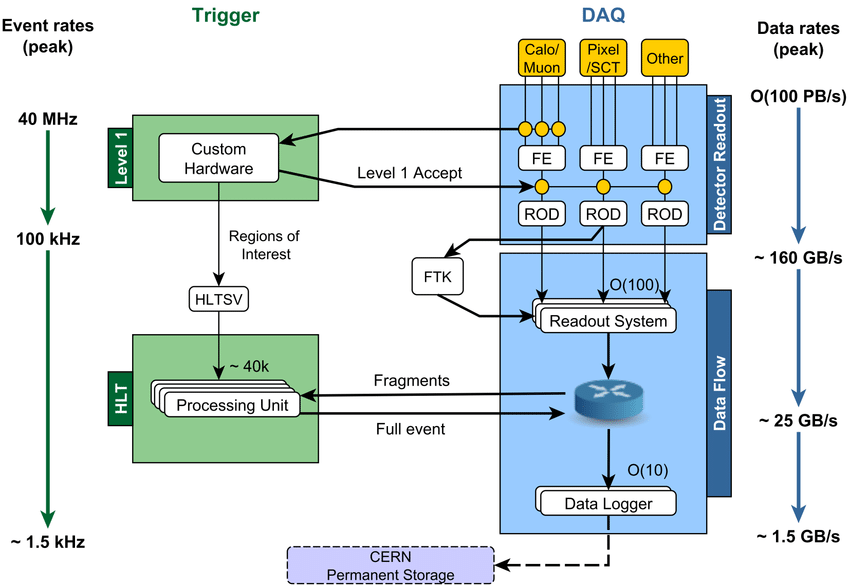}
\caption{ATLAS TDAQ System in Run~2.}
\label{fig:ATLASTDAQ}
\end{figure}

Collisions happen at every bunch crossing with a frequency of 40 MHz. The average pile-up during Run~2 is around 33 collisions with an instantaneous luminosity of  $\approx 2\times 10^{34}$ cm$^{-2}$s$^{-1}$. This large amount of data can not be processed or stored. Therefore triggers are necessary to reduce this rate to approximately 1 kHz, almost a factor $10^5$. The ATLAS triggers are designed to decide what events are worth keeping for the analysis in the shortest time possible.\\
This is implemented by the Trigger and Data AcQisition (TDAQ) systems \cite{ATLAS-TDAQ1,ATLAS-TDAQ2} and the Detector Control System (DCS). TDAQ and DCS are divided into sub-systems associated to the various sub-detectors.\\
During Run~1 the trigger had a system composed of three different levels more and more selective: Level 1 (L1), Level 2 (L2) and Event Filter. The first was hardware-based while the other two were software based. Due to the higher luminosity in Run~2, the triggers have been upgraded and the new trigger system for the Run~2 L2 and Event Filter have been merged into a single High Level Trigger (HLT). A scheme of the trigger flow during Run~2 is shown in Fig.~\ref{fig:ATLASTDAQ}.

The TDAQ workflow can be summarized as follow:
\begin{itemize}
\item[1.] L1 makes a decision based on the information from just a part of the detector with reduced granularity, the calorimeters and the MS. These detectors are able to identify high $p_T$ muons, electron/photons, jets, taus decaying into hadrons and missing energy in less than 2.5 $\mu$s, reducing the rate to 100 kHz. L1 is composed of different subtriggers: L1 Calorimeter (L1Calo), L1 Muon (L1Muon), Central Trigger Processors (CTP) and the L1 Topological (L1Topo) trigger modules. The L1Topo part of the trigger calculates event topological quantities between L1 objects within the L1 latency time ($\approx 2\mu$s) and uses these to perform selections. For example, it is possible to compute invariant masses or angular distances between objects.

\item[2.] When an event passes L1 a signal is sent back to the rest of the detector causing the data associated with the event to be read out for all components of the detector. Only the data for the events selected by L1 is then transferred from the Front-End memories (FE) into the detector ReadOut Drivers (RODs). The data contained in the RODs are sent for storage to the ReadOut Buffers (ROBs) contained in the ReadOut System units (ROSs), where it is temporarily stored and provided on request to the following stages of event selection.

\item[3.] For every accepted event, the L1 system produces the \textit{Region of Interest} (RoI) information, which includes the positions of all the identified interesting objects in units of $\eta$ and $\phi$. This information is sent by the different elements of the L1 trigger system to the RoI Builder (RoIB), which assembles it into a unique data fragment and sends it to the HLT SuperVisor (HLTSV).

\item[4.] HLT uses fast algorithms to access RoIs to read information such as coordinates, energy and type of signature to reduce once again the rate of events. The HLT output rate is approximately 1.5 kHz with a processing time of 0.2s on average. Events passing also the HLT are permanently moved to the CERN storage.
\end{itemize}

The DCS is active during all data taking periods, monitoring every aspect of the ATLAS detector, from the value of magnetic fields to the humidity and temperature. It also checks for abnormal behaviour and permits to safely operate all the hardware components.

\subsection{Run~3 detector upgrade}
\label{ssec:run3detector}
\begin{figure}[!htb]
\centering
\includegraphics[width=0.9\textwidth]{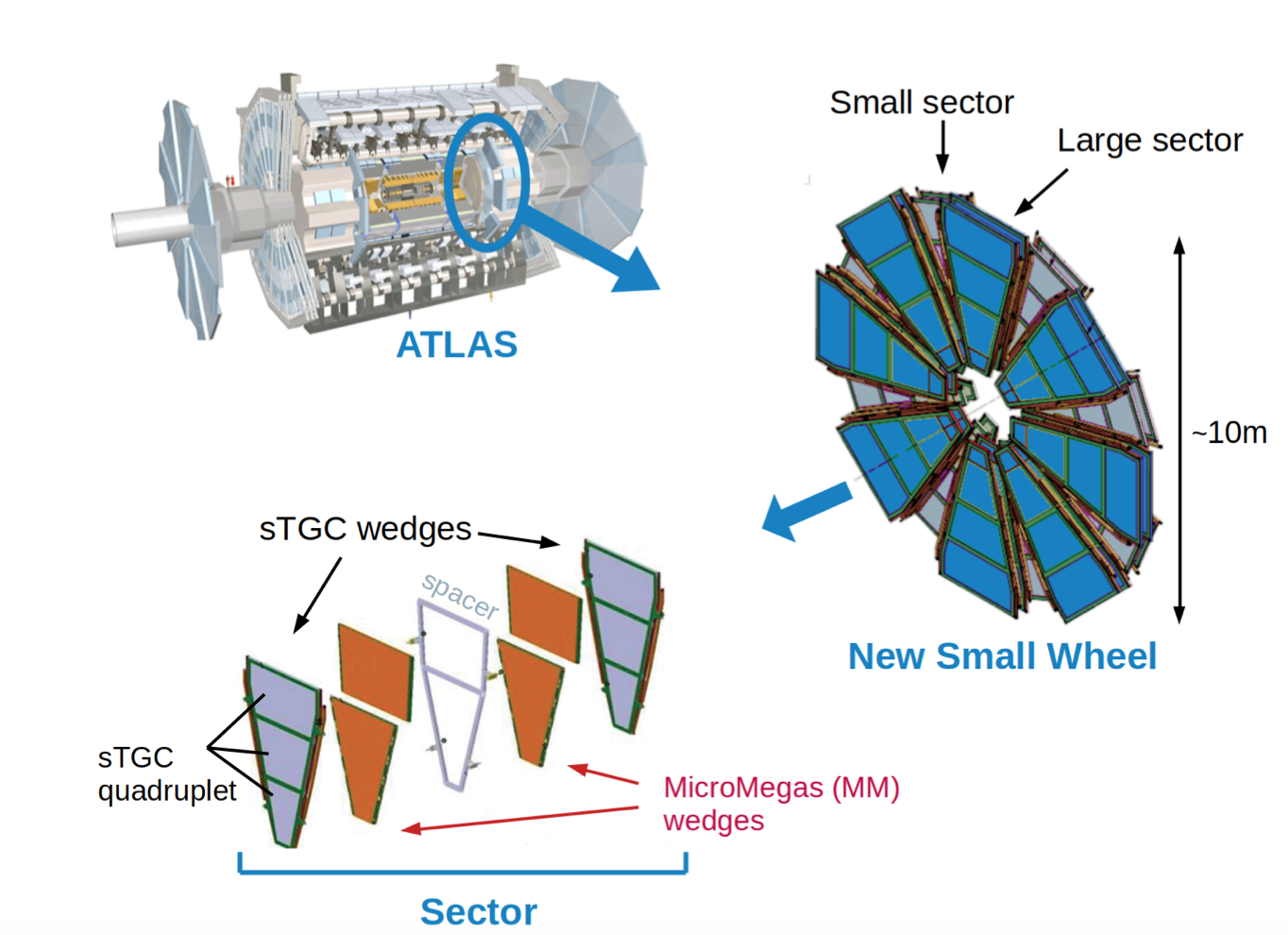}
\caption{Schematic view of the ATLAS NSW.}
\label{fig:ATLASNSW}
\end{figure}

Among the new systems which have been installed in the detector upgrade plan, there are the 10-metre diameter New Small Wheels (NSWs). The installation of the NSWs has been carried out during LS2 to have them available during Run~3. The NSWs employ two detector technologies: small-strip Thin Gap Chambers (sTGC) and MicroMegas (MM). The sTGC are optimised for triggering while MM detectors are optimised for precision tracking. Both technologies can withstand the higher flux of neutrons and photons expected in future LHC interactions. A schematic diagram is shown in Fig. \ref{fig:ATLASNSW}. One wheel is made of 16 sectors and each sector is composed of two sTGC wedges and one MM double-wedge. The sTGC wedges are made of three quadruplets modules, each composed of four sTGC layers.

Additional improvements to the ATLAS MS include 16 new chambers featuring Small Monitored Drift Tubes (sMDT) and Resistive Plate Chambers (RPCs) that have been installed in the barrel of the experiment, thus improving the overall trigger coverage of the detector. The smaller diameter tubes of the sMDTs provide an order of magnitude higher rate capability.
LS2 has also seen the enhancement of the ATLAS LAr calorimeter with new FE electronics and optical-fibre cabling. This improves the resolution of the detector at the trigger level, providing 4 times higher granularity to allow jets to be better differentiated from electrons and photons, thus refining the first decision level where collision events are accepted for offline storage or dismissed. ATLAS trigger and data-acquisition systems have also been upgraded during LS2 with new electronics boards, further improving the overall resolution and fake rejection of the experiment, while preparing for the HL-LHC.
\chapter{Radiation Damage Effects on the ATLAS Pixel Detector}

The ATLAS pixel detector is the closest one to the interaction point and so the most affected by the effects of radiation damage. In this Chapter, some aspects of the ATLAS Pixel detector are presented and the effects of the radiation damage are discussed. The simulation of the radiation damage effects in ATLAS is presented and data to Monte Carlo (MC) comparisons are shown for validation.

\minitoc
\medskip

\section{The ATLAS Pixel Detector}
\label{sec:pixel}
In this Section, the geometry and layout of the detector are described in more detail than Section \ref{ssec:ID}. The sensor technology, the electronics that read the signal, and the calibration procedure are also described.

\subsection{General Layout}
\begin{figure}[!htb]
\centering
\includegraphics[width=0.7\textwidth]{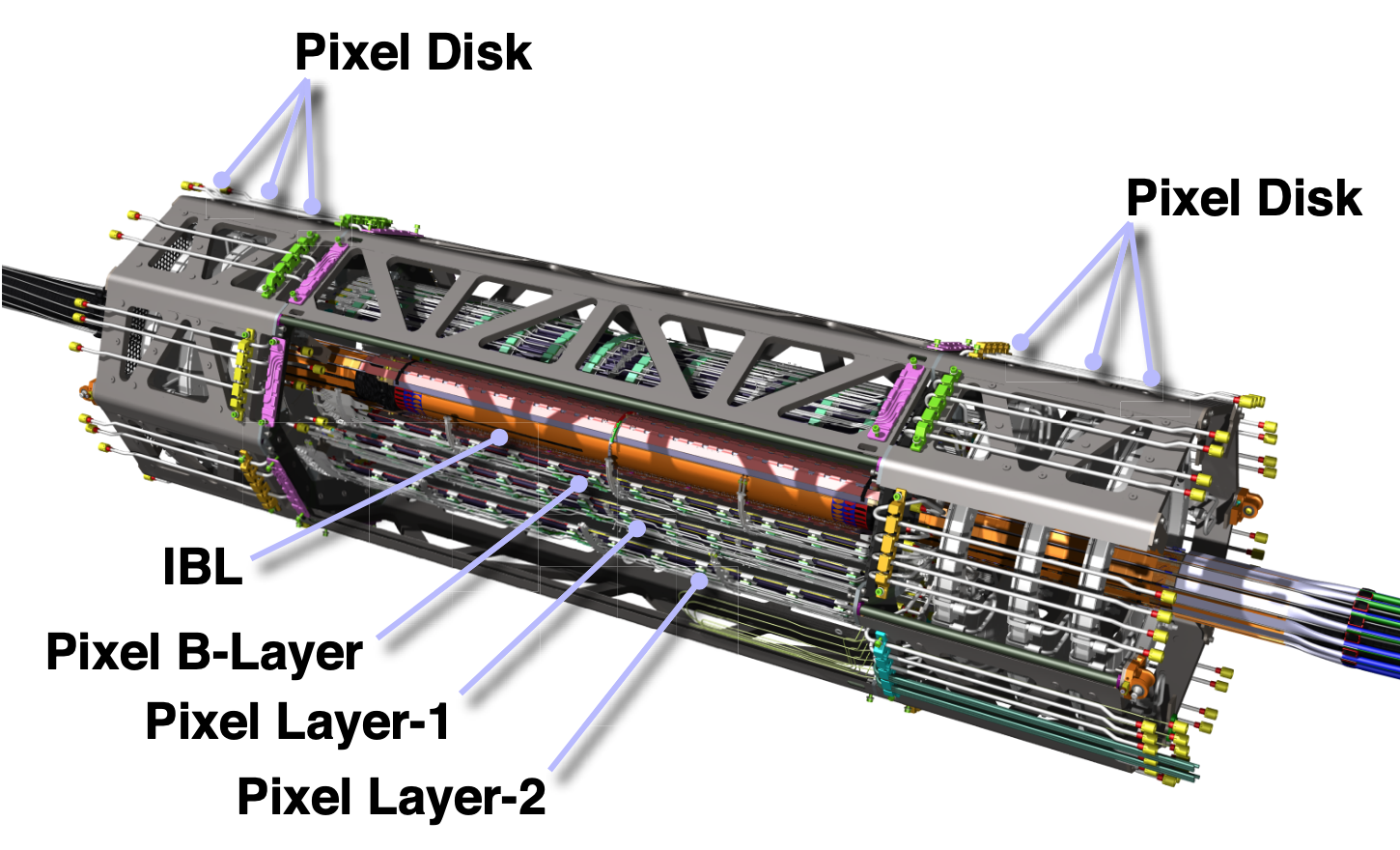}
\caption{Schematic view of the ATLAS pixel detector.}
\label{fig:PixelDetector}
\end{figure}

The Pixel Detector \cite{ATLAS-ID} is made of four barrel layers and three disk layers per end cap, six end cap disks in total, as shown in Fig.~\ref{fig:PixelDetector}. With the addition of the IBL \cite{ATLAS-IBL} the first layer is now at 33.5 mm from the beam axis, where the beam pipe has been reduced inner radius size of 23.5 mm to accommodate the new layer of pixels. 

\begin{figure}[!htb]
\centering
\includegraphics[width=0.6\textwidth]{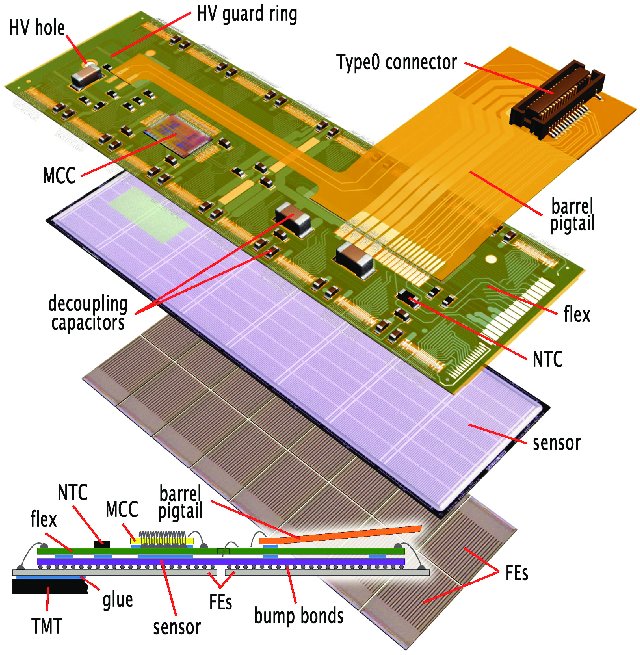}
\caption{Assembly view and cross-section of an ATLAS Pixel Detector module. Sixteen front-end chips are bump bonded to the silicon pixel sensor. Interconnections are done on a flexible Kapton PCB, which is connected by wire bonds to the electronic chips.}
\label{fig:PixelDetector-Module}
\end{figure}

Its coverage goes up to $|\eta| <2.5$, while it covers the full azimuthal angle ($\phi$). The nominal pixel size is 50 $\mu$m in $\phi$ direction and 250 $\mu$m in $z$ direction for the IBL, with a depth of 200 $\mu$m for the planar sensors and 230 $\mu$m for the 3D, while the size is 50 $\mu$m in $\phi$ direction and 400 $\mu$m in $z$ direction, with a depth of 250 $\mu$m for the other layers (B-Layer, Layer 1 and Layer 2). The base element of the detector is the pixel module, composed of the silicon sensor itself, the front-end electronics, and the flex-hybrids with control circuits. There are 46,080 pixel electronic channels in a pixel module and 43,000 in a IBL module. Pixel modules are connected with FE-I3 front-end chip for the readout, while the IBL modules use a FE-I4B, which allow also for a larger active area (Fig. \ref{fig:PixelDetector-Module}).

\begin{figure}[!htb]
\centering
\includegraphics[width=0.7\textwidth]{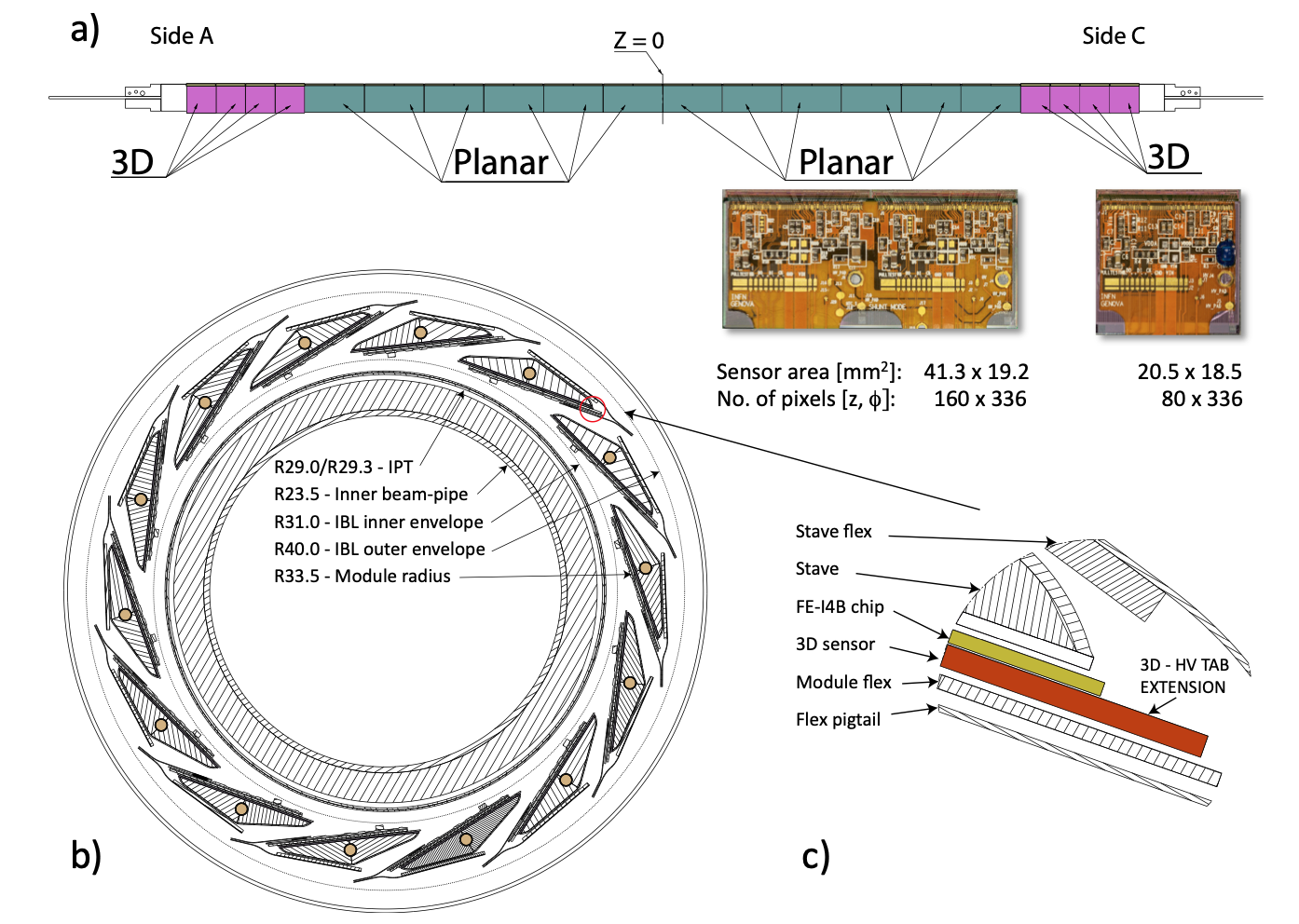}
\caption{IBL detector layout: (a) Longitudinal layout of planar and 3D modules on a stave. (b) An $r-\phi$ section showing the beam pipe, the inner positioning tube (IPT), the staves of the IBL detector and the inner support tube (IST), as viewed from the C-side. (c) An expanded $r-\phi$ view of the corner of a 3D module fixed to the stave. From Ref.~\cite{ATLAS-IBL-Appendix}.}
\label{fig:ATLAS-IBL-geometry}
\end{figure}

The pixel system is then composed of sub-elements called staves (in the barrel) and sectors (in the disks) that contain pixel modules, the mechanics and the cooling system. Staves and sectors are then mounted together on supporting structures to form the barrel and the disks. The barrel modules are mounted on the staves overlapping in $z$ to avoid any gap in particle detection, at least for particles with $p_{\mathrm{T}} > 1$ GeV, and facing the beam pipe. Moreover, the pixel modules are tilted by an angle in the azimuthal direction achieving overlap in the active area, and also helping in compensating for the Lorentz angle drift of the charges inside the detector. Each IBL stave is composed of 12 modules with planar sensors in the middle and 4 with 3D sensors at each end of the stave, as shown in Fig.~\ref{fig:ATLAS-IBL-geometry}. Barrel and endcap disks are mounted on carbon-fibre support. Services, such as electronics, optics, and cooling, are connected within service panels from patch panels (Patch Panel 0-PP0) at the ends of the supporting spaceframe to the end of the Pixel Support Tube. Services connections are made at the end of the Pixel Support Tube at Patch Panel 1 (PP1), while connections of external services are at additional patch panels (PP2, PP2, and PP4), situated outside the ID.

The main parameters of the Pixel Detector are summarized in Table \ref{tab:Pixel-Numbers}. 

\begin{table}[!htb]
\scalebox{0.6}{
\centering
\begin{tabular}{lcccccccc}
\noalign{\smallskip}\hline\noalign{\smallskip}
Layer & Mean        & Number    & Number     & Number       &  Pixel         & Sensor             & Active       & Pixel array \\
Name  & Radius [mm] & of Staves & of Modules & of Channels  &  Size [$\mu$m] & Thickness [$\mu$m] & Area [m$^2$] & (columns rows) \\
\noalign{\smallskip}\hline\noalign{\smallskip}
IBL     & 33.5 & 14 & 12+8 (p/3D) & 43,000 & $50\times 250$      & 200/230 (p/3D) & 0.15 & $336\times 80$\\
B-Layer & 50.5 & 22 & 13               & 46,080 & $50\times 400$ & 250 & 0.28 & $160\times 18$\\
Layer 1 & 33.5 & 38 & 13               & 46,080 & $50\times 400$ & 250 & 0.49 & $160\times 18$\\ 
Layer 2 & 33.5 & 52 & 13               & 46,080 & $50\times 400$ & 250 & 0.67 & $160\times 18$\\
\noalign{\smallskip}\hline\noalign{\smallskip}
Layer & Mean & Number & Number & Number & Pixel & Sensor & Active & Pixel array\\
Name & $z$ [mm] & of Sectors & of Modules & of Channels  & Size [$\mu$m] & Thickness [$\mu$m] & Area [m$^2$] & (columns rows)\\
\noalign{\smallskip}\hline\noalign{\smallskip}
Disk 1 & 495 & 8 & 6  & 46,080 & $50\times 400$ & 250 & 0.0475 & $160\times 18$ \\
Disk 2 & 580 & 8 & 6  & 46,080 & $50\times 400$ & 250 & 0.0475 & $160\times 18$ \\
Disk 3 & 650 & 8 & 6  & 46,080 & $50\times 400$ & 250 & 0.0475 & $160\times 18$ \\
\noalign{\smallskip}\hline\noalign{\smallskip}
\end{tabular}}
\caption{Basic parameters for pixel layers and disks.}
\label{tab:Pixel-Numbers}
\end{table}

An important aspect that has been considered during the production of the pixel detector is the material budget, e.g. the amount of material that composes the detector. This must be optimized to reduce the effect of multiple scattering of incoming particles and to ensure good performance for tracking and vertex reconstruction. For the IBL the radiation length averaged over is 1.88\% $X_0$ for tracks perpendicular to the beam axis originating in $z=0$, while it is 30\% more for the other pixel layers. IBL small radiation length was achieved with new technology: staves with lower density and the CO$_2$ evaporative cooling instead of C$_3$F$_8$, allowing for more efficient cooling in terms of mass and pipe size; new modules with lower mass; and using aluminium conductors for the electrical power services.

\subsection{Sensors}
Sensors are the sensitive part of the pixel detector and work as a solid-state ionization chamber for charged particles. Sensors must have high geometry precision and granularity. Another important requirement is a high charge collection efficiency, while at the same time being resistant to the high dose of radiation damage from ionizing and non-ionizing particles. The requirements are met with a careful design of the structure of the sensor and the bulk material. In the ATLAS pixel detector, there are two different technologies implemented: planar and 3D.

\subsubsection{Planar Sensors}
\begin{figure}[!htb]
\centering
\includegraphics[width=1.\textwidth]{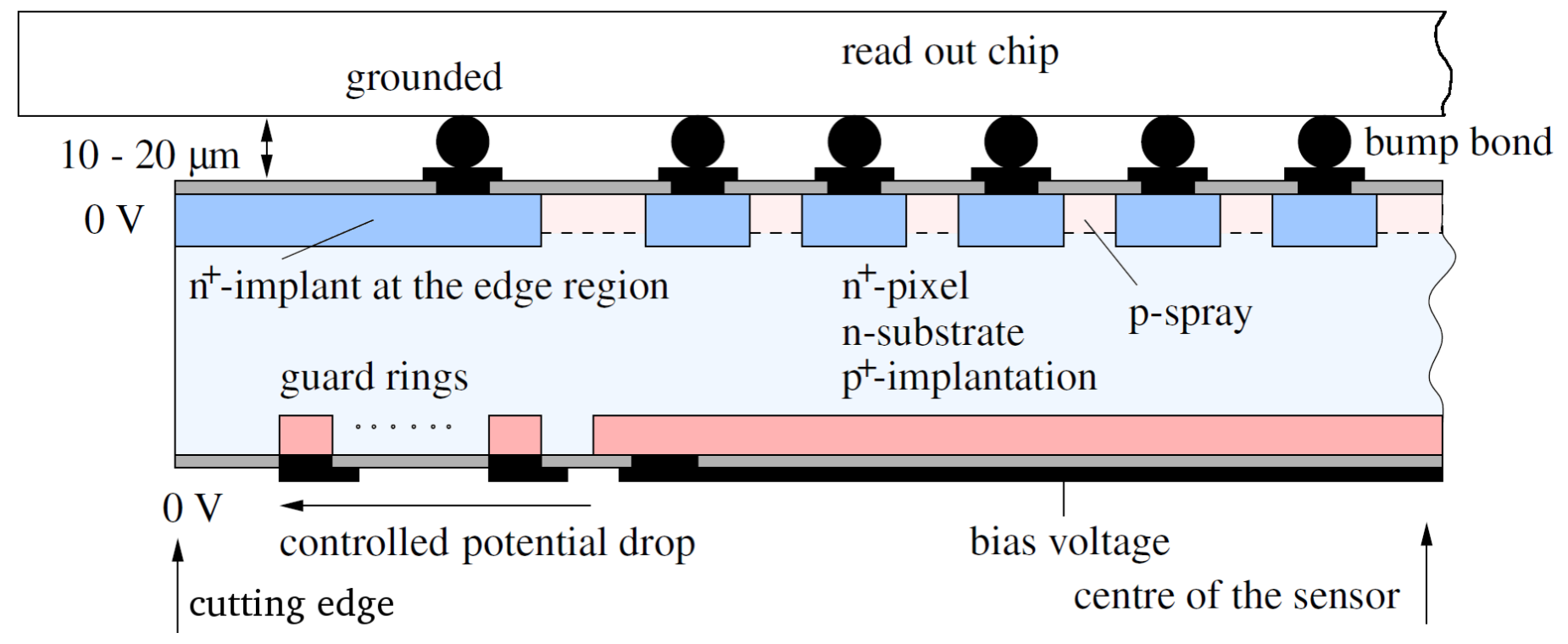}
\caption{Cross-section sketch of the ATLAS pixel module.}
\label{fig:ATLAS-pixel-principle}
\end{figure}

The pixel planar sensors are arrays of bipolar diodes placed on a high resistivity $n$-type bulk close to the intrinsic charge concentration. Planar sensors are $n^{+}$-in-$n$ sensors with a high positive ($p^+$) implantation and negative ($n^+$) dose regions implanted on each side of the wafer. Due to the concentration gradient, electrons and holes from the $n$ and $p$ type sides are recombined, forming a small depletion region. This is expanded with a reverse bias applied to the sensor. The voltage at which point the sensor is fully depleted in the bulk is called \textit{depletion voltage}. This is proportional to the doping concentration and depends also on the thickness of the bulk, and its resistivity. Charges and holes produced by the ionization of charged particles passing through the sensor are then free to reach the electrodes and can be detected by the electronics. Fig.~\ref{fig:ATLAS-pixel-principle} shows a sketch of the ATLAS pixel module.

The $n^+$ implants are on the readout side and the $p$-$n$ junction is on the backside. The $n$-side is made to match the FE-I4B \cite{GARCIASCIVERES2011S155} readout electronics, for IBL, while FE-I3 \cite{PERIC2006178} for the other layers. Guard-rings are positioned on the $p$-side: 13 in the IBL and 16 on the other pixel layers. The module's edges are kept at ground to avoid discharge in the air causing the sensors at the edges being not depleted. Inactive edge has also been reduced from 1100 $\mu$m to 200 $\mu$m between the outermost pixels and IBL. Pitch size in IBL is 250 $\mu$m by 50 $\mu$m, while in the two central columns of the double-chip sensor they are extended to 450 $\mu$m instead of 250 $\mu$m to cover the gap between the two adjacent FE-I4B chip.
In the other layers, the nominal pitch is 400 $\mu$m and 50 $\mu$m against long pixels with a pitch of 600 $\mu$m and 50 $\mu$m. All the readout channels in a tile are connected to a common structure for the bias voltage via a punch-through connection that provides DC-coupled bias to each channel. This allows bringing the bias voltage to the sensors without individual connection, but still having isolation between pixels.

\subsubsection{3D Sensors}
\begin{figure}[!htb]
\centering
\includegraphics[width=1.\textwidth]{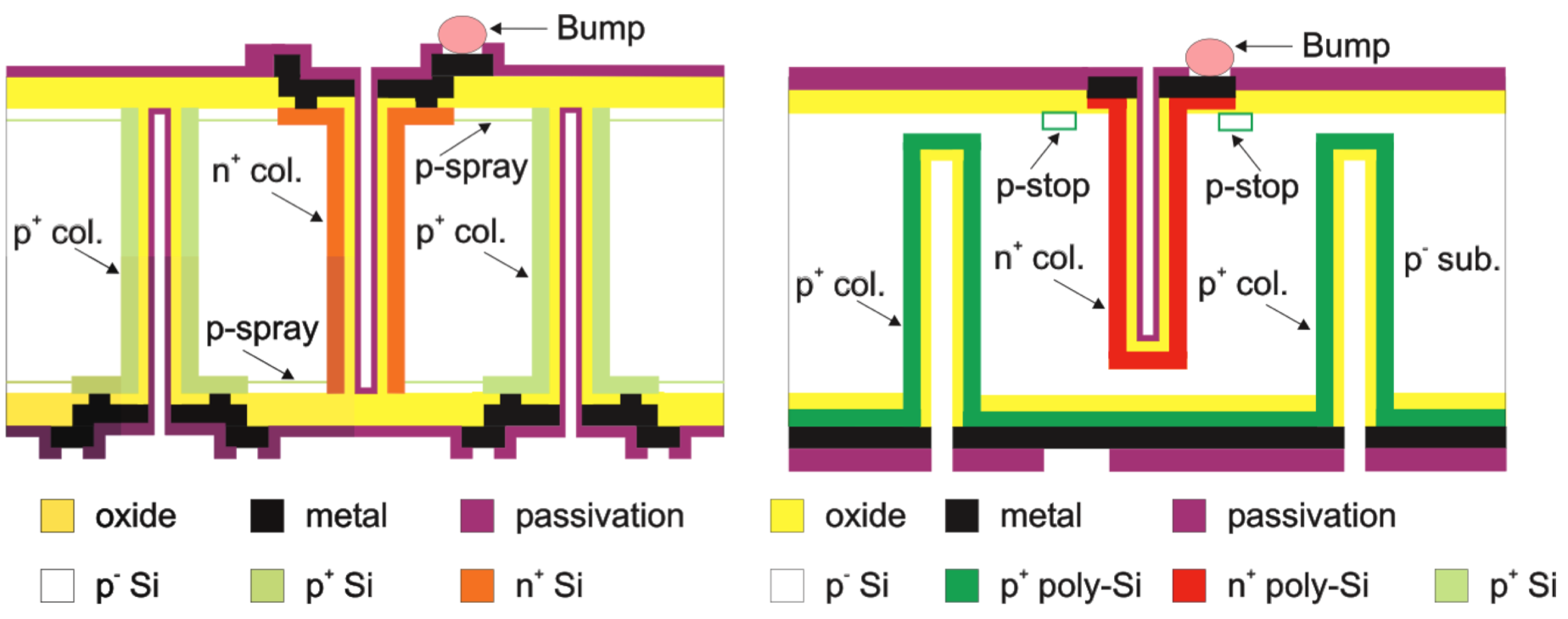}
\caption{Schematic cross-section of the 3D detector with passing-through columns from FBK (left) and with partial columns from CNM (right) fabricated on a $p$-type substrate. From Ref.~\cite{ATLAS-IBL-3D}.}
\label{fig:IBL3DSensors}
\end{figure}

\begin{figure}[!htb]
\centering
\includegraphics[width=1.\textwidth]{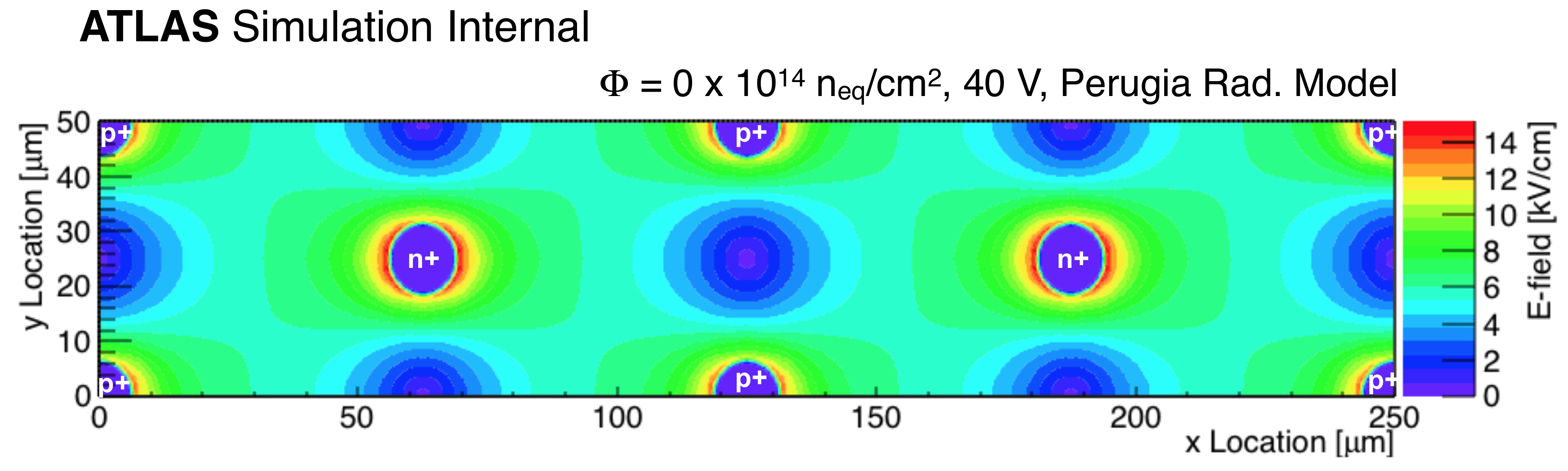}
\caption{Electric field in a IBL 3D sensors. From Ref \cite{Aaboud:2019wgd}.}
\label{fig:IBL3DEfields}
\end{figure}

New technologies have been developed to sustain the high dose of radiation that the pixel has receive. 3D sensors \cite{PARKER1997328} have been developed to sustain these problems, and also to keep low power consumption even after irradiation.\\
Fig.~\ref{fig:IBL3DSensors} shows the schematics of 3D sensors built by two main manufacturers.
3D sensors are silicon sensors with electrodes that are columns that penetrate the bulk, reducing the drifting path while keeping the same signal size. This means that the electric field is parallel to the surface instead of perpendicular. Each column is $\sim12\, \mu$m wide and closer to each other, dramatically reducing the depletion voltage (it can be as low as 20 V) and charge collection distance, reducing the probability of charge trapping. The low depletion voltage also implies a low leakage current, therefore requiring less cooling. Signal size is still determined by the thickness of the sensor, meaning signals with more similar amplitudes than the planar sensors. However signal is much faster.\\
The 3D sensors in the IBL were produced with a double-sided technology: columns of electrodes were implanted on both sides in a $p$-bulk sensor. Sensor bias is applied on the backside ($p^+$) as in planar sensors. Each pixel has two readout columns ($n^+$) with an inter-electrode spacing between $n^+$ and $p^+$ columns of 67 $\mu$m. Fig.~\ref{fig:IBL3DEfields} shows the electric field in a 3D pixel grid.

\subsection{Electronics}
\begin{figure}[!htb]
\centering
\includegraphics[width=0.9\textwidth]{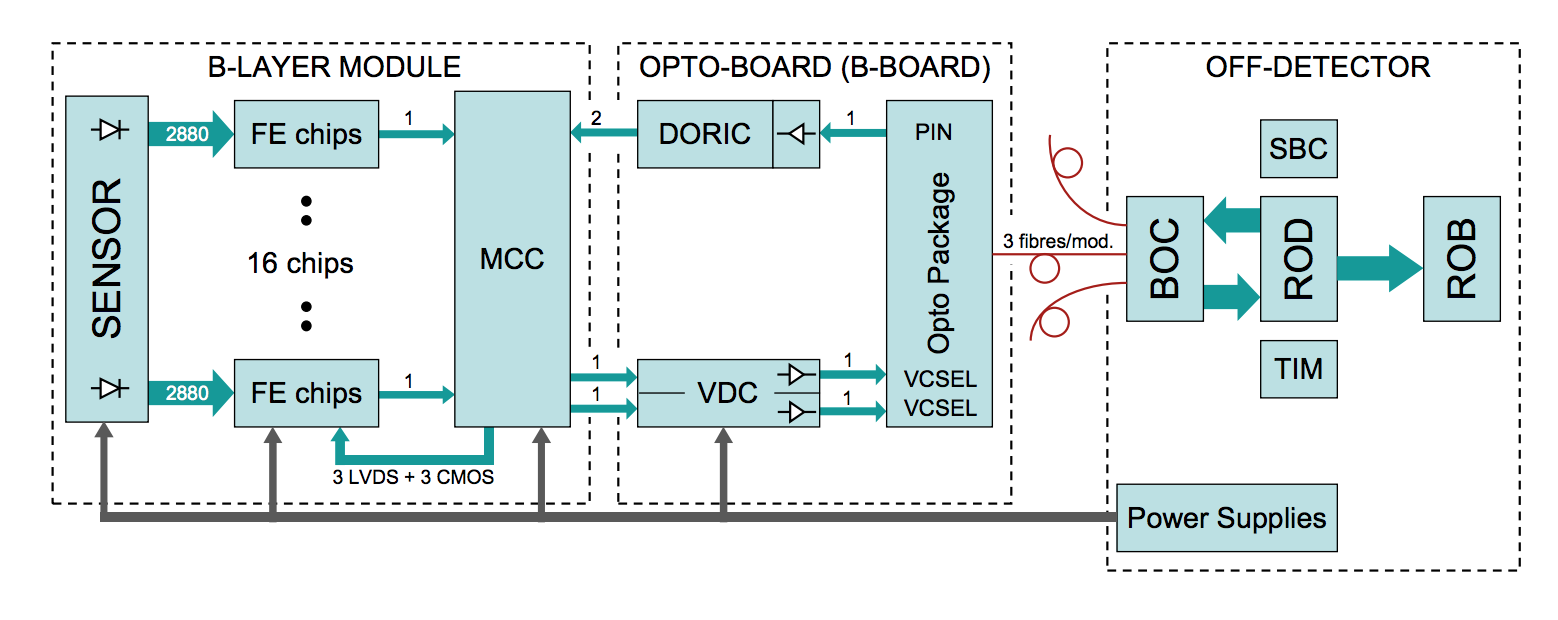}
\caption{Pixel Electronics schematics.}
\label{fig:PixElectrSchem}
\end{figure}

A schematic of the electronics in the Pixel layers is shown in Fig. \ref{fig:PixElectrSchem}. For each module, there are 16 Front-End (FE) chips, FE-I3, arranged in two rows of 8 chips each. IBL, instead, uses FE-I4B. The read-out of the chip is done by the Module Control Chip (MCC), and data are transferred between chips and MCC through Low Voltage Differential Signaling (LVDS) serial links. Modules are then connected with optical fibre links (opto-links) to the Read-Out Drivers (RODs) of the off-detector. The power supply is provided from a single DC supply over long cables, requiring low-voltage regulators boards.

\subsubsection*{FE chips}
There are two types of FE chips in the ATLAS pixel detector: FE-I4B for the IBL modules and FE-I3 for the modules of the other layers. FE-I3 readout chip contains 2880 pixel cells of $50\times 400\;\mu $m$^2$ arranged in a $18\time 160$ matrix, while FE-I4B contains 26880 pixel cells of $50\times 250\;\mu $m$^2$ ordered in a $336 \times 80$ matrix. Each cell contains an analog block where the signal is amplified. The digital readout saves a timestamp at the Leading Edge (LE) and at the Trailing Edge (TE) and sent it to the buffers and uses the difference to evaluate the ToT. This is proportional to the signal charge and can therefore be used to estimate amplitude.\\
Information can be stored in the chip for latency up to 255 LHC clock cycles of 25 ns, and it is saved if a trigger is supplied within this time. FE-I3 chips have a readout in 8 bits, giving ToT signals from 0 to 255, while FE-I4B chips have 4 bit readout, giving a range in ToT from 0 to 15.

\subsubsection*{MCC}
The MCC works on three different aspects: first, it loads the parameters and configuration settings into the FE chips, it distributes timing signal, L1 triggers and resets, then it reads out the FE chips. MCC must be set up at the beginning of each run of data taking, and when a L1 trigger command arrives at the MCC, a trigger is sent to the FE, as long as there are less than 16 events stored, otherwise the event is not saved. Information is sent back with also the number of missed events to keep the synchronization.

\subsubsection*{Opto-links}
Opto-links make possible communication between the modules of the detector and the off-detector electronics. This is made with optical fibres designed to have electrical decoupling and to minimize the material budget. The two main components are the opto-boards, mounted on the module side, and the Back of Crate Card (BCC), on the off-detector side. Signal transmission from the modules to the opto-boards is made with LVDS electrical connections. Readout bandwidth depends on the LHC instantaneous luminosity, the L1 trigger, and the distance from the interaction point. Electrical-to-optical conversion happens on the opto-board. The bandwidth of the opto-link has been improved during Run~2 to keep up with the increase in luminosity.

\subsubsection*{TDAQ}
The Pixel ROD is a 9U-VME module whose purpose is to transfer the data from the detector electronics to the ROS system. ROD modules are organized in 9 crates, each one can contain 16 ROD for a total of 14 RODs modules for the IBL (one crate), 44 modules (three crates) for B-Layer, 38 modules for Layer-1, 28 modules for Layer-2 (together in four crates), and 24 modules (two crates) for the disk. In the crates is also present the interface with the trigger system. Data are routed directly from the ROD to the ROS with optical links. Commands, trigger and clock are transmitted with one down link, while event readout goes through another one (or two) up-link. The readout is done with \say{data-push}, meaning that when buffers are full there is no mechanism to stop transmission (busy). This means also that each component of the chain (from FE to ROD) always transmits at the maximum rate. Each step also monitors the number of events received and the triggers sent, if these are different by a certain amount, triggers are blocked and empty events are generated.

\subsection{Calibration}
\label{ssec:Calibration}
Pixel calibration is needed to convert the ToT (the output of the sensors) into charge. Three consecutive steps are needed: calibration of the time walk (assigning the event to the correct bunch crossing), threshold calibration, and finally the ToT-to-charge conversion.\\

\subsubsection*{Time Walk}
\begin{figure}[!htb]
\centering
\subfloat{\includegraphics[width=0.45\textwidth]{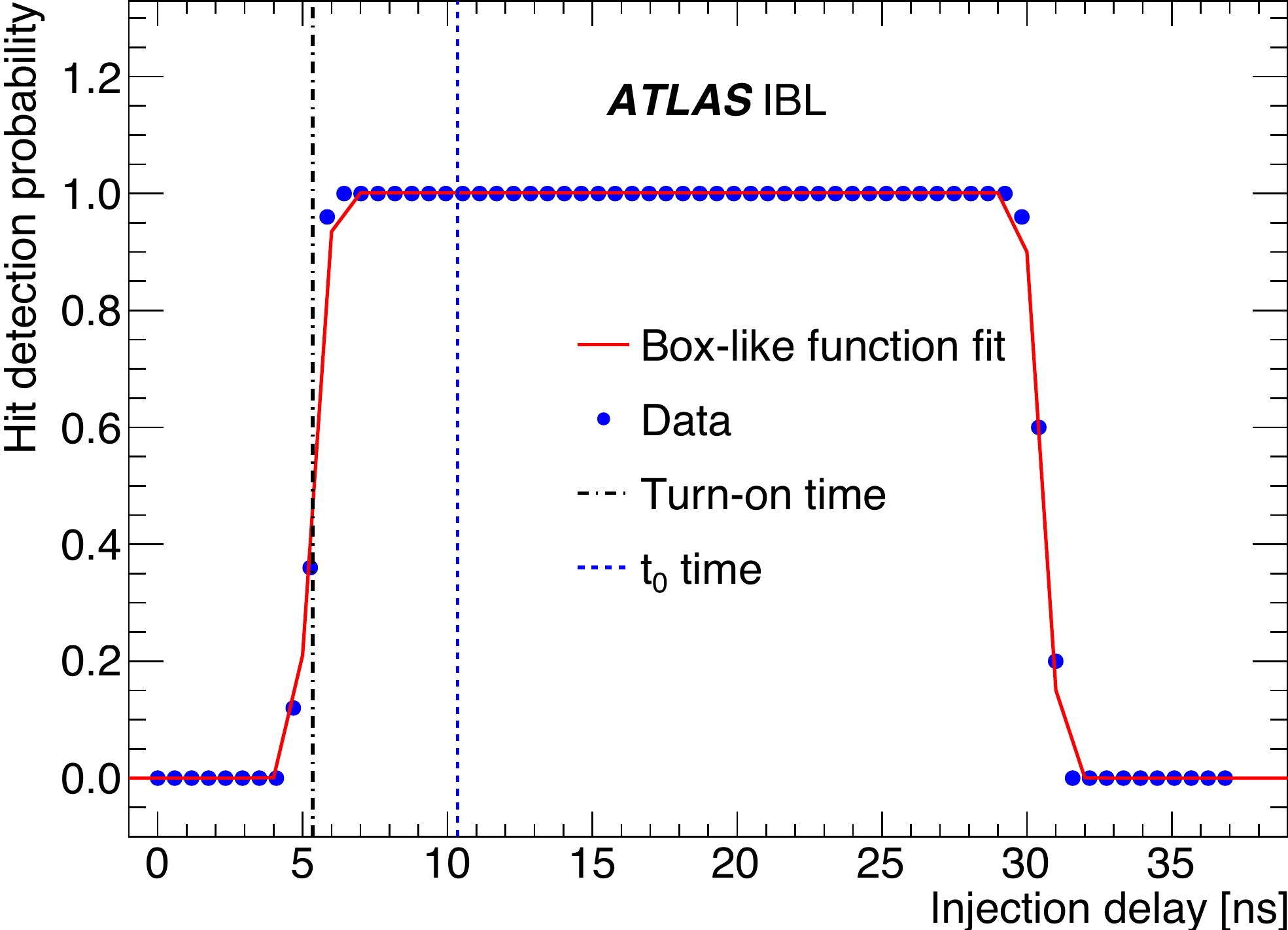}}
\subfloat{\includegraphics[width=0.45\textwidth]{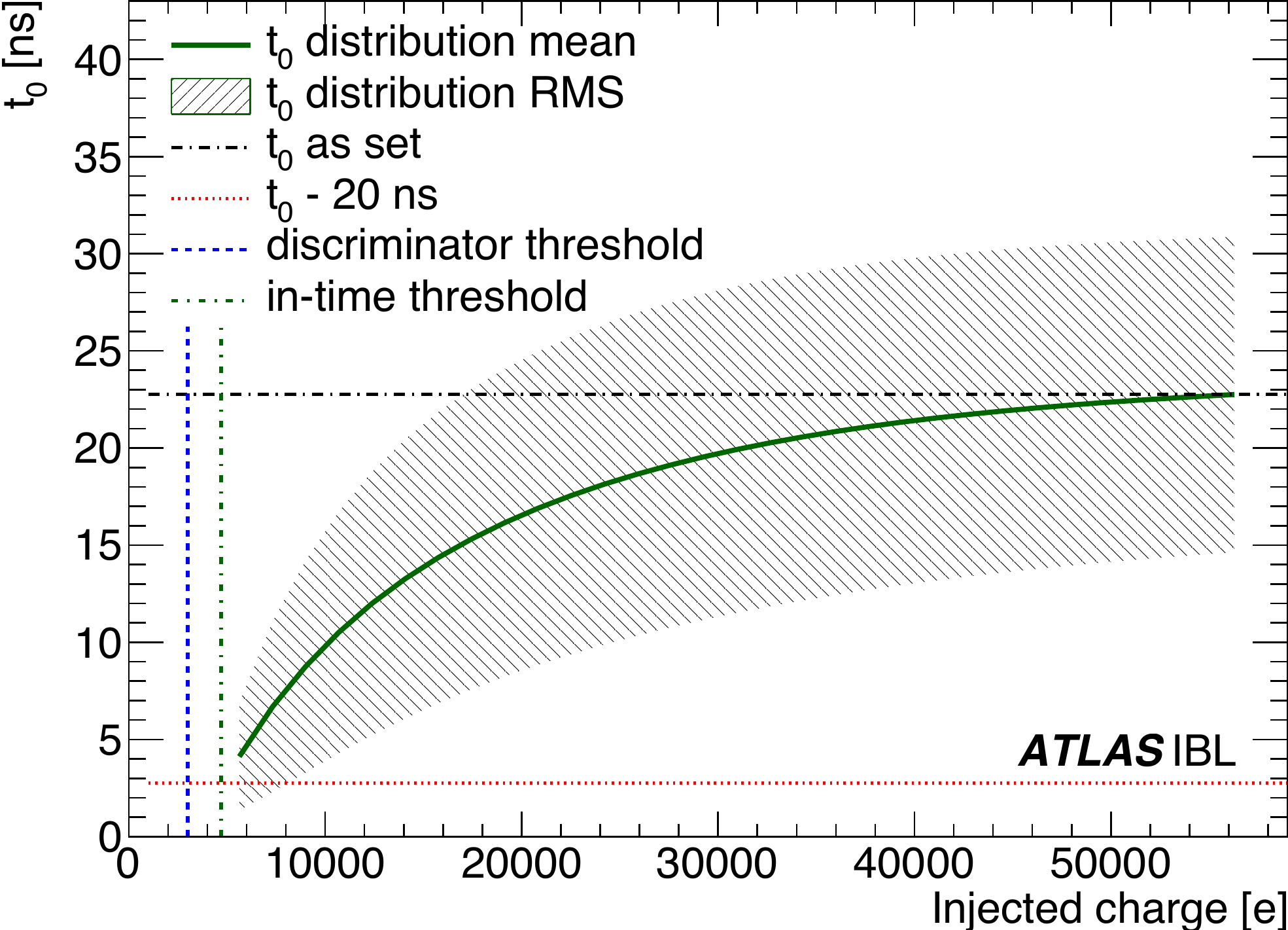}}
\caption{Single pixel hit detection probability for one pixel during $t_0$ scan with an injected charge of 10 $\mathrm{ke^{-}}$ (left) and distribution of the measured $t_0$ as a function of the injected charge for an entire chip (right). The solid line corresponds to the average of all measurements. From Ref. \cite{Abbott:2018ikt}.}
\label{fig:PixTimeWalk}
\end{figure}

Hits are digitized by the pixel modules relative to the master clock, which is also synchronized to the LHC clock. Hits are saved only if they happen within one clock cycle, meaning in a time interval of 25 ns. It can happen that small signal charges when passing through the amplifier cross the discriminator threshold with a time delay with respect to the signal reference that might be a large signal. This small signal charge will then be assigned to the wrong bunch crossing. This effect is the time walk. To estimate the time walk it is therefore needed to measure the difference in time between when the signal charge arrives at the input of the amplifier and the time when amplifier output crosses the discriminator threshold. This is measured by injecting a known charge directly from the FE chip with an adjustable delay, that allows changing globally the injection time with respect to the chip master clock. In this way, it is possible to decrease the time difference between the charge injection and the digitization window. A scan of the delay is performed and the hit detection probability is measured. The $t_0$ time is defined as the time for a 50\% hit detection probability plus a safety margin of 5 ns. A small $t_0$ means that the time between the charge injection and the digitization time window is larger. Fig.~\ref{fig:PixTimeWalk} (left) shows the hit detection probability for one pixel as a function of the delay time for a fixed charge of 10 k$e^-$.
Fig.~\ref{fig:PixTimeWalk} (right) shows the distribution of $t_0$ for all the pixel arrays as a function of injected charge. Time walk is visible for small charges, where $t_0$ is small, meaning that there is a large difference in time. To recover hits of small charges being out of time, all hits below a certain ToT value are duplicated in the previous bunch crossing in the pixel FE-I3 while in the FE-I4B hits with a ToT of 1 or 2 are duplicated in the previous bunch crossing only if they are near-by a larger hit (ToT$>3$). This is because small charge hits are usually due to charge sharing among pixels.

\subsubsection*{Threshold}
\begin{figure}[!htb]
\centering
\includegraphics[width=1.\textwidth]{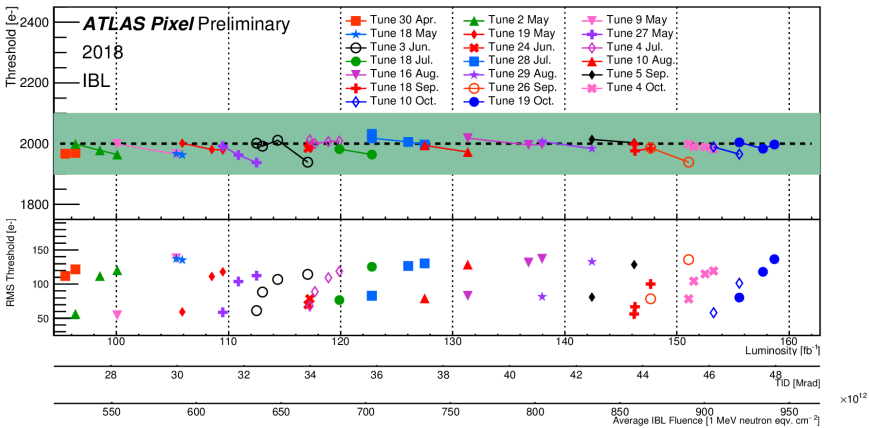}
\caption{The evolution of the measured threshold over all pixels in the IBL detector as a function of the integrated luminosity, the corresponding TID, and the average IBL fluence as measured in calibration scans estimated considering the data-taking periods from 2015 to 2018. The threshold was tuned to 2000 electrons. Each colour/symbol series corresponds to a single tuning of the detector. The shade indicates $\pm 100$ electrons of the tuning point. From Ref.~\cite{IBLCalibrationPlots2016}.}
\label{fig:ThresholdShift}
\end{figure}

Signal charge is saved when it passes a module threshold. This threshold has to be measured and it is done by injecting a known charge in the FE and measuring where there is a 50\% hit efficiency, both at the global and the pixel level. Tuning of the threshold has been done constantly during Run~2 operations. In fact, due to radiation damage effects on the electronics, the actual threshold drifts away from the nominal value. This is possible to be seen in Fig.~\ref{fig:ThresholdShift}, where it is shown the measured threshold over all the pixels in IBL as a function of integrated luminosity and the corresponding Total Ionizing Dose (TID) of radiation sustained. Radiation effects caused the measured threshold to drift and its standard deviation to increase with integrated luminosity, but regular re-tunings brought the mean threshold back to the tuning point and the reduction of root mean square (r.m.s.).

\subsubsection*{ToT Tuning and Calibration} 
\begin{figure}[!htb]
\centering
\subfloat{\includegraphics[width=0.45\textwidth]{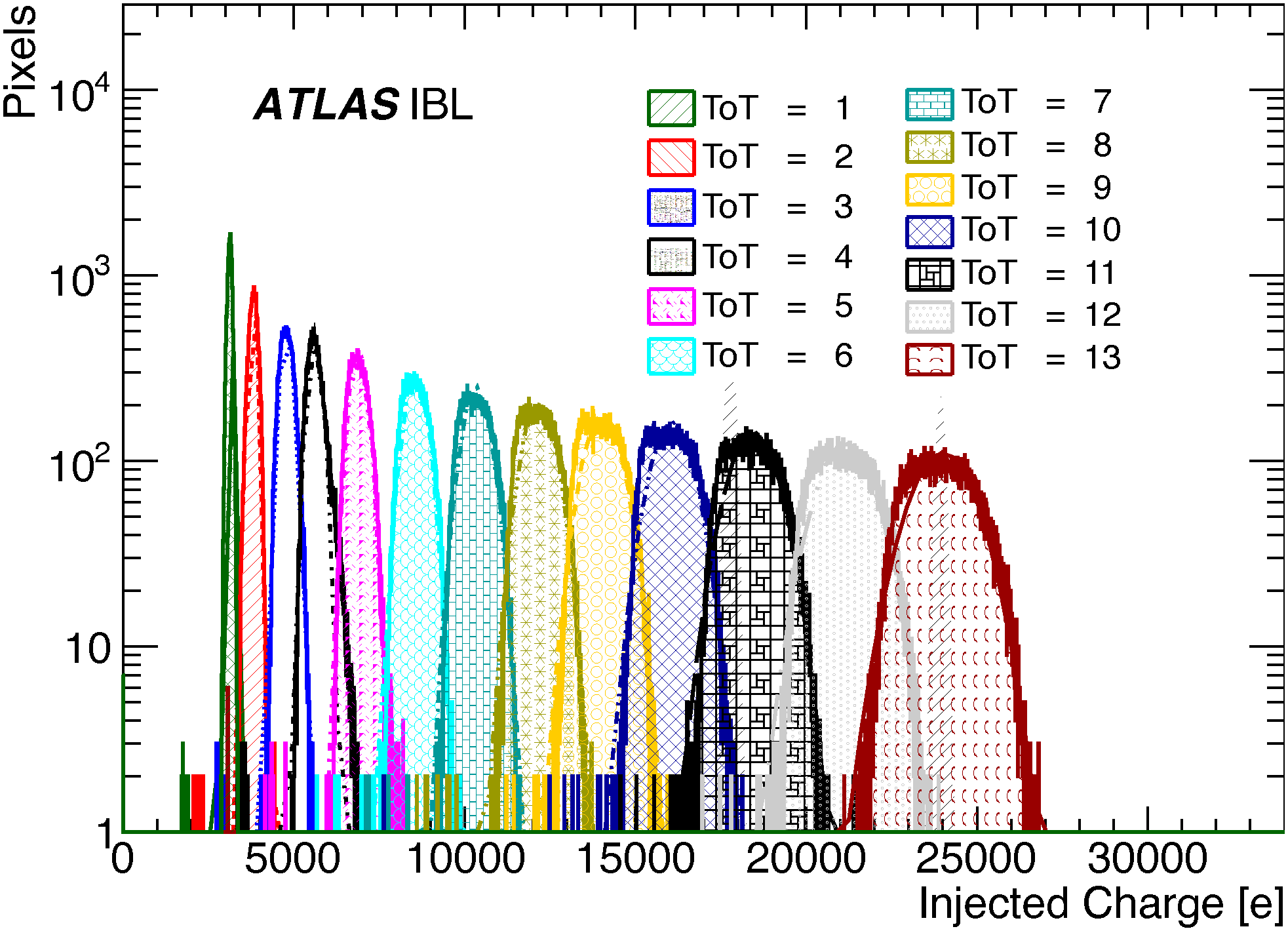}}
\subfloat{\includegraphics[width=0.45\textwidth]{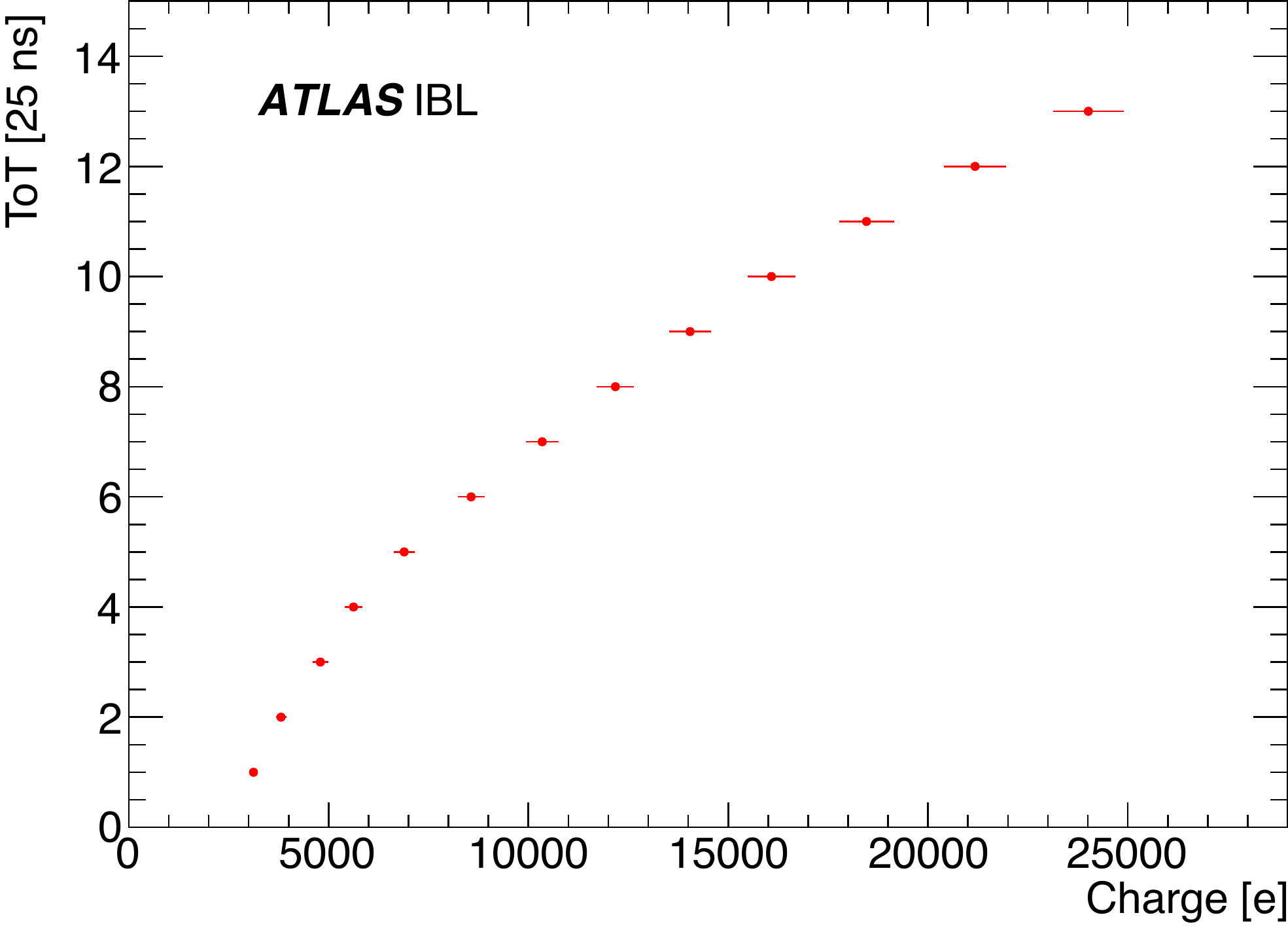}}
\caption{Number of pixels with a certain ToT given a certain charge injected (left) and charge to ToT calibration curve (right). From Ref.~\cite{Abbott:2018ikt}.}
\label{fig:ToTResponseIBL}
\end{figure}

The ToT is the final output of the digitizer. It is an integer number ranging from 0 to 15 for IBL and from 0 to 255 for the other pixel layers. ToT must be first tuned at some value to have a coherent response between all the pixels, then it must be calibrated locally to give the same results for all the signal charges above the threshold. Calibration is done using a fit function and a tuning of the threshold is repeated after the ToT one as the latter changes the threshold.\\
Tuning is done by injecting a known charge and asking that the output is a given value of ToT. For the three outermost pixel layers, a 20 $\mathrm{ke^{-}}$ charge is injected and ToT is set to 30. Tuning is done in three steps: first, an algorithm tunes the average ToT for each FE chip; then, another algorithm is used to tune the ToT at pixel level, while keeping the same average ToT for each FE chip, allowing to reduce the r.m.s. of all the responses; finally, any badly tuned FE chip is identified and the tuning procedure repeated.\\
After tuning, the response is the same for all the pixels, but only for the value of the injected charge. In general, the response to the charge of the sensor is not linear and the ToT needs to be calibrated with the charge. The response is almost linear around the tuning point, but it is more quadratic at low ToT and reaches a plateau at high ToT. Calibration is done at module level. A fit function is used to have a map of the values, and the ToT is the obtained as
\begin{equation}
\mathrm{ToT} = a_0 \frac{a_1+ Q}{a_2 + Q}
\end{equation}
where $Q$ is the charge, and $a_0$, $a_1$, and $a_2$ are the fit parameters. The procedure starts by injecting charge and dividing the pixels by the response in ToT, as shown in Fig.~\ref{fig:ToTResponseIBL} (left). The mean of each peak is then reported and an error is assigned as the width of the distribution. These values are then graphed as shown in Fig.~\ref{fig:ToTResponseIBL} (right) and used to fit the function and obtain the parameters $a_0$, $a_1$, and $a_2$.

\section{Radiation Damage Simulation}
\label{sec:RadDam}
As the detector component closest to the interaction point, the ATLAS Pixel detector has been exposed to a significant amount of radiation over its lifetime. Radiation creates defects in the sensors and, over time, reduces their efficiency, causing a degradation of the performance of the whole detector to reconstruct physical quantities. Starting from Run~3, the ATLAS simulations will take into account the effects of the radiation damage to the pixel detector in reconstruction. To this end, a \textit{digitizer} has been implemented in the digitization step of the reconstruction, e.g. the step where the energy deposits from Geant4 are converted into digital signals. In the digitizer, the charge collected converted is reduced to account for the loss of efficiency.\\
In Section~\ref{ssection:RadDam} it is discussed how radiation damage degrades silicon detectors and what are the consequences on the performance of a detector. In Section~\ref{ssection:ATLASRadDam}, the current status at the end of Run~2 of the ATLAS pixel detector and its level of damage sustained is presented, while the description of the digitizer and how it was implemented is described in Section~\ref{ssection:DigiModel} and its predictions are compared to data in Section~\ref{ssection:ModelValidation}.\\

\subsection{Radiation Damage effects}
\label{ssection:RadDam}
Radiation passing through the detector damages the sensors. These effects are caused by the appearance of defects in either the bulk structure, where crystal atoms are displaced, or the surface \cite{PixelDetectorsBook2006}. Surface defects imply an increase of the interface oxide region charge, which saturates after $\sim$ kGy of the ionizing dose. The macroscopic effects due to damage in the bulk are: increase of leakage current, change in the depletion voltage and charge trapping. A brief description of the nature of the effects due to bulk damage will be presented here.

\subsubsection{Microscopic Nature of Radiation Damage}
\begin{figure}[!htb]
\centering
\includegraphics[width=0.7\textwidth]{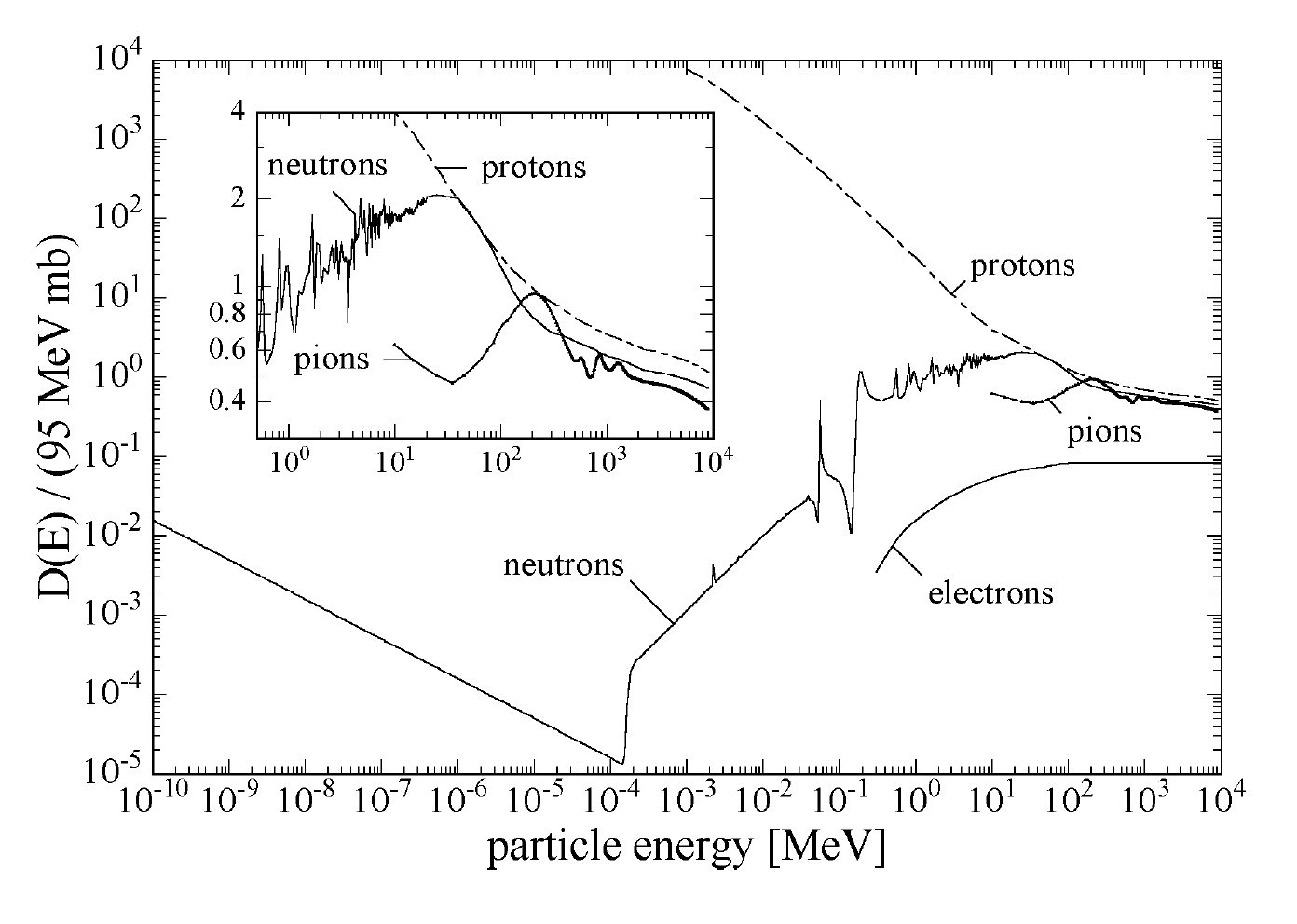}
\caption{Displacement Damage function ($D(E)$) normalized to 95 MeV mb, for neutrons, protons, pions, and electrons. Because of the normalization to 95 MeV mb the ordinate axis represents the damage equivalent to 1 MeV neutrons. The insert is a zoom. From Ref.~\cite{Moll:1999kv}.}
\label{fig:ConversionNIEL}
\end{figure}

An initial concentration of defects is first introduced in the crystal of the silicon sensors depending on the purity of the initial wafer. These defects introduce localized energy levels that can, if within the forbidden band gap, change the electrical characteristic of the bulk. If a particle with a high enough energy (around $\mathcal{O}(10)$ keV) \cite{Moll:1999kv} crosses the detector, it may collide with an atom of the lattice and, if there is enough energy, it will remove it from its position, leaving an empty space called \textit{vacancy}, while the atom will end on a position non in the lattice, called \textit{interstitial} defect. This pair of displaced atom-hole is called a \textit{Frenkel Pair}. If the atom itself has enough energy it might release energy by ionization first, and then by nuclear interaction, generating more defects. If many Frenkel pairs are grouped together they can form clusters \cite{LINDSTROM2001308}.\\
Different particles interact in different ways with the lattice of the sensor: charged particles tend to scatter via electromagnetic interaction with the atoms, while neutrons interact directly with the nuclei. In order to compare the damage from different types of particles with different energies, radiation damage is scaled to the \textit{Non-Ionizing Energy Loss} (NIEL), which is the component that causes the damages to the lattice. In semiconductor \textit{ionizing} damage is generally fully recovered. Particles can lose energy with both non-ionizing and ionizing processes, in different ratios depending on the type of radiation involved. NIEL then summarizes the energy deposited from only the processes that cause non-reversible damage on the lattice of the sensor, using as reference 1 MeV neutrons. In this way a fluence $\phi_{\text{phys}}$ from an arbitrary particles is equivalent to the fluence $\phi_{\text{eq}}$ of a 1 MeV neutron. The conversion factor $k$ (called also \textit{hardness factor}) between $\phi_{\text{phys}}$  and $\phi_{\text{eq}}$ must then be calculated for each specific particle and energy, and are provided as look-up table \cite{Vasilescu:1997}. Fig.~\ref{fig:ConversionNIEL} shows the displacement damage function $D(E)$, normalized to 95 MeV mb, which correspond to the $D(E)$ value for 1 MeV neutron, because of this the ordinate axis actually represent the damage equivalent to 1 MeV neutrons.

Silicon interstitial, vacancy, and primary defects can move inside the crystal (they are not stable), and if they meet inside the crystal this can lead to the creation of a cluster defect or secondary defects. This process of travelling and combination is called \textit{defects annealing} and impacts the energy levels in the band gaps and the space charge in the depletion zone. The mobility of the defects is strongly dependent on the temperature, and therefore the changes in the detector will have a complex annealing behaviour due to the many possible secondary defects.\\

\subsubsection{Leakage Current}
\begin{figure}[!htb]
\centering
\includegraphics[width=0.7\textwidth]{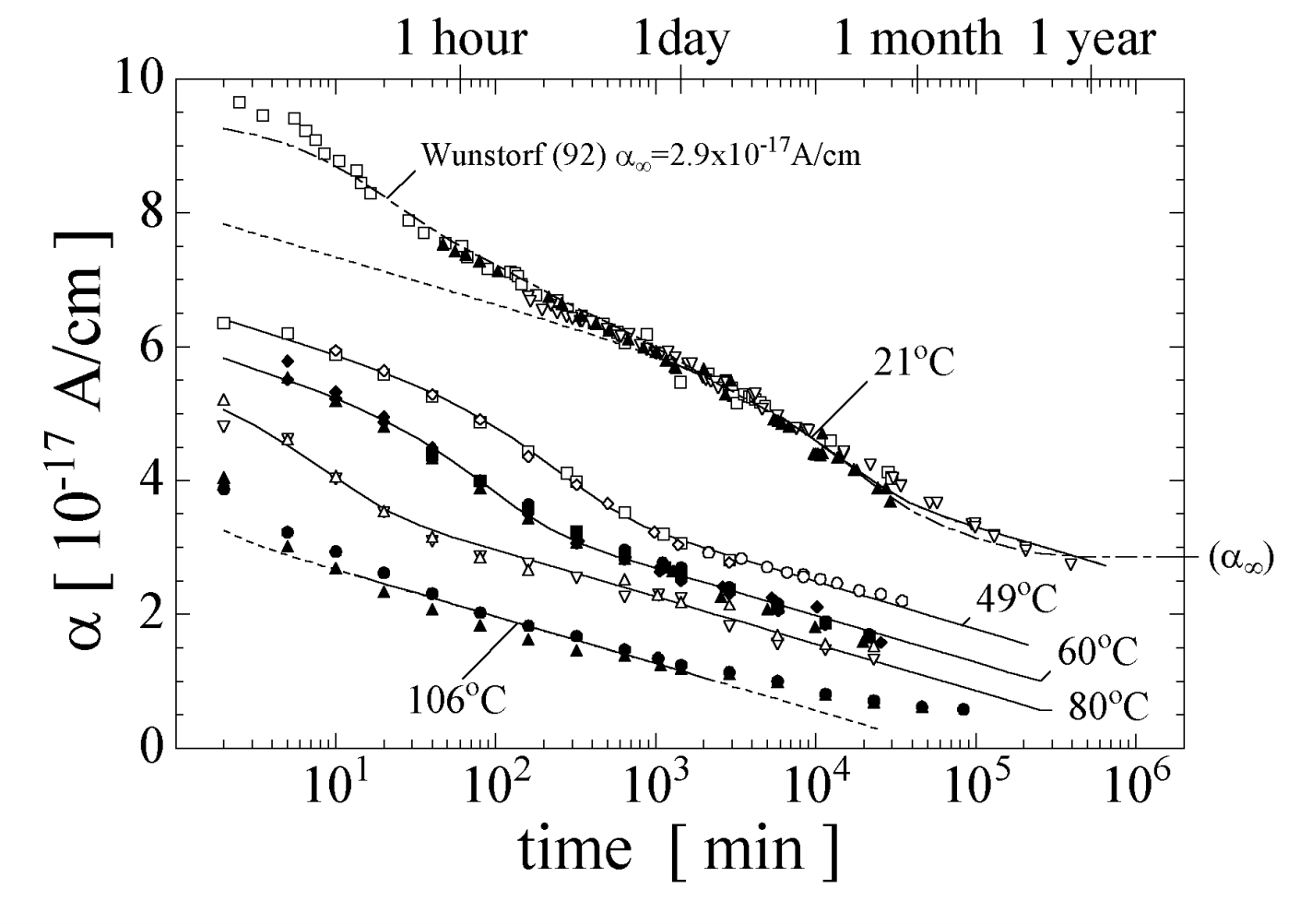}
\caption{Rate of increase of the leakage current $\alpha$ as a function of the annealing time. Top and bottom axis represent time, the bottom one in minutes, while the top one shows the  time but in hours/days/months/years. From Ref.~\cite{Moll:1999kv}.}
\label{fig:LeakCurrentAlpha}
\end{figure}

The presence of energy levels in the band gap caused by defects in the crystal helps generating additional electron-hole pairs within the depleted region of the sensor. This leads to a decrease of the generation lifetime $\tau_g$ and an increase of the volume generation current $I_{\text{vol}}$ proportional to the fluence $\phi$:
\begin{linenomath*}
\begin{align}
\frac{1}{\tau_g}  = &  \frac{1}{\tau_{g,\phi=0}}+k_{\tau}\phi\\
\frac{I_{\text{vol}}}{V} = & \frac{I_{\text{vol},\phi=0}}{V}  + \alpha\cdot\phi
\label{Eq:LeakageCurrent}
\end{align}
\end{linenomath*}

with $V$ being the volume, defined as junction area times detector thickness. Instead $k_{\tau}$ is the \textit{life time related damage rate} and $\alpha$ the \textit{current related damage rate}. An increase of the leakage current is a problem because it increases the noise and it heats up the sensor, therefore needing a cooling system to avoid a thermal runaway.\\
The two constants can also be related between them by the relation: $\alpha = en_i k_{\tau}$, where $n_i$ is \textit{intrinsic carrier concentration}. Also, it is possible to reformulate as $\alpha = \Delta  I_{\text{vol}} / \phi\cdot V$.\\
It is important to note that $\alpha$ is independent of the initial resistivity of the silicon and the production method of the sensor.\\
Fig.~\ref{fig:LeakCurrentAlpha} shows that with time, after irradiation, the leakage current will anneal and that this process is strongly dependent on the temperature \cite{Chilingarov_2013}.

For long annealing times or high annealing temperature, it is possible to parametrize the evolution of $\alpha$ with a function such as:
\begin{linenomath*}
\begin{equation}
\alpha (t) = \alpha_i \cdot \exp\left( -\frac{t}{\tau_i}\right)   + \alpha_0 -\beta\cdot \ln \left( \frac{t}{t_0}\right)
\label{eq:evLeakCurrent}
\end{equation}	
\end{linenomath*}

with $t_0$ an arbitrary time (e.g. set at 1 min), and $\tau_i$ hides the dependence on the temperature in the following way:
\begin{linenomath*}
\begin{equation}
\frac{1}{\tau_i} =  k_{0,i}\cdot \exp \left( -\frac{E_i}{kT_a}\right).
\end{equation}	
\end{linenomath*}
where $E_i$ is a parameter set to $E_i=(1.11\pm0.05) $ eV. It is also worth noticing that $T_a$ is evaluated at the temperature at which the sensor was annealed, and not the current one.

\subsubsection{Effective doping concentration}
\label{subsec:EffDop}

\begin{figure}[!htb]
\centering
\includegraphics[width=0.7\textwidth]{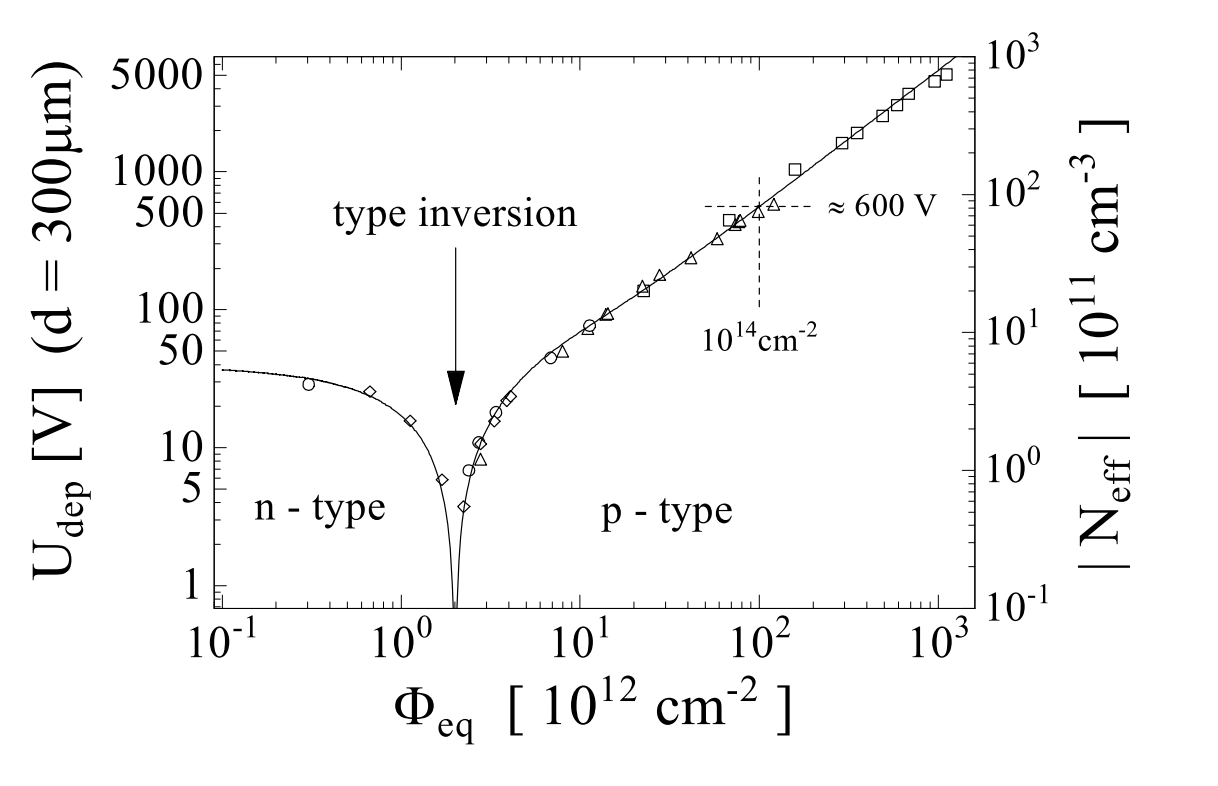}
\caption{Evolution of the depletion voltage for a 300 $\mu$m thick silicon sensor as a function of the fluence. From \cite{Moll:1999kv}. }
\label{fig:VDepFluence}
\end{figure}

A sensor is considered fully depleted when there are virtually no free carriers in the bulk, and there is an electric field that collects the charge created by ionizing particles passing through it.\\
For biases below the depletion voltage $V_{\mathrm{depl}}$, charges are recombined inside the non-depleted region and can't reach the electrode. This means that only the depleted part of the sensor is sensible and, if this is on the side of the electrode, the sensor works as if it was thinner. \\
The depletion voltage is also related to the \textit{net doping} or \textit{effective doping} $N_{\mathrm{eff}}$, which is the difference of all donor-like states and all acceptor-like states, via the following formula:
\begin{linenomath*}
\begin{equation}
|N_{\mathrm{eff}}|  = \frac{2\epsilon_0\epsilon_{\mathrm{Si}}V_{\mathrm{depl}}}{ed^2}
\label{eq:VdepNeff}
\end{equation}
\end{linenomath*}
where $d$ is the depth of the sensor. $N_{\mathrm{eff}}$ can be both negative or positive, depending on whether acceptors or donors are dominating. In general, defects caused by irradiation are responsible for a change in the $N_{\mathrm{eff}}$. This is due to the removal or formation of acceptor or donor, caused by the formation of either defects cluster containing acceptor or donor, or cluster assuming positive/negative charge states in the space-charge region. When irradiated, a $n$-doped material will decrease its $N_{\mathrm{eff}}$ up until a certain fluence, depending on the initial concentration, where the material becomes intrinsic. The material then with increasing the dose increases the absolute value of $N_{\mathrm{eff}}$, dominated by acceptor-like defects with a negative space charge, showing the behaviours of a $p$-type material \cite{Lutz:411172}. This change in nature is called \textit{type inversion}. Fig.~\ref{fig:VDepFluence} shows the evolution of the depletion voltage for a 300 $\mu$m thick silicon sensor as a function of the fluence. This cause the shift of the $n$ side of the sensor to the $p$ side.

The \textit{Hamburg Model} gives an empirical expression of the evolution of $N_{\mathrm{eff}}$ after irradiation,
\begin{linenomath*}
\begin{equation}
N_{\mathrm{eff}}  = N_{\mathrm{eff},\phi=0} - \left(  N_C(\phi) + N_a(\phi,T_a,t) + N_Y(\phi,T_a,t) \right),
\label{eq:HamburgModel}
\end{equation}
\end{linenomath*}
where:
\begin{itemize}
    \item $N_C$ describes the \textit{stable damage} that does not depend on time or temperature. This term has an exponential dependency on the fluence, $N_{C}(\phi)  = N_{C,\phi=0}  \left( \-e^{-e\phi} \right) + g_e \phi$;
    \item $N_a$ describes the short-term or \textit{beneficial} annealing and is parametrised as $N_{a}(\phi)  = \phi \sum_i g_{a,i} e^{-t/\tau_{a,i}(T_a)} \sim \phi g_ae^{-t/\tau_a(T_a)}$;
    \item $N_Y$ describes the \textit{reverse annealing} which describes the increase of the full depletion voltage after a few weeks at room temperature. It can be parametrised as $N_Y  = g_Y \phi \left( 1- \frac{1}{1+t/\tau_Y} \right)$.
\end{itemize}

\subsubsection{Trapping}
\begin{figure}[!htb]
\centering
\includegraphics[width=0.7\textwidth]{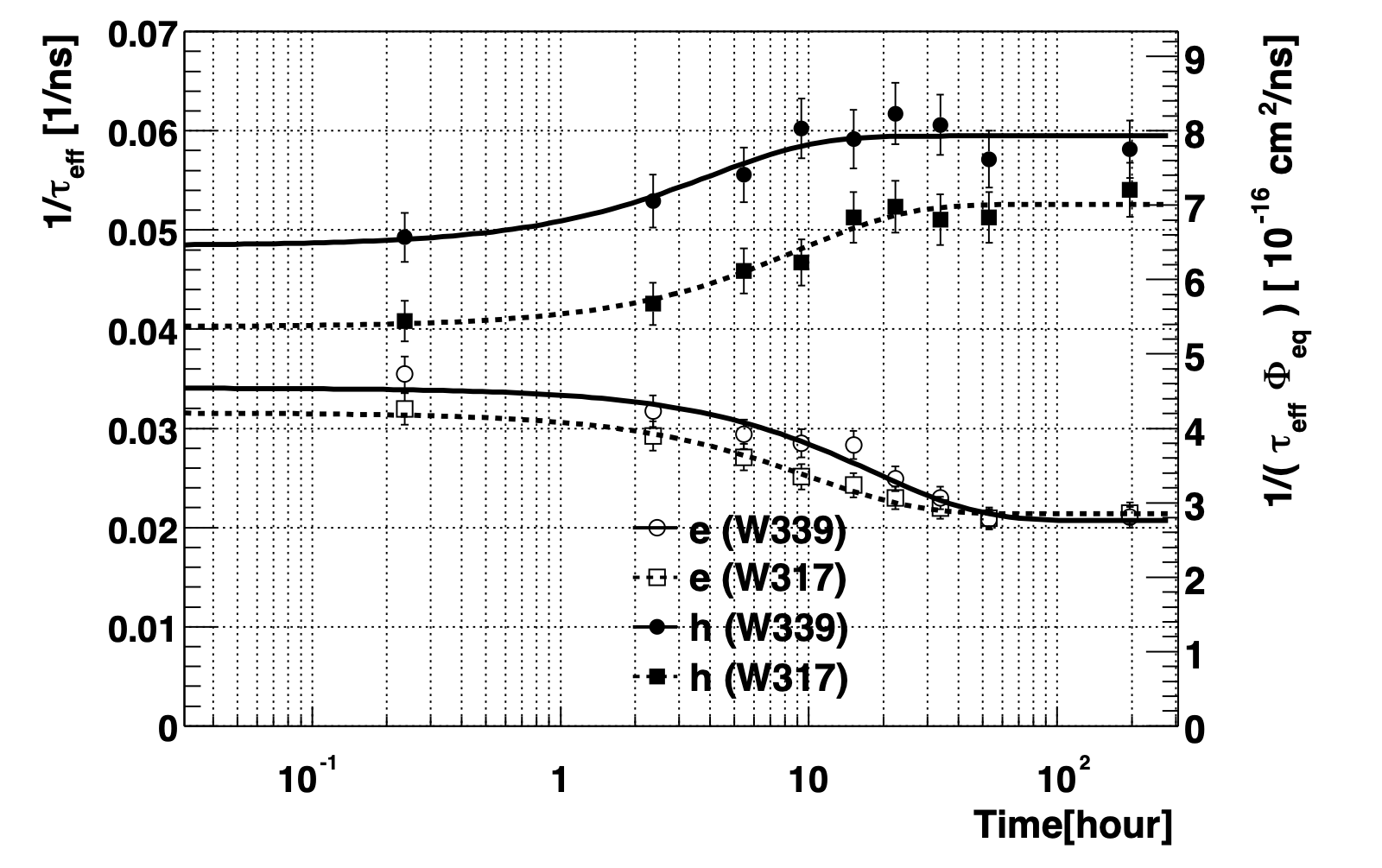}
\caption{Annealing of effective trapping probability at 60 °C. Measurements were taken at $T=10$ °C. From Ref.~\cite{Kramberger:2002zb}.}
\label{fig:Kramberger}
\end{figure}

Another important effect of radiation damage is the creation inside the sensor of trapping centres. Crystalline defects introduce localized energy levels in the bulk with high capture cross-section. A charge carrier trapped inside one of these levels has a re-emission time that is far larger than the charge collection time needed for a tracking detector, and therefore its signal is lost, reducing the total amplitude of the signal. \\
An important parameter is the \textit{trapping time} $\tau_i$, which describes the (inverse of the) probability of a charge to be trapped: 
\begin{linenomath*}
\begin{equation}
\frac{1}{\tau_t(\phi)} = \frac{1}{\tau_{t,\phi=0}} +\gamma\phi
\end{equation}
\end{linenomath*}
where $\gamma$ is a coefficient that has been measured and it is $0.41\times10^{-6}$cm$^2$s$^{-1}$ for electrons and  $0.60\times10^{-6}$cm$^2$s$^{-1}$ for holes. Smaller values of the trapping time for the electrons than the holes means also that holes are more likely to be trapped. This is why, in general, sensors that collect electrons instead of holes are more used in applications. These values however have a dependence on the annealing time, as found in Ref \cite{Kramberger:2002zb}. Fig.~\ref{fig:Kramberger} shows the trapping constant as a function of the annealing time for two different sensors using neutrons for irradiation and showing results for both electrons (empty marker) and holes (full marker).

\subsection{ATLAS Pixel Detector Conditions}
\label{ssection:ATLASRadDam}
\begin{figure}[!htb]
\centering
\begin{minipage}{0.54\textwidth}
\includegraphics[width=1.\linewidth]{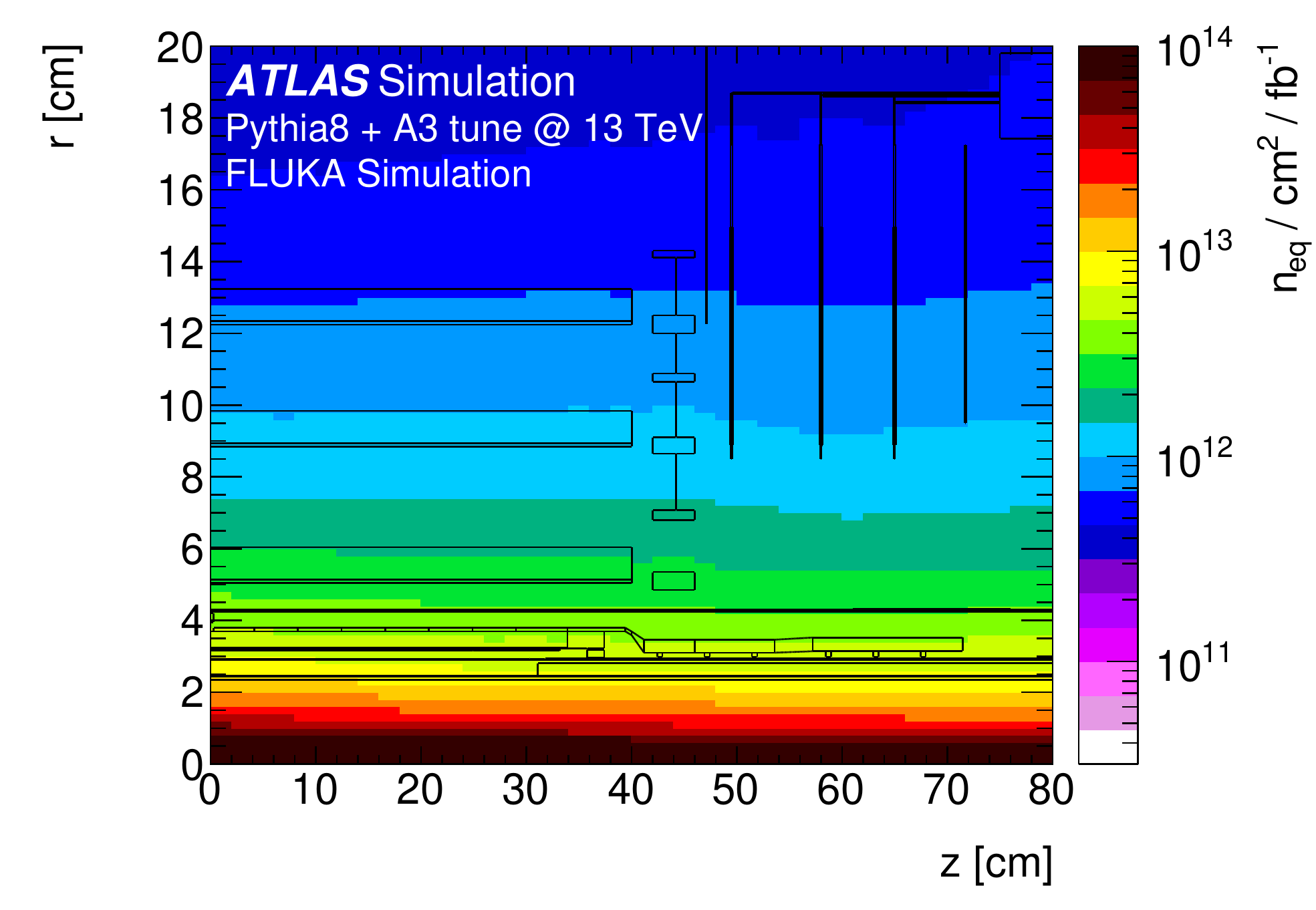}
\end{minipage}
\begin{minipage}{0.45\textwidth}
\includegraphics[width=1.1\linewidth]{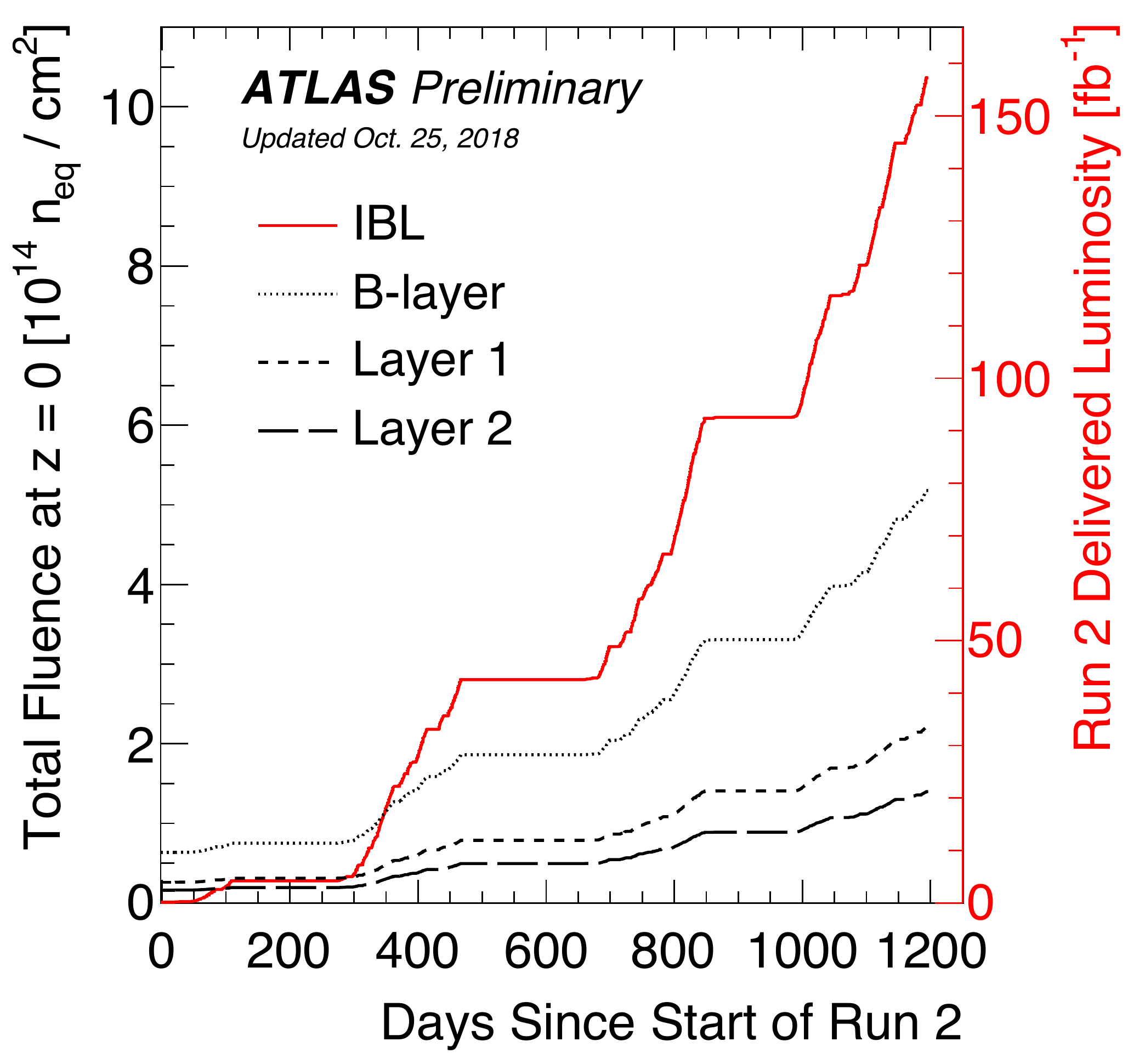}
\end{minipage}
\caption{Simulated 1 MeV $\mathrm{n_{eq}}$ fluence prediction in the Pixel detector as a function of $r$ and $z$ position using FLUKA for 1 $\mathrm{fb}^{-1}$ (left). Predictions for the lifetime fluence experienced by the four layers of the current ATLAS pixel detector as a function of time since the start of Run~2 at $z \sim 0$ up to the end of 2017 (right). For the IBL, the lifetime fluence is only due to Run~2 and for the other layers, the fluence includes all of Run~1. The IBL curve represents both the fluence on the IBL (left axis) as well as the delivered integrated luminosity in Run~2 (right axis). (left) from Ref.~\cite{RadiationFlukaSim}, (right) from Ref.~\cite{PixelLumiAndFLuence}.}
\label{fig:FluenceFLUKA}
\end{figure}

As it has been shown in the previous section, radiation-induced defects change the characteristic of the sensors, in particular the voltage needed to fully deplete the sensor. Predictions of the radiation fluence that will impact the detector are then important for the performance of the detector itself.\\
The estimate of the fluence depends on two different key aspects: first is the modelling of the secondary particles produced in the collisions, and second, their interactions with the detector. In ATLAS this estimation is done using a combination of different simulations. Pythia 8 \cite{Sjostrand:2014zea,Sjostrand:2006za} generates inelastic proton-proton scattering using the MSTW2008LO parton distribution with the tuned set of parameters A2 \cite{ATL-PHYS-PUB-2016-017}. The produced particles are then propagated through the detector using the particle transport software FLUKA \cite{doi:10.1063/1.2720455,Ferrari:898301}. Particles are transported down to an energy of 30 keV for photons, thermal energy for neutrons, and 100 keV for everything else.\\
It is very important to model correctly the geometry of the Inner Detector because of secondary particles generated in the high energy hadronic interactions in the detector. Fig.~\ref{fig:FluenceFLUKA} (left) shows the estimated 1 MeV neutron-equivalence fluence per $\mathrm{fb^{-1}}$, while Fig.~\ref{fig:FluenceFLUKA} (right) shows the same information but as a function of time, divided for the 4 layers of the pixel detectors. The most important contribution comes from charged pions coming directly from the proton-proton collision. As it is possible to see, there is a $z$ dependence that can be as high as 10\%. From Fig.~\ref{fig:FluenceFLUKA} (right) instead is possible to notice how, even if the IBL was installed after the other pixel layers, it has received more fluence, due to its proximity to the interaction point.

\subsubsection{Luminosity to fluence validation}
Fluence is also an important input for the simulation. To compare simulation to data, it is needed to know the correct fluence corresponding to the luminosity of the data sample. A conversion factor from luminosity to fluence is then needed. FLUKA is used to get the conversion factor and estimate the systematic effects. Fluence is then converted into leakage current. \\

The leakage current can be predicted using Eq.~(\ref{Eq:LeakageCurrent}) knowing the fluence and the temperature and the Hamburg model parametrization of $\alpha=\alpha(t,T)$ as shown in Eq.~(\ref{eq:evLeakCurrent}). The leakage current for $n$ time intervals can be written as \cite{LeakageCurrent}

\begin{equation}
I_{\text{leak} } = \frac{\phi}{L_{\text{int}}} \cdot \sum_i^n V_i L_{\text{int},i}\left[ \alpha_1 e^{\left( -\sum_{j=i}^n \frac{t_j}{\tau(T_j)}\right)}+\alpha_0^{\star}-\beta \log\left( \frac{\Theta\cdot t_j}{t_0}\right) \right]
\label{eq:leakPred}
\end{equation}
where
\begin{itemize}
    \item $L_{\text{int},i}$ is the luminosity in the $i$-th time interval;
    \item $T_i$ is the temperature in the $i$-th time interval;
    \item $V_i$ is the depleted volume in the $i$-th time interval;
    \item $t_0$ is a constant term, $t_0 = 1$ min;
    \item $\alpha_1$, $\alpha^{\star}_{0}$, $\beta$ are empirical constant.
    In Run~2, they were set to $\alpha_1 = (1.23 \, \pm \, 0.06) \times 10^{-17} \mathrm{A/{cm}}$, $\alpha^{\star}_{0} = 7.07 \times 10^{-17} \mathrm{A/{cm}}$, and $\beta = (3.29 \, \pm \, 0.18) \times 10^{-18} \mathrm{A/{cm}}$;
    \item $\tau$ is a time constant that follows the Arrhenius distribution:  $\tau^{-1}=k_0\cdot\exp\left(-\frac{E}{k_BT_a} \right)$, where $k_{0} = 1.2^{+5.3}_{-1.0} \times 10^{13} \mathrm{s}^{-1}$ and $E = (1.11 \pm 0.05)$ eV.
    \item $\Theta$ is a function that scales the temperature to a reference value ($T_{\text{ref}}=20^{\circ}$ C), and is described as:
\begin{equation}
\Theta(T) = \exp \left[  -\frac{E_I^{\star}}{k_B}\left( \frac{1}{T}  -\frac{1}{T_{\text{ref}}} \right) \right]
\end{equation}
where $E_I^{\star}=(1.30\pm 0.14)$ eV and $T_{\mathrm{ref}}$ is a reference temperature, typically 20 °C.
\end{itemize}

\begin{figure}[!htb]
\centering
\includegraphics[width=0.8\textwidth]{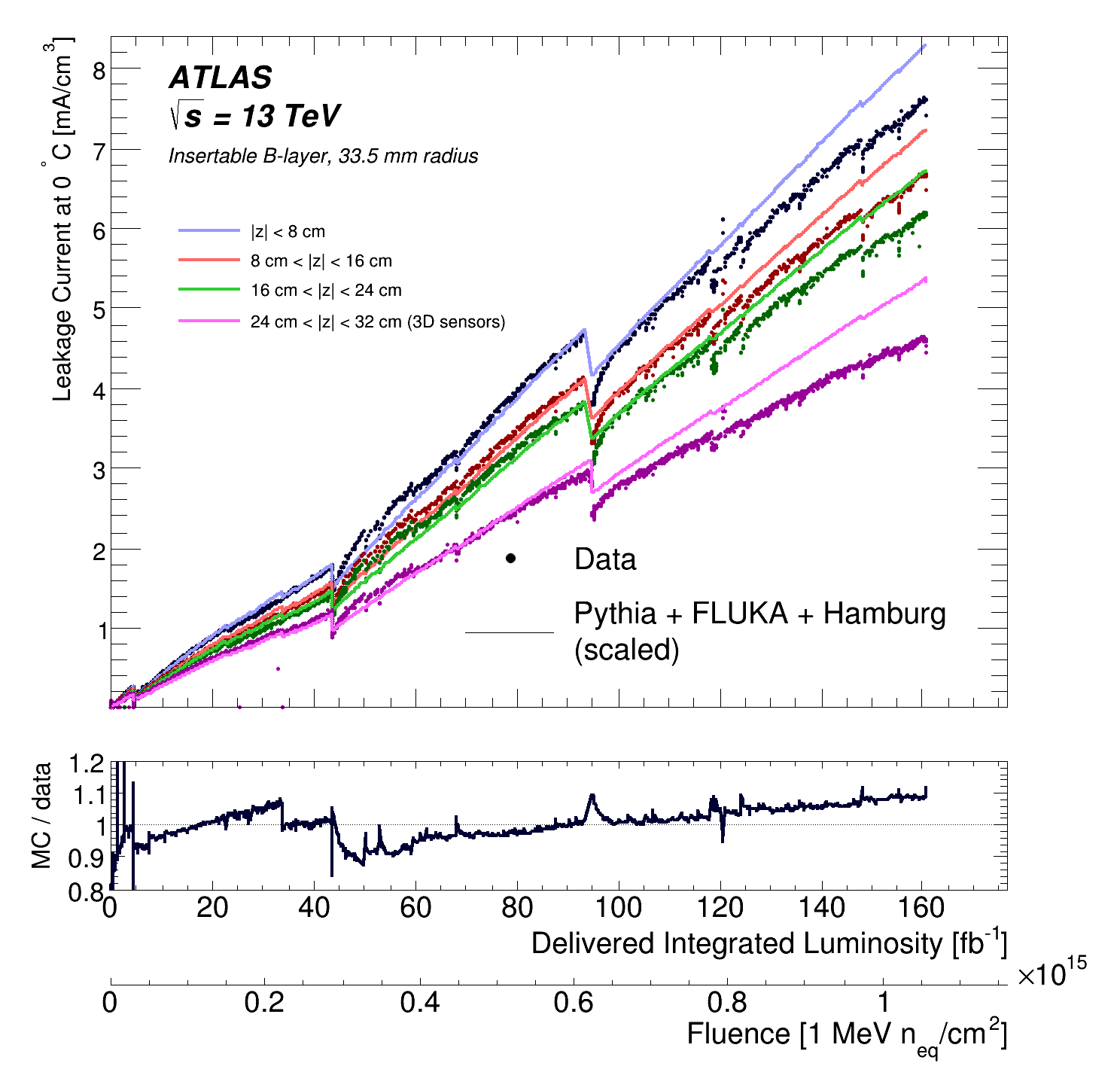}
\caption{The measured and predicted leakage current for sensors on the IBL, both normalized to 0 degrees celsius for different $|z|$ regions. The lower panel shows the ratio between the prediction and the data for the innermost module group. From Ref. \cite{LeakageCurrent}.}
\label{fig:LeakCurrFit}
\end{figure}

Properties were predicted and measured in time intervals of 10 minutes during the whole Run~2, and in each time period predictions were fitted to the data, divided per group of modules. This is because different modules have different distances from the interaction point along the beam axis. Predictions were also scaled to a reference temperature of 20 $^{\circ}$C. 
Leakage current measurements are done with two different subsystems: the high voltage patch panel subsystem (HVPP4) and the multi-module power supply subsystem. The HVPP4 monitors the current using a Current Monitoring Board system, at the pixel level from the high voltage power supplies, while the multi-module power supply system uses a mix of custom components and commercially available components for high and low voltage for the readout electronics and sensor bias.\\
Fig.~\ref{fig:LeakCurrFit} shows the measured and predicted leakage current for the IBL. The prediction is based on the thermal history of the modules combined with the Hamburg model \cite{Moll:1999kv} for modelling changes to the leakage current and Pythia + Fluka for simulating the overall fluence. For all four predictions, the overall scale is normalized based on a fit with the data across the entire range. Similar MC/data trends are observed for the $|z|$ regions.

\subsubsection{Annealing and depletion voltage}
\begin{figure}[!htb]
\begin{minipage}{0.49\textwidth} \centering
\includegraphics[width=1.\textwidth]{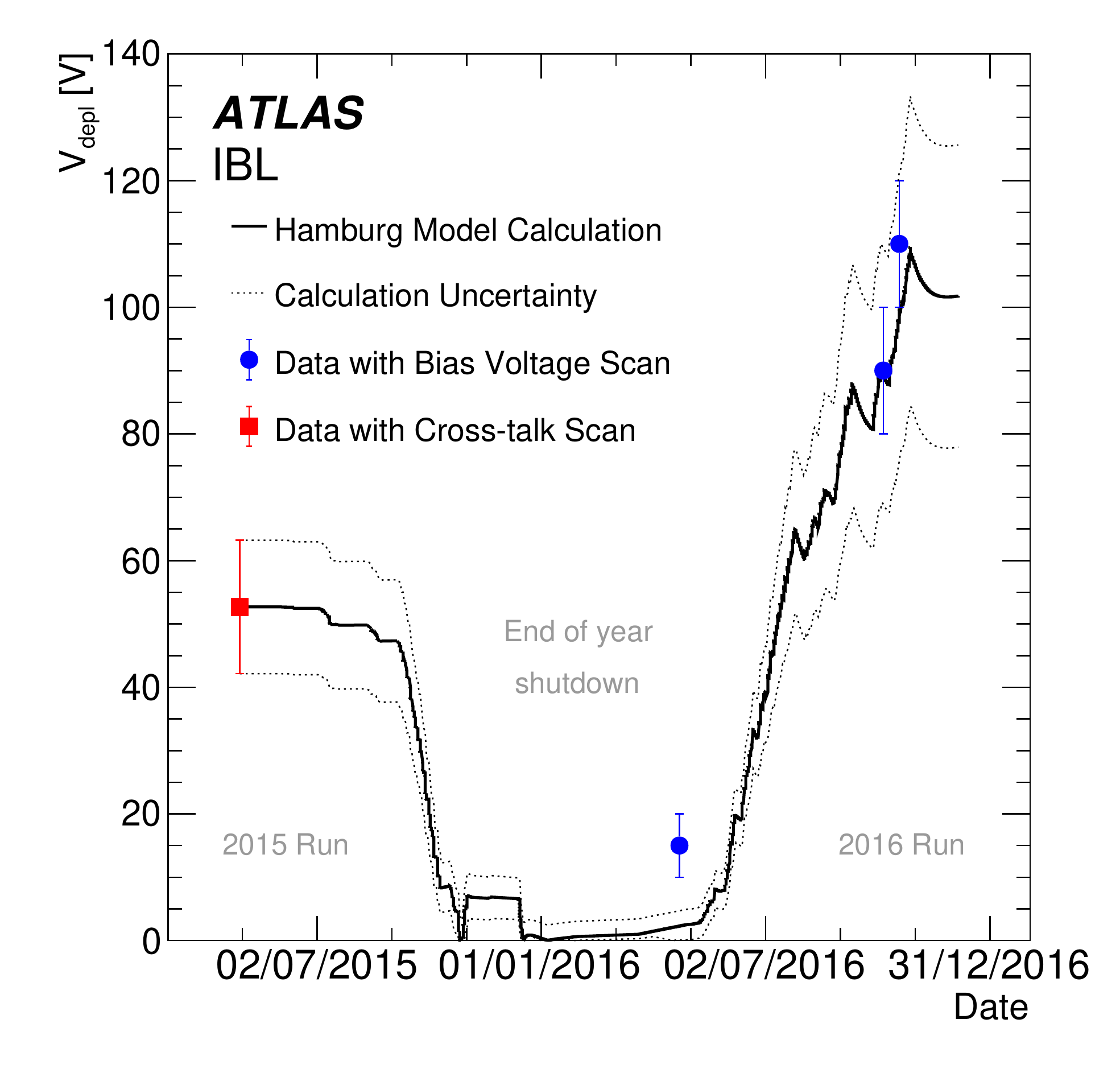}
\end{minipage}
\begin{minipage}{0.49\textwidth} \centering
\includegraphics[width=1.\textwidth]{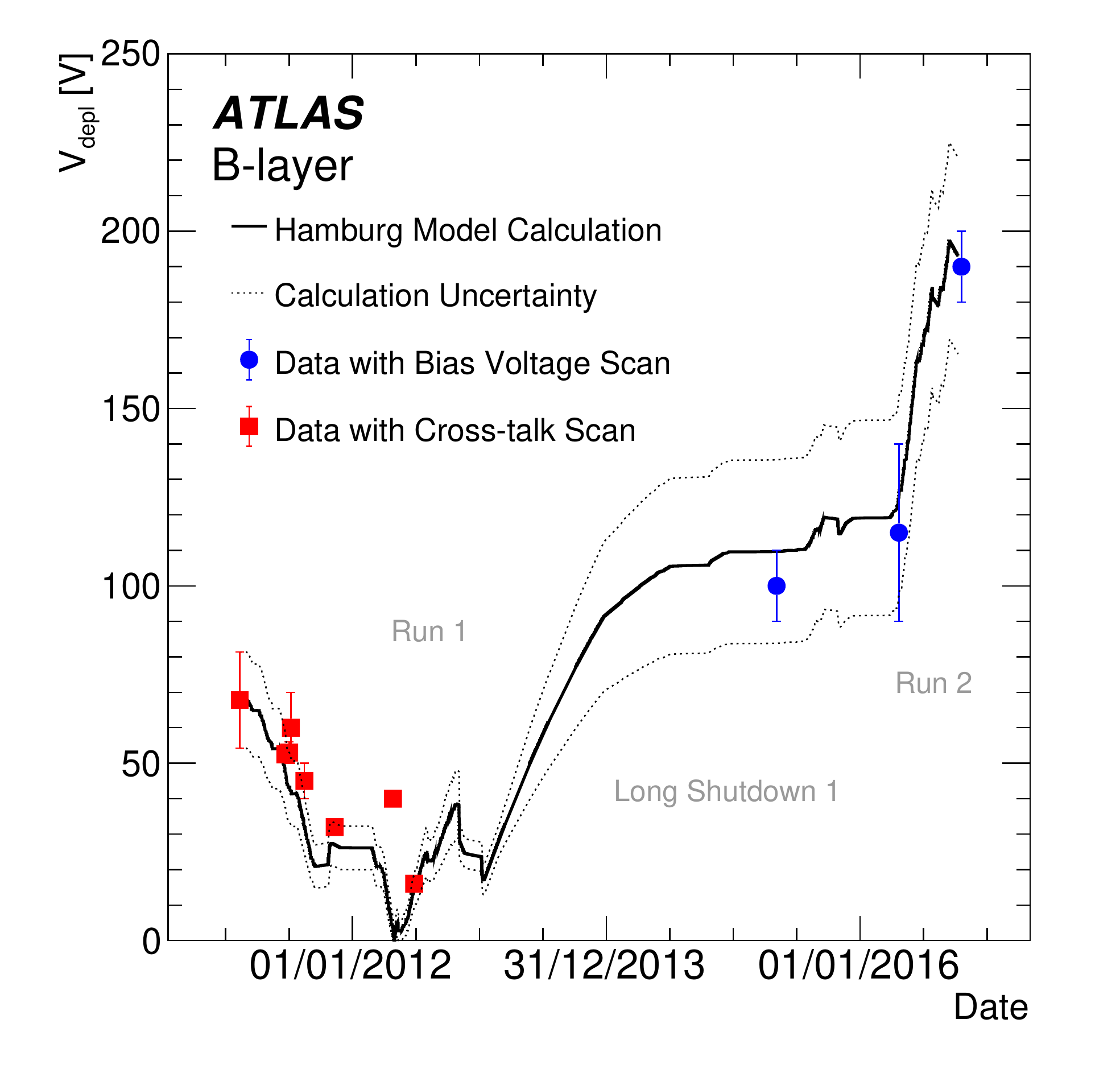}
\end{minipage}
\caption{Calculated depletion voltage of (left) IBL and (right) B-layer according to the Hamburg model as a function of time from the date of their installation until the end of 2016. Circular points indicate measurements of the depletion voltage using the bias voltage scan method while square points display earlier measurements using cross-talk scans. From Ref. \cite{Aaboud:2019wgd}.}
\label{fig:EvolutionVDep}
\end{figure}

An important aspect of the detector that must be kept under control is the annealing and the depletion voltage of the sensors. From Eqs.~(\ref{eq:VdepNeff}) and (\ref{eq:HamburgModel}) it is possible to predict the evolution of the depletion voltage with fluence. This can be measured in two different methods. The first one consists of using cross-talk of adjacent pixels since pixels are only isolated when fully depleted. However, this is true only before type inversion, after that point pixels are isolated even before full depletion. The second one is the bias voltage scan and can be used after type inversion. The operating voltage of the sensor is raised in steps and, at each step, the collected charge is measured. 
At high fluence, the depletion voltage loses its meaning. It is however used as an operational parameter to indicate the bias voltage needed to recover most of the charge. This is done by fitting with two curves, a straight line and squared function that parametrize the two different behaviours, to the mean collected charge as a function of the bias voltage. The depletion voltage is then defined as the operating voltage where the two curves cross. Fig.~\ref{fig:EvolutionVDep} shows the evolution of the depletion voltage as a function of the days of operation of the LHC during 2015 and 2016 for IBL (left) and B-Layer (right). The points are data collected with both the cross-talk and the voltage scan. Prediction are from Eq.~(\ref{eq:HamburgModel}). Uncertainty contains variations of the initial parameter of the equation and an additional 20\% in the initial doping concentration.

\subsection{Digitizer Model}
\label{ssection:DigiModel}
A description of the effect of radiation damage on the detector response and performance is needed in order to correctly predict the behaviour of the detector, decide the operation conditions (like electronics threshold, temperature and bias voltage) that minimize the impact on the performance, and also have a good description of physical quantities. 

\subsubsection{Overview}
\begin{figure}[!htb]
\centering
\includegraphics[width=1.\textwidth]{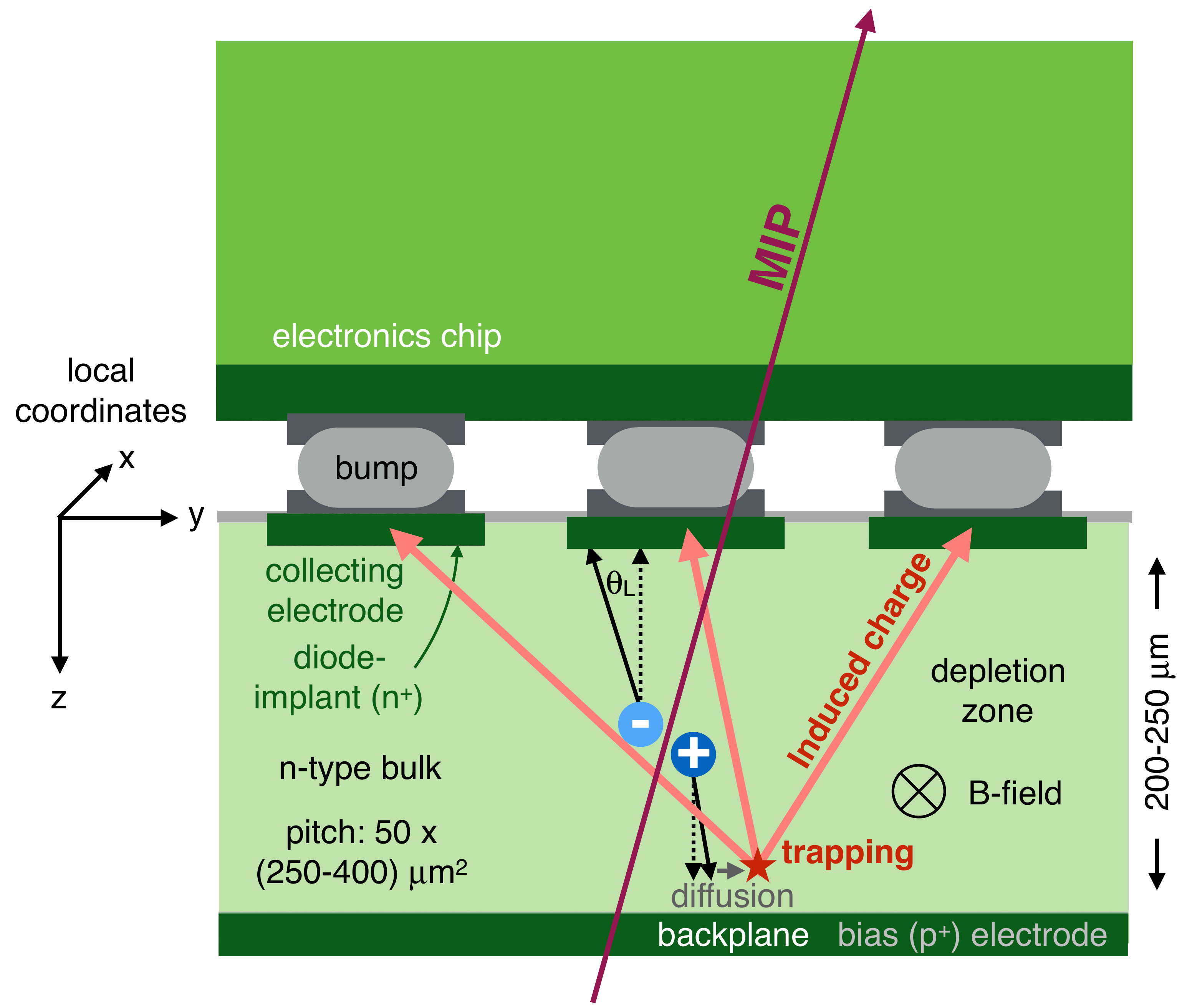}
\caption{Schematics of the planar digitizer. From Ref. \cite{Aaboud:2019wgd}.}
\label{fig:RadDamDigitizer}
\end{figure}

The implementation of these effects is done in the \textit{digitization} step, where the energy deposits of charged particles are converted to digital signals sent from the front ends to the detector readout system. Energy deposits are obtained from Geant4, a software that evaluates the trajectories of particles inside the detector and their interaction with the material, and whose output is a list of energy deposits and their position in the sensitive material, the hits. The radiation damage digitizer is implemented in the ATLAS common software framework, Athena \cite{Athena}, which describes the whole ATLAS detector and has been used the production of MC samples so far. A schematic of the planar digitizer is shown in Fig.~\ref{fig:RadDamDigitizer}.\\

\begin{figure}[!htb]
\centering
\includegraphics[width=0.8\textwidth]{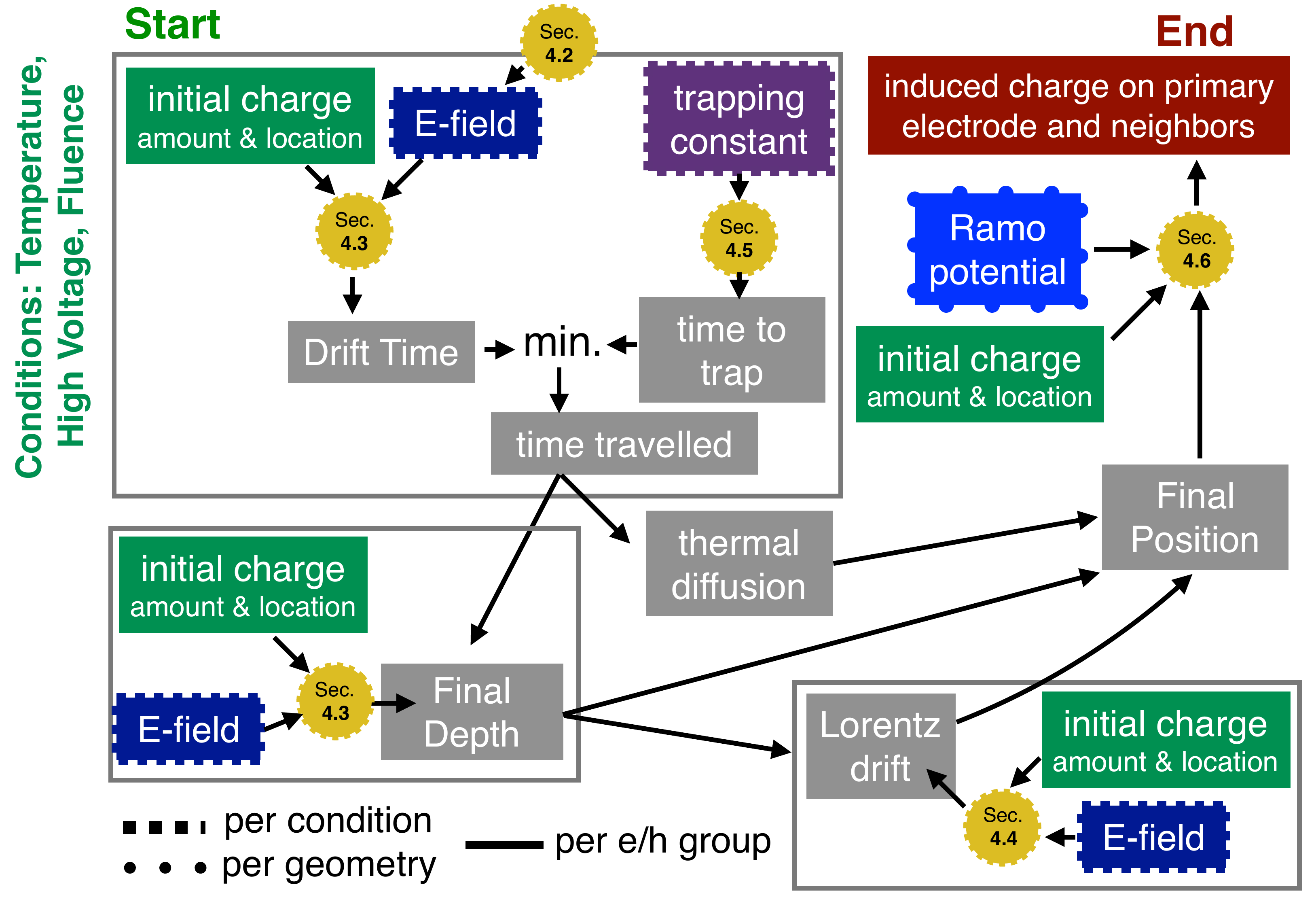}
\caption{Workflow of the digitizer. From Ref.~\cite{Aaboud:2019wgd}.}
\label{fig:RadDamDigitizer-Workflow}
\end{figure}

A standalone software outside Athena, called Allpix \cite{Allpix}, is used to set the parameters concerning the geometry of the pixel module (thickness, pitch, number of rows and columns, and tilt relative to the beam) and to send them to Geant4 for generating the energy deposits. This way, Allpix allows to easily validate the effects of the radiation damage without the reconstruction of Athena, which takes longer and is generally harder. Other constant values, such as fluence, trapping time for electrons and holes, temperature, and B field strength, are instead set in the digitizer. Still in the initialization, Ramo potential maps and Electric field maps corresponding to the correct bias voltage and fluence are loaded and stored in root histograms, ready to be used as look-up table. These maps are obtained from separate simulations with the TCAD (Technology Computer Aided Design) tool, containing the radiation damage effects. Secondary maps are built from all the inputs and the maps, such as: Lorentz angle values, and final position maps. \\
The digitizer reads the ionization energy deposits of the hits created by Geant4 and converts them into electron-hole pairs. The energy needed for a particle to create an electron/hole pair is $\sim 3.6$ eV. Electrons and holes are then drifted towards the opposite electrodes using the information from the lookup tables. In order to speed up the software, charges are grouped in chunks of $\sim 10$, although this is a settable parameter. For each charge then the probability of being trapped is evaluated, and charge carriers are considered trapped if the time needed to reach the electrode is larger than a random trapping time $\tau$ exponentially distributed as $1/\beta\phi$, where $\phi$ is the fluence and $\beta$ the trapping constant. In case the charge carrier is trapped, it is necessary to evaluate how much charge is induced in the neighbouring pixel. This is done with \textit{Ramo} maps, as it will be explained in Section~\ref{sec:RamoMaps}. The workflow of the digitizer is shown in Fig.~\ref{fig:RadDamDigitizer-Workflow}.\\

During Run~2, the Athena simulation has been performed without the inclusion of the radiation damage digitizer. For Run~3, instead, the effects of the radiation damage will be accounted for in the Athena simulation by the digitizer. For this reason, during my PhD, I compared several simulated distributions related to the properties of the collected charge into pixels with the digitizer turned on or off. This allowed to understand the effects of the radiation damage on the collected charge. As a second step, the simulation with the digitizer turned on was compared to data to understand if the radiation damage corrections were leading to a better agreement to data, this way testing the implementation of the digitizer. This led to some improvements in the digitizer code, such as the implementation of scale factors accounting for different pixel sizes at the borders or the improvement of the granularity of the maps used to estimate to time travelled by the electrons before being trapped. Once the validation was concluded, it was possible to tune the simulation parameters to data, focusing in particular on the trapping constants and using various available measurements including high voltage scans. In the following paragraphs, the most important aspects of the radiation damage digitizer are presented.

\subsubsection{Electric Field Simulation}
\label{ssec:EField}
\begin{figure}[!htb]
\centering
\includegraphics[width=0.6\textwidth]{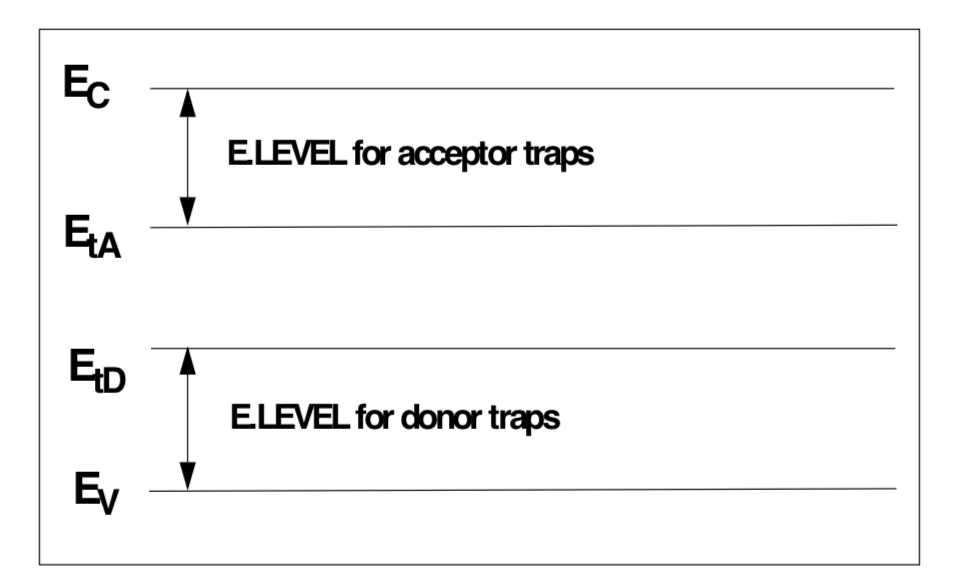}
\caption{Energy band and trapping levels for acceptor and donor.}
\label{fig:EnBandTrap}
\end{figure}

In the presence of a constant doping inside the bulk, as in the case of unirradiated sensors, the electric field is linear. The Hamburg \cite{Moll:1999kv} model predicts the change in time and temperature of the effective doping concentration ($N_{\mathrm{eff}}$), but not the change in concentration within the sensor, which is responsible for the non-trivial shape of the electric field. The electric field shape is instead simulated with software based on TCAD, which is a type of automation for electronic design that models the fabrication and operation of semiconductor devices. The commercial TCAD products used were Silvaco Atlas (for planar modules) \cite{silvaco} and Synopsys (for 3D) \cite{synopsys}. These tools allow us to model the diffusion and ion implantation, and to see the effects on the electrical devices according to the doping profiles. Another important aspect of TCAD is the possibility of modelling the radiation damage effects. This is done by adding trap centres to the energy band gap, which is positioned between the valence band ($E_V$) and conduction band ($E_C$), influencing the density of space charge. Since there are two technologies in the pixel sensors ($n^+$-in-$n$ for the planar and $p$-in-$n$ for the 3D), two sets of simulations are used. Fig.~\ref{fig:EnBandTrap} shows the positions and name convention for the energy bands and the acceptor and donor traps.

The radiation damage model used for planar sensors was proposed in \cite{Chiochia:2004qh} by Chiochia et al. and has been found to give better modelling than the alternative Petasecca model \cite{Petasecca:1710302}. The Chiochia model uses a double trap, with one acceptor and one donor trapping centre, with energy levels at $E_C - 0.525$ eV and $E_V + 0.48$ eV for the conduction and valence band respectively. This model was first developed for CMS sensors, which are also $n^+$-in-$n$ pixel modules. The Petasecca model was found to not correctly model some of the important variables, such as the Lorentz angle as a function of fluence.\\
Simulations are performed for only one quarter of the pixel sensors due to the symmetry of the pixel geometry. The $z$ direction is defined as the direction of the depth of the sensor, while $x$ and $y$ correspond to the $\phi$ and $\eta$ directions, respectively. 
The values of the main parameters used in the TCAD simulation for the planar modules are reported in Table~\ref{tab:ChiochiaNumbers}, where $N_{A/D}$ is the concentration of the acceptor/donor defects and $\sigma_{e/h}^{A/D}$ are the electrons/holes capture cross-section for acceptor and donor defects. The introduction rate, $g_{int}^{A/D} = N_{A/D}/\phi$, is also reported as the density of traps increase with fluence.

\begin{table}[!htb]
\begin{center}
\scalebox{0.75}{\begin{tabular}{cccccccc}
\noalign{\smallskip}\hline\noalign{\smallskip}
Fluence & $N_A$  & $N_D$ & $\sigma_e^{A/D}$ &  $\sigma_h^{A}$ & $\sigma_h^{D}$ & $g_{int}^A$ & $g_{int}^D$ \\
$[10^{14}\mathrm{n_{eq/cm^{2}}}]$ & $[10^{-15}\text{cm}^{-3}]$ &  $ [10^{-15}\text{cm}^{-3}]$ &  $[ 10^{15}\text{cm}^{2}]$  &  $[10^{15}\text{cm}^{2}]$  &  $[ 10^{15}\text{cm}^{2}]$  & $[\text{cm}^{-1}]$ & $[\text{cm}^{-1}]$ \\
\noalign{\smallskip}\hline\noalign{\smallskip}
1 & 0.36 & 0.5  & 6.60 & 1.65 & 6.60 & 3.6 & 5\\
2 & 0.68 &   1  & 6.60 & 1.65 & 6.60 & 3.4 & 5\\
5 & 1.4  &  3.4 & 6.60 & 1.65 & 1.65 & 2.8 & 6.8\\
\noalign{\smallskip}\hline\noalign{\smallskip}
\end{tabular}}
\end{center}
\caption{Values used in TCAD simulations for acceptor (donor) defect concentrations $N_A$ ($N_D$), for their electron (hole) capture cross-sections ($\sigma_{e,h}^{A/D}$) and introduction rates $g_{int}^A/D$ for three different fluences. Values are derived from the Chiochia model ~\cite{Chiochia:2004qh} for temperature $T=-10^{\circ}$C.}
\label{tab:ChiochiaNumbers}
\end{table}

Radiation damage effects in the 3D sensors are instead implemented with the Perugia model \cite{7542192} using the Synopsys TCAD package. In this model there are instead three trap levels: two acceptor and one donor trap levels with energies: $E_C -0.42$ eV,  $E_C -0.46$ eV, and $E_V + 0.36$ eV.

\begin{figure}[!htb]
\begin{minipage}{0.49\textwidth} \centering
\includegraphics[width=1.\textwidth]{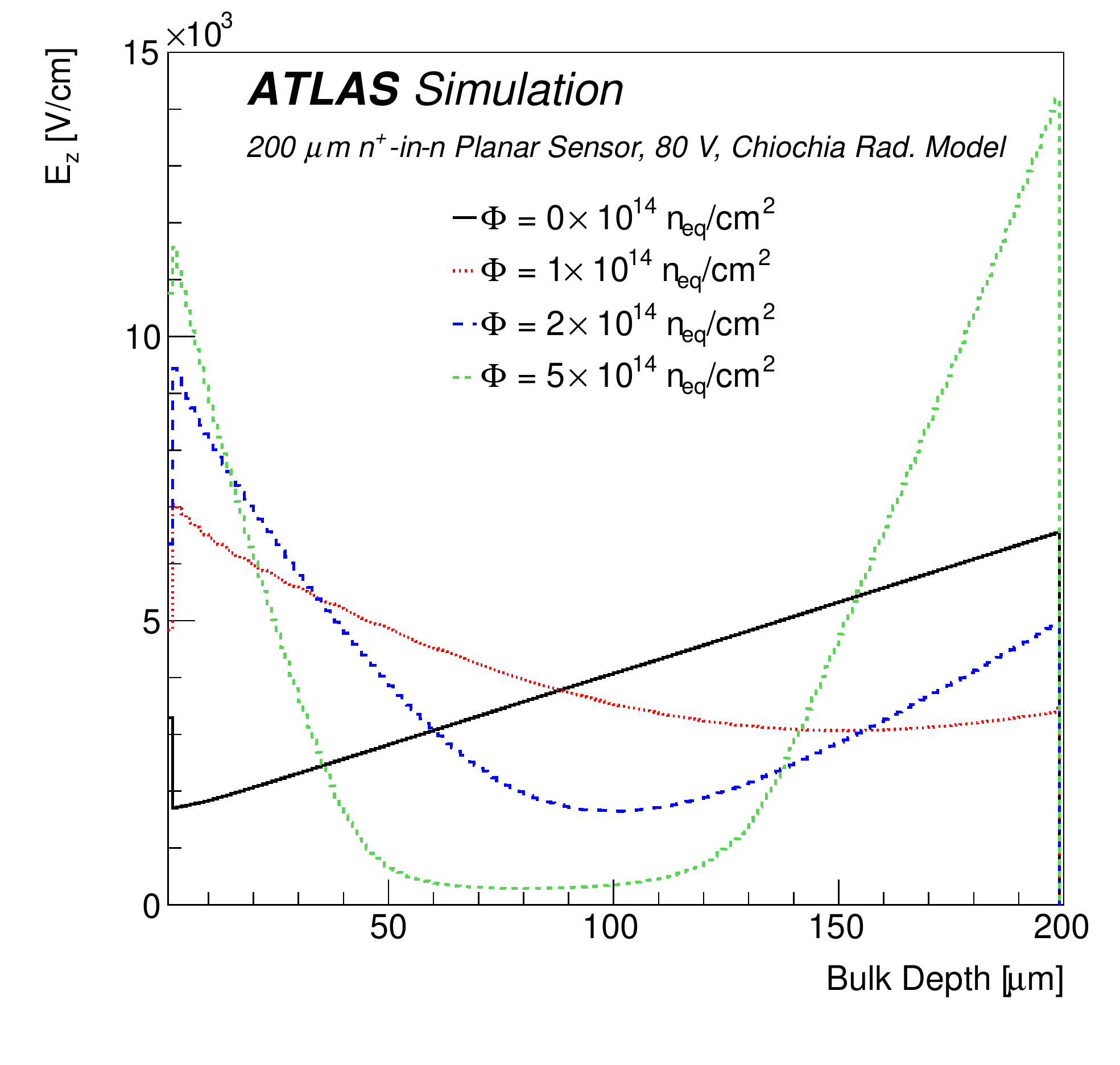}
\end{minipage}
\begin{minipage}{0.49\textwidth} \centering
\includegraphics[width=1.\textwidth]{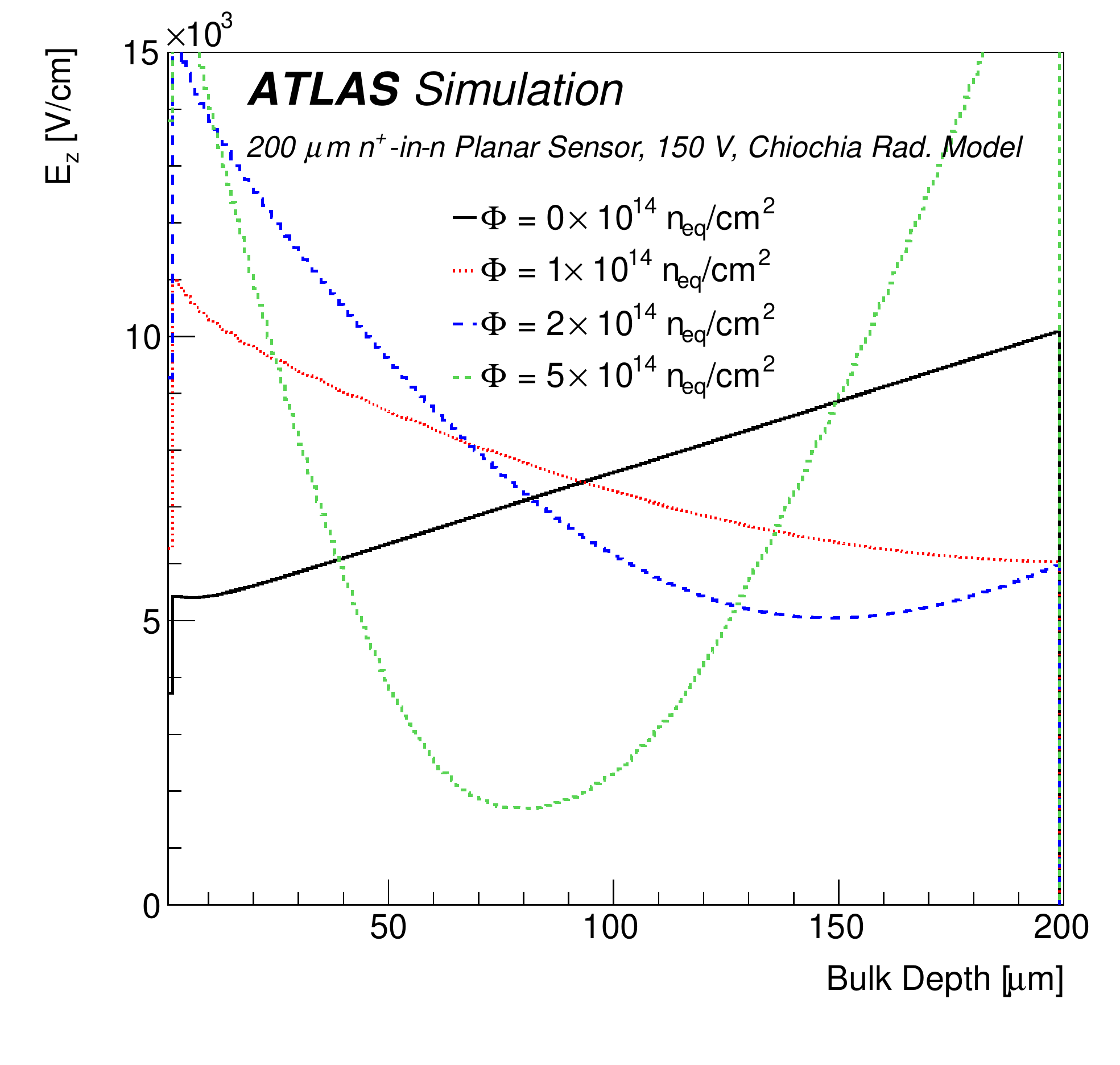}
\end{minipage}
\caption{Simulation of the electric field profile for ATLAS planar pixel modules along the $z$ axis. From Ref. \cite{Aaboud:2019wgd}.}
\label{fig:EfieldTCAD}
\end{figure}

For planar sensors, the electric field profile is rather independent of the $x$ and $y$ positions. Fig.~\ref{fig:EfieldTCAD} shows the shape of the electric field as a function of the bulk depth ($z$ dependency) for different fluences and two bias voltages, 80 V (left) and 150 V (right). It is possible to see that for low fluences the electric field is almost linear, but after type inversion the field is almost all shifted on the other side. After even more fluence a minimum appears in the centre and the electric field has a typical \textit{U-shaped} profile with the characteristic \textit{double peak} \cite{bib:DP} structure.
In the low electric field region, the charges move slowly and are more likely to be trapped at small distances, therefore the charge in this region is not collected efficiently. However the electric field is not zero, at high enough fluence, and therefore the meaning of depletion depth is not valid anymore.

\subsubsection{Time-to-electrode, position-at-trap}

\begin{figure}[!htb]
\centering
\includegraphics[width=0.6\textwidth]{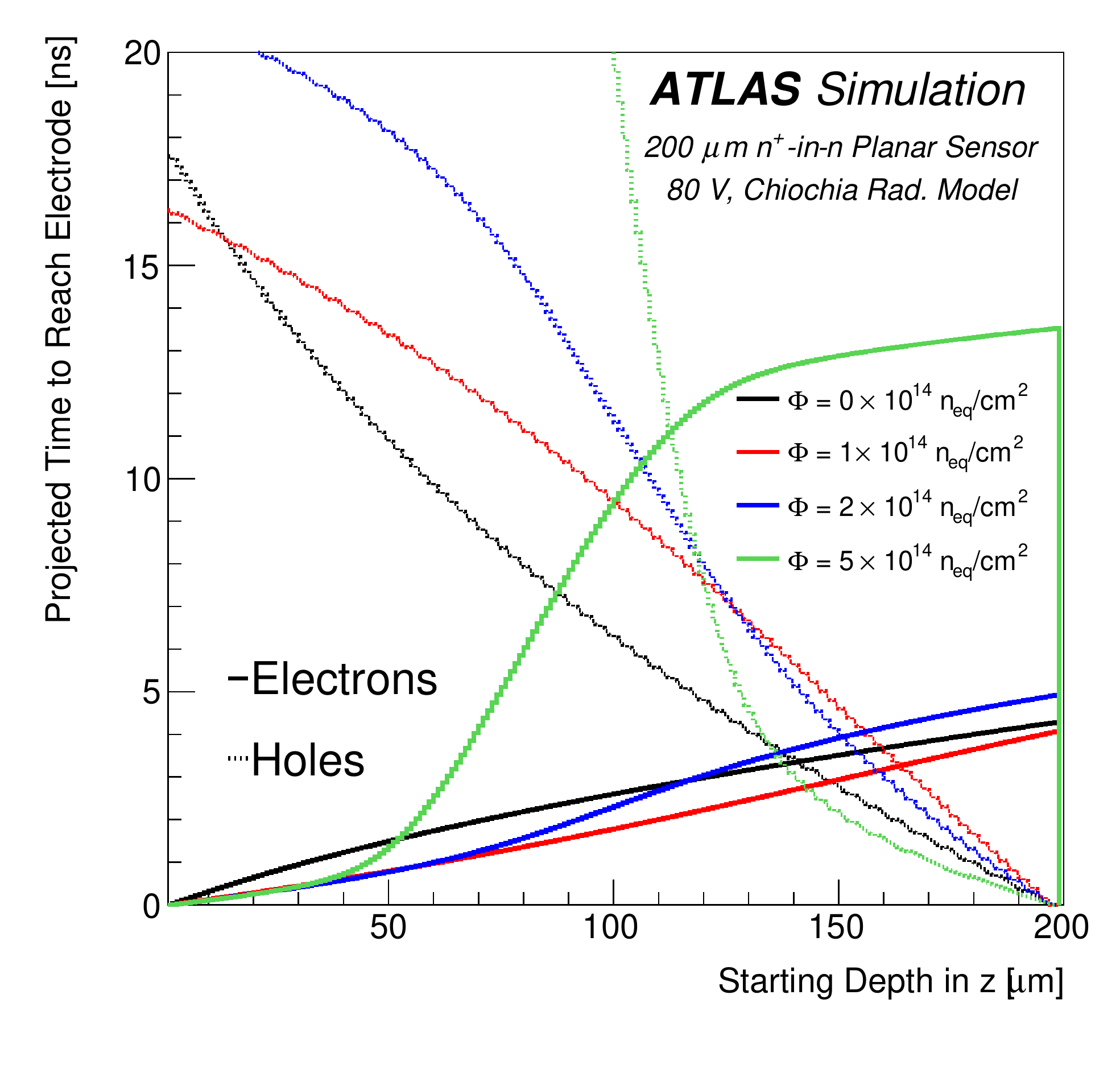}
\caption{Time to reach the electrodes as a function of the initial position of the charge carrier. Electrons go toward the 0 $\mu$m side, while holes toward the 200 $\mu$m one. From Ref.~\cite{Aaboud:2019wgd}.}
\label{fig:TimeCollection}
\end{figure}

Another important input in the digitizer is the time needed by the charge carriers to reach the electrodes, and what is the final position. In fact, due to trapping, if the time is too long, the electron/hole will be trapped, and the final position will define where the charge ends, and thus the charge induced on each pixel cell. A map is computed with the final position as a function of both the initial position and time of drift and is used in each loop of the digitizer. The maps are computed once per geometry and conditions (fluence, bias voltage and temperature).
Electrons and holes drift towards the opposite electrodes with a mobility $\mu$ that depends  on the nature of the charge carrier, the electric field and also the temperature \cite{LUTZ1996}, following the equation from Jacoboni-Canali \cite{JACOBONI197777}
\begin{linenomath*}
\begin{equation}
\begin{split}
\mu_e(T) =& 1533.7 \text{cm}^2/(V\cdot s)\times T_n^{-2.42}\\
\mu_h(T) =& 463.9 \text{cm}^2/(V\cdot s)\times T_n^{-2.20}.
\end{split}
\end{equation}
\end{linenomath*}
Drift velocity is then given by $\vec{v}(E)\sim r\mu(E)\vec{E} $, where $r$ is the Hall scattering factor. From this, the estimated time for collecting the charge is given by
\begin{linenomath*}
\begin{equation}
t_{\text{collection}} (\vec{x}_{\text{initial}}) = \int_C \frac{ds}{r\mu(E)E}
\end{equation}
\end{linenomath*}
where $C$ is the path from the initial to the final position. The integration is done in the $z$ direction since the field is nearly independent of $x$ and $y$. Fig.~\ref{fig:TimeCollection} shows the time to reach the electrode for both electrons and holes, for different fluences, for an IBL planar sensor with a bias voltage of 80 V. Holes drift toward the 200 $\mu$m side, while electrons toward the zero side. Also, it is possible to see that the times go from a few ns to tens of ns.\\
Electrons are collected in few ns, except for very high fluences, where the electric field is quite low in the central part of the sensor; in this case, most of the charges are trapped before reaching the electrodes. Holes instead are slower. Signal formation is still, in general, faster than the LHC clock of 25 ns, but could be a problem for very high fluences.\\
It is important to know the position where the charges are trapped to evaluate the induced charge. As previously explained, a charge carrier is trapped if its time to reach the electrode is larger than a random number distributed as an exponential function with the trapping time as mean value. Therefore it is possible to evaluate the final position as:
\begin{linenomath*}
\begin{equation}
\vec{x}_{\text{trap}} = \int^{t_{\text{trap}}}_{0} r\mu(E)\vec{E}dt.
\end{equation} 
\end{linenomath*}
where $t_{\text{trap}}$ is the random time of trapping of the charge carrier considered.

\subsubsection{Lorentz angle}
\begin{figure}[!htb]
\centering
\includegraphics[width=0.45\textwidth]{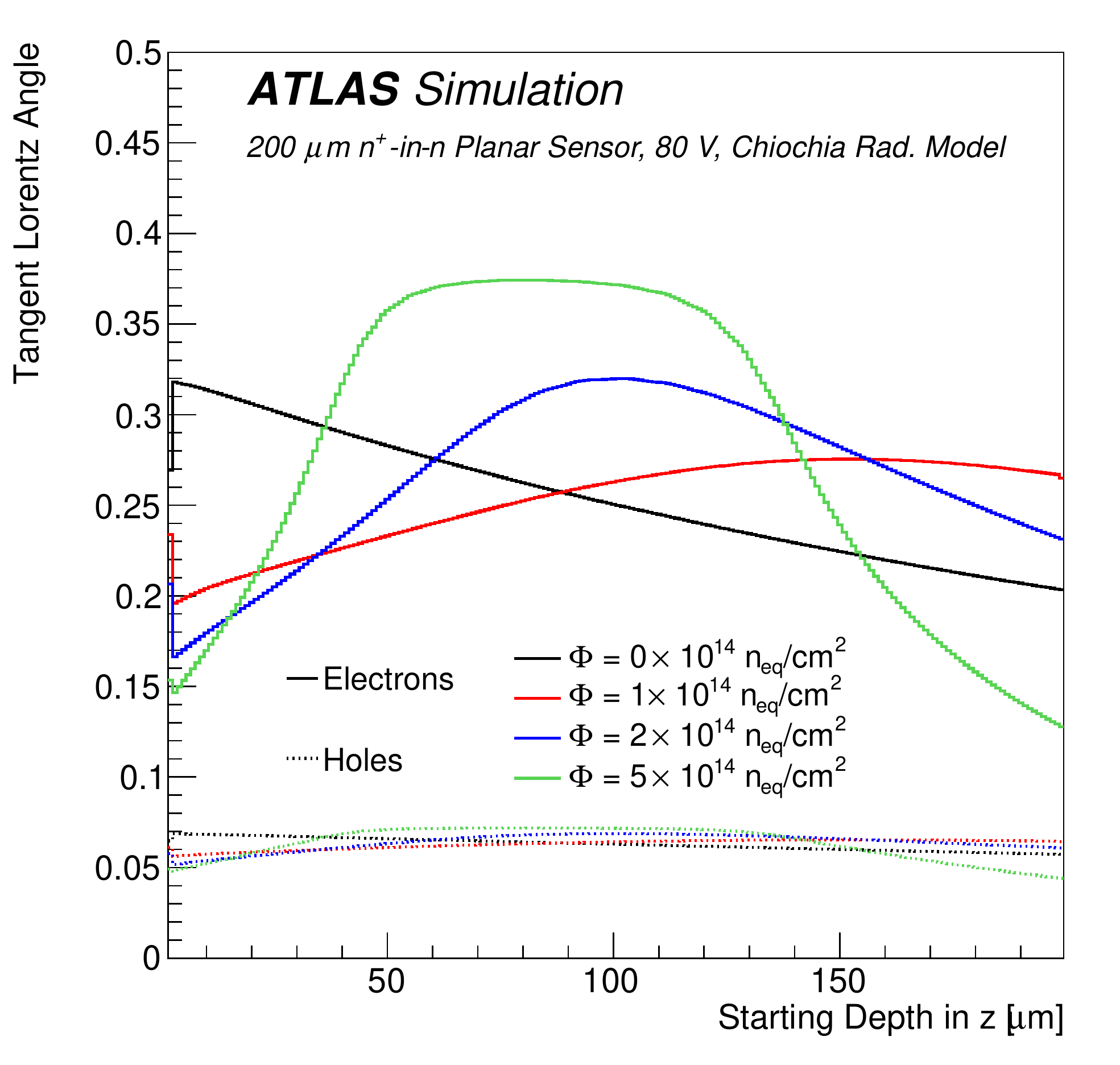}
\includegraphics[width=0.45\textwidth]{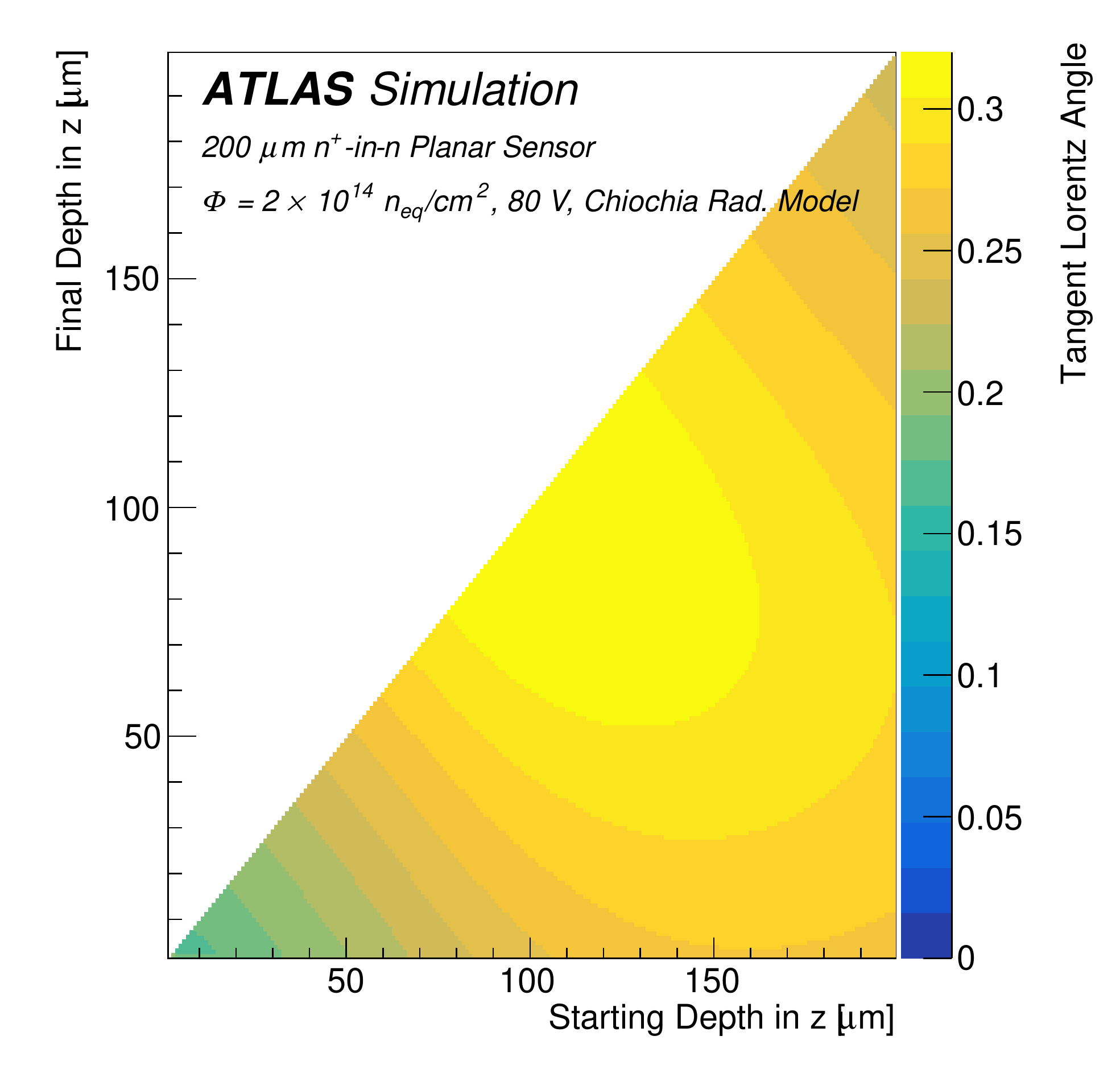}
\caption{Tangent of the Lorentz angle as a function of the starting position in $z$ of the charge carrier (left) or as a function of both the initial and final position (right) in an ATLAS IBL planar module, with a bias voltage of 80 V. From Ref. \cite{Aaboud:2019wgd}.}
\label{fig:LorentzAngle}
\end{figure}

Charge carriers in the silicon sensors drift towards the electrodes due to the electric field. The presence of a magnetic field deviates the path of the charge carriers from straight lines. The Lorentz angle ($\theta_L$) is defined as the angle between the drift direction and the electric field, and this causes the minimum of the cluster size to be obtained for particles entering the sensor with an incident angle equal to the Lorentz angle. In a given point inside the bulk of the sensor, the Lorentz angle is given by
\begin{linenomath*}
\begin{equation}
\tan \theta_L = rB \mu(E(z))
\end{equation}
\end{linenomath*}
where $\mu$ is the mobility. The mobility depends on the electric field and it diminishes for a very high field, thus also the Lorentz angle depends on the electric field. This also means that the total effects, as the incident angle corresponding to the minimum cluster size, depends not only on the average electric field (which is the bias voltage divided by the depth of the sensor) but also on the profile of the electric field which is modified by the radiation damage. It is possible to write the Lorentz angle with the following formula:
\begin{linenomath*}
\begin{equation}
\tan\theta_L (z_i,z_f) = \frac{rB}{|z_f - z_i |}\int^{z_f}_{z_i}\mu(E(z))dz.
\end{equation}
\end{linenomath*}
where $z_{i/f}$ is the initial/final position of the charge carrier. \\
In the digitizer code, the Lorentz angle maps are saved at the beginning for each geometry and condition setup (fluence, bias voltage, and temperature). The final position of the charge carrier when adding together the drift and the Lorentz angle is then given by
\begin{linenomath*}
\begin{equation}
\begin{split}
x_f =& x_i + |z_f - z_i|\cdot \tan\theta_L + d_x \\
y_f =& y_i + d_y
\end{split}
\end{equation}
\end{linenomath*}
where the $y$ direction is the direction perpendicular to the magnetic field, while the $x$ is parallel to it. $d_{x/y}$ instead is the thermal diffusion in the $x$ and $y$ direction, and it is given by
\begin{linenomath*}
\begin{equation}
d = \epsilon\cdot d_0 \sqrt{\frac{|z_f - z_i|\cdot \cot\theta }{0.3}} 
\end{equation}
\end{linenomath*}
with $d_0$ a diffusion constant and $\epsilon$ a random number. The Lorentz angle for electrons is larger than for holes because of the larger mobility. Fig.~\ref{fig:LorentzAngle} (left) shows the tangent of the Lorentz angle as a function of the initial position in a planar module with a bias voltage of 80 V for different fluences. Fig.~\ref{fig:LorentzAngle} (right) shows the same plot but as a function of both the initial and final position, for a fluence of $2\times 10^{14} \mathrm{n_{eq}/cm^{2}}$.

\subsubsection{Charge Trapping}
In the digitizer, the charge carriers are declared trapped if their time to reach the electrodes is larger than a random time distributed as an exponential with mean value $1/\phi\beta $, where $\beta$ is the trapping constant.\\
This constant is set at the beginning of the digitizer, and it is taken from literature. From different measurements, $\beta$ has been found to depend on the type of irradiation, the temperature, and the annealing history, and also if the charge carrier is an electron or a hole. In the digitizer, an average of different measurements is used. Values are reported in Table~\ref{tab:trapping} for the different values of $\beta$, together with the method of irradiation, the level of annealing and the type of irradiation. 
\begin{table}[!htbp]
\footnotesize
\centering
\begin{tabular}{cccccc} %
\noalign{\smallskip}\hline\noalign{\smallskip}
Irradiation   & Annealing & $\beta_e$  & $\beta_h$ & Reference & Method  \\
  &  &  ($10^{-16} \text{cm}^2$/ns) & ($10^{-16} \text{cm}^2$/ns) &  &   \\
\noalign{\smallskip}\hline\noalign{\smallskip}
Neutrons      & minimum $V_{\text{depl}}$ & $4.0 \pm 0.1$ & $5.7 \pm 0.2$ & \cite{Kramberger:2002zb} & TCT \\
Pions         & minimum $V_{\text{depl}}$ & $5.5 \pm 0.2$ & $7.3 \pm 0.2$ & \cite{Kramberger:2002zb} & TCT\\
Protons         & minimum $V_{\text{depl}}$ & $5.13 \pm 0.16$ & $5.04 \pm 0.18$ & \cite{Krasel:2004mi} & TCT\\
\noalign{\smallskip}\hline\noalign{\smallskip}
Neutrons      & $> 50$~hours at 60$^{\circ}$C & $2.6 \pm 0.1$ & $7.0 \pm 0.2$ & \cite{Kramberger:2002zb}& TCT \\
Protons         & $> 10$~hours at 60$^{\circ}$C & $3.2 \pm 0.1$ & $5.2 \pm 0.3$ & \cite{Krasel:2004mi}& TCT \\
\noalign{\smallskip}\hline\noalign{\smallskip}
Protons         & minimum $V_{\text{depl}}$ & $4.0 \pm 1.4$ &   -    & \cite{Aad:2008zz,Alimonti:2003laa} & Test beam\\
Protons         & 25h at 60$^{\circ}$C    & $2.2 \pm 0.4$  &  -    & \cite{Aad:2008zz,Alimonti:2003laa} & Test beam\\
\noalign{\smallskip}\hline\noalign{\smallskip}
\end{tabular}
\caption{Measurements of the trapping constant $\beta$ are summarized, normalized to a temperature of 0$^{\circ}$C. Some measurements are reported after annealing to the minimum in the full depletion voltage $V_{\text{depl}}$ (reached in about 80 minutes at 60$^{\circ}$C) while others correspond to the asymptotic values observed after long annealing times.}
\label{tab:trapping}
\end{table}
Measurements were obtained with two different techniques: a transient current technique (TCT) was used for Refs.~\cite{Krasel:2004mi} and \cite{Kramberger:2002zb}, while results from test beam were used for Ref.~\cite{Alimonti:2003laa}. Measurements were also performed at different temperatures, between $-10^{\circ}$C and $10^{\circ}$C, and a significant dependence of $\beta$ on temperature was found. Therefore all the results were scaled to $0^{\circ}$C to be comparable. Both results with TCT find a $\beta$ increasing with annealing for electrons while decreasing for holes.\\
In the digitizer the values used were:
\begin{linenomath*}
\begin{equation}
\begin{split}
\beta_e = & (4.5 \pm 1.5) \times 10^{-16} \text{cm}^2/\text{ns}\\
\beta_h = & (6.5 \pm 1.5) \times 10^{-16} \text{cm}^2/\text{ns}
\end{split}
\end{equation}
\end{linenomath*}
These values were chosen to be representative of the conditions of the ATLAS Pixel detector during Run~2. The uncertainty instead was set to cover differences between all the references used.

\subsubsection{Ramo potential and induced charge}
\label{sec:RamoMaps}
\begin{figure}[!htb]
\centering
\includegraphics[width=0.7\textwidth]{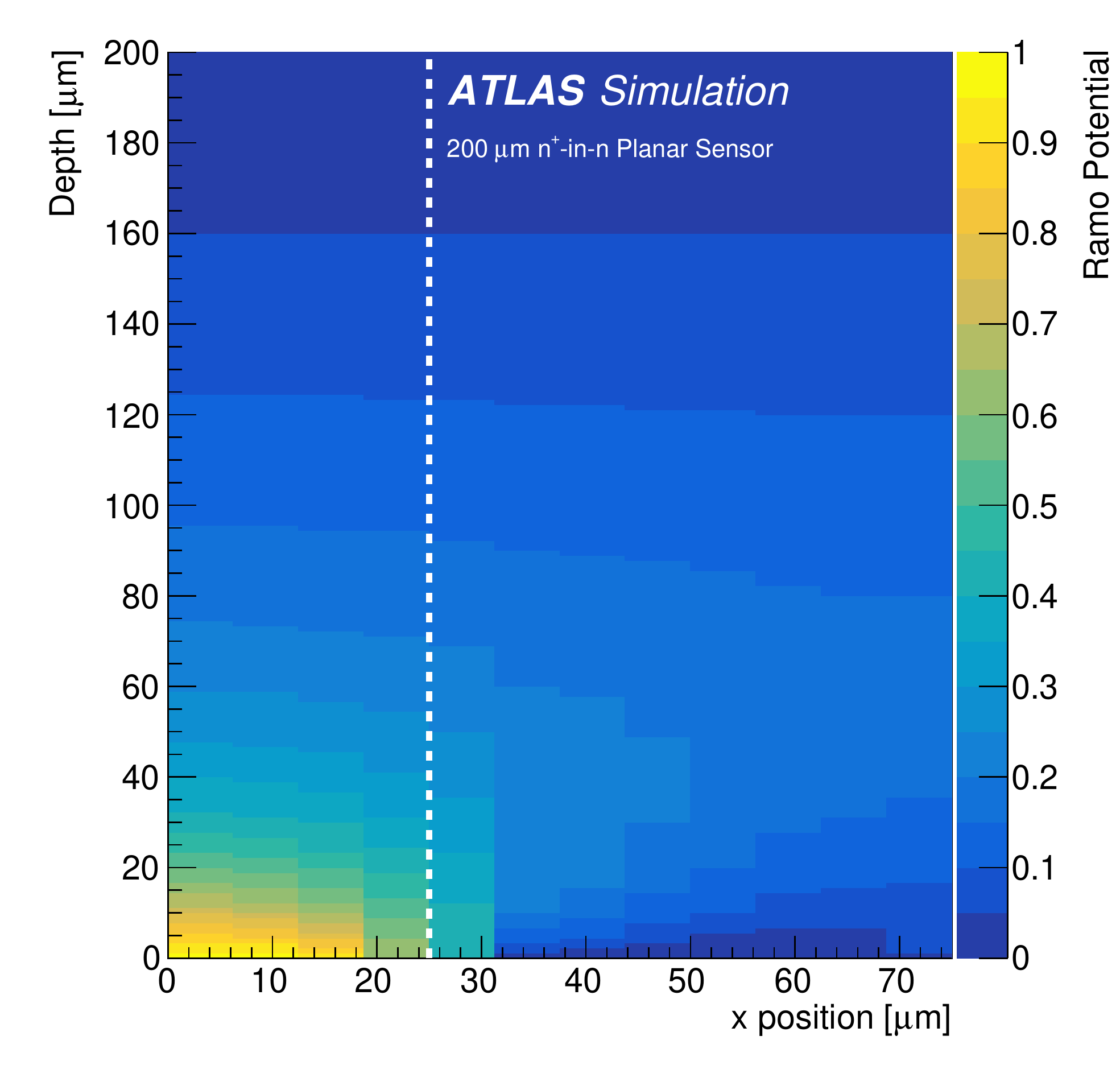}
\caption{Ramo potential map of a quarter of a ATLAS IBL planar module in the $z-x$ plane. From Ref. \cite{Aaboud:2019wgd}.}
\label{fig:RamoMapsPlanar}
\end{figure}

\begin{figure}[!htb]
\centering
\includegraphics[width=0.6\textwidth]{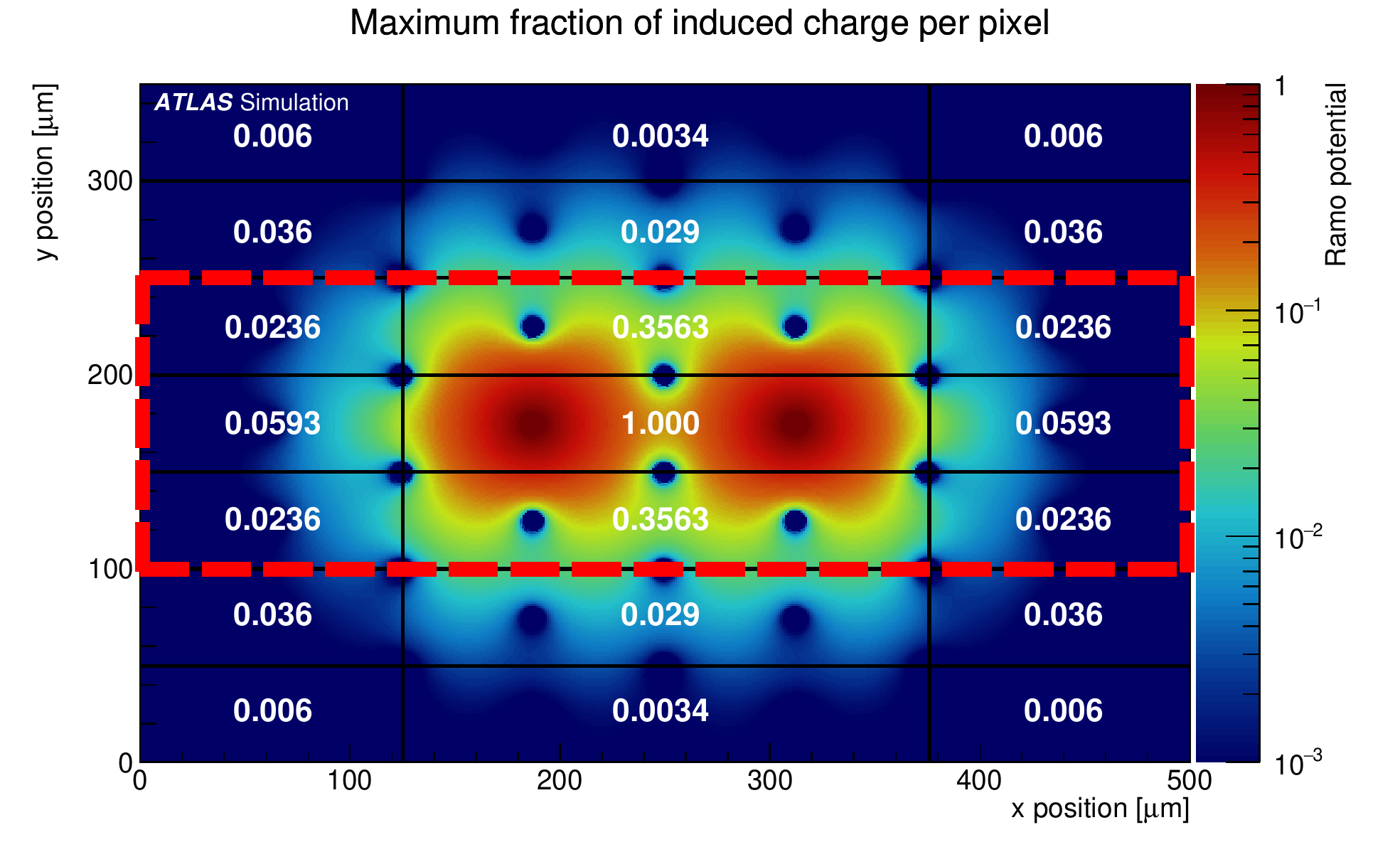}
\caption{Ramo potential map of an ATLAS IBL 3D module. Numbers indicate the maximum induced charged (normalized to one). Blue holes are the p$^+$ columns. The red dashed lines illustrate which pixels are put in the simulations. From Ref. \cite{Aaboud:2019wgd}.}
\label{fig:RamoMaps3D}
\end{figure}

Drifting charges inside the bulk of the sensors towards the electrodes induce a signal that is then read by the electronics. This signal can be analytically calculated, by using the Shockley-Ramo theorem \cite{Ramo}. The theorem states that the instantaneous current $i$ induced on an electrode by a moving charge $q$ is given by
\begin{linenomath*}
\begin{equation}
i(t) =  q \vec{v}\cdot \vec{E}_w(\vec{r})
\label{eq:ramotheorem}
\end{equation}
\end{linenomath*}
where $\vec{v}$ is the instantaneous velocity of the charge. Instead, $E_w$ is the electric field generated at the position $r$ by $q$ on the electrode considered and removing all other charges and electrodes. $E_w$ is called \textit{weighting field} or \textit{Ramo field}. Integrating Eq.~(\ref{eq:ramotheorem}) over time it is obtained:
\begin{linenomath*}
\begin{equation}
Q_{\text{induced}} = -q[\phi_{\text{w}}(\vec{x}_{\text{f}})-\phi_{\text{w}}(\vec{x}_{\text{i}})],
\label{eq:ramopotential}
\end{equation}
\end{linenomath*}
where $\phi_w$ is the Ramo potential $\vec{E}_w = \nabla \phi_w $. The Ramo potential depends only on the geometry of the electrodes, and therefore it is possible to be evaluated in advance. In presence of a pair of electrons-holes formed in the position $x_i$ that drifts towards their respective electrodes and they both arrive at the end, the induced charge is $q$, the charge of the electrons. However, if one charge carrier is trapped, the charge is not zero but can be evaluated using Eq.~(\ref{eq:ramopotential}), and it is always smaller than the charge $q$. \\
In the digitizer, the Ramo maps are loaded in the initialization process and are used in each loop whenever a charge is trapped to estimate the induced charge in all the pixels in a $3\times3$ matrix around the closest pixel to the trapping position. These maps are evaluated with TCAD to solve the Poisson equation. For planar sensors, there is a small $x$ and $y$ dependence, while the main changes are in the $z$ direction. However $x$ and $y$ directions are important to evaluate the charge induced on the neighbouring pixels. Fig.~\ref{fig:RamoMapsPlanar} shows the Ramo potential of a quarter of an IBL planar sensor. The white dashed line indicates the edge of the electrode. It is then possible to see that indeed the potential is not zero outside the pixel area. \\
The Ramo potential for 3D sensors is slightly more complex, due to not only the 3D geometry, but also to the fact that the two $n^+$ columns are connected, and so they must be kept at ground together when doing the calculation, and this requires a relatively large simulation area. Fig.~\ref{fig:RamoMaps3D} shows the Ramo potential map for a 3D sensor.

\subsection{Model Validation on data}
\label{ssection:ModelValidation}

The digitizer presented is tested by comparing its prediction to data taken by the ATLAS detector during Run~2. Simulations are obtained using the Allpix software and here referred to as \textit{Standalone simulation}. Events from data or simulations passing di-muon or di-jet trigger are considered and charged-particle tracks are reconstructed from hits in the pixel detector, silicon strip detector, and transition radiation tracker. Clusters on the innermost pixel layer associated with tracks are considered for further analysis. Two key observables are chosen to study the correct modelling of the radiation damage effects: the Charge Collection Efficiency (CCE) and the Lorentz angle. Predictions from the simulation with the ATHENA common software are used to validate the radiation damage digitizer using a data sample collected at the start of Run~3.

\subsubsection{Charge Collection Efficiency}
\begin{figure}[!htb]
\centering
\includegraphics[width=0.65\textwidth]{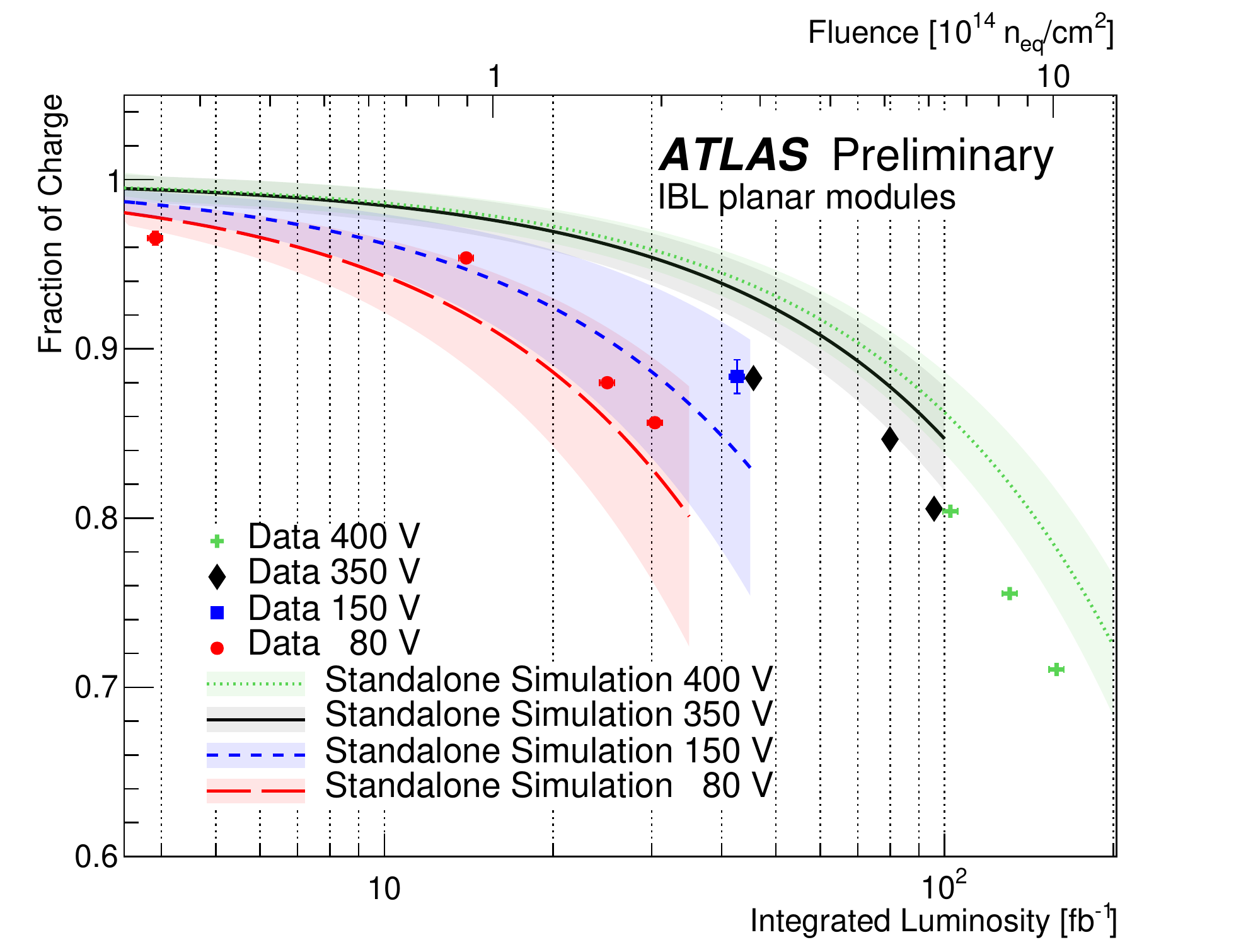}
\caption{Charge Collection Efficiency as a function of the integrated Luminosity  (bottom axis) and fluence (top axis) for IBL modules with $|\eta|<0.2$. For the simulation, the vertical bars include radiation damage parameter variations and the horizontal error bars reflect uncertainty in the luminosity-to-fluence conversion. From Ref. \cite{CCEPlotsPixelAtalsPublic}.}
\label{fig:CCEPlanar}
\end{figure}

The collected charge is one of the most important parameters to monitor as it is directly affected by radiation damage. A decrease in the collected charge implies a decrease in the cluster size, with the pixels with low charge can end up being below threshold. Also, it is possible that due to the reduction of collected charges whole clusters might disappear, therefore reducing the efficiency of the tracking performance. It is then clear why an accurate description of this phenomenon is essential.  \\
Charge deposited in the pixel cluster is well described by a Landau distribution \cite{Landau:1944if}, and from this is possible to define a CCE value as the ratio of the Most Probable Value (MPV) of the Landau distribution of the sensor at one fluence and the MPV from an unirradiated sensor in over-depletion. \\
Fig.~\ref{fig:CCEPlanar} shows the CCE for IBL planar modules with $|\eta| < 0.2 $ as a function of luminosity (bottom axis) and fluence (top axis). 

Data are selected in $Z\rightarrow\mu\mu$ events with track with $3.5 < p_{\mathrm{T}} < 150$ GeV, $ 0<\phi$ on surface $<0.5$ , $|\theta$ on surface $|<0.2$. No clusters with pixels with ToT$=1$ or $>14$ are used. Results are obtained from ToT. \\
As expected the CCE decreases with luminosity, and therefore fluence. At the end of 2016 (around 30 $\mathrm{fb}^{-1}$) the IBL detector was under-depleted, and the CCE was quickly decreasing. Increasing the bias voltage from 80 V to 150 V then was needed to recover this trend. From mid-2017 to the beginning of 2018 the bias voltage was increased again to 350 V, and in 2018 increased again to 400 V. \\
Error bands on the $y$ axis on the simulations account for all the systematic variations presented in Section \ref{ssec:EField}, and also variation (of 1 $\sigma$) of the trapping constants. On the $x$ axis there is a 15 \% uncertainty from the fluence-to-luminosity conversion. Data have instead an uncertainty that accounts for the shift in ToT (described in Section \ref{ssec:Calibration}) along the $y$ axis, and a 2\% uncertainty on the $x$ due to the luminosity measurement uncertainty. Data and simulation are in agreement within the uncertainty, even if the last data points seem to be systematically lower than the prediction.\\

\begin{figure}[!htb]
\centering
\includegraphics[width=0.7\textwidth]{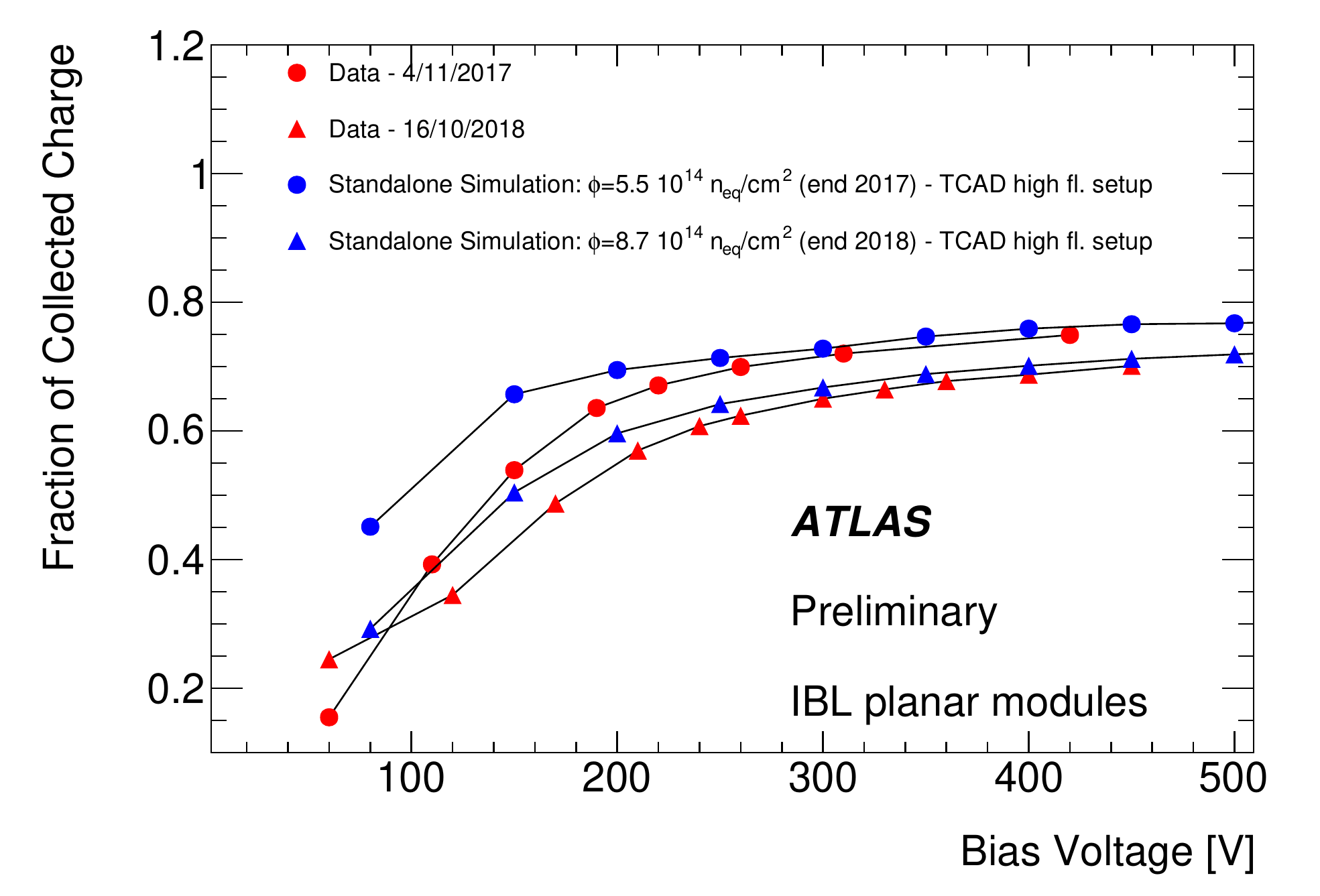}
\caption{Charge Collection Efficiency as a function of the Bias voltage applied to IBL modules. From Ref.~\cite{CCEPlotsPixelAtalsPublic}.}
\label{fig:CCEHVScan}
\end{figure}

It is also possible to evaluate the charge collection efficiency as a function of the bias voltage applied, instead of as a function of the luminosity. This is possible because the bias voltage was varied in special runs, called \textit{Voltage Scan}, during the data taking. In the whole Run~2, eight different voltage scans were taken, one at the beginning and one at the end of each year. This was done to monitor the depletion voltage of the sensors. Fig.~\ref{fig:CCEHVScan} shows the fraction of collected charge as a function of the bias voltage in IBL planar modules for two data runs, one at the end of 2017 and one at the end of 2018, compared with the corresponding simulations.

Simulations agree with data at high bias voltage, while they are higher than data at low bias voltages. The change in the slope of the curves with the increase of the bias voltage can be then used to estimate the depletion voltage. The obtained depletion voltages are reported in Table~\ref{tab:BiasVoltageCalc}. 

\begin{table}
\begin{center}
\begin{tabular}{ccc} %
\noalign{\smallskip}\hline\noalign{\smallskip}
Sample		& Fluence $[10^{14} \mathrm{n_{eq}/{cm^2}}]$  & Depletion Voltage [V]\\
\noalign{\smallskip}\hline\noalign{\smallskip}
Data & 5.5 & $240\pm4$\\
Data & 8.7 & $278\pm4$\\
\noalign{\smallskip}\hline\noalign{\smallskip}
Simulation & 5.5  & $250\pm4$\\
Simulation & 8.7  & $268\pm4$\\
\noalign{\smallskip}\hline\noalign{\smallskip}
\end{tabular}
\caption{Depletion voltage obtained from fit on data and simulation.}
\label{tab:BiasVoltageCalc}
\end{center}
\end{table}

The depletion voltage can be regarded as the voltage value where the electric field is high enough to efficiently collect most of the charge from the whole sensor. The discrepancy at low bias voltage does not compromise the ability to emulate the behaviour of the detector during Run~2, since the operational bias voltage has always been kept at levels higher than the depletion voltage.

\subsubsection{Lorentz angle}
\begin{figure}[!htb]
\centering
\begin{minipage}{0.45\textwidth}\centering
\includegraphics[width=1.\textwidth]{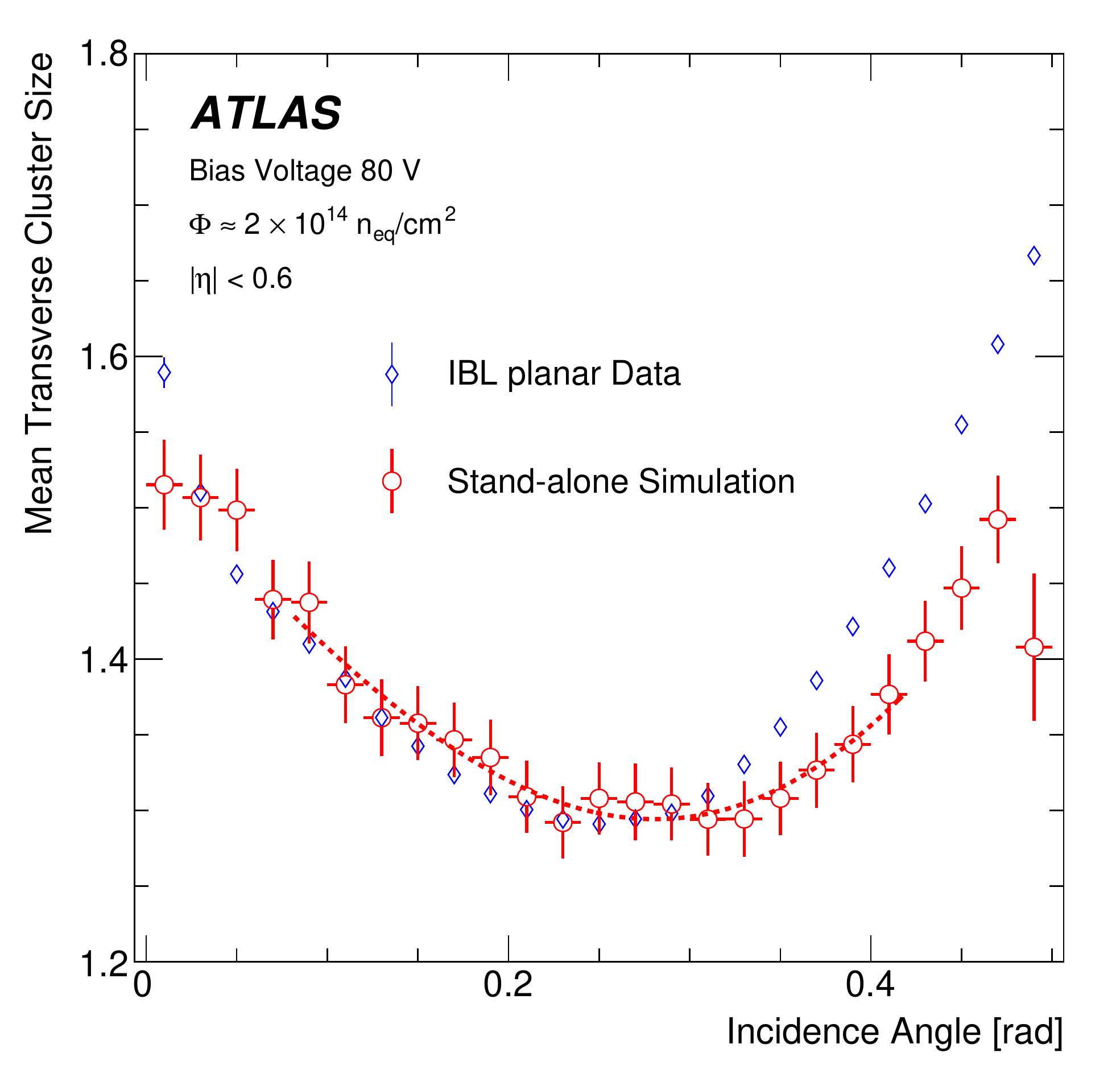}
\end{minipage}
\begin{minipage}{0.45\textwidth} \centering
\includegraphics[width=1.\textwidth]{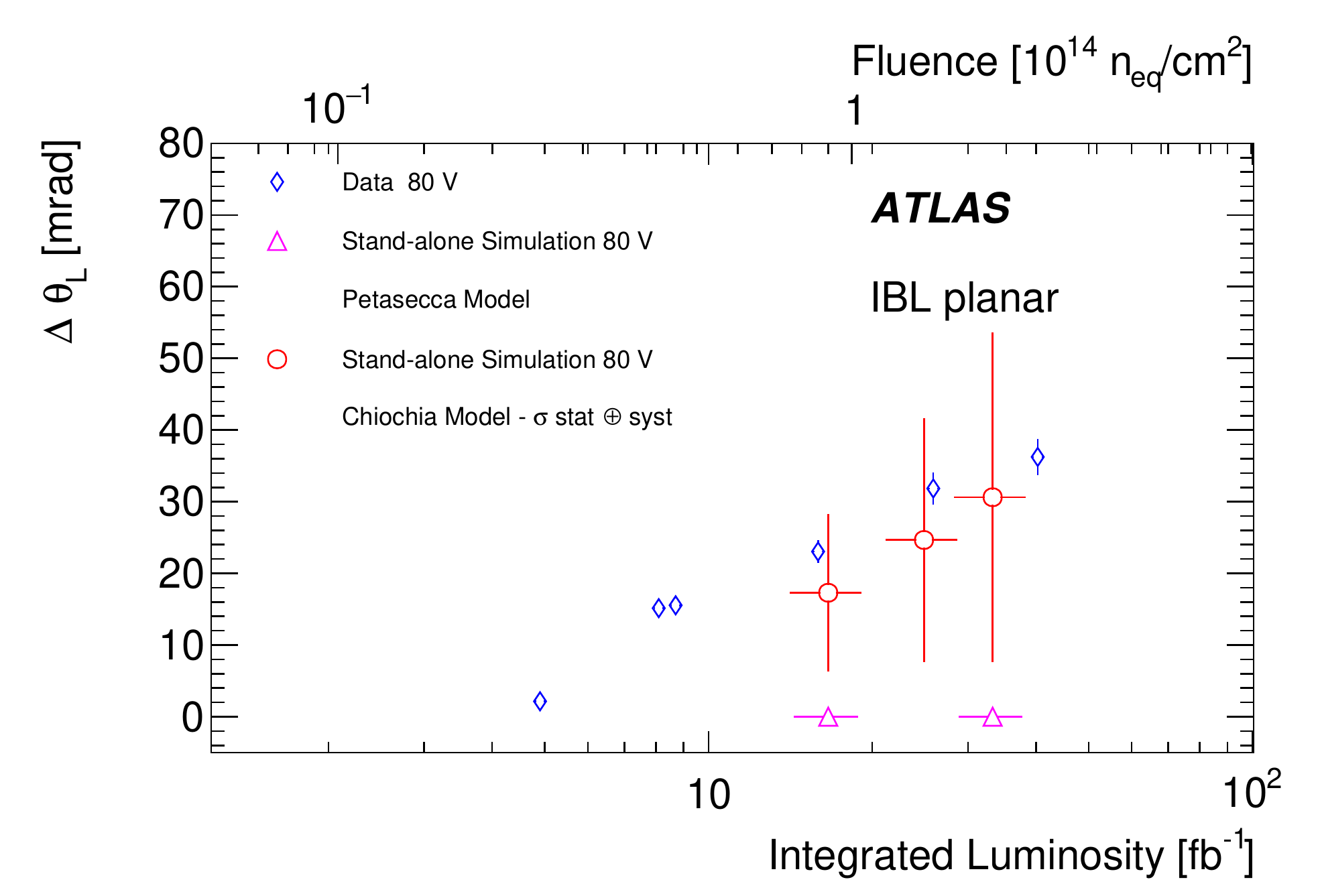}
\end{minipage}
\caption{The mean transverse cluster size versus transverse incidence angle near the end of the 2016 Run ($\sim2\times 10^{14}$ $\mathrm{n_{eq}/{cm}^{2}}$) with a bias voltage of 80~V (left). The change in the Lorentz angle ($\theta_{\text{L}}$) from the unirradiated case as a function of the integrated luminosity in 2015-2016 (right). Two TCAD radiation damage models are considered, Chiochia and Petasecca. Chiochia model points have both statistical and systematic uncertainties, while Petasecca only the statistical uncertainties. From Ref. \cite{Aaboud:2019wgd}.}
\label{fig:LA2:Run2}
\end{figure}

The Lorentz angle is determined by performing a fit to the transverse cluster size $F$ as a function of the incidence angle of the associated track using the following functional form:
\begin{linenomath}
\begin{align*}
F(\alpha)=[a\times |\tan\alpha-\tan\theta_{\text{L}}|+b/\sqrt{\cos\alpha}]\otimes G(\alpha|\mu=0,\sigma),
\end{align*}
\end{linenomath}
where $\alpha$ is the incidence angle with respect to the normal direction of the sensor in the plane perpendicular to the magnetic field. $\theta_{\text{L}}$ is the fitted Lorentz angle, $G$ is a Gaussian probability distribution evaluated at $\alpha$ with mean $0$ and standard deviation $\sigma$, and $a$ and $b$ are two additional fit parameters related to the depletion depth and the minimum cluster size, respectively. An example input to the fit is shown in Fig.~\ref{fig:LA2:Run2} (left).
Cluster size depends on many effects that are not included in the simulations, nonetheless, the position of the minimum should still correspond to the Lorentz angle. For example, the geometry used for this simulation is simplified and the extreme incidence angles are likely more impacted in the actual geometry. The simulation in Fig.~\ref{fig:LA2:Run2} (left) matches the low incidence angles well, but this is not seen for all fluences; it could be due in part to the uncertainty in the fluence.\\

The fitted Lorentz angle as a function of integrated luminosity is shown in Fig.~\ref{fig:LA2:Run2} (right). Due to the degradation in the electric field, the mobility and thus the Lorentz angle increase with fluence. This is not true for the Petasecca model, which does not predict regions of low electric field.
Charge trapping does not play a significant role in the Lorentz angle prediction.  The overall normalisation of the simulation prediction is highly sensitive to the radiation damage model parameters, but the increasing trend is robust. An overall offset (not shown) is consistent with previous studies and appears even without radiation damage (zero fluence)~\cite{ATL-INDET-PUB-2018-001}, which is why only the difference in the angle is presented.

\begin{figure}[!htb]
\centering
\begin{minipage}{0.8\textwidth}\centering
\includegraphics[width=1.\textwidth]{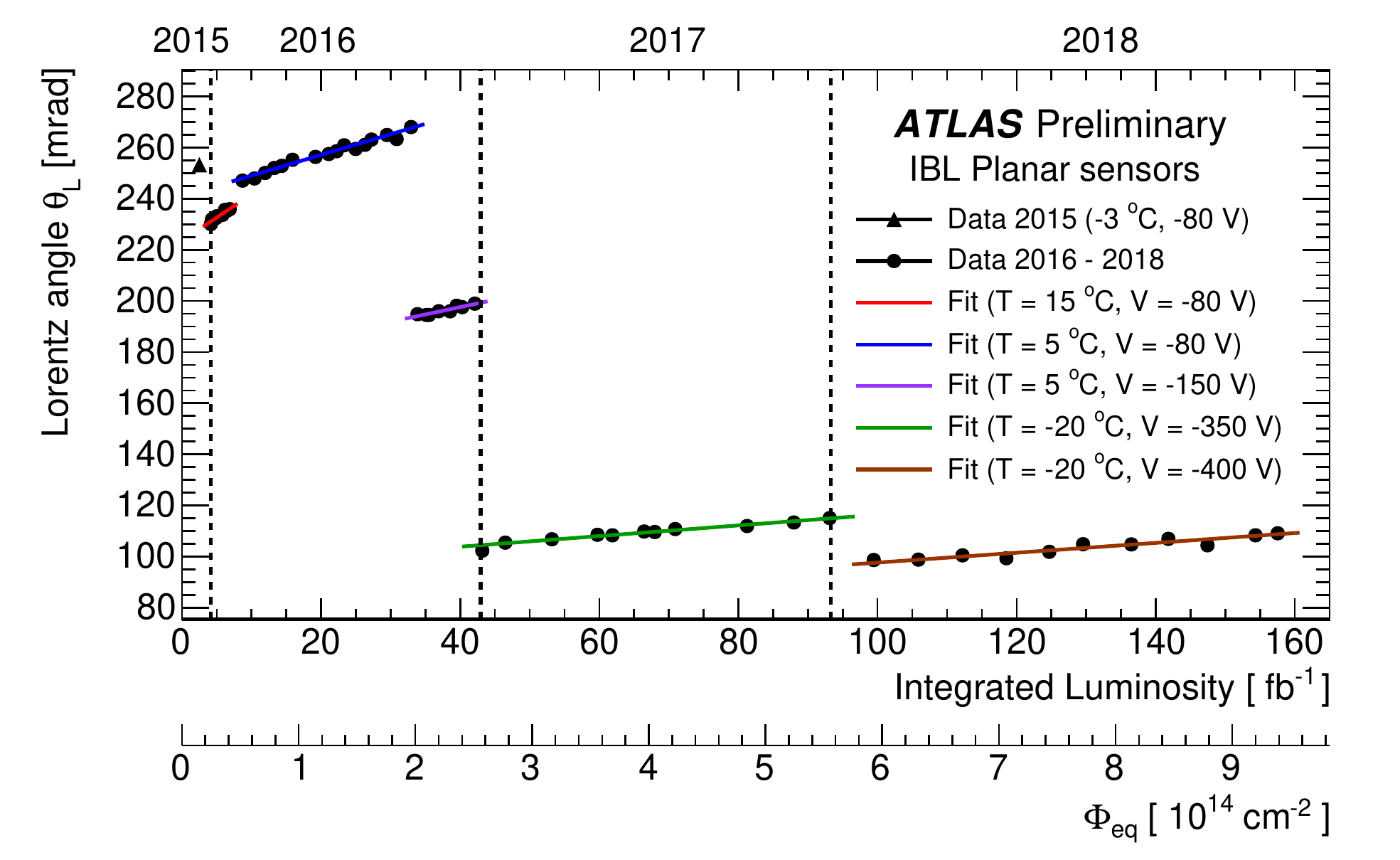}
\end{minipage}
\begin{minipage}{0.8\textwidth}\centering
\includegraphics[width=1.\textwidth]{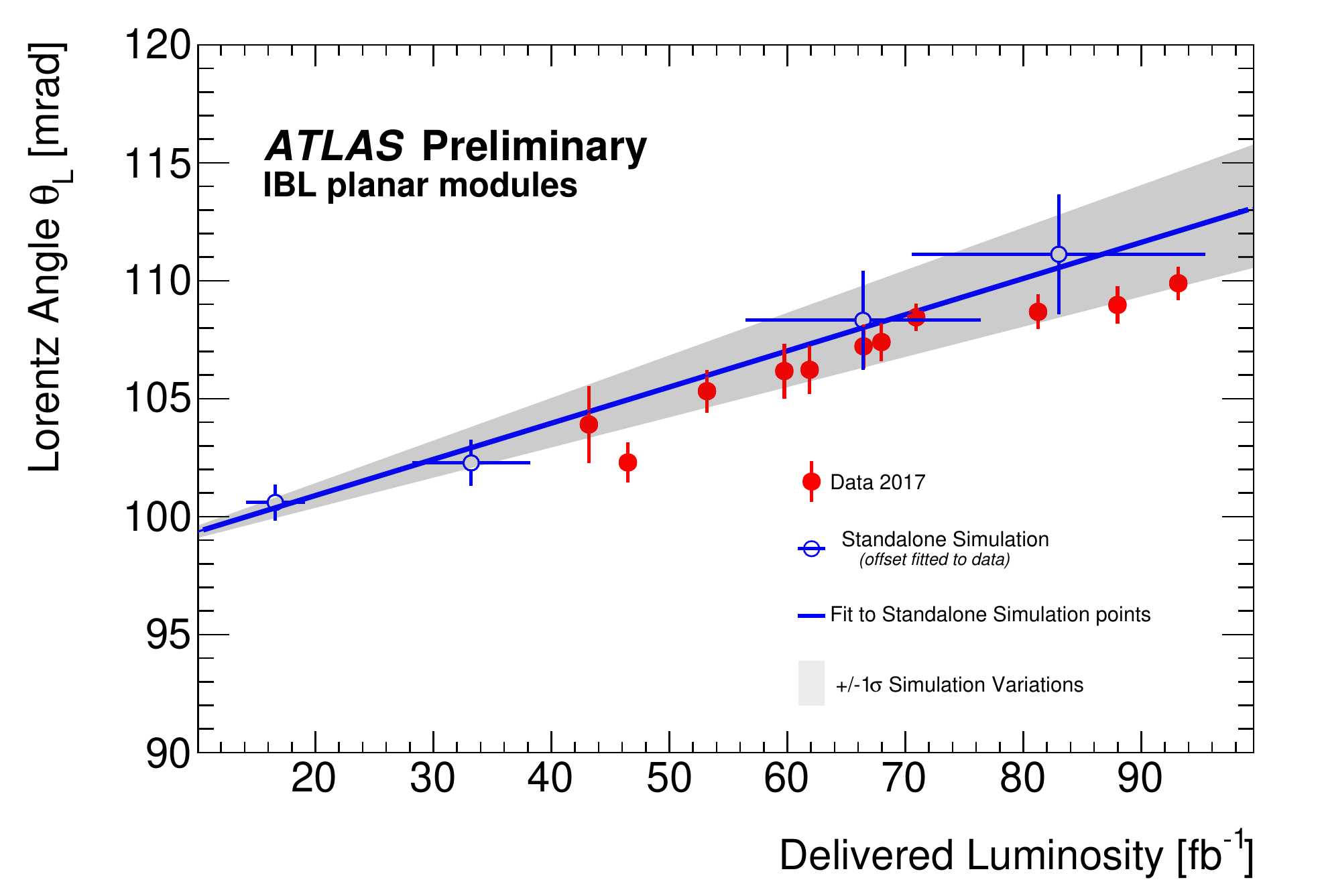}
\end{minipage}
\caption{The evolution of the Lorentz angle ($\theta_{\text{L}}$) during the full Run~2 with only data (top), and during 2017 compared with simulations from Allpix using the Chiochia model (bottom). From Ref. \cite{RadiationLorentzAngle}.}
\label{fig:LAFullRun2Comp}
\end{figure}

Fig.~\ref{fig:LAFullRun2Comp} (top) shows the evolution of the Lorentz angle during Run~2. Different fits for data are performed for different conditions of temperature and bias voltage. Differences in Lorentz angle are due to changes in bias voltage and temperature. Fig.~\ref{fig:LAFullRun2Comp} (bottom) shows the evolution of the Lorentz angle during 2017 compared with the prediction from the Allpix simulations. The simulated points are also fitted with a straight line that has an offset fixed to match the one from the data. Error bands account for all the systematic variations.

\subsubsection{Model validation with Run~3 data}
\begin{figure}[!htb]
\centering
\includegraphics[width=0.65\textwidth]{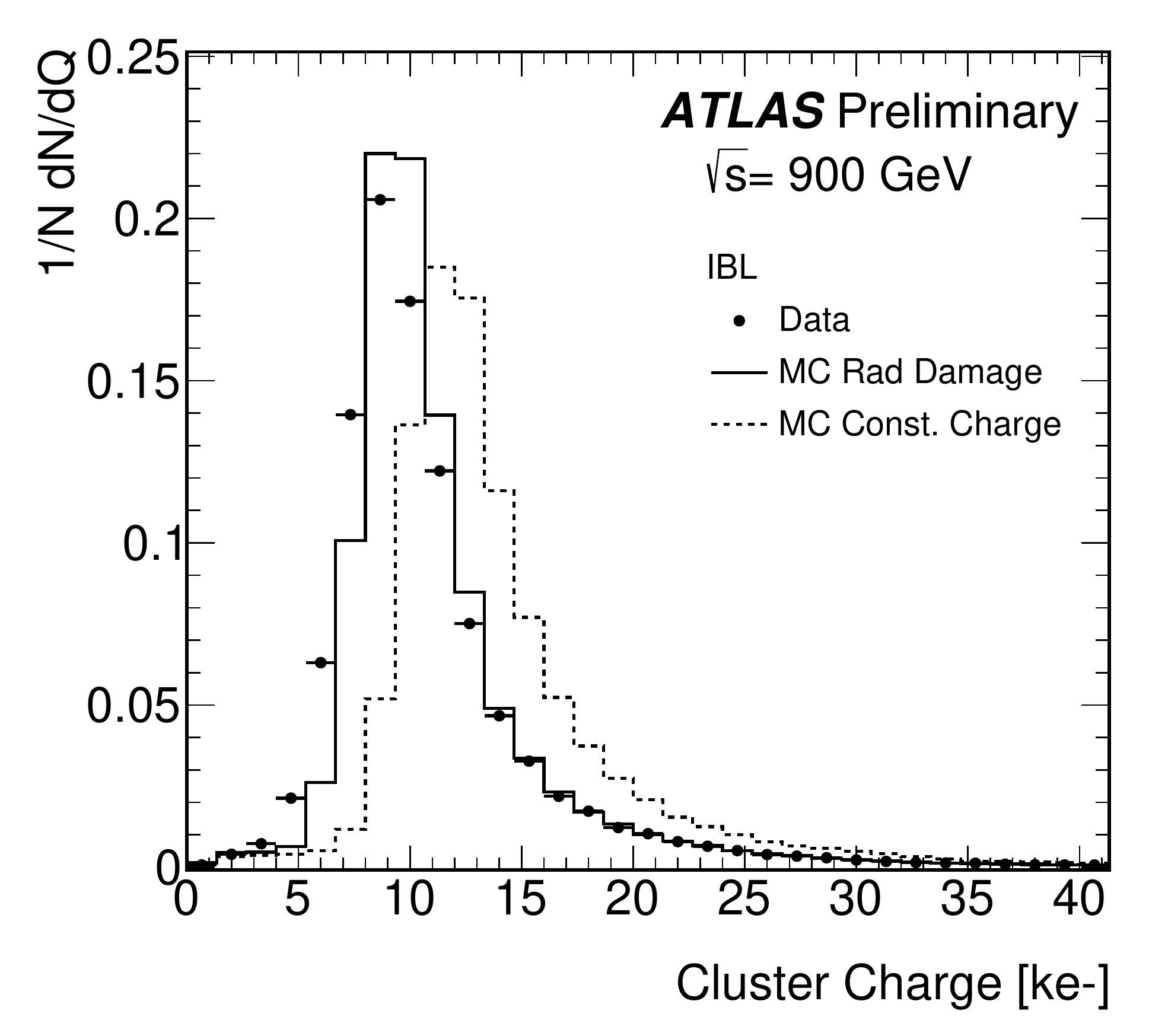}
\caption{Cluster charge for the IBL as measured in data and obtained from simulations with radiation damage (continuous line histogram) and constant charge (dashed line histogram). From Ref.~\cite{ATL-PHYS-PUB-2022-033}.}
\label{fig:data22-RadDam}
\end{figure}

\begin{figure}[!htb]
\centering
\begin{minipage}{0.45\textwidth}
\includegraphics[width=1.\textwidth]{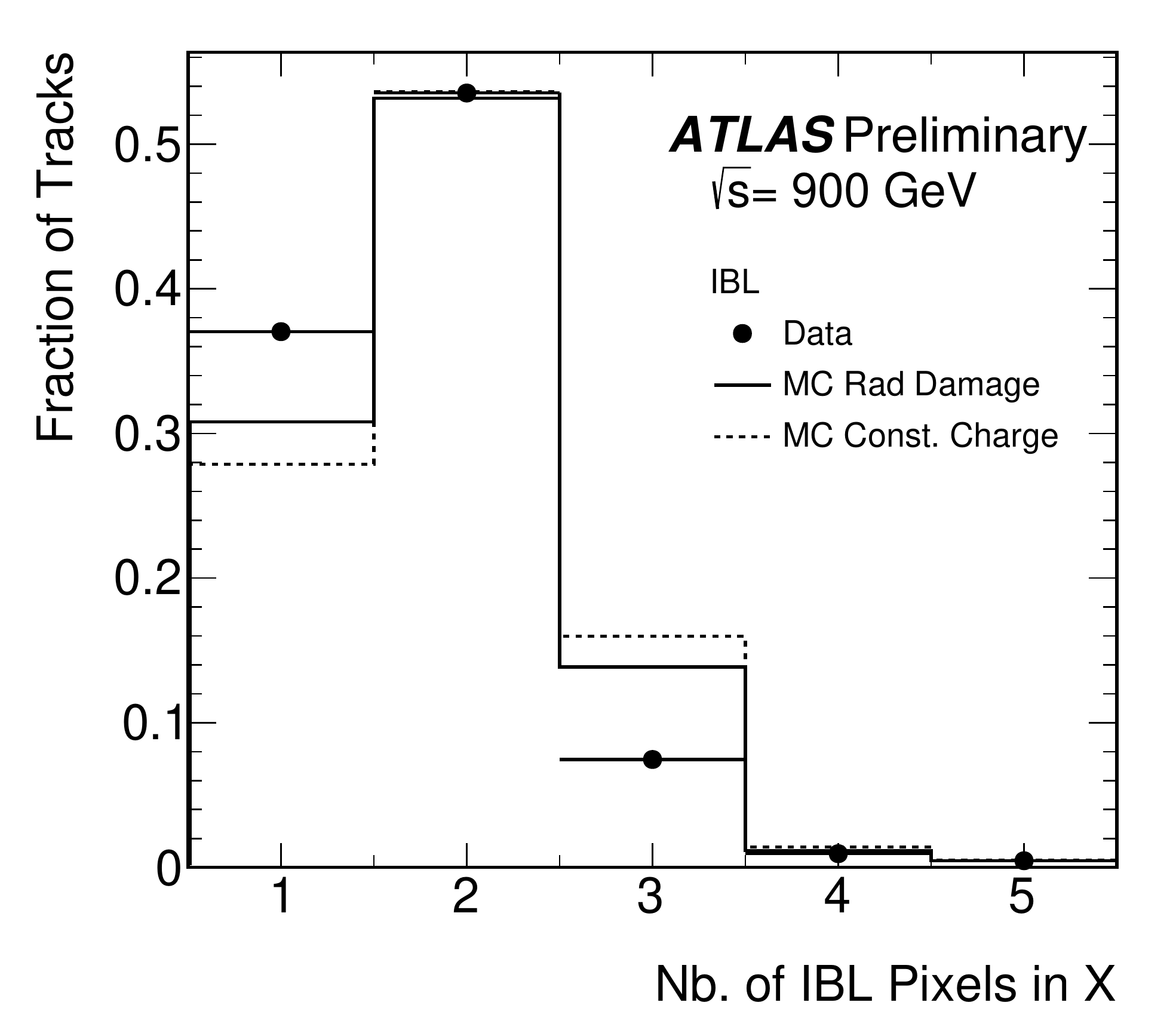}
\end{minipage}
\begin{minipage}{0.45\textwidth}
\includegraphics[width=1.\textwidth]{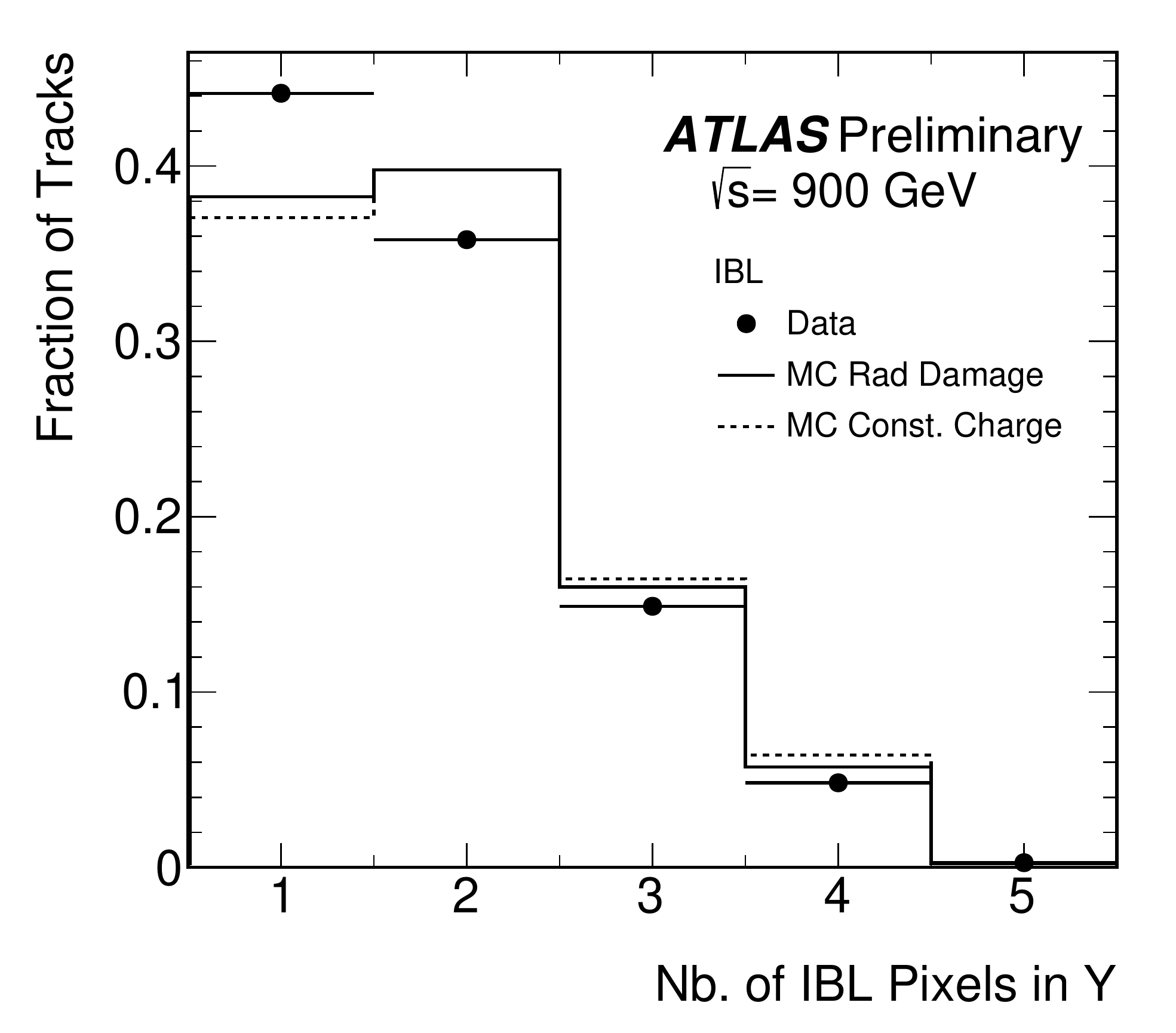}
\end{minipage}
\caption{Number of pixels in clusters in the transverse (left) and longitudinal (right) projections to the beam axis for the IBL. From Ref.~\cite{ATL-PHYS-PUB-2022-033}.}
\label{fig:data22-RadDam-cluster}
\end{figure}

\begin{figure}[htb]
\centering
\begin{tabular}{cc}
\includegraphics[width=0.87\textwidth]{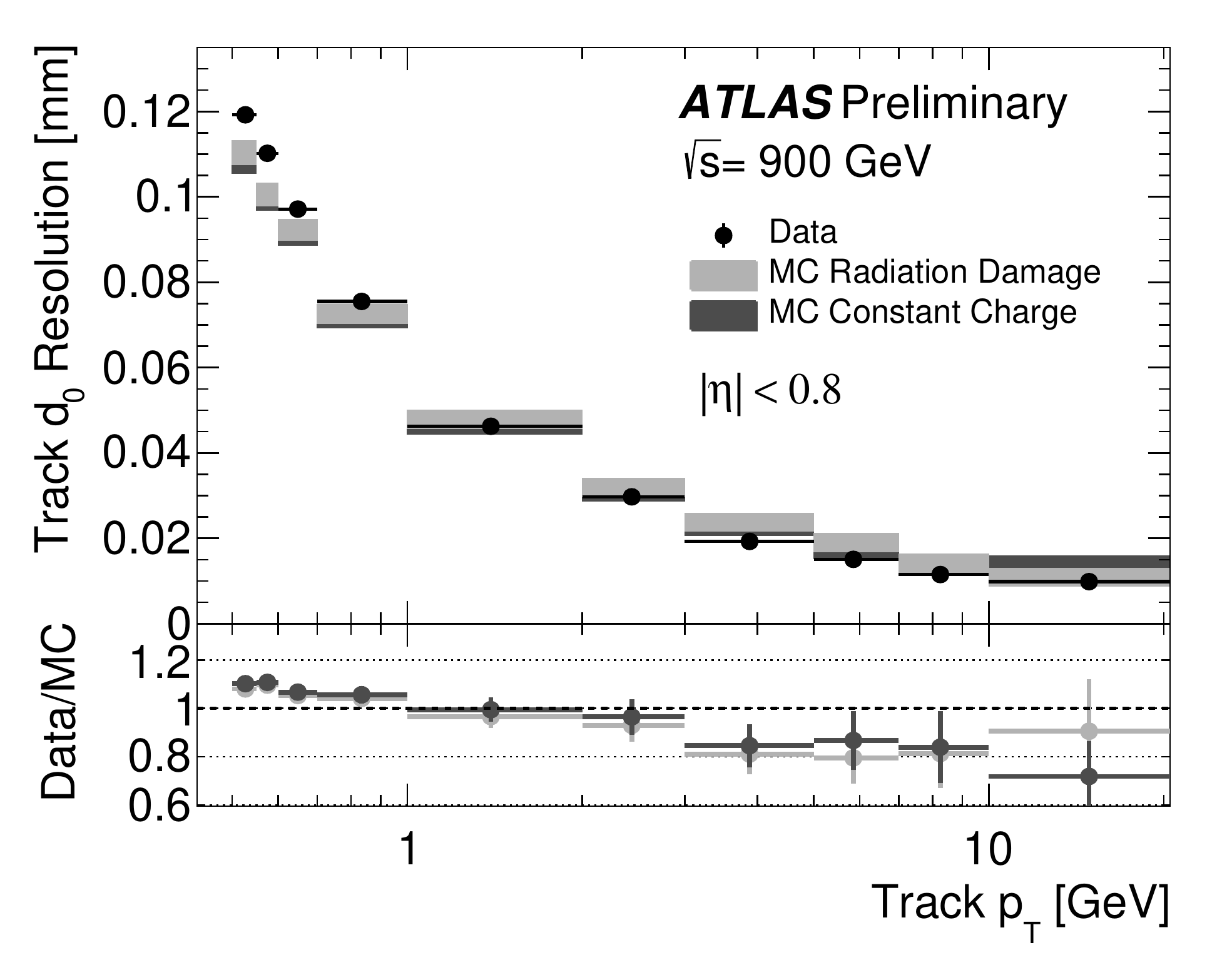}
\end{tabular}
\caption{Transverse $d_0$ resolution measured in data as a function of the track $p_{T}$ for $|\eta|<$0.8 in comparison to the predictions of radiation damage and constant charge simulation. The lower panel shows the ratio of the resolution measured in data to the MC predictions. From Ref.~\cite{ATL-PHYS-PUB-2022-033}.}
\label{fig:d0ResVsPt}
\end{figure}

The radiation damage ditigizer is further studied using a data sample corresponding to events recorded during the special LHC operation at a centre-of-mass energy of 900 GeV in June 2022 at the start of Run~3~\cite{ATL-PHYS-PUB-2022-033}. Simulated samples containing a total of 82 k events are generated both with and without the inclusion of the radiation damage effects and referred to as \textit{Radiation Damage MC} and \textit{Constant Charge MC}, respectively. An estimated average fluence of $7.5 \times 10^{14} \, \mathrm{n_{eq}/cm^{2}}$, expected on the IBL in the 2022 run after 12.5 $\mathrm{fb^{-1}}$ of data collection, is used in the electric field maps.\\
As a first measure of the detector response and its sensitivity to radiation damage, the charge collected in pixel clusters is shown in Fig.~\ref{fig:data22-RadDam}.
The collected charge with the inclusion of the radiation damage is reduced, as expected, and the simulation with the inclusion of the radiation damage provides a better agreement to data. The differences between data and Radiation Damage MC can be explained in terms of uncertainties in the radiation damage model parameters, as well as the uncertainty in the charge calibration and luminosity-to-fluence conversion. As a second measure of the pixel detector response, the distribution of the number of pixels in clusters associated to tracks is shown in Fig.~\ref{fig:data22-RadDam-cluster} (left) for the transverse ($r-\phi$ or local X) projection to the beam axis and in Fig.~\ref{fig:data22-RadDam-cluster} (right) for the longitudinal ($z$ or local Y) projection to the beam axis.
These distributions depend on the incidence angle of the particle and the position of impact of the particle on the detector plane but also on the charge diffusion and the pixel thresholds. The inclusion of the radiation damage effects clearly improves the modelling of the cluster shape. Differences between the two MC models in the transverse projection are due to the change in the collected charge and in the calculated Lorentz angle that drives the cluster multiplicity along the small pixel pitch projection.\\
Finally, the resolution of the impact parameter $d_{0}$ is studied for a subset of reconstructed tracks. The impact parameter $d_{0}$ is defined as the point of closest approach of the track to the primary vertex in the plane transverse to the beam axis. Its resolution is measured as a r.m.s. extracted from a fit on the central portion of the $d_{0}$ peak contained within a $\pm \, 1.5 \sigma$ interval centred at the maximum value of the distribution. The track $d_{0}$ resolution is shown in Fig.~\ref{fig:d0ResVsPt} as a function of the track $p_{\mathrm{T}}$. 
The track $d_{0}$ resolution measured in data is close to that obtained with data at the start of Run~2~\cite{ATL-PHYS-PUB-2015-018} and the agreement to simulation is improved with the inclusion of the radiation damage effects.

\chapter{Assembly and quality control of ITk pixel modules}
\label{sec:ModuleAssembly}

In this Chapter, the assembly and quality control of pixel modules for the Phase-II upgrade of the ATLAS Inner Tracker (ITk) is presented. The assembly procedure is the production stage of pixel modules when different components of a module are joined together, while the quality control is the set of procedures aimed at assuring that the pixel module components reflect the design requirements and that the assembled module can be operated as expected.

\minitoc
\medskip

\section{ATLAS Phase-II detector upgrade}
\label{ssec:phaseIIdetector}

The ATLAS detector will undergo a significant set of changes in preparation for the HL-LHC. First, there are changes to detector systems that are related to radiation damage. This comes either from the damage the existing systems will have already suffered or from the fact that these existing systems were not designed to accept the fluences that will result from HL-LHC. Secondly, there are changes related to the increases in trigger rates and detector occupancy that comes about when large numbers of interactions occur within each beam crossing.\\

\subsubsection{ITk}
\begin{figure}[!htb]
\centering
\includegraphics[width=1.\textwidth]{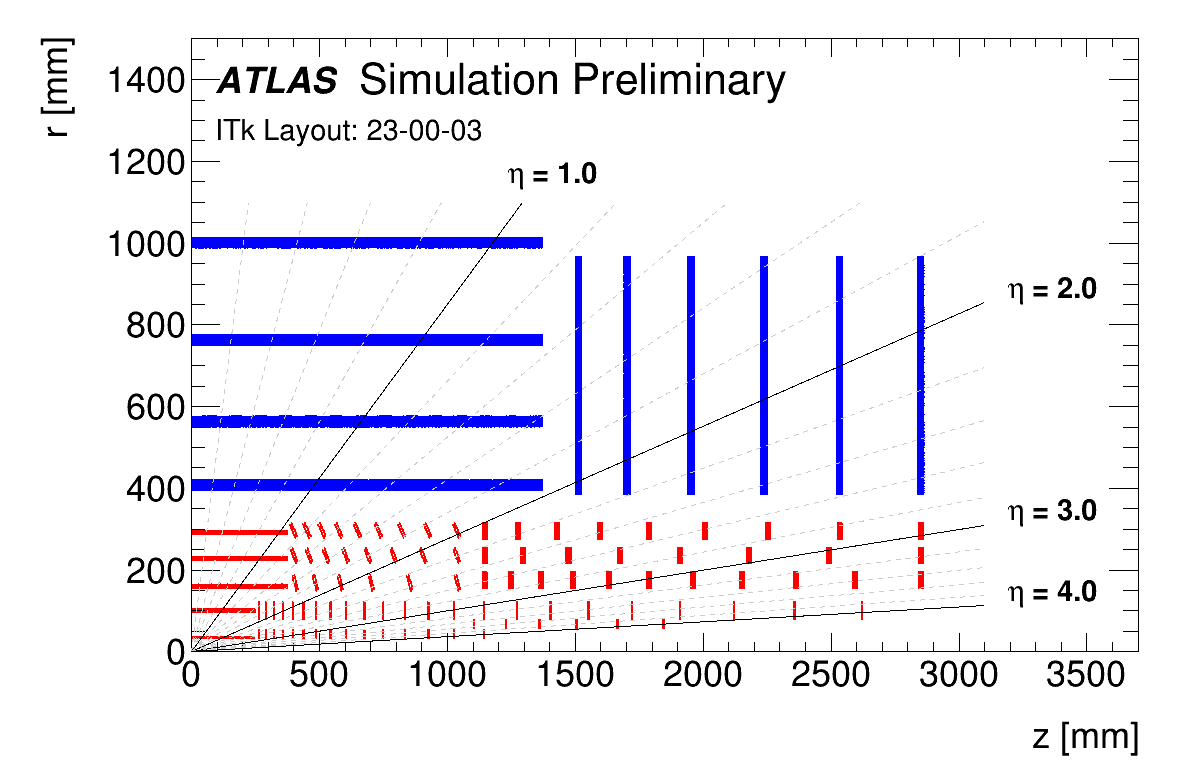}
\includegraphics[width=1.\textwidth]{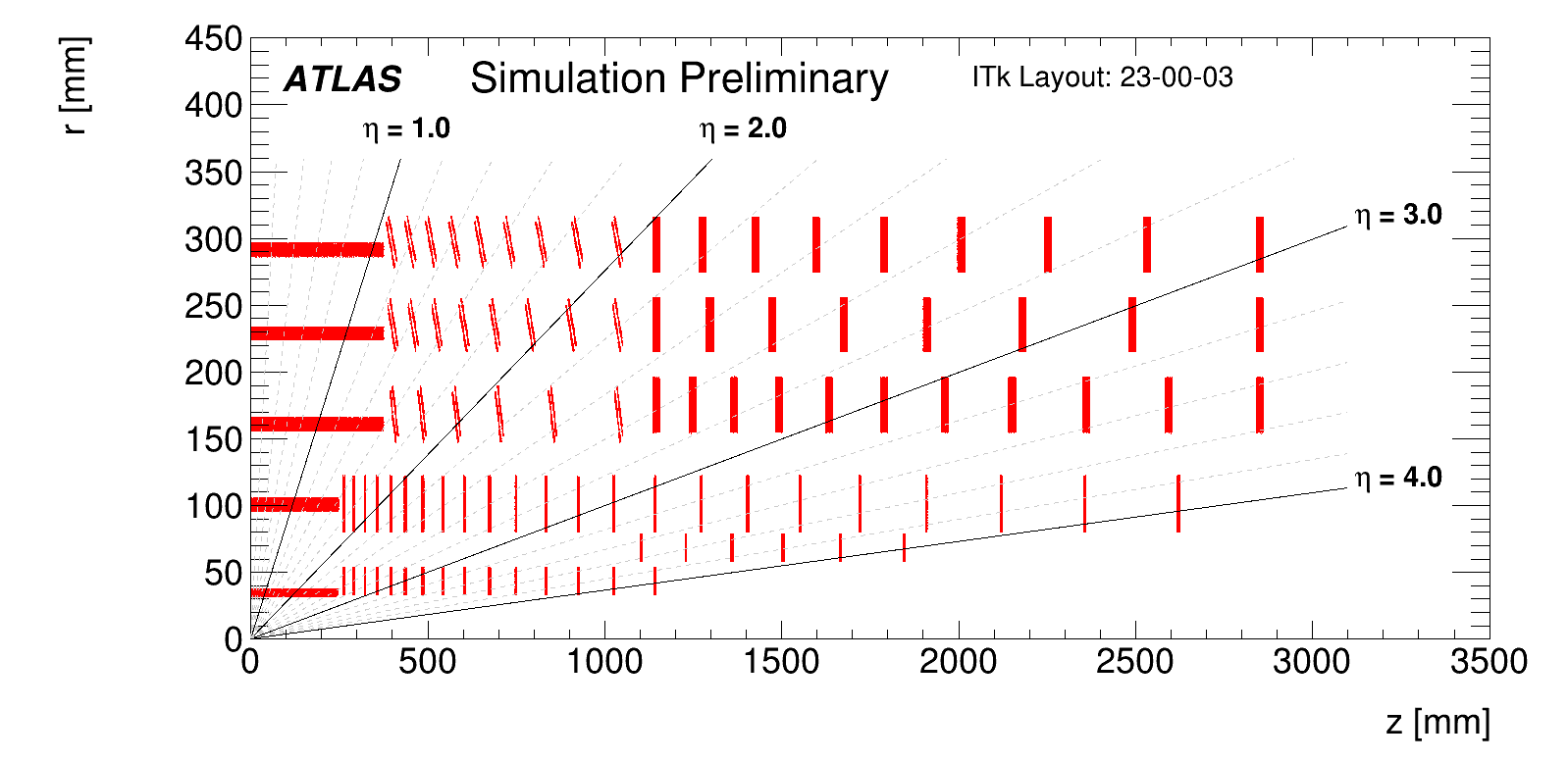}
\caption{Top: A schematic layout of the ITk Layout 23-00-03 as presented in \cite{ATL-PHYS-PUB-2021-024}. A zoomed-in view of the pixel detector. In each case, only one quadrant and only active detector elements are shown. The active elements of the strip detector are shown in blue, and those of the pixel detector are shown in red. The horizontal axis is along the beam line with zero being the nominal interaction point. The vertical axis is the radius measured from the interaction point.}
\label{fig:ATLASITkLayout}
\end{figure}

The Phase-II ITk will be a silicon detector \cite{ATLAS-ITkTDR-30,ATL-PHYS-PUB-2021-024}. The whole of the inner tracker (silicon and TRT) will be removed, and it will be replaced with an all-silicon tracker which fills the existing tracking volume and will increase the $|\eta|$ coverage to 4. A schematic of the baseline layout for the ITk is shown in Fig.~\ref{fig:ATLASITkLayout}. It will consist of an inner part made of pixel detectors, with 5 barrel layers and multiple inclined or vertical ring-shaped end-cap disks, and an outer part made of strip detectors, with 4 barrel layers and 6 endcap rings.

The innermost pixel layer will use ${n}^{+}$-in-$n$ 3D pixel sensor technology with a pixel size of $25\times100 \, \mathrm{\mu m^{2}}$.
The pixels in the outer layers and in the end-cap disks will be fabricated using the more standard $n$-in-$p$ technology which does not require double-sided processing and therefore helps to reduce costs as the sensor areas increase. The size for these pixels is $50\times50 \, \mathrm{\mu m^{2}}$. Both inner and outer pixel sensors will be read out electrically with a RD53 chip with a high-speed differential serial signal along the beam line where the digital signal will be converted to optical pulses by the Versatile Link for transmission off detector. The Phase-II pixel detector will have 8.2 $\mathrm{m^{2}}$ of active silicon area and 638 million channels.\\
The Phase-II silicon strip detector will have 165 $\mathrm{m^{2}}$ of active silicon area and 60 million channels. The central barrel region extends up to $z = \pm 1.4$ m and two end-caps extend the length of the strip detector up to $z = \pm 3$ m. The barrel layers consist of 392 staves with modules on both sides (196 staves on each side of $z = 0$). Each barrel stave is populated with 28 modules (14 on each stave side). The strips on the inner two cylinders are 24.1 mm long (short-strips) and those on the outer two cylinders are 48.2 mm long (long-strips). One side of the stave will have sensors with strips oriented axially while the other side will have the same sensors oriented such that they form a small stereo angle with the axial sensors to improve $z$ resolution. The strip modules on each side of the stave are rotated with respect to the $z$-axis by $\pm$ 26 mrad such that there is a total rotation between the strips on each side of the stave of 52 mrad.
The strip end-caps will consist of 6 disks using a \textit{petal} concept, with each end-caps disk composed of 32 identical petals. Each petal has nine modules on each side with six different sensor geometries to cover the wedge shaped petal surface. A stereo angle of 20 mrad is directly implemented in the end-cap sensors to achieve a total stereo angle of 40 mrad.

\subsubsection{The trigger upgrades}
A consequence of higher luminosity running apart from radiation fluence is the increased trigger rate. As a result, there are significant changes planned to the trigger architecture for the Phase-II upgrade. 
The design of the upgraded architecture of the TDAQ is a single-level hardware trigger that features a maximum rate of 1 MHz and 10 $\mu$s latency \cite{ATLAS-TDR-029}. It will be an hardware-based L0 trigger system, composed of the L0 Calorimeter Trigger (L0Calo), the L0 Muon Trigger (L0Muon), the Global Trigger and the Central Trigger sub-systems. In the L0Calo sub-system, a new forward Feature EXtractor (fFEX) will be added to reconstruct forward jets and electrons. The new L0Muon sub-system will use an upgraded barrel and end-cap sector logic for the reconstruction of muon candidates in the barrel RPCs and in the endcap TGCs, respectively. In addition, MDT information will be used in new dedicated processors to improve the robustness and efficiency of the L0Muon, its $p_\mathrm{T}$ resolution and selectivity. The Global Trigger will replace and extend the Run~2 and Phase-I Topological Processor by accessing full-granularity calorimeter information to refine the trigger objects calculated by L0Calo, perform offline-like algorithms, and calculate event-level quantities before applying topological selections. The final trigger decision is made by the Central Trigger Processor (CTP), which can apply flexible prescales and vetoes to the trigger items. The CTP also drives the trigger, timing and control system network to start the readout process of the detectors.

\section{ITk pixel modules}
\begin{figure}[!htb]
\begin{center}
\includegraphics[scale=0.25]{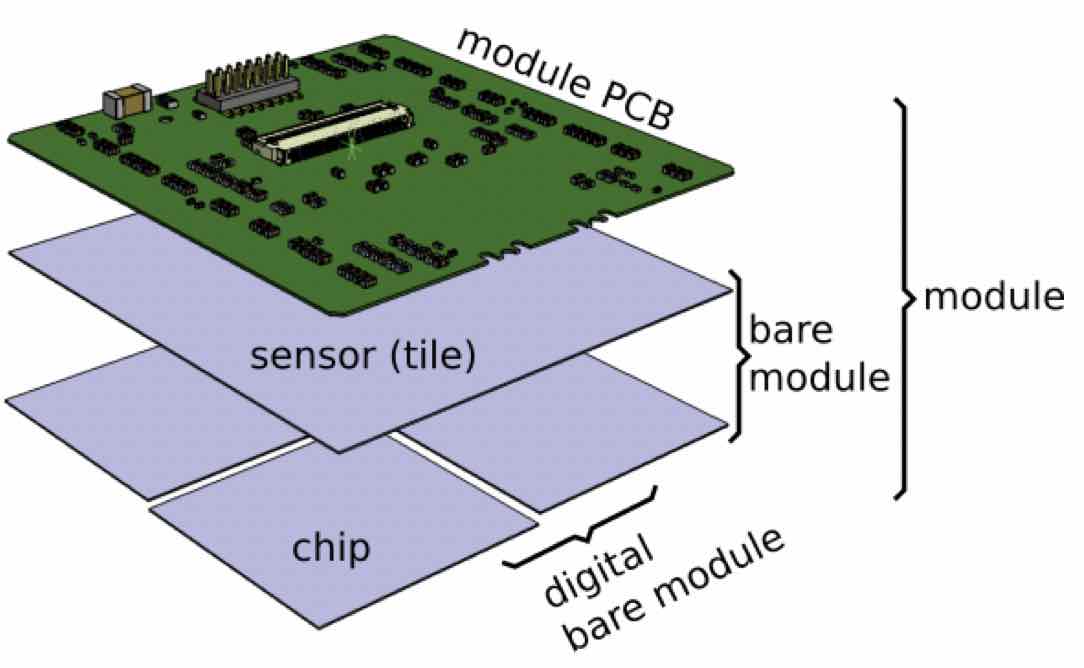}
\caption{Schematics of a pixel module.}
\label{fig:sketch-pixelmodule}
\end{center}
\end{figure}

The design of the hybrid pixel module is similar to the one adopted for the Run~2 Pixel Detector~ \cite{ATLAS-ITkTDR-30}. The hybrid pixel module is shown in Fig.~\ref{fig:sketch-pixelmodule}

and is made of two parts: 
\begin{itemize}
    \item a bare module, consisting of a passive high resistivity silicon sensor and FE read-out chips fabricated in CMOS technology, as depicted in Fig.~\ref{fig:sketch-baremodule}. 
    \begin{figure}[!htb]
    \begin{center}
    \includegraphics[scale=0.3]{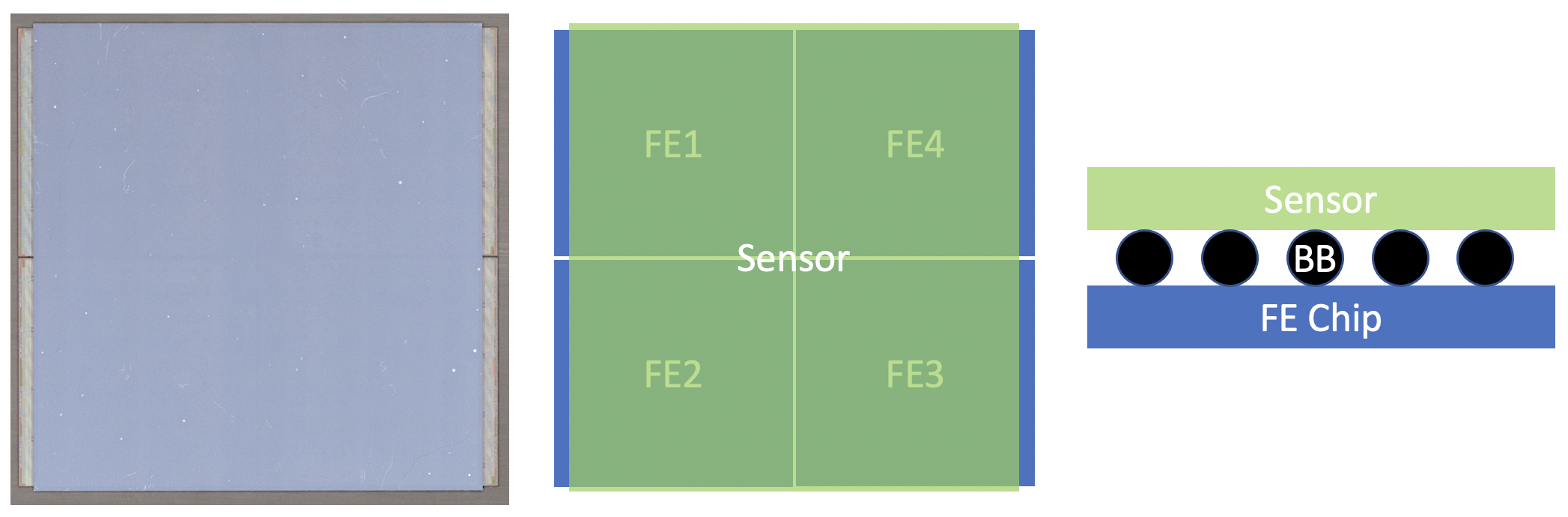}
    \caption{Image of real quad bare module (left), top view diagram of quad bare module components (centre), side view of components including bump bonds (right).}
    \label{fig:sketch-baremodule}
    \end{center}
    \end{figure}
    The silicon sensor and the FE read-out chips are joined using a high-density connection technique called \textit{flip-chip bump-bonding (BB)};
   \item a module flex, consisting of a flexible PCB.
\end{itemize}

All connections to the bare modules are routed to the active elements via the module flex, which is glued to the backside of the sensor. Sometimes we refer to a pixel module without a sensor tile, e.g. FE chips attached to a module PCB, as a digital module.\\
There will be two main types of hybrid pixel modules: 
\begin{itemize}
    \item \textit{quad modules}, consisting of four chips bump-bonded to a single sensor, around $4 \times 4 \, \mathrm{cm^{2}}$ in area, which are used in the outer flat barrel layers and in the outer end-cap rings;
    \item \textit{single-chip modules}, consisting of one FE chip bump-bonded to a sensor, around $2 \times 2 \, \mathrm{cm^{2}}$ in area, which will be arranged into triplets and used in the innermost barrel layer and in the first two end-cap ring layers.
\end{itemize}

\begin{figure}[!htb]
\begin{center}
\includegraphics[scale=0.5]{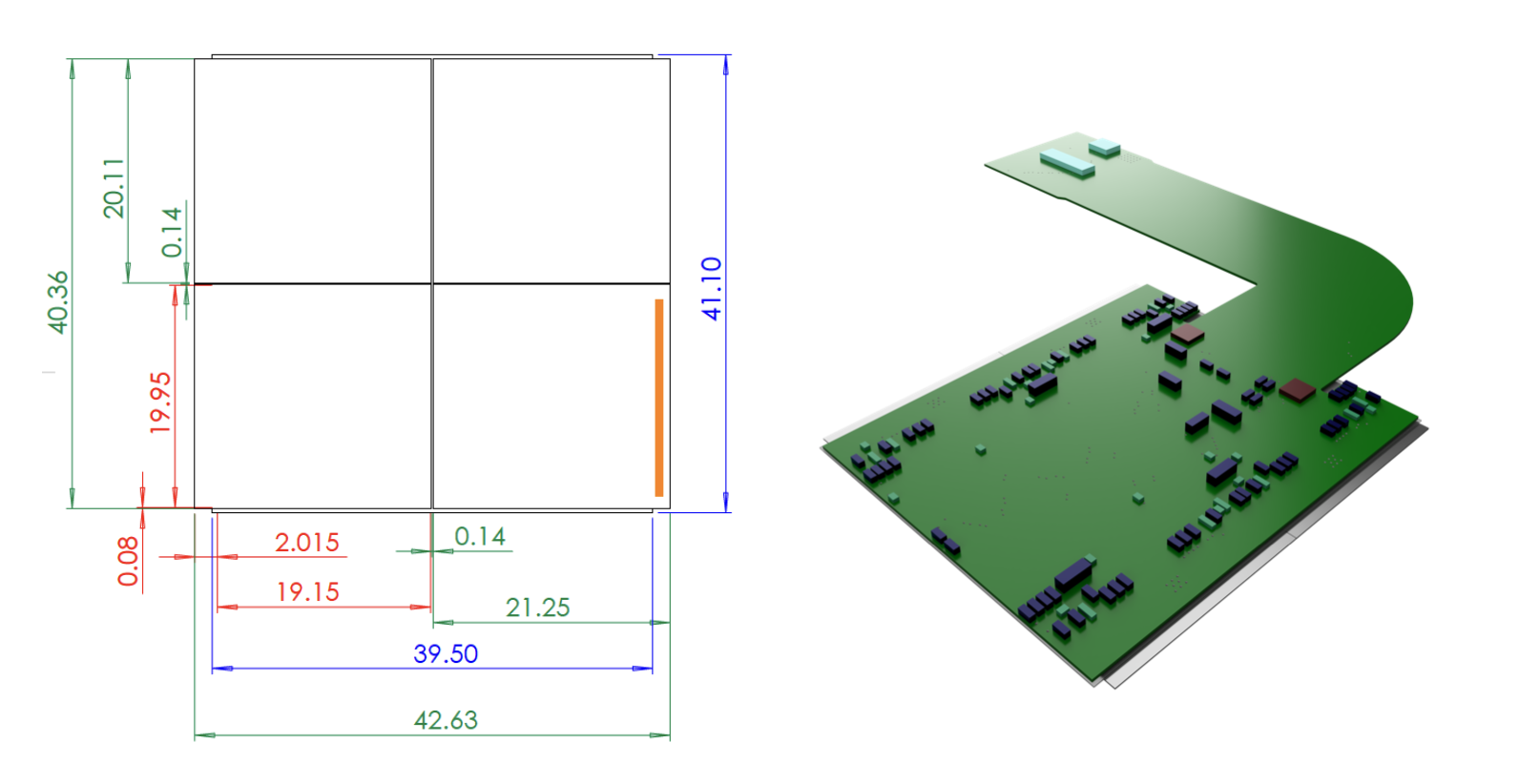}
\caption{Left: Drawing of a quad module with four pixel FE chips viewed from the FE chip side, all dimensions in mm. FE chip dimensions are green with the wire-bond pad area indicated in gold on the lower right chip. Sensor dimensions are in blue and in red the distances from and to the outermost bump-bond pads of the lower left chip are given. Right: 3D view of a quad module for the outer barrel flat section viewed from the module flex side.}
\label{fig:ITk-Module-Picture}
\end{center}
\end{figure}

Fig.~\ref{fig:ITk-Module-Picture} shows a drawing and a three-dimensional rendering of a quad module based on the ATLAS pixel FE chip with $50 \times 50 \, \mathrm{\mu m^{2}}$ pixel size. All connections (clock and command input, data output, low voltage and sensor high voltage) to the modules are routed to the active elements via the module flex which is glued to the backside of the sensor. Internal to the module, the module flex connects the high voltage bias to the sensor’s backside and the low voltage supply to the FE chip. For multi-chip modules, the FE chip chips are connected in parallel for powering from a common low voltage input on the module flex. A single downlink data line is connected to the module and the signal is routed in parallel to each chip in the module. The clock and command signals are extracted from this data by the FE chip. The uplink data streams from the FE chip are multiplexed together into high speed electrical data cables, which are routed to the opto-converters. The data signals to and from the module are transmitted on differential pairs and are AC coupled. For module temperature monitoring a temperature sensor (NTC) is connected to the Pixel DCS chip. The connections to the FE chips and the sensor are made with wire bonds and passive components, such as decoupling capacitors and termination resistors, are mounted on the module flex.
The power consumption of a module is around 7 W and this must be removed to prevent thermal run-away and a per pixel leakage current above 10 nA. The modules will be placed on the local support, with the backside of the FE chips in contact with the support. This interface is part of the thermal path between the module and the cooling fluid.

\subsection{Bare Module}
\begin{figure}[!htb]
\centering
\includegraphics[width=0.55\textwidth]{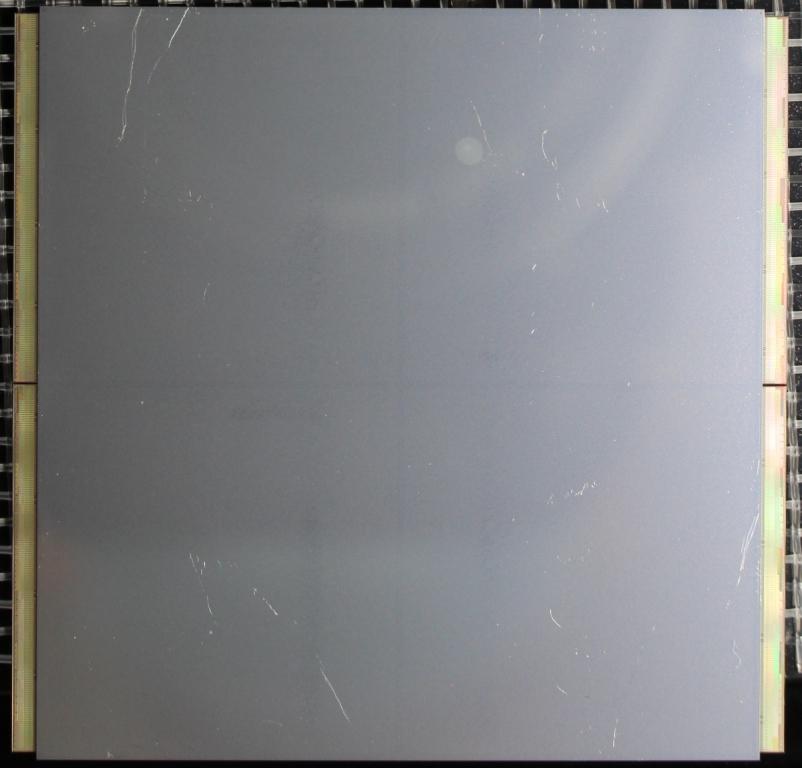}
\caption{Picture of a RD53A bare module.}
\label{fig:RD53A-module}
\end{figure}

The pixel bare modules for ITk are designed in the so-called RD53B format. The first pixel detector chip produced and used during the pre-production stage is called \textit{ITkPix-V1}, while \textit{ITkPix-V2} is currently in a production stage. 
As a prototype chip, the RD53A chip is used for testing purposes in the initial stages. Considering that most of the functionalities needed for the ITk chips are already available in RD53A, there is considerable confidence in working with RD53A chips while waiting for the final \textit{ITkPix} chips produced in the RD53B format. A picture of a RD53A bare module is shown in Fig.~\ref{fig:RD53A-module}.

The nominal dimensions are $(41.15 \pm 0.05) \times (42.25\pm 0.05)\,$mm$^2$ and ($325^{+55}_{-40}$)\,\textmu m thickness. \\

\begin{figure}[!htb]
\centering
\includegraphics[width=0.8\textwidth]{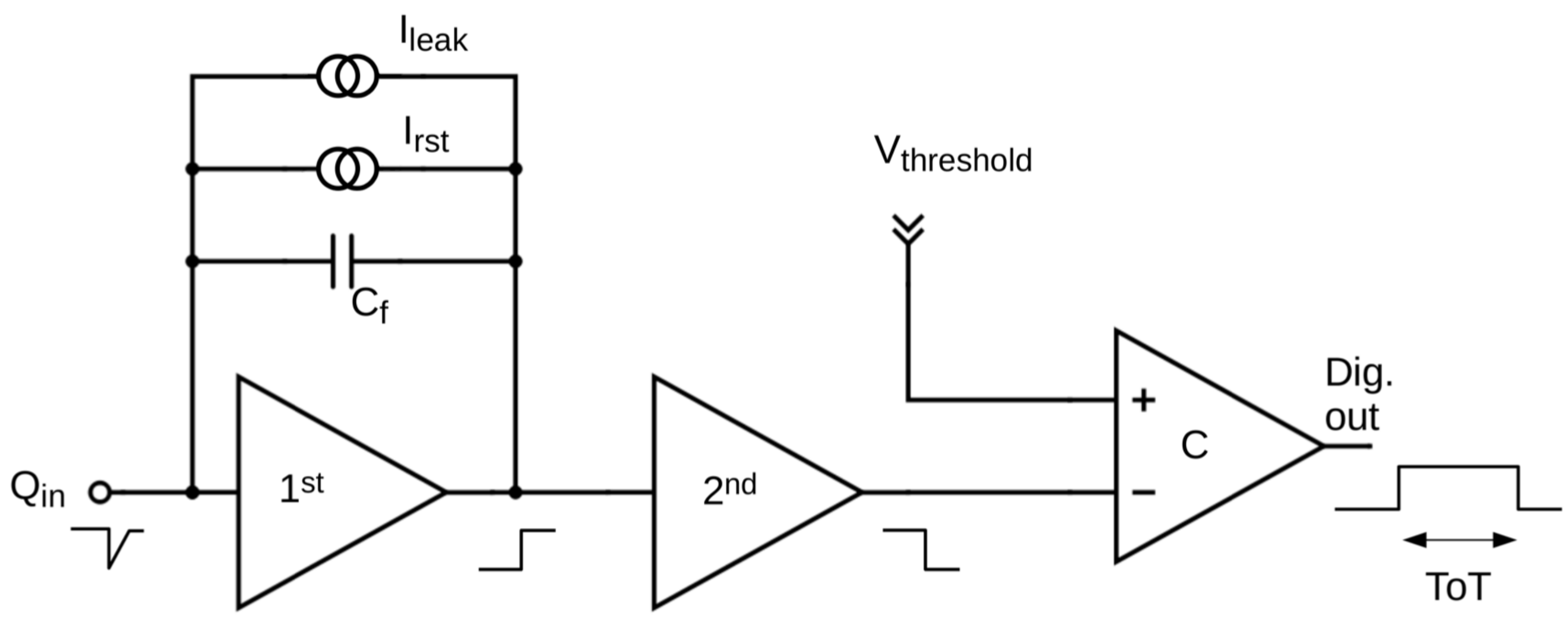}
\caption{Simplified diagram of the analog FE chip.}
\label{fig:AnalogFE}
\end{figure}

The RD53A FE chip is a half-size prototype chip. Pixels are arranged in cores of 8 $\times$ 8 pixels and the readout is performed through 4 $\times$ 1.28 Gbit/s lanes, with 400 cols $\times$ 192 rows of 50 $\times$ 50 $\mathrm{{\mu m}^2}$ pixels. The wire bond pads are daisy chained together to allow the mechanical modules to be used for electrical connectivity testing. A simplified diagram of the circuit of an analog FE chip is shown in Fig.~\ref{fig:AnalogFE}.

The circuit processes the charge $Q_{in}$ received from the bump pad of the sensor and converts it into a digital output. The first signal processing stage is represented by a pre-amplifier with a feedback capability $C_{f}$: this device is generally a Charge Sensitive Amplifier (CSA), which is an element that integrates the current flowing through the inverting terminal generating an amplitude voltage proportional to the input charge. A second stage provides signal gain right before a comparator which is used to discriminate the hit threshold.\\

The RD53A FE chip comprises 3 different analog FE flavours that are substantially different from each other to allow detailed performance comparisons when testing \cite{ATLAS-RD53A-chip}:
\begin{itemize}
    \item \textit{Synchronous} FE uses a baseline \say{auto-zeroing} scheme that requires periodic acquisition of a baseline instead of pixel-by-pixel threshold trimming;
    \item \textit{Linear} FE implements a linear pulse amplification in front of the discriminator, which compares the pulse to a threshold voltage;
    \item \textit{Differential} FE uses a differential gain stage in front of the discriminator and implements a threshold by unbalancing the two circuit branches.
\end{itemize}

\begin{figure}[!htb]
\centering
\includegraphics[width=0.65\textwidth]{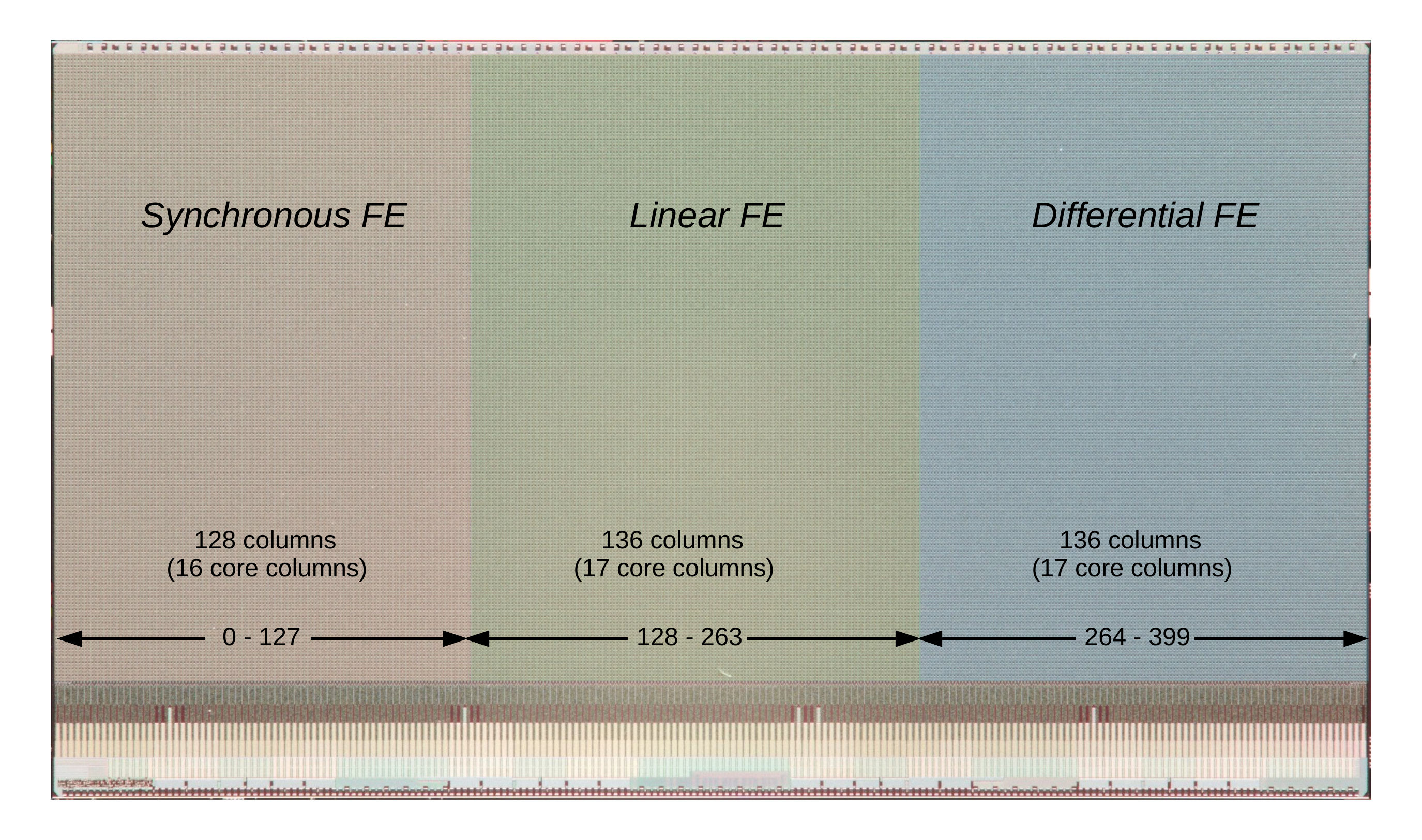}
\caption{Picture of a RD53A FE chip flavours.}
\label{fig:RD53A-FEchip}
\end{figure}

The analog FE flavours are shown in Fig.~\ref{fig:RD53A-FEchip}: synchronous FE in core columns 1-16, linear FE in core columns 17-33 and differential in core columns 34-50. The linear FE flavour has been chosen by the CMS collaboration, while the ATLAS collaboration has chosen to use the differential FE flavour.

\begin{figure}[!htb]
\centering
\includegraphics[width=0.6\textwidth]{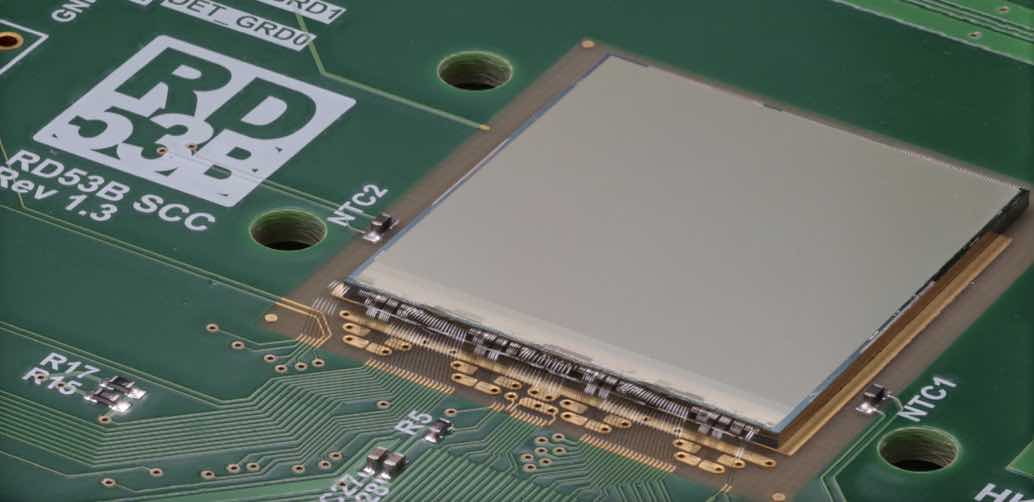}
\caption{ITkPixV1 chip on Single Chip Card.}
\label{fig:RD53B-module}
\end{figure}

A picture of ITkPixV1 chip on Single Chip Card (SCC) is shown in Fig.~\ref{fig:RD53B-module}.
It consists of 400 columns $\times$ 384 rows of 50 $\times$ 50 $\mathrm{{\mu m}^2}$ pixels and differential FE. The chip size is $20 \times 21\,$mm$^2$, with a sustained hit rate of 3 GHz/$\mathrm{cm^{2}}$, a data rate up to 5.12 Gbit/s per chip, and a radiation tolerance of 500 Mrad.

\subsection{Module Flex}
\begin{figure}[!htb]
\centering
\includegraphics[height=0.6\textwidth]{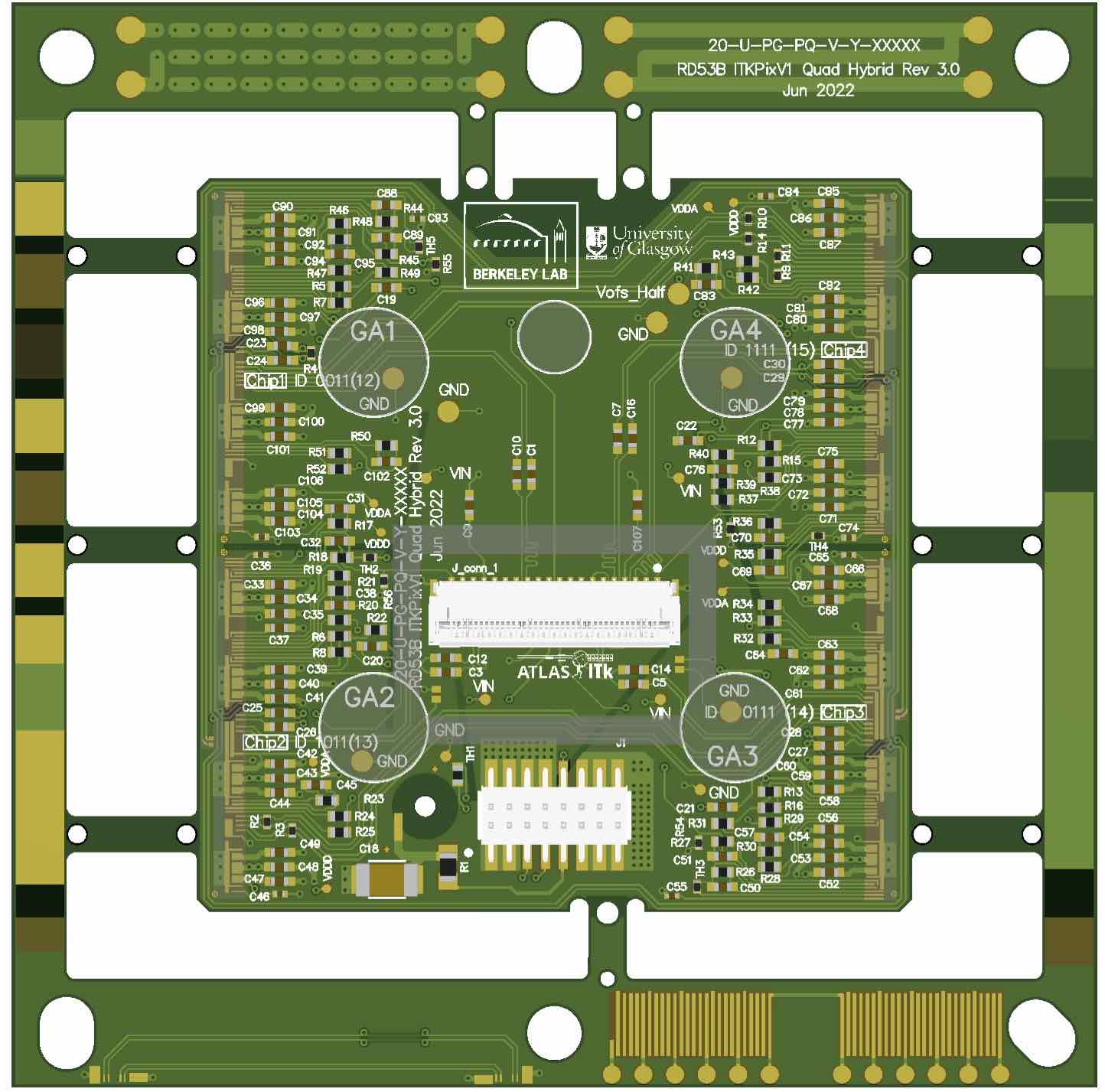}
\caption{Module flex for RD53B ITkPixV1 quad modules.}
\label{fig:flex}
\end{figure}

Each bare module is glued and wire bonded to a module flex PCB. A module flex for a quad module in the RD53B format (ITkPixV1) is shown in Fig.~\ref{fig:flex}.

The module flex contains the necessary Low Voltage (LV), High Voltage (HV), DCS, data and command lines, and several passive and active components that are needed to operate and read out the FE chips. The module flex consists of two metal layers, where one of the layers serves as the line for the supply current, while the second layer represents the local module ground. Data lines, the supply line for the sensor bias voltage and the lines for the DCS system are placed within these layers. The sensor bias voltage is routed to one or more holes on the module flex, from where it is supplied to the sensor backside via wire bonds. Termination resistors and coupling capacitors are mounted on the flex for the clock and trigger signals, while bias resistors and filter capacitors are mounted for LV and HV supply lines.

\section{Stages of the module assembly and quality control}
ATLAS is currently in the pre-production phase, a period during which tests are conducted in the laboratory to prepare for the final production phase of the hybrid modules. During pre-production, a standard operating routine is defined and several studies are conducted to verify that each production stage is carried out optimally, by considering a set of criteria involving precision, needed time, and cost. At the end of the pre-production phase, a set of guidelines ensuring the sustained and reliable production rate of hybrid pixel modules is developed and the readiness of the laboratory for the next phase is certified. As part of the ATLAS pixel module team at CERN, I was involved in several key activities that start from the reception of the pixel module components at CERN and end with the submission of the assembled module for cell-loading, mainly involving the assembly and the quality control of the components to produce the assembled hybrid module.\\

The module assembly and quality control comprises various stages:
\begin{itemize}
\item Visual inspection, metrology and initial probing
\item Module assembly, including gluing the module flex to the bare module
\item Assembly metrology
\item Wire bonding
\item Plasma cleaning
\item Electrical tests including X-ray scans.
\end{itemize}

In the following sections, the most important aspects of these stages are discussed.

\section{Visual Inspection}
The visual inspection of components forms part of the quality checking stage to ensure that anomalies or damages are identified early to remove components that do not meet the specifications from the assembly process. This applies to all components on reception and prior to assembly.

\subsection{Visual inspection of bare modules}
\begin{figure}[!ht]
\centering
\includegraphics[width=12cm]{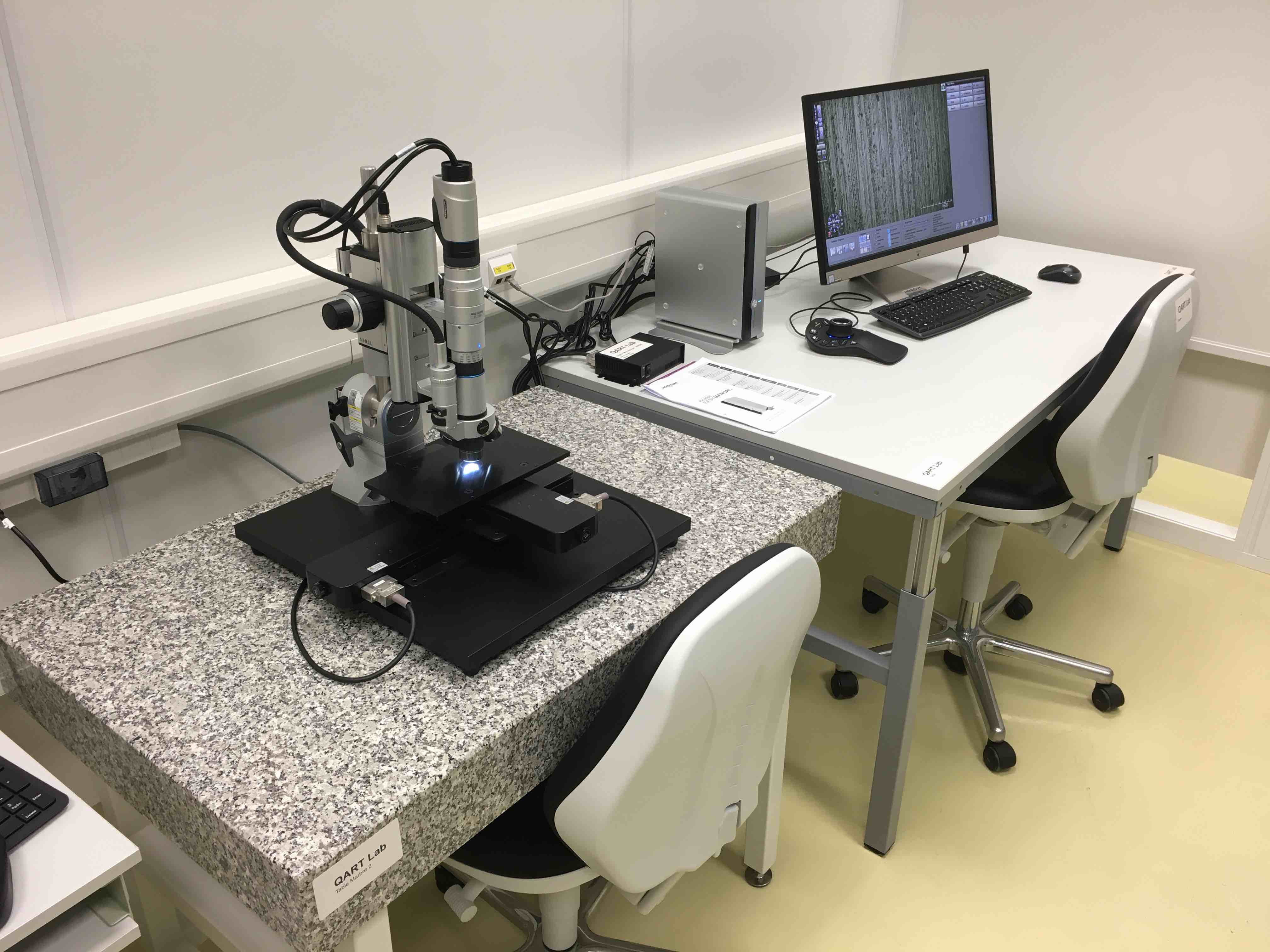}
\caption{The QART lab Hirox MXB-5000REZ optical system.}
\label{fig:HiroxMicroscope}
\end{figure}

Detailed microscope inspection at different lighting and angles is needed to find production imperfections and features that are not obvious, e.g. loose wire bonds, corrosions and excess encapsulant. Sometimes other means might be needed, e.g. use dry air to check loose wire bonds or to blow away debris. At CERN, the visual inspection of the modules is carried out using the QART lab Hirox MXB-5000REZ optical system with a high magnification capability to image the different regions of the module (see Fig.~\ref{fig:HiroxMicroscope}). Due to the nature of the silicon surface of the bare modules, the need for polarised light along with a polarisation filter is essential. The bare modules are placed directly on a microscope table.

In order to produce systematically comparable inspection images of each bare module, the lighting and optical settings have to be consistent for the entire batch. To this end, the following list details the optimum settings adopted for each inspection:
\begin{itemize}
    \item Magnification: for large area features, such as chip corners or general surface features, low magnification lens at $100\times$. For closer inspections, such as wire bonds or bond pads, magnification lens at $600\times$;
    \item Polarisation filter: the filter must be kept to the second notch from the right;
    \item Directional Lighting Adapter: the knob must be set away from the user, to its maximum point.
    \item Brightness: this value is set to 50\% on the software user interface.
\end{itemize}

\begin{figure}[!ht]
\centering
\includegraphics[width=12cm]{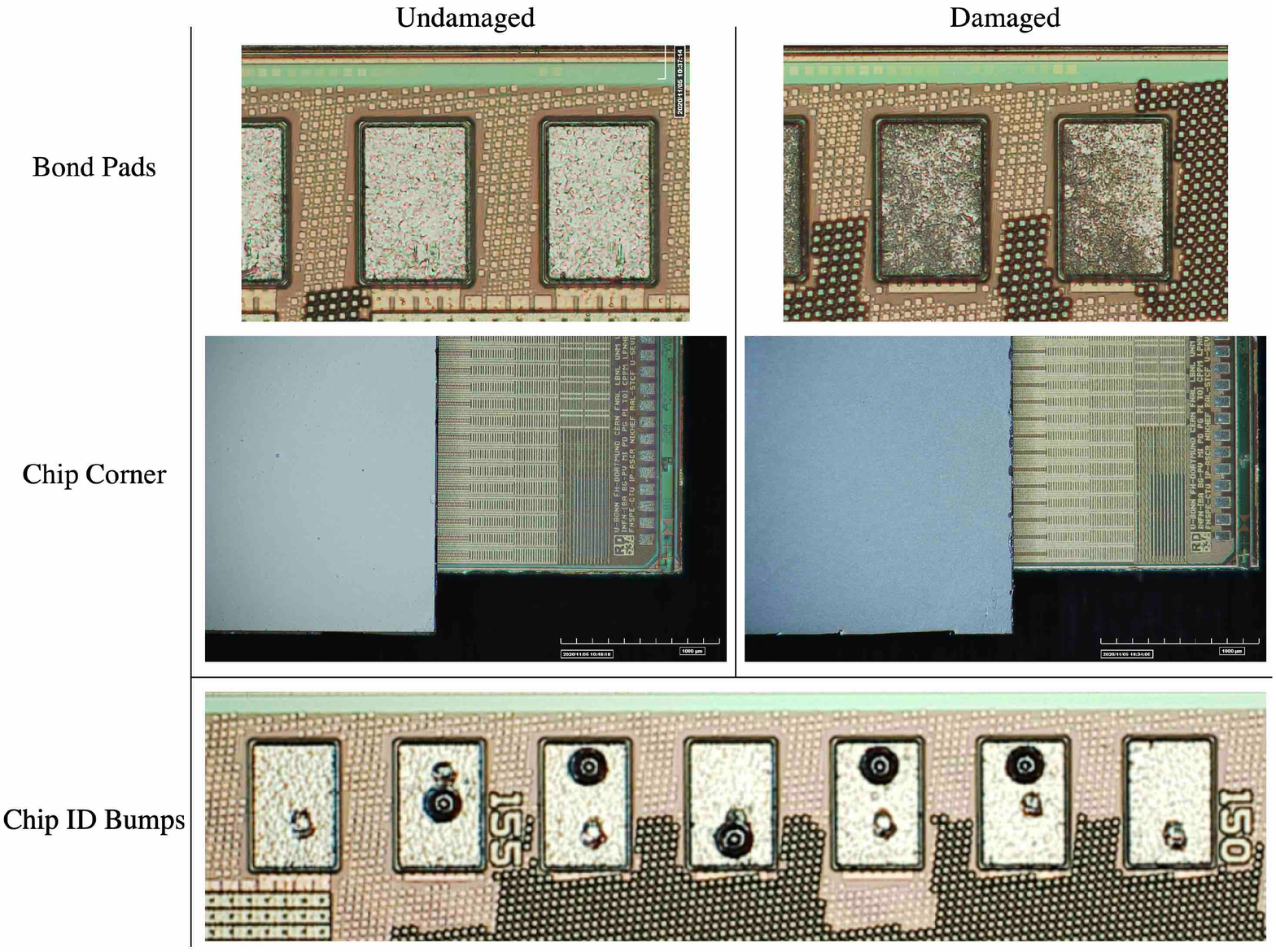}
\caption{Examples of the different features and their corresponding potential damaged state.}
\label{fig:DamageExamples}
\end{figure}

Fig.~\ref{fig:DamageExamples} illustrates different types of features and damages that are inspected on bare modules. 

Wire bond pads are inspected at high magnification aiming to identify scratches or discolourations of the pads, the four chip corners are checked at low magnification for mechanical damage and finally chip ID bond pad bumps are inspected at high magnification on each FE chip.

\subsection{Visual inspection of flexes}
\begin{figure}[!ht]
\centering
\includegraphics[width=6cm]{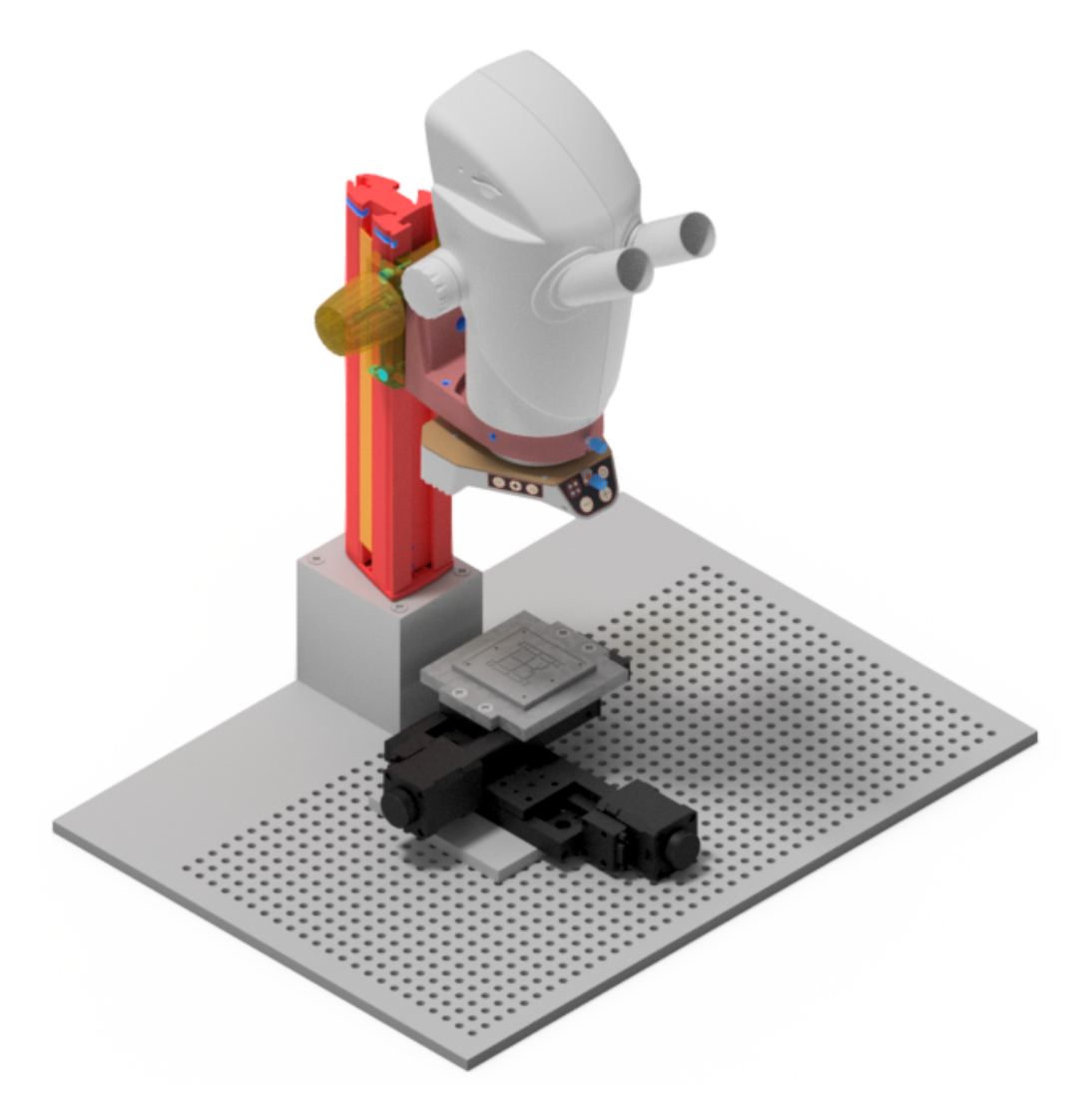}
\includegraphics[width=6cm]{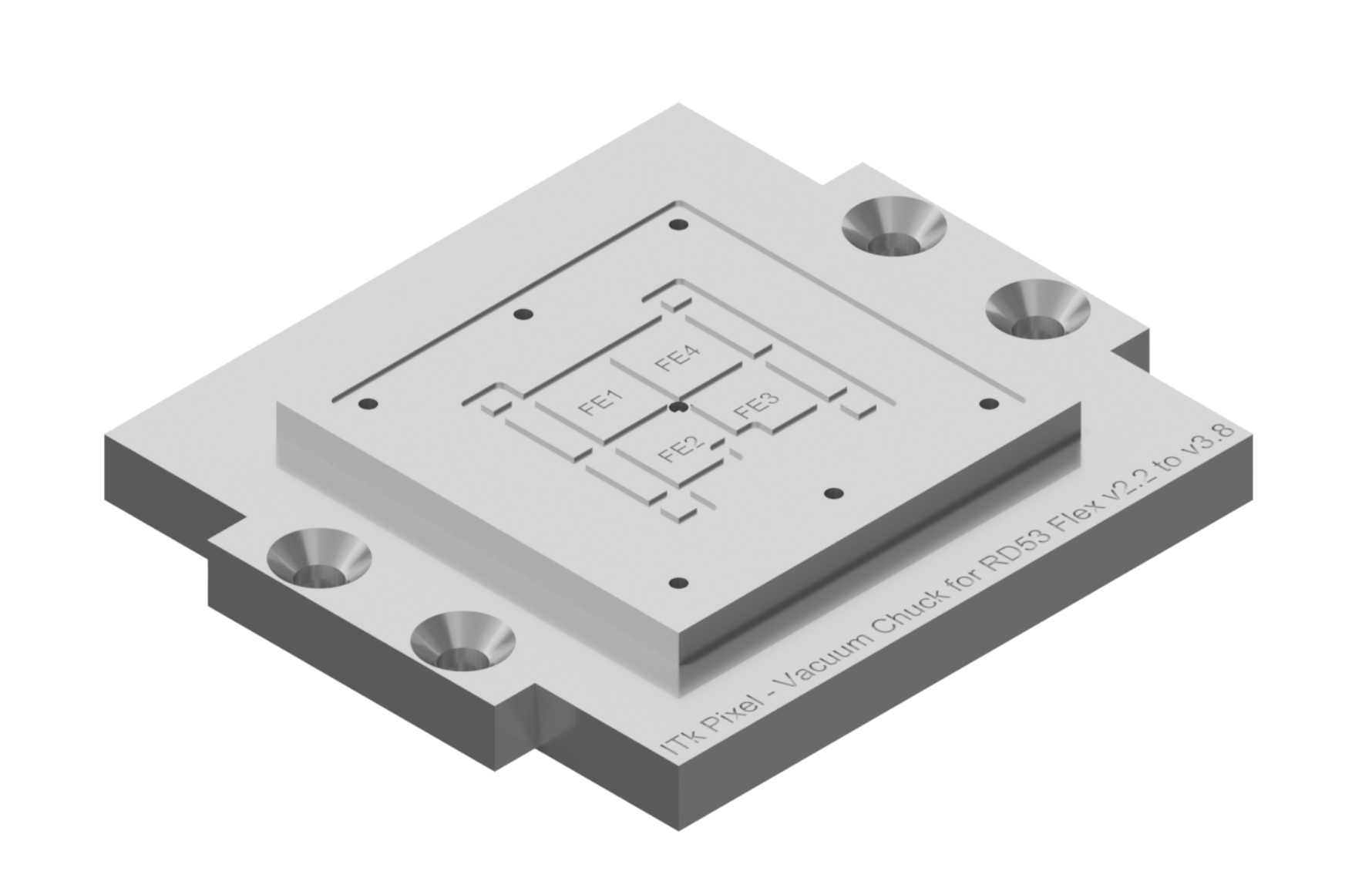}
\caption{Leica S9i microscope and vacuum chuck used for the visual inspection of the flexes.}
\label{LeicaMicroscopeVacuumChuck}
\end{figure}

At CERN, a Leica S9i microscope is used to conduct the visual inspection of pixel flexes. The microscope has a magnification range of 6.1 to 55 times fitted with a polarised light source and filter lens as well as a 10 MP camera. The systematic positioning of the sample being inspected is ensured through the use of an X-Y linear stage system which can be configured from a computer. The Leica S9i microscope is shown in Fig.~\ref{LeicaMicroscopeVacuumChuck}.
During the visual inspection, the flex needs to be placed in a stable and repeatable position. This is achieved through the use of an aluminium vacuum chuck, which is also shown in Fig.~\ref{LeicaMicroscopeVacuumChuck}. The dedicated chuck has been designed to hold the flex horizontally flat and counter any intrinsic bending or warping during the inspection. The chuck designed for this purpose has been made compatible with RD53A flex versions 2.2 and 3.8.

\begin{figure}[!htb]
\centering
\includegraphics[width=0.68\textwidth]{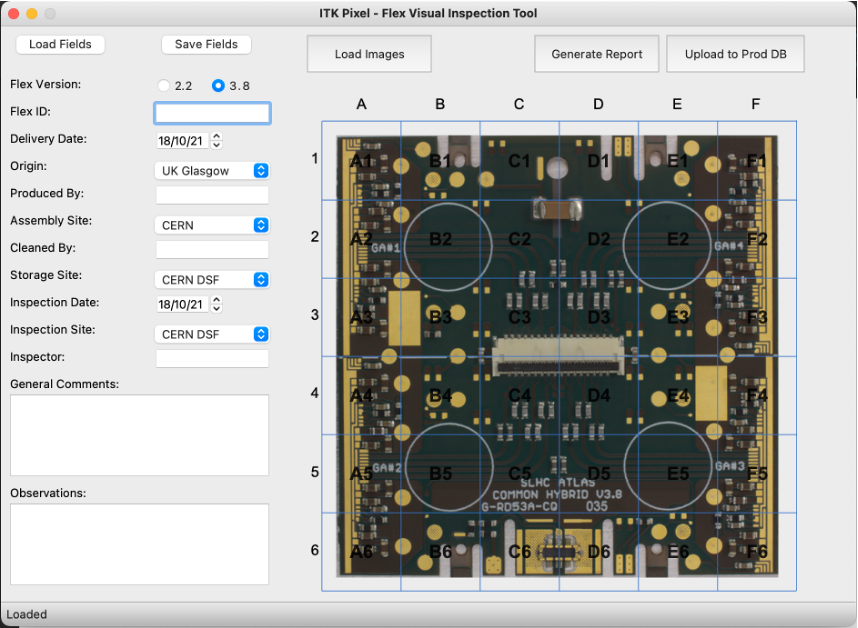}
\includegraphics[width=0.68\textwidth]{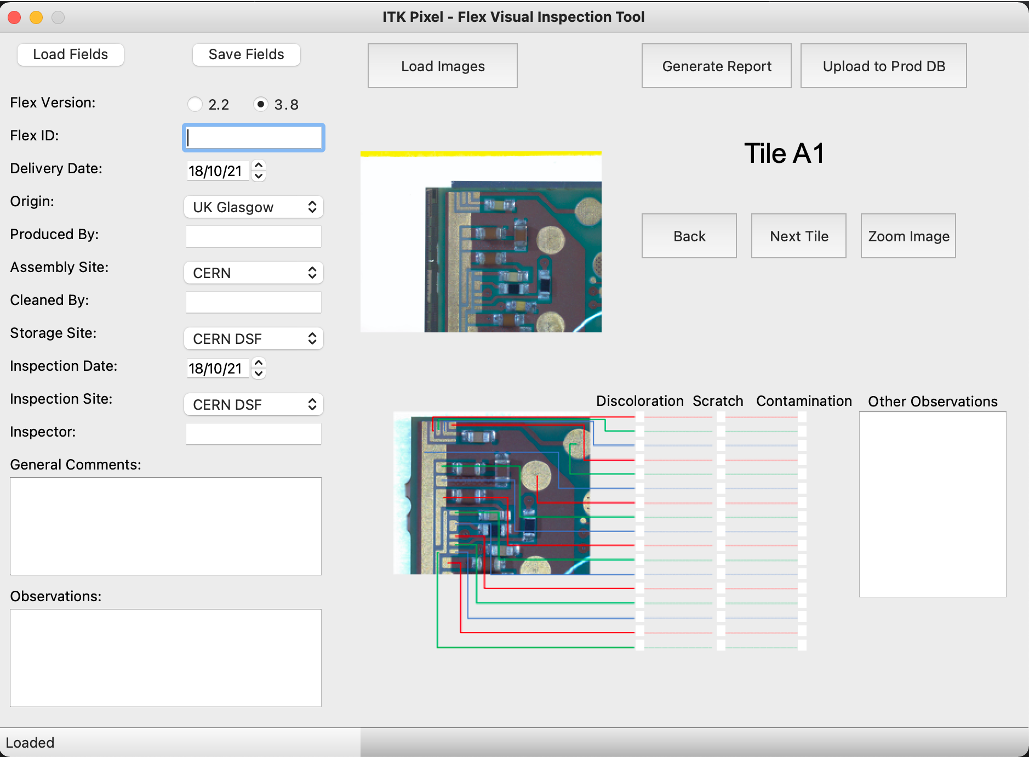}
\caption{ITk pixel visual inspection tool.}
\label{fig:VisualInspectionTool}
\end{figure}

An automated tool has been developed to speed up the visual inspection of flexes, using pyQt as programming software for the GUI \cite{pyqt}. The tool is used to inspect all the images acquired from the microscope in an automated way: an image is acquired for each flex tile and the tool allows to mark down the defects for each wire bond pad and to generate a report with all the relevant information. Two screenshots are shown in Fig.~\ref{fig:VisualInspectionTool}. 

First, on the left section, it is required to set the flex version for the visual inspection (currently 2 versions are available: 2.2 and 3.8). Then, some fields about the flex ID, its origin and inspection need to be filled or they can be loaded if previously saved, along with any general remarks or observations. The right section displays a reference image of an entire flex which is subdivided into a 6 × 6 grid, where each tile is labelled from A1 to F6. Pressing the \textit{Load Image} button imports all 36 tile images in the working directory that have the provided flex ID appended in their file name. Once all information is filled in on the left-hand side of the GUI, the inspection can begin. The tool allows to select a flex tile, e.g. A1, and start the visual inspection for it. A new window will appear, showing the acquired image of the flex tile being investigated on the top side and a reference image of the same tile with no anomalies on the bottom. On the bottom, each wire bond pad is connected with a line to 3 different key defects that can be selected: discolouration, scratch, or contamination. Additional observations can also be registered per tile. The tool allows to move quickly through each tile sequentially and to easily compare the image acquired from the microscope to the reference one by possibly zooming the acquired image. Once all the flex tiles are inspected, it is possible to generate a report with all the information inserted using the \textit{Generate Report} button and to upload it to the central production database using the \textit{Upload to Prod DB} button.

\section{Bare module to flex assembly}
\begin{figure}[!htb]
\centering
\includegraphics[width=1.\textwidth]{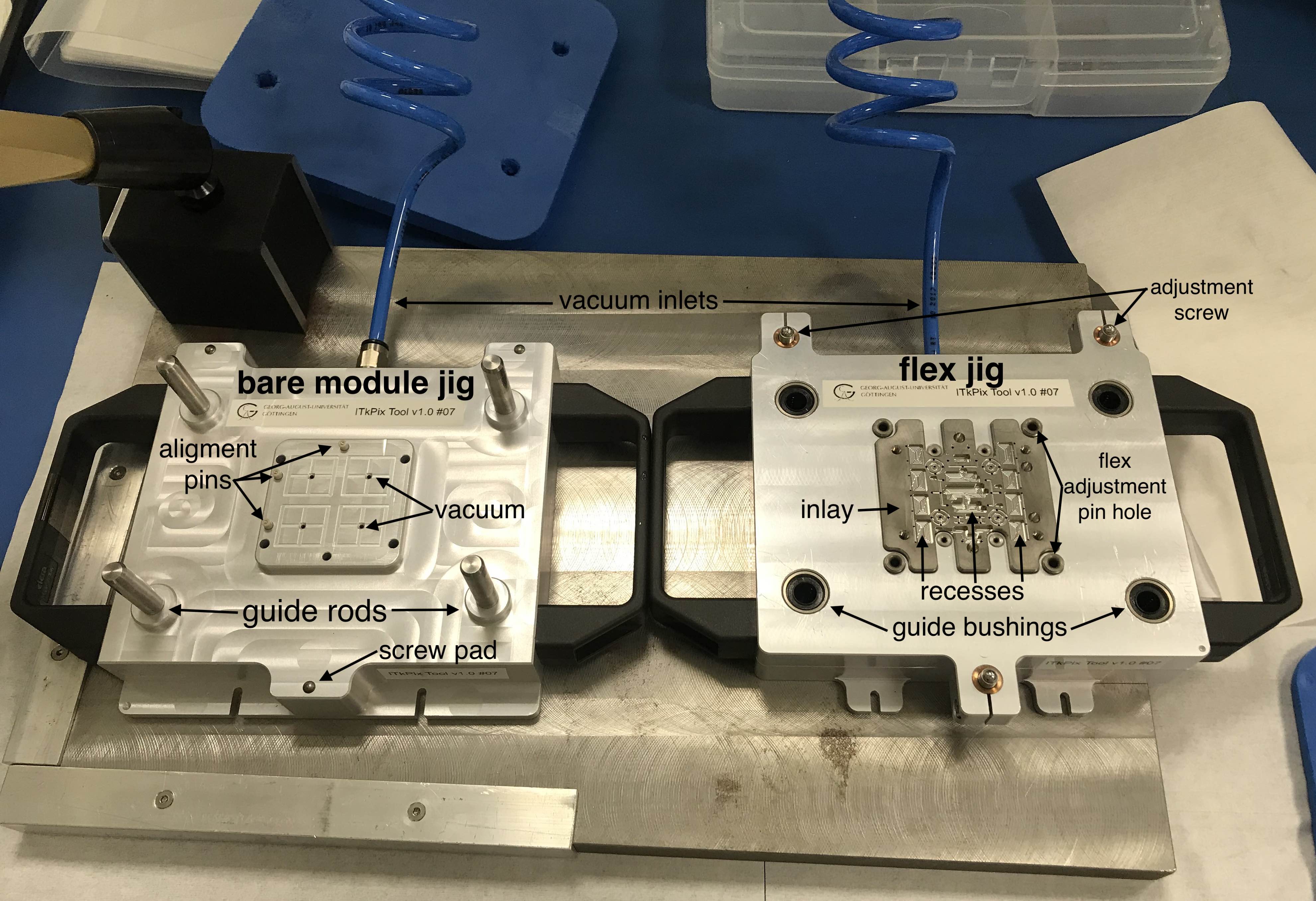}\hfill
\caption{ITPix v1 assembly tooling, showing the bare module jig (left) and flex jig (right).}
\label{fig:ITkPixAssemblyTooling}
\end{figure}

The tooling used for module assembly is shown in Fig~\ref{fig:ITkPixAssemblyTooling}. 

It consists of two jigs for alignment, one for the bare module and one for the flex. Several key components are highlighted in the bare module jig:
\begin{itemize}
\item Alignment pins are used to position the bare module such that it aligns properly with the flex. This ensures that the module can be wire bonded and that the adhesive is localised to its proper locations. The alignment pins are made of a metal core coated in PEEK which avoids damage to the bare modules.
\item Vacuum is used for securing the bare module into position and to keep the bare module as flat as possible. There are four vacuum holes which are centred on each chip. Good experiences have been made with a pump that supplies 70 mbar and 2.0-2.2 $\mathrm{m^{3}/h}$ vacuum pressure.
\item The vacuum inlet provides the external vacuum connection to the bare module jig. The inlet requires external tubing of 6 mm.
\item Guide rods are used to align the flex jig to the bare module jig by guiding the guide bushings in the flex jigs over the guide rods on the bare module jig. The rods and bushings are spaced such that the bare module and flex jigs can only be brought into contact in the correct orientation.
\item Screw pads are where adjustment screws from the flex jig make contact. The adjustment screws are used to adjust the spacing between the bare module and flex jigs to set the glue height.
\end{itemize}
The flex jig is composed of two pieces, the aluminium jig used for alignment to the bare module jig and the inlay with recesses for the SMD components of the flex. Several key components are highlighted in the flex jig:
\begin{itemize}
\item The inlay is where the flex is placed during the assembly process. It contains vacuum holes to hold the flex in place, as well as to keep it flat.
\item Recesses are \say{cutouts} in the inlay that house various components on the flex (resistors, connectors, etc.) in order to allow the flex to lay flat relative to the jig and provide more uniform force over the surface area of the bare module when the flex jig is placed on the bare module jig.
\item The vacuum inlet is where the external vacuum is connected to the flex jig. The inlet requires external tubing of 6 mm.
\item Guide bushings are where the guide rods from the bare module jig are inserted into the flex jig at glue-time. The rods and bushings are spaced such that the bare module and flex jigs can only be brought into contact in the correct orientation, as well as to ensure the proper alignment of the flex and the bare module.
\item Flex adjustment pin holes are used to align the flex frame. Alignment pins are inserted into the flex adjustment pin holes to position the frame into place.
\item Adjustment screws control the spacing between the bare module and flex jig when they are brought into contact during assembly. The spacing between the two jigs corresponds to the desired adhesive height. Use of adjustable screws in three points also accounts for non-parallelism between the surface of the flex jig and the bare module jig, thus providing a uniform glue height along the module.
\end{itemize}

\begin{figure}[!htb]
\centering
\includegraphics[width=0.4\textwidth]{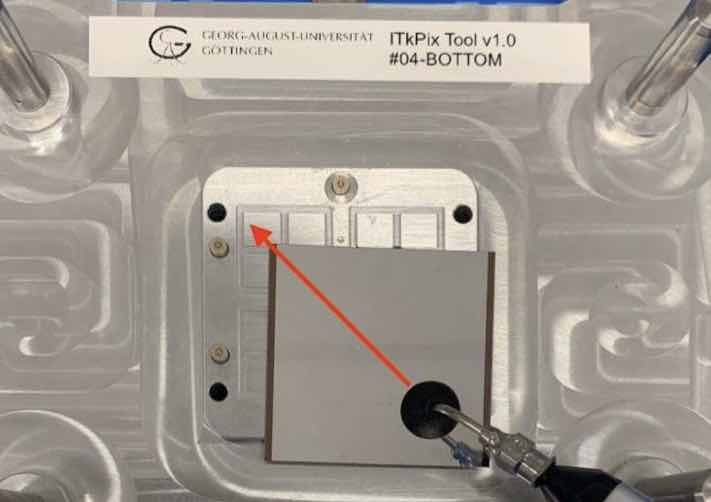}
\includegraphics[width=0.4\textwidth]{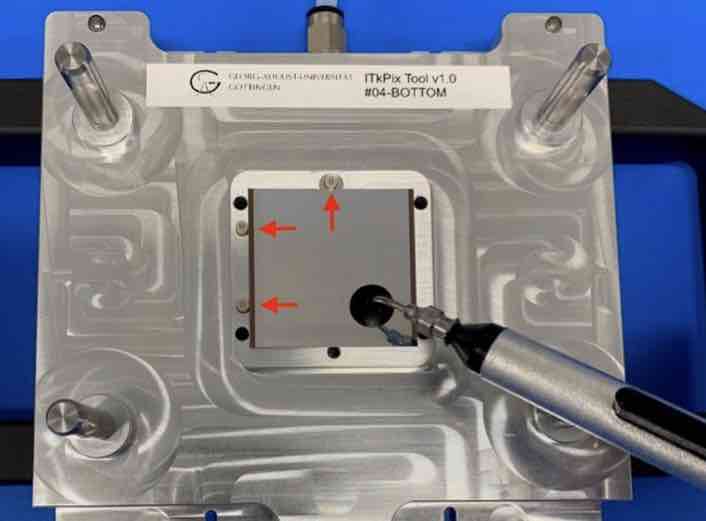}
\includegraphics[width=0.4\textwidth]{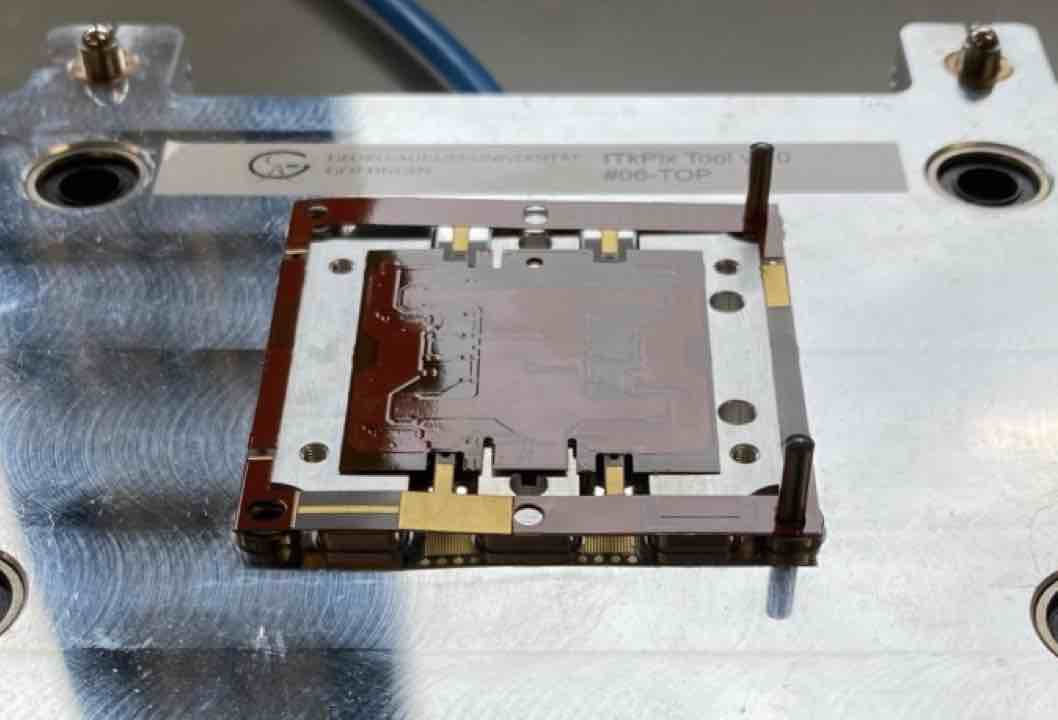}
\includegraphics[width=0.4\textwidth]{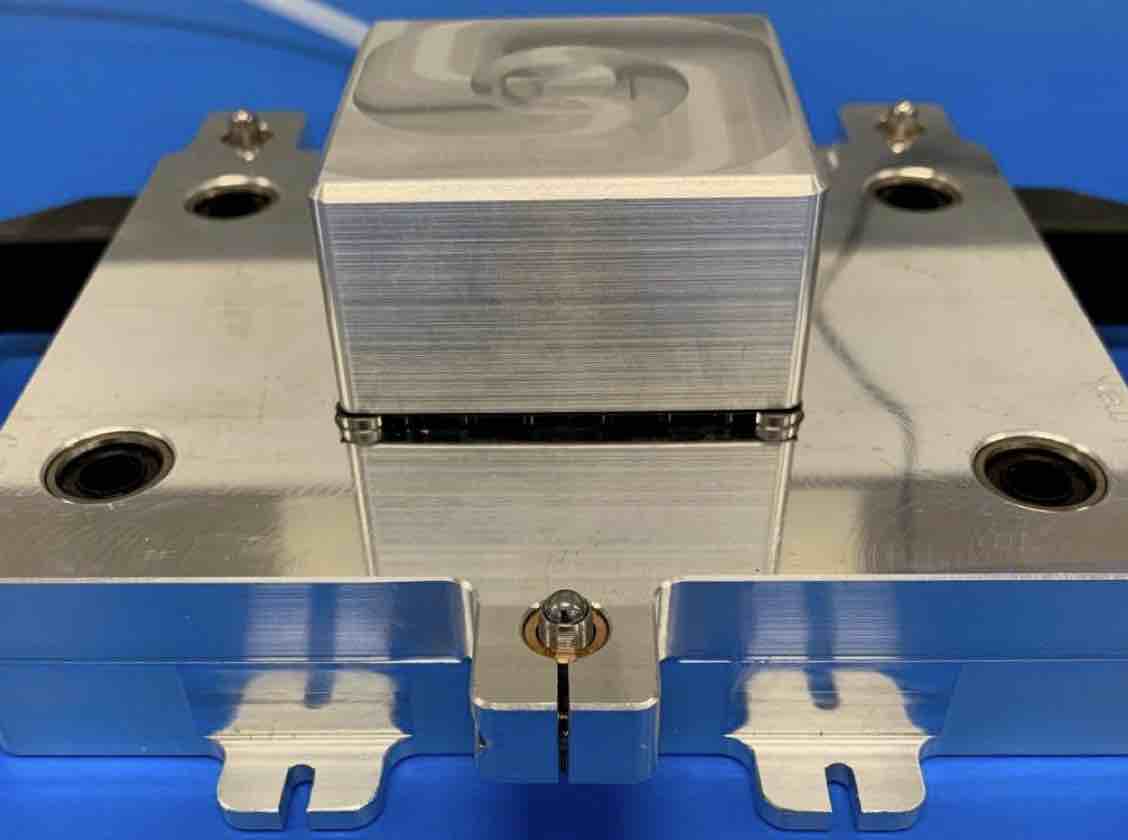}
\includegraphics[width=0.4\textwidth]{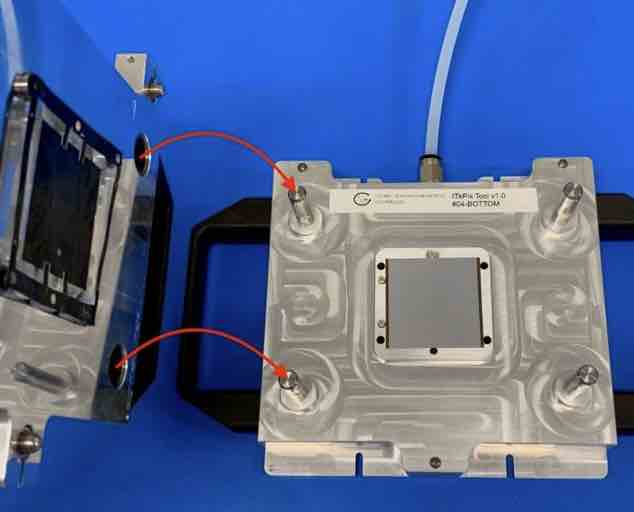}
\caption{The various steps of the ITPix v1 assembly procedure.}
\label{fig:ITkPixAssembly}
\end{figure}

The basic steps of the bare module to flex assembly (see Fig.~\ref{fig:ITkPixAssembly}), are:
\begin{itemize}
\item[1.] Align bare module on jig: place the bare module on the bare module jig and align it by pushing the bare module against the alignment pins. Turn on the vacuum.
\item[2.] Align flex on jig: place flex on the flex jig, insert the frame alignment pins according to the version of flex, use an aluminium block to flatten the flex and make sure that the flex does not move relative to the inlay. Turn on the vacuum. Lift the aluminium block and remove frame-aligning pins from the flex jig.
\item[3.] Set the glue gap between the module components. This can be achieved by placing shims on top of the bare module with the help of a vacuum pin and placing the flex jig onto the bare module jig by mating the guide rods and the guide bushings. Screw the adjustable screws to the appropriate level with the help of a hex wrench until a slight resistance can be felt. Screw down all three screws and remove the shims from the bare module. 
\item[4.] Apply glue on the flex jig using a stencil or a glue robot. This is described in the following section.
\item[5.] Place flex jig on bare module jig.
\end{itemize}

\section{Bare module to flex attach}
\begin{figure}[!htb]
\centering
\includegraphics[width=0.7\textwidth]{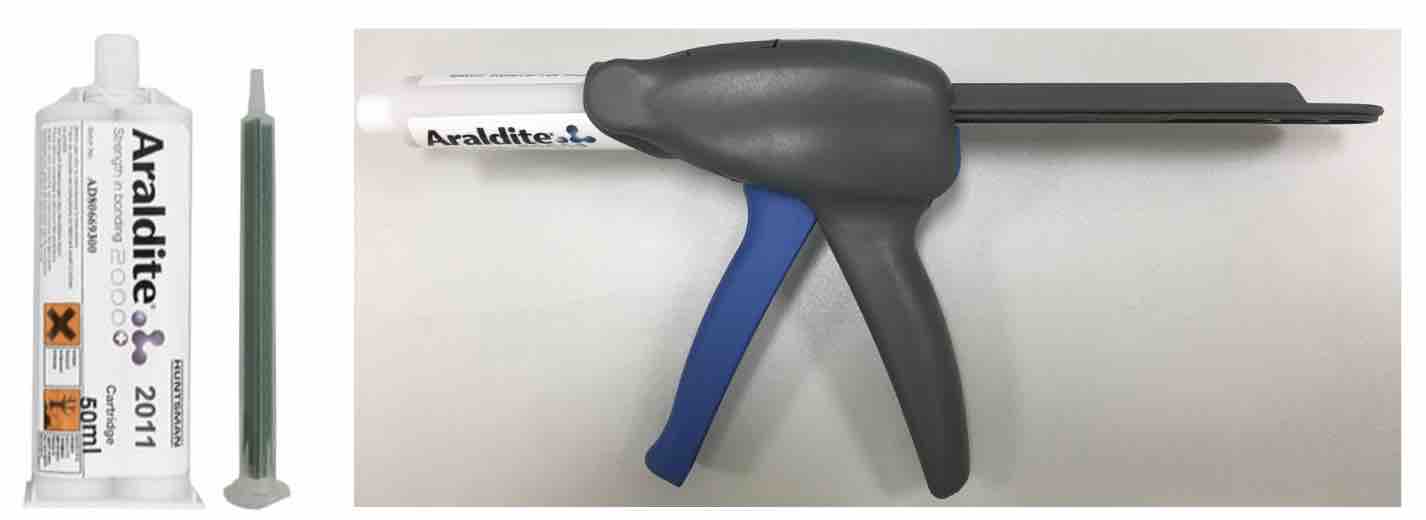}
\caption{Araldite 2011 barrels with mixing nozzle (left) and glue gun assembly (right).}
\label{fig:AralditeGlue}
\end{figure}

The assembly of the RD53A modules includes gluing the bare module to the module flex. The glue selected for this purpose is Araldite 2011A1 (shown in Fig.~\ref{fig:AralditeGlue}). Araldite is available in dual 25 mL tubes at the CERN stores and is to be used with its corresponding nozzle and glue-gun for dispensing.

\begin{figure}[!htb]
\centering
\includegraphics[width=0.7\textwidth]{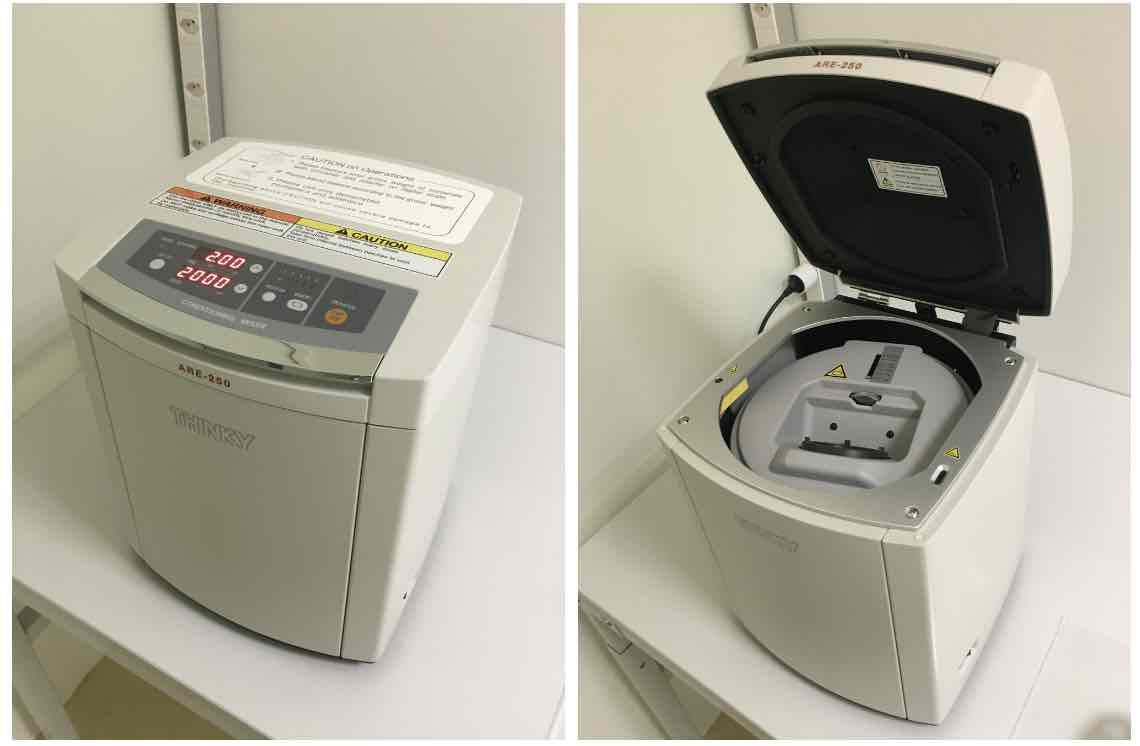}
\caption{Thinky ARE-250-CE mixer and degasser.}
\label{fig:ThinkyMixer}
\end{figure}

This two-component epoxypaste adhesive requires to be mixed such that it is bubble-free and of uniform consistency to yield a homogeneous mechanical performance. In order to obtain a highly uniform mixture of both components, a dedicated mixing and defoaming instrument, the Thinky ARE-250-CE is used at CERN, as shown in Fig.~\ref{fig:ThinkyMixer}.

\subsection{Gluing using the stencil}
\begin{figure}[!htb]
\centering
\includegraphics[height=0.5\textwidth]{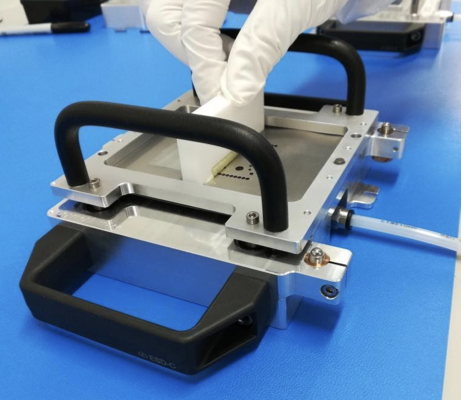}
\includegraphics[height=0.5\textwidth]{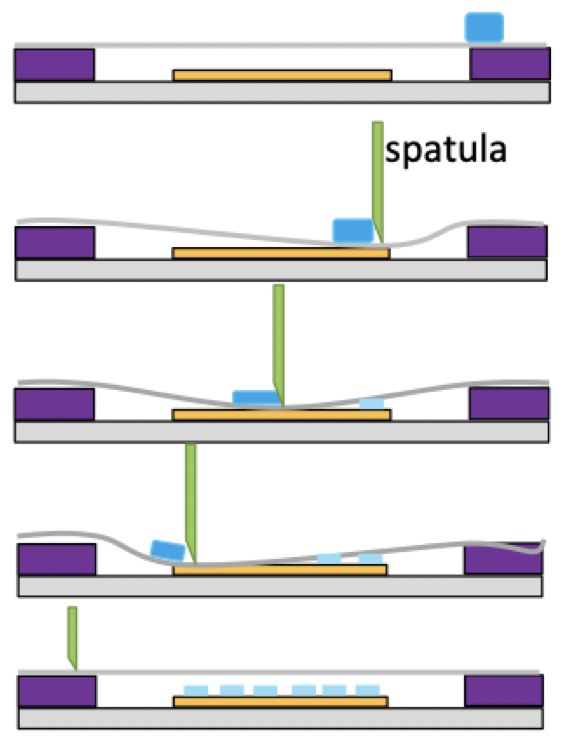}
\caption{Glue being distributed into the stencil pattern using the spatula and diagram showing that the height of the stencil determines the quantity of deposited glue on the flex when being applied with the help of the spatula.}
\label{fig:StencilTooling}
\end{figure}

The bare module to flex attach is usually performed by adopting a stencil (see Fig.~\ref{fig:StencilTooling}). The stencil tooling has to be placed onto the flex jig and the glue has to be placed on the stencil using the mixing nozzle. Pushing down on the stencil, a spatula is adopted to apply adhesive into the stencil holes in a controlled motion, as shown in Fig.~\ref{fig:StencilTooling}. The best performance is achieved when the spatula is moved in a constant manner over the stencil in 7-10 s.

\subsection{Gluing using the glue robot}
\begin{figure}[!htb]
\centering
\includegraphics[width=0.4\textwidth]{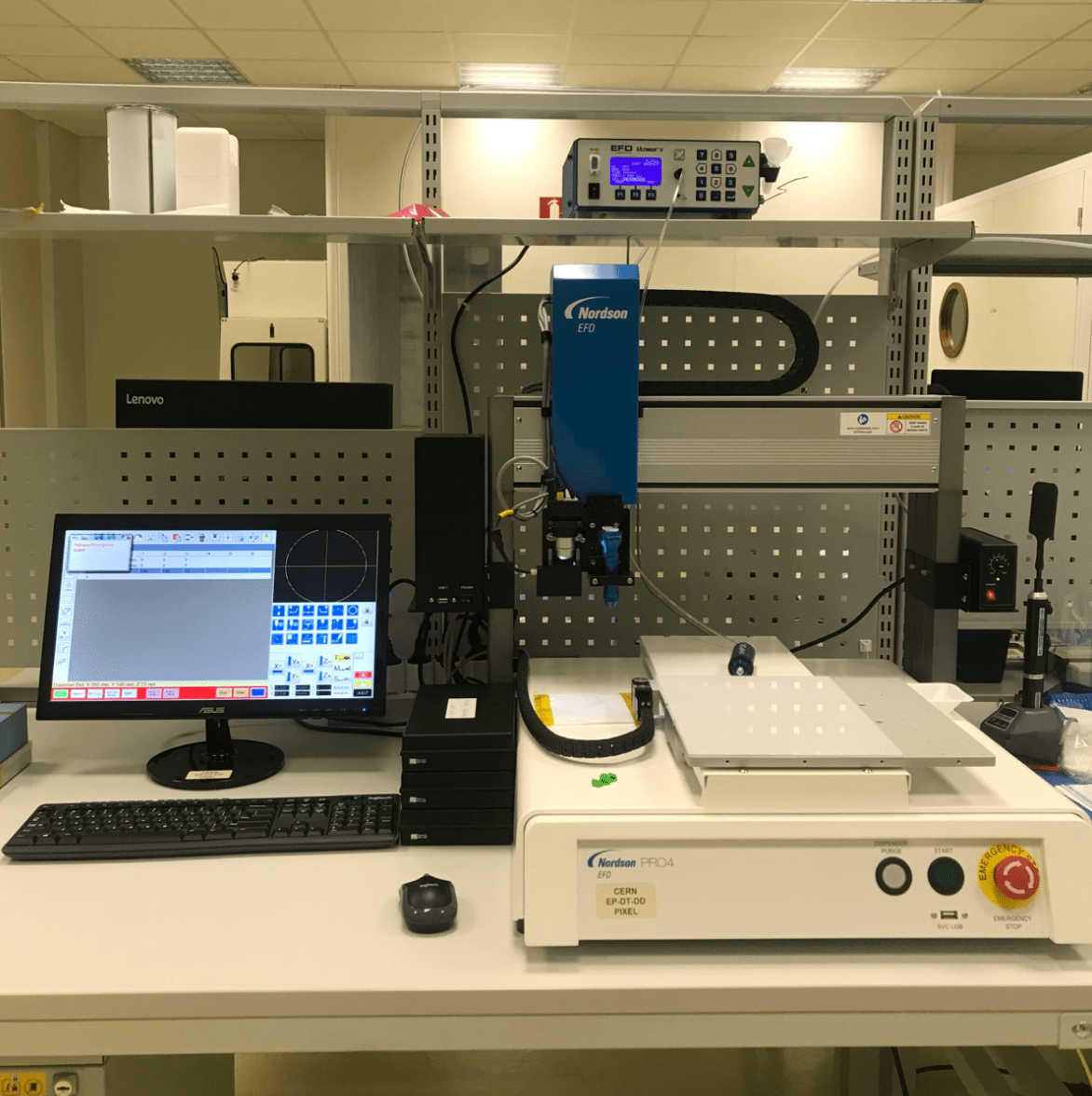}
\includegraphics[width=0.52\textwidth]{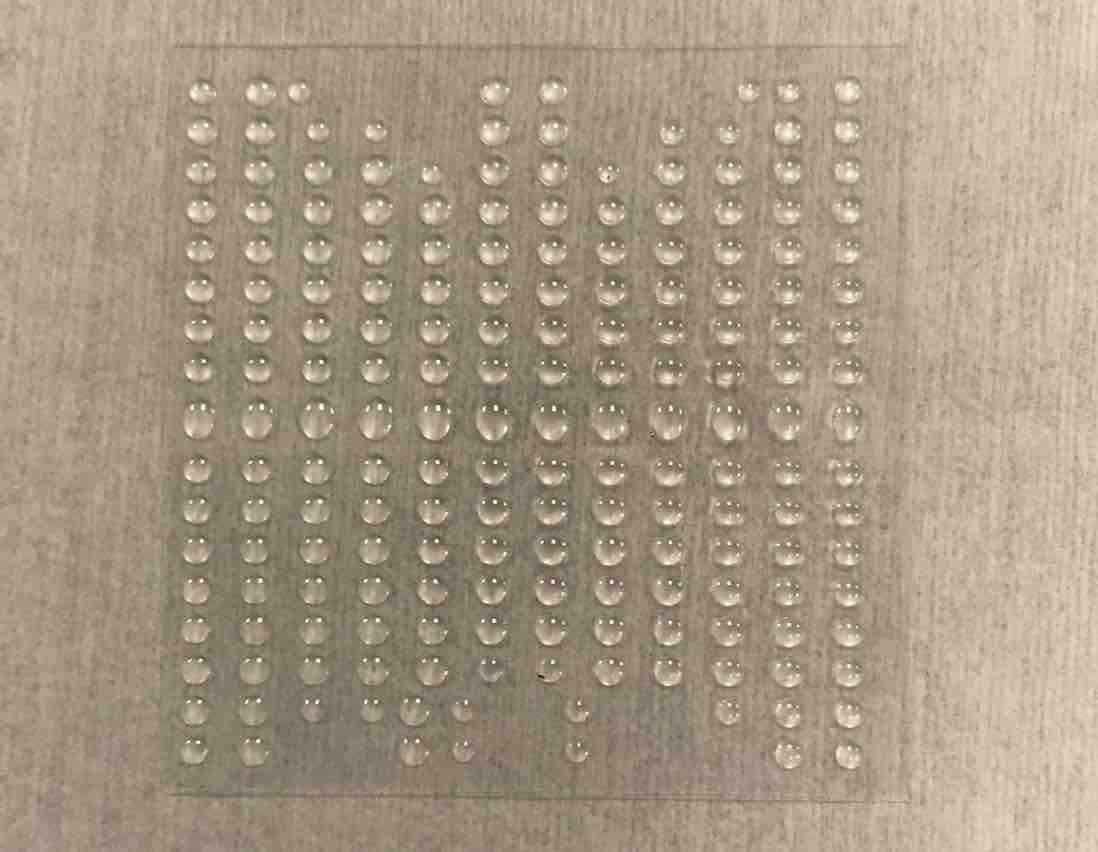}
\caption{Glue robot used for the reproduction of the glue pattern and an example of a pattern produced on a glass plate.}
\label{fig:GlueRobot}
\end{figure}

Another possibility is to use a glue robot to increase repeatability and reduce operator dependency (see Fig.~\ref{fig:GlueRobot}). A glue dispensing program for the glue robot has been developed to reproduce the design pattern, and some parameters were tuned to improve the quality of the shape of the dots as measured using a microscope and to reproduce the pattern as quickly as possible. An example of a pattern produced on a glass plate is also shown in Fig.~\ref{fig:GlueRobot}. 

\begin{figure}[!htb]
\centering
\includegraphics[width=1.\textwidth]{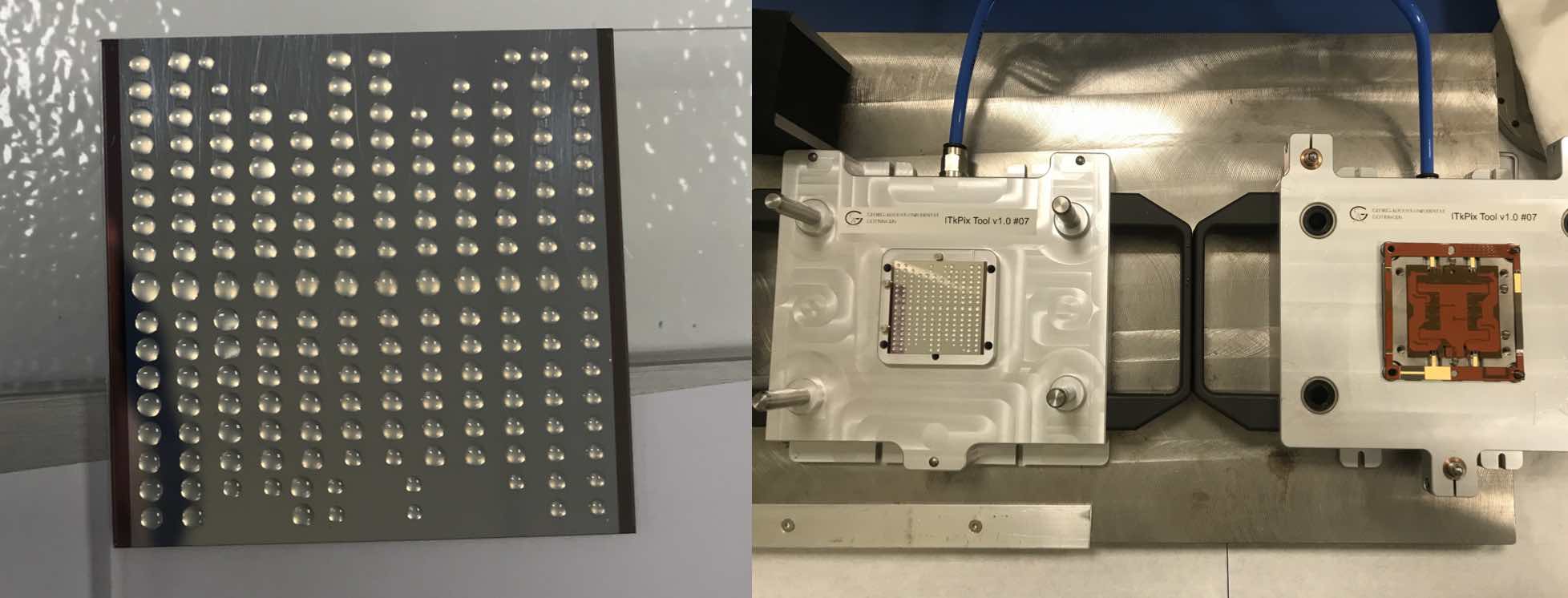}
\caption{An example of a pattern produced on a silicon dummy.}
\label{fig:GlueRobotSiliconDummy}
\end{figure}

The production of the glue pattern has been repeated 10 times to estimate the deviation in the glue weight. The mean value of the deposited glue is found to be 55 mg, with a sample standard deviation of 1.72 mg and a mean standard deviation of 0.544 mg. This test showed that it is possible to produce a much more repeatable pattern using the glue robot which has no operator dependency instead of the manual stencil. \\
Tests have also been conducted to see the glue spread after the bare module is attached to the flex. For this purpose, the pattern has been produced on a silicon dummy instead of the bare module, which was then glued on the backside of the flex. The realized pattern is shown in Fig.~\ref{fig:GlueRobotSiliconDummy}. Optimal glue spread is realised when the proper separation between jigs is set in the assembly tooling.

\section{Plasma cleaning}
\begin{figure}[!ht]
\centering
\includegraphics[width=1.\textwidth]{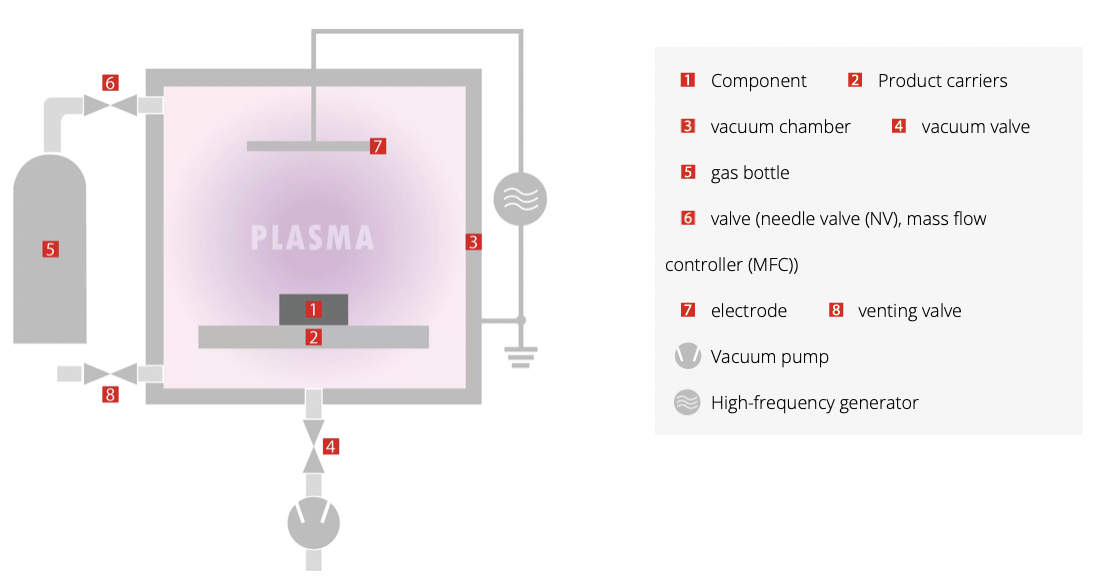}
\caption{Henniker HPT-500 Plasma cleaner and the plasma cleaning process. From Ref.~\cite{Diener}.}
\label{fig:PlasmaCleaning}
\end{figure}

Plasma cleaning is the process of modifying the surface characteristics in a targeted manner, including fine-cleaning of contaminated components, plasma activation of plastic parts, etching of polytetrafluorethylene (PTFE) or silicon, and coating of plastic parts with PTFE-like layers.\\
At CERN, plasma cleaning of pixel modules is performed using the Henniker HPT-500 Plasma cleaner. The plasma cleaning process is shown in Fig.~\ref{fig:PlasmaCleaning}. 

The component (1) lies down on a supporting plate (2) in the chamber (3). A vacuum valve (4) to the pump is initially opened and the chamber can thus be evacuated. The process gas (5) is then supplied to the chamber via a valve (6). When the pressure is stabilised, the generator is ignited and the plasma treatment can start. At the end of the treatment, the gas supply (6) is terminated, the angle valve (4) is closed and the chamber is vented (8). The treated components can then be removed.
To generate plasma, a gas is supplied with sufficient energy to make a critical number of electrons leave their atomic shell. The positively charged ions are responsible for the plasma cleaning following 3 different mechanisms:

\begin{itemize}
\item {\textit{Micro-cleaning} – Degreasing in oxygen plasma} \\
Hydrocarbons are present as residues of greases, oils, or release agents on nearly all surfaces. These coatings drastically reduce the adhesion of other materials during subsequent surface treatment. The plasma reactions during this cleaning process are shown in Fig.~\ref{fig:PlasmaCleaning-Degreasing}. 
\begin{figure}[!ht]
\centering
\includegraphics[keepaspectratio=true,height=5cm]{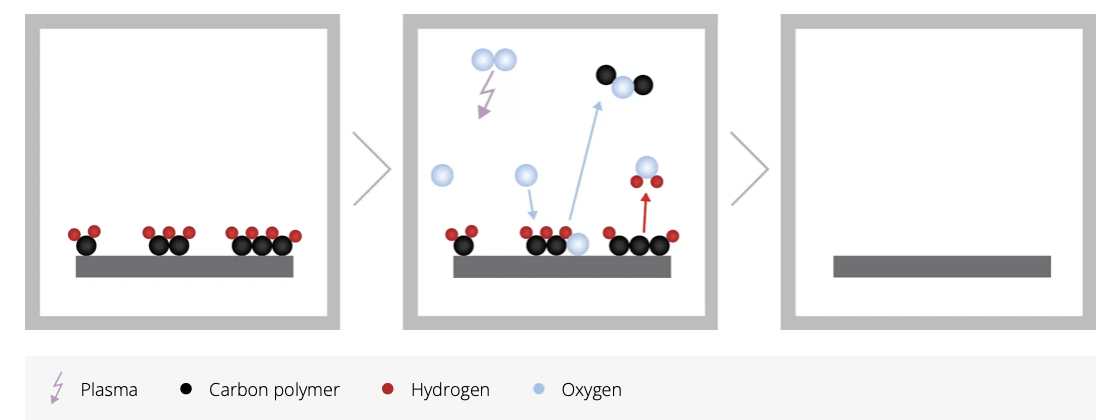}%
\caption{The removal of chemical hydrocarbons through the formation of $\mathrm{CO_{2}}$ and $\mathrm{H_{2}O}$ molecules from oxygen ions leaving a clean surface with good wettability. From Ref.~\cite{Diener}.}
\label{fig:PlasmaCleaning-Degreasing}
\end{figure}
On the left, the component before plasma treatment is shown with the surface contaminated with hydrocarbons. High-energy UV radiation splits macromolecules into ions. Ions, oxygen radicals and hydrogen radicals occupy free chain ends of the polymer chains to $\mathrm{H_{2}O}$ and $\mathrm{CO_{2}}$. The products of the degradation of the hydrocarbons are gaseous in the low-pressure plasma and are siphoned off. Oil, grease or release agents containing additives cannot always be completely removed in oxygen plasma. Solid oxides may form which adhere to the substrate. These can be purified in additional downstream purification processes if needed.

\item  {Mechanical cleaning by \textit{micro-sandblasting}}\\
Noble gas plasma is a particularly simple plasma. It consists only of ions, electrons, and noble gas atoms. As the gas is always atomic, there are no radicals and, since noble gases do not react chemically, there are also no reaction products. Argon plasma is nevertheless active due to the kinetic energy of the heavy ions. The kinetic energy of the impacting ions chips away at the atoms and molecules forming the coat, gradually decomposing them.

\item  {\textit{Reduction} - Removing oxide layers}\\
Plasma can be also adopted to remove oxide coats. Pure hydrogen or a mixture of argon and nitrogen can be used as a process gas (see Fig.~\ref{fig:PlasmaCleaning-Reduction}). The hydrogen plasma is able to chemically reduce the oxidised surface. This generates water which is simply discharged via a continuous gas flow. These processes can also be run in two stages: for example, the items to be treated are first oxidised with oxygen and then can be reduced with the argon-hydrogen mixture.

\begin{figure}[!ht]
\centering
\includegraphics[keepaspectratio=true,height=5cm]{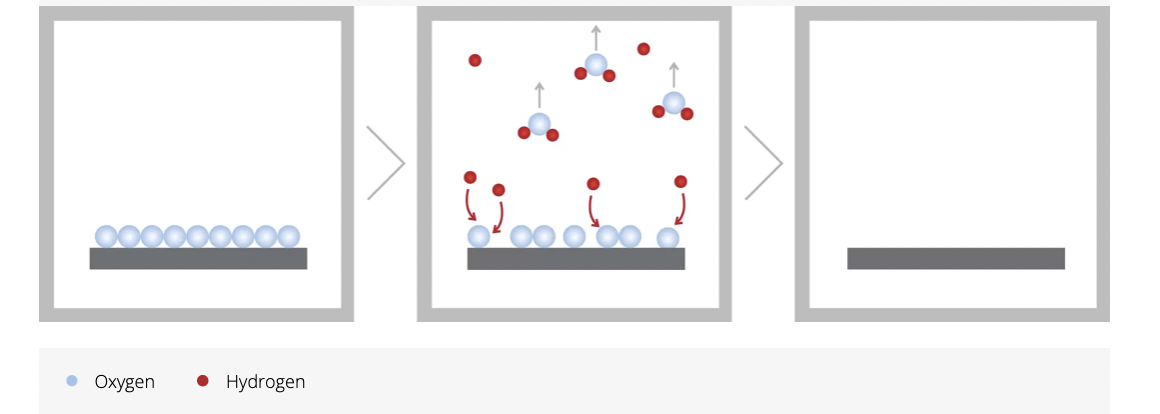}%
\caption{The removal of oxide layers through the formation of $\mathrm{H_{2}O}$ molecules from pure hydrogen leaving an oxide-free surface with good wettability. From Ref.~\cite{Diener}.}
\label{fig:PlasmaCleaning-Reduction}
\end{figure}

\end{itemize}

The plasma cleaning can be performed with two cleaning gas at CERN: a gas mixture of Ar (80\%) and $\mathrm{O_{2}}$ (20\%) or pure $\mathrm{O_{2}}$. 

A study varying the time under which the flex is treated with the gas plasma, while keeping other parameters fixed (power, gas flow, evacuating time and gas stabilisation time) is conducted. The set of parameters chosen for the plasma cleaning can be summarised as follows:
\begin{itemize}
    \item Cleaning time: 1 min, 2 min, 3 min, 5 min, 10 min
    \item Power: 300 W
    \item Gas flow: 5 sccm
    \item Evacuating time: 120 s
    \item Gas stabilisation time: 240 s
\end{itemize}

Because each plasma cleaning session is performed after the previous ones, in the following we will refer to the plasma cleaning time as the total cleaning time considering the time of the last cleaning session summed with the ones of the previous sessions, e.g. 1 min, 3 min, 6 min, 11 min, 21 min.

\begin{figure}[!htb]
\centering
\includegraphics[width=0.8\textwidth]{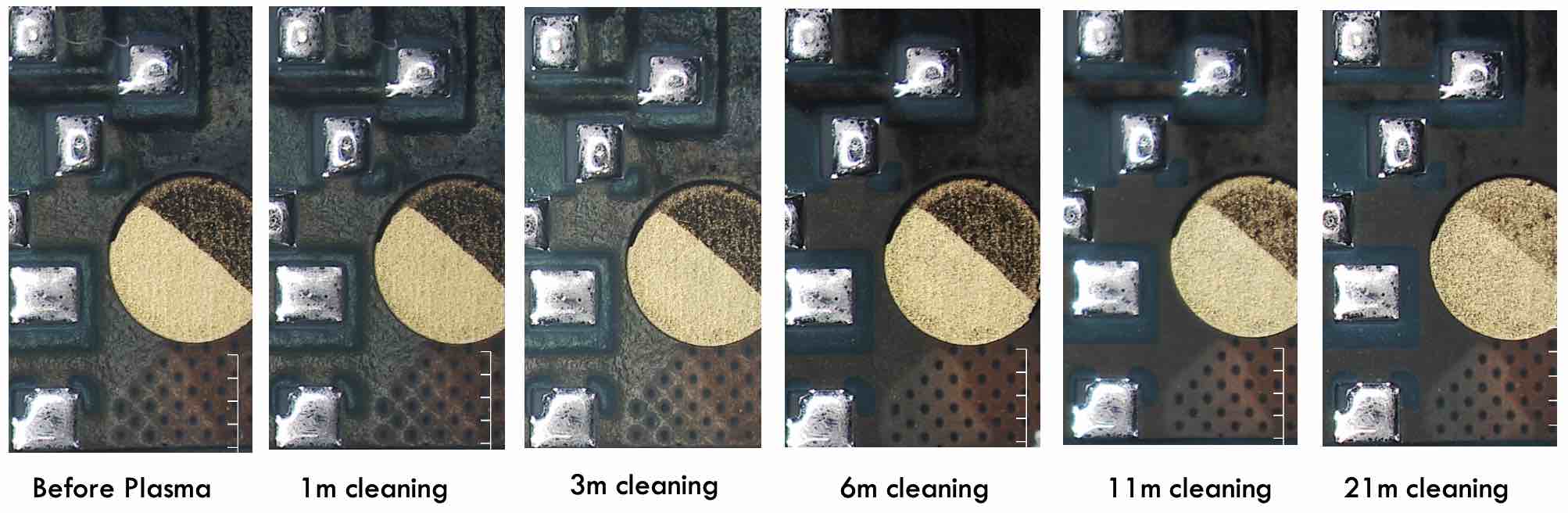}
\includegraphics[width=0.8\textwidth]{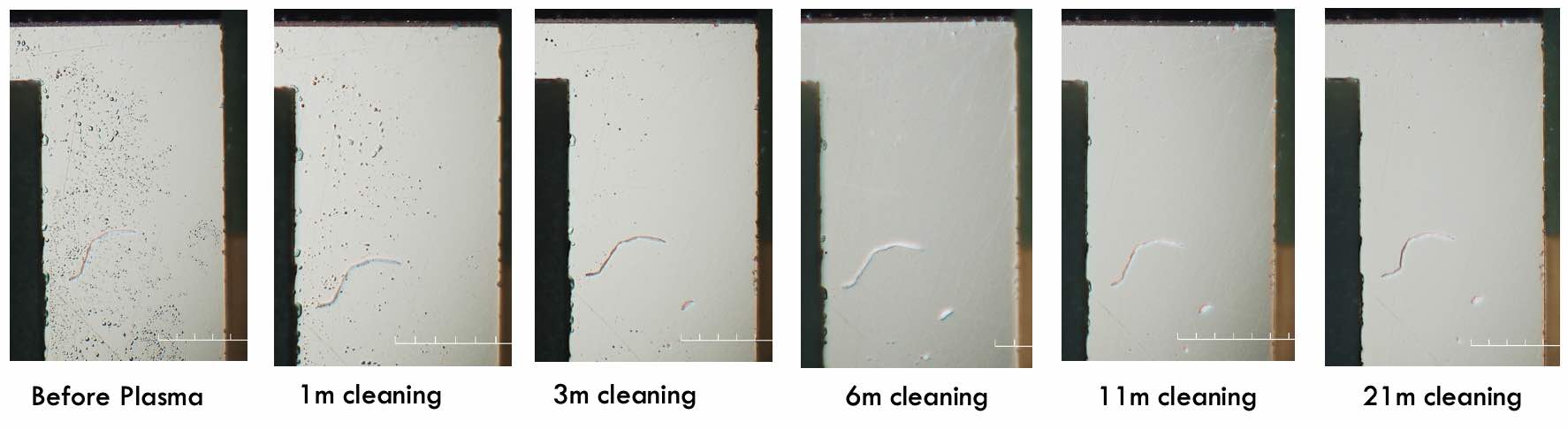}
\centering
\includegraphics[width=0.8\textwidth]{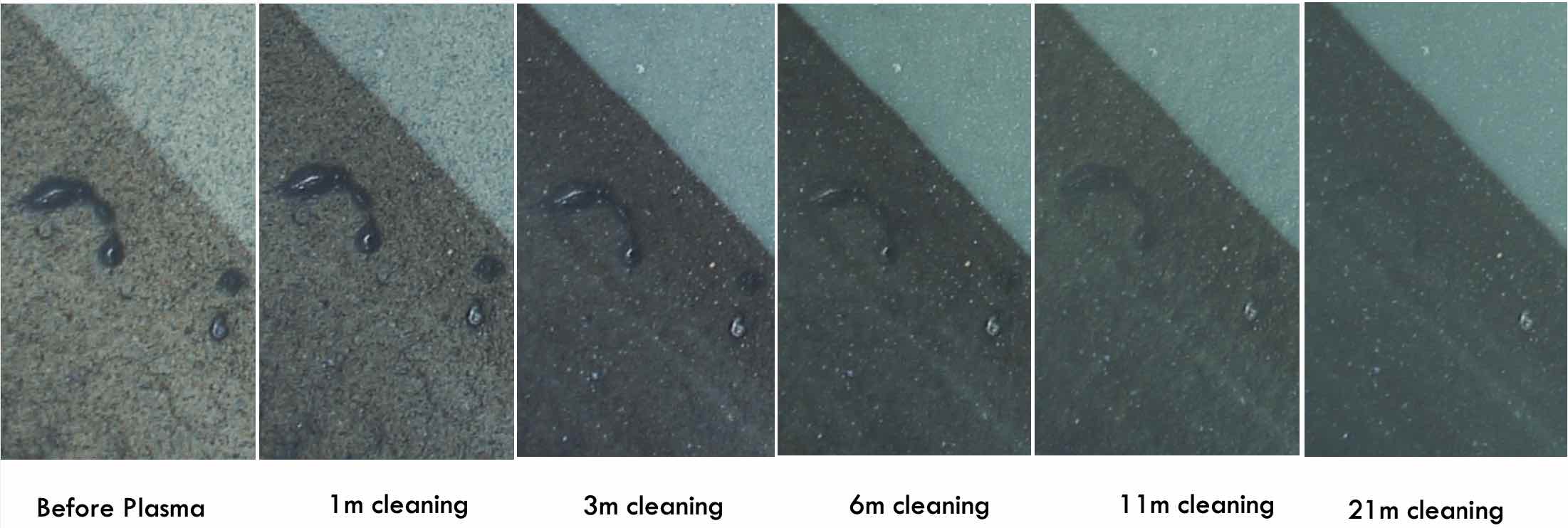}
\includegraphics[width=0.8\textwidth]{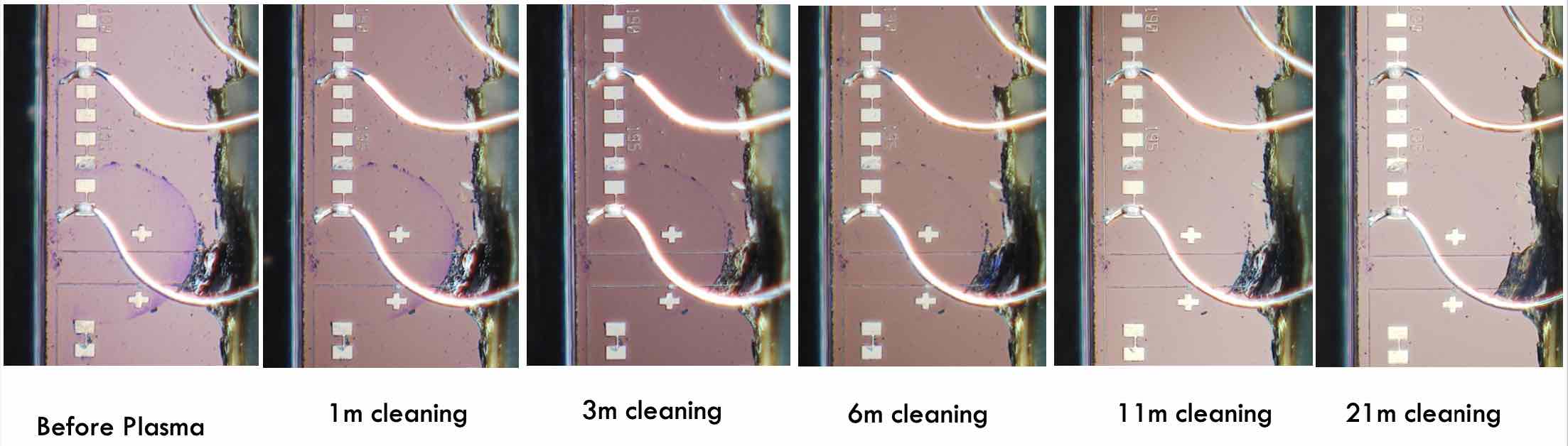}
\caption{Visual Inspection of flex tiles and silicon wire bonds as a function of the cleaning gas time.}
\label{fig:VisualInspectionImagesAr02}
\end{figure}

A systematic visual inspection is performed through the HIROX MXB-5000REZ microscope, available at the CERN QART Lab. The bare modules are placed directly on a microscope table. Fig.~\ref{fig:VisualInspectionImagesAr02} shows the cleaning evolution of the hybrid module as a function of the cleaning time.

A diminishing level of different types of contamination on the surface material of the flex is observed. In particular, the first row shows that a black ink \textit{QC-reject mark} needs 21 min to be removed from the wire bond pad. In the second row, we observe a considerable reduction in droplet-like contaminants from the surface of the dummy silicon, which appears very clean after 6 minutes, while the third shows a reduction of contaminants from the surface of the flex. The last row shows a diminishing level of pink contamination on the silicon dummy.

\section{Electrical testing}

Electrical testing is a crucial step of the quality control of pixel modules, aimed at ensuring that assembled modules work properly. A sequence of action needs to be followed:

\begin{itemize}
    \item Get module information: check out the module on the production database
    \item Verification: visual inspection, initial probing, soldering if necessary, verify functionality
    \item Electrical tests at 30 °C
    \item Electrical tests at 20 °C
    \item Cold startup to qualify module’s cold behaviour
\end{itemize}

Electrical tests include:

\begin{itemize}
    \item Sensor IV (FE chips powered off)
    \item Shunt-LDO VI - verify powering functionality
    \item Probing, trimming, ADC calibration
    \item Chip tuning with fully depleted sensor
    \item Crosstalk/disconnected bump scans
    \item Chip masking (digital, analog, noise)
    \item Source scan
\end{itemize}

\subsection{Experimental setup}
\begin{figure}[!ht]
\begin{minipage}[c]{\textwidth}
\centering
\includegraphics[width=0.8\textwidth]{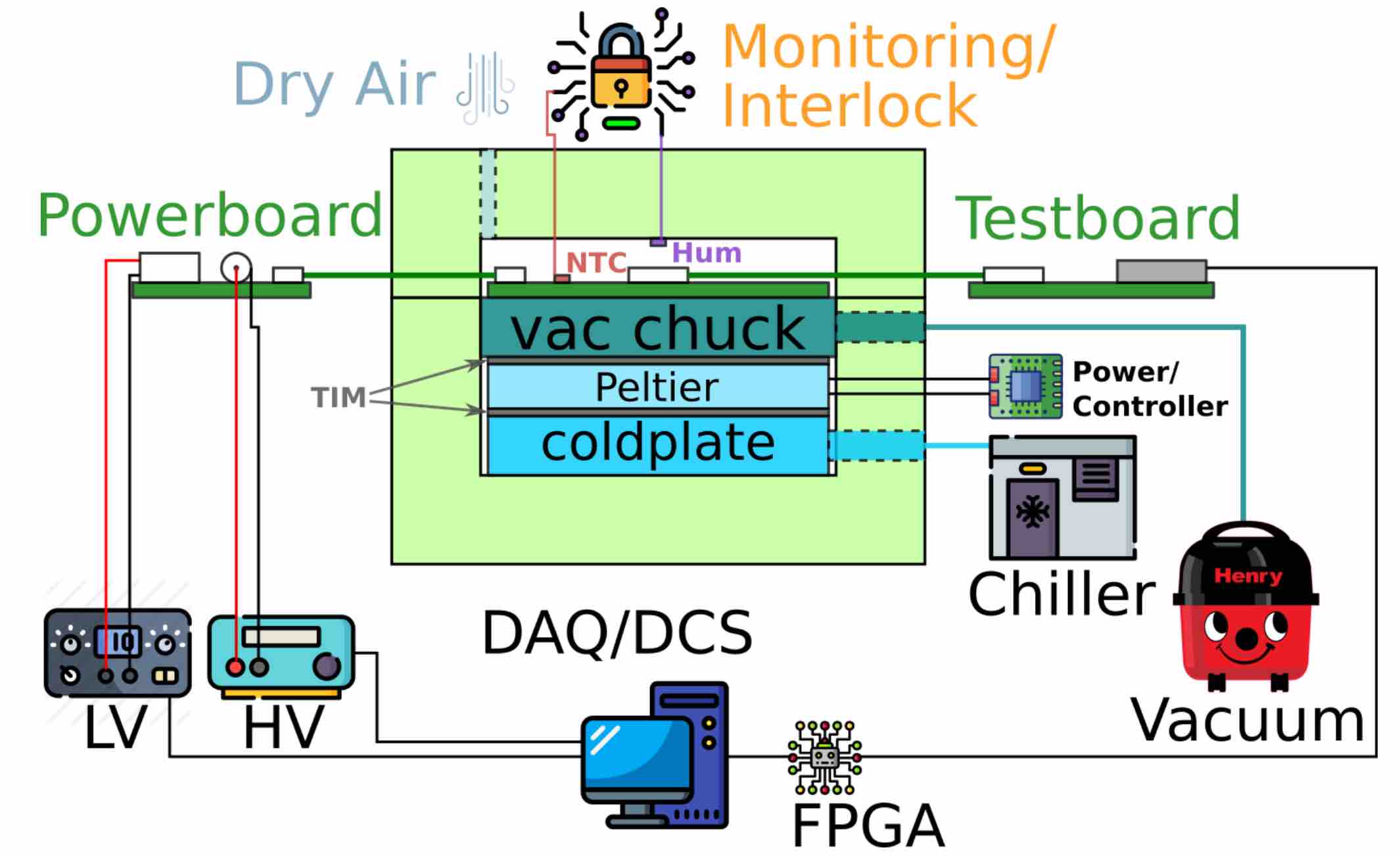}
\end{minipage}
\begin{minipage}[c]{\textwidth}
\centering
\includegraphics[width=0.85\textwidth]{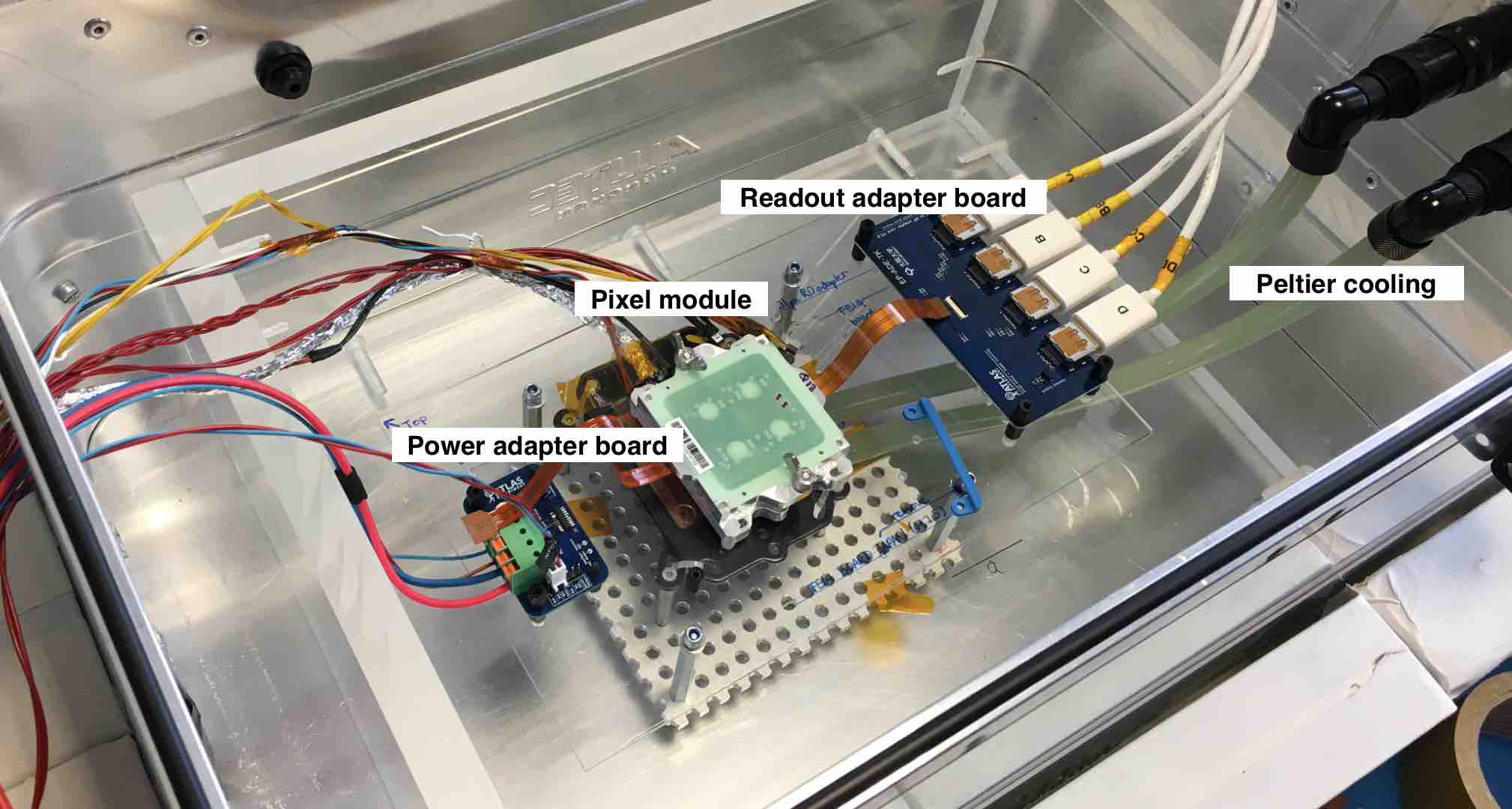}
\end{minipage}
\caption{Schematic of hardware needed for electrical testing and single module setup in 161/1-024 at CERN.}
\label{fig:ElectricalTesting-Setup}
\end{figure}

A schematic of the hardware needed for electrical testing of pixel modules and a picture of the experimental setup in 161/1-024 at CERN are shown in Fig.~\ref{fig:ElectricalTesting-Setup}.

The central green box is a cooling unit whose volume is flushed with dry air and contains the pixel module within its carrier sitting on top of a stack consisting of a vacuum chuck, a Peltier cooling system and a cold plate. The layers are connected to an external vacuum pump, control and power elements, and a chiller, respectively. The thermal interface between the layers in the stack is ensured by the use of Thermal Interface Material (TIM) sheets.
The pixel module is connected to the power adapter board, a rigid PCB which works as a Power Supply (PS) for the pixel module and is needed to power it on. The pixel module needs a LV PS of 1.9 V to operate the FE chips, and a HV PS to deplete the silicon sensor. The power adapter board adapts HV and LV cables from the PS to the power pigtail connected to the flex PCB of the pixel module. The pixel module is also connected to the readout adapter board, a rigid PCB adapting the data pigtail (Molex ZIF connector) to 4 lines of Display Port (DP) connection. 

A generic DAQ/DCS system using power supplies and sensors for monitoring is implemented in commercial computers. The DAQ system of pixel readout chips is based on the Yet Another Rapid Readout (YARR) software~\cite{YARR}. The YARR software reads data from the pixel module and sends commands to the pixel readout chip via a Peripheral Component Interconnect Express (PCI-e) FPGA board that is installed in the host computer. The software implements a kernel driver to communicate with the firmware via PCIe, an engine to drive the scans of the pixel matrix, and processors which analyse the received data. The FPGA is directly connected to the computer through the YARR system and no longer contained in cards connected to the individual detector modules, acts as a simple I/O readout interface, aggregating the data read from the FE chips and sending them to the processor. The latter takes care of the management and effective execution of the scans necessary for the calibration of the chip and also of the organisation of the data received in histograms. In this way, the CPU contains all the information coming from the FE electronics and this makes it possible to carry out a more detailed analysis. The YARR system, therefore, benefits from the remarkable advancement of technology that allows the transfer of a greater amount of data between the FPGA and the CPU via the PCIe serial bus, developed over the last few years to manage video cards of the last generations.

\subsection{Sensor IV}
\begin{figure}[!htb]
\centering
\includegraphics[width=0.5\textwidth]{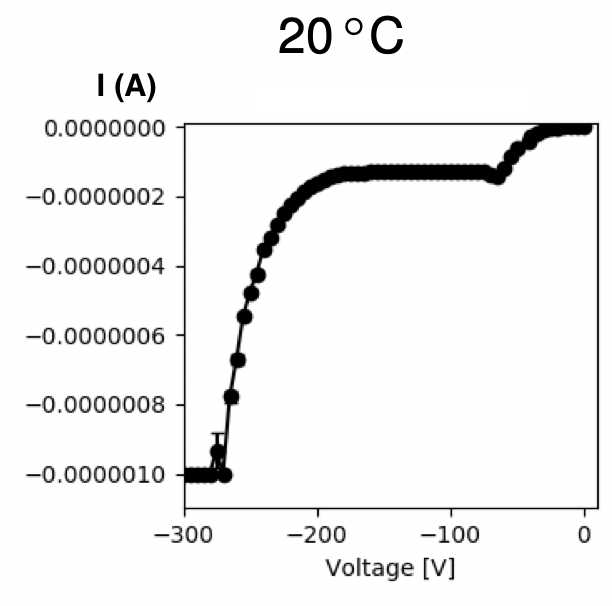}
\caption{An example of a typical sensor IV curve.}
\label{fig:SensorIV}
\end{figure}

Sensor IV curve is measured to check for sensor damages and changes in response to the bias voltage. For each fixed value of voltage, the current is measured. The FE chips are not powered during this measurement. The test is conducted in a light-tight environment at 20°C with a tolerance of $\pm 2$°C and relative humidity of <50\%. The dew point should never exceed $T-20$°C. Fig.~\ref{fig:SensorIV} shows an example of a sensor IV curve, with the depletion voltage lying between 0 and $-100$ V, while the breakdown below $- 200$ V.

\subsection{Shunt-LDO VI}
A Shunt-LDO regulator is a device used to maintain a constant voltage level by diverting any excess current towards ground. LDO refers to the fact that the device works at Low DropOut, that is with a small difference between input and output voltage. A Shunt-LDO VI curve is measured to verify that FE chips are functional when the voltage is supplied to FE chips through regulators which alter the voltage received from the external power supply. 
Shunt-LDO VI curve is measured by powering up the chip with a 1.25 A/FE current applied and measuring the voltage at 0.1 A current steps from 5 A for a quad module down to 0 A. The measurement is done by switching the power supply off and on at each current step. There is no exact temperature required because as the power of the chip changes at each step, the temperature also changes, so a temperature requirement close to room temperature to an upper limit of 40°C is acceptable.

\subsection{Probing, trimming, ADC calibration}
In the module flex, probing pads are to be used to check some parameters are consistent with wafer probing data. An example of a parameter to check is that of the reference current used to operate the DAC that supplies power to the FE chips; this, usually indicated as $I_{\mathrm{ref}}$, must be equal to 4 $\mu$A. Trimming is instead the procedure of changing some register values of FE chips to get the correct values of the probing parameters. The chip has also an ADC which has to be calibrated to ensure that the voltage generated from the DAC is correct and to check other analog inputs such as the temperature.

\subsection{Chip tuning}
The readout system of the chips must be tuned before any data taking. The main purpose of chip tuning the readout chips is to ensure a uniform value of threshold and ToT gain over the entire chip. First, a sequence of scans has to be performed to test the functionality of the readout chip. While in the case of \text{tuning} the registers that define the characteristic parameters of the chip are modified, a \textit{scan} is instead used to measure the actual value of a parameter, without altering the configuration of the chip. A sequence of tunings for the threshold, ToT and noise is also performed.

\subsubsection{Digital scan}
\begin{figure}[!htb]
\centering
\includegraphics[width=0.48\textwidth]{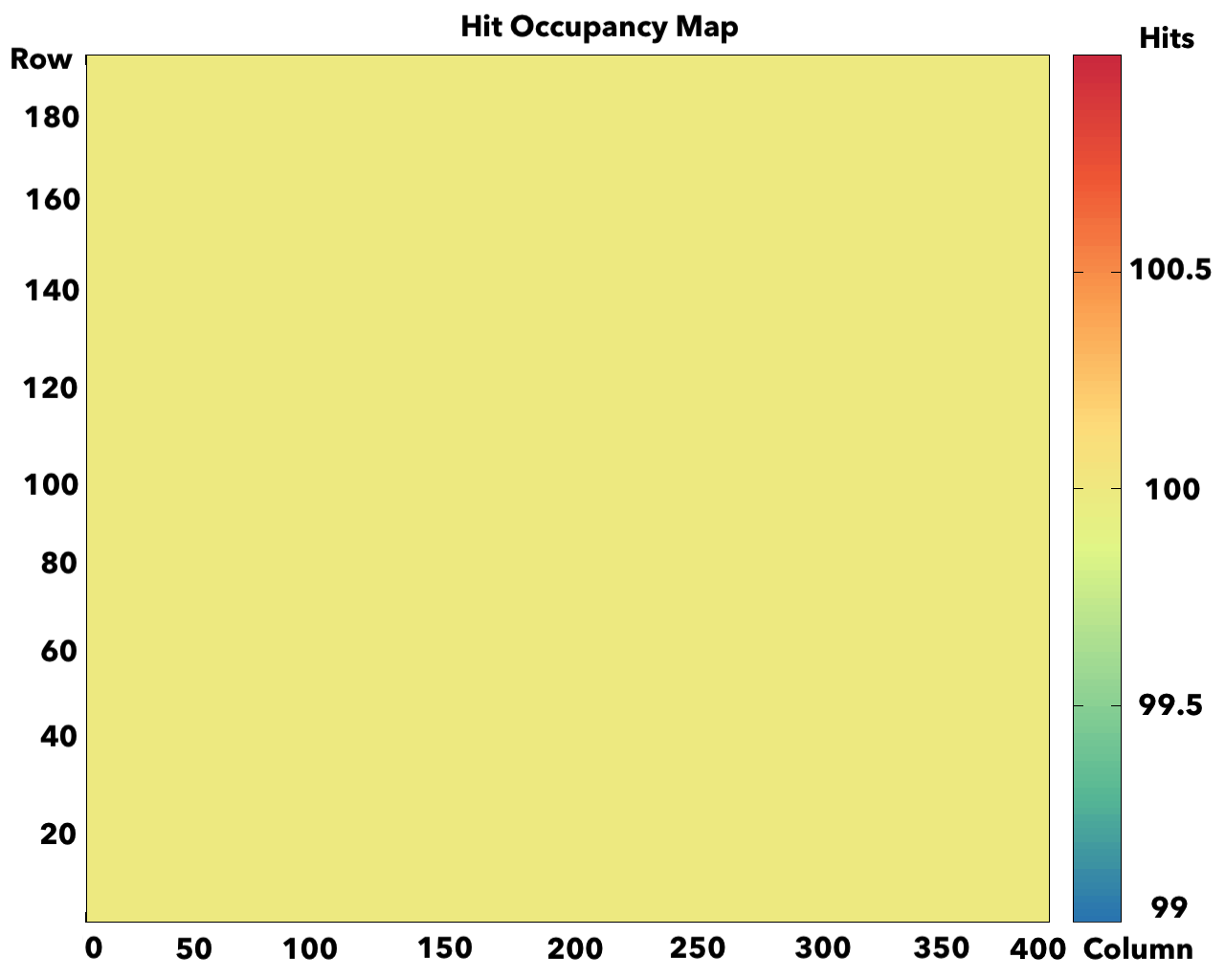}
\includegraphics[width=0.48\textwidth]{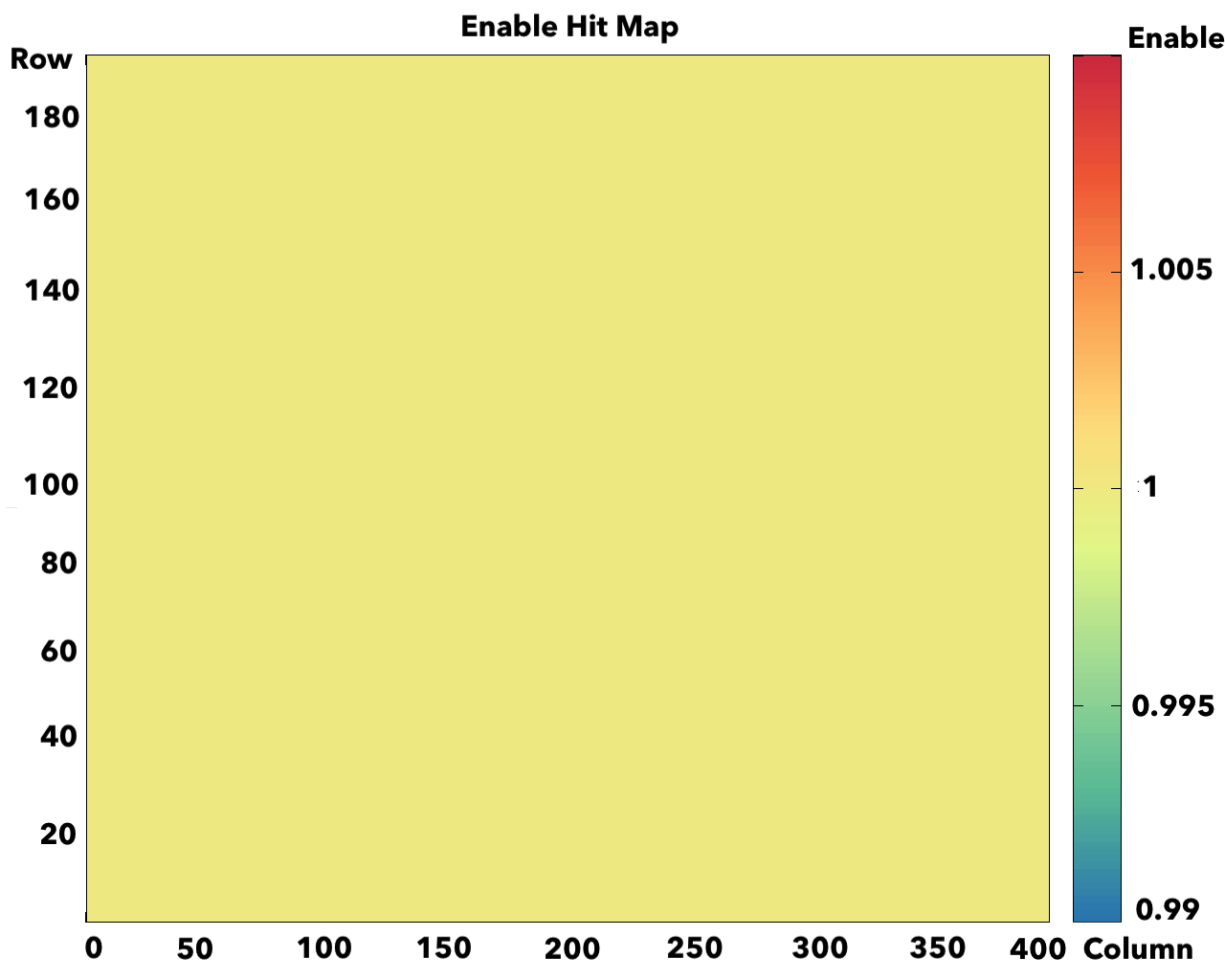}
\caption{The hit occupancy map (left) and enable map (right) for a perfectly working chip.}
\label{fig:HitMaps}
\end{figure}

In a digital scan, electrical pulses simulating the reception of a signal above the threshold are injected into the output of the discriminator of the chips to check the correct functioning of the digital component of the FE. To decouple the digital component under examination from the analog one, the threshold of the discriminators in the three different FE chips is set at a sufficiently high value to avoid a hit may be generated as a consequence of the noise and not of the digital injection.
The digital scan produces two two-dimensional maps, the \textit{hit occupancy map} and the \textit{enable hit map}. The hit occupancy map shows the number of hits recorded by each pixel for a fixed value of injections (100 by default), while the enable hit map is similar to the hit occupancy map but associates a value of 1 if the number of hits recorded is equal to that of the injections performed, 0 otherwise. The enable hit map is basically a map of working pixels and can possibly be used to disable pixels that have responded inadequately in subsequent scans. Fig.~\ref{fig:HitMaps} (left) shows the hit occupancy map for a perfectly functional chip where all pixels have recorded 100 hits, while Fig.~\ref{fig:HitMaps} (right) the corresponding enable hit map where a value of 1 is associated to each pixel.

\subsubsection{Analog scan}
In an analog scan, a certain amount of charge is injected at the amplifier input of each FE and the number of hits recorded by each pixel of the chip is recorded. The injected charge is sufficiently high to be above the threshold determined by the value of the register configuration saved in the memory. The analog scan produces maps similar to those provided by the digital scan. By combining the results of the analog scan with those of the digital scan, it is possible to identify the pixels for which the analog component of the FE chip exhibits incorrect behaviour; in fact, a malfunction of the analog component affects the response of the digital component, but not vice versa.

\subsubsection{Threshold tuning}
The goal of threshold tuning is to set the discriminator threshold to the desired value. The register containing the bits converted by the DAC into a threshold voltage of the discriminator is changed in each iteration. In correspondence of each iteration, an amount of charge equal to the desired threshold is injected and the number of hits recorded by each pixel is measured. The iterations are configured to gradually decrease the threshold, starting from a presumably higher value than that of the injected charge; consequently, the number of hits recorded by each pixel is initially zero. Assuming a Gaussian distribution of the thresholds for each pixel around the average value, the desired threshold is found to be in correspondence with a threshold value where the number of recorded hits is equal to 50\% of that expected from the injected charge. The tuning procedure ends when this condition occurs. Two types of threshold tuning can be performed, a global one to tune the threshold value for the entire pixel matrix and a local threshold one. 

\subsubsection{ToT tuning}
The goal of ToT tuning is to modify the proportionality relationship between the charge at the input of the amplifier of each FE chip and the corresponding ToT. In each iteration, the parameters that determine the discharge current of the capacitor in the integrated circuit are changed and the average of the ToT values recorded by each pixel is calculated. The tuning procedure ends when the averages of these distributions coincide with the desired ToT value, specified at the moment in which the scan is started. The ToT is expressed in units of bunch crossing (bc), with 1 bc = 25 ns.

\subsubsection{Crosstalk/disconnected bump scans}
The crosstalk refers to the phenomenon for which the charge released in a pixel induces a hit also in the adjacent ones due to parasitic capacities between them. Since these capacities are inevitably present in the module, the absence of crosstalk in a pixel is interpreted as damage to the corresponding bump bond and therefore a non-connection to the sensor. This conclusion is based on the fact that the crosstalk induced by the chip electronics is negligible and so this phenomenon is attributable solely to the sensor.
The crosstalk scan is evaluated by injecting a determined value of charge in the neighbouring pixels of a central pixel and checking the occupancy in the central pixel. The disconnected bump scan uses the crosstalk scan to identify pixels without any crosstalk which are likely due to disconnected bumps.

\subsection{Source scan}
In the previous steps, the behaviour of the chips has been verified by injecting a quantity of charge into the input of the FE circuits. To investigate the behaviour of the chips in response to the signal generated by the interaction with a particle, the module is exposed to a radioactive source.\\
At CERN, a $\mathrm{^{90}Sr}$ radioactive source is used. $\mathrm{^{90}Sr}$ is a radioactive isotope of strontium subject to $\beta$ decay into yttrium with a half-life of 28.79 years, $\mathrm{^{90}Sr \rightarrow e^{-} \, \bar{\nu}}_{e} \,\, \mathrm{^{90}Y}$. The isotope $\mathrm{^{90}Y}$ is also unstable and $\beta$ decays with a half-life of 64 hours into $\mathrm{^{90}Zr}$ which is a stable isotope of zirconium, $\mathrm{^{90}Y} \rightarrow e^{-} \, \bar{\nu}_{e} \,\, \mathrm{^{90}Zr}$.\\
Signals from the sensor can be transmitted only when the bump bonding connection between the sensor and the readout chip is functioning. These scans allow to investigate the quality of the bump bonds and ultimately verify the proper module functionality. They are particularly useful at the early stages of module building where the hybridization procedure needs to be qualified.

\chapter{Object reconstruction}
\label{sec:objreco}

The raw events provided by the data acquisition system are basically a list of detector element identifiers, with information on the signal registered by each such as energy or time. These events are subject to a procedure called object reconstruction, which processes the raw information to produce a higher level of data: a list of charged particle trajectories (tracks), jets, electrons, photons, muons, hadronically decaying $\tau$s, along with information such as their trajectory and energy. These are the objects which are the input to the subsequent physics analyses. \\

In this Chapter, the reconstruction of the relevant objects for the analysis is presented: charged particle trajectories (tracks) and vertices, electrons, muons, jets, $b$-jets, and missing transverse energy. \minitoc

\medskip

During object reconstruction, calibration of momentum and energy is also applied. The identification of the nature of an object (such as an electron or a jet) is never unambiguous: tighter selections lead to smaller misidentification rates but lower efficiency, and the optimal choices are analysis-dependent. For this reason, object reconstruction often provides candidates with a quality score, and the same detector signature can be often reconstructed as different objects (for example, all electron candidates are also jet candidates). Each analysis selects objects with the desired quality level and removes overlaps.\\

\section{Tracks and vertices}
\label{ssec:reco-Tvtx}

\subsubsection{Definitions}
This section presents the definition of items that are used in the following to build the objects:
\begin{itemize}
\item \textit{Hits} are the space points where charged particles cross the detector, measured by Pixel, SCT and TRT and they are the inputs to the tracks reconstruction.
\item \textit{Tracks} in the ID are obtained by fitting sets of hits in the various layers of the detector. The first algorithm searches for a set of three hits compatible with a helicoidal track. The three hits are then the seed for the Kalman filter, which is used to build complete track candidates by incorporating additional hits to the track seeds. A second fit is performed on the obtained track candidates and an ambiguity-solving procedure is applied for the tracks that have hits in common.
\item \textit{Vertices} are the points where two or more tracks start from, with no other hits behind. In addition to vertices coming from pile-up of the proton-proton collisions, they can be categorized into: 
\begin{itemize}
\item \textit{Primary vertex} where the hard scattering process occurs, located by the tracks with the highest $p_{\mathrm{T}}$.
\item \textit{Secondary vertex} where a particle with a relatively long lifetime decays into other charged particles.
\end{itemize}
\item \textit{Impact parameters of tracks} are defined with respect to the interaction point, $d_{0}$ is the transverse impact parameter and $z_{0}$ denotes the longitudinal one. They correspond to the distance of the closest approach to the primary vertex in the $r-\phi$ plane ($d_0$) and in the longitudinal plane ($z_0$), respectively. $\sigma(d_{0})$ and $\sigma(z_{0})$ are their corresponding uncertainties.
\item \textit{Energy clusters} are obtained by grouping calorimeter cells around a seed cell with an energy deposition above a certain threshold. Different methods are used in ATLAS to form energy clusters: the sliding window algorithm is used for the reconstruction of electrons and photons, while the topological or particle flow clustering is used for the reconstruction of jets.
\end{itemize}

\subsubsection{Track reconstruction}
\label{Section:TrackReconstruction}
The tracks of charged particles are reconstructed from the hits in many steps \cite{Cornelissen:2008zzc}.\\
\begin{itemize}
    \item[1.] \textit{Clusterization} of the raw measurements from the pixel and SCT detectors. A particle leaves a signal on more than one adjacent pixel or strip, depending on the position and the incidence angle on the sensor. Therefore, a Connected Component Analysis (CCA) algorithm \cite{Rosenfeld:1966:SOD:321356.321357} is applied to cluster neighbouring pixels or strips into a single object that represents the position of passage of the particle in the layer. Merged clusters from several charged particles are then split into sub-clusters using a Neural Network (NN) \cite{NNclustering} to measure more accurately the position of the clusters, the impact parameter resolution in both the transverse and longitudinal planes and to reduce the number of clusters shared between tracks in highly energetic jets.
    \item[2.] \textit{Iterative combinatorial track finding} algorithm, producing track seeds from three space points. The algorithm starts with SCT-only seeds, then adds pixel-only seeds, and finally mixed seeds, in order of purity. Purity is then improved with an additional requirement on $d_0$ and $p_{\mathrm{T}}$, and by requiring an additional space point to be compatible with the track extrapolation from the seed.
    \item[3.] \textit{Kalman filter} \cite{Fruhwirth:1987fm} to build tracks candidates including additional space points in the remaining layers.
    \item[4.] \textit{Ambiguity solver} algorithm, scoring the tracks according to the likelihood of being a good candidate, using information such as the number of clusters associated with it, the number of holes, and the $\chi^2$ of the track fit. Tracks are then processed in increasing order of track score and later the ambiguity solver deals with clusters assigned to multiple tracks, requiring that clusters are not shared among more than two tracks, there are no more than two shared clusters in the same track and other quality requirements.
    \item[5.] \textit{Extension to the TRT}, where tracks in the silicon detector are used as input to search for matching measurements in the TRT. The silicon-only track is not modified by this process and the association with the TRT hits are only extensions. The first step is to perform a fit between the TRT hits and the silicon tracks and then, as for the silicon hits, try to match onwards. A fit is performed again to try to improve the momentum resolution.
\end{itemize}

Sometimes it might happen that a candidate track in the TRT does not match any tracks in the SCT. This can happen when ambiguous hits shadow the tracks or when tracks come from a secondary vertex with few hits in the silicon. In this case, the algorithm will start a second sequence starting from the TRT and moving inside towards the silicon.

\subsubsection{Vertex Reconstruction}
Vertices can be identified starting from the reconstructed tracks \cite{Aaboud:2016rmg,Fruhwirth:2007hz} using two different approaches:

\begin{itemize}
    \item \textit{finding-through-fitting} approach, which reconstructs the vertices starting from tracks which are likely to be originated from the same area and fit them together with only one vertex candidate. Outlier tracks are removed and refitted using them as a seed for an additional vertex. The fit is redone and the process is repeated.
    \item \textit{fitting-after-finding} approach, which reconstructs the vertices searching for a cluster of tracks in the longitudinal projection. The cluster is fitted and the outlier rejected and never used in any other cluster. The maximal number of vertices is then decided at the seeding stage.
\end{itemize}

In a single collision, different vertices are identified since on average per bunch crossing there are $13$ interactions in 2015 data, 25 in 2016, 37 in 2017, and 36 in 2018, for a total average of 33.

\section{Electrons}
\label{ssec:reco-ele}
\subsubsection{Reconstruction}
An electron can lose a significant amount of its energy due to bremsstrahlung when interacting with the material it traverses. The radiated photon may convert into an electron-positron pair which itself can interact with the detector material. These positrons, electrons, and photons are usually emitted in a very collimated fashion and are normally reconstructed as part of the same electromagnetic cluster. The electron reconstruction procedure is based on clusters in the electromagnetic calorimeter that are matched with a reconstructed track inside the ID \cite{Aaboud:2019ynx,Ereconstruction1908}. The algorithm is built to allow for optimal reconstruction of the momentum and energy of the electrons in the whole pseudorapidity range and can be summarized in the following steps: 

\begin{itemize}
    \item[1.] It selects clusters of energy deposits measured in topologically connected EM and hadronic calorimeter cells, denoted \textit{topo-clusters}, reconstructed using a set of noise thresholds for the cell initiating the cluster and the neighbouring cells. During Run~2, the ATLAS reconstruction of clusters was improved from fixed-size topo-clusters of calorimeter cells to dynamic, variable-size topo-clusters, also called \textit{superclusters}. While fixed-size clusters naturally provide a linear energy response and good stability as a function of pile-up, dynamic clusters change in size is needed to recover energy from bremsstrahlung photons or electrons from photon conversions.
    \item[2.] The topo-clusters are matched to ID tracks, which are re-fitted accounting for bremsstrahlung. The algorithm also builds conversion vertices and matches them to the selected topo-clusters. A supercluster-building algorithm constructs electron and photon superclusters separately using the matched clusters as input. 
   \item[3.] After applying the initial position corrections and the energy calibrations to the resulting superclusters, the supercluster-building algorithm matches tracks to the electron superclusters and conversion vertices to the photon superclusters.
\end{itemize}

The electron and photon objects to be used for analyses are then built, discriminating variables used to identify electrons or photons from the backgrounds are defined, and the energies need to be calibrated.

\subsubsection{Identification}
Electron identification relies on a likelihood-based discriminant, whose inputs are variables with high discriminating power between isolated electrons and jets signatures. These variables include the information from the tracker and the matching, the information from the electromagnetic calorimeter, and hadronic leakage. The combination of all these variables is put together in a likelihood $L_{s/b}$, for both signal $s$ and background $b$. The discriminant is $d_L = \frac{L_s}{L_s + L_b}$ and a value is attributed to each electron candidate.
Electrons are identified by different sets of criteria. Different thresholds of the discriminant correspond to different benchmark working points (WPs): \textit{Loose}, \textit{Medium}, \textit{Tight}, and are chosen to have efficiency for electrons with $E_{\mathrm{T}}>$ 40 GeV of 93\%, 88\%, and 80\% respectively. This means that they are inclusive, and each one is a subset of the others. All of these WPs have fixed requirements on tracking criteria, they all require at least two hits in the pixel detector and at least 7 hits in the pixel and SCT detectors combined. Medium and Tight WPs additionally require that one of the pixel hits must be in the IBL to reduce backgrounds from photon conversions. 

\begin{figure}[!htb]
\centering
\includegraphics[width=0.7\textwidth]{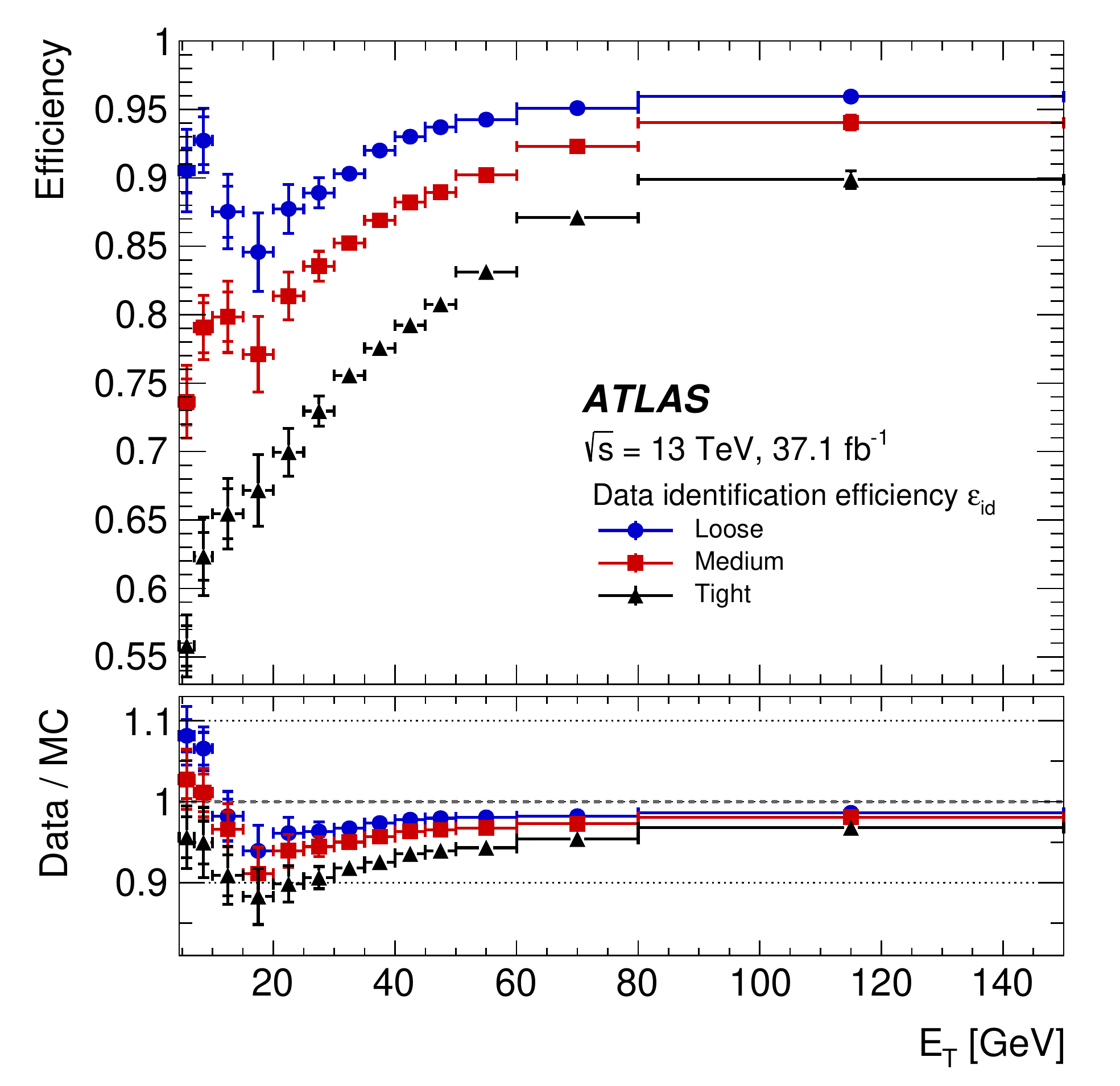}
\caption{The reconstruction and identification efficiency as a function of the $E_{\mathrm{T}}$ for MC and data (2016) for three different WPs. From Ref. \cite{Aaboud:2019ynx}.}
\label{fig:eleRecoEff}
\end{figure}

Fig.~\ref{fig:eleRecoEff} shows the combined reconstruction and identification of electrons for both data and MC, using $Z\rightarrow ee$ and $J/\Psi \rightarrow ee$ events and covering both high and low $E_{\mathrm{T}}$ of the electrons. The minimum $E_{\mathrm{T}}>$ of the electron identification was reduced from 7.5 GeV in Run~1 to 4.5 GeV in Run~2, which is a huge improvement. Even if electrons are identified only for $E_{\mathrm{T}}>4.5$ GeV, they can be reconstructed for lower $E_{\mathrm{T}}$, although with lower efficiency.

\subsubsection{Calibration}
Electron energy needs to be calibrated to deal with effects such as energy losses in passive materials, EM shower leakages and fluctuations in the deposited energy \cite{Aaboud:2018ugz}. These corrections are evaluated by comparing data and MC simulations for a well know Standard Model process, i.e. $Z\rightarrow e^{+}e^{-}$, $W\rightarrow e \nu$, and $J/\Psi \rightarrow e^{+}e^{-}$. The correct parameters are obtained after a global fit on the invariant mass of the $e^{+}e^{-}$ couple. Any residual miscalibration is corrected by the scale factor defined by
\begin{linenomath*}
\begin{equation}
\alpha = \frac{E_{\mathrm{measured}}-E_{\mathrm{truth}} }{E_{\mathrm{truth}}}
\end{equation}
\end{linenomath*}
where $E_{\mathrm{measured}}$ is the energy measured by the calorimeters after the MC-based correction, and $E_{\mathrm{truth}}$ is the energy at the truth level of the electrons.

\subsubsection{Isolation}
Prompt electrons coming from signal processes (from the hard-scattering vertex or the decay of heavy resonances such as Higgs, $W$, and $Z$ bosons) need to be separated from background processes such as semileptonic decays of heavy quarks, hadrons misidentified as leptons and photons, and photons converting into electron-positron pairs in the detector material upstream of the electromagnetic calorimeter.
A characteristic signature of such a signal is represented by little activity (both in the calorimeter and in the inner detector) in an area of $\Delta\eta\times\Delta\phi$ surrounding the candidate object. However, the production of boosted particles decaying, for example, into collimated electron-positron pairs or the production of prompt electrons, muons, and photons within a busy experimental environment such as in $t\bar{t}$ production can obscure the picture.\\
Calorimetric isolation variables are built by summing the transverse energy of positive energy deposit clusters whose barycenter falls within a cone centered around the selected electron. Then the energy of the electron itself is removed, and other corrections are done to account for pile-up. This variable is called $E_\mathrm{T}^{\mathrm{coneXX}} $ where XX depends on the size of the cone. Tracking isolation is similar to calorimetric isolation, with the difference that the $ p_{\mathrm{T}}$ of the tracks is used, summing together the tracks falling inside a cone of a given size around the candidate electrons. To compensate for the very busy environment at high $p_{\mathrm{T}}$, the cone is of variable size
\begin{equation}
\Delta R = \min \left( \frac{k_{\mathrm{T}}}{p_\mathrm{T}},R_{\mathrm{max}}\right).
\end{equation}

Different WPs can be defined using a combination of the calorimetric and tracking isolation variables (see Table \ref{tab:EleIsoWP} for the definition of the WPs).

\begin{table}[!htb]
\begin{center}
\begin{tabular}{lcc}
\noalign{\smallskip}\hline\noalign{\smallskip}
Working Point & Calorimetric Isolation & Track Isolation\\
\noalign{\smallskip}\hline\noalign{\smallskip}
Gradient & $\epsilon = 0.1143\times p_\mathrm{T}+92.14\% $ & $\epsilon = 0.1143\times p_\mathrm{T}+92.14\% $ \\ 
FCHighPtCaloOnly & $E_{\mathrm{T}}^{\mathrm{cone20}} <\max (0.015\times p_\mathrm{T},3.5\,\mathrm{GeV})$ & -\\
FCLoose & $E_{\mathrm{T}}^{\mathrm{cone20}}/ p_\mathrm{T} <0.20$ &  $ p_\mathrm{T}^{\mathrm{varcone20}}/ p_\mathrm{T} <0.15$	\\
FCTight & $E_{\mathrm{T}}^{\mathrm{cone20}}/ p_\mathrm{T} <0.06$ & $ p_\mathrm{T}^{\mathrm{varcone20}}/ p_\mathrm{T} <0.06$ \\
\noalign{\smallskip}\hline\noalign{\smallskip}
\end{tabular}
\end{center}
\caption{Definition of the electron isolation WPs. For the Gradient WP, the units of $p_{\mathrm{T}}$ are GeV. All operating points use a cone size of $\Delta R = 0.2$ for calorimeter isolation and $R_{\mathrm{max}} = 0.2$ for track isolation. The values are obtained from a simulated $Z \rightarrow ee$ sample where electrons satisfy Tight identification requirements.}
\label{tab:EleIsoWP}
\end{table}

The WPs are obtained by asking for either a fixed value of efficiency or a fixed cut on the isolation variables. \textit{Gradient} WP is built by asking that the efficiency is 90\% at $ p_\mathrm{T}=25$ GeV and 99\% at $ p_\mathrm{T}=60$ GeV, and uniform in $\eta$. Instead, the other WPs have fixed cuts on track and calorimeter isolation information. \textit{FCHighPtCaloOnly} does not use tracking information to reduce the contribution of fake leptons and high $p_{\mathrm{T}}$ events from multijet processes.

\section{Muons}
\label{ssec:reco-muon}

\subsubsection{Reconstruction}
Muons are first reconstructed independently by both the MS and the ID, and then the information from these two subdetectors is combined \cite{Aad:2016jkr}.\\
Reconstruction in the ID is the same as for any other particle. The reconstruction in the MS starts with the search in the muon chamber of hit patterns forming segments in the bending plane of the detector. Muon track candidates are built from a fit with segments from different layers. The algorithm starts from the segments generated in the middle layers of the detector where more trigger hits are available and the search is then extended to use the segments from the outer and inner layers using criteria such as hit multiplicity and fit quality. Hits associated with a track are fitted and the track candidates are accepted if the $\chi ^2 $ of the fit satisfies the selection criteria. Outlier hits in the $\chi ^2 $ are removed.\\
The next step in the reconstruction is matching the information between the ID and the MS. Four muon \textit{types} are defined according to which subdetectors are used in reconstruction:
\begin{itemize}
\item \textit{Combined (CB)} muons. Tracks are reconstructed independently by both the ID and MS and a global fit is performed. Hits may be removed or added to have a better fit. Muons reconstruction follows an outside-in approach, starting from the MS and searching for a match in the ID, while inside-out approaches are used as complementary.
\item \textit{Segment-tagged (ST)} muons. Tracks are reconstructed by the ID and, when extrapolated to the MS, at least one track segment in the MDT or CSC chambers is required. This is helpful for low $p_\mathrm{T}$ muons or for muons that fall in regions with reduced MS acceptance.
\item \textit{Calorimeter-tagged (CT)} muons. Tracks are reconstructed by the ID and associated with energy deposits in the calorimeter which are compatible with minimum ionizing particles. This type of muons have a low purity but they are helpful in the less instrumented parts of the MS detector, where the cabling and services of the calorimeter and the ID are positioned. CT reconstruction criteria are optimised for a range $15 <  p_\mathrm{T} < 100$ GeV.
\item \textit{Extrapolated (ME)} muons. Tracks are reconstructed by the MS and the muon trajectories are extrapolated to the ID by requiring loose criteria on the compatibility with the interaction point. Energy loss in the calorimeter is also estimated. Tracks are required to have traversed at least two layers of MS chambers in the barrel and at least three in the forward region. This type of muons is helpful to recover the muons outside the ID acceptance, in the pseudorapidity range 2.5 $ < | \eta | < $ 2.7.
\end{itemize}

Overlaps between different types are resolved by giving different priorities. For muons sharing the same ID tracks, preference is given to the CB category, then ST and finally CT. ME overlaps are resolved by selecting tracks with better fit quality.\\

\subsubsection{Identification}
Muon identification is done by applying quality requirements. This suppresses background from pions or kaons decays. Several variables that have high discriminating power are studied using a $t\bar{t}$ sample. 

Some of these variables are: 
\begin{itemize}
\item $q/p $ significance: difference between the ratio of the charge and momentum of muons candidates in the ID and the MS divided by the sum in quadrature of the corresponding uncertainties;
\item $\rho '$: difference between the transverse momentum measured in the ID and MS over the combined $p_\mathrm{T}$;
\item normalised $\chi^2$ of the combined track.
\end{itemize}

Then five different identification categories are defined with different sets of requirements. This corresponds to five different WPs, with different background rejection rates and identification efficiencies.
\begin{itemize}
\item \textit{Medium muons.} This is the default selection, it minimises the systematic uncertainties. Only CB and ME tracks are used. The CB muons are required to have at least three hits in at least two MDT layers, unless in $ | \eta | < $ 0.1 where also tracks with up to one MDT hole are allowed. The ME are used in the 2.5  $ < | \eta | < $ 2.7. Requirements on ID and MS momentum compatibility are added to suppress hadrons misidentified.
\item \textit{Loose muons.} This WP is designed to maximise the reconstruction efficiency, with all types of muons being used. CB and ME medium muons are included, while CT and ST are restricted to the $|\eta|<$ 0.1 region. This selection is optimised to provide good quality muon track, specifically for Higgs searches in the four-lepton channel.
\item \textit{Tight muons.} This set of cuts is chosen to increase the purity of muons at the cost of efficiency. Only CB medium muons with hits in at least two stations of the MS are selected. Cuts on the $\chi^2$ and the $p_\mathrm{T}$ are also applied. 
\item \textit{High $p_\mathrm{T}$ muons.} This WP aims to maximise the $p_\mathrm{T}$ resolution for tracks with high transverse momentum (over 100 GeV). CB medium muons with at least three hits in three MS stations are selected. Tracks in specific regions of the MS where there are discontinuities are vetoed. This procedure reduces the efficiency by almost 20 \% but improves the $p_\mathrm{T}$ resolution by almost 30 \% for muons up to 1.5 TeV. This WP is helpful for searches for high-mass $Z'$ and $W'$.
\item \textit{Low $p_\mathrm{T}$ muons.} This WP is used to reconstruct events with very low momentum: down to 4 GeV in the 2015-2016 data-taking period, and then down to 3 GeV. Only CB tracks are used, and at least one MS station for $|\eta| < 1.3$, while at least two MS stations for $1.3 < |\eta| < 1.55$. Medium WP is required for $|\eta|>1.55$. Additional variables are used to discriminate prompt and fake muons. Compared to the other WPs, the low $p_\mathrm{T}$ WP allows for higher efficiency in the barrel at the expense of a higher fake rate. 
\end{itemize}

\subsubsection{Calibration}
A measurement of the muon reconstruction efficiency in the region $ | \eta | < $ 2.5 is obtained with a tag-and-probe method. This method is similar to the one used for the electrons, it selects a pure sample of $J/ \Psi \rightarrow \mu \mu$ and $Z\rightarrow \mu \mu$ events. The difference between data and MC prediction is used to compute scale factors for compensating energy losses in the materials or distortions in the magnetic fields not optimally simulated.

\subsubsection{Isolation}
Similarly to the electrons, different isolation WPs are defined to reduce the contribution of non-prompt muons. In the same way as for electrons, they use tracking and calorimetric isolation variables based on $p_\mathrm{T}^{\mathrm{varconeXX}}$ and $E_{\mathrm{T}}^{\mathrm{cone20}}$, and share the same definitions.

\section{Jets}
\label{ssec:reco-jets}
Coloured particles arising from the hard scattering (gluons and quarks) can not stay in a free state, therefore they create other particles to a have colourless state \cite{Forshaw:1997dc}. Initial partons involved in the hard scattering may radiate further gluons, which then may split into further quark anti-quark pairs, and so on until partons are confined in a colourless state, i.e. hadrons. This process is called hadronization. This happens in a time of the order of $\Lambda_{\text{QCD}}^{-1}$, which for the time scale considered in the colliders, is almost instantaneous, and therefore happens inside the beam pipe, in the collision point. The produced particles (typically $K$, $\pi$, neutrons, and very few protons) will then reach the detector and consequently interact in the matter, creating a chain reaction, called a \textit{shower}, which generates many other particles. Moreover, $\pi^0$ generates photons couples that generate more compact electromagnetic showers inside the hadronic one. \\
The QCD radiation associated with an initial quark and gluons with a momentum above 10-20 GeV, and then the resulting hadrons, is usually at close angles with the direction of the initial parton, therefore it is possible to define a \textit{jet} as an object formed by the vector sum of all the particles generated by the hadronization process inside a given cone. A jet contains information about the properties of the initial parton, which otherwise would be unknown.\\

\subsubsection{Reconstruction}
Hadronic particles deposit most of their energy in the calorimeter system, and it is possible to see the jet as a \textit{local maximum} of deposited energy. Clustering together the inputs from the cells makes it possible to construct the jets and point to the original coloured particle \cite{Lampl:1099735,Aad:2016upy}. The reconstruction of a jet evolves through a series of steps.\\
The first step in the reconstruction of a jet is to cluster the proto-clusters in the calorimeter and sum together their energy. This process starts with a seed cell and then adds the neighbouring cells if the energy in these cells is over a certain threshold. This algorithm is called \textit{TopoCluster} and it is divided into two separate steps: \textit{cluster maker} and \textit{cluster splitter}.

\begin{itemize}
    \item \textit{cluster maker.} Initially, all the cells with a signal-to-noise ratio over a certain threshold $t_{\text{seed}}$ are identified. The noise here is the RMS of the electronics noise, while the signal is the cell energy. These cells are the seed around which to build the cluster, called now \textit{proto-cluster}. Now all the neighbouring cells are considered and if their signal-to-noise ratio is above a $t_{\text{neighbor}}$ threshold, the cell is added to the proto-cluster. If a cell is adjacent to more than one proto-cluster, these proto-clusters are merged together. This process is repeated until all the cells are in a proto-cluster or below threshold. Clusters are selected according to the transverse energy, $E_{\mathrm{T}}$. A cluster is removed if it has a $E_{\mathrm{T}}$ less than a certain threshold. This is useful to remove pure noise proto-clusters.
    \item \textit{cluster splitter.} The cluster is separated by finding a set of local maximum cells satisfying: 
    \begin{itemize}
    \item $E>500$ MeV
    \item Energy greater than any adjacent cell
    \item At least 4 neighbouring cells with energy over threshold.
    \end{itemize}
     Clusters are then grown around this set of local maxima as before, except that only the cells originally clustered are used, no thresholds are applied, and no cluster merging occurs. Cells shared by multiple proto-clusters are added to the two most energetic proto-clusters with a weight $w_{1,2}$ defined by:
     \begin{equation}
     w_1=\frac{E_1}{E_1+rE_2}, \qquad w_2=1-w_1, \qquad r=\text{exp}(d_1-d_2),
     \end{equation} 
     where $E_{1,2}$ are the energies of the proto-clusters and $d_{1,2}$ are the distances between the cell and the proto-cluster centres.
\end{itemize}

The next step is to cluster the proto-clusters in jets, which is done by the anti-$k_{\mathrm{T}}$ jet clustering algorithm \cite{Cacciari:2008gp}. The algorithm works by first defining a distance $d_{ij}$ between the objects $i$ and $j$, and the distance $d_{iB} $ between the object $i$ and the beam ($B$). These quantities are computed as
\begin{linenomath*}
\begin{equation}
\begin{split}
d_{ij} & = \min ( k_{{\mathrm{T}}i}^{2p},k_{{\mathrm{T}}j}^{2p} ) \, \frac{R_{ij}^2}{R^2} \\
d_{iB} & = k_{{\mathrm{T}}i}^{2p}
\end{split}
\end{equation}
\end{linenomath*}
where $k_{{\mathrm{T}}i}$ is the transverse momentum of the particle $i$ and $R_{ij}^2=(\eta_i - \eta_j)^2 + (\phi_i - \phi_j)^2 $ where $\eta_{i,j}$ and $\phi_{i,j}$ are the pseudorapidity and azimuth of the particle $i$, respectively. $R$ is the radius jet parameter (set to $R =$ 0.4), while $p$ is a parameter that governs the relative power of the energy versus geometrical scales $R_{ij}^2$ (set to $p=-1$ for the anti-$k_{\mathrm{T}}$ algorithm). \\
The algorithm then works by identifying the smallest of the distances and, if it is a $d_{ij}$, the objects $i$ and $j$ are added together so that $k^{\text{new}}=k_i + k_j$; instead, if it is $d_{iB}$, $i$ is declared as jet and removed from the list. The process is then repeated until no more objects are left. \\
The choice of the parameter $p=-1$ is performed so that for equally separated particles, the $d_{ij}$ for a hard particle $i$ and a soft particle $j$ is smaller than the $d_{ij}$ between two soft terms, therefore the algorithm clusters the soft and the hard particle before. This way, the algorithm is more stable with respect to soft particles and improves the ability to converge. In other algorithms, soft particles tend to destabilise the process of convergence, while in the anti-$k_{\mathrm{T}}$ soft terms do not modify the shape of the jet, while hard particles do.\\
In Run~1 of the LHC, the ATLAS experiment used either solely the calorimeter or solely the tracker to reconstruct hadronic jets and soft particle activity. The vast majority of analyses utilised jets that were built from topological clusters of calorimeter cells, the topo-clusters discussed. An alternative approach, called \textit{Particle flow} \cite{PFlowJets}, reconstruct \texttt{AntiKt4EMPFlowJets} by combining signal measurements from both the tracker and the calorimeter. Jet reconstruction is performed on an ensemble of \textit{particle flow objects} consisting of the calorimeter energies and tracks which are matched to the hard interaction. The advantage of using particle flow objects is that, for low-energy charged particles, the momentum resolution of the tracker is significantly better than the energy resolution of the calorimeter, as well as the acceptance of the detector is extended to softer particles, as tracks are reconstructed for charged particles with a low transverse momentum, whose energy deposits often do not pass the noise thresholds required to seed topo-clusters. The capabilities of the tracker in reconstructing charged particles are complemented by the ability of the calorimeter to reconstruct both the charged and neutral particles. At high energies, instead, the energy resolution of the calorimeter is superior to the momentum tracker resolution. Outside the geometrical acceptance of the tracker, only the calorimeter information is available. Hence, the topo-clusters in the forward region are used alone as inputs to the particle flow jet reconstruction. Thus a combination of the two subsystems is preferred for optimal event reconstruction. Additionally, when a track is reconstructed in the tracker, one can ascertain whether it is associated with a vertex, and if so the vertex from which it originates. Therefore, in the presence of multiple in-time pile-up interactions, the effect of additional particles on the hard-scatter interaction signal can be mitigated by rejecting signals originating from pile-up vertices.

\subsubsection{Jet Energy Calibration}
\begin{figure}[!htb]
\centering
\includegraphics[width=1.\textwidth]{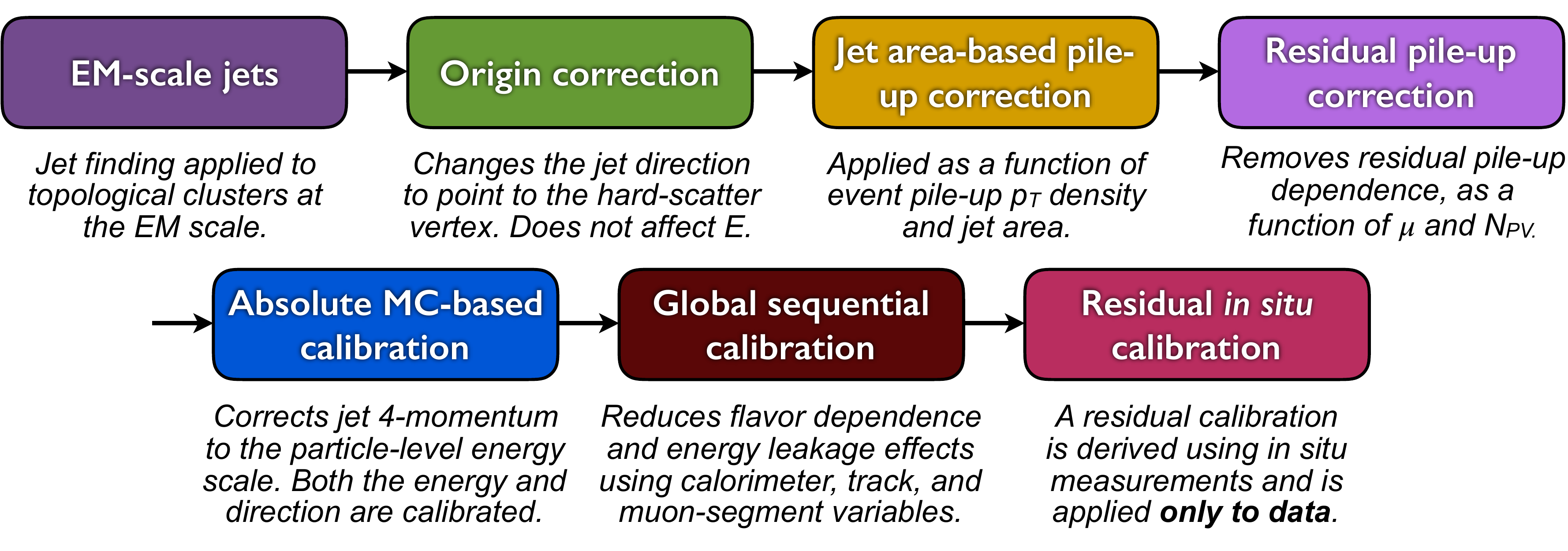}
\caption{Schematic diagram of the algorithms used for the Jet Energy Calibration.}
\label{fig:SchemJETCal}
\end{figure}

Jets are built by clustering energy deposits in the calorimeter. This energy is measured at the electromagnetic scale (EM-scale), which is the signal scale that electromagnetic showers deposit in the calorimeter. This means that for hadrons the energy measurement is underestimated by $15-55$ \% because hadronic and electromagnetic particles interact differently in material and the ATLAS calorimeter does not compensate for this effect. Variable electromagnetic content and energy losses in the dead material lead to a worse resolution on the jet energy measurement in comparison to particles interacting only electromagnetically (electrons and photons). Therefore, jet energy calibration is needed to correct the bias in the reconstructed energy and reduce as much as possible the spread in the response. The calibration corrections are obtained by trying to unify the response of the jets by applying corrections obtained from MC simulations and data-driven methods \cite{Aaboud:2017jcu}. This process defines the \textit{jet energy scale} (JES). Fig. \ref{fig:SchemJETCal} shows a schematic diagram of the different steps used in the calibration.

\begin{itemize}

\item \textit{Origin correction} The first step is to change the origin direction of the four-momentum of the jet so that it will point to the hard-scatter primary vertex, rather than the centre of the detector. The jet energy is kept constant. This step improves the resolution in $\eta$.
\item \textit{Pile-up correction} Two other steps are used to reduce the effects of in-time and out-of-time pile-up \cite{Cacciari:2007fd,ATLAS-CONF-2013-083}. In the first part of the procedure (\textit{jet area-based pile-up correction}) the average pile-up contribution in each event is removed from the $p_{\mathrm{T}}$ of each jet, according to an area-based method. The pile-up contribution is obtained from the $p_{\mathrm{T}}$ density of jets ($\rho$) in the $\eta-\phi$ plane. The density of each jet is defined as $p_\mathrm{T}/A$. The second part of the procedure takes care instead of the residual $p_{\mathrm{T}}$ dependence on the number of reconstructed primary vertices $N_{PV}$ and the number of interactions per bunch crossing $\mu$ (\textit{Residual pile-up correction}). These dependencies are found to be linear and independent of one another and coefficients are fitted. After these corrections, the $p_{\mathrm{T}}$ is
\begin{equation}
p_\mathrm{T}^{\text{corr}} = p_\mathrm{T}^{\text{reco}} - \rho\times A - \alpha\times(N_{PV}-1) -\beta\times\mu.
\end{equation}
\item \textit{Absolute calibration} The absolute jet calibration corrects the reconstructed jet four-momentum to the particle-level energy scale and accounts for biases in the jet $\eta$ reconstruction, caused by the transition between different parts of the calorimeter. The correction is derived from MC, matching jets to truth particles within $\Delta R=0.3$, and using only isolated jets (no further jets of $p_\mathrm{T} > 7$ GeV within $\Delta R=0.6$). The response is then defined as $E^{\mathrm{reco}}/E^{\mathrm{truth}} $ and binned in $\eta$.
\item \textit{Global sequential calibration} This calibration scheme is based on the jet structure to try to compensate for the energy fluctuation \cite{Aaboud:2017jcu}. This method uses the topology of the jet (number of tracks in the jets, or muons segments) and its energy deposit to characterize the energy fluctuations. For each observable used, the four-momentum is corrected, as a function of $p_\mathrm{T}^{\mathrm{truth}}$ and $\eta$, but with an overall constant in order to leave unchanged the average energy of the jets at each step.
\item \textit{In-situ calibration} The last step of the calibration accounts for differences in the response between the data and the MC, due to imperfect description in the simulations: from detector material to hard scatter and pile-up. This is done by balancing the $p_{\mathrm{T}}$ of the jet against well-known objects. Central jets ($|\eta|<0.8$) use $Z/\gamma$+jets events, where the jets are balanced against the $Z$ boson or the $\gamma$. Multijet events are instead used for high $p_{\mathrm{T}}$ central jets (300 GeV $< p_\mathrm{T} <$ 2000 GeV), where the high $p_{\mathrm{T}}$ jets are balanced against well-known central low $p_{\mathrm{T}}$ ones. Dijet events are instead used for forward jets ($0.8<|\eta|<4.5$), where the jets are balanced against the central jets.
\end{itemize}

\subsubsection{Jet Calibration Systematic Uncertainties}
\begin{figure}[!htb]
    \centering
\subfloat[ ]{\includegraphics[width=0.49\textwidth]{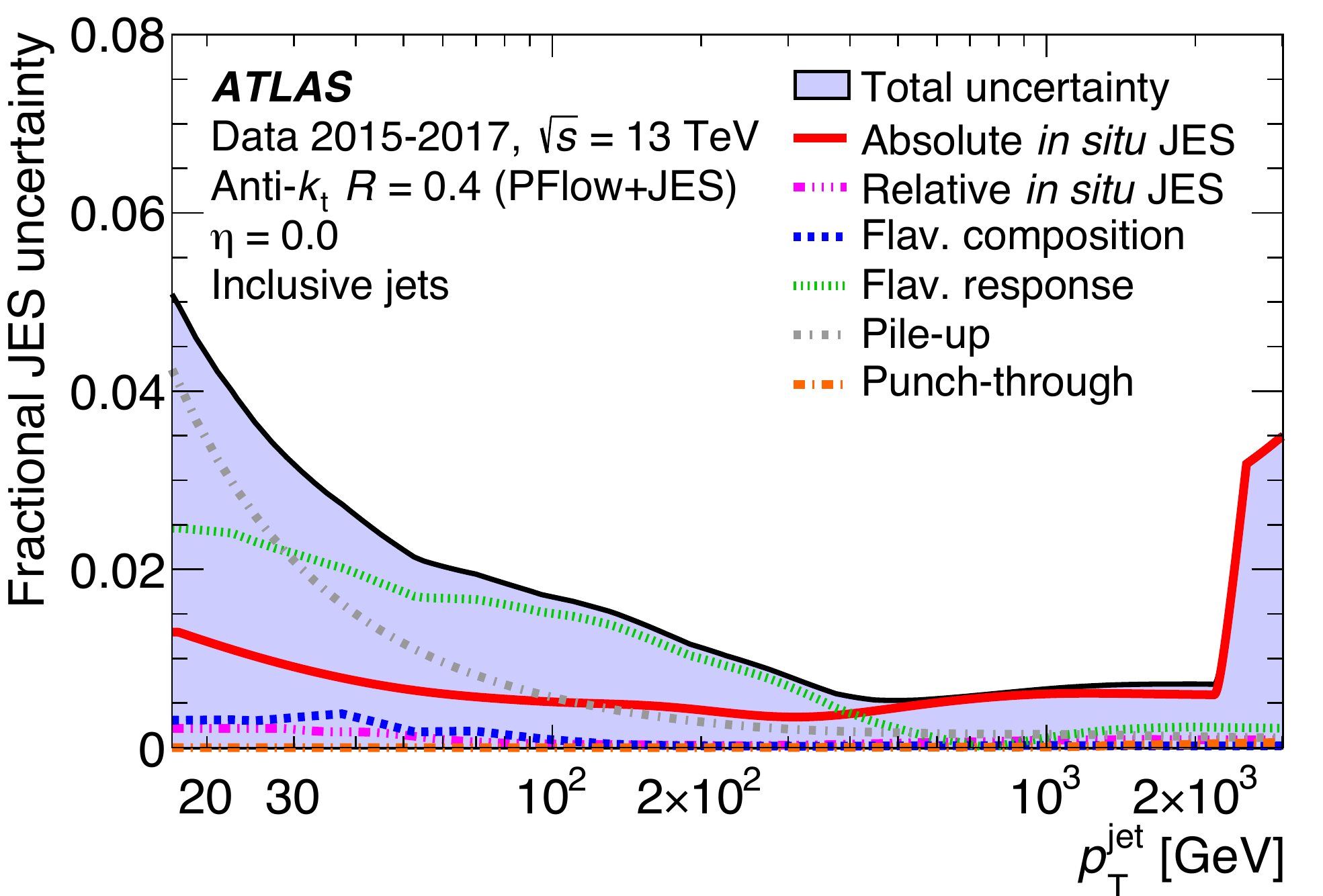}}
\subfloat[ ]{\includegraphics[width=0.49\textwidth]{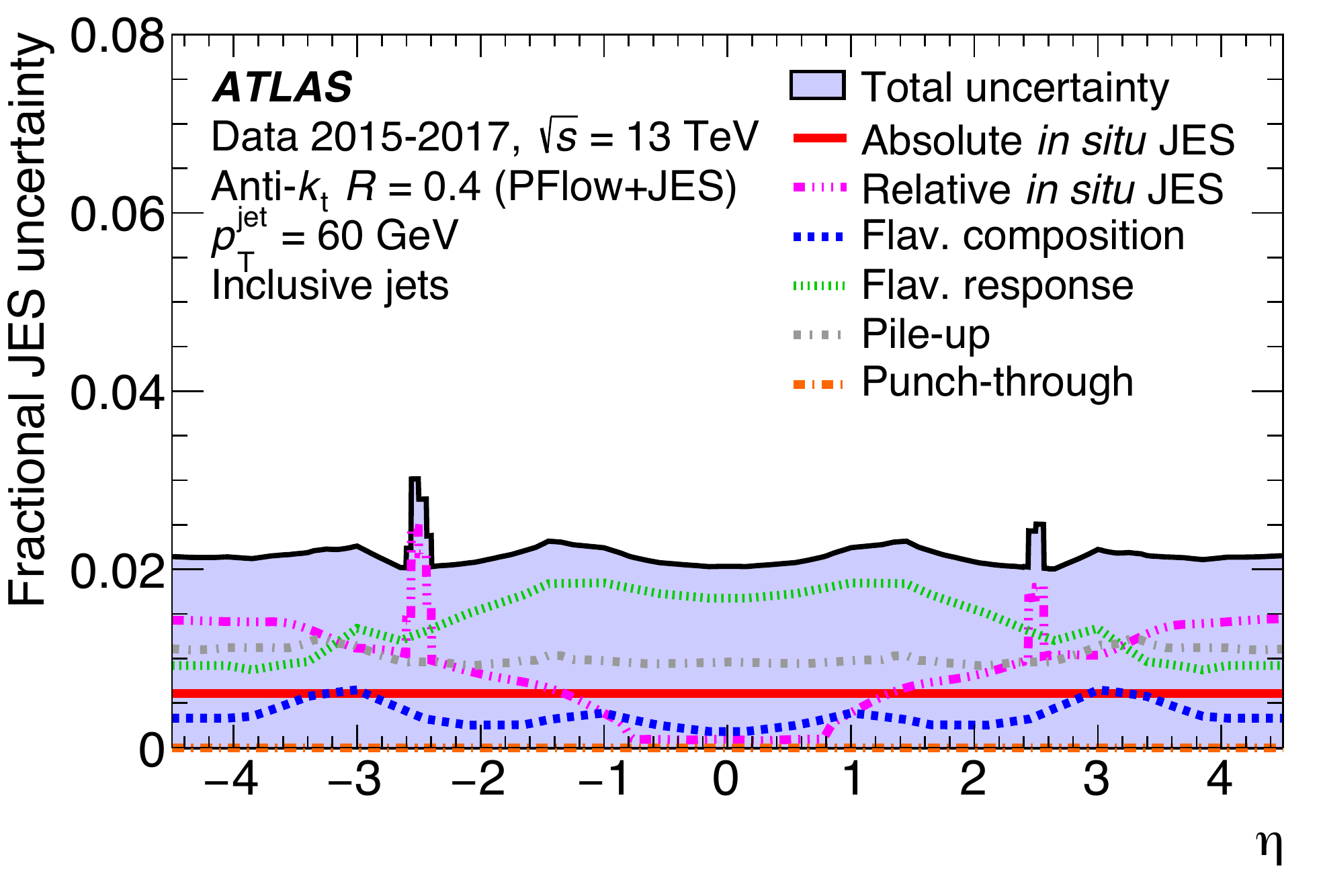}}
\caption{Fractional jet energy scale systematic uncertainty components for $R=0.4$ anti-$k_{\mathrm{t}}$ jets for: (a) $\eta=0$ as a function of $p_{\mathrm{T}}^{\mathrm{jet}}$, (b) $p_\mathrm{T}^{\mathrm{jet}} = 60$ GeV as a function of $\eta$, reconstructed from electromagnetic-scale topo-clusters. The total uncertainty (all components summed in quadrature) is shown as a filled region topped by a solid black line. From Ref.~\cite{ATLASJesJer}.}
\label{fig:JetJESunc}
\end{figure}

The calibration procedure brings with it a set of uncertainties that are propagated from the individual calibration to the final jet \cite{Aaboud:2017jcu}. There is a total of 80 JES systematic uncertainties: 67 come from $Z/\gamma+$jets in situ calibration and account for topology assumption, MC simulation and statistic, and propagated electrons/muons/photon energy scale. The other 13 systematic uncertainties come from pile-up (4), $\eta$-intercalibration in the region with $2.0<|\eta|<2.6$ region (3), and difference in response of light-quark, $b$-quark, and gluon initiated jets (3). Another uncertainty comes from the Global Sequential Calibration (GSC) punch-through correction. For jets outside the in-situ methods (with a $p_\mathrm{T}> 2$ TeV) an additional uncertainty is applied. For fast simulation, an AFII modelling uncertainty is also considered for non-closure in the JES calibration. Fig.~\ref{fig:JetJESunc} shows the total uncertainty as a function of jet $p_{\mathrm{T}}$ and $\eta$.
However, most of the physics analyses do not need to evaluate and propagate each one of the systematic uncertainties because most of the information might be unnecessary. Therefore, a reduced set of nuisance parameters (NPs) is produced, trying to preserve as precisely as possible the correlation across $p_{\mathrm{T}}$ and $\eta$.\\ 
The set of in-situ systematic uncertainties is reduced from 67 to the 5 most relevant and the others are combined in a single parameter. Then these remaining 19 NPs (6 from the in-situ plus 13) are combined into four reduced NPs. This reduction of course reduces the correlations between most of the uncertainties, but the loss of information is indeed small for most of the analyses.

\subsubsection{Jet Energy Resolution}
\begin{figure}[!htb]
    \centering
\subfloat[ ]{\includegraphics[width=0.49\textwidth]{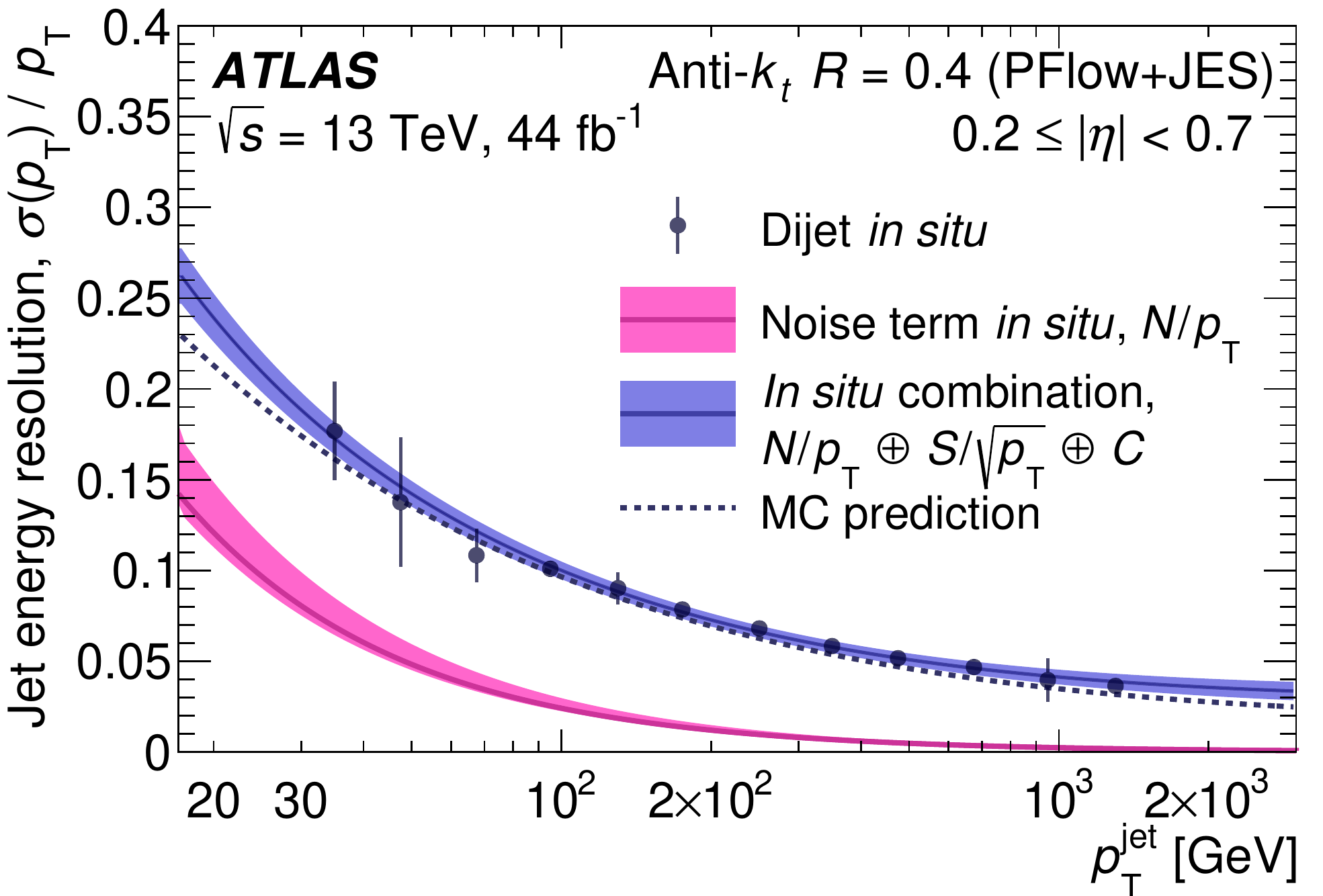}}
\subfloat[ ]{\includegraphics[width=0.49\textwidth]{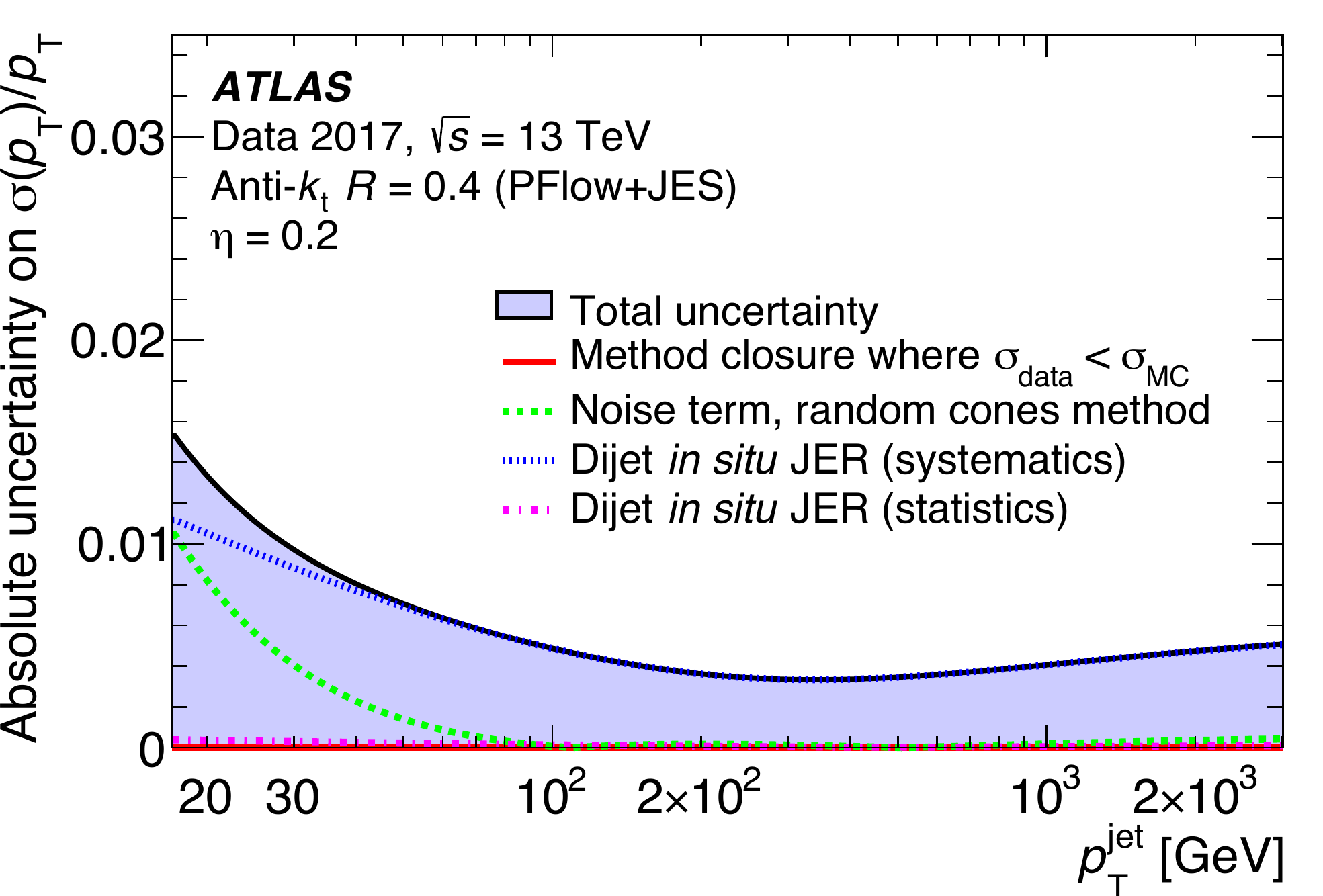}}
\caption{(a) The jet energy resolution $\sigma(p_\mathrm{T})/p_\mathrm{T}$ and (b) the absolute uncertainty on the jet energy resolution, as a function of $p_{\mathrm{T}}^{\mathrm{jet}}$ for anti-$k_{\mathrm{t}}$ jets with a radius parameter of $R = 0.4$ and inputs jets calibrated with the PFlow+JES scheme followed by a residual in situ calibration and using the 2017 dataset. From Ref.~\cite{ATLASJesJer}.}
\label{fig:JETEMJERSyst}
\end{figure}

After the jet energy scale calibration, it is also measured the energy resolution (JER). This can be parametrized as
\begin{equation}
\frac{\sigma(p_\mathrm{T})}{p_\mathrm{T}} = \frac{N}{p_\mathrm{T}}\oplus\frac{S}{\sqrt{p_\mathrm{T}}} \oplus C,
\end{equation}
where $N$ is a \textit{noise} term that contains effects from pile-up and electronic noise, that enters at very low $p_{\mathrm{T}}$. $S$ is the statistical Poisson fluctuations due to the sampling nature of the calorimeter. The last term, $C$, is due to the passive material inside the detector. JER is measured in data and MC by balancing the jet $p_{\mathrm{T}}$ with dijet events, $Z+$jets and $\gamma+$jets in a similar way as for the JES. Again, this procedure results in $\sim100$ uncertainties, which should be propagated to the analysis level, and as for the JES they are combined together in NPs, two sets are possible: 7 NPs and 12 NPs, depending on the needs of the single analysis. The uncertainties are also constrained with respect to the inputs by the use of a fit function that constrains $N$, $S$, and $C$. Fig.~\ref{fig:JETEMJERSyst}(a) shows the JER as a function of jet $p_{\mathrm{T}}$ for 2017 data and MC, while Fig.~\ref{fig:JETEMJERSyst} (b) shows its absolute uncertainty, divided by type.

\subsubsection{Jet Vertex Tagger}
\begin{figure}[!htb]
    \centering
\subfloat[]{\includegraphics[width=0.3\textwidth]{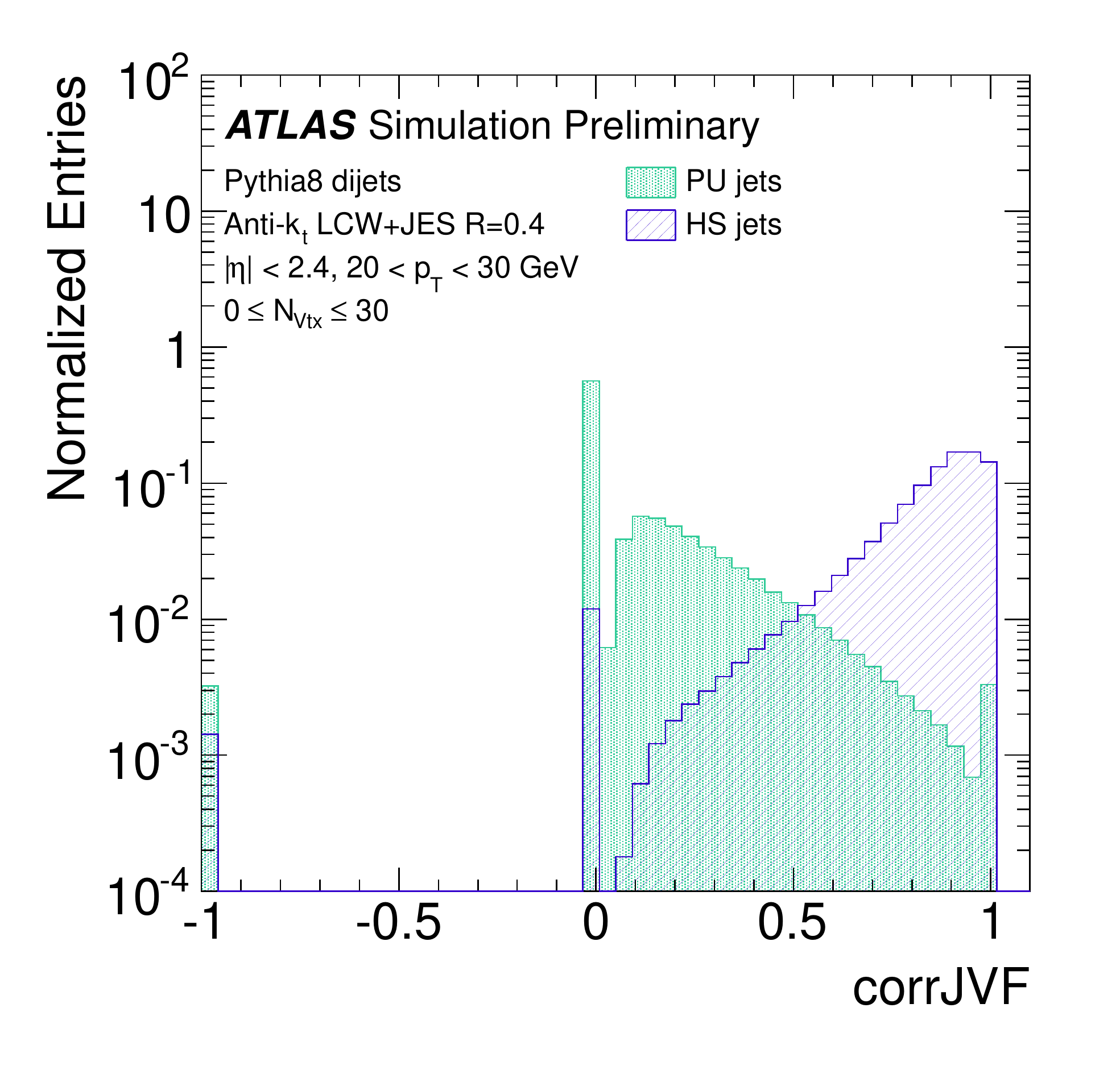}}
\subfloat[]{\includegraphics[width=0.3\textwidth]{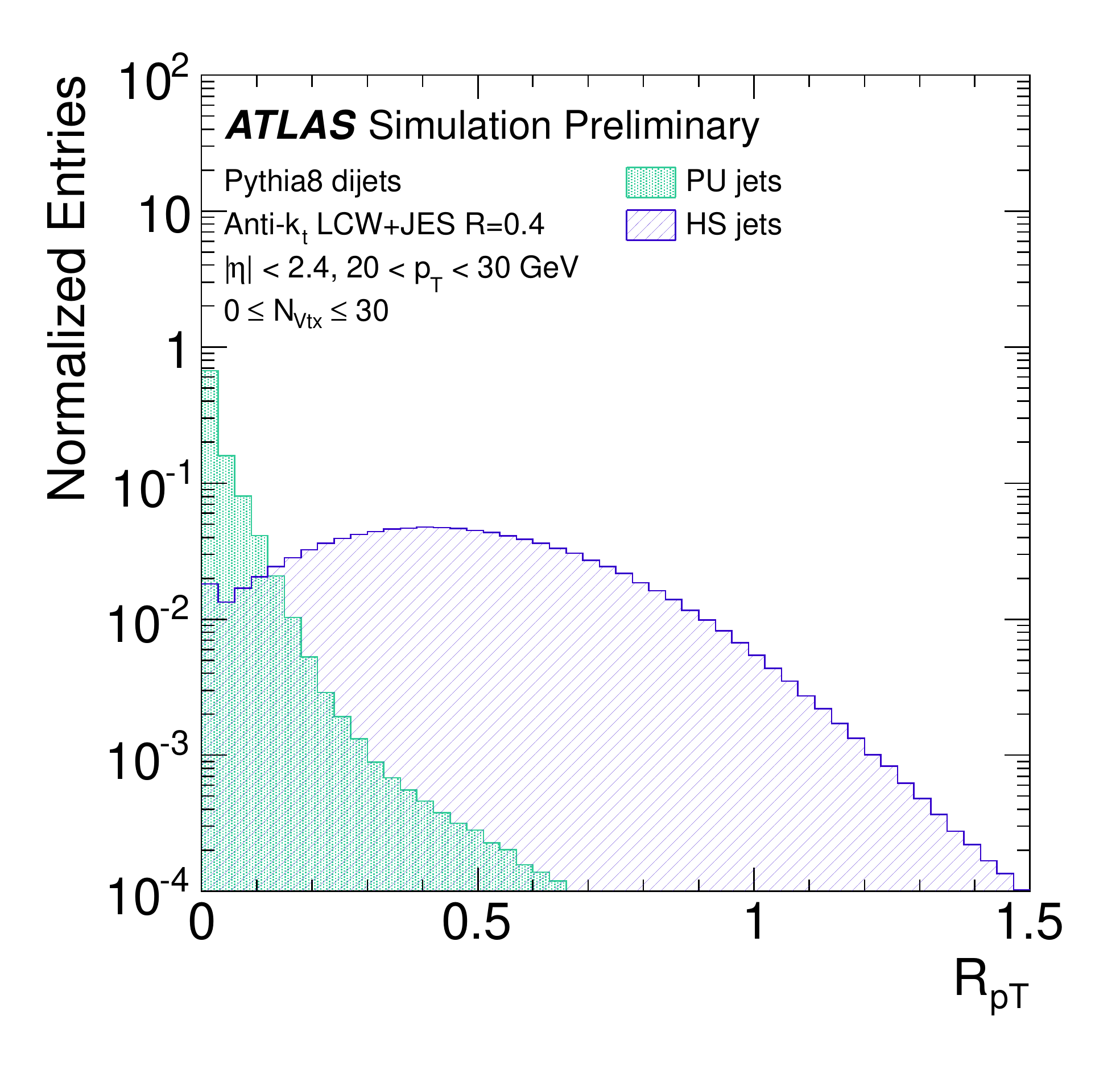}}
\subfloat[]{\includegraphics[width=0.3\textwidth]{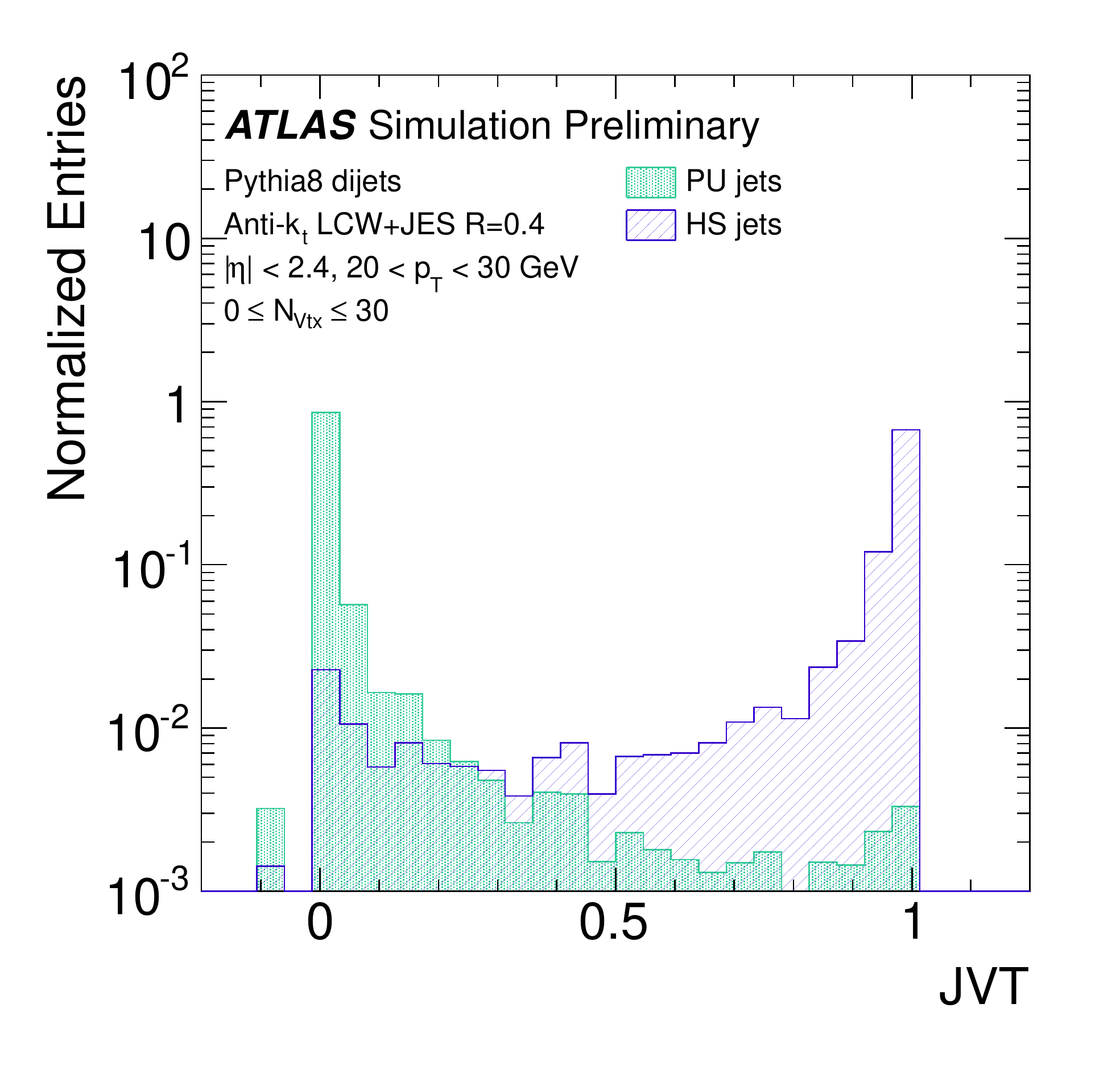}}
\caption{Distribution of (a) corrJVF , (b) $R_{p_\mathrm{T}}$, and (c) JVT for pile-up and hard-scatter jets with 20 GeV $< p_\mathrm{T} < 30$ GeV. From Ref. \cite{ATLAS-CONF-2014-018}.}
\label{fig:corrJVFRPT}
\end{figure}

Pile-up can be a problem not only because it can bias the energy of the jets, but also because it can lead to the reconstruction of jets that are actually not originating from the hard scattering interaction. Most of the pile-up jets however can be removed using the \textit{Jet-Vertex-Fraction} (JVF) \cite{ATLAS-CONF-2014-018,Aad:2015ina,ATLAS-CONF-2011-102}. This variable is the ratio between the scalar sum of the tracks $p_{\mathrm{T}}$ associated to the jet and to the vertex, and the scalar sum of the $p_{\mathrm{T}}$ of all the tracks:
\begin{linenomath*}
\begin{equation}
\text{JVF} = \frac{\sum_k p_\mathrm{T}^{\text{trk}_k}(\text{PV}_0)}{\sum_l p_\mathrm{T}^{\text{trk}_l}(\text{PV}_0) +  \sum_{n\geq1} \sum_l p_\mathrm{T}^{\text{trk}_l}(\text{PV}_n)}
\end{equation}
\end{linenomath*}
where PV$_0$ is the primary vertex of hard scatter and PV$_j$ are the primary vertices of pile-up events. JVF is bound between 0 and 1, but $-1$ is assigned to jets with no associated tracks. With increasing pile-up, however, this variable is less efficient due to its dependence on the scalar sum of $p_{\mathrm{T}}$ on the number of vertexes. For this reason, an additional variable called corrJVF is introduced, defined as:
\begin{linenomath*}
\begin{equation}
\text{corrJVF} = \frac{ \sum_k p_\mathrm{T}^{\text{trk}_k}(\text{PV}_0 )}{\sum_l p_\mathrm{T}^{\text{trk}_l}(\text{PV}_0) + \frac{ \sum_{n\geq1} \sum_l p_\mathrm{T}^{\text{trk}_l}(\text{PV}_n)}{(k\cdot n_{\text{trk}}^{\text{PU}})}}
\end{equation}
\end{linenomath*}
where $n_{\mathrm{trk}}^{\text{PU}}$ is the number of tracks per event and $k=0.01$, and should be the slope of $\langle p_\mathrm{T}^{\text{PU}}\rangle $. Another important variable used to discriminate hard scattering events and pile-up ones is $R_{p_\mathrm{T}}$, defined as the ratio between the scalar sum of the tracks $p_{\mathrm{T}}$ associated with the PV$_0$ and the $p_{\mathrm{T}}$ of the jet,
\begin{linenomath*}
\begin{equation}
R_{p_\mathrm{T}} = \frac{\sum_k p_\mathrm{T}^{\text{trk}_k}(\text{PV}_0 )}{p_\mathrm{T}^{\mathrm{jet}}}.
\end{equation}
\end{linenomath*}
Figs.~\ref{fig:corrJVFRPT} (a) and (b) show the distribution of corrJVF and $R_{p_\mathrm{T}}$, respectively. These two variables are fed into another discriminant, called \textit{Jet Vertex Tagger} (JVT), with a 2-dimensional likelihood based on a k-Nearest Neighbour (kNN) algorithm. Fig.~\ref{fig:corrJVFRPT} (c) shows the JVT distribution for hard scattering and pile-up jets.

\section{b-tagging}
\label{ssec:reco-btag}
\begin{figure}[!htb]
\centering
\includegraphics[width=0.55\textwidth]{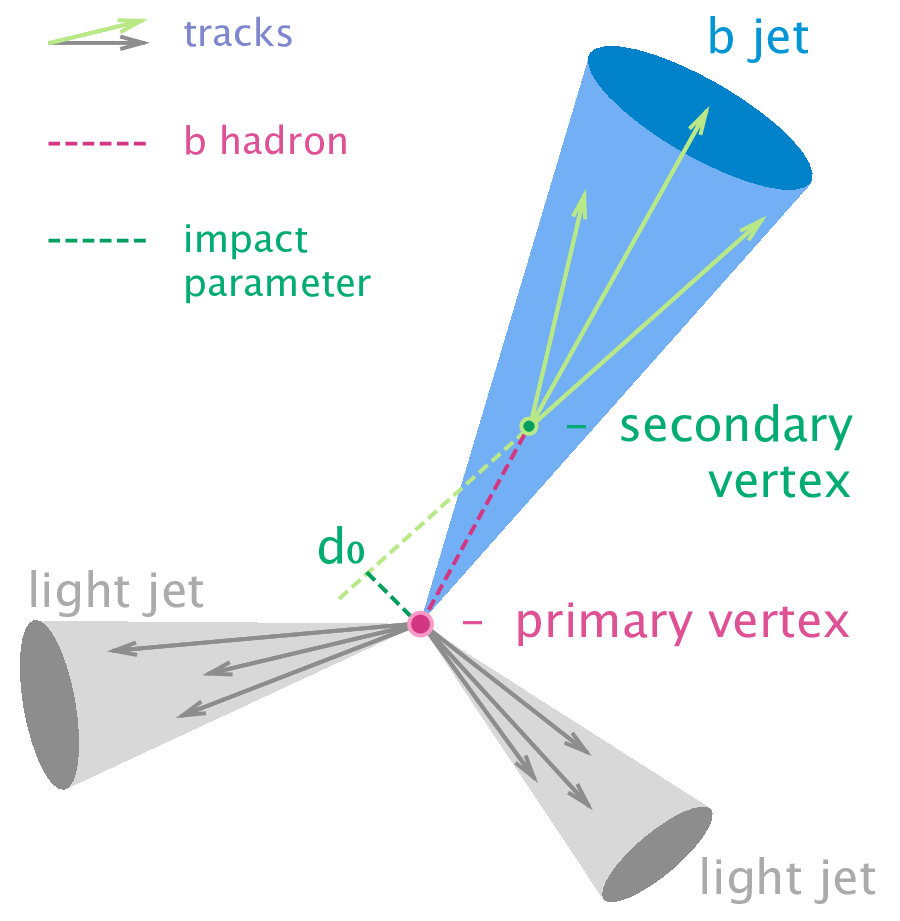}
\caption{Picture representing the decay of a hadron containing a $b$-quark.}
\label{fig:b-tagging}
\end{figure}

The identification of jets as coming from $b$-quarks, called \textit{b-tagging}, is a vital aspect of the ATLAS experiment. Due to the high energy involved, hadrons containing $b$-quarks have relatively long lifetimes ($\approx10^{-12}$ s) and can travel long distances ($\approx$ 1 mm) before decaying, therefore leaving a secondary vertex in the ID (see Fig.~\ref{fig:b-tagging}).

Different \textit{low-level tagging} algorithms are adopted to discriminate between $b$-quark jets and light-quark ($u$, $d$, $s$) jets \cite{ATLASbtagging}, using track and impact parameters and different quality selection criteria. The output of the low-level tagging algorithms is used by higher-level algorithms which are chosen at the analysis level.

\subsubsection{Low-level tagging algorithms}
\begin{itemize}
\item \textit{IP3D algorithm} uses the significance of the impact parameters (both longitudinal and transverse) of the tracks associated with a jet. The significance of the impact parameter $S$ is the ratio between the impact parameter and its error: $S(d_0) = d_0/\sigma(d_0)$ for the transverse plane and $S(z_0) = z_0/\sigma(z_0)$ for the longitudinal one. Probability density functions for the impact parameter are used to define ratios for the $b$- and light-jet hypotheses and combined together in a Log Likelihood Ratio discriminant (LLR). LLR can be constructed with different sets of PDF for different track categories. During Run~2 these categories have been refined. 
\item \textit{SV1 algorithm} tries to reconstruct the displaced secondary vertex within the jet. The first step is to reconstruct two-track vertices, fake vertices that can be suppressed by noticing they do not have associated detector hits with radii smaller than the radius of the secondary vertex found by the algorithm. Then, tracks compatible with $K_S$ or $\Lambda$ are rejected by exploiting the invariant masses of the particles produced by their decays and searching for peaks in the invariant mass distributions of $\pi^{+}\pi^{-}$ and $p\pi$. Also, photon conversions or hadronic interactions with the detector material. A discriminant is constructed with the decay length significance: $L/\sigma(L)$.
\item \textit{JetFitter algorithm} uses the topological structure of $b$ and $c$ hadrons to reconstruct the whole chain decay PV$\rightarrow b \rightarrow c$ decay. The algorithm tries to find a common line between the primary vertex and the bottom and charm vertices, as well as their position on the line, approximating the $b$-hadron flight path. The discrimination between $b$-, $c$- and light-jets is based on a neural network using similar variables.
\end{itemize}

\subsubsection{MV2 and DL1 algorithms}
The output scores of IP3D, SV1 and JetFitter algorithms are combined together to better discriminate $b$-jets from light- and $c$-jets \cite{MV2DL1}. Two different algorithms are obtained, the \textit{MV2 algorithm} and the \textit{DL1 algorithm}.

\begin{itemize}
\item \textit{MV2 algorithm} uses a Boosted Decision Tree (BDT) classifier to combine the outputs of the low-level tagging algorithms. The BDT algorithm is trained using the ROOT Toolkit for Multivariate Data Analysis (TMVA) \cite{TMVA} on a $t\bar{t}$ sample. 
The algorithm treats $b$-jets as signal and light and $c$-jets as background and includes the kinematic properties of the jets ($p_{\mathrm{T}}$ and $|\eta|$) to take advantage of the correlations with the other input variables. The $b$-jets and $c$-jets are reweighted in $p_\mathrm{T}$ and $|\eta|$ to match the spectrum of the light-flavour jets.
\item \textit{DL1 algorithm} uses a deep feed-forward Neural Network (NN) trained using Keras \cite{Keras} with the Theano \cite{Theano} backend and the Adam optimiser \cite{Adam}. The DL1 NN has a multidimensional output corresponding to the probabilities for a jet to be a $b$-jet, a $c$-jet or a light-flavour jet. The topology of the output consists of a mixture of fully connected hidden and Maxout layers \cite{MaxoutNetworks}. The input variables to DL1 consist of those used for the MV2 algorithm with the addition of some JetFitter $c$-tagging variables exploiting the secondary and tertiary vertices (distance to the primary vertex, invariant mass and number of tracks, energy, energy fraction, and rapidity of the tracks associated with the secondary and tertiary vertices). A jet $p_\mathrm{T}$ and $|\eta|$ reweighting similar to the one used for MV2 is performed. Since all flavours are treated equally during training, the trained network can be used for both $b$-jet and $c$-jet tagging. The final DL1 $b$-tagging discriminant is defined as
\begin{equation}
    D_{DL1} = \ln \Big( \frac{p_{b}}{f_{c} \cdot p_{c} + (1-f_{c})\cdot p_{\text{light}}} \Big)
\end{equation}
where $p_{b}$, $p_{c}$, $p_{\text{light}}$ and $f_{c}$ represent the $b$-jet, $c$-jet, light-flavour jet probabilities, and the effective $c$-jet fraction in the background training sample, respectively. Using this approach, the $c$-jet fraction in the background can be chosen a posteriori in order to optimise the performance of the algorithm.
\end{itemize}

\begin{figure}[!htb]
\centering
\subfloat[]{\includegraphics[width=0.45\textwidth]{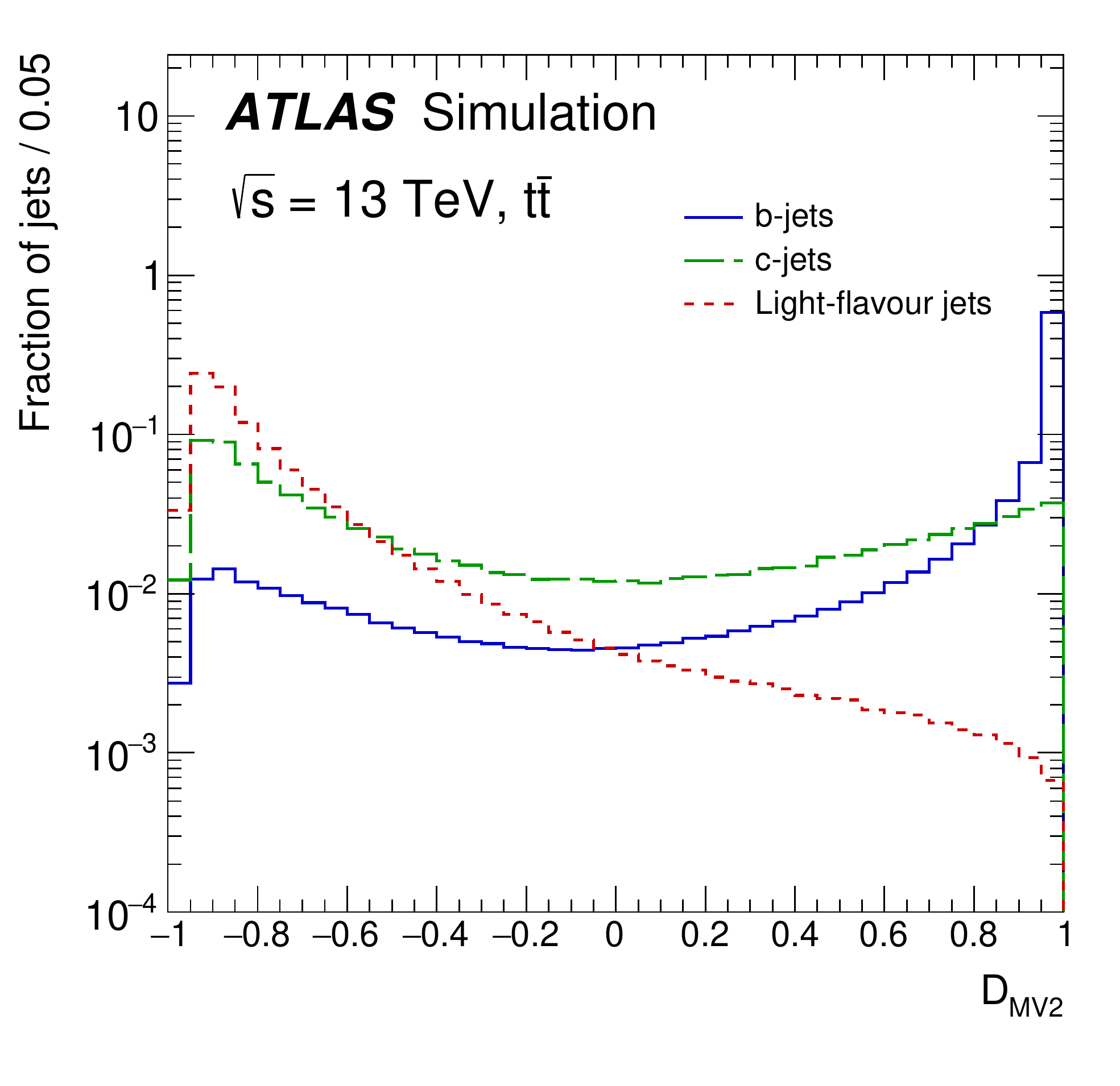}}
\subfloat[]{\includegraphics[width=0.45\textwidth]{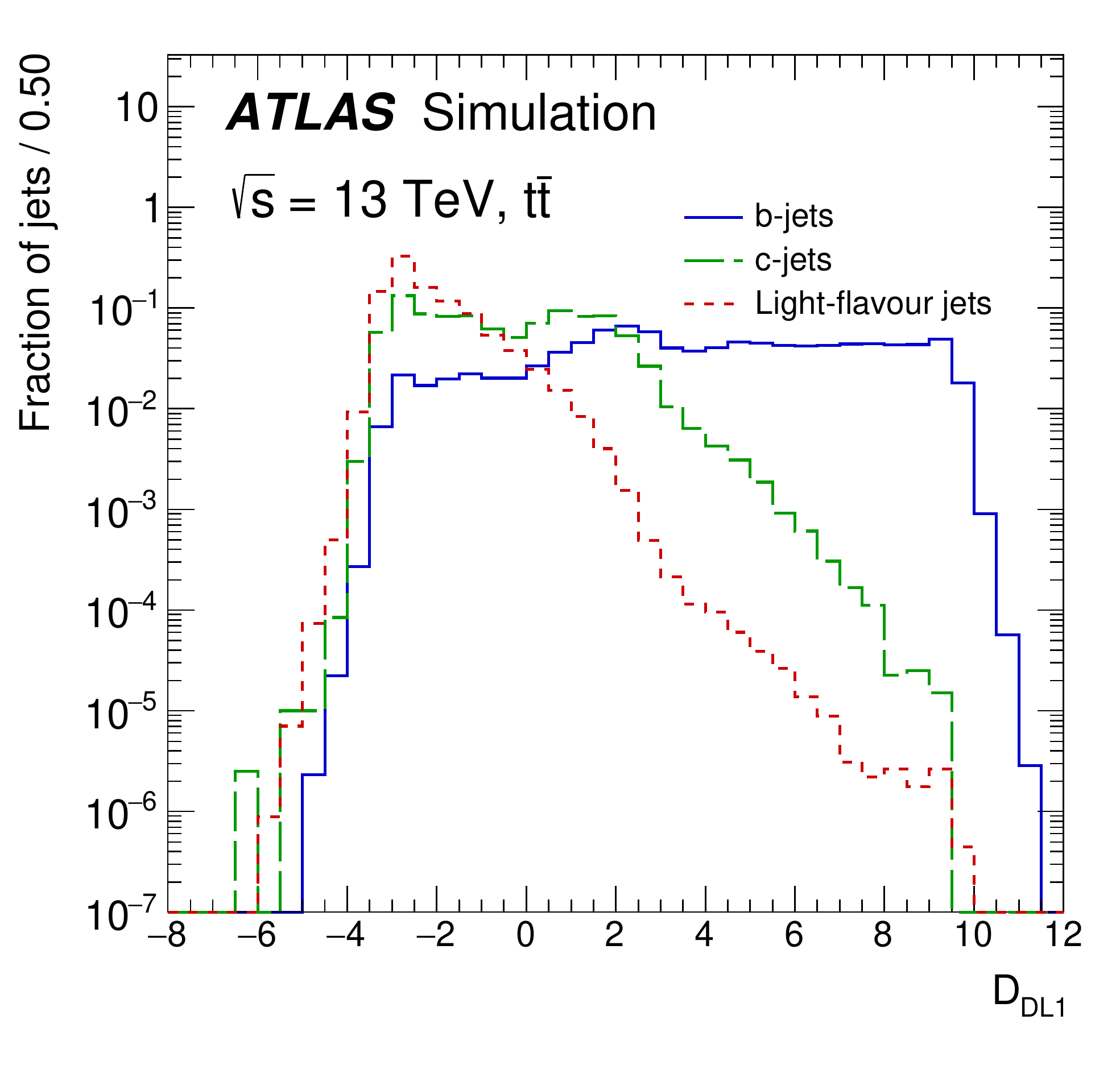}}
\caption{Distribution of the output discriminant of the (a) MV2 and (b) DL1 b-tagging algorithms for $b$-jets (dashed blue), $c$-jets (solid green)and light-jets (dotted red) in $t\bar{t}$ events. From Ref. \cite{ATLASbtagging}.}
\label{fig:MV2c10}
\end{figure}

A comparison of the distributions of the two output discriminants is shown in Fig.~\ref{fig:MV2c10}.

The evaluation of the performance of the algorithms is carried out using different WPs defined by a fixed selection on the $b-$tagging output discriminant and ensuring a specific $b$-jet tagging efficiency, $\epsilon_{b}$, for the $b$-jets present in the baseline $t\bar{t}$ simulated sample. The selections used to define the single-cut WPs of the MV2 and the DL1 algorithms, as well as the corresponding $c$-jet, $\tau$-jet and light-flavour jet rejections, are shown in Table \ref{tab:btaggingOPs}.

\begin{table}[!htb]
\centering
\begin{tabular}{c|cccc|cccc}
\noalign{\smallskip}\hline\noalign{\smallskip}
\multirow{2}{*}{$\epsilon_{b}$} & \multicolumn{4}{c|}{MV2}   & \multicolumn{4}{c}{DL1} \\
 & Selection & $c$-jet & $\tau$-jet & light-jet & Selection & $c$-jet & $\tau$-jet & light-jet \\ \noalign{\smallskip}\hline\noalign{\smallskip}
60\% & \textgreater 0.94 & 23 & 140  & 1200 & \textgreater 2.74 & 27 & 220 & 1300 \\
70\% & \textgreater 0.83 & 8.9 & 36  & 300  & \textgreater 2.02 & 9.4 & 43 & 390  \\
77\% & \textgreater 0.64 & 4.9 & 15  & 110  & \textgreater 1.45 & 4.9 & 14 & 130  \\
85\% & \textgreater 0.11 & 2.7 & 6.1 & 25   & \textgreater 0.46 & 2.6 & 3.9 & 29  \\
\noalign{\smallskip}\hline\noalign{\smallskip}
\end{tabular}
\caption{Selection and $c$-jet, $\tau$-jet and light-flavour jet rejections corresponding to the different $b$-jet tagging efficiency single-cut WPs for the MV2 and the DL1 b-tagging algorithms, evaluated on the baseline $t\bar{t}$ events.}
\label{tab:btaggingOPs}
\end{table}

\begin{figure}[!htb]
\centering
\subfloat[]{\includegraphics[width=0.45\textwidth]{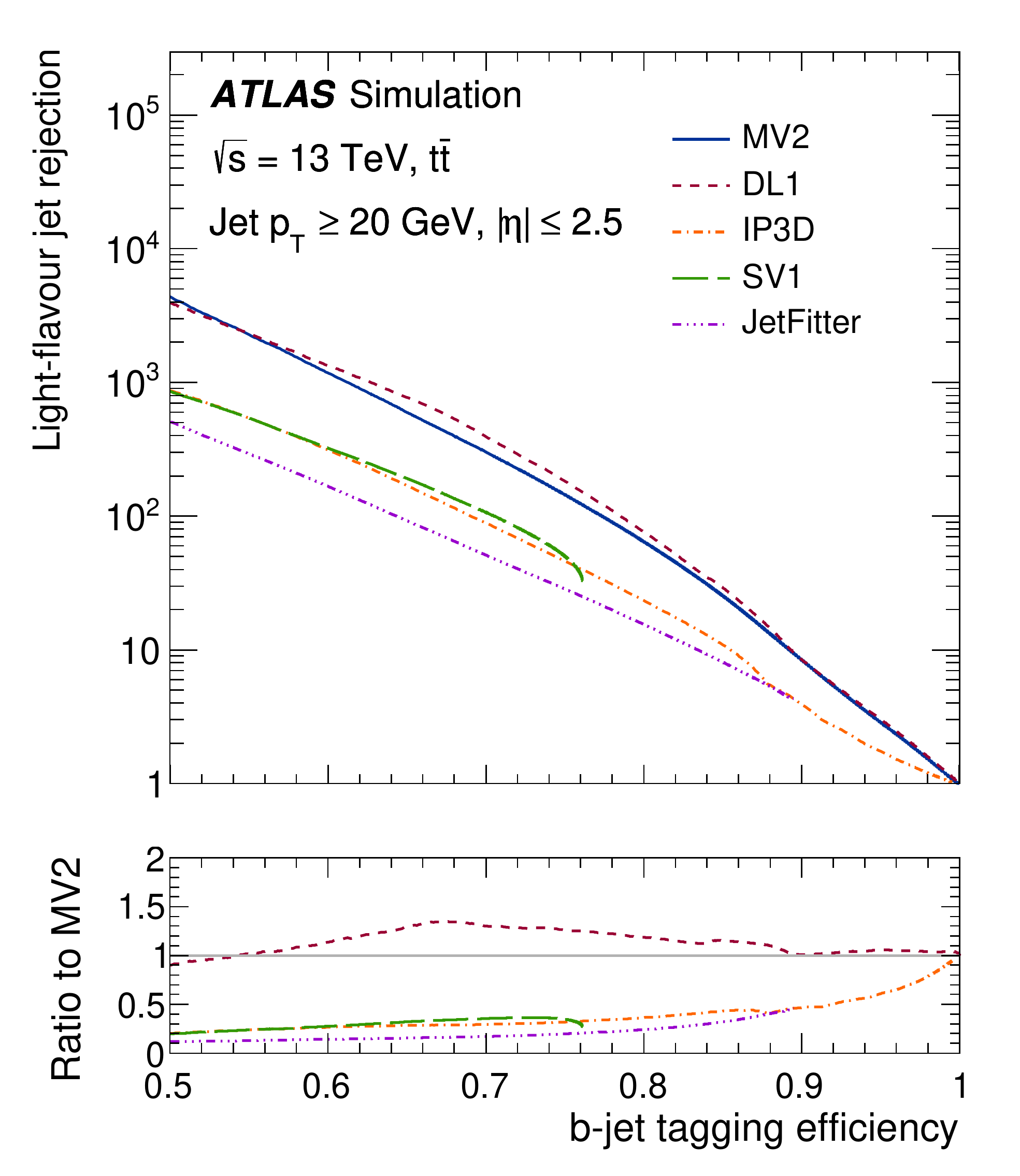}}
\subfloat[]{\includegraphics[width=0.45\textwidth]{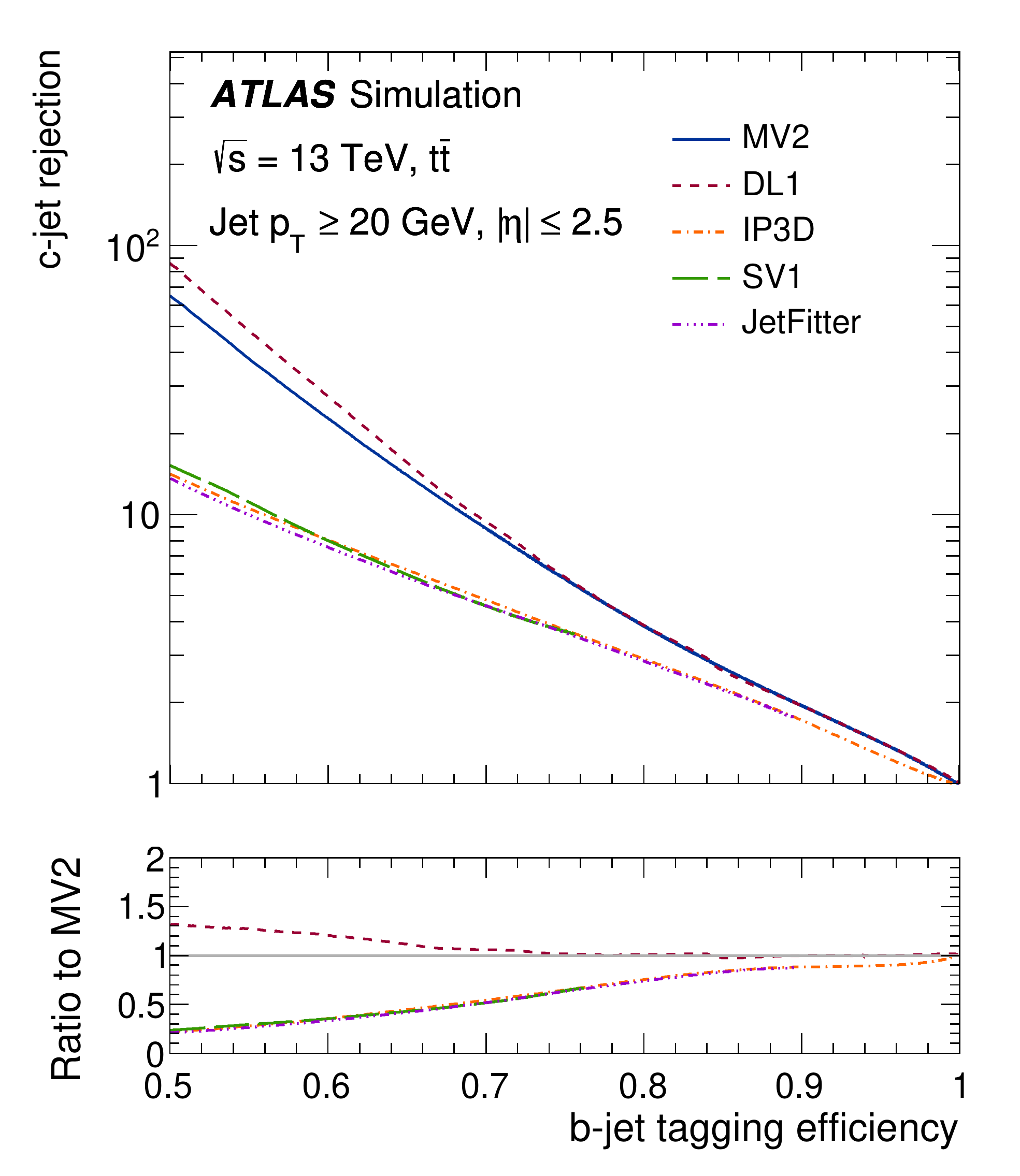}}
\caption{The (a) light-flavour jet and (b) c-jet rejections versus the $b$-jet tagging efficiency for the IP3D, SV1, JetFitter, MV2 and DL1 b-tagging algorithms evaluated on the baseline $t\bar{t}$ events.}
\label{fig:bTaggingRejections}
\end{figure}

The light-flavour jet and $c$-jet rejections as a function of the $b$-jet tagging efficiency are shown in Fig.~\ref{fig:bTaggingRejections} for the various low- and high-level $b$-tagging algorithms. 

This demonstrates the advantage of combining the information provided by the low-level taggers, where improvements in the light-flavour jet and $c$-jet rejections by factors of around 10 and 2.5, respectively, are observed at the $\epsilon_{b}$ = 70\% single-cut WP of the high-level algorithms compared to low-level algorithms. This figure also illustrates the different $b$-jet tagging efficiency ranges accessible with each low-level algorithm and thereby their complementarity in the multivariate combinations, with the performance of the DL1 found to be better than the MV2 discriminant. The two algorithms tag a highly correlated sample of $b$-jets, where the relative fraction of jet exclusively tagged by each algorithm is around 3\% at the $\epsilon_{b}$ = 70\% single-cut WP. The relative fractions of light-flavour jets exclusively mistagged by the MV2 or the DL1 algorithms at the $\epsilon_{b}$ = 70\% single-cut WP reach 0.2\% and 0.1\%, respectively. \\

\subsubsection{DL1r algorithm}
\begin{figure}[!htb]
\centering
\includegraphics[width=0.8\textwidth]{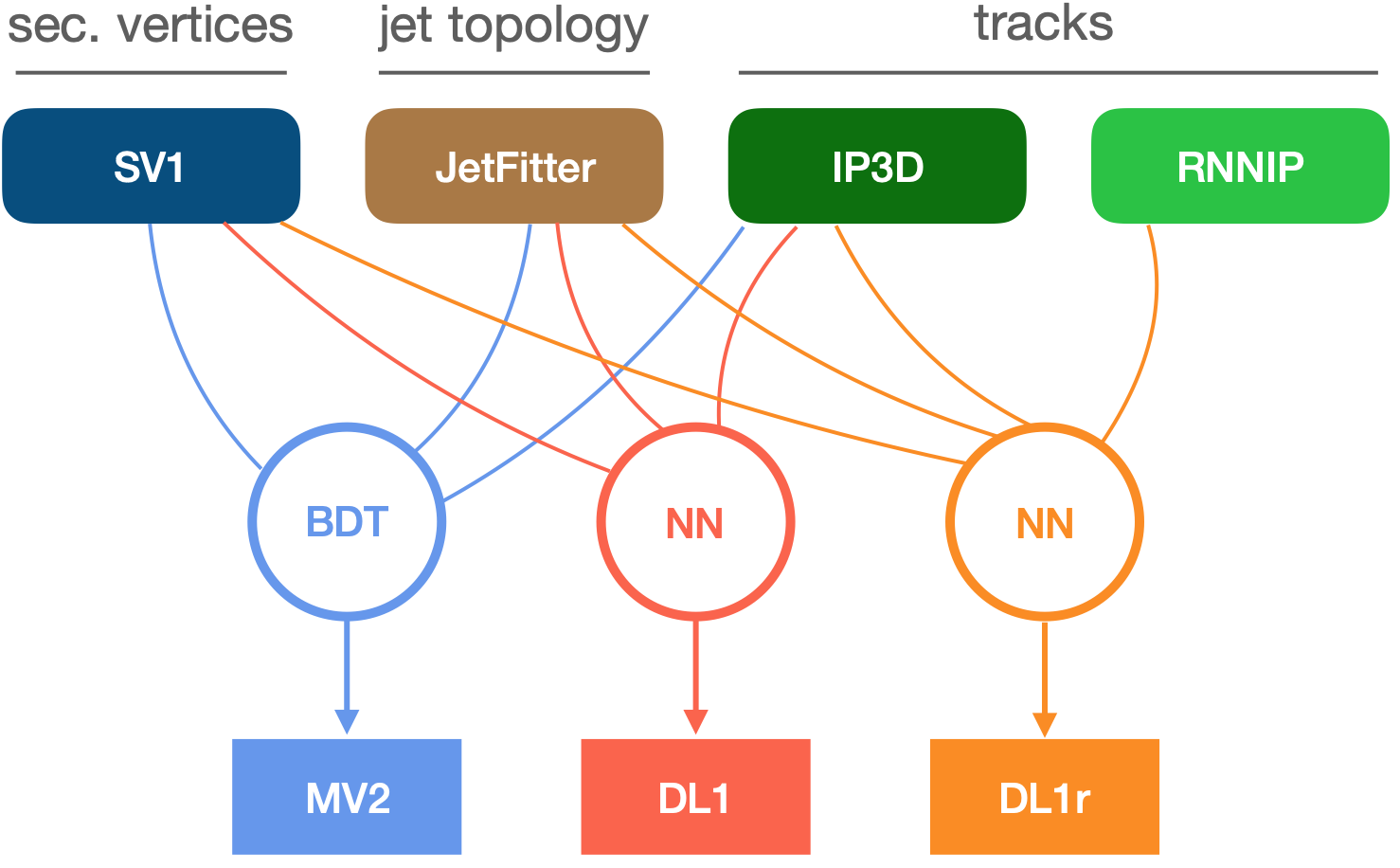}
\caption{The inputs used in the MV2, DL1 and DL1r algorithms.}
\label{fig:DL1r}
\end{figure}

An improved version of the DL1 tagging algorithm, called \textit{DL1r} \cite{DL1r}, uses a Recursive Neural Network (RNN) with the Impact Parameters of the tracks (RNNIP) in addition to the SV1, JetFitter, IP3D low-level taggers as shown in Fig.~\ref{fig:DL1r}.
The RNNIP network architecture learns the track multiplicity and impact parameters of the B-hadron decays and matches the tracks with larger IPs to $b$-jets decays which tend to have them harder and wider.

\section{Missing Transverse Energy}
\label{ssec:reco-met}
In proton-proton colliders the fraction of energy the partons have in the collision is unknown, therefore we do not know the initial energy and cannot use the conservation of the momentum. However, it is known the initial energy in the transverse plane is zero and it is possible to use the conservation of momentum in this plane. The missing transverse energy ($E_{\mathrm{T}}^{\mathrm{miss}}$) accounts for all the particles that are invisible to the detector, mainly neutrinos or particles beyond the Standard Model, such as neutralinos. $E_{\mathrm{T}}^{\mathrm{miss}}$ is defined as
\begin{linenomath*}
\begin{equation}
E_{\mathrm{T}}^{\mathrm{miss}} = \sqrt{(E_x^{\mathrm{miss}})^2 + (E_y^{\mathrm{miss}})^2}
\end{equation}
\end{linenomath*}
where $E_{x(y)}^{\mathrm{miss}} = -\sum E_{x(y)}$, and $E_{x(y)}$ is the energy deposited in the detector. Contribution for the $E_{\mathrm{T}}^{\mathrm{miss}}$ comes from the calorimeters, the MS, and also the ID. $E_{\mathrm{T}}^{\mathrm{miss}}$ reconstruction uses calorimeter cells calibrated for the different reconstructed objects (electrons, muons, photons, hadronically decaying $\tau$-leptons, jets) and tracks ($|\eta|<2.5$) and cells in the calorimeter ($|\eta| > 2.5$) with no object attached \cite{Aaboud:2018tkc,ATLAS-CONF-2018-023}.\\
More specifically, the $E_{\mathrm{T}}^{\mathrm{miss}}$ is calculated as
\begin{linenomath*}
\begin{equation}
\begin{split}
E_{x(y)}^{\mathrm{miss}} & =E_{x(y)}^{\mathrm{miss},e}+E_{x(y)}^{\mathrm{miss},\mu}+E_{x(y)}^{\mathrm{miss},\gamma}+E_{x(y)}^{\mathrm{miss},\tau}+E_{x(y)}^{\mathrm{miss},\mathrm{jets}}\\
&+E_{x(y)}^{\mathrm{miss},\mathrm{softjets}}+E_{x(y)}^{\mathrm{miss},\mathrm{calo}\mu}+E_{x(y)}^{\mathrm{miss},\mathrm{tracks}}+E_{x(y)}^{\mathrm{miss},\mathrm{softcalo}}
\end{split}
\end{equation}
\end{linenomath*}
where, in particular:
\begin{itemize}
    \item $E_{x(y)}^{\mathrm{miss},\mathrm{calo}\mu}$ is the energy of muons in the calorimeters;
    \item $E_{x(y)}^{\mathrm{miss},\mathrm{tracks}}$ is the energy of the tracks associated with the hard-scatter vertex but not with any hard object. Tracks are taken from the ID so they can be matched to the primary vertex and can help to reconstruct low $p_{\mathrm{T}}$ particles;
    \item $E_{x(y)}^{\mathrm{miss},\mathrm{softcalo}}$ is the sum of the terms in the calorimeter that don't match any object.
\end{itemize}
The convention adopted is that every $\textit{miss}$ term is the negative sum of the measured energy for a specific category. The sum of these terms is done in the pseudorapidity range $|\eta| < 4.9$. The azimuthal coordinate is evaluated as
\begin{equation}
\phi^{\mathrm{miss}} =\arctan(E_y^{\mathrm{miss}}/E_x^{\mathrm{miss}})
\end{equation}
Reducing noise contamination is crucial, therefore, only the cells belonging to the topological clusters are considered. An overlap removal between calorimeter clusters with high-$p_\mathrm{T}$ and tracks is requested to avoid double counting. Tracks with more than 40\% of uncertainty on the $p_\mathrm{T}$ are removed.\\
Three different WPs are provided to satisfy the needs at the analysis level:
\begin{itemize}
\item \textit{Loose} WP, using all jets with $p_\mathrm{T} > 20$ GeV that pass the JVT cut for $|\eta|<2.4 $ and $p_\mathrm{T} < 60$ GeV;
\item \textit{Tight} WP, reducing the $E_{\mathrm{T}}^{\text{miss}}$ dependence on the pile-up by additionally vetoing the forward jets with $|\eta| > 2.4$ and 20 GeV $< p_\mathrm{T} < 30$ GeV;
\item \textit{Forward JVT} WP, vetoing the forward jets with $|\eta|>2.4$ and 20 GeV $ < p_\mathrm{T} < 30$ GeV and additionally failing the 'Loose' fJVT criteria \cite{fJVTcriteria}.
\end{itemize}

Performance studies of events with no real $E_{\mathrm{T}}^{\text{miss}}$ are done by looking at $Z\rightarrow \mu \mu$ events while for events with real $E_{\mathrm{T}}^{\text{miss}}$, $W\rightarrow \ell \nu$ is studied. The resolution of the Soft Track Term (TST) in $E_{\mathrm{T}}^{\text{miss}}$ is estimated by comparing data to simulated $Z\rightarrow \mu \mu$ events, where no real $E_{\mathrm{T}}^{\text{miss}}$ is expected. The resolution is evaluated as the r.m.s. of the combined $E_x^{\text{miss}}$ and $E_x^{\text{miss}}$. \\
\chapter{Analysis regions and statistical interpretation}
\label{sec:statistics}

Before diving into the analyses, this Chapter is aimed at illustrating the general strategy for defining different analysis regions and providing an illustration of the statistical interpretation of analysed events.

\minitoc
\medskip

\section{Definition of Analysis regions}
The overall analysis strategy of a search relies on the definition of dedicated analysis regions, which can be categorised as follows:
\begin{itemize}
\item Control Regions (CRs), used to control and estimate the main background contributions;
\item Validation Regions (VRs), used to validate the background estimation with respect to data;
\item Signal Regions (SRs), used to test the signal model with respect to data.
\end{itemize}

\begin{figure}[!htb]
\centering
\includegraphics[width=0.8\textwidth]{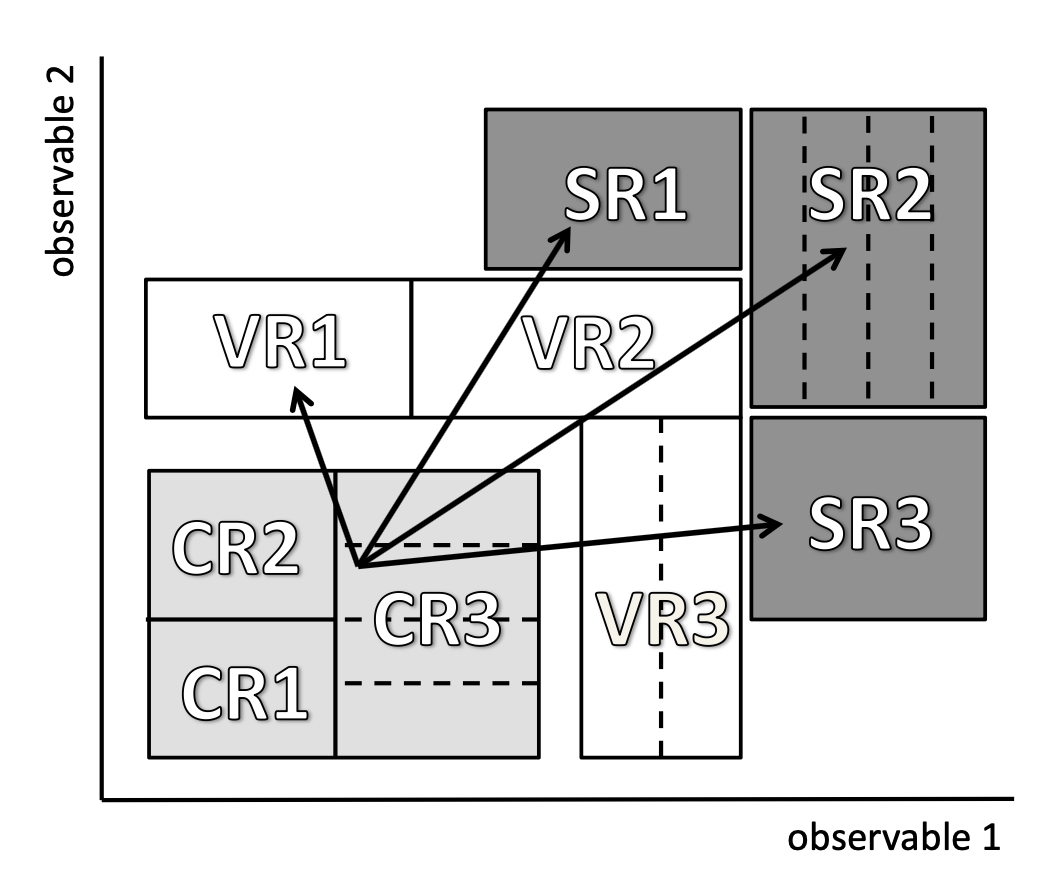}
\caption{Schematic view of how SRs, CRs, and VRs are defined as a function of two arbitrary variables.}
\label{fig:AnalysisRegions}
\end{figure}

Fig.~\ref{fig:AnalysisRegions} shows a schematic representation of how the CRs, VRs, and SRs are defined in a 2D plane of two arbitrary variables.

The SM background is estimated either from MC samples or from data-driven techniques. It is possible to take these MC samples at face value, or it is possible to improve the prediction by using special regions where to constrain the background and therefore have a better prediction. This is done in CRs: phase space regions, close to the other analysis regions but with a negligible expected rate of signal events, where the backgrounds are checked and normalized against the observed data. Generally, a specific CR is defined for each main background, by inverting one or more kinematic cuts that are applied in the SR. This allows also to have non overlapping CRs and SRs.\\
Usually, normalization factors for the different background samples are extrapolated from the CRs to the VRs and SRs. The fit to observed data in CRs allows to eliminate any mismodelling in the normalization of the MC. Another advantage is that when evaluating systematic uncertainties in the SR, these depend only on the changes in the ratio of the expected background yields in SR and CR.\\
The normalization obtained in the CR is checked by extrapolating the results in the VRs. These are kinematic regions that are close to the SR but still have a small expected signal contamination. Only after the modelling of the background is checked against data in the VRs it is possible to look at data inside the SR, a procedure which is called \textit{unblinding}. SRs initially are blinded, e.g. it is not allowed to look at data inside these regions, to avoid any biases in the definition of the SRs and the test statistic that might come from a first observation of the data before the background estimation strategy is optimized.

\section{Likelihood fit}
\label{sec:likelihoodfit}
A binned likelihood function $L(\mu, \boldsymbol \theta)$ is constructed from the product of Poisson probabilities for all $N$ bins,
\begin{equation}
    L(\mu, \boldsymbol{\theta}) = \prod_{i=1}^{N} \frac{(\mu s_{i} + b_{i})^{n_{i}}}{n_{i}!} e^{-(\mu s_{i}+b_{i})} \prod_{j=1}^{M} \frac{{u_{j}}^{m_{j}}}{m_{j}!}e^{-{u_{j}}},
\end{equation}
with $s_{i}$ and $b_{i}$ representing the number of expected signal and background events in each bin and $n_{i}$ the observed number of events whose expected value is $E[n_{i}]=\mu s_{i} + b_{i}$. More precisely, $s_{i}=s_{\mathrm{tot}}\int_{\mathrm{bin}\,i} f_{s}(x, \boldsymbol{\theta}_{s})$ and $b_{i}=b_{\mathrm{tot}}\int_{\mathrm{bin}\,i} f_{b}(x, \boldsymbol{\theta}_{b})$ where $s_{\mathrm{tot}}$ and $b_{\mathrm{tot}}$ are the total mean number of signal and backgrounds events, $f_{s}(x, \boldsymbol{\theta}_{s})$ and $f_{b}(x, \boldsymbol{\theta}_{b})$ are the probability density functions (pdfs) of the variable $x$ for signal and background events, and $\boldsymbol{\theta}_{s}$ and $\boldsymbol{\theta}_{b}$ represent nuisance parameters that might affect the shape of the pdfs. \\
In addition to the $N$ values, one often makes further subsidiary measurements that help constrain the set of nuisance parameters $\boldsymbol{\theta}$, selecting some kinematic variables. This then gives a set of values $m_{j}$ for the number of entries in each of the $M$ bins of these variables whose expectation values is $E[m_{j}]=u_{j}(\boldsymbol{\theta})$. \\
In multi-bin fits, several regions need to be combined. In this case, a likelihood function $L_{c}=(\mu,\boldsymbol{\theta}_{c})$ is defined for each region $c$, with $\mu$ representing the signal strength and $\boldsymbol{\theta}_{c}$ representing the set of nuisance parameters for the $c-$th region. The signal strength is assumed to be the same for all regions but in general the set of nuisance parameters can vary among regions. Assuming the regions to be statistically independent, the combined likelihood function is given by the product over all of them, 
\begin{equation}
L(\mu, \boldsymbol{\theta}) = \prod_{c} L_{c}(\mu, \boldsymbol{\theta}_{c}).
\end{equation}
To test a hypothesised value of $\mu$, the profile likelihood ratio, 
\begin{equation}
    \lambda(\mu) = \frac{L(\mu, \hat{\hat{\boldsymbol{\theta}}})}{L(\hat{\mu},\hat{\boldsymbol{\theta}})}
\end{equation}
is considered, where $\hat{\hat{\boldsymbol{\theta}}}$ in the numerator denotes the value of $\boldsymbol{\theta}$ that maximises $L$ for the specified $\mu$, i.e. it is conditional maximum-likelihood (ML) estimator of $\boldsymbol{\theta}$ (and thus is a function of $\mu$). The denominator is the maximised likelihood function, i.e. $\hat{\mu}$ and $\hat{\boldsymbol{\theta}}$ are their ML estimators. The profile likelihood ratio $\lambda(\mu)$ assumes values between 0 and 1 (at $\mu = \hat{\mu}$), with $\lambda$ close to 1 implying a good agreement between data and the hypothesised value of $\mu$. The presence of the nuisance parameters broadens the profile likelihood as a function of $\mu$ relative to what one would have if their values were fixed. This reflects the loss of information about $\mu$ due to the systematic uncertainties. In our analyses, the contribution of the signal process to the mean number of events is assumed to be non-negative. However, it is convenient to define an effective estimator $\hat{\mu} < 0$, but providing that the Poisson mean values, $\mu s_{i} + b_{i}$, remain non-negative. This will allow us to model $\hat{\mu}$ as a Gaussian-distributed variable and, in this way, we can determine the distributions of the test statistics that we consider.\\
To set exclusion limits and set upper limits on the strength parameter $\mu$, we consider the test statistic $\tilde{q}_{\mu}$, defined as
\begin{equation}
    \tilde{q}_{\mu} = 
    \begin{cases}
  -2 \, \mathrm{ln} \, \tilde{\lambda}(\mu) & \hat{\mu}\leq \mu\\ 0 & \hat{\mu} > \mu
    \end{cases}
\end{equation}
where $\tilde{\lambda}(\mu)$ is the profile likelihood ratio. The reason for setting $\tilde{q}_{\mu} = 0$ for $\hat{\mu} > \mu$ is to avoid considering, as signal evidence, upward fluctuations of the data guaranteeing a one-sided confidence interval, and therefore this is not taken as part of the rejection region of the test. From the definition of the test statistic, one sees that higher values of $\tilde{q}_{\mu}$ represent greater incompatibility between the data and the hypothesised value of $\mu$.\\
In our case, $\mu > 0$, so considering $\tilde{\lambda}(\mu)$, we have
\begin{equation}
    \tilde{q}_{\mu} = 
    \begin{cases}
  -2 \, \mathrm{ln} \, \frac{L(\mu,\hat{\hat{\boldsymbol{\theta}}}(\mu))}{L(0,\hat{\hat{\boldsymbol{\theta}}}(0))} & \hat{\mu} < 0\\ 
   -2 \, \mathrm{ln} \, \frac{L(\mu,\hat{\hat{\boldsymbol{\theta}}}(\mu))}{L(\hat{\mu},\hat{\boldsymbol{\theta}})} & 0 \leq \hat{\mu} \leq \mu\\
   0 & \hat{\mu} > \mu.\\
    \end{cases}
\end{equation}
Assuming the Wald approximation, we find that
\begin{equation}
    \tilde{q}_{\mu} = 
    \begin{cases}
  \frac{\mu^2}{\sigma^2} - \frac{2\mu\hat{\mu}}{\sigma^2} & \hat{\mu} < 0\\ 
  \frac{{(\mu-\hat{\mu})}^2}{\sigma^2} & 0 \leq \hat{\mu} \leq \mu\\
   0 & \hat{\mu} > \mu\\
    \end{cases}    
\end{equation}
where $\hat{\mu}$ follows a Gaussian distribution centred at $\mu'$ with a standard deviation $\sigma$. The pdf $f(\tilde{q}_{\mu}|\mu')$ is found to be 
\begin{equation}
\begin{split}
  f(\tilde{q}_{\mu}|\mu') = & \Phi\Big(\frac{\mu'-\mu}{\sigma}\Big) \, \delta(\tilde{q}_{\mu}) + \\
  & +
    \begin{cases}
  \frac{1}{2}\frac{1}{\sqrt{2\pi}}\frac{1}{\sqrt{\tilde{q}_{\mu}}} \, \exp \, \Big[ - \frac{1}{2} \Big( \sqrt{\tilde{q}_{\mu}} - \frac{\mu - \mu'}{\sigma} \Big)^2 \Big] & 0 < \tilde{q}_{\mu} < \mu^2/\sigma^2\\
   \frac{1}{\sqrt{2\pi}\,(2\mu/\sigma)} \, \exp \, \Big[ - \frac{1}{2} \frac{(\tilde{q}_{\mu}-(\mu^2-2\mu\mu')/\sigma^2)^2}{(2\mu/\sigma)^2} \Big] & \tilde{q}_{\mu} > \mu^2/\sigma^2\\
    \end{cases}     
\end{split}
\end{equation}
where $\Phi$ is the cumulative distribution function of the standard normal distribution. \\
The special case $\mu = \mu'$ is therefore 
\begin{equation}
    f(\tilde{q}_{\mu}|\mu') = \frac{1}{2}\,
    \delta(\tilde{q}_{\mu})+
    \begin{cases}
  \frac{1}{2}\frac{1}{\sqrt{2\pi}}\frac{1}{\sqrt{\tilde{q}_{\mu}}} \, \exp \, \Big[ - \frac{1}{2}  \tilde{q}_{\mu} \, \Big] & 0 < \tilde{q}_{\mu} < \mu^2/\sigma^2\\
   \frac{1}{\sqrt{2\pi}\,(2\mu/\sigma)} \, \exp \, \Big[ - \frac{1}{2} \frac{(\tilde{q}_{\mu}+\mu^2/\sigma^2)2}{(2\mu/\sigma)^2} \Big] & \tilde{q}_{\mu} > \mu^2/\sigma^2\\
    \end{cases}    
\end{equation}

The cumulative distribution function $F(\tilde{q}_{\mu}|\mu')$ corresponding to the  $f(\tilde{q}_{\mu}|\mu')$ pdf is given by
\begin{equation}
    F(\tilde{q}_{\mu}|\mu') = \begin{cases}
  \Phi \Big(\sqrt{\tilde{q}_{\mu}} - \frac{\mu - \mu'}{\sigma} \Big) & 0 < \tilde{q}_{\mu} < \mu^2/\sigma^2\\
  \Phi \Big(\frac{\tilde{q}_{\mu} - (\mu^2-2\mu\mu')/\sigma^2}{2\mu/\sigma}\Big) & \tilde{q}_{\mu} > \mu^2/\sigma^2.\\
    \end{cases} 
\end{equation}
The special case $\mu = \mu'$ is therefore 
\begin{equation}
    F(\tilde{q}_{\mu}|\mu') = \begin{cases}
  \Phi \Big(\sqrt{\tilde{q}_{\mu}} \Big) & 0 < \tilde{q}_{\mu} < \mu^2/\sigma^2\\
  \Phi \Big(\frac{\tilde{q}_{\mu} + \mu^2/\sigma^2}{2\mu/\sigma}\Big) & \tilde{q}_{\mu} > \mu^2/\sigma^2.\\
    \end{cases} 
\end{equation}
The $p$-value of the hypothesised $\mu$ is given by the formula 
\begin{equation}
\label{Equation:pvalue}
    p_{\mu} = 1 - F(\tilde{q}_{\mu}|\mu)
\end{equation}
with a corresponding significance 
\begin{equation}
Z_{\mu} = \begin{cases}
   \sqrt{\tilde{q}_{\mu}} & 0 < \tilde{q}_{\mu} < \mu^2/\sigma^2,\\
   \frac{\tilde{q}_{\mu} + \mu^2/\sigma^2}{2\mu/\sigma} & \tilde{q}_{\mu} > \mu^2/\sigma^2.\\
    \end{cases} 
\end{equation}
If the $p$-value is found below a specific threshold $\alpha$ (often taken as $\alpha$ = 0.05), then the value of $\mu$ is said to be excluded at a confidence level (CL) of 1-$\alpha$. The observed upper limit on $\mu$ is the smallest $\mu$ such that $p_{\mu} < \alpha$. Therefore, the observed upper limit on $\mu$ at CL $1-\alpha$ is found by setting $p_{\mu}=\alpha$ and solving Eq.~(\ref{Equation:pvalue}) for $\mu$, that is
\begin{equation}
    \mu_{up} = \hat{\mu}\,+\,\sigma\Phi^{-1}(1-\alpha).
\end{equation}
If $\alpha=0.05$, then $\Phi^{-1}(1-\alpha)=1.64$. Moreover, $\sigma$ depends in general on the hypothesised value of $\mu$. Upper limits closer to the hypothesised value of $\mu$ correspond to stricter constraints on the hypothesised value for $\mu$ and thus higher significance.

\subsection{The CL technique}
\label{sec:CLtechnique}
By using the statistical test $\tilde{q}_{\mu}$, it is possible to exclude a specific signal model or to compute the upper limit on the visible cross-section using the $CL_{s}$ method. This method is introduced to not exclude signals for which an analysis has little or no sensitivity; for example in cases where the expected number of signal events is much smaller than the background and, consequently, the $\tilde{q}_{\mu}$ distributions under the background-only and signal+background hypotheses almost overlap each other. Assuming signal plus background hypothesis as null hypothesis, the observed value $\tilde{q}_{\mu}^{\mathrm{obs}}$ of the statistical test $\tilde{q}_{\mu}$ for the given signal strength ($\mu=1$) is computed and, subsequently, the $p-$values for the signal+background ($p_{s+b}$) and background-only ($p_{b}$) hypotheses are estimated as
\begin{equation}
\label{Eq:CLs+b}
    p_{s+b} = CL_{s+b} =\int_{\tilde{q}_{\mu=1}^{\mathrm{obs}}}^{+\infty} f(\tilde{q}_{\mu}| \, \mu=1, \hat{\hat{\boldsymbol{\theta}}}_{\mu}) \, \mathrm{d} \tilde{q}_{\mu}
\end{equation}
\begin{equation}
\label{Eq:CLb}
    p_{b} = 1 - CL_{b} =\int_{-\infty}^{\tilde{q}_{\mu=1}^{\mathrm{obs}}} f(\tilde{q}_{\mu}| \, \mu=0, \hat{\hat{\boldsymbol{\theta}}}_{\mu}) \, \mathrm{d} \tilde{q}_{\mu}
\end{equation}
where $f(\tilde{q}_{\mu}| \, \mu=1, \, \hat{\hat{\boldsymbol{\theta}}}_{\mu})$ and $f(\tilde{q}_{\mu}| \, \mu=0, \, \hat{\hat{\boldsymbol{\theta}}}_{\mu})$ are the distributions of the test statistic $\tilde{q}_{\mu}$ for the signal+background and background-only hypothesis. \\
The $CL_{s}$ is therefore calculated as 
\begin{equation}
    CL_{s} = \frac{CL_{s+b}}{CL_{b}} = \frac{p_{s+b}}{1-p_{b}}
\end{equation}
and the signal+background hypothesis is rejected for all models having a $CL_{s}$ value lower than 0.05, thus with a 95\% CL. \\
An expected $CL_{s}$, referred as $CL_{s}^{\mathrm{exp}}$, can be computed using the same Eqs.(\ref{Eq:CLs+b})-(\ref{Eq:CLb}) but replacing $\tilde{q}_{\mu}^{\mathrm{obs}}$ with the median $\tilde{q}_{\mu}^{\mathrm{med}}$ of the $f(\tilde{q}_{\mu}| \, \mu=0, \hat{\hat{\boldsymbol{\theta}}}_{\mu})$ distribution.\\
For the exclusion limits, the method just described is applied for each signal model, building a likelihood that takes into account both CRs and SRs. A contour exclusion plot, selecting the mass region with a CLs value lower than 0.05, is then obtained. For the purpose of setting upper limits on a model, a scan varying the hypothesised $\mu$ is performed.
Given the estimated total SM with its error and the observed data, an upper limit on the number of observed signal events ($S_{\mathrm{obs}}^{95}$) and expected signal events ($S_{\mathrm{exp}}^{95}$) is estimated. An upper limit on the visible cross-section $\sigma_{\mathrm{vis}}$ by taking the ratio between $S_{\mathrm{obs}}^{95}$ and the integrated luminosity considered in the analysis.

\subsection{Nuisance parameters in the likelihood}
\label{Sec:nuisanceparameters}

The nuisance parameters, generically referred to as $\boldsymbol{\theta}$, may enter the likelihood function in different ways. For illustrating this, a parametrized probability density function is constructed by HistFactory. This likelihood pdf is sufficiently flexible to describe many analyses based on template histograms. The general form of the HistFactory pdf can be written as
\begin{equation}
\begin{split}
L(n_{c}, x_{e}, a_{p} | \phi_{p}, \alpha_{p}, \gamma_{b}) & = \prod_{c \, \in \, \{\mathrm{channels\}}} \Big[ \mathrm{Pois}(n_{c} | \nu_{c}) \prod_{e = 1}^{n_{c}} f_{c} (x_{e} | \mathbf{\alpha}) \Big] \cdot \\
& \cdot G(L_{0} | \lambda, \Delta_{L}) \cdot \prod_{p \in \mathbb{S}+\Gamma} f_{p} (a_{p} | \alpha_{p})
\end{split}
\end{equation}

where:
\begin{itemize}
    \item $c$ is the channel index, $e$ is an event index, $p$ is the parameter index;
    \item $n_{c}$ and $\nu_{c}$ are the total observed and expected mean number of events in channel $c$, respectively;
    \item $f_{c} (x_{e} | \alpha)$  is the pdf of channel $c$, assuming $x_{e}$ as the number of events in a generic bin $b_{e}$ and $\alpha$ as a nuisance parameter;
    \item $G(L_{0} | \lambda, \Delta_{L})$  is a Gaussian constraint term referred to the luminosity parameter $\lambda_{cs}$. In principle, it might depend on the sample and channel, but it is typically assumed constant in a given channel for all the samples that are normalized with a normalization factor $\lambda_{cs}=L_{0}$;
    \item $f_{p} (a_{p} | \alpha_{p})$ is a constraint term describing an auxiliary measurement $a_{p}$ that constrains the nuisance parameter $\alpha_{p}$.
\end{itemize}

The likelihood pdf can be more easily rewritten in terms of the individual bins as
\begin{equation}
\begin{split}
L(n_{cb}, a_{p} | \phi_{p}, \alpha_{p}, \gamma_{b}) = & \prod_{c \, \in \, \{\mathrm{channels\}}} \prod_{b \, \in \, \{\mathrm{bins\}}} \mathrm{Pois}(n_{cb} | \nu_{cb}) \\
& \cdot G(L_{0} | \lambda, \Delta_{L}) \cdot \prod_{p \in \mathbb{S}+\Gamma} f_{p} (a_{p} | \alpha_{p})
\end{split}
\end{equation}

where the expected mean of events in a given bin $b$, $\nu_{cb}$, is given by
\begin{equation}
    \nu_{cb}(\phi_{p}, \alpha_{p}, \gamma_{b}) = \lambda_{cs}\, \gamma_{cb} \, \phi_{cs}(\alpha) \, \eta_{cs} (\alpha) \, \sigma_{csb} (\alpha).
\end{equation}

The nuisance parameters that enter in the pdf are listed in Table \ref{Tab:HistFactoryParameters}. 
\begin{table}[!htb]
\centering
\begin{tabular}{c|c|c} & Constrained & Unconstrained \\
\noalign{\smallskip}\hline\noalign{\smallskip}
Normalization Variation  & overallSys ($\eta_{cs}$)              & normFactor ($\phi_{p}$)     \\
Coherent Shape Variation & histoSys ($\sigma_{csb}$)             & -                           \\
Bin-by-bin variation     & shapeSys \& statError ($\gamma_{cb}$) & shapeFactor ($\gamma_{csb}$)
\end{tabular}
\caption{Parameters implemented in the HistFactory pdf.}
\label{Tab:HistFactoryParameters}
\end{table}

There are two sets of nuisance parameters, constrained and unconstrained. The set of unconstrained normalization factors as \textit{normFactor} is indicated with $\mathbb{N} = \{\phi_{p}\}$, the set of parameters which have external constraints such as \textit{histoSys} and \textit{overallSys} with $\mathbb{S} = \{\alpha_{p}\}=\{\sigma_{csb}, \eta_{cs}\}$, the set of bin-by-bin uncertainties with constraints such as \textit{shapeSys} and statistical errors but not those associated to an unconstrained \textit{shapeFactor} with $\Gamma = \{\gamma_{cb}\}$.\\
A \textit{histoSys} parameter describes a correlated uncertainty of the shape and normalization while an \textit{overallSys} parameter describes an uncertainty of the global normalization of the sample, the latter not affecting the shape of the distributions of the sample. A \textit{shapeSys} parameter describes an uncertainty of statistical nature, typically arising from limited MC statistics. In HistFactory, \textit{shapeSys} is modelled with an independent parameter for each bin of each channel and shared among all samples having statistical uncertainty associated. Starting from these native HistFactory types of nuisance parameters, others are defined. For example, \textit{normHistoSys} is defined as a histoSys systematic type where \textit{norm} indicates that the total event count is required to remain invariant in a user-specified list of normalization regions when constructing up/down variations. Such a systematic uncertainty is therefore transformed from an uncertainty on event counts in each region into a systematic uncertainty on the transfer factors. \textit{OneSide} and \textit{Sym} types indicate that an uncertainty is constructed as a one-sided or a symmetrized one, respectively, from the inputs.

In the RooFit workspace produced by HistFactory, $\alpha_{p}$ can be assumed as the mean of a Gaussian distribution
\begin{equation}
    G(a_{p} | \alpha_{p}, \sigma_{p}) = \frac{1}{\sqrt{2\pi\sigma_{p}^2}} \exp \Big[ - \frac{(a_{p}-\alpha_{p})^2}{2\,\sigma_{p}^2} \Big]
\end{equation}
with $\sigma_{p} = 1$ by default. It is possible, however, to constrain these parameters with a Poisson distribution or a Log-normal one.

\subsection{Systematic interpolation}
The treatment of systematic uncertainties is subtle, particularly when one wishes to take into account the correlated effect of multiple sources of systematic uncertainty across many signal and background samples. The most important conceptual issue is that we separate the source of the uncertainty (for instance the uncertainty in the response of the calorimeter to jets) from its effect on an individual signal or background sample (e.g. the change in the acceptance and shape of a $W$+jets background). In particular, the same source of uncertainty has a different effect on the various signal and background samples 1. The effect of these $\pm 1 \sigma$ variations about the nominal predictions $\eta_{0} = 1$ and $\sigma_{sb}^{0}$ is quantified by dedicated studies that provide $\eta^{\pm}_{sp}$ and $\sigma^{\pm}_{sp}$.\\
Once one has estimated the effects of the individual sources of systematic uncertainty for each sample, one must address two related issues to form a likelihood parametrized with continuous nuisance parameters. First, one must provide an interpolation algorithm to interpolate to define $\eta_{s}(\mathbf{\alpha})$ and $\sigma_{sb}(\mathbf{\alpha})$. Secondly, one must incorporate constraint terms on the $\alpha_{p}$ to reflect that the uncertain parameter has been estimated with some uncertainty by an auxiliary measurement. A strength of the histogram template based approach (compared to parametrized analytic functions) is that the effects of individual systematics are tracked explicitly; however, the ambiguities associated with the interpolation and constraints are a weakness.

For each sample, one can interpolate and extrapolate from the nominal prediction $\eta^{0}_{s} = 1$ and the variations $\eta^{\pm}_{sp}$ to produce a parametrized $\eta_{s}(\mathbf{\alpha})$. Similarly, one can interpolate and extrapolate from the nominal shape $\sigma^{0}_{sb}$ and the variations $\sigma_{psb}^{\pm}$ to produce a parametrized $\sigma_{psb}(\mathbf{\alpha})$. We choose to parametrize $\alpha_{p}$ such that $\alpha_{p} = 0$ is the nominal value of this parameter, $\alpha_{p} = \pm 1$ are the $\pm 1 \sigma$ variations. Needless to say, there is a significant amount of ambiguity in these interpolation and extrapolation procedures and they must be handled with care.

\subsubsection{Polynomial Interpolation and Exponential Extrapolation}

The strategy is to use the piecewise exponential extrapolation with a polynomial interpolation that matches $\eta(\alpha = \pm \alpha_{0})$, $\mathrm{d}\eta/\mathrm{d}\alpha_{|\alpha = \pm \alpha_{0}}$, and $\mathrm{d}^2\eta/\mathrm{d}\alpha^2_{|\alpha = \pm \alpha_{0}}$ and the boundary $\pm \alpha_{0}$ is defined by the user (with default $\alpha_{0} = 1$),
\begin{equation}
    \eta_{s} (\alpha) = \prod_{p \in \mathrm{Syst}} I_{\mathrm{poly|exp}} (\alpha_{p}; 1, \eta^{+}_{sp}, \eta^{-}_{sp}, \alpha_{0}),
\end{equation}
with
\begin{equation}
    I_{\mathrm{poly|exp}} (\alpha_{p}; I_{0}, I^{+}, I^{-}, \alpha_{0}) =
    \begin{cases}
    (I^{+}/I_{0})^{\alpha}           &= \alpha \geq \alpha_{0} \\
    1+\sum_{i=1}^{6} a_{i}\alpha^{i} &= |\alpha| < \alpha_{0} \\
    (I^{-}/I_{0})^{-\alpha}          &= \alpha \leq \alpha_{0}
    \end{cases}
\end{equation}

\begin{figure}[!htb]
\centering
\includegraphics[width=0.5\textwidth]{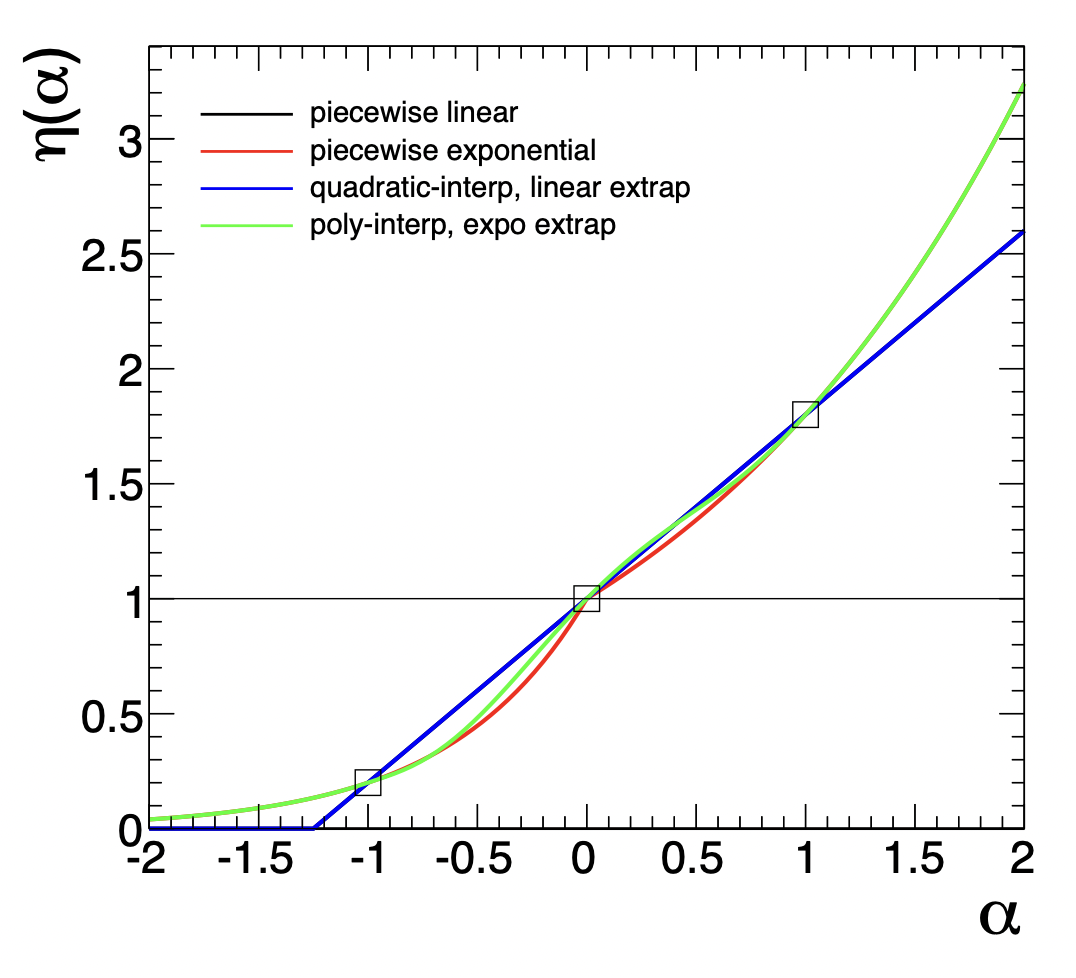}
\caption{Comparison of the four interpolation options choosing $\eta^{+} = 1.2$ and $\eta^{-} = 0.8$.}
\label{fig:InterpolationCodes}
\end{figure}

This approach avoids the kink (discontinuous first and second derivatives) at $\alpha = 0$, which can cause some difficulties for numerical minimization packages such as Minuit and ensures that $\eta(\alpha) \geq 0$, as shown in Fig. \ref{fig:InterpolationCodes}.

\subsubsection{Systematic pulling and profiling}
The fits do not only change normalizations at the post-fit level, but they also \textit{profile} uncertainties. Two important effects can be distinguished: 

\begin{itemize}
    \item \textit{Pulling} shifts the nominal value of a nuisance parameter and causes a change in the yield prediction within its uncertainty to better match the data;
    \item \textit{Profiling} reduces the uncertainty of a nuisance parameter associated with a systematic when testing its compatibility to data.
\end{itemize}

\begin{figure}[!htb]
\centering
\includegraphics[width=0.7\textwidth]{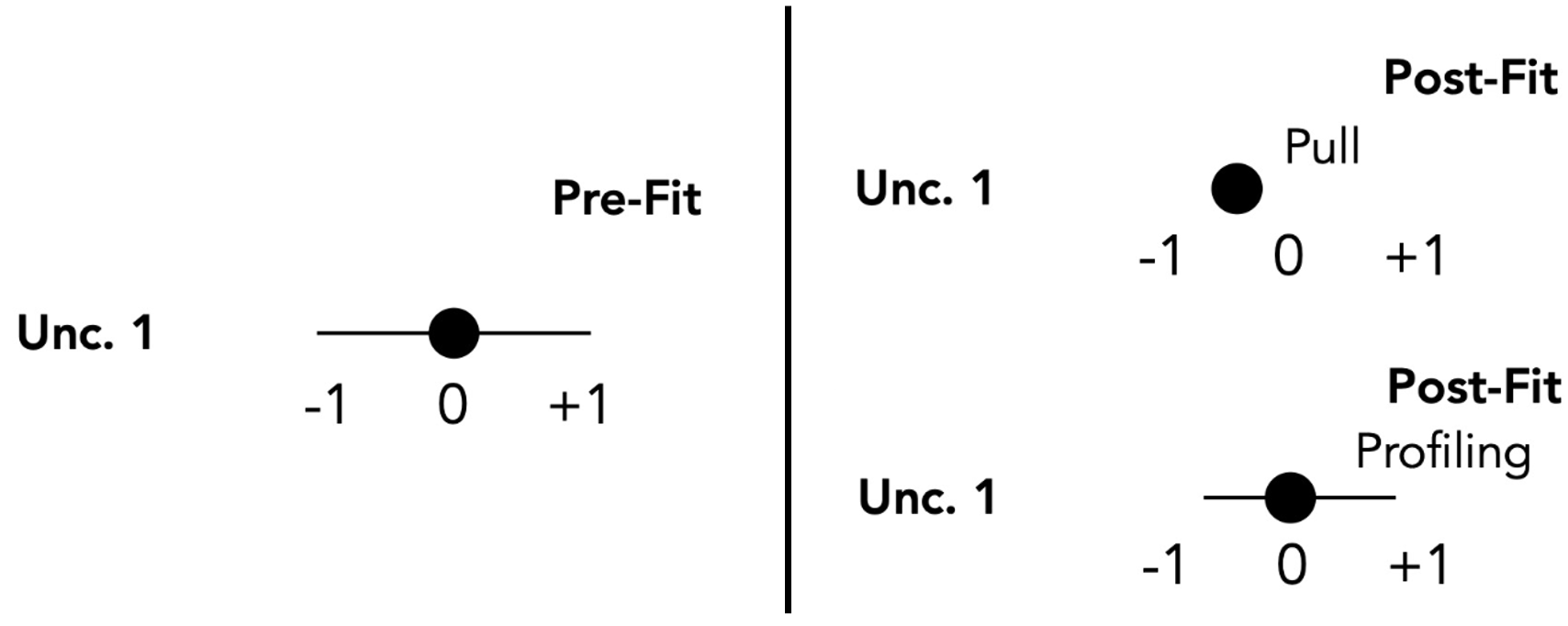}
\caption{The effects of the systematic pulling and profiling for an uncertainty of $\pm 1$ set to a nominal value of 0 at the pre-fit level.}
\label{fig:SystematicPullingProfiling}
\end{figure}

These effects are shown in Fig.~\ref{fig:SystematicPullingProfiling} for an uncertainty of $\pm 1$ set to a nominal value of 0 at the pre-fit level.

At the post-fit level, its nominal value can be shifted towards one side of the initial interval due to pulling, or constrained due to profiling. These effects need to be considered very carefully in our fits, especially if their entity is considered to be large. Pulling is particularly dangerous in shape fits because it can tune a background yield in a specific bin, which usually happens if the bin has high statistics, to a level that is incompatible with the other bins, while profiling is dangerous because it can reduce too much a systematic uncertainty, e.g. to a level that is was not designed for with the object reconstruction. The opposite of profiling, e.g. the expansion of the interval of an uncertainty around its range, is also possible in fits, however this is considered a less worrisome issue compared to the other effects discussed before since an increase of the uncertainty can reduce the sensitivity but does not bias the results, yet it still has to be monitored.

\section{Types of fit}
There are three most commonly used fit strategies: the \textit{background-only fit}, the \textit{exclusion} fit (also called \textit{model-dependent signal fit}), and the \textit{discovery fit} (also called \textit{model-independent signal fit}:
\begin{itemize}
\item Background-only fit. The purpose of this fit strategy is to estimate the total background in SRs and VRs, without making assumptions on any signal model. As the name suggests, only background samples are used in the model. The CRs are assumed to be free of signal contamination. The fit is only performed in the CR(s), and the dominant background processes are normalized to the observed event counts in these regions. As the background parameters of the PDF are shared in all different regions, the result of this fit is used to predict the background event levels in the SRs and VRs.
The background predictions from the background-only fit are independent of the observed number of events in each SR and VR, as only the CR(s) are used in the fit. This allows for an unbiased comparison between the predicted and observed number of events in each region. The background-only fit predictions are used to present the validation of the transfer factor-based background level predictions.
Another important use case for background-only fit results in the SR(s) is for external groups to perform a hypothesis test on an untested signal model, which has not been studied by the experiment. With the complex fits currently performed at the LHC, it may be difficult (if not impossible) for outsiders to reconstruct these. An independent background estimate in the SR, as provided by the background-only fit, is then the correct estimate to use as input to any hypothesis test.
\item Exclusion fit. This fit strategy is used to study a specific signal model. In the absence of a significant event excess in the SR(s), as concluded with the background-only fit configuration, exclusion limits can be set on the signal models under study. In case of excess, the model-dependent signal fit can be used to measure properties such as the signal strength. The fit is performed in the CRs and SRs simultaneously. Along with the background samples, a signal sample is included in all regions, not just the SR(s), to correctly account for possible signal contamination in the CRs. A normalization factor, the signal strength parameter $\mu_{sig}$, is assigned to the signal sample. Note that this fit strategy can be run with multiple SRs (and CRs) simultaneously, as long as these are statistically independent, non-overlapping regions. If multiple SRs are sensitive to the same signal model, performing the model-dependent signal fit on the statistical combination of these regions shall, in general, give better (or equal) exclusion sensitivity than obtained in the individual analyses.
\item Discovery fit. An analysis searching for new physics phenomena typically sets model-independent upper limits on the number of events beyond the expected number of events in each SR. In this way, for any signal model of interest, anyone can estimate the number of signal events predicted in a particular signal region and check if the model has been excluded by current measurements or not.
Setting the upper limit is accomplished by performing a model-independent signal fit. For this fit strategy, both the CRs and SRs are used, in the same manner as for the model-dependent signal fit. Signal contamination is not allowed in the CRs, but no other assumptions are made for the signal model, also called a \say{dummy signal} prediction. The SR in this fit configuration is constructed as a single-bin region, since having more bins requires assumptions on the signal spread over these bins. The number of signal events in the signal region is added as a parameter to the fit. Otherwise, the fit proceeds in the same way as the model-dependent signal fit.
The model-independent signal fit strategy, fitting both the CRs and each SR, is also used to perform the background-only hypothesis test, which quantifies the significance of any observed excess of events in a SR, again in a manner that is independent of any particular signal model. One key difference between the model-independent signal hypothesis test and the background-only hypothesis test is that the signal strength parameter is set to one or zero in the profile likelihood numerator respectively.
\end{itemize}

\section{Statistical significance}
\label{sec:StatisticalSignificance}
In order to optimise a search, or estimate how much data deviate from the SM prediction, it is useful to introduce the statistical significance. The definition of the statistical significance used in the analyses is:
\begin{equation}
Z_{n} = \sqrt{ 2 \Big[n \log \left( \frac{n(b+\sigma^2)}{b^2+n\sigma^2} \right) - \frac{b^2}{\sigma^2}\log \left( 1+\frac{\sigma^2 (n-b)}{b(b+\sigma^2)} \right) \Big]}
\label{eqn:SignificanceMeasure}
\end{equation}
where $n$ is the number of observed events ($n = s+b$) given a background prediction of $b \pm \sigma$ events. In order to exclude a model at 95\% CL, a significance of $Z_{n} = 1.64$ is needed (the $p$-value drops below $0.05$ for $Z_{n} = 1.64$). The formula reproduces well the significance calculated through a toy (frequentist) approach for a corresponding model. \footnote{Historically, the signal (s) over background (b) $s$/$b$ or $s$/$\sqrt{b}$ was used in searches that were systematics or statistics dominated, respectively. However, it has been shown that when the statistics is very low, both indicators fail.} \\
The derivation of Eq.~(\ref{eqn:SignificanceMeasure}) makes use of the asymptotic formulae for the distributions of profile likelihood test statistics. A likelihood function $L(s)$ can be constructed from the product of two Poisson distributions
\begin{equation}
    L(s) = P(n|s+b)P(m|\tau b).
\end{equation}
The term $P(m|\tau b)$ for this likelihood comes from the fact that the background rate $b$ is supposed to be a true but unknown parameter which is constrained by an auxiliary measurement $m$. The factor $\tau$ relates the event rates in the region where we perform the auxiliary measurement $m$ to the region where we measure $n$, and it is a function of the background $b$ and its uncertainty $\sigma$ through $\tau=b\/\sigma^2$. For the purpose of computing significance, the observed value of $m$ is taken to be $\tau b$. This is, however, not required to be an integer, and so the second Poisson distribution given here is actually described by gamma functions.
The likelihood is maximised for $\hat{s} = n-m/\tau$, $\hat{b} = m/\tau$. For the background-only hypothesis, the maximum likelihood is for $\hat{\hat{b}} = (n + m)/(1 + \tau)$. The profile likelihood ratio $0 \leq \lambda(s, b) = L(s,b)/L(\hat{s},\hat{b}) \leq 1$ becomes for the background-only hypothesis
\begin{equation}
    \lambda(0) = \frac{L(0, \hat{\hat{b}})}{L(\hat{s},\hat{b})} = \Big( \frac{n+m}{1+\tau} \Big)^{n+m} \frac{\tau^{m}}{n^{m}m^{m}}.
\end{equation}
Using the test statistic $q_{0} = -2 \ln \lambda(0)$ and Wilk’s approximation for the test statistic distribution we can write
\begin{equation}
    Z = \sqrt{-2 \ln \lambda(0)}
\end{equation}
which, setting $m = \tau b$, yields Eq.~(\ref{eqn:SignificanceMeasure}).

\chapter{Compressed chargino search}
\label{sec:compressedscharginos}
In this Chapter, the compressed chargino search is presented. This search targets the direct production of chargino pairs decaying into the LSP neutralinos through the emission of $W$ bosons, which further decay into leptons and neutrinos. A signature with two leptons, $E_{\mathrm{T}}^{\text{miss}}$ and no hadronic activity is considered and a compressed mass spectrum where the difference in mass between chargino and neutralino is close to the mass of the $W$ boson is targeted. \\

\minitoc
\medskip

\section{Analysis overview}
\label{sec:compressedscharginos-analysisoverview}

\begin{figure}[!htb]
\centering
\includegraphics[width=0.5\textwidth]{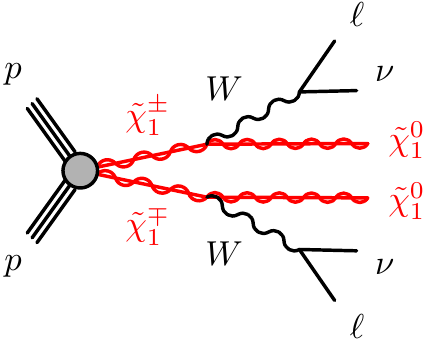}
\caption{Diagram of the supersymmetric production of charginos, with two leptons and weakly interacting particles in the final state.  In this model with intermediate $W$ bosons, two flavour $\tilde{e}, \tilde{\mu}$ are considered.}
\label{fig:c1c1viawwfeynman}
\end{figure}

The analysis targets the direct production of chargino pairs $\tilde{\chi}_1^+\tilde{\chi}_1^-$ in the wino-bino scenario, where each chargino decays into the LSP neutralino via the emission of a SM $W$ boson, as shown in Fig.~\ref{fig:c1c1viawwfeynman}. The topology of the signal is close to the SM $WW$ process and it is characterised by 2 leptons that can be both of the Same Flavour (SF), namely $ee$ and $\mu\mu$, or Different Flavour (DF), namely $e\mu$.\\

A previous search \cite{SUSY-2018-32} considering the same model and signature was performed. The search exploited the full ATLAS Run 2 data set too but it was optimized to target the phase space where a large mass difference between chargino and the LSP is present and resulted in the exclusion of $\tilde{\chi}_1^{\pm}$ mass up 420 GeV in the case of a massless $\tilde{\chi}_{1}^{0}$. The search used the $m_{\mathrm{T2}}$ variable as a key discriminating variable. The analysis strategy is not effective in scenarios with a compressed mass spectrum where the SUSY particles are at the scale of the SM bosons.\\

In order to explore the compressed phase space, Machine Learning (ML) techniques are exploited in this new effort \cite{ATLAS-CONF-2022-006,SUSY-2019-02}. The analysis strategy relies on Boosted Decision Trees (BDTs), developed in the LightGBM framework \cite{Ke:2017LightGBM} designed for both speed and performance on large data sets. The classifier is trained to separate into 4 classes: signal, diboson, Top ($t\bar{t}$ and $Wt$) and all other backgrounds ($Z(ee/\mu\mu)$+jets, $Z(\tau\tau)$+jets, $VVV$ and all other minor backgrounds), with the output score of each class corresponding to the probability for the event of being in each class. The four output scores sum to one. The technique is found to be more effective at discriminating the signal and backgrounds than a binary signal vs background classifier, and the scores corresponding to each background can be used to isolate that background for control and validation regions. 

The main SM backgrounds are the irreducible diboson ($VV=WW/WZ/ZZ$) and the Top processes, and dedicated CRs and VRs are designed for them. For the $VV$ background, a selection is performed in BDT-signal score cut and two intervals are used for the CR and VR. Cuts on BDT-VV score are applied to define the VV CRs and VRs. A selection on BDT-top score is instead used to reduce the top contamination in VV CRs and VRs. The top processes ($t\bar{t}$ and $Wt$) contain $b$-jets in the final state, so by requiring events with a $b$-jet, regions rich in top processes can be defined. This has the additional benefit of orthogonality to all the signal regions, which have a $b$-veto. The BDT-signal score is used to select a region close to the SR, while a selection on the BDT-top score ensures a large purity.\\

A SR is defined by taking the region in BDT-signal score with the high significance, and a shape fit as a function of BDT-signal score is performed, by considering the DF and SF channels separately. No significant excess of events is observed in the SRs and exclusion limits at 95\% of CL are reported as results.
The results supersede the previous ATLAS results in particularly interesting regions where the difference in mass between the chargino and neutralino is close to the mass of the $W$ boson and the chargino pair production could have hidden behind the looking-alike $WW$ background.

\section{Event samples}
\label{sec:compressedscharginos-eventsamples}

\subsection{Signal samples}
The MC signal samples with $\tilde{\chi}_1^+\tilde{\chi}_1^-\rightarrow W\tilde{\chi}_1^0W\tilde{\chi}_1^0$ are generated from leading-order (LO) matrix elements with up to two extra partons using {\textsc{MadGraph5}\_aMC@NLO} \cite{Alwall:2014hca} v2.6.2 interfaced with \textsc{PYTHIA} 8.2 \cite{Sjostrand:2014zea} with the A14 tune \cite{ATL-PHYS-PUB-2014-021}, for the modelling of the SUSY decay chain, parton showering, hadronisation and the description of the underlying event. Parton luminosities are provided by the NNPDF23LO PDF set \cite{Ball:2012cx}. Jet-parton matching has been done following the CKKW-L prescription \cite{Lonnblad:2012ix}, with a matching scale set to one quarter of the pair-produced superpartner mass. In order to allow for the off-shell $W$ decay to keep track of the spin correlations effects, the MadSpin code \cite{Artoisenet:2012st} has been used in generation for mass splitting between chargino and neutralino lower than 100 GeV. Signal cross-sections were calculated assuming the wino-bino mass hierarchy to next-to-leading order (NLO) in $\alpha_S$ adding the resummation of soft gluon emission at next-to-leading-logarithm accuracy (NLO+NLL) \cite{Debove,Fuks13040790,xsec13TeV-5}. The nominal cross-sections and their uncertainties were taken from an envelope of cross-section predictions using different PDF sets and factorisation and renormalisation scales, as described in Ref.~\cite{xsec13TeV-1}.
The samples used, filtered in generation by requiring 2 leptons with $p_{\mathrm{T}} > 3$ GeV and the leading lepton with $p_{\mathrm{T}} > 20$ GeV, are listed in Table \ref{Table:SleptonCharginosGrid}.

\begin{table}[!htb]
\centering
\scalebox{0.58}{
\centering
\begin{tabular}{c|ccccccccc}
\noalign{\smallskip}\hline\noalign{\smallskip}
$\Delta m(\tilde{\chi}_{1}^{\pm}, \tilde{\chi}_{1}^{0})$ [GeV] & \multicolumn{9}{c}{$m(\tilde{\chi}_{1}^{\pm}$, $\tilde{\chi}_{1}^{0})$ [GeV]}\\
\noalign{\smallskip}\hline\noalign{\smallskip}
10& (100, 90) & (125, 115) & (150, 140) & (175, 165) & (200, 190) &            & (250, 240) &            &            \\
20& (100, 80) & (125, 105) & (150, 130) & (175, 155) & (200, 180) &            & (250, 230) &            &            \\
30& (100, 70) & (125, 95)  & (150, 120) & (175, 145) & (200, 170) &            & (250, 220) &            &            \\
40& (100, 60) & (125, 85)  & (150, 110) & (175, 135) & (200, 160) &            & (250, 210) &            &            \\
50& (100, 50) & (125, 75)  & (150, 100) & (175, 125) & (200, 150) &            & (250, 200) &            &            \\
60& (100, 40) & (125, 65)  & (150, 90)  & (175, 115) & (200, 140) &            & (250, 190) & (300, 240) &            \\
70& (100, 30) & (125, 55)  & (150, 80)  & (175, 105) & (200, 130) &            & (250, 180) & (300, 230) &            \\
80& (100, 20) & (125, 45)  & (150, 70)  & (175, 95)  & (200, 120)  &            & (250, 170) & (300, 220) &            \\
90& (100, 10) & (125, 35)  & (150, 60)  & (175, 85)  & (200, 110)   &            & (250, 160) & (300, 210)           &            \\
100&              & (125, 25)  & (150, 50)  & (175, 75)  & (200, 100)   & (225, 125) & (250, 150) & (300, 200) &            \\
125&              &                  & (150, 25)  & (175, 50)  & (200, 75)    & (225, 100) & (250, 125) & (300, 175) &            \\
150&              &                  &                 &  (175, 25)  &  (200, 50)   &            &            & (300, 150) & (350, 200)  \\
175&              &                  &                 &                   &  (200, 25)   &            &            & (300, 125) & (350, 175) \\
200&              &                  &                 &                   &                   &            &            &            & (350, 150)  \\
225&              &                  &                 &                    &                  &            &            &            & (350, 125) \\ 
\noalign{\smallskip}\hline\noalign{\smallskip}
\end{tabular}}
\caption{Signal samples produced for the chargino search. Grid points are distinguished as a function of their mass splitting $\Delta m(\tilde{\chi}_{1}^{\pm}, \tilde{\chi}_{1}^{0})$ and separated in order of  $m_{\tilde{\chi}_{1}^{\pm}}$ in the different columns.}
\label{Table:SleptonCharginosGrid}
\end{table}

\subsection{Background samples}

The SM background processes considered in the search are top (pair production $t\bar{t}$ and single top $Wt$), diboson ($VV$), triboson ($VVV$), $W/\gamma+$jets, $Z/\gamma+$jets, and other processes. \\

\begin{itemize}
    \item Top processes. The production of $t\bar{t}$ and $t\bar{t}H$ events is modelled using the \textsc{POWHEG-BOX-v2} \cite{Nason:2004rx,Frixione:2007vw,Alioli:2010xd} generator at NLO with the NNPDF3.0NLO \cite{Ball:2014uwa} set of PDFs and the hdamp parameter\footnote{The hdamp parameter is a resummation damping factor and one of the parameters that controls the matching of Powheg matrix elements to the parton shower and thus effectively regulates the high-$p_{\mathrm{T}}$ radiation against which the $p_{\mathrm{T}}$ system recoils.} set to 1.5 $m_{t}$~\cite{ATL-PHYS-PUB-2016-020}. The events are interfaced to \textsc{PYTHIA8} ~\cite{Sjostrand:2014zea} to model the parton shower, hadronisation, and underlying event, with parameters set according to the A14 tune ~\cite{ATL-PHYS-PUB-2014-021} and using the NNPDF2.3LO~\cite{Ball:2012cx} set of PDFs. The decays of bottom and charm hadrons are performed by \textsc{EvtGen~v1.6.0} \cite{Lange:2001uf}. \\
    The associated production of top quarks with $W$ bosons ($Wt$) is modelled using the \textsc{POWHEG-BOX-v2} generator at NLO in QCD using the five-flavor scheme and the NNPDF3.0NLO set of PDFs. The Diagram Removal (DR) scheme \cite{Frixione:2008yi} is used to remove interference and overlap with $t\bar{t}$ production. The events are interfaced to \textsc{PYTHIA} v8 using the A14 tune and the NNPDF2.3LO set of PDFs. The  {\textsc{MadGraph5}\_aMC@NLO} v2.3.3 \cite{Alwall:2014hca} generator at NLO with the NNPDF3.0NLO PDF is used to model the production of $t\bar{t}V$, $tWZ$ and $tZq$ events. The events are interfaced to \textsc{PYTHIA} v8.2 \cite{Sjostrand:2014zea} using the A14 tune and the NNPDF2.3LO PDF set. The {\textsc{MadGraph5}\_aMC@NLO} v2.3.3 generator at LO with the NNPDF2.3LO PDF is used for the production of $t\bar{t} \gamma$ events. The events are interfaced with \textsc{8.2} using the A14 tune and the NNPDF2.3LO PDF set.
    
    \item $VV$ processes. Samples of $VV$ processes are simulated with the \textsc{SHERPA} v2.2.1 or v2.2.2 \cite{Bothmann:2019yzt} generator depending on the process, including off-shell effects and Higgs-boson contributions, where appropriate. Fully leptonic final states and semileptonic final states, where one boson decays leptonically and the other hadronically, are generated using matrix elements at NLO accuracy in QCD for up to one additional parton and at LO accuracy for up to three additional parton emissions. Samples for the loop-induced processes $gg \to VV$ are generated using LO-accurate matrix elements for up to one additional parton emission for both cases of fully leptonic and semileptonic final states. The matrix element calculations are matched and merged with the \textsc{SHERPA} parton shower based on Catani-Seymour dipole \cite{Gleisberg:2008fv,Schumann:2007mg} using the MEPS@NLO prescription \cite{Hoeche:2012yf,Catani:2001cc,Hoeche:2009rj}. The virtual QCD correction are provided by the {\textsc{openloops}} library~\cite{Cascioli:2011va}. The set of PDFs used is NNPDF3.0NLO, along with the dedicated set of tuned parton-shower parameters developed by the \textsc{SHERPA} authors. Electroweak production of diboson in association with two jets ($VVjj$) is simulated with the \textsc{SHERPA}~v2.2.2 generator. The LO-accurate matrix elements are matched to a parton shower based on Catani-Seymour dipoles using the MEPS@LO~\cite{Hoeche:2012yf,Catani:2001cc,Hoeche:2009rj} prescription. Samples are generated using the NNPDF3.0NLO set, along with the dedicated set of tuned parton-shower parameters developed by the \textsc{SHERPA} authors. 

    \item $VVV$ processes. The production of $VVV$ events is simulated with the \textsc{SHERPA}~v2.2.2 generator using factorised gauge boson decays. Matrix elements, accurate at NLO for the inclusive process and at LO for up to two additional parton emissions, are matched and merged with the \textsc{SHERPA} parton shower based on Catani-Seymour dipoles using the MEPS@NLO prescription. The virtual QCD correction for matrix elements at NLO accuracy are provided by the {\textsc{openloops}} library. Samples are generated using the NNPDF3.0NLO set, along with the dedicated set of tuned parton-shower parameters developed by the \textsc{SHERPA} authors.
    
    \item $V(W,Z)+$jets processes. The production of $V(W,Z)+$jets is simulated with the \textsc{SHERPA}~v2.2.1 generator using NLO-accurate matrix elements for up to two jets, and LO-accurate matrix elements for up to four jets calculated with the \textsc{COMIX} and \textsc{OPENLOOPS} libraries. They are matched with the \textsc{SHERPA} parton shower using the MEPS@NLO prescription using the set of tuned parameters developed by the \textsc{SHERPA} authors. The NNPDF3.0NLO set of PDFs is used and the samples are normalised to a Next-to-Next-to-Leading Order (NNLO) prediction~\cite{Anastasiou:2003ds}. Electroweak production of $\ell\ell jj$, $\ell\nu jj$ and $\nu\nu jj$ final states are generated using \textsc{SHERPA} v2.2.1 using leading order matrix elements with up to two additional parton emissions. The matrix elements are merged with the \textsc{SHERPA} parton shower following the MEPS@NLO prescription and using the set of tuned parameters developed by the \textsc{SHERPA} authors. The NNPDF3.0NLO set of PDFs is employed. The samples are produced using the VBF approximation, which avoids the overlap with semi-leptonic diboson topologies by requiring a $t$-channel colour singlet exchange.
\end{itemize}

\subsection{Data samples}
The full Run~2 ATLAS data set is exploited, corresponding to 3.2 fb$^{-1}$ of data collected in 2015, 33.0 fb$^{-1}$ of data collected in 2016, 44.3 fb$^{-1}$ of data collected in 2017 and 58.45 fb$^{-1}$ of data collected in 2018. The dataset considered corresponds then to a combined total integrated luminosity of 138.95~fb$^{-1}$. The uncertainty in the combined 2015–2018 integrated luminosity is 1.7\% \cite{LuminosityUncertainty}, obtained using the LUCID-2 detector \cite{Avoni} for the primary luminosity measurements.

\section{Object definition}
\label{sec:compressedscharginos-objectdefinition}
This Section is dedicated to specify the objects used for the analysis: electrons, muons, jets and $E_{\mathrm{T}}^{\text{miss}}$. Hadronic taus are not selected but there is no explicit veto on them. The set of quality cuts and the trigger selection required for the events is first discussed. The object definition criteria for electrons, muons and jets can be found in Tables~\ref{tab:eledef},~\ref{tab:muondef} and~\ref{tab:jetsdef}, respectively.

\subsection{Event Quality Cuts}
\label{sec:techintro}
The $pp$ collision events recorded by the ATLAS experiment are on top of a background that is due to both collision debris and non-collision components. The non-collision background comprises three types: beam-induced backgrounds (BIB), cosmic particles and detector noise. BIB is due to proton losses upstream of the interaction point with the proton losses inducing secondary cascades which can reach the ATLAS detector and become a source of background for physics analyses. In general, this background is difficult to model in simulation and a set of selections on top of the analysis level is applied to reduce it and to veto inactive regions of the detector. The set of requirements applied at event level is:

\begin{itemize}
\item {ATLAS Good Run List (GRL)} (data only): data events must satisfy the GRL, which selects good luminosity blocks within the data runs (spanning 1-2 minutes of data-taking). The decision of including a luminosity block in the GRL is taken online during data-taking, while the other requirements for data from the errors received by subdetectors are usually performed in a subsequent step, in the analysis of the RAW data. Luminosity blocks that are absent in the GRL are considered bad, thus they are not considered in the analysis.
\item {LAr/Tile error} (data only): events with noise bursts and data integrity errors in the LAr calorimeter/Tile are removed;
\item {Tile Trip} (data only): events with Tile trips are removed;
\item {SCT error} (data only): events affected by the recovery procedure for single event upsets in the SCT are removed;
\item {Cosmic or bad muons} (data and MC): fake muons are reconstructed muons not corresponding to true objects coming from $pp$ collisions. They can sometimes be created from high hit multiplicities in the muon spectrometer in events where some particles from very energetic jets punch through the calorimeter into the muon system, or from badly measured inner detector tracks in jets wrongly matched to muon spectrometer segments. They can also be caused by the cavern background creating hits in the muon spectrometer. Cosmic muons are also a source of muons unrelated to the hard scattering. If an event contains at least one cosmic muon after overlap removal (defined as having $abs(z_{0}^{PV})>1$ mm or $d_{0}^{PV}>0.2$ mm) or at least one bad baseline muon before overlap removal (satisfying $\sigma(q/p)/(|q/p|)>0.4$) then the event is rejected.
\item {Bad jets} (data and MC): non-collision background processes can lead to (fake or real) energy deposits in the calorimeters. These energy deposits are reconstructed as jets. Jet properties can be used to distinguish background jet candidates not originating from hard scattering events from jets produced in $pp$ collisions. Events with a \texttt{BadLoose} jet \cite{JetSelectionBadLoose} with $p_\mathrm{T}>$ 20 GeV are rejected.
\item {Primary Vertex} (data and MC): events must have a Primary Vertex (PV), selected as the one with the highest $\sum p_{\mathrm{T}}^{2}$ of associated tracks, with at least two tracks.
\end{itemize}

\subsubsection{Electrons}
Electrons are reconstructed as described in Section~\ref{ssec:reco-ele} and are required to reside within $|\eta|$ <2.47. At baseline level, electrons must have $p_\mathrm{T} > 9$ GeV, satisfy the \texttt{LooseAndBLayerLLH} Particle Identification (PID) quality criteria and also satisfy the Interaction Point (IP) condition $|z_0 \sin \theta| < 0.5$~mm.
Signal electrons must have $p_\mathrm{T} > 9$ GeV and be isolated with respect to other high-$p_{\mathrm{T}}$ charged particles satisfying the \texttt{FCLoose} isolation criteria. Moreover signal electrons must pass \texttt{TightLLH} quality criteria and also satisfy the IP condition $S(d_0)<5$. The electron selection is summarised in Table~\ref{tab:eledef}.

\begin{table}[!htb]
\begin{center}
\begin{tabular}{l|c}
\noalign{\smallskip}\hline\noalign{\smallskip}
\multicolumn{2}{c}{Baseline electron}\\
\noalign{\smallskip}\hline\noalign{\smallskip}
Acceptance       & $p_{\mathrm{T}}$ > 9 GeV, $|\eta^\mathrm{clust}|$ < 2.47 \\
PID Quality      & \texttt{LooseAndBLayerLLH} \\
Impact parameter & $|z_0 \sin\theta|< 0.5$ mm \\
\noalign{\smallskip}\hline\noalign{\smallskip}
\noalign{\smallskip}\hline\noalign{\smallskip}
\multicolumn{2}{c}{Signal electron}\\
\noalign{\smallskip}\hline\noalign{\smallskip}
Acceptance       & $p_{\mathrm{T}}$ > 9 GeV, $|\eta^\mathrm{clust}|$ < 2.47 \\
PID Quality      & \texttt{TightLLH}  \\
Isolation        & \texttt{FCLoose} \\ 
Impact parameter & $S(d_0)< 5$ \\ 
\noalign{\smallskip}\hline\noalign{\smallskip}
\end{tabular}
\end{center}
\caption{Summary of the electron selection criteria. The signal selection requirements are applied on top of the baseline selection and after Overlap Removal has been performed.}             
\label{tab:eledef}
\end{table}

\subsubsection{Muons}
Muons used in this analysis must have $p_\mathrm{T} > 9$ GeV and reside within $|\eta|<2.6$. Baseline muons must pass \texttt{Medium} quality requirements and also satisfy the IP condition $|z_0 \sin \theta|<0.5$~mm. Signal muons must have $p_\mathrm{T} > 9$ GeV, pass the \texttt{Medium} quality criteria, be isolated with respect to other high-$p_{\mathrm{T}}$ charged particles, satisfying the \texttt{FCLoose} isolation criteria and additionally having $S(d_0)<3$ constraint on the IP. The muon selection criteria are summarised in Table~\ref{tab:muondef}.

\begin{table}[!htb]
\begin{center}\renewcommand\arraystretch{1.2}
\begin{tabular}{l|c}
\noalign{\smallskip}\hline\noalign{\smallskip}
\multicolumn{2}{c}{Baseline muon}\\
\noalign{\smallskip}\hline\noalign{\smallskip}
Acceptance        & $p_{\mathrm{T}}$ > 9 GeV, $|\eta|$ < 2.6  \\
PID Quality       & \texttt{Medium}    \\
Impact parameter  & $|z_0 \sin\theta|$ < 0.5 mm \\
\noalign{\smallskip}\hline\noalign{\smallskip}
\noalign{\smallskip}\hline\noalign{\smallskip}
\multicolumn{2}{c}{Signal muon}\\
\noalign{\smallskip}\hline\noalign{\smallskip}
Acceptance        & $p_{\mathrm{T}}$ > 9 GeV, $|\eta|$ < 2.6  \\
PID Quality       & \texttt{Medium}    \\
Isolation         & \texttt{FCLoose} \\
Impact parameter  & $S(d_0)$ < 3 \\
\noalign{\smallskip}\hline\noalign{\smallskip}             
\end{tabular}
\caption{Summary of the muon selection criteria. The signal selection requirements are applied on top of the baseline selection after Overlap Removal has been performed.} 
\label{tab:muondef}
\end{center}
\end{table}

\subsubsection{Jets}
This analysis uses \texttt{PFlow} jets reconstructed using the anti-$k_{\mathrm{T}}$ algorithm with distance parameter $D=0.4$. At baseline level these jets are required to have $p_\mathrm{T} > 20$ GeV and fulfill the pseudorapidity requirement of $|\eta|<2.8$.
To reduce the effects of pile-up, signal jets are further required to pass a cut of $JVT>0.5$ on the $JVT$ ~\cite{ATLAS-CONF-2014-018}, if their $p_{\mathrm{T}}$ is in the 20-60 GeV range and they reside within $|\eta|<2.4$. Only jet candidates with $p_\mathrm{T} > 20$ GeV and $|\eta|<2.4$ are finally considered,\footnote{Hadronic $\tau$-lepton decay products are treated as jets.} although jets with $|\eta|<4.9$ are included in the missing transverse momentum calculation and are considered when applying the procedure to remove reconstruction ambiguities, which is described later in this Section.

The DL1r algorithm \cite{DL1r} identifies $b$-jets. A selection that provides 85\% efficiency for tagging $b$-jets in simulated $t\bar{t}$ events is used. The choice of 85\% WP ensured a stronger $t\bar{t}$ and single top rejection, without a significant loss of signal statistics. The jet selection criteria are summarised in Table~\ref{tab:jetsdef}.

\begin{table}[!htb]
\begin{center}
\begin{tabular}{l|c}
\noalign{\smallskip}\hline\noalign{\smallskip}
\multicolumn{2}{c}{Baseline jet} \\
\noalign{\smallskip}\hline\noalign{\smallskip}
Collection     & \texttt{AntiKt4EMPFlowJets} \\
Acceptance     & $p_{\mathrm{T}}$ > 20 GeV, $|\eta|$ <2.8 \\
\noalign{\smallskip}\hline\noalign{\smallskip}
\noalign{\smallskip}\hline\noalign{\smallskip}
\multicolumn{2}{c}{Signal jet} \\
\noalign{\smallskip}\hline\noalign{\smallskip}
JVT                        & \texttt{Tight} \\
Acceptance                 & $p_\mathrm{T} > 20$ GeV, $|\eta | < 2.4$ \\ 
\noalign{\smallskip}\hline\noalign{\smallskip}
\noalign{\smallskip}\hline\noalign{\smallskip}
\multicolumn{2}{c}{Signal $b$-jet} \\
\noalign{\smallskip}\hline\noalign{\smallskip}
$b$-tagger Algorithm      & DL1r \\
Efficiency                & \texttt{FixedCutBEff\_85} \\
Acceptance                & $p_{\mathrm{T}}$ > 20 GeV, $|\eta|$ < 2.4 \\ 
\noalign{\smallskip}\hline\noalign{\smallskip}
\end{tabular}
\end{center}
\caption{Summary of the jet and $b$-jet selection criteria. The signal selection requirements are applied on top of the baseline requirements after Overlap Removal has been performed. }
\label{tab:jetsdef}
\end{table}

\subsubsection{Missing transverse momentum}
The missing transverse energy is built from the transverse momenta of all physics objects considered in the analysis (jets, muons and electrons), as well as photons and all tracks matched to the primary vertex not associated with these objects. The $E_{\mathrm{T}}^{\text{miss}}$ is reconstructed using the \texttt{Tight} working point, where the jets with $|\eta|>$~2.4 are required to have $p_{\text{T}}>$~30~GeV. 

Associated to the $E_{\mathrm{T}}^{\text{miss}}$ value is the $E_{\mathrm{T}}^{\text{miss}}$ significance value, obtained by also considering the resolution of each physics object considered in the analysis. $E_{\mathrm{T}}^{\text{miss}}$ significance helps to separate events with true $E_{\mathrm{T}}^{\text{miss}}$ (arising from weakly interacting particles) from those where it is consistent with particle mismeasurement, resolution or identification inefficiencies. On an event-by-event basis, given the full event composition, $E_{\mathrm{T}}^{\text{miss}}$ significance evaluates the p-value that the observed $E_{\mathrm{T}}^{\text{miss}}$ is consistent with the null hypothesis of zero real $E_{\mathrm{T}}^{\text{miss}}$, as further detailed in Ref.~\cite{ATLAS-CONF-2018-038}.

\subsection{Trigger selection}
Single lepton triggers requiring the presence of at least an electron or a muon in each event are adopted. Events are required to satisfy a logical OR of the triggers listed in Table~\ref{tab:trigger}. The triggers are divided by year and ordered by $p_{\mathrm{T}}$ threshold, with the first triggers listed on top of each year corresponding to lower $p_{\mathrm{T}}$ thresholds but also having isolation requirements (\textit{ivar}) while the other triggers having higher $p_{\mathrm{T}}$ thresholds allowing high efficiency at high $p_{\mathrm{T}}$ \cite{ATLASTrigger1,ATLASTrigger2}.

\begin{table}[!htb]
\begin{center}
\begin{tabular}{lll}
\noalign{\smallskip}\hline\noalign{\smallskip}
 & Single electron      & Single muon \\
\noalign{\smallskip}\hline\noalign{\smallskip}
2015       & \texttt{HLT\_e24\_lhmedium\_L1EM20VH}    & \texttt{HLT\_mu20\_iloose\_L1MU15}    \\
           & \texttt{HLT\_e60\_lhmedium}              & \texttt{HLT\_mu60\_0eta105\_msonly}   \\
           & \texttt{HLT\_e120\_lhloose}              & \texttt{HLT\_mu50}                  \\
\noalign{\smallskip}\hline\noalign{\smallskip}
2016       & \texttt{HLT\_e24\_lhtight\_nod0\_ivarloose}   & \texttt{HLT\_mu26\_ivarmedium}   \\
           & \texttt{HLT\_e26\_lhtight\_nod0\_ivarloose}   & \texttt{HLT\_mu50}              \\
           & \texttt{HLT\_e60\_lhmedium\_nod0}             & \\
           & \texttt{HLT\_e140\_lhloose\_nod0}             & \\
           & \texttt{HLT\_e300\_etcut}             &    \\
\noalign{\smallskip}\hline\noalign{\smallskip}
2017-2018       & \texttt{HLT\_e26\_lhtight\_nod0\_ivarloose}    & \texttt{HLT\_mu26\_ivarmedium} \\
                & \texttt{HLT\_e60\_lhmedium\_nod0}             & \texttt{HLT\_mu50}        \\
                & \texttt{HLT\_e140\_lhloose\_nod0}              & \texttt{HLT\_mu60\_0eta105\_msonly} \\
                & \texttt{HLT\_e300\_etcutP}    &         \\
\noalign{\smallskip}\hline\noalign{\smallskip}
\end{tabular}
\end{center}
\caption{Summary of the triggers used in the analysis. Further details on the triggers are provided in Refs.~\cite{ATLASTrigger1,ATLASTrigger2}.}  
\label{tab:trigger}
\end{table}

\subsection{Overlap Removal}\label{sec:overlapRemoval}
The Overlap Removal (OR) procedure is performed with baseline objects (electrons, muons and jets) to avoid the double counting of analysis baseline objects. This configuration is described from the following steps:
\begin{itemize}
    \item jet candidates within $\Delta R =\sqrt{\Delta y^2+\Delta\phi^2}$ = 0.2  of an electron candidate, or jets with fewer than three tracks that lie within $\Delta R = 0.4$ of a muon candidate are removed, as they mostly originate from calorimeter energy deposits from electron shower or muon bremsstrahlung;
    \item electrons and muons within $\Delta R' = \min(0.4, 0.04 + 10/p_{\mathrm{T}})$ of the remaining jets are discarded, to reject leptons from the decay of $b$- or $c$-hadrons;
    \item calo-tagged muon candidates sharing an ID track with an electron are removed. Electrons sharing an ID track with remaining muons are removed.
\end{itemize}

\subsection{Correction to the MC event weight}
To account for differences between data and simulation, MC event weights are corrected by using several scale factors. These factors are related to lepton identification and reconstruction efficiency, to the efficiency of the lepton isolation requirements, to $JVT$ cut, to jet $b$-tagging efficiencies and trigger efficiencies. Additionally, a pile-up weight is applied to the MC samples, so that their distribution of the average number of interactions per bunch crossing reproduces the observed distribution in the data.

\section{The Fake and Non Prompt background}
One of the backgrounds that have to be estimated is the one coming from Fake and Non Prompt (FNP) leptons. These come from QCD or conversion processes and their contribution is estimated through the \textit{Matrix Method} (MM)~\cite{TOPQ-2010-01}. 
The MM uses two sets of lepton identification criteria: i) the standard set used in the analysis (called \textit{tight}) and ii) a looser selection (called \textit{loose}) achieved by relaxing or removing some of the requirements in the signal definitions. The definition of the loose and tight leptons used in the MM follow exactly the \textit{Baseline} and \textit{Signal} selections listed in Table \ref{tab:eledef} and \ref{tab:muondef} for electrons and muons, respectively.

\begin{figure}[!htb]
\centering                     
\includegraphics[width=0.55\textwidth]{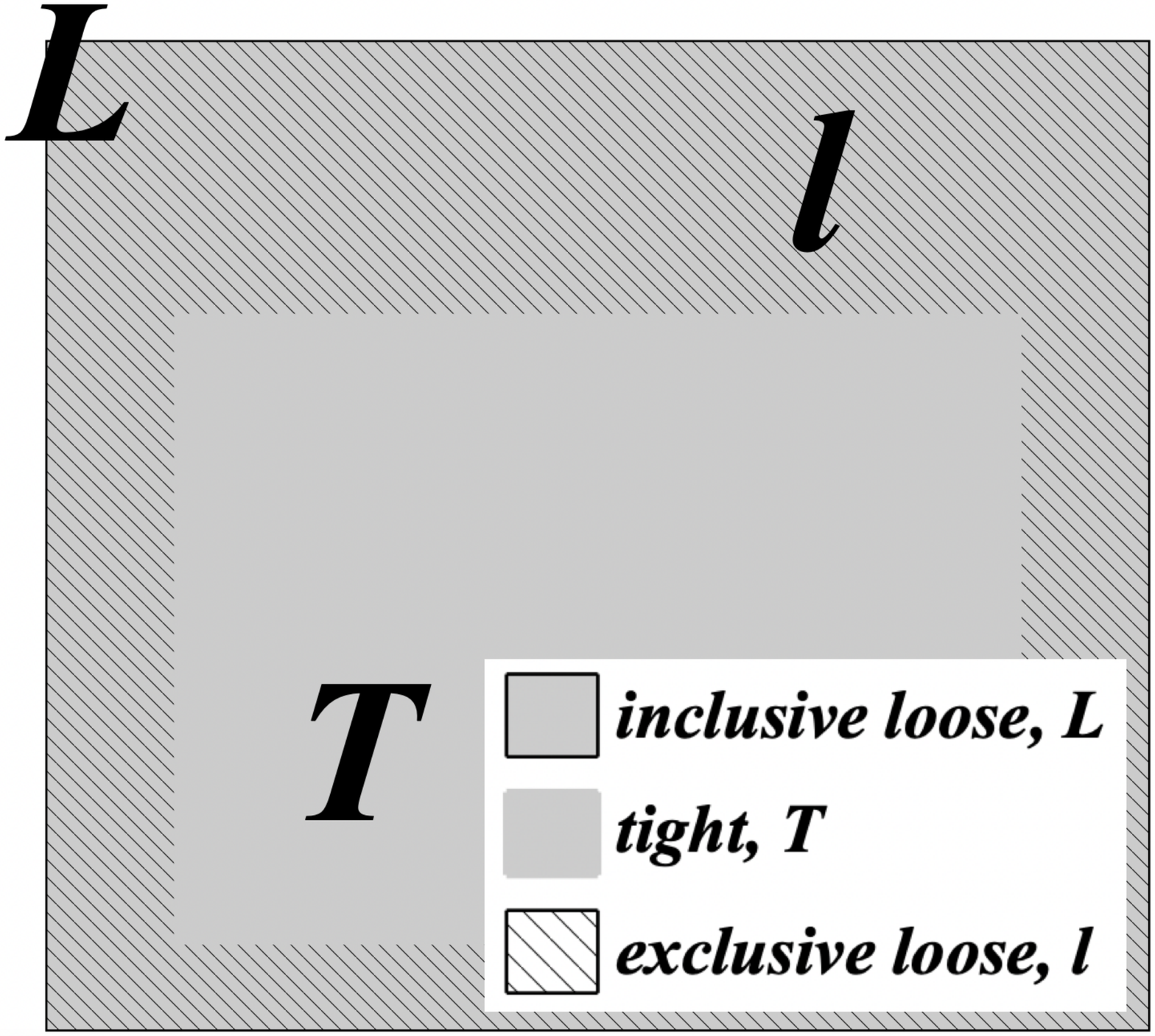}
\caption[The classification of leptons used in the MM]{A schematic view of the lepton categorisation used in the MM. \textit{L} denotes all leptons that pass the loose criteria, \textit{T} all leptons passing both loose and tight and \textit{l} leptons passing loose but not tight.}
\label{fig:T_l_L}
\end{figure}

Let \textit{T} denote the leptons passing the tight identification criteria and \textit{L} leptons that at least pass the loose criteria (\textit{inclusive loose}). Leptons passing loose but not tight are called \textit{exclusive loose} and are denoted by denoted \textit{l}. A schematic diagram of these three sets is shown in Fig.~\ref{fig:T_l_L}.

All the observed events containing two inclusive loose leptons are counted, ordered in $p_{\mathrm{T}}$, and classified into four different categories: $N_{TT}$, $N_{Tl}$, $N_{lT}$ and $N_{ll}$, where the first letter in the subscript indicates the lepton with the highest $p_{\mathrm{T}}$. 
In addition two probabilities, $r$ and $f$, are defined. The real efficiency, $r$, is the probability that a real prompt lepton passing the loose identification criteria also passes the tight. The fake rate, $f$, is the corresponding probability for an FNP lepton that passes loose to also pass tight. 
If the event contains two leptons of opposite flavour ($e\mu$ or $\mu e$) four probabilities are needed since the electron and muon probabilities are in general different. 
Using $p_{\mathrm{T}}$ and/or $\eta$ dependent efficiencies demands also four different efficiencies since the two leptons in an event often have different values of $p_{\mathrm{T}}$ and $\eta$ and thus different probabilities. 
The probabilities used in the MM are therefore $r_1$, $r_2$, $f_1$ and $f_2$ where the subscripts, $1$ and $2$, indicate the hardest and second hardest lepton in $p_{\mathrm{T}}$, respectively. 
With these probabilities and the number of events in each of the categories, $N_{TT}$, $N_{Tl}$, $N_{lT}$ and $N_{ll}$, the estimated number of events with two real ($N_{LL}^{RR}$), one real and one fake ($N_{LL}^{RF}$ and $N_{LL}^{FR}$) and two fake ($N_{LL}^{FF}$) leptons is achieved by inverting the following matrix
\begin{equation}
\small
\begin{bmatrix} N_{TT}\\ N_{Tl}\\ N_{lT}\\ N_{ll}\end{bmatrix} = \begin{bmatrix} r_1r_2& r_1f_2& f_1r_2& f_1f_2\\ r_1(1-r_2)& r_1(1-f_2)& f_1(1-r_2)& f_1(1-f_2)\\(1-r_1)r_2& (1-r_1)f_2& (1-f_1)r_2& (1-f_1)f_2\\(1-r_1)(1-r_2)& (1-r_1)(1-f_2)& (1-f_1)(1-r_2)& (1-f_1)(1-f_2)\end{bmatrix}
\begin{bmatrix} N^{RR}_{LL}\\ N^{RF}_{LL}\\ N^{FR}_{LL}\\ N^{FF}_{LL}\end{bmatrix},
\label{eq:mm_matrix}
\end{equation}
obtaining
\begin{equation}
\small
\begin{split}
N_{LL}^{RR} &= \phantom{-}(1-f_1)(1-f_2)N_{TT} - \left[f_2(1-f_1)\right]N_{Tl} - \left[f_1(1-f_2)\right]N_{lT} + f_1f_2N_{ll}\\
N_{LL}^{RF} &= -(1-f_1)(1-r_2)N_{TT} + \left[r_2(1-f_1)\right]N_{Tl} + \left[f_1(1-r_2)\right]N_{lT} + f_1r_2N_{ll}\\
N_{LL}^{RF} &= -(1-f_2)(1-r_1)N_{TT} + \left[f_2(1-r_1)\right]N_{Tl} + \left[r_1(1-f_2)\right]N_{lT} + f_2r_1N_{ll}\\
N_{LL}^{FF} &= \phantom{-}(1-r_1)(1-r_2)N_{TT} - \left[r_2(1-r_1)\right]N_{Tl} - \left[r_1(1-r_2)\right]N_{lT} + r_1r_2N_{ll}.
\end{split}
\label{eq:NLL}
\end{equation}

These are the expected number of events with two, one and zero prompt real leptons in a sample of two inclusive loose leptons (thus the $LL$ subscript). 
Since we are interested in regions with two tight leptons in the final state the above estimates need to be translated into the corresponding yields for a sample containing two tight leptons by multiplying Eq.~(\ref{eq:NLL}) by the appropriate probabilities
\begin{align}
N_{TT}^{RR} &= r_1r_2N_{LL}^{RR}\label{eq:NTTRR}\\
N_{TT}^{RF} &= r_1f_2N_{LL}^{RF}\label{eq:NTTRF}\\
N_{TT}^{FR} &= f_1r_2N_{LL}^{FR}\label{eq:NTTFR}\\
N_{TT}^{FF} &= f_1f_2N_{LL}^{FF}\label{eq:NTTFF}.
\end{align}

\subsection{Real and fake efficiencies}
The efficiencies, $r$, and fake rates, $f$, are in general measured in data using dedicated CRs, as will be discussed in detail in Section~\ref{sec:FNP_r_and_f}. Once these are defined, events with exactly two inclusive loose leptons are used to calculate the real efficiencies and fake rates. First, the hardest lepton is tagged while the other lepton acts as a probe and it is checked whether it passes tight or not. The procedure is then repeated on the same event, but now tagging the softest lepton and probing the hardest one whether it is tight. To find the efficiencies/rates as a function of some kinematic variable, like $\eta$ or $p_{\mathrm{T}}$, one histogram for the numerator and one for the denominator is filled using $\eta$ or $p_{\mathrm{T}}$ of the probe lepton. Then, to get the final efficiency/rate, the two histograms are divided by each other, bin-by-bin. \\
The FNP leptons originate in general from different sources and thus the final fake rates used in the MM is a linear combination of the different sources given by Eq.~(\ref{eq:fake_total_comb})
\begin{align}\label{eq:fake_total_comb}
f_{\mathrm{total}}(p_{\mathrm{T}}) = \sum_{i}f_i(p_{\mathrm{T}})w_i(p_{\mathrm{T}}),
\end{align}
where $i$ runs over the different FNP sources and $f_i$ is the corresponding fake rate for that source.

\subsection{Lepton Classification}
In ATLAS, leptons are classified into different classes by the \textit{IFFTruthClassifier} algorithm. The possible sources leptons come from are:
\begin{itemize}
    \item \texttt{Unknown}, leptons that cannot be attributed to any of the classes listed below;
    \item \texttt{KnownUnknown}, leptons which, in principle, could be classified, but the tool fails with the classification due to missing information;
    \item \texttt{IsoElectron}, electrons classified as prompt or isolated electrons;
    \item \texttt{ChargeFlipIsoElectron}, electrons with correctly-assigned and mis-identified charge (charge-flip) identified from the charge of the mother-particle being different from the reconstructed charge;
    \item \texttt{PromptMuon}, muons classified as prompt or isolated 	muons;
    \item \texttt{PromptPhotonConversion}, electrons originating from the conversion of prompt photons or an electromagnetic process;
    \item \texttt{ElectronFromMuon}, electrons classified as muons but reconstructed as electrons;
    \item \texttt{TauDecay}, non-isolated electrons and muons from hadronic tau-decays;
    \item \texttt{BHadronDecay}, electrons and muons originating from heavy-flavor decays can of $b$-hadrons;
    \item \texttt{CHadronDecay}, electrons and muons originating from heavy-flavor decays can of $c$-hadrons;
    \item \texttt{LightFlavorDecay}, leptons produced by light-flavor jets.
\end{itemize}
The MM estimates the contribution from leptons believed to come from \texttt{PromptPhotonConversion}, \texttt{BHadronDecay}, \texttt{CHadronDecay} and \texttt{LightFlavorDecay}. The remaining sources of FNP leptons are taken from MC.
\subsection{CRs for efficiency measurements}
\label{sec:FNP_r_and_f}
Table~\ref{tab:CR_def} shows the CRs used to extract the real efficiencies and fake rates for the FNP estimates.

\begin{table}[!htb]
\def\arraystretch{1.1}
\resizebox{\textwidth}{!}{
\begin{tabular}{l|c|p{1.3cm}|p{2.5cm}|c|c}
\noalign{\smallskip}\hline\noalign{\smallskip}
Variable & $\mathrm{CR^{REAL}}$ & \multicolumn{2}{c}{$\mathrm{CR_{HF}^{FAKE}}$} & $\mathrm{CR_{CO}^{FAKE}}$ & $\mathrm{CR_{LF}^{FAKE}}$\\
\noalign{\smallskip}\hline\noalign{\smallskip}
Data/MC & MC & \multicolumn{3}{c}{Data} & MC \\
\noalign{\smallskip}\hline\noalign{\smallskip}
$p_{\mathrm{T}}^{\ell}$ [GeV] & \multicolumn{5}{c}{$p_{\mathrm{T}}^{\ell_{1}}>27$, $p_{\mathrm{T}}^{\ell_{2}}>9$}\\
$n_{\mathrm{leptons}}$ & $2$ & \multicolumn{2}{c|}{$2$} & $3$ & $2$ \\
Type & ($e^{\pm}e^{\mp}$) or ($\mu^{\pm}\mu^{\mp}$) & \multicolumn{1}{c|}{$\mu$($e$)} & \multicolumn{1}{c|}{($\mu\mu$)} & $\mu^{+}\mu^{-}$($e^{\pm}$) & ($e^{\pm}e^{\pm}$) \\
$|m_{\ell\ell} - m_{Z}|$ [GeV] & - & \multicolumn{1}{c|}{-}  & \multicolumn{1}{c|}{$>10$} & $<10$ & - \\
$E_{\mathrm{T}}^{\text{miss}}$ [GeV] & Significance $>3$ & \multicolumn{2}{c|}{$<50$} & $<50$ & $<40$ \\
\ensuremath{M_{\text{T}}(\text{tag},\, E_{\mathrm{T}}^{\text{miss}})} [GeV] & - & \multicolumn{2}{c|}{$<50$} & - & -\\
$n_{b\mathrm{-jets}}$ & $\leq 1$ & \multicolumn{2}{c|}{$1$} & $0$ & - \\ 
$\Delta R(\text{lep},\,b-$jet$)$ & passOR & \multicolumn{2}{c|}{tag: $<0.3$} & passOR & passOR\\
$\Delta R(\text{lep},\,$jet$)$   & passOR & \multicolumn{2}{c|}{probe: $>0.4$} & passOR & passOR\\
\noalign{\smallskip}\hline\noalign{\smallskip}
\end{tabular}}
\caption{The cuts used to define the CRs for extracting the real efficiencies and fake rates used as input to the MM. The leptons in parenthesis indicate the ones used as probe to calculate the real efficiency/fake rate. If two leptons are in parenthesis, they are both used as probes.}
\label{tab:CR_def}
\end{table}

\begin{itemize}
    \item {$\mathrm{CR^{REAL}}$} The real efficiencies are extracted from MC after applying all the relevant scale factors with a SR-like selection. 
    \item {$\mathrm{CR_{HF}^{FAKE}}$} The fake rate for leptons originating from decays of heavy flavoured (HF) jets are estimated from a $b$ control region. The region is defined by requiring exactly one tag muon passing the baseline selection, before overlap removal is applied, within $0.3$ in $\Delta R$ of a reconstructed $b$-jet. The probe lepton (either electron or muon) is required to be separated by $0.4$ in $\Delta R$ from any jet in the event. By requiring exactly one $b$-jet (the one overlapping with the tag muon) the probe is believed to come from a second, unreconstructed, $b$-jet in the event. In order to minimize the real lepton contamination additional requirements of $E_{\mathrm{T}}^{\text{miss}} < 50$ GeV and $M_{\text{T}}(\text{tag},E_{\mathrm{T}}^{\text{miss}})<50$ GeV are added. Events from $Z$ boson decays in the $\mu\mu$ region are suppressed by applying an additional $Z$ boson veto.
    \item {$\mathrm{CR_{CO}^{FAKE}}$} The CR used to estimate the fake rate for electrons from converted photons (CO) targets events with $Z$ boson decaying into two muons where one of the muons radiates a photon which again converts into an electron/positron pair for which only one of them are reconstructed. The region require two muons of opposite sign together with an electron where the invariant mass of the three leptons is within $10$ GeV of the $Z$ boson mass. In order to reduce the prompt lepton contamination an additional $b$-jet veto and $E_{\mathrm{T}}^{\text{miss}}<50$ GeV are required.
    \item {$\mathrm{CR_{\text{LF}}^{FAKE}}$} The fake rate for electrons coming from decays of light-flavoured (LF) jets or jets reconstructed as electrons in the detector is taken from MC as it is difficult to construct a CR in data pure enough in FNP electrons. The fake rate is calculated by using an inclusive $ee$ region and selecting electrons from light flavoured sources using the truth information.
\end{itemize}

\section{Preselection}
Candidate events are firstly selected by applying a preselection, reported in Table \ref{tab:preselCuts_c1c1ww}.

\begin{table}[!htb]
\begin{center}
\begin{tabular}{l | c }
\noalign{\smallskip}\noalign{\smallskip}\noalign{\smallskip}\hline
\noalign{\smallskip}\noalign{\smallskip}\noalign{\smallskip}
Variable & Cut \\
\noalign{\smallskip}\noalign{\smallskip}\noalign{\smallskip}\hline
\noalign{\smallskip}\noalign{\smallskip}\noalign{\smallskip}
$N_\textrm{OS leptons}$ &  $= 2$ \\
$p_{\mathrm{T}}^{\ell_{1}}$ &  $> 27$ GeV \\
$p_{\mathrm{T}}^{\ell_{2}}$ &  $>  9$ GeV \\
$m_{\ell\ell}$     &  $> 11$ GeV  \\
$n_{\mathrm{jet}}$ & $=0$ \\
$E_{\mathrm{T}}^{\text{miss}}$ significance &   $> 3$  \\
$|m_{\ell\ell} - m_{Z}|$ & $ >  15~$ GeV  (for SF only)\\
\noalign{\smallskip}\noalign{\smallskip}\hline
\noalign{\smallskip}\noalign{\smallskip}
\end{tabular}
\end{center}
\caption{Preselection cuts on SF and DF events for the training of the chargino signals via $WW$ analysis.}
\label{tab:preselCuts_c1c1ww}
\end{table}

Exactly 2 Opposite Signed (OS) charged leptons (electrons and/or muons) are required. The leading and sub-leading lepton transverse momenta, $p_{\mathrm{T}}^{\ell_{1}}$ and $p_{\mathrm{T}}^{\ell_{2}}$, are required to be $>27$ GeV and $>9$ GeV, respectively. An invariant mass ($m_{\ell\ell}> 11$ GeV) cut is applied in order to remove low mass resonances ($J/\Psi,\phi',\Upsilon,...$). A veto on jets with $p_\mathrm{T}>20$ GeV and $|\eta|<2.4$ removes the bulk of the top backgrounds. The cut on $E_{\mathrm{T}}^{\text{miss}}$ significance $>3$ further reduces $Z$+jets event contamination. Then, the cut $|m_{\ell\ell} - m_{Z}| > 15$ GeV on SF leptons excludes a mass window close to the $Z$ mass.

\begin{figure}[!p]
\centering
\includegraphics[width=0.45\linewidth]{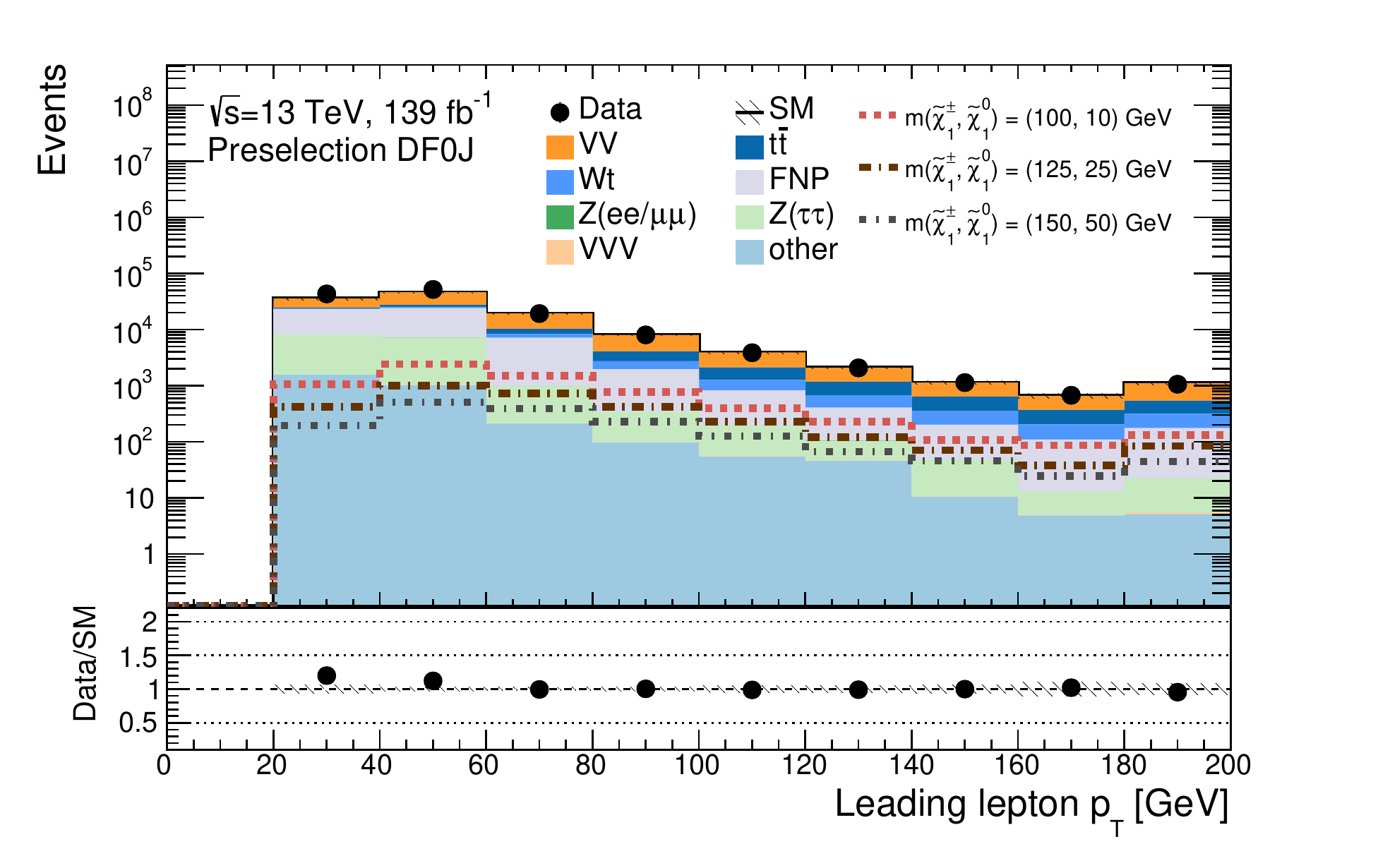}
\includegraphics[width=0.45\linewidth]{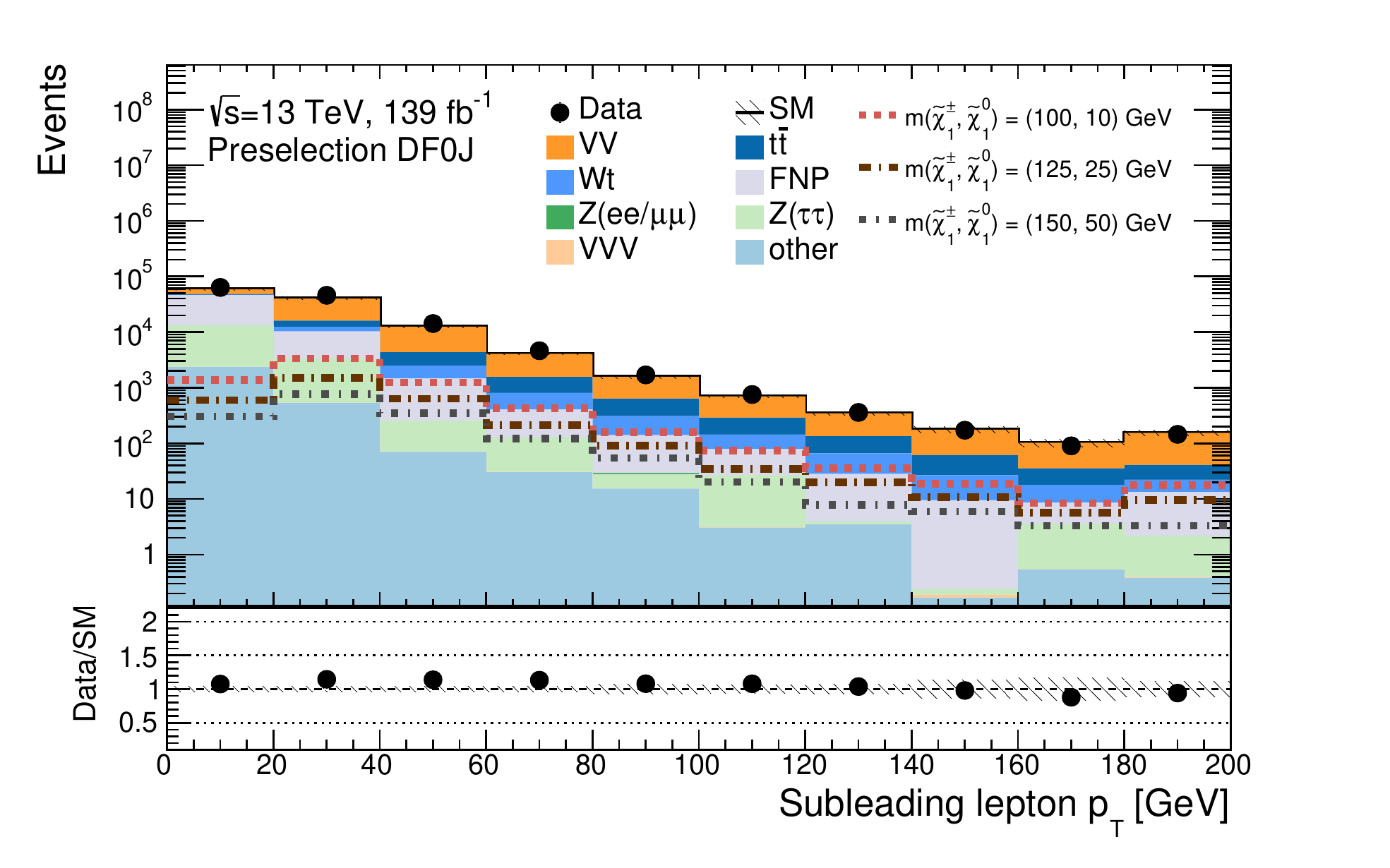}
\includegraphics[width=0.45\linewidth]{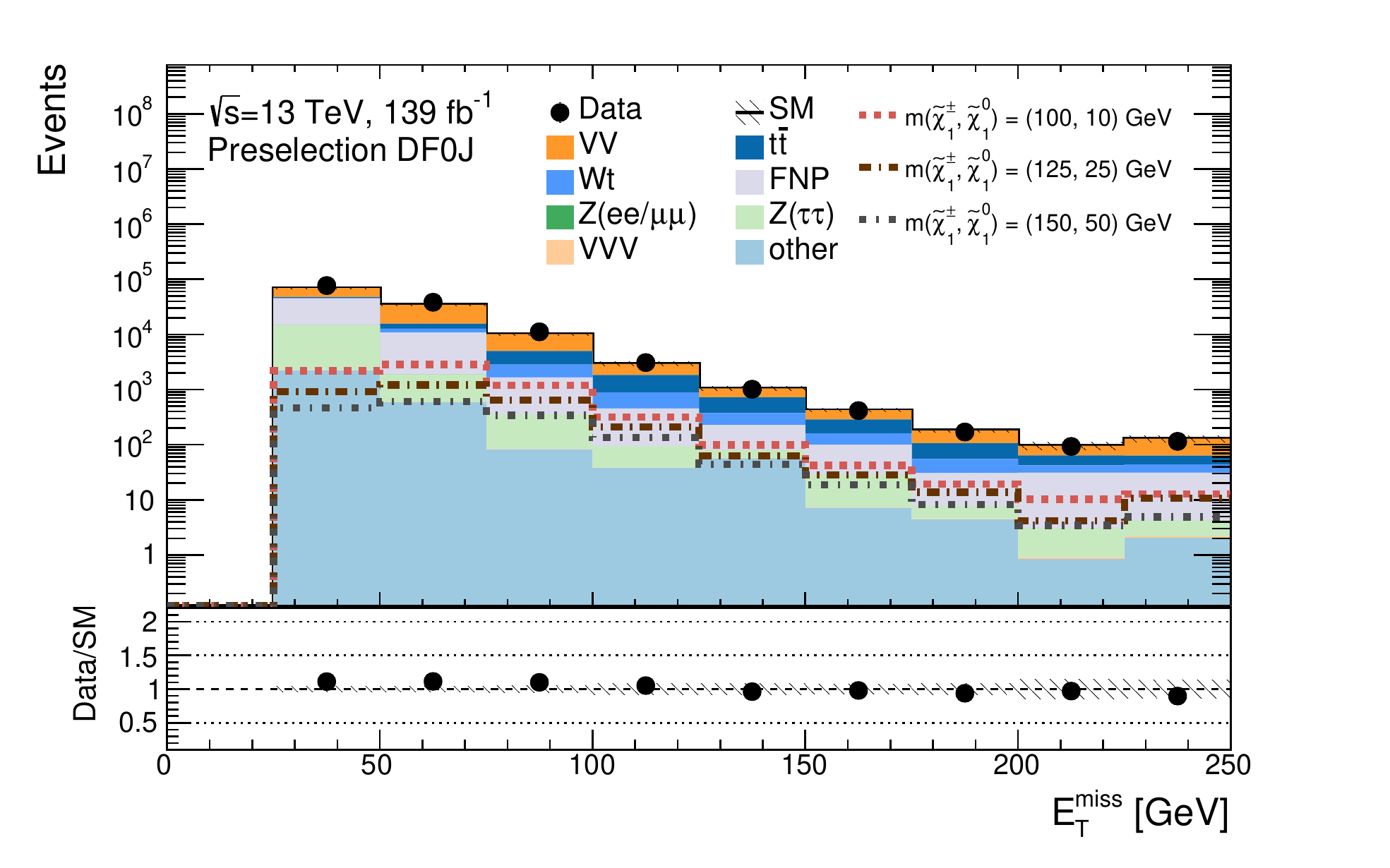}
\includegraphics[width=0.45\linewidth]{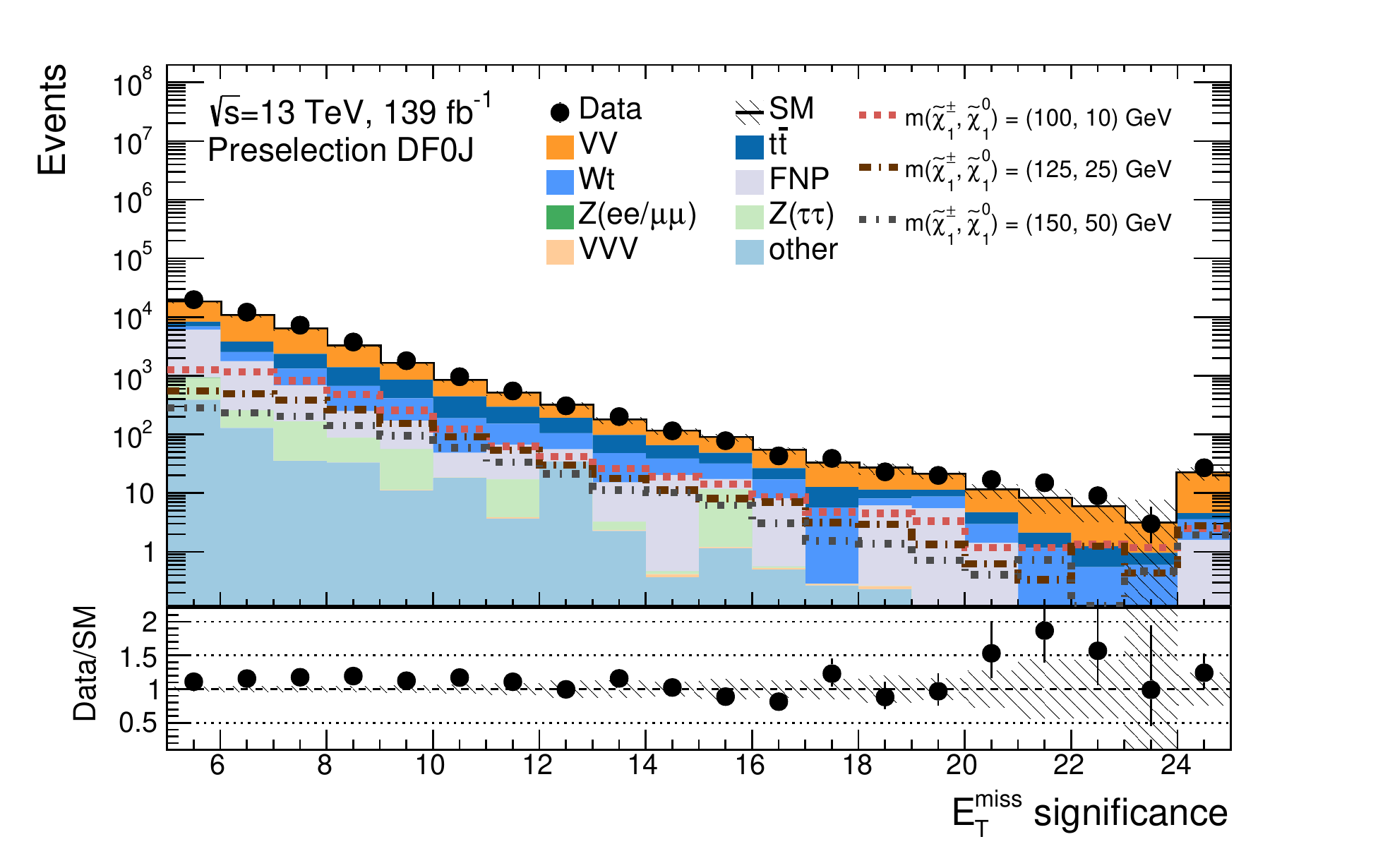}
\includegraphics[width=0.45\linewidth]{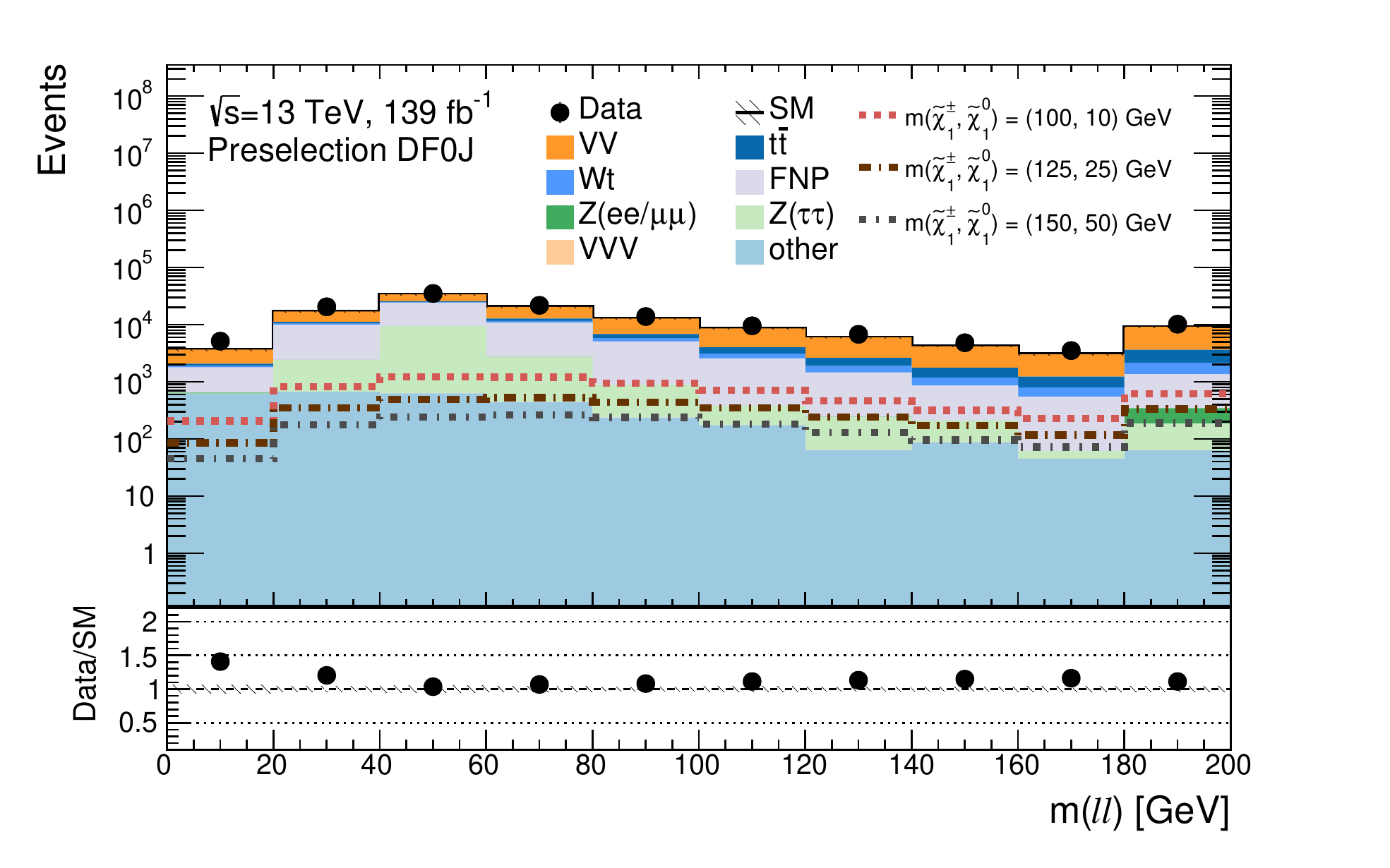}
\includegraphics[width=0.45\linewidth]{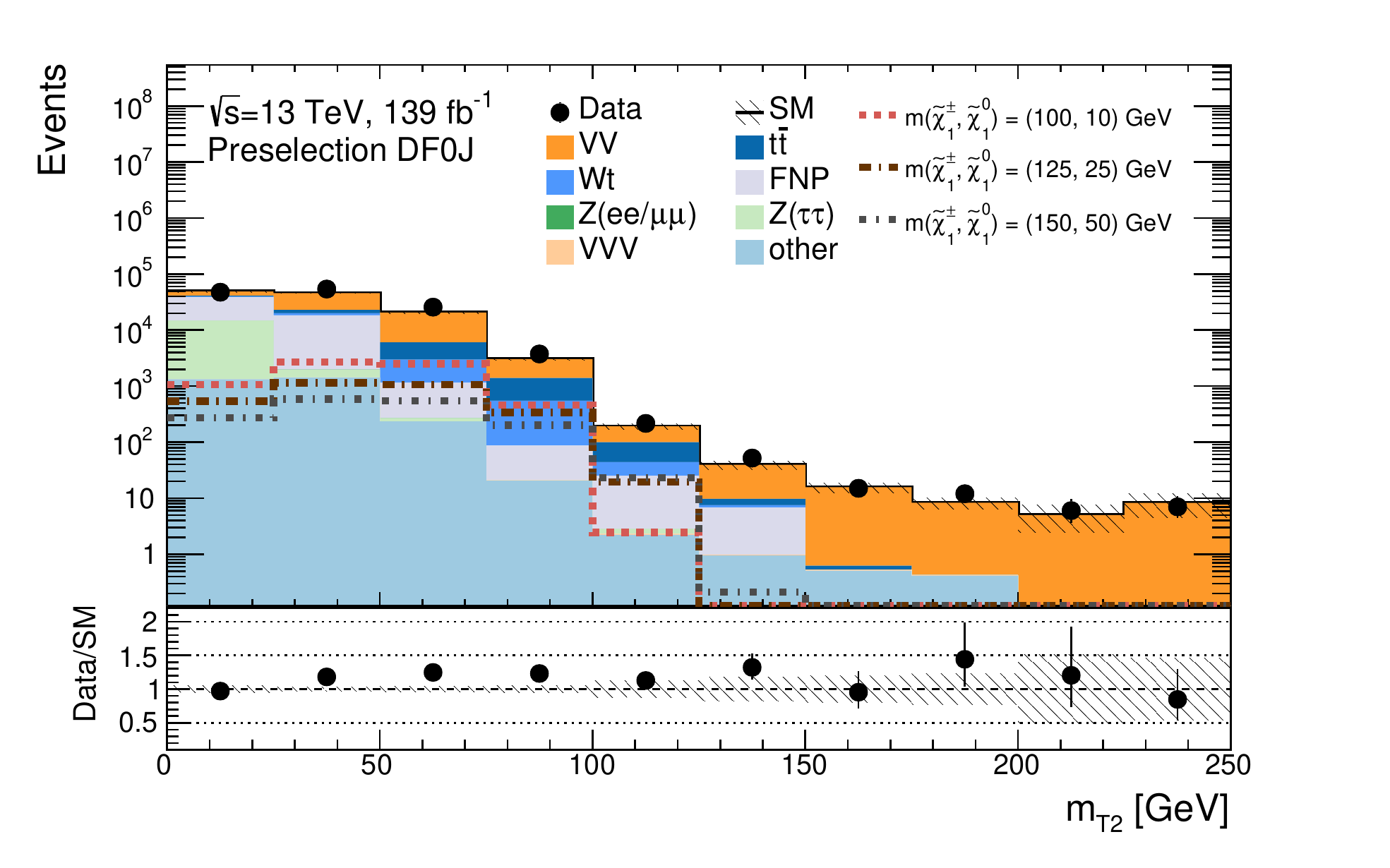}
\includegraphics[width=0.45\linewidth]{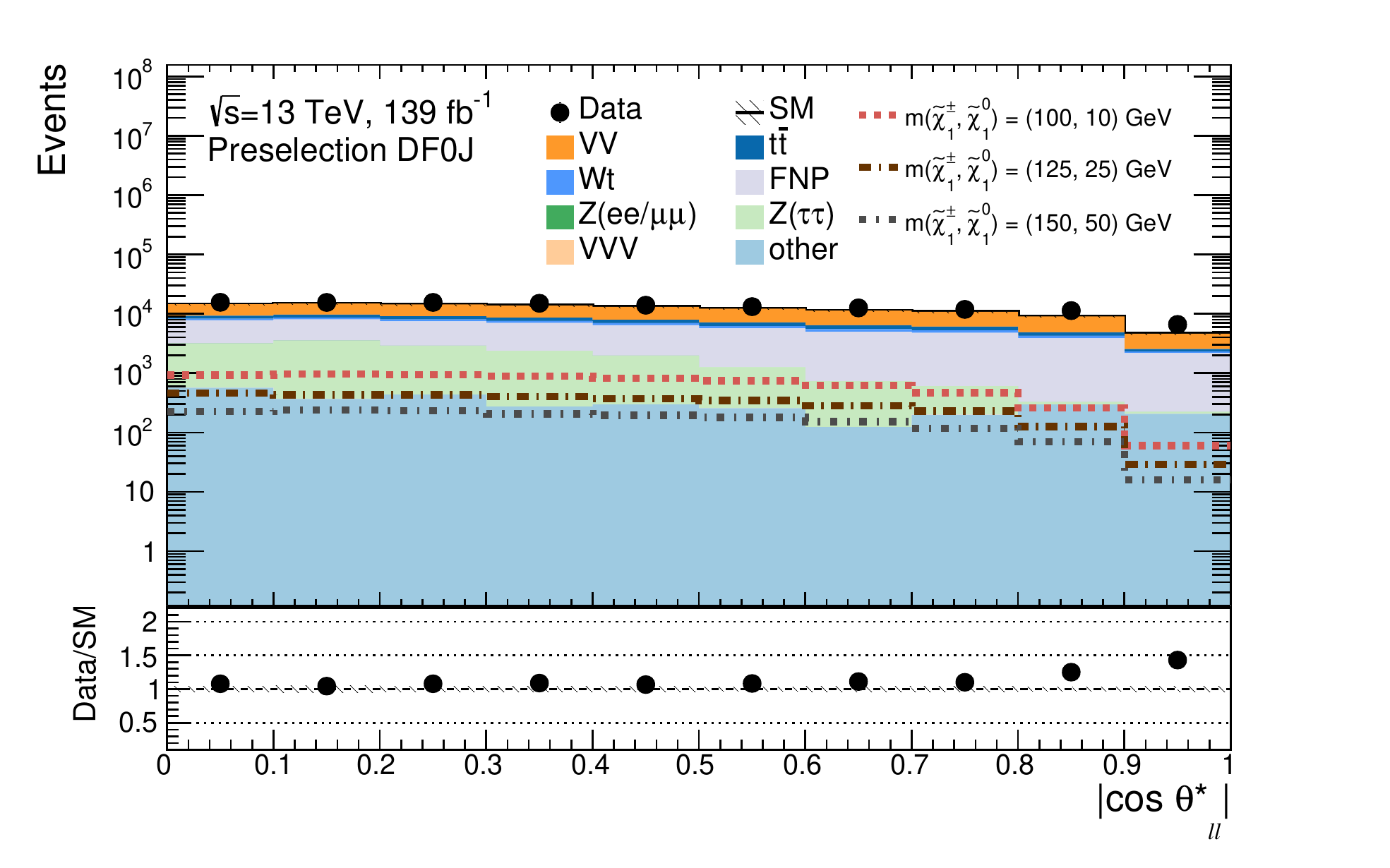}
\includegraphics[width=0.45\linewidth]{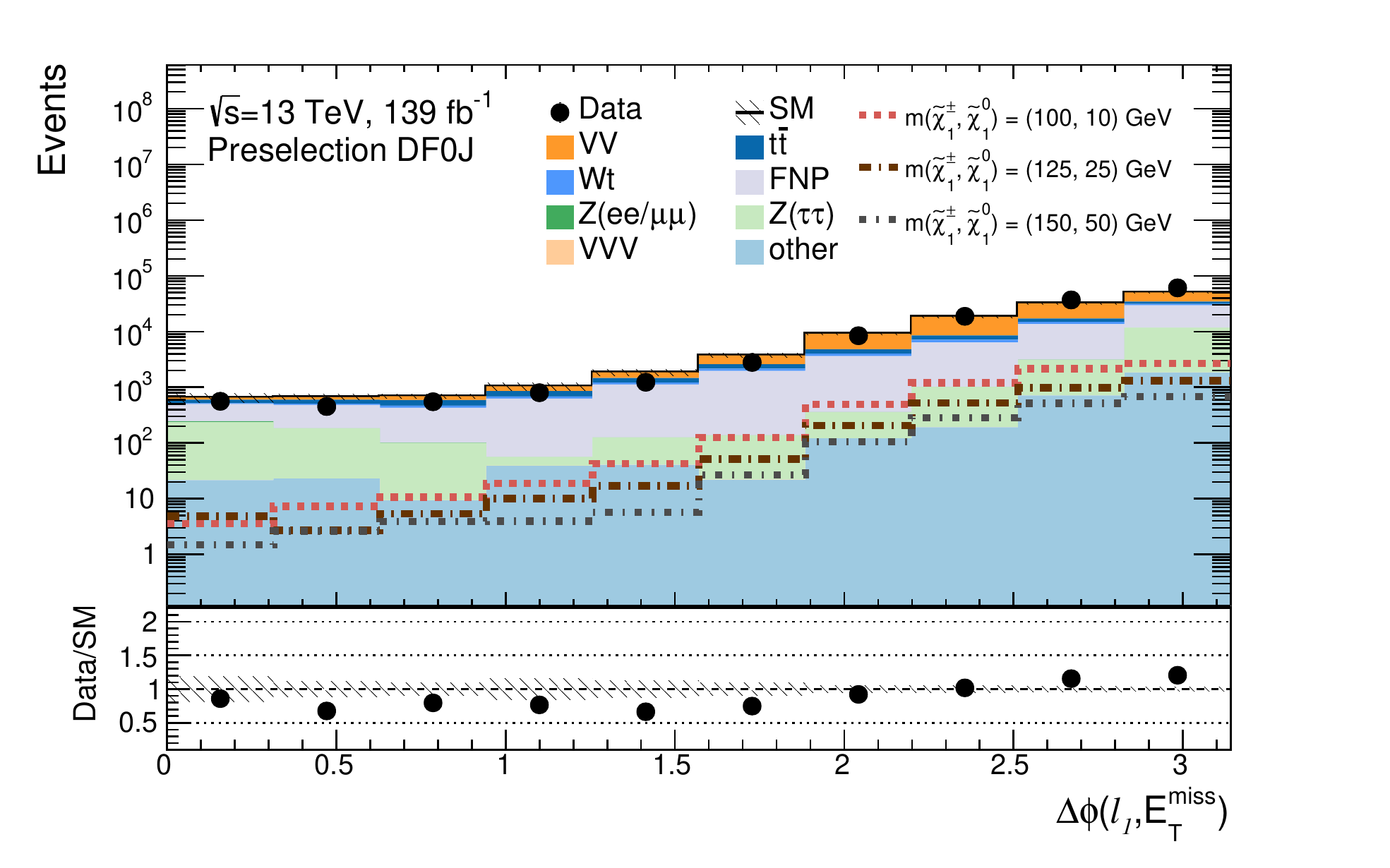}
\includegraphics[width=0.45\linewidth]{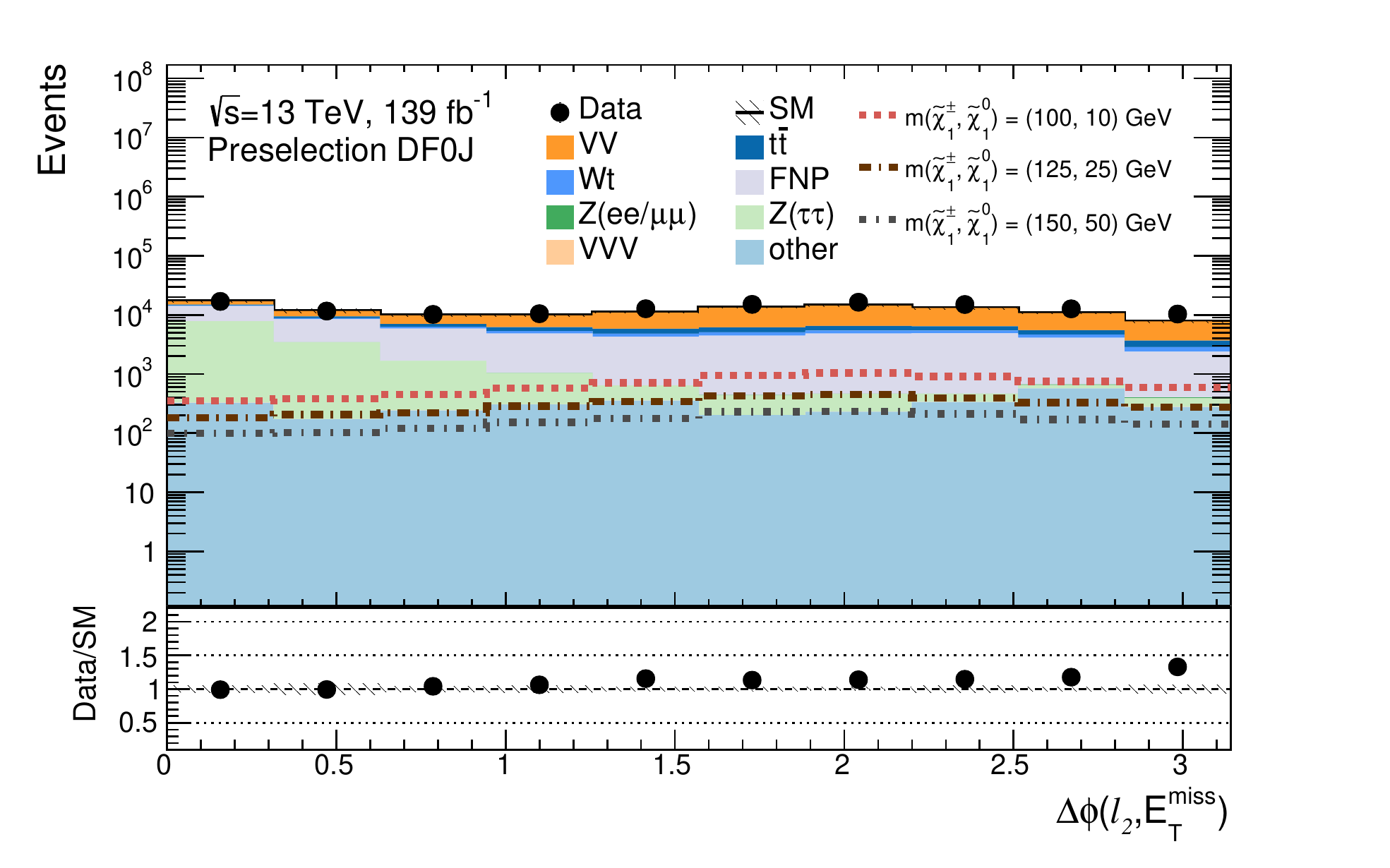}
\includegraphics[width=0.45\linewidth]{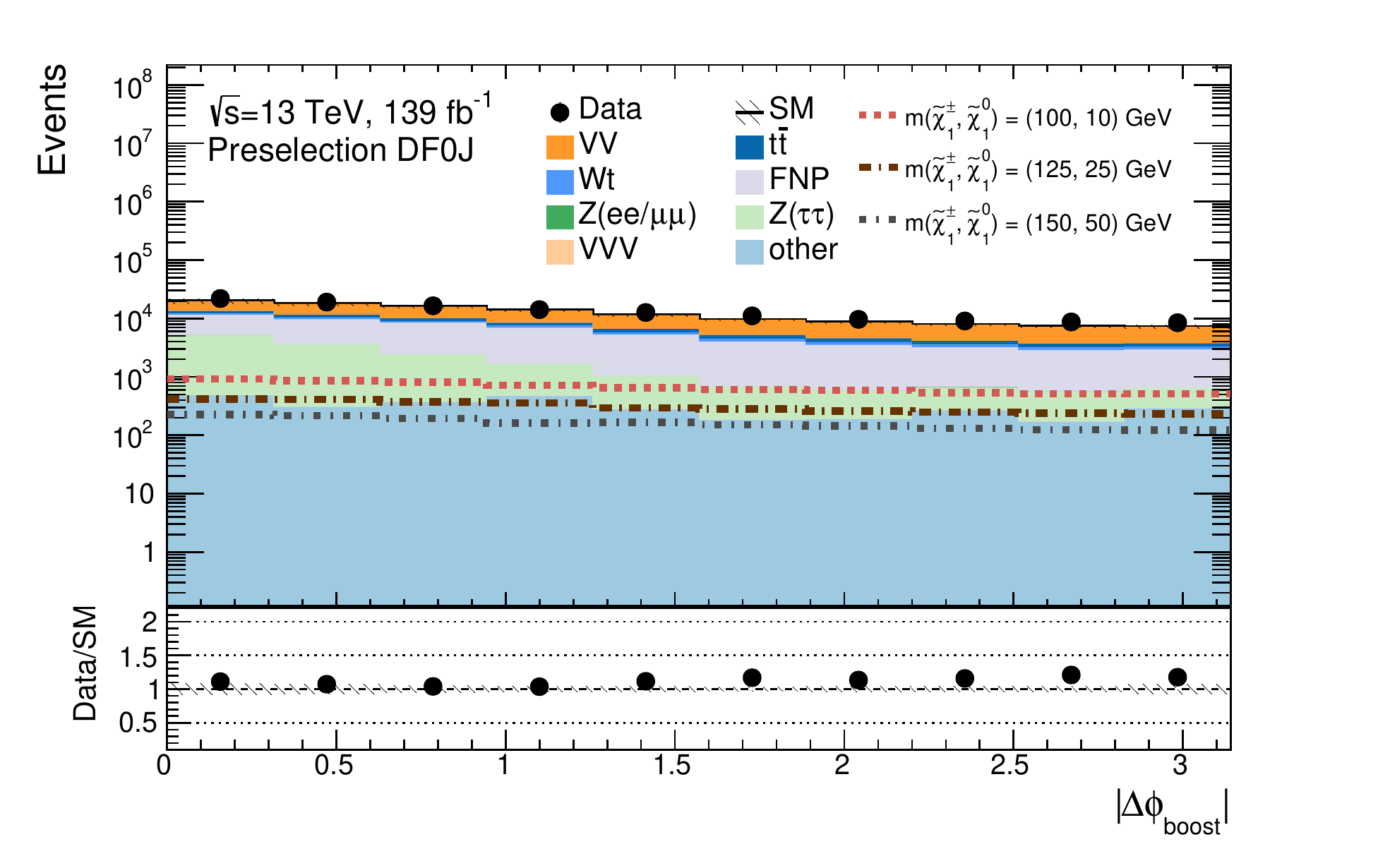}
\caption{The data and MC distributions of the variables trained over, after the preselection and DF 0-jet selections. Both statistical and systematic uncertainties are shown.}
\label{fig:Dists_presel}
\end{figure}

\begin{figure}[!p]
\centering
\includegraphics[width=0.45\linewidth]{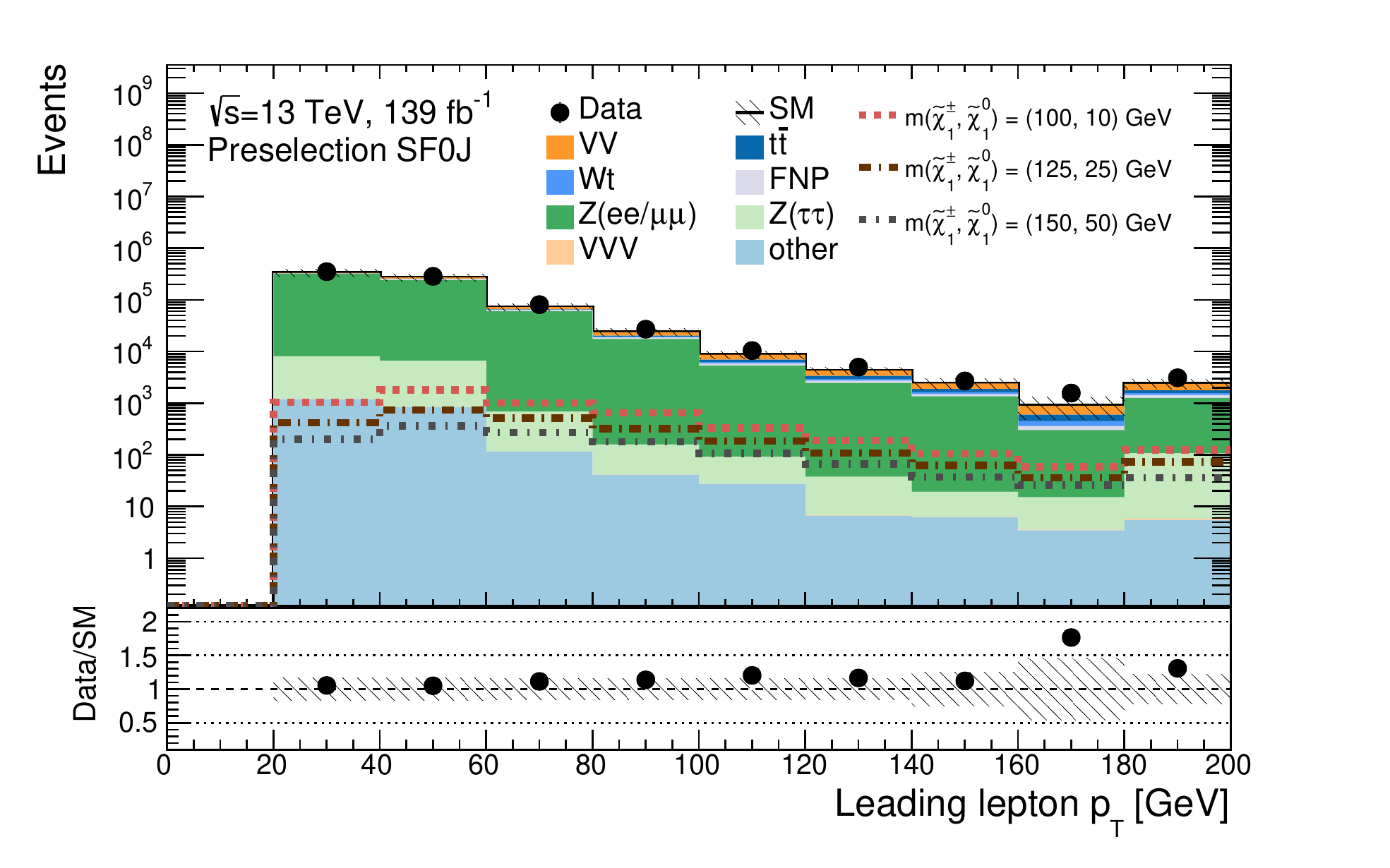}
\includegraphics[width=0.45\linewidth]{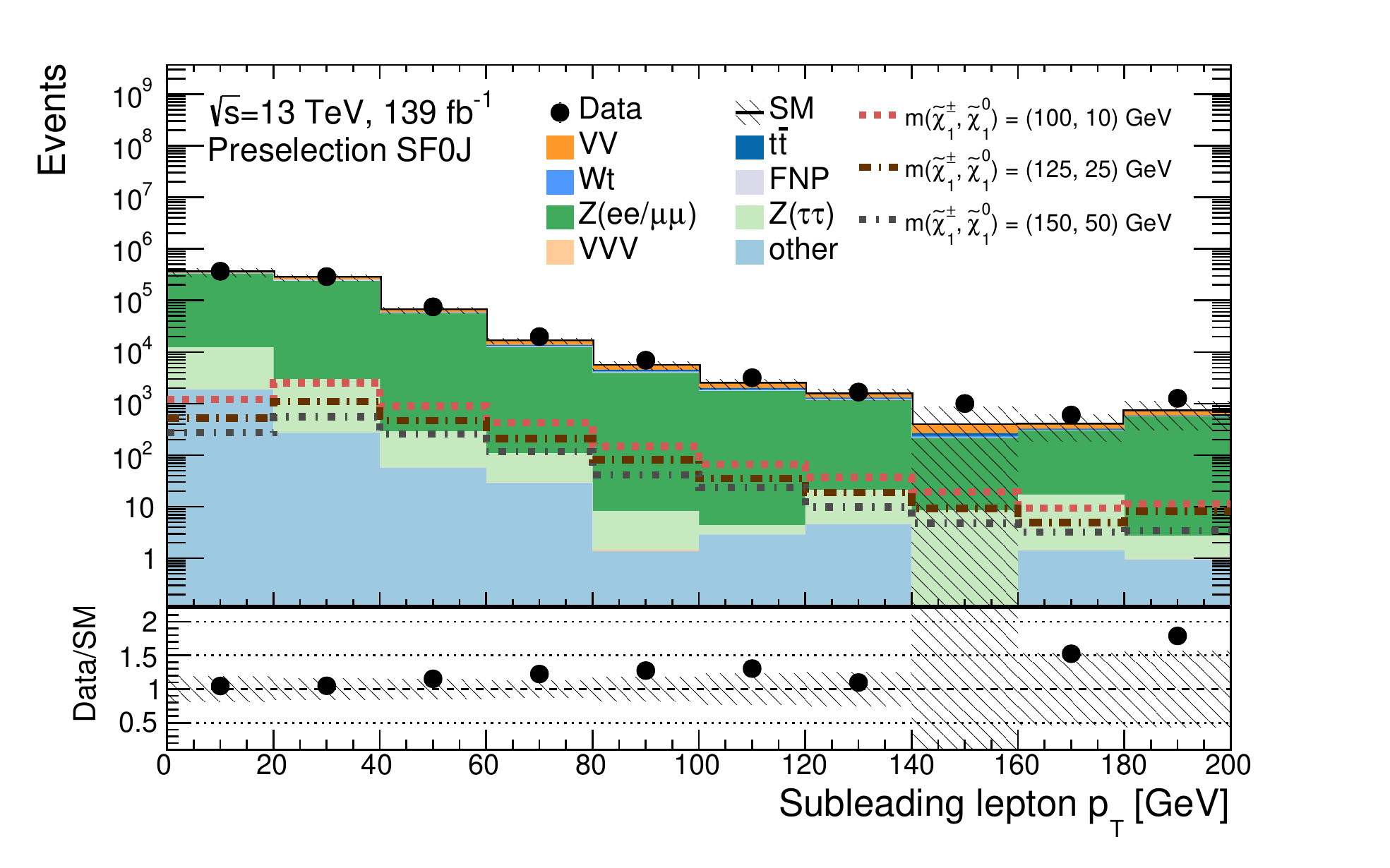}
\includegraphics[width=0.45\linewidth]{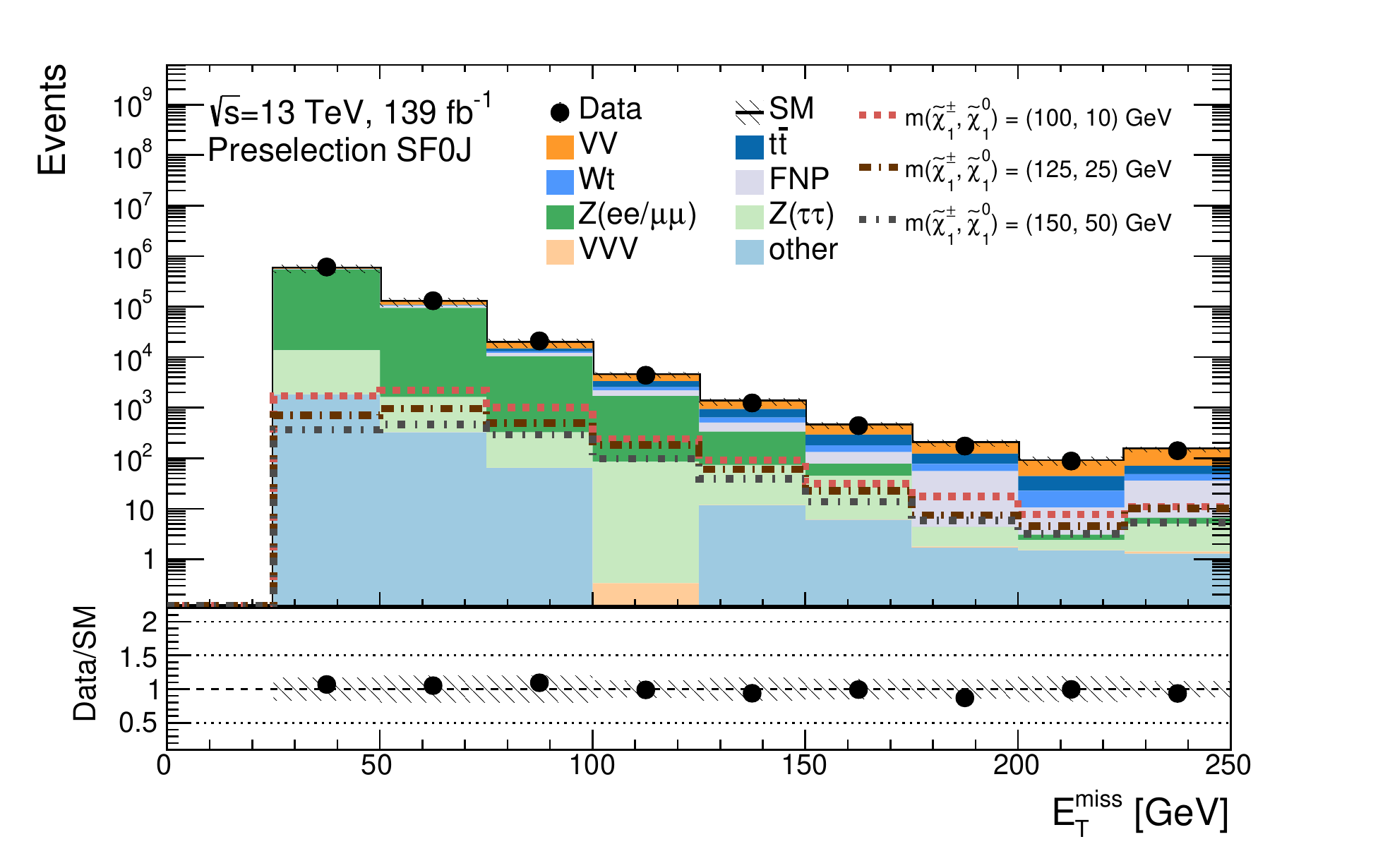}
\includegraphics[width=0.45\linewidth]{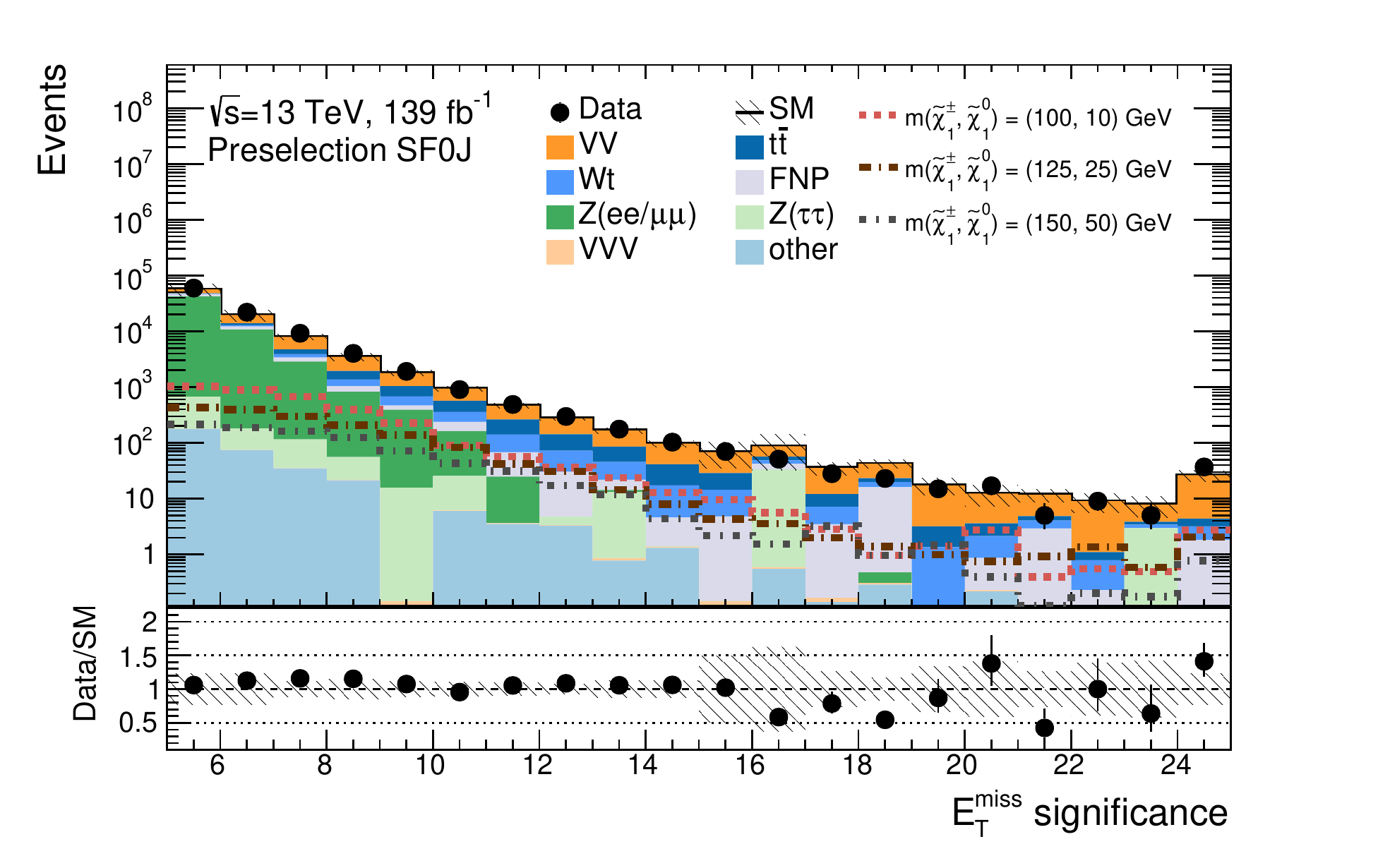}
\includegraphics[width=0.45\linewidth]{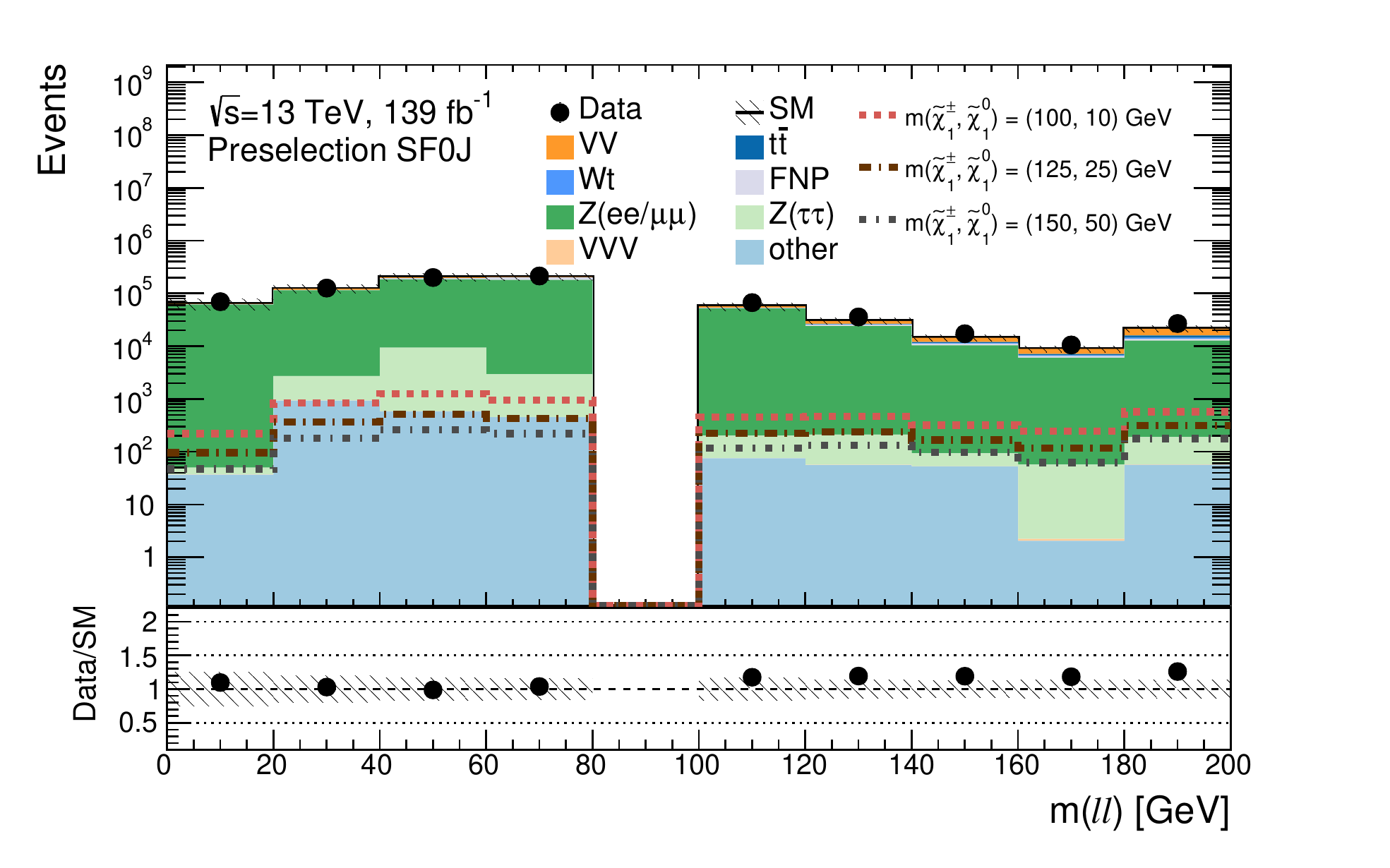}
\includegraphics[width=0.45\linewidth]{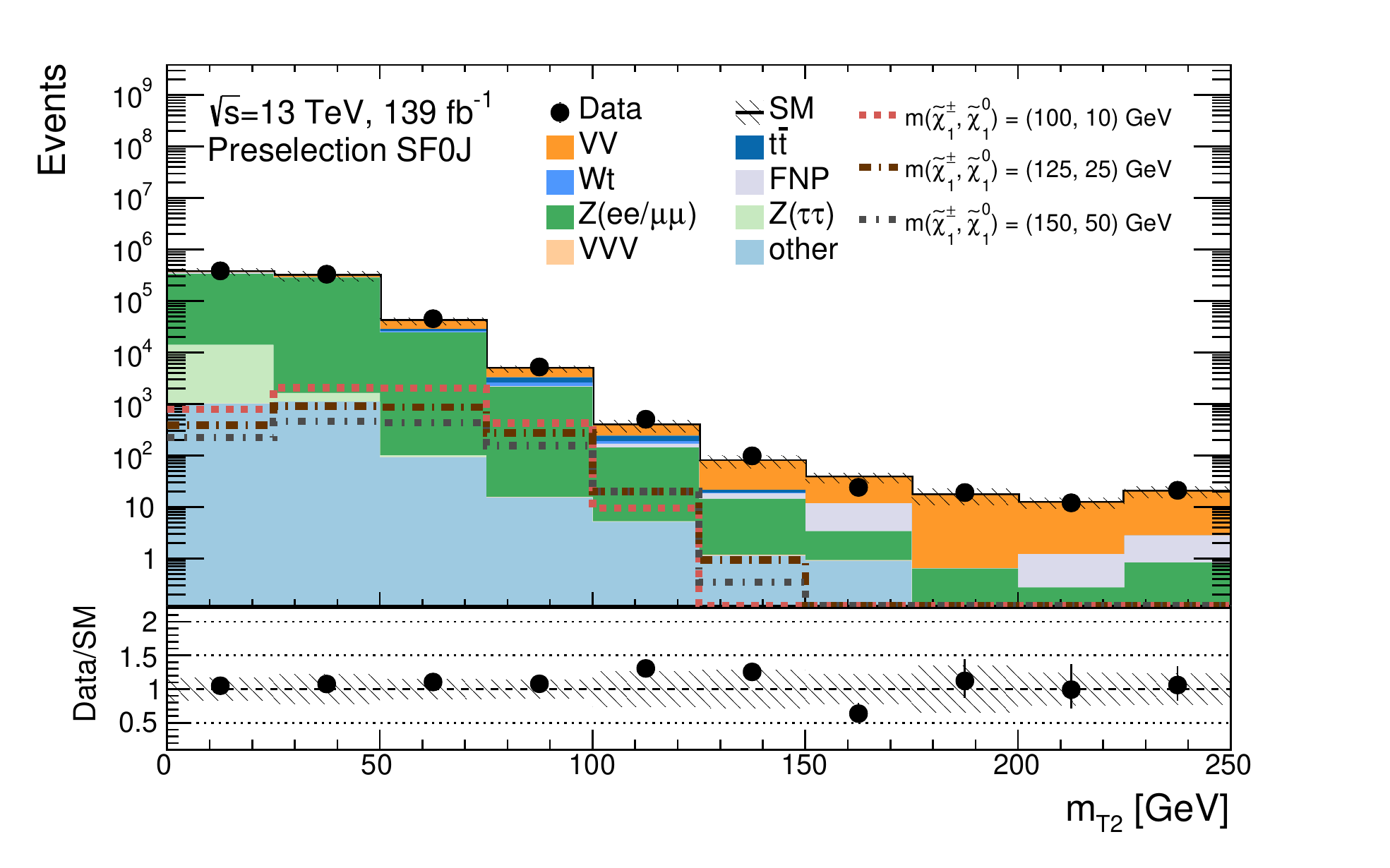}
\includegraphics[width=0.45\linewidth]{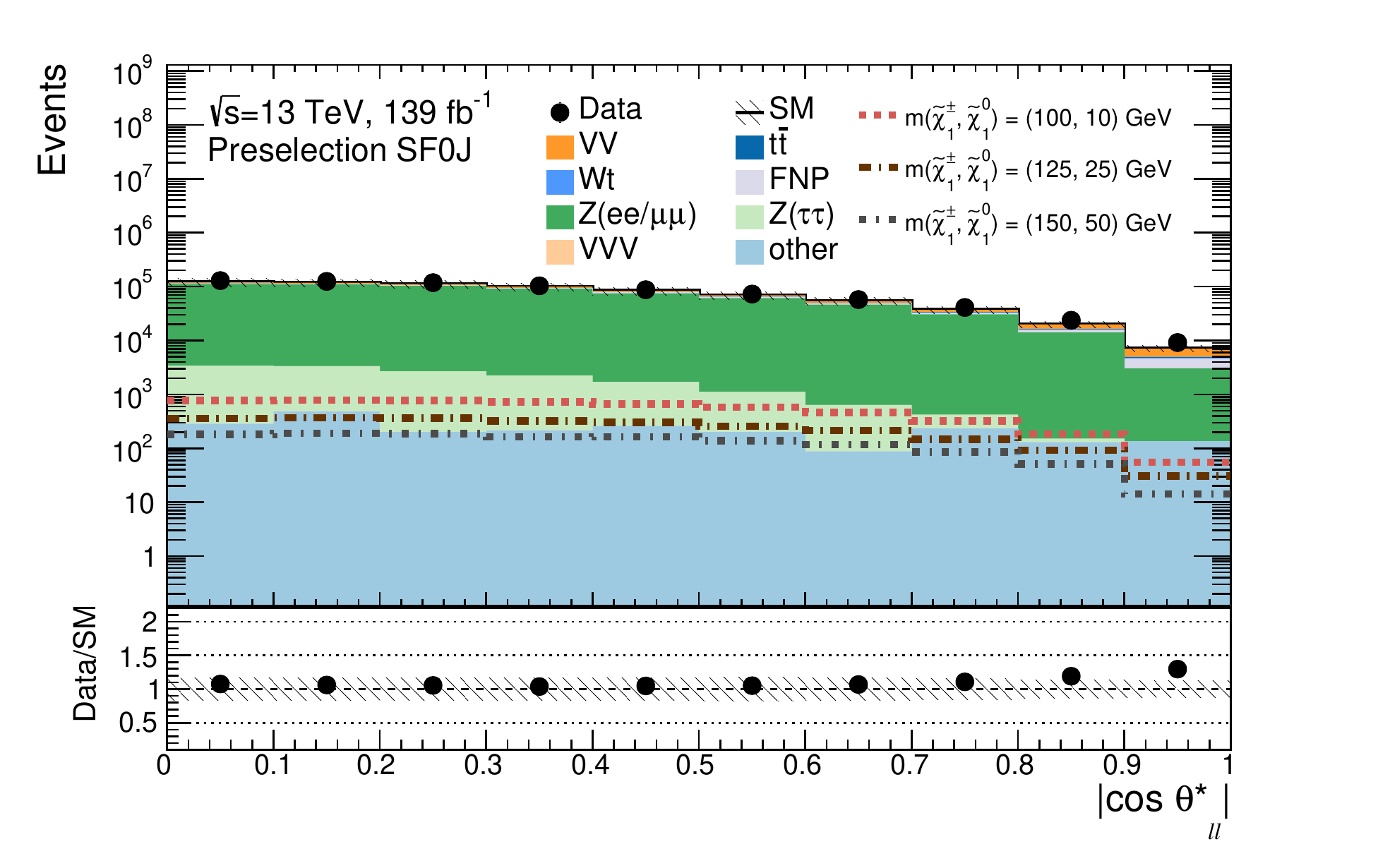}
\includegraphics[width=0.45\linewidth]{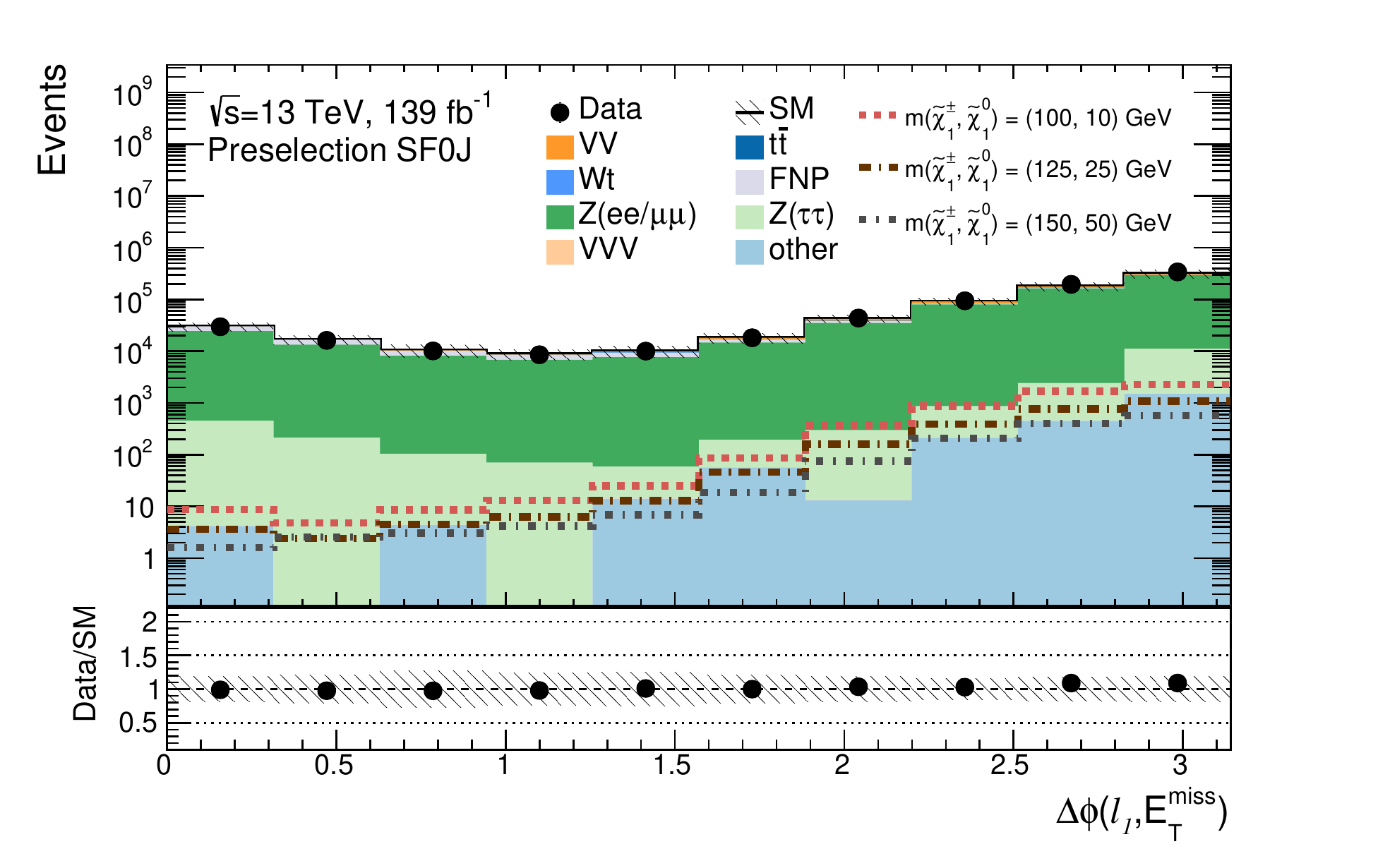}
\includegraphics[width=0.45\linewidth]{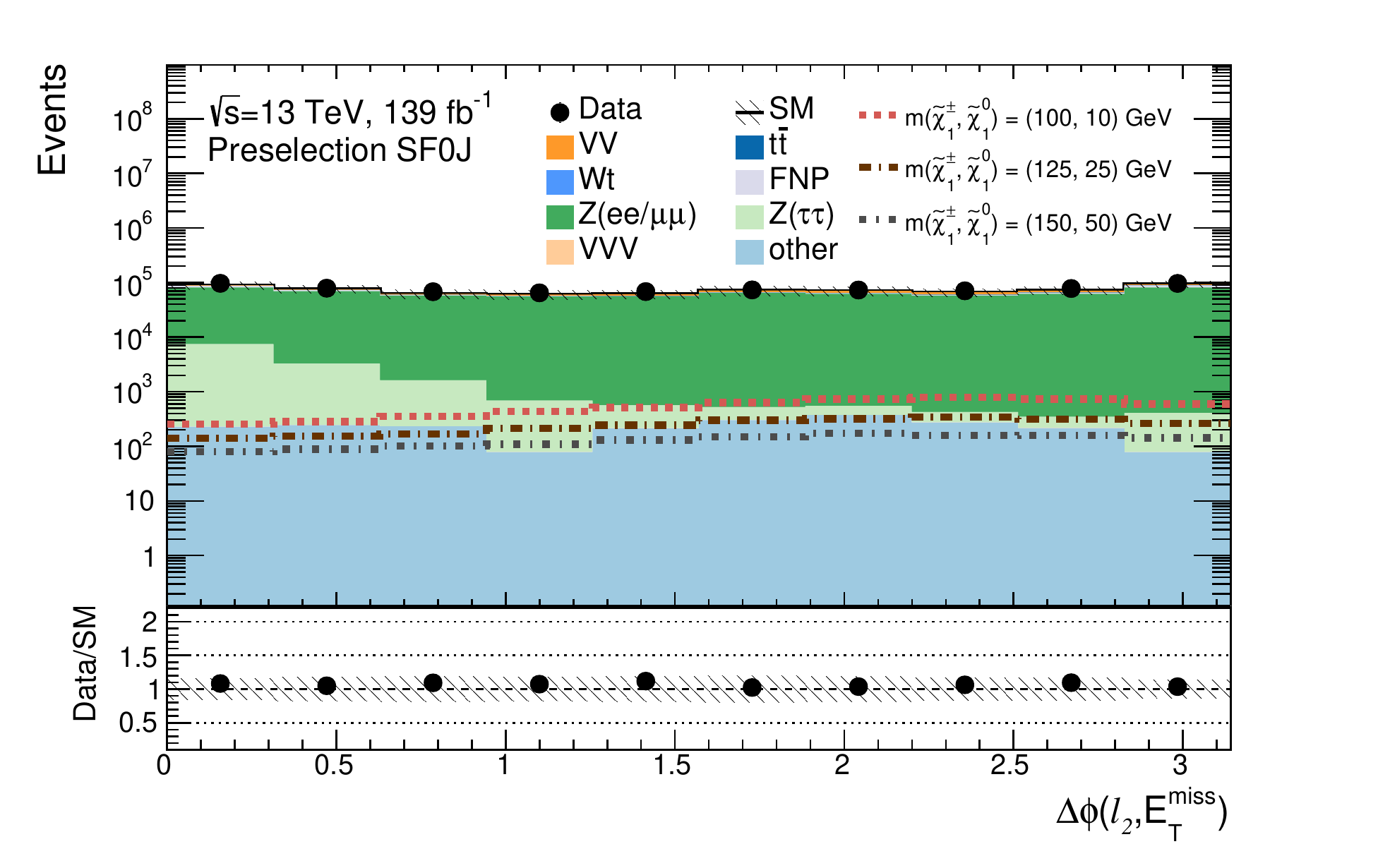}
\includegraphics[width=0.45\linewidth]{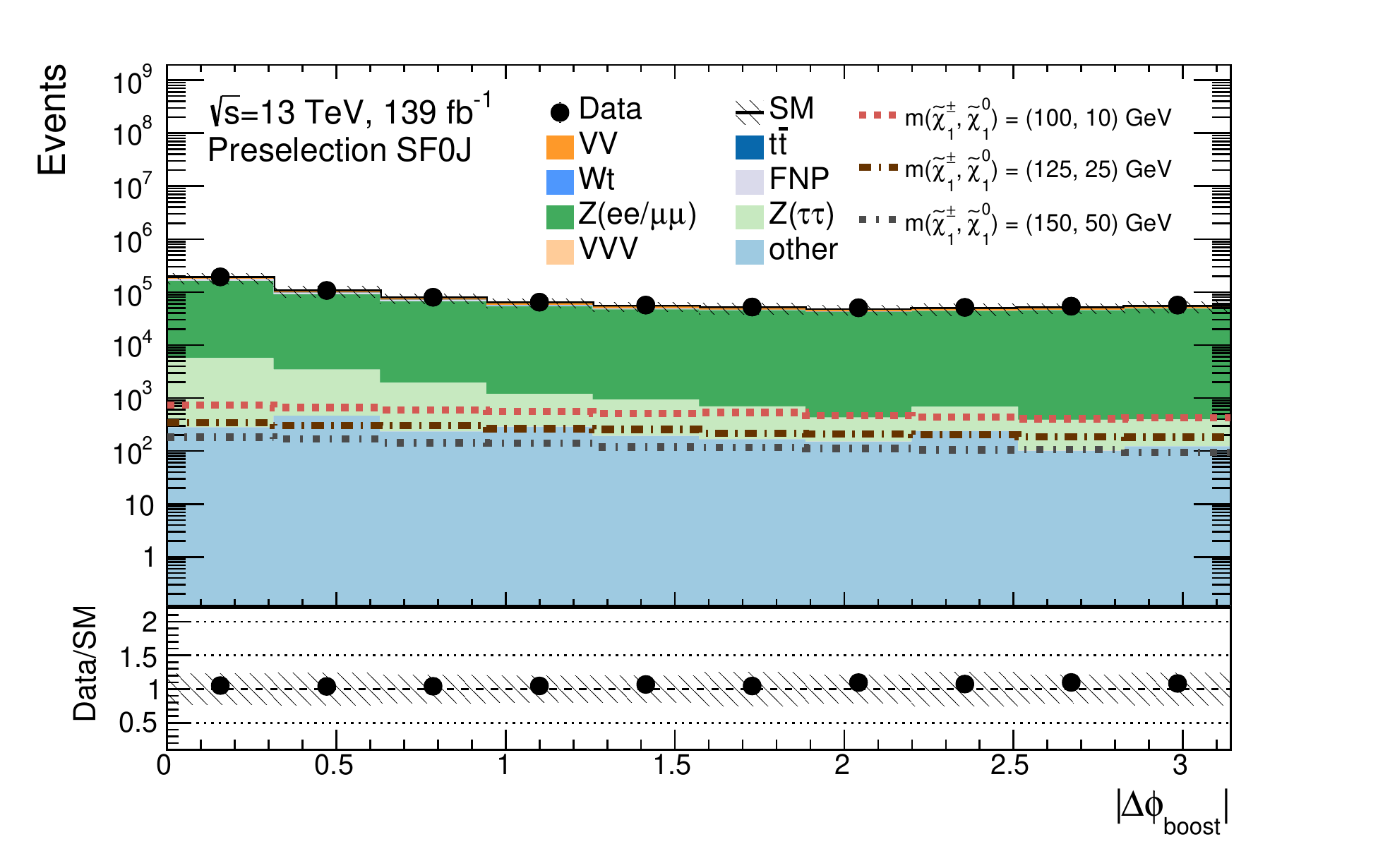}
\caption{The data and MC distributions of the variables trained over, after the preselection and SF 0-jet selections. Both statistical and systematic uncertainties are shown.}
\label{fig:Dists_presel_C1C1_SF0J}
\end{figure}

Figs.~\ref{fig:Dists_presel} and \ref{fig:Dists_presel_C1C1_SF0J} show the data and MC distributions of the variables trained over at the preselection level for DF0J and SF0J events, respectively.

\section{BDT classification}
\label{sec:compressedscharginos-BDT}

\subsection{Training setup}
The signal and background MC samples are split into two sets: the \textit{training} and \textit{test} sets. A classifier is trained over the training set, and the classifier tests the testing set. The two sets are then inverted, and the test set is used for training a second classifier which tests the original training set. The training and test sets are split such that they both contain half of the total MC samples, according to the \say{EventNumber} being odd or even. In this way, the entire signal and background samples get tested and we can use the entire MC statistics for the fit, whilst always testing over a statistically independent data sample.

Two regions are trained over, defined by requirements on the lepton flavour combination SF or DF, with 0 jets and on top of the preselection cuts in Table~\ref{tab:preselCuts_c1c1ww}. These two regions that are trained over (DF 0-jets and SF 0-jets) consist of a loose selection, such that the machine learning classifier has lots of information for learning. Moreover, the two regions are different from the kinematic and background composition point of view, so trainings are separated. Training was also performed over the 1-jet channels, however no sensitivity can currently be attained so these channels are not further considered.

\subsection{Signal samples}
Concerning the $\tilde{\chi}_{1}^{\pm}\tilde{\chi}_{1}^{\mp}$ signal samples used for the training, different training strategies were tested. Depending on the mass splitting between chargino and neutralino, $\Delta m(\tilde{\chi}_{1}^{\pm}, \tilde{\chi}_{1}^{0})$, being equal or below the $W$ boson mass, the kinematic of the signal is different and classifiers trained on one mass region have no sensitivity for signals in another region. Three different trainings were tested:
\begin{itemize}
\item for the \textit{more-compressed} region with $\Delta m(\tilde{\chi}_{1}^{\pm}, \tilde{\chi}_{1}^{0}) < 80$ GeV, a classifier was trained with signal samples having $\Delta m(\tilde{\chi}_{1}^{\pm}, \tilde{\chi}_{1}^{0}) = $ 30 GeV. For signals with low mass splitting, the 1-jet selection was found to have a better performance compared to the 0-jet one, but this training is no longer considered due to the very limited sensitivity.
\item for the region where the $\tilde{\chi}_{1}^{\pm}\tilde{\chi}_{1}^{\mp}$ are very $WW$-like with a $\Delta m(\tilde{\chi}_{1}^{\pm}, \tilde{\chi}_{1}^{0}) = 80$ GeV, little sensitivity can be achieved and it is not considered further.
\item for the \textit{less-compressed} region with $\Delta m(\tilde{\chi}_{1}^{\pm}, \tilde{\chi}_{1}^{0}) > 80$ GeV, a classifier was trained with the samples with $\Delta m(\tilde{\chi}_{1}^{\pm}, \tilde{\chi}_{1}^{0}) =$ 90 and 100 GeV combined together. This is found to provide the best sensitivity in this less-compressed region.
\end{itemize}
The classifier trained over $\Delta m(\tilde{\chi}_{1}^{\pm}, \tilde{\chi}_{1}^{0})$ = 90 and 100 GeV signal samples was found to be the one driving the sensitivity of the search, therefore it is the only classifier considered in the search.

\subsection{Input features}
\label{sec:compressedscharginos-inputfeature}

The choice of variables is a key consideration when training a BDT. Having poor discriminating input features reduces the performance of a classifier, while adding too many input features makes the classifiers more likely to overfit. Choosing an optimal set of input features is thus of utmost importance.

Different variables are considered as input features. The performance when each variable is removed and the BDT retrained is assessed. A variable which, when removed, resulted in an increase of performance is removed and this process is repeated until no gain in performance is gained by removing any of the variables. The best variable set is found to be: $p_{\mathrm{T}}^{\ell_{1}}$, $p_{\mathrm{T}}^{\ell_{2}}$, $E_{\mathrm{T}}^{\mathrm{miss}}$, $E_{\mathrm{T}}^{\mathrm{miss}}$ significance, $m_{\ell \ell}$, the \textit{stransverse mass} $m_{\mathrm{T2}}$ \footnote{The \textit{stransverse mass}~\cite{Lester:1999tx,Barr:2003rg} is a kinematic variable used to bound the masses of a pair of particles that are assumed to have each decayed into one visible and one invisible particle. It is defined as
\begin{equation*} 
m_{\mathrm{T2}}^{m_{\chi}}
( \mathbf p_{\mathrm{T,1}}, \mathbf p_{\mathrm{T,2}},\mathbf p^\mathrm{miss}_\mathrm{T}) = \min_{\mathbf q_{\mathrm{T,1}} + \mathbf q_{\mathrm{T,2}} = 
\mathbf p^\mathrm{miss}_\mathrm{T} }
\left\{ \max [\; 
    m_{\mathrm{T}}( \mathbf p_{\mathrm{T,1}}, \mathbf q_{\mathrm{T,1}};m_{\chi} ),  
    m_{\mathrm{T}}( \mathbf p_{\mathrm{T,2}}, \mathbf q_{\mathrm{T,2}};m_{\chi} ) 
\;] \right\},
\label{mt2eq}
\end{equation*}
\noindent where $\mathbf p_{\mathrm{T,1}}$ and $\mathbf p_{\mathrm{T,2}}$ indicate the transverse-momentum vectors of the two leptons, $\mathbf q_{\mathrm{T,1}}$, $\mathbf q_{\mathrm{T,2}}$ are vectors with $\mathbf p^\mathrm{miss}_\mathrm{T}= \mathbf q_{\mathrm{T,1}} + \mathbf q_{\mathrm{T,2}}$, $m_{\chi}$ is the mass of the invisible particle and $m_{\mathrm T}$ indicates the transverse mass, defined as 
\begin{equation*} 
m_{\mathrm{T}}(\mathbf p_{\mathrm{T}},\mathbf q_{\mathrm{T}},{m_{\chi}}) = \sqrt{m_{\ell}^2+m_{\chi}^2+2(E_{\mathrm{T}}^{\ell}E_{\mathrm{T}}^{q}-\mathbf p_{\mathrm{T}}\cdot\mathbf q_{\mathrm{T}})}.
\label{mteq}
\end{equation*}
The minimisation in $m_{\mathrm{T2}}^{m_{\chi}}$ is performed over all the possible decompositions of $\mathbf p^\mathrm{miss}_\mathrm{T}$. $m_{\chi}$ is a free parameter, which has been set to 0 GeV since this choice improves the sensitivity to several chargino signals. The interesting property of $m_{\mathrm{T2}}^{m_{\chi}}$ is that, for $m_{\chi} = m_{\mathrm{inv}}$, the value of $m_{\mathrm{T2}}$ has a kinematic endpoint at the mass $M$ of the mother particle.}, the angular variable $\cos{\theta^{*}}$ \footnote{The angular variable $\cos{\theta^{*}}$, where $\theta^{*}$ is the polar angle between the incoming quark in one of the protons
and the produced sparticle, is sensitive to the spin of the produced particles \cite{Barr:2005dz}. Since $\theta^{*}$ is not directly measurable, $\cos{\theta_{\ell\ell}^{*}} = \cos \left( 2\tan^{-1} e^{\Delta\eta_{\ell\ell}/2} \right) = \tanh(\Delta\eta_{\ell\ell}/2)$ is defined in terms of the difference in pseudorapidity between the two leptons. The leptons \say{inherit} some knowledge of the rapidity of their supersymmetric parents, and the two variables $\cos{\theta^{*}}$ and $\cos{\theta_{\ell\ell}^{*}}$ are well correlated to each other.}, the azimuthal separation between $E_{\mathrm{T}}^{\mathrm{miss}}$ and $\ell_{1}$ denoted by $\Delta \phi_{E_{\mathrm{T}}^{\mathrm{miss}},\ell_{1}}$, the azimuthal separation between $E_{\mathrm{T}}^{\mathrm{miss}}$ and $\ell_{2}$ denoted by $\Delta\phi_{E_{\mathrm{T}}^{\mathrm{miss}},\ell_{2}}$, and the azimuthal separation between $E_{\mathrm{T}}^{\mathrm{miss}}$ and the vector sum of the $p_{\mathrm{T}}$ of the two leptons and $E_{\mathrm{T}}^{\mathrm{miss}}$ denoted by $\Delta\phi_{\mathrm{boost}}$.

\begin{figure}[!htb]
\begin{center} 
\includegraphics[width=0.7\textwidth]{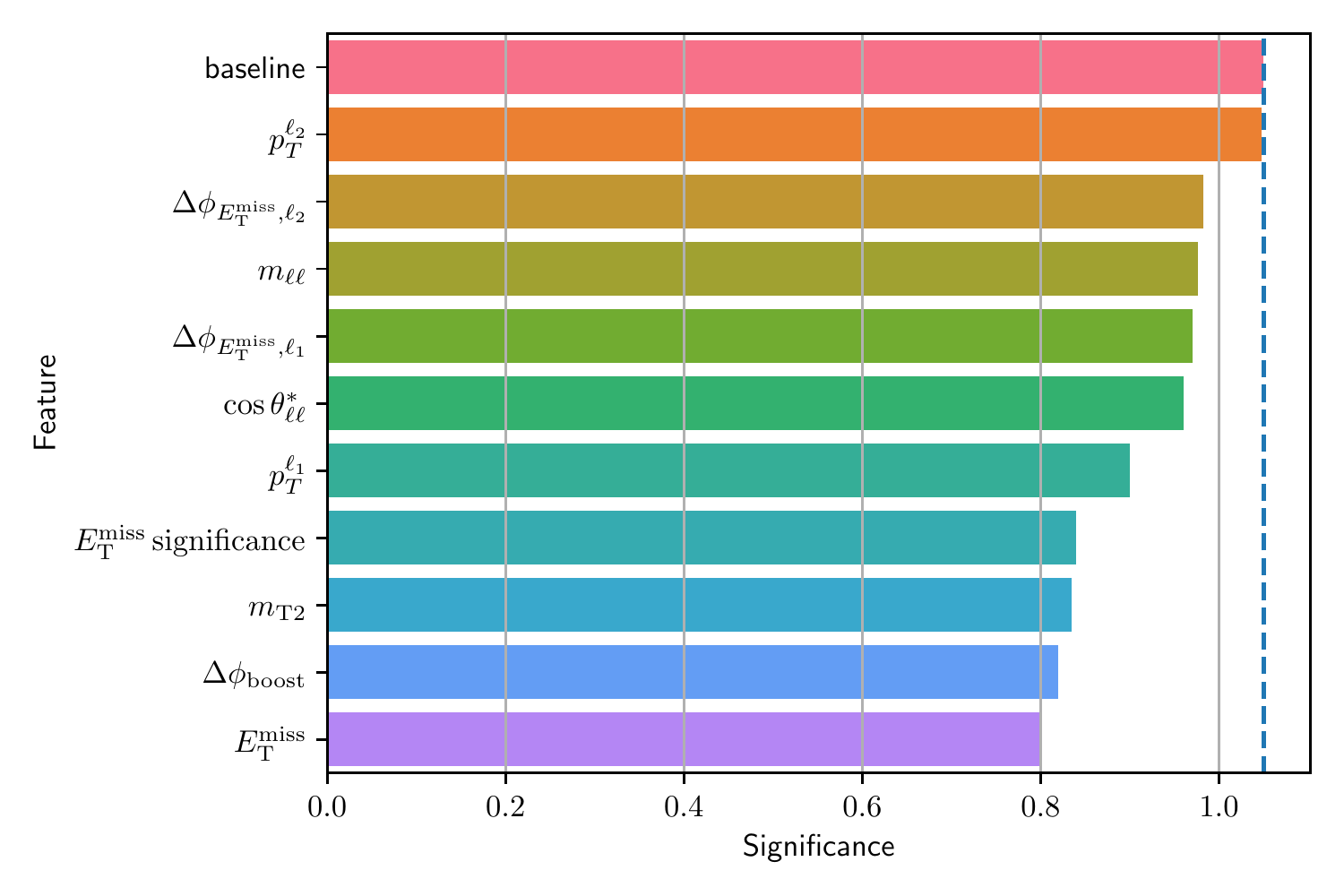}
\end{center}
\caption{The significance of the $m(\tilde{\chi}_{1}^{\pm}, \tilde{\chi}_{1}^{0})$ = (150, 50) GeV signal sample when removing each variable and retraining the classifier in the DF channel. The blue vertical line indicates our current significance.}
\label{fig:c1c1_variable_removing}
\end{figure}

Fig.~\ref{fig:c1c1_variable_removing} shows the significance of the $m(\tilde{\chi}_{1}^{\pm}, \tilde{\chi}_{1}^{0})$ = (150, 50) GeV signal sample when each variable is removed from the training in the DF channel, which is found to be the most sensitive one.

In all cases, removing a variable determines a loss in the baseline significance, meaning that all the variables in the chosen set are beneficial. The input features that, when removed, determine the most drastic loss in significance are the ones which turn out to be the most beneficial for gaining sensitivity. Furthermore, we cannot gain any significant sensitivity by reducing this set of variables trained over and training with the extra variables showed no increase in sensitivity.

\subsection{ROC and PR curves}
The performance of the classifier can be quantified using \textit{Receiver Operating Characteristic} (ROC) and \textit{Precision-Recall} (PR) curves. These take the signal and background BDT score distribution and apply cuts on the BDT score, classifying events with BDT score less than the cut value as background and events with BDT score above the cut value as signal. 
The signal events correctly classified as signal are called \textit{True Positives} (TP), whereas the background events that have been incorrectly classified as signal are called \textit{False Positives} (FP). Similarly, the background events correctly classified as background are called \textit{True Negatives} (TN), whereas the signal events that have been incorrectly classified as background are called \textit{False Negatives} (FN). These four categories form a \textit{confusion matrix}. 

\begin{figure}[!htb]
\centering
\includegraphics[width=0.7\linewidth]{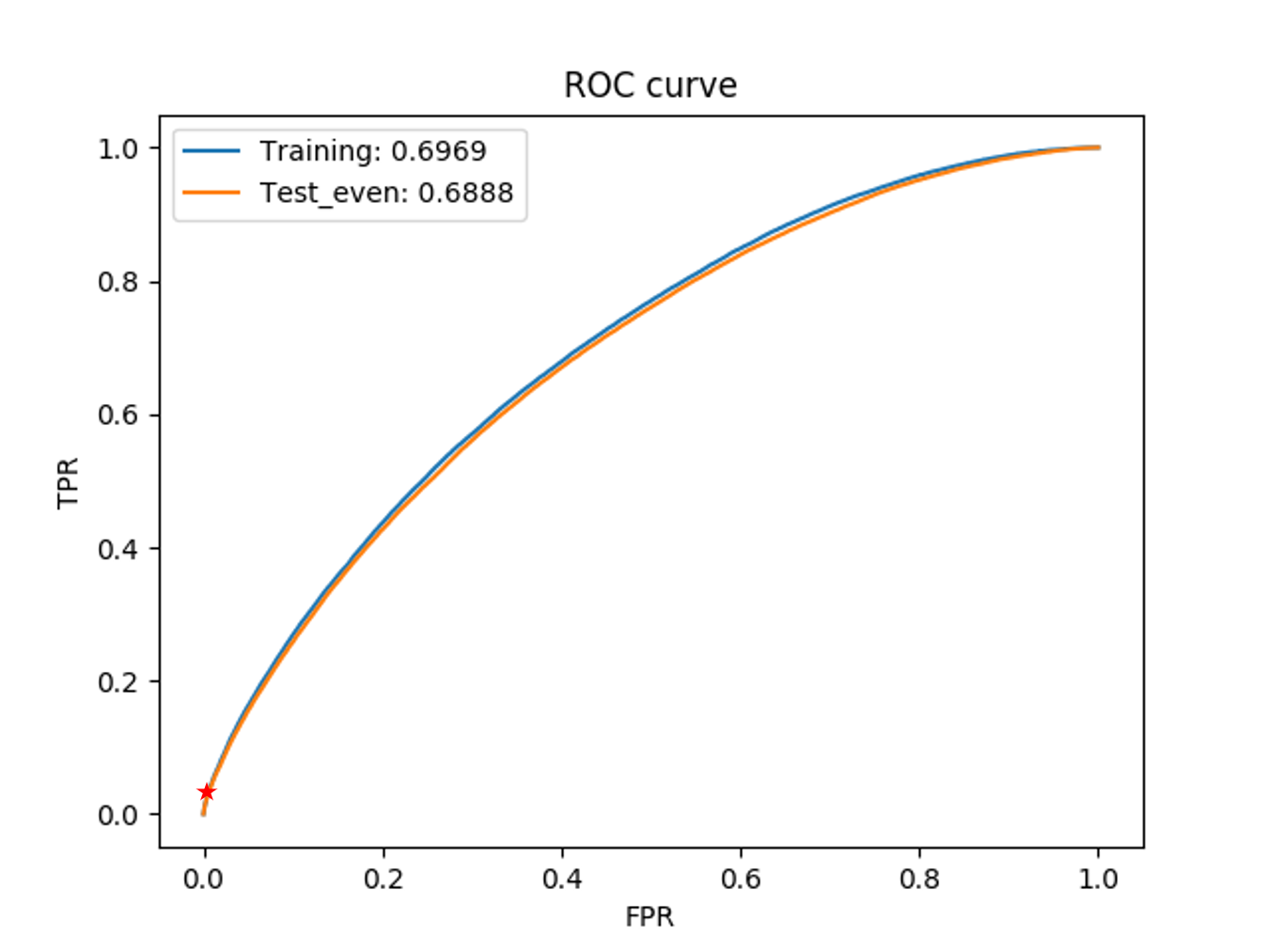}
\caption{The ROC curve for the training and testing sets in the DF channel. These have been calculated considering the signal against all the other backgrounds. The AUC values are included in the legend. The star indicates where the SRs begin.}
\label{fig:ROCcurve}
\end{figure}

ROC curves plot the True Positive Rate (TPR = TP / TP + FN) as a function of the False Positive Rate (FPR = FP / FP + TN), with a better classifier maximising the TPR whilst minimising the FPR. The Area Under the Curve (AUC) is a numerical measure of the quality of the classifier, which corresponds to the probability that the classifier would rank a randomly chosen signal event higher than a randomly chosen background event. AUC = 1 represents perfect signal/background separation and AUC = 0.5 represents what is expected from random guessing. The ROC curves for the training and testing sets in the DF channel can be seen in Fig.~\ref{fig:ROCcurve}, with the AUC values for each set listed in the legend.

\begin{figure}[!htb]
\centering
\includegraphics[width=0.7\linewidth]{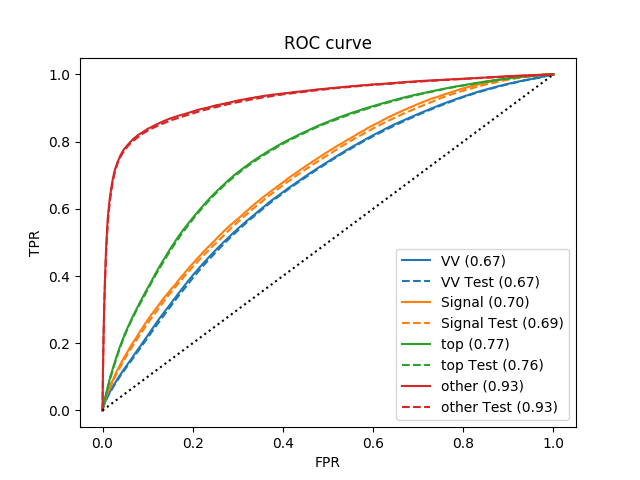}
\caption{The ROC for each of the four categories ($VV$, signal, top, other) in the multiclass classifier in the DF channel. The curves use the respective BDT score and measure the separation to the other categories. The AUC are included in the legend, for the training and test sets.}
\label{fig:ROCmulti}
\end{figure}

Also, the ROC curves for each of the four categories ($VV$, signal, top, other) are shown in Fig.~\ref{fig:ROCmulti}, which indicates which of the categories are easiest to distinguish from the rest. Clearly, the other more minor backgrounds are the easiest to separate from the rest of the backgrounds, with a very high AUC of 0.93, due to very different event kinematics. 

\begin{figure}[!htb]
\centering
\includegraphics[width=0.7\linewidth]{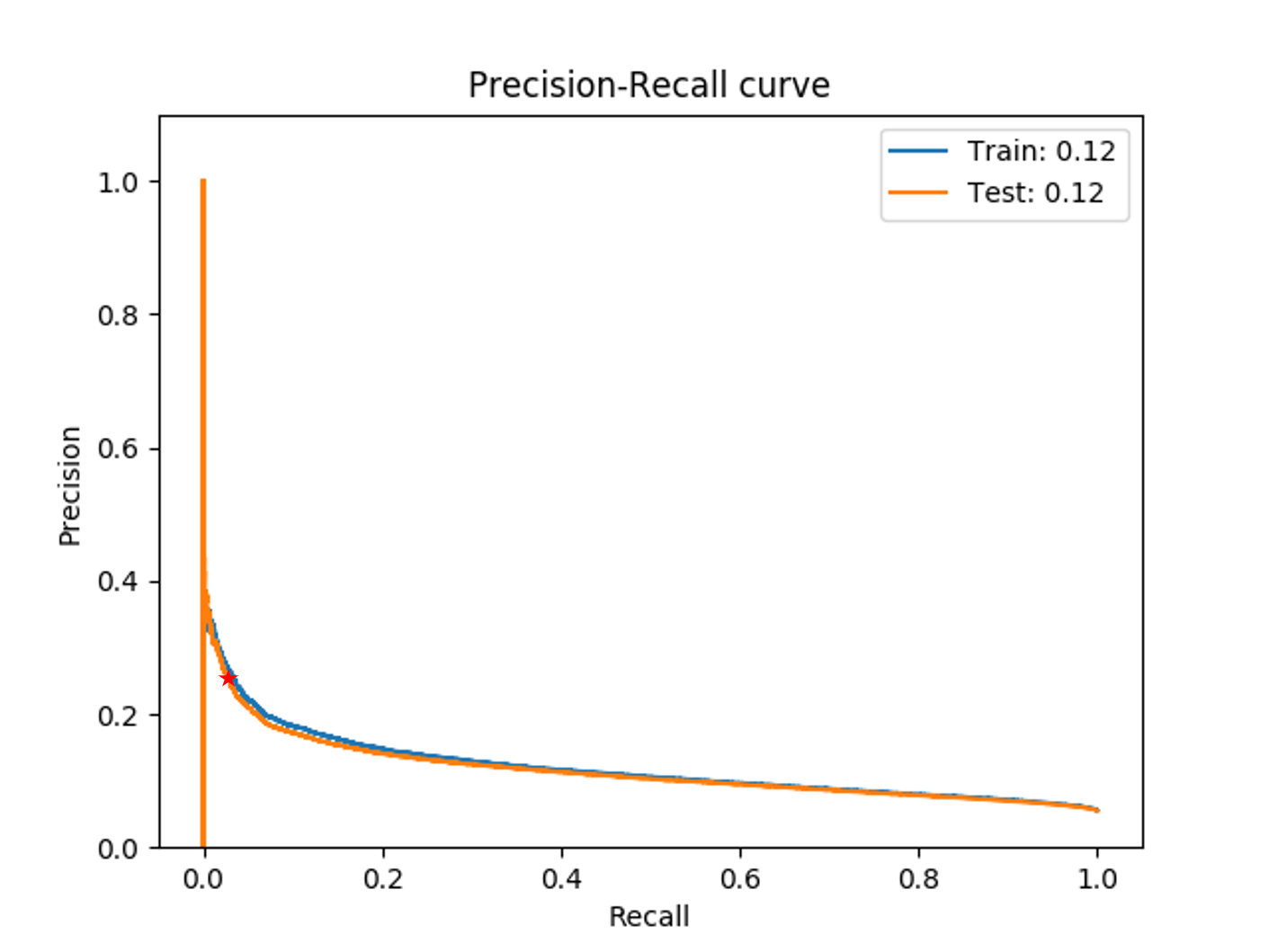}
\caption{The PR curves for the training and testing sets in the DF channel. These have been calculated considering the signal against all the other backgrounds. The average-precision values are included in the legend. The star indicates where the SRs begin.}
\label{fig:PRcurve}
\end{figure}

One drawback of ROC curves for datasets which have a large mismatch in statistics between signal and background samples, with few small signal events among a large background as in the case here considered, is that they can be overly sensitive to large numbers of true negatives, that is background correctly identified as background. For this case, PR curves can be more illustrative of performance: they show the precision (precision = TP / TP + FP) against the recall (recall = TPR = TP / TP + FN). Intuitively, precision is the ability of the classifier not to label as positive an event that is negative, and recall is the ability of the classifier to find all the positive events. The PR curves in the DF channel are shown in Fig.~\ref{fig:PRcurve}, with the area under the PR curve (the average precision) for each set shown in the legend.

Both the AUC and average precision illustrate the performance on the BDT score. Because these estimators are computed across the entire range of the BDT score, it can happen that a classifier has a lower AUC or average precision but actually provides better sensitivity to a specific signal.

\subsection{Feature importance}
\label{sec:compressedcharginos-featureimportance}

\begin{figure}[!htb]
\centering
\includegraphics[width=0.7\linewidth]{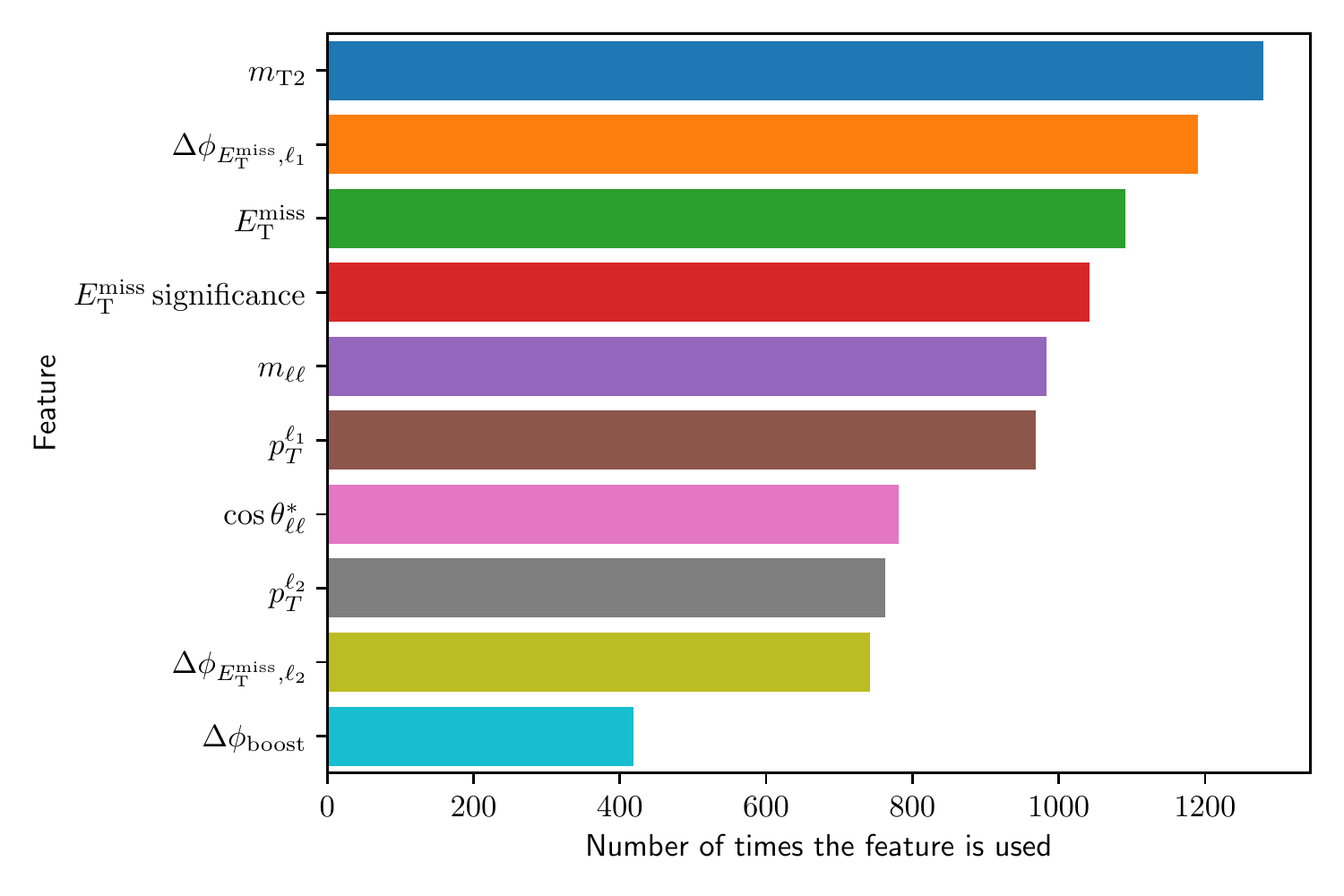}
\caption{The number of times each feature is used to split in the BDT.}
\label{fig:featureimportance}
\end{figure}

Understanding how important each input feature is to the classifier gives a good insight into interpreting the classifier output. Firstly, we can consider the number of times each feature is used in building the model. As can be seen in Fig.~\ref{fig:featureimportance}, the most used variables when constructing the BDT are $m_{\mathrm{T2}}$, $\Delta \phi_{E_{\mathrm{T}}^{\text{miss}},\ell_{1}}$ and $E_{\mathrm{T}}^{\text{miss}}$. 

\begin{figure}[!htb]
\centering
\includegraphics[width=0.75\linewidth]{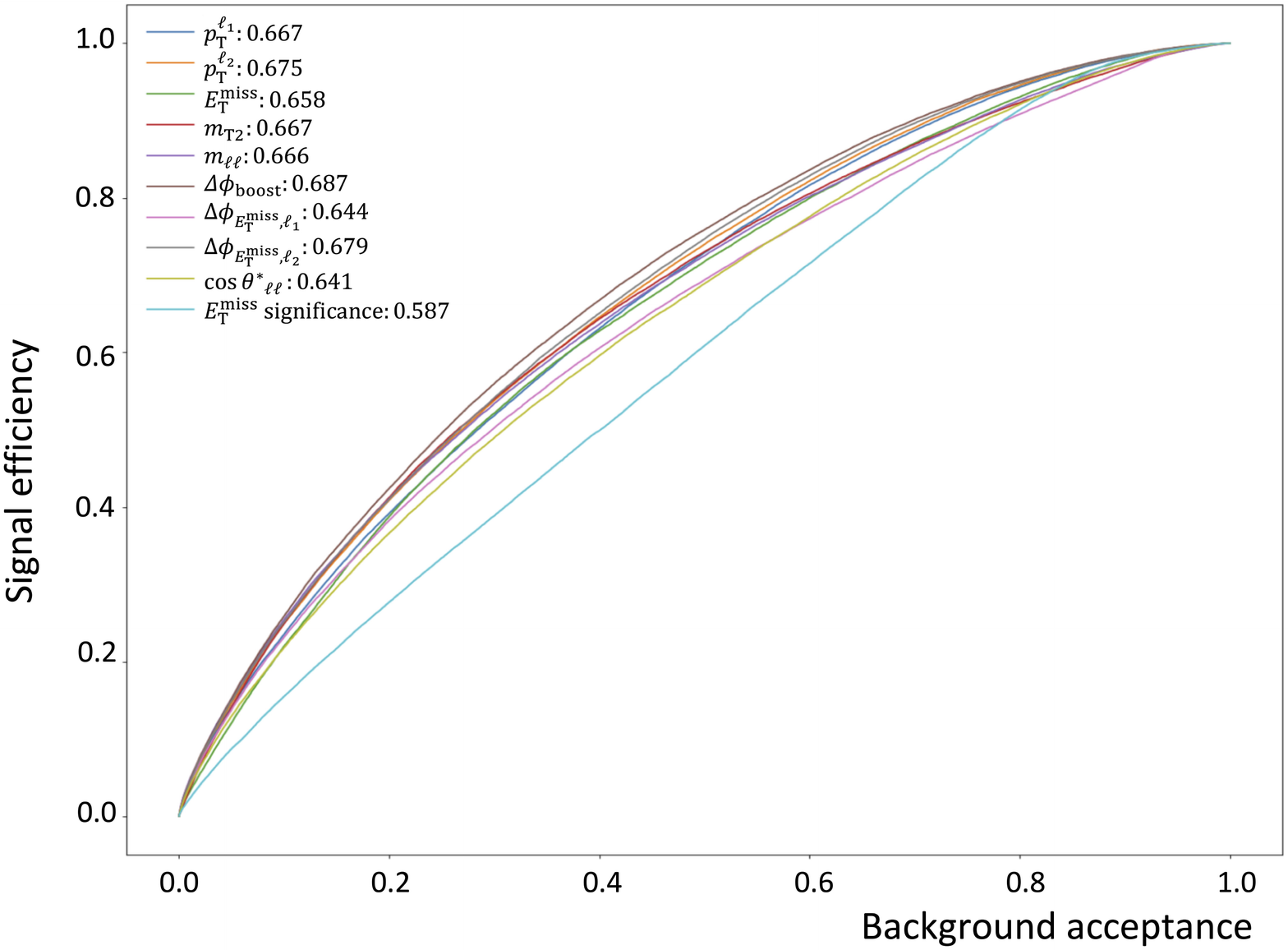}
\caption{The ROC curves for the test set with one variable randomly shuffled across all events, as indicated by the legend. Lower values of the AUC indicate a more important variable.}
\label{fig:permutationimportance}
\end{figure}

A further method to determine the feature importance is through \textit{permutation importance}.
A trained classifier is considered and the performance is tested with one variable removed in order to explicitly see the effect of each variable in the classification, with the removal of a more important variable causing the performance to decrease more. Since the classifier will not work if a variable is removed, the variable is instead replaced with random noise drawn from the same distribution as the original values for that variable. No useful information can be gained from that variable and the effect is the same as removing the variable. The variable can be replaced with noise by randomly shuffling, or permuting, its values across all events. This both saves time in re-training the classifier, and more importantly allows us to assess the classifier. 
Fig.~\ref{fig:permutationimportance} reports the ROC curves for the test set with one variable shuffled to assess the importance. 

Interestingly, this permutation importance leads to the conclusion that $E_{\mathrm{T}}^{\text{miss}}$ significance is the most important variable since its removal leads to worse performance. The $E_{\mathrm{T}}^{\text{miss}}$ significance is used fewer times for splitting the trees, however, clearly is important when doing the splitting. Since these two feature importances metrics are inherently different - the number of times each variable is used compared to the AUC - it is not unexpected that there are some differences in ordering. Both methods provide complementary information and help us understand the classifier.

\subsection{Hyperparameters}
When training a machine learning classifier, certain parameters are defined before the training procedure and do not change throughout. These are known as hyperparameters, the choice of which is essential for good performance. For example, with too few trees the classifier will not be able to generate the required complexity to describe the data, whereas if there are too many trees the classifier can tune to fluctuations in the training set and not generalise to the underlying distributions. These two scenarios are typically known as underfitting and overfitting, respectively. The optimal set of hyperparameters for each situation is not known before training, and so must be determined by hyperparameter optimisation methods.

For the BDTs, four hyperparameters are varied whilst keeping the others at their default value. These hyperparameters are: the number of trees in the BDT, the maximum number of leaves of each tree, the learning rate (how often the loss is updated during training), and the minimum number of samples per each leaf. Cross validation is used to find the best significance with the simplest classifier. The significance used for this optimisation is calculated with Eq.~(\ref{eqn:SignificanceMeasure}), using a flat 20$\%$ systematic uncertainty. The classifiers in the DF and in the SF channels are independently optimised in this way.

The optimisation of the hyperparameters does not make too much difference to the overall results, with many hyperparameter configurations giving a very similar performance. We prefer simpler classifiers, such that we are less likely to be overfitting, so when two classifiers have the same performance we prefer the simpler one. We measure the complexity of the BDT as the number of trees multiplied by the number of leaves in each tree.

In order to optimise, we use 5-fold cross validation and look at the significance value. We then choose the simplest classifier with significance within one standard error of the best classifier.

We observed that having a high number of trees with a small number of leaves gives a good performance. With this setup, the sensitivity is comparable with a large range of hyperparameter selections. Table~\ref{tab:hyperparameter_study} illustrates the final significance on the representative $m(\tilde{\chi}_{1}^{\pm}, \tilde{\chi}_{1}^{0})$ = (150, 50) GeV sample as well as the ROC AUC for the chosen simpler classifiers, as well as the classifier which was seen to maximise the significance. Interestingly, the classifiers which had the highest sensitivity tend to have more trees, however we can see that the sensitivity is very comparable between this and the simpler classifiers used in the final analysis, since the significance values are observed to be very stable across a wide range of hyperparameter choices.

\begin{table}[!htb]
\begin{center}
\begin{tabular}{l|cc|cc}
\noalign{\smallskip}\hline\noalign{\smallskip}
Classifier & DF chosen & DF best Z & SF chosen & SF best Z \\ \noalign{\smallskip}\hline\noalign{\smallskip}
$n_{\mathrm{trees}}$ & 100 & 752 & 150 & 771 \\
$n_{\mathrm{leaves}}$ & 24 & 40 & 20 & 40 \\
$\min n_{\mathrm{data}}$ in trees & 11 & 482 & 177 & 207 \\
learning rate & 0.06 & 0.03 & 0.05 & 0.04 \\
\noalign{\smallskip}\hline\noalign{\smallskip}
$Z$ & 1.02 & 1.05 & 0.45 & 0.48 \\
AUC-train & 0.697 & 0.731 & 0.729 & 0.775 \\
AUC-test & 0.689 & 0.694 & 0.721 & 0.726 \\
\noalign{\smallskip}\hline\noalign{\smallskip}
\end{tabular}
\caption{Various metrics for the classifiers. The significance $Z$ of the $m(\tilde{\chi}_{1}^{\pm}, \tilde{\chi}_{1}^{0})$ = (150, 50) GeV sample when calculated with a single cut on the BDT score, and the ROC area under the curve (AUC) calculated on the train and test sets.}
\label{tab:hyperparameter_study}
\end{center}
\end{table}

The AUC values on the training and test set show a clear mismatch for the more complicated classifiers, indicating that if we did use a more complicated classifier we would be at risk of overfitting. The hyperparameter selection is simple enough to not overfit to the training set and to ensure enough complexity to effectively separate the signal and backgrounds, thus providing good sensitivity.

\subsection{SHAP values}
SHAP (SHapley Additive exPlanations) values \cite{SHAPpaper} are a novel way of explaining machine learning models. These help us improve in model explainability and transparency, which are very important for machine learning models. 

For each event, SHAP values are calculated for each input feature to indicate the marginal contribution of each variable to the output score from the BDT. For example, the value of $p_{\mathrm{T}}^{\ell_{1}}$ can, averaged over all permutations of adding it, make the event more signal-like and increase the BDT score by 0.1: $p_{\mathrm{T}}^{\ell_{1}}$ for this event would have a SHAP value of 0.1.

In our case, since we are doing multi-class classification, the BDT has four output scores: BDT-VV, BDT-signal, BDT-top and BDT-others. In this case, we have four BDT output scores for which we can obtain SHAP values, i.e. we can obtain the contributions of the variables to each of the four scores. 

\begin{figure}[!htb]
\begin{center}
\includegraphics[width=0.8\textwidth]{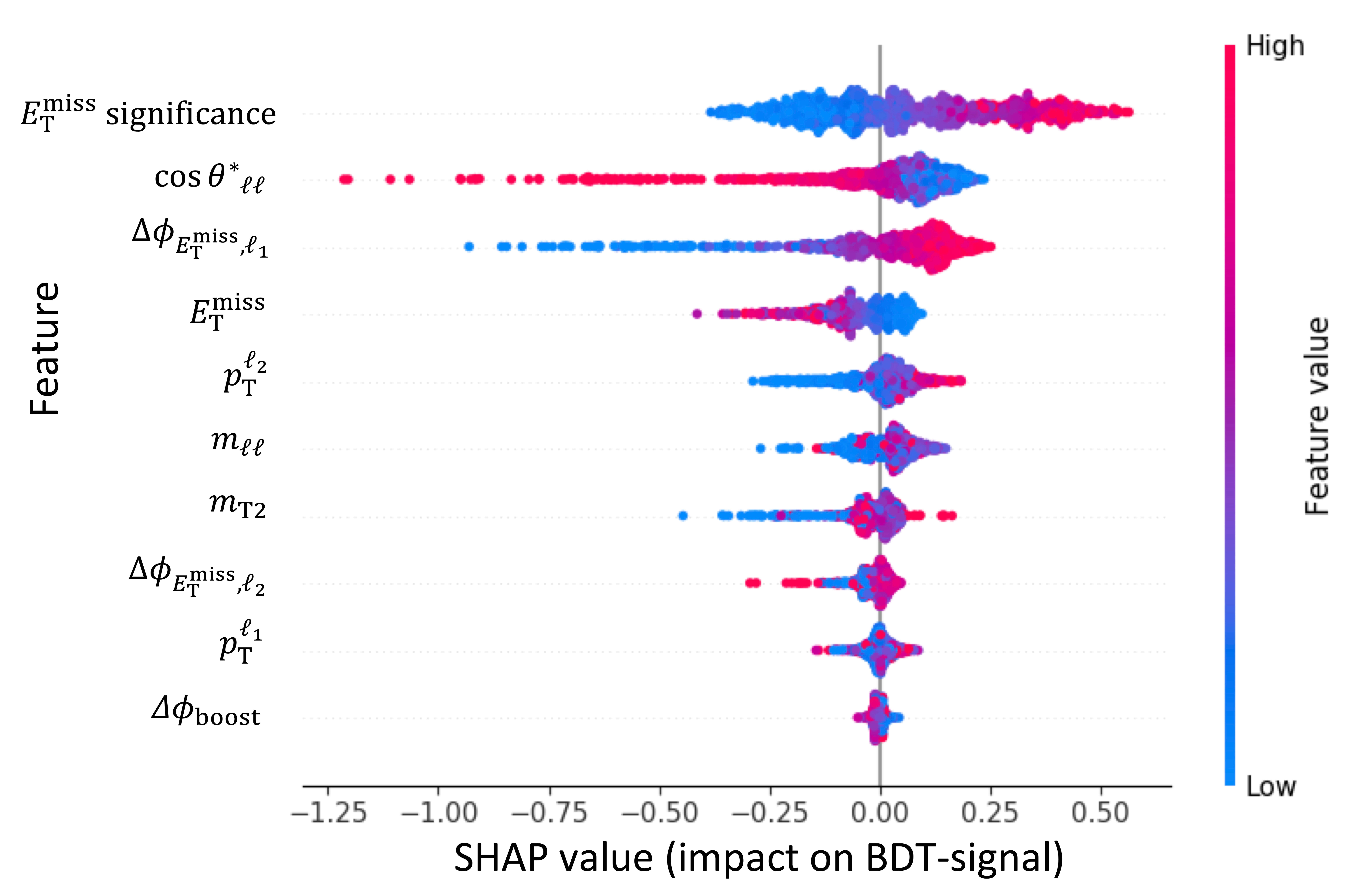}
\caption{The SHAP values for the BDT-signal score. Points to the right are more signal-like and to the left are more background-like. The colour of the point indicated the value of the corresponding variable.}
\label{fig:SHAP signal}
\end{center}
\end{figure}

We can firstly look at the SHAP values for the BDT-signal score in Fig.~\ref{fig:SHAP signal}. 

\begin{figure}[!htb]
\begin{center}
\includegraphics[width=0.8\textwidth]{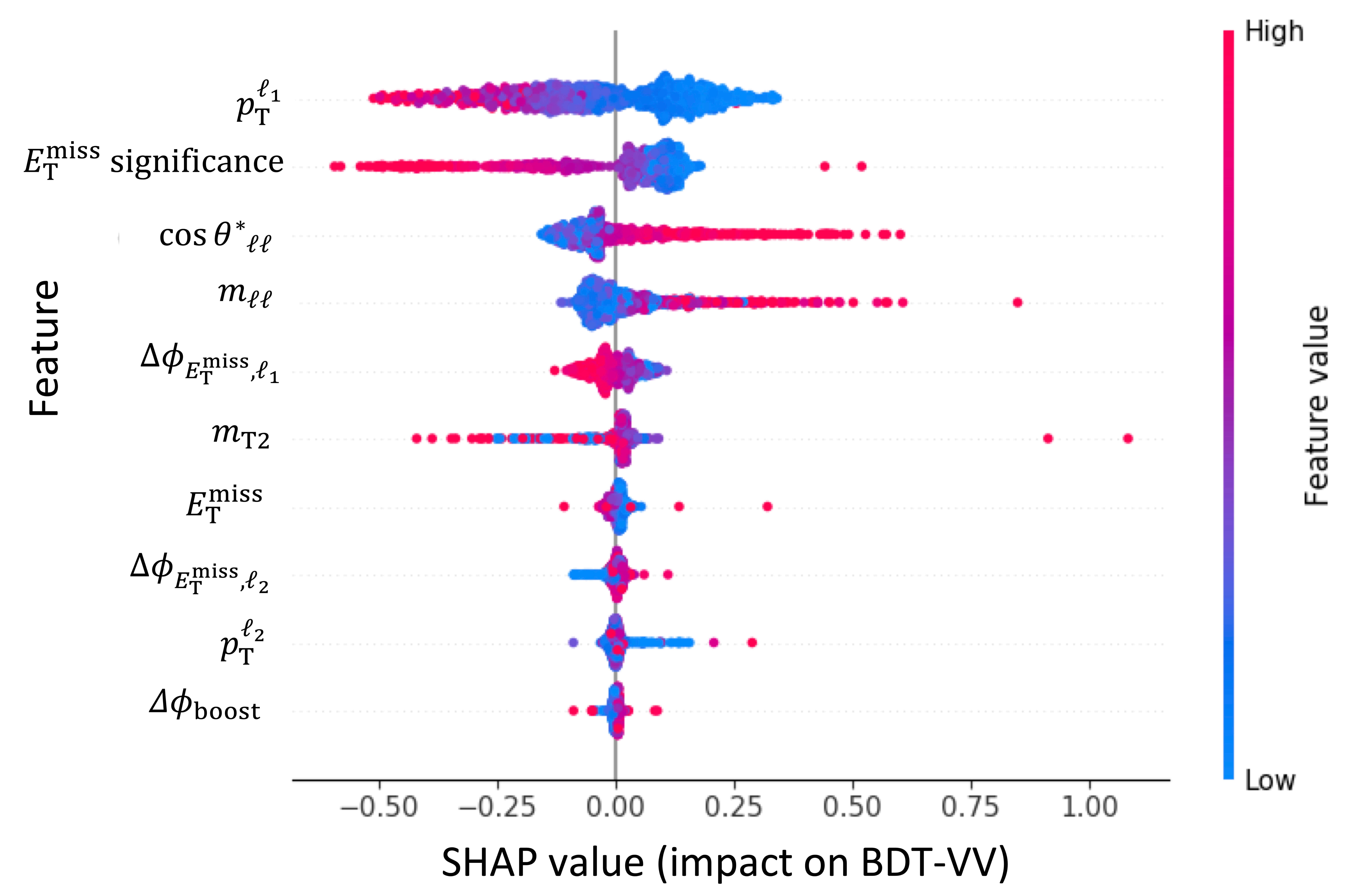}
\caption{The SHAP values for the BDT-VV score. Points to the right are more VV-like. The colour of the point indicated the value of the corresponding variable.}
\label{fig:SHAP VV}
\end{center}
\end{figure}

\begin{figure}[!htb]
\begin{center}
\includegraphics[width=0.8\textwidth]{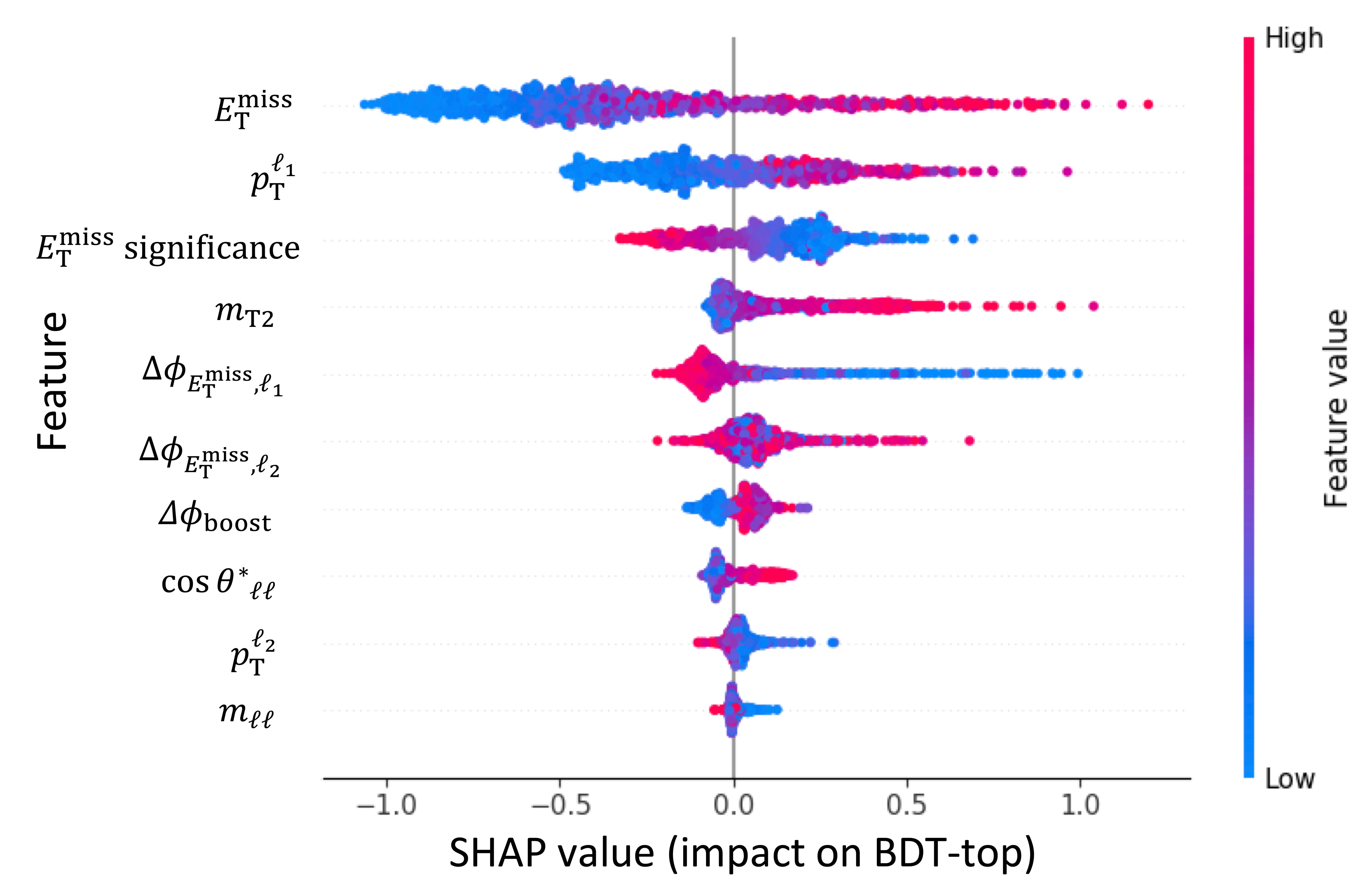}
\caption{The SHAP values for the BDT-top score. Points to the right are more top-like. The colour of the point indicated the value of the corresponding variable.}
\label{fig:SHAP top}
\end{center}
\end{figure}

\begin{figure}[!htb]
\begin{center}
\includegraphics[width=0.8\textwidth]{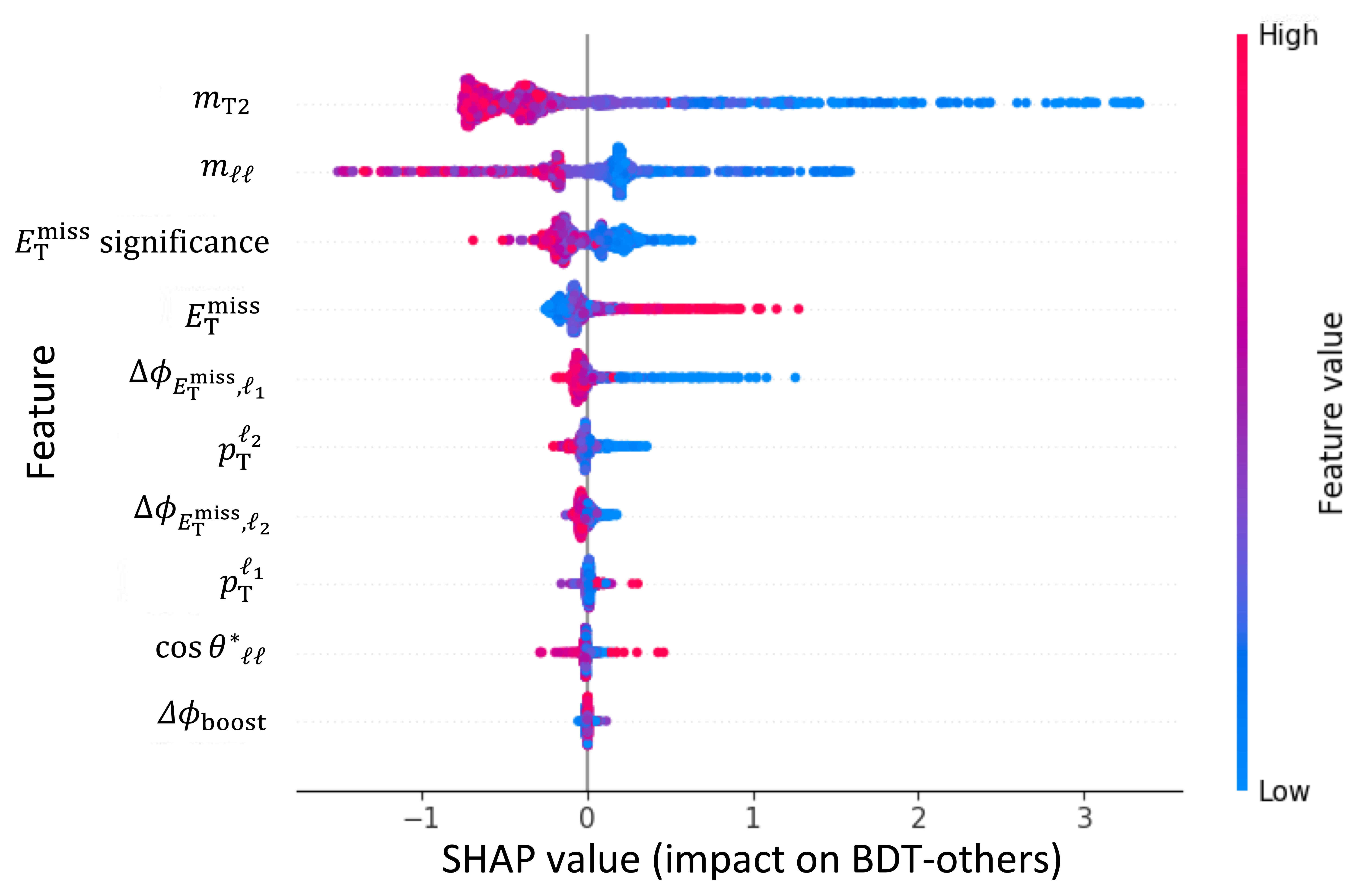}
\caption{The SHAP values for the BDT-others score. Points to the right are more others-like. The colour of the point indicated the value of the corresponding variable.}
\label{fig:SHAP others}
\end{center}
\end{figure}

The $x$-axis indicates the SHAP value, e.g. the impact on the model output.  Higher SHAP values mean more signal-like for the event while lower SHAP values mean more background-like for the event. Each dot on the plot corresponds to one event, and its colour (on the scale blue to red) indicates the value of the corresponding variable the SHAP value is calculated for, labelled on the $y$-axis. For example, at the top of the $y$-axis is the $E_{\mathrm{T}}^{\text{miss}}$ significance variable. This plot indicates that high (red dots) values of $E_{\mathrm{T}}^{\text{miss}}$ significance are more signal-like events (higher SHAP values). The corresponding plots for the other classes (VV, top and others) are shown in Figs.~\ref{fig:SHAP VV},~\ref{fig:SHAP top}~and~\ref{fig:SHAP others}.


The variables on the $y$-axis are in descending feature importance order, that is the variables at the top ($E_{\mathrm{T}}^{\text{miss}}$ significance, $\cos{\theta_{\ell\ell}^{*}}$, $\Delta\phi_{E_{\mathrm{T}}^{\mathrm{miss}},\ell_{1}}$) affect the BDT-signal score the most. Interestingly, this is not seen when using simpler feature importance metrics such as the number of times the variable is used for splitting in the tree. 

\begin{figure}[!htb]
\begin{center}
\includegraphics[width=12cm]{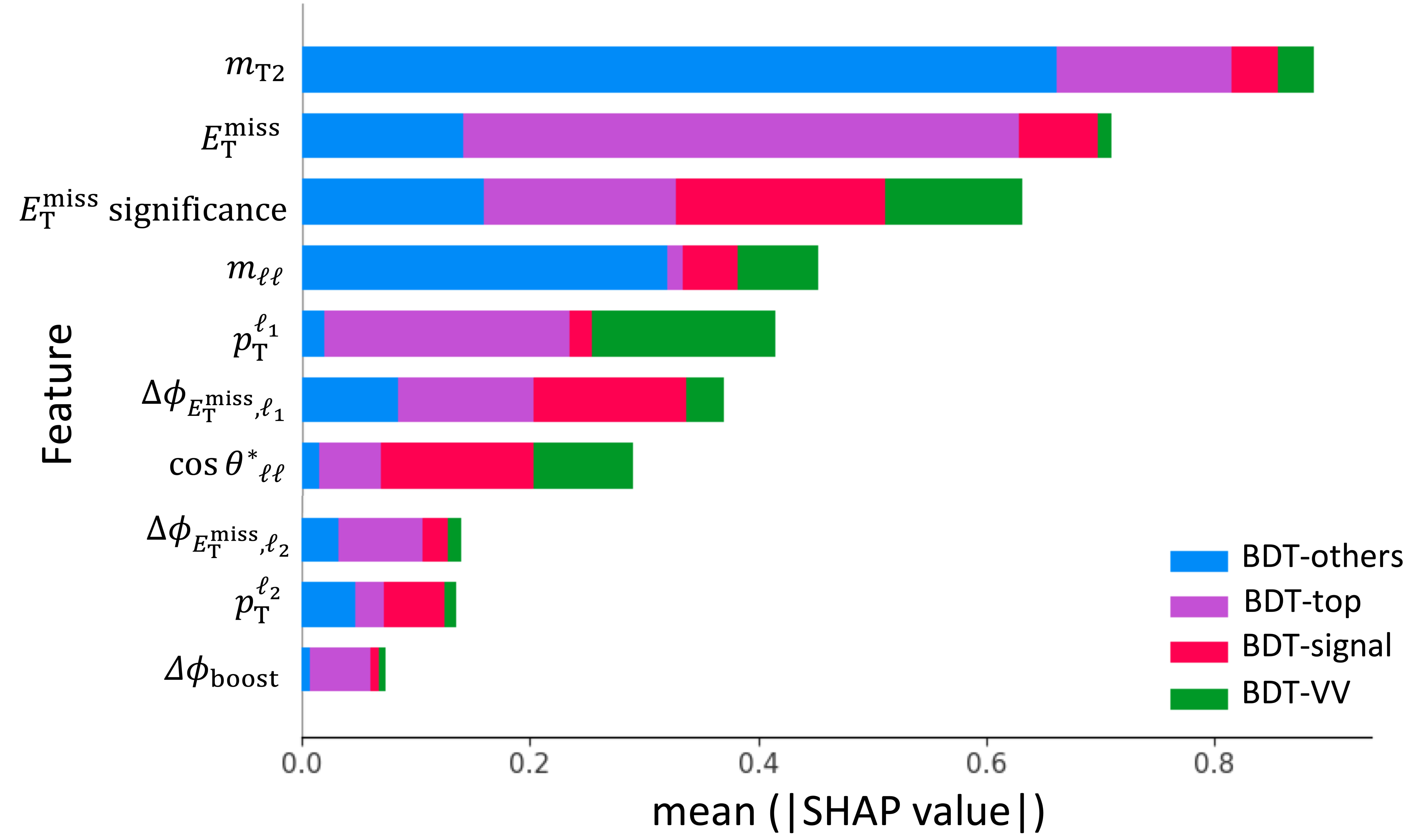}
\caption{SHAP bar chart indicating the average impact (magnitude of SHAP value) of each variable on each specific BDT score.}
\label{fig:SHAP bar}
\end{center}
\end{figure}

Secondly, we can take the mean of the absolute values of the SHAP values for each score, and plot it as a bar chart in Fig.~\ref{fig:SHAP bar}. 

Here, we can see which variables are best for distinguishing each class by the relative size of the coloured bar. For example, as shown earlier $E_{\mathrm{T}}^{\text{miss}}$ significance makes the largest contribution to the BDT-signal score, whereas $E_{\mathrm{T}}^{\text{miss}}$ makes the largest contribution to the BDT-top score. This gives us a much clearer insight into how the classifier works than the simple feature importances previously considered.

\begin{figure}[!htb]
\begin{center}
\includegraphics[width=0.49\textwidth]{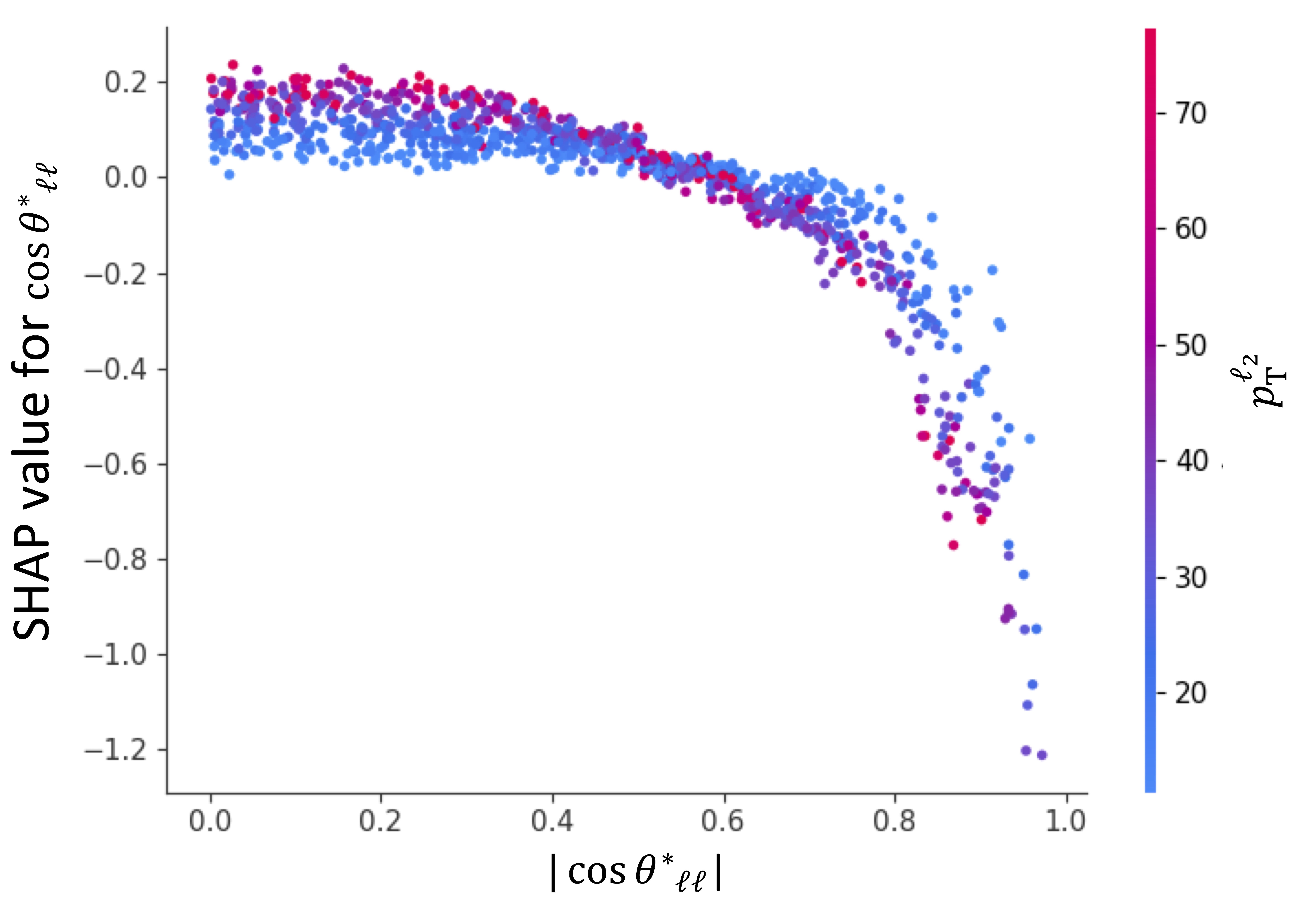}
\includegraphics[width=0.49\textwidth]{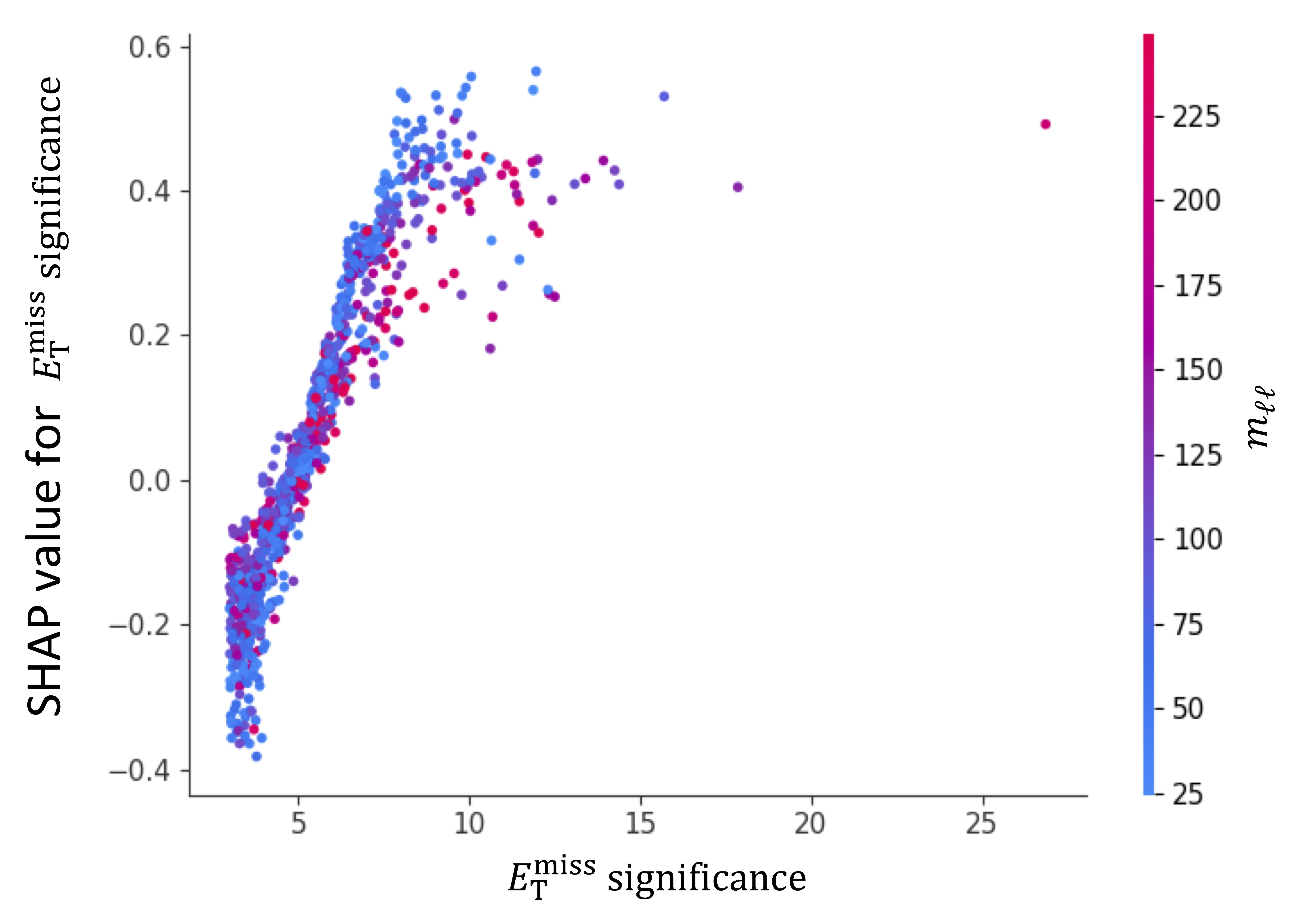}
\end{center}
\caption{SHAP interaction values. Here the SHAP values are plotted as a function of a variable such that we can explicitly see how the SHAP values depend on $\cos{\theta_{\ell\ell}^{*}}$ and $\Delta\phi_{E_{\mathrm{T}}^{\mathrm{miss}},\ell_{1}}$. The interaction between these variables and the $p_{\mathrm{T}}^{\ell_{1}}$ variable can also be seen by the colour of the dots.}
\label{fig:SHAP-interaction}
\end{figure}

Finally, and perhaps most interestingly, we can plot \textit{SHAP dependency} plots, in Fig.~\ref{fig:SHAP-interaction}. 
These have the value of the variable considered on the $x$-axis, with its corresponding SHAP value on the $y$-axis. Again each dot corresponds to an event, however in this case the colour of the dots corresponds to the value of a \textit{different} variable. This allows us to see how the interaction between two variables can affect the SHAP value, that is whether two variables interact to make the event more signal or background-like. Fig.~\ref{fig:SHAP-interaction}, illustrates that high $\cos{\theta_{\ell\ell}^{*}}$ values are more background-like (low SHAP), as we ascertained previously. Moreover, we can see an interaction such that at low $\cos{\theta_{\ell\ell}^{*}}$ values high $p_{\mathrm{T}}^{\ell_{2}}$ (red dots) are more signal-like, whereas at high $\cos{\theta_{\ell\ell}^{*}}$ values the reverse is true - low $p_{\mathrm{T}}^{\ell_{2}}$ (blue dots) values are more signal-like. On the other hand, there is no clear interaction between $E_{\mathrm{T}}^{\text{miss}}$ significance and $m_{\ell\ell}$ as can be seen in Fig.~\ref{fig:SHAP-interaction}. We can still see that higher values of $E_{\mathrm{T}}^{\text{miss}}$ significance indicate a more signal-like event as we would expect.

\subsection{BDT-signal score}
\begin{figure}[!htb]
\centering
\includegraphics[width=0.8\linewidth]{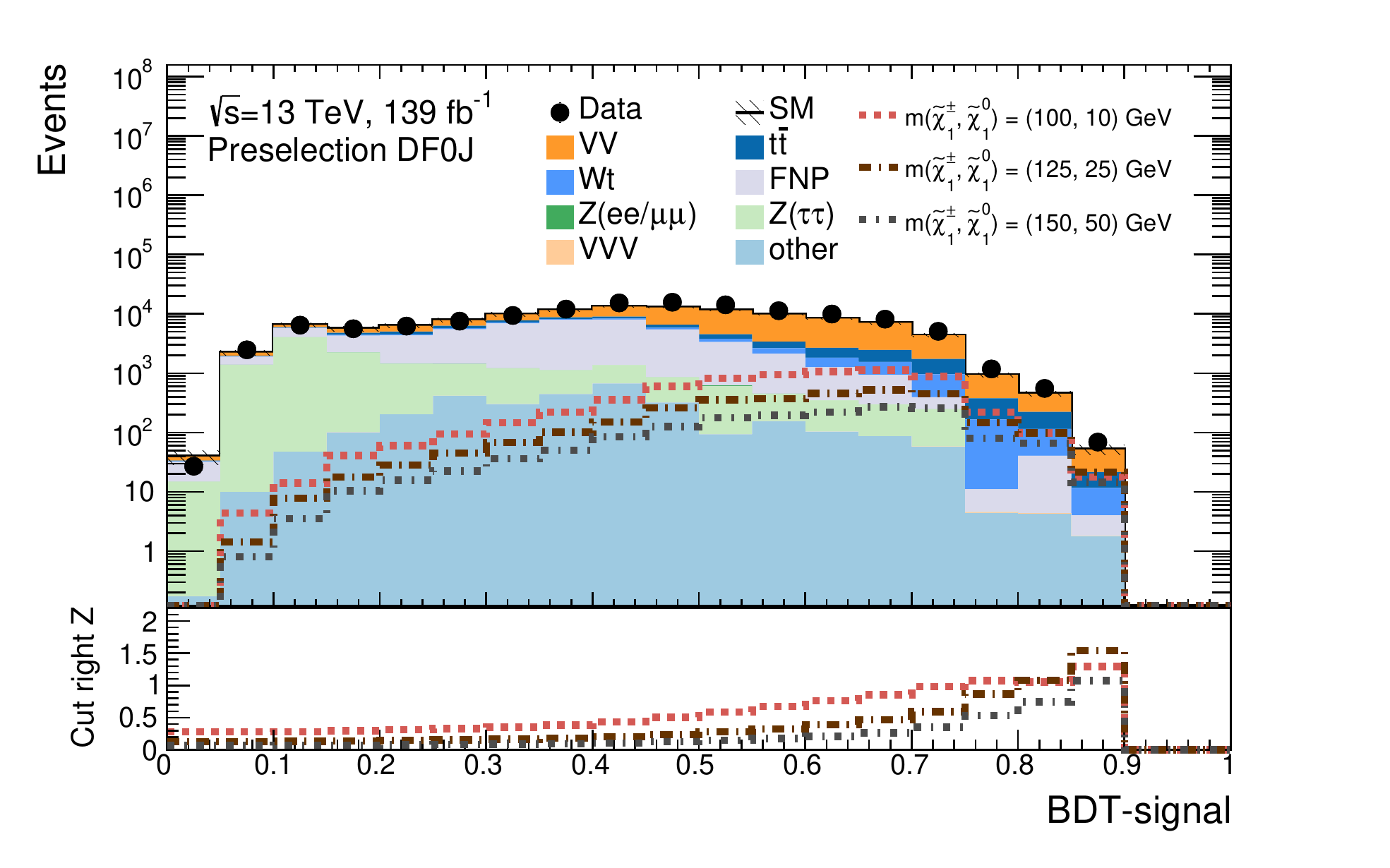}
\includegraphics[width=0.8\linewidth]{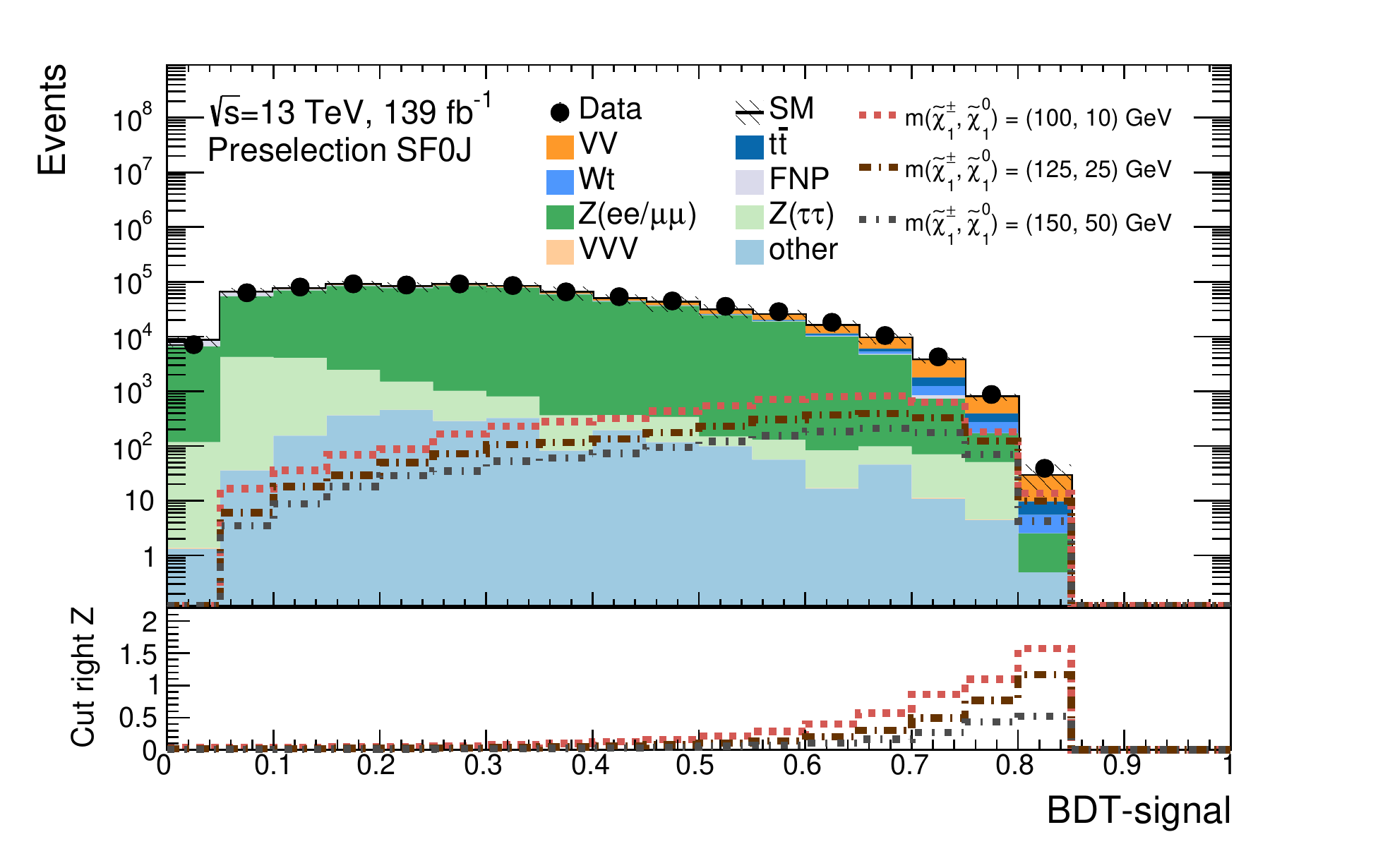}
\caption{The BDT-signal score for the backgrounds and the signal samples, in the DF0J and SF0J channels.}
\label{fig:BDTscore}
\end{figure}

The BDT-signal score for the chargino signals and the backgrounds is shown in Fig.~\ref{fig:BDTscore}. As we can see, we gain signal sensitivity with $\Delta m(\tilde{\chi}_{1}^{\pm}, \tilde{\chi}_{1}^{0})$ = 100 GeV and 90 GeV at higher BDT-signal scores, where the BDT has identified the events to be more signal-like.

\section{SR definition}
The definition of the SR proceeds by first applying two baseline cuts, then cuts on the BDT output score.

\subsection{Baseline cuts}
Considering the $E_{\mathrm{T}}^{\text{miss}}$ significance distribution, it can be noticed that the BDT-signal requires high $E_{\mathrm{T}}^{\text{miss}}$ significance for the more signal-like events, as we would expect. We apply a cut of $E_{\mathrm{T}}^{\text{miss}}$ significance $>$~8 as well as BDT-signal score cuts to define the SRs, and we also validate the BDT scores with orthogonal events that have $E_{\mathrm{T}}^{\text{miss}}$ significance $< 8$. This general validation region is used to verify the data/MC shape agreement for the BDT score.\\
A further selection of $m_{\mathrm{T2}}>50$ GeV is applied to the SR. The cut is designed to ensure good data/MC modelling in the VRs and therefore applied to all the region relevant for the search. The impact of the $m_{\mathrm{T2}}>50$ GeV cut on the signal regions is very small at high BDT-signal score, as expected from the correlation between $m_{\mathrm{T2}}$ and BDT-signal score.

\subsection{BDT-signal score cut}
Different ways of defining the SR and its binning are investigated to obtain the best possible sensitivity. First, the starting point in BDT-signal of the SR has to be chosen, in both the DF and SF channels, this way fixing the amount of events entering in the SR. Secondly, one can improve the sensitivity in the SR by subdividing it into bins and performing simultaneous measurements for the combination of the SR bins in the likelihood fit. The choice of the binning for the DF and SF channels is performed separately, due to differing BDT score shapes. It was observed that the full fit (with all the systematic uncertainties included) with SR bins with high statistics becomes unstable, with systematic pulling and profiling. To recover for these effects, one can start by adding bins from the end of the BDT-signal score, ensuring that the background yield of each SR bin remains below a certain threshold, and keep adding SR bins going back in the BDT-signal until no pathological behaviour is observed. When performing the following binning studies it was ensured that the background yield remained below roughly 30 background events, since in the end the fit was observed to be stable with this threshold.

For these studies, the expected CLs values for the $m(\tilde{\chi}_{1}^{\pm}, \tilde{\chi}_{1}^{0})$ = (125, 25), (150, 50), (125, 35) GeV signal samples are considered. The $m(\tilde{\chi}_{1}^{\pm}, \tilde{\chi}_{1}^{0})$ = (150, 50) and (125, 35) GeV signal samples are on the edge of the expected exclusion sensitivity. Table~\ref{tab:binning studies} shows these CLs values obtained. 
\begin{itemize}
\item Firstly, a single bin signal region is considered with cuts of BDT-signal > 0.84 and BDT-signal > 0.785 considered for DF and SF respectively. These are the rows \textit{DF one bin above 0.84} and \textit{SF one bin 0.785}.

\item Secondly, splitting this one bin into multiple bins (5 bins for SF and 4 bins for DF) in the rows \textit{DF binned above 0.84 (4 bins)} and \textit{SF binned above 0.785 (5 bins)}, we see a sensitivity improvement over the one bin case.

\item Thirdly, including more bins with around 30 background events to the values of BDT-signal of 0.81 and 0.77 for DF and SF in the rows \textit{DF binned above 0.81 (16 bins)} and \textit{SF binned above 0.77 (8 bins)}, respectively. We see a further noticeable improvement in the CLs values.

\item Lastly, fitting the DF and SF bins together, we obtain the best sensitivity on the top line of the table. This is the binning used for the final results.
\end{itemize}

\begin{table}[!htb]
\begin{center}
\begin{tabular}{l|rrr}
\noalign{\smallskip}\hline\noalign{\smallskip}
Binning & \multicolumn{3}{c}{CLs for $m(\tilde{\chi}_{1}^{\pm}, \tilde{\chi}_{1}^{0})$ [GeV]}  \\
        & (125,25) &  (150,50) &  (125,35) \\
\noalign{\smallskip}\hline\noalign{\smallskip}
DF one bin above 0.84 & 0.032 & 0.13 & 0.13 \\ 
DF binned above 0.84 (4 bins) & 0.021 & 0.10 & 0.098 \\
DF binned above 0.81 (16 bins) & 0.0056 & 0.046 & 0.043 \\ \hline
SF one bin 0.785  & 0.099 & 0.44 & 0.25 \\
SF binned above 0.785 (5 bins) & 0.087 & 0.43 & 0.22 \\
SF binned above 0.77 (8 bins) & 0.0512 & 0.26 & 0.17 \\\hline
DF and SF binned combined & 0.0028 & 0.038 & 0.030 \\
\noalign{\smallskip}\hline\noalign{\smallskip}
\end{tabular}
\caption{The expected CLs values with statistical and a flat 20$\%$ uncertainty on three representative $m(\tilde{\chi}_{1}^{\pm}, \tilde{\chi}_{1}^{0})$ = (125,25), (150,50) and (125,35) GeV signal samples. Different binning setups are considered.}
\label{tab:binning studies}
\end{center}
\end{table}

The SR is defined to start by taking BDT-signal score > 0.81 for DF and BDT-signal score > 0.77 for SF.

\subsection{BDT-others cut}
\begin{figure}[!htb]
\centering
\includegraphics[width=0.75\linewidth]{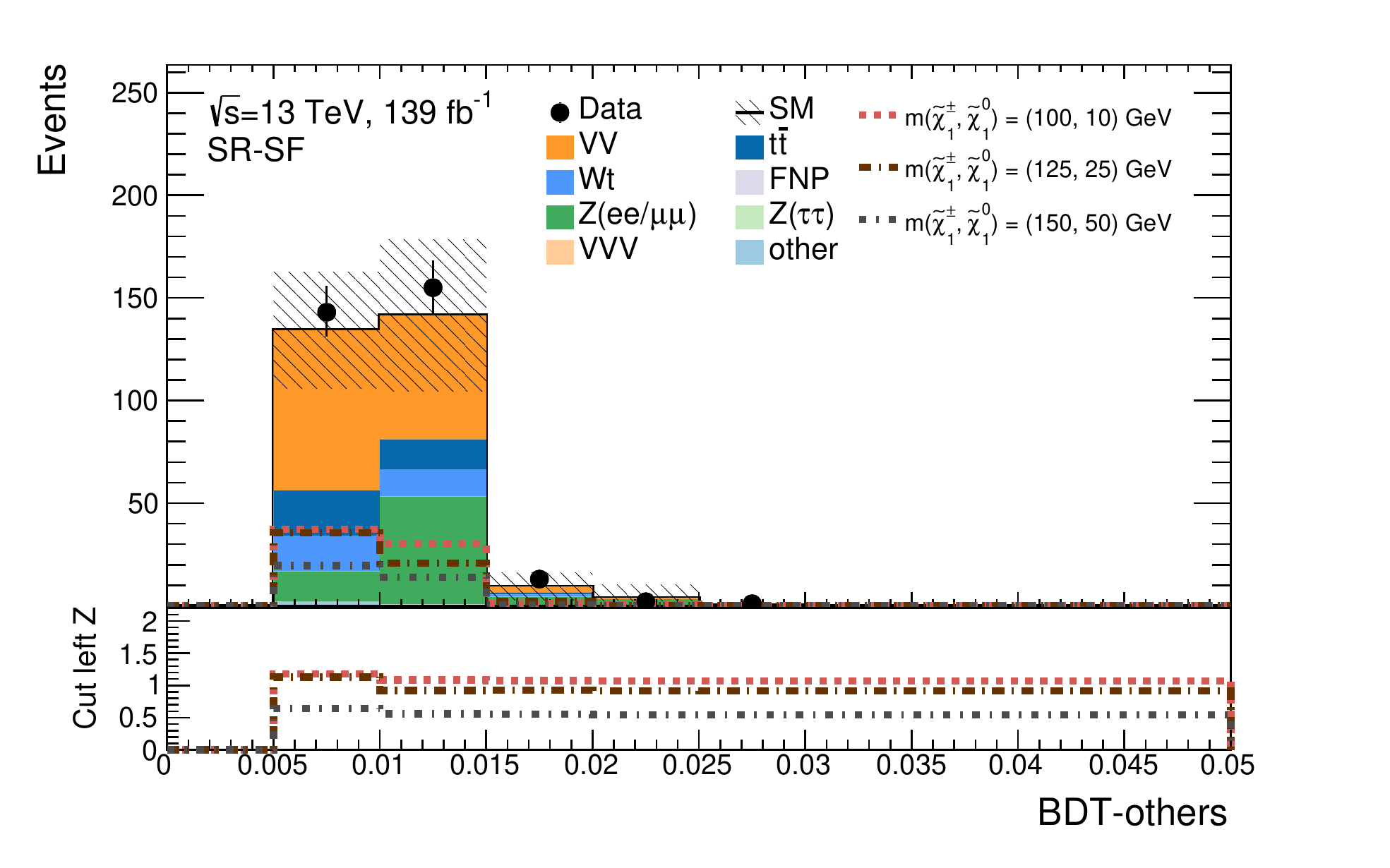}
\caption{The signal and background BDT others distribution in the SF 0-jet channel with $E_{\mathrm{T}}^{\text{miss}}$ significance $>$ 8 and BDT-signal $>$ 0.77.}
\label{fig:SF_SR_BDTothers}
\end{figure}

For the SR in the SF channel, the $Z+$jets contribution is much greater than in the DF channel even when requiring high BDT-signal. Looking at the BDT-others distribution with $E_{\mathrm{T}}^{\text{miss}}$ significance $> 8$ and BDT-signal $> 0.77$ in Fig.~\ref{fig:SF_SR_BDTothers} an additional cut of BDT-others $< 0.01$ is made to further reduce the $Z$+jets in the SRs. This is essentially requiring our SRs events to be very unlike $Z+$jets events. 

\subsection{Shape fit}
Binned SRs are used for the model-dependent fit. For the model-independent fit, we define multiple inclusive SRs for each channel as described below. In the region of high BDT-signal score, as the BDT-signal score increases the backgrounds decrease whilst the signals remain relatively constant from bin to bin. This shape difference can be exploited to perform a shape fit on these bins to achieve a larger sensitivity. The sensitivity can be estimated by combining the significance in quadrature across the bins used for the shape fit. For the shape fit, we use bins with less than approximately 30 background events. Therefore, for the DF0J 16 signal region bins are defined starting from a BDT-signal score of 0.81. For the SF0J, 8 signal region bins are defined starting from a BDT-signal score of 0.77. The definitions of the SR bins in both channels can be seen in Table~\ref{tab:Signal_region_cuts_c1c1ww}.

\clearpage
\begin{table}[!htb]
\begin{center}
\footnotesize
\begin{tabular*}{0.7\textwidth}{@{\extracolsep{\fill}}l|ll}
\noalign{\smallskip}\hline\noalign{\smallskip} 
          \multicolumn{1}{l}{} & &   \\[\dimexpr-\normalbaselineskip-\arrayrulewidth] Signal region (SR)           & SR-DF & SR-SF  \\
\noalign{\smallskip}\hline\noalign{\smallskip}
			 $n_{b \text{-tagged jets}}$     & \multicolumn{2}{c}{= 0} \\ 
			 $n_{\text{non-} b \text{-tagged jets}}$ 	& \multicolumn{2}{c}{= 0} \\ 
			 $E^{miss}_T$ significance	& \multicolumn{2}{c}{$>$8} \\ 
			 $m_{\mathrm{T2}}$ [GeV]            	& \multicolumn{2}{c}{$>$50} \\ 
			 BDT-other					& 			& $<$ 0.01 \\

\noalign{\smallskip}\hline\noalign{\smallskip}
  		Binned SRs &   \multicolumn{2}{c}{} \\ 
\noalign{\smallskip}\hline\noalign{\smallskip}
	           
\multirow{6}{*}{BDT-signal} & {\footnotesize $\in$(0.81,\,0.8125] } & {\footnotesize $\in$(0.77,\,0.775] }\\
        & {\footnotesize $\in$(0.8125,\,0.815] } & {\footnotesize $\in$(0.775,\,0.78] }\\
        & {\footnotesize $\in$(0.815,\,0.8175] } & {\footnotesize $\in$(0.78,\,0.785] }\\
        & {\footnotesize $\in$(0.8175,\,0.82] }  & {\footnotesize $\in$(0.785,\,0.79] }\\
        & {\footnotesize $\in$(0.82,\,0.8225] }  & {\footnotesize $\in$(0.79,\,0.795] }\\
        & {\footnotesize $\in$(0.8225,\,0.825] } & {\footnotesize $\in$(0.795,\,0.80] } \\
        & {\footnotesize $\in$(0.825,\,0.8275] } & {\footnotesize $\in$(0.80,\,0.81] } \\
        & {\footnotesize $\in$(0.8275,\,0.83] }  & {\footnotesize $\in$(0.81,\,1] } \\
        & {\footnotesize $\in$(0.83,\,0.8325] }  & \\
        & {\footnotesize $\in$(0.8325,\,0.835] } & \\
        & {\footnotesize $\in$(0.835,\,0.8375] } & \\
        & {\footnotesize $\in$(0.8375,\,0.84] }  & \\
        & {\footnotesize $\in$(0.84,\,0.845] }  & \\
        & {\footnotesize $\in$(0.845,\,0.85] }  & \\
        & {\footnotesize $\in$(0.85,\,0.86] }  & \\
        & {\footnotesize $\in$(0.86,\,1] }  & \\
\noalign{\smallskip}\hline\noalign{\smallskip}
  		Inclusive SRs &   \multicolumn{2}{c}{} \\ 
\noalign{\smallskip}\hline\noalign{\smallskip}
     & {\footnotesize $\in$(0.81,\,1] }  & {\footnotesize $\in$(0.77,\,1] }\\
     & {\footnotesize $\in$(0.81,\,1] }  & 				\\
     BDT-signal & {\footnotesize $\in$(0.82,\,1] }  & 				\\
     & {\footnotesize $\in$(0.83,\,1] }  & 				\\
     & {\footnotesize $\in$(0.84,\,1] }  & 				\\
     & {\footnotesize $\in$(0.85,\,1] }  & 				\\
     &   					  & {\footnotesize $\in$(0.77,\,1] }\\
     &   					  & {\footnotesize $\in$(0.78,\,1] }\\
     &   					  & {\footnotesize $\in$(0.79,\,1] }\\
     &   					  & {\footnotesize $\in$(0.80,\,1] }\\
     \noalign{\smallskip}\hline\noalign{\smallskip}

\end{tabular*}
\caption{Signal regions definition for the chargino analysis. \label{tab:Signal_region_cuts_c1c1ww}}
\end{center}
\end{table}

For the inclusive SR, we define overlapping inclusive SRs for each channel. These have the previously described cuts of $E_{\mathrm{T}}^{\mathrm{miss}}$ significance $> 8$ and BDT-others $<$ 0.01 (SF only). The looser selections of BDT-signal $>$ 0.81 and BDT-signal $>$ 0.77 for DF0J and SF0J respectively are used, with tighter selections defined in Table~\ref{tab:Signal_region_cuts_c1c1ww}. The looser regions are defined to have the same start point as the binned SRs, and tighter selections are made to maximise the significance.

\section{Background estimation}
The main SM backgrounds in the SRs are the irreducible diboson ($VV$) and the top processes ($t\bar{t}$ and $Wt$). The contribution of these backgrounds is estimated by data-driven normalised scale factors extracted from a likelihood fit to data in a set of dedicated CRs. VRs are used to validate the extrapolation of the fit results to the SRs. VRs lie in-between the CRs and SRs.

\begin{figure}[!htb]
\centering
\includegraphics[width=0.6\textwidth]{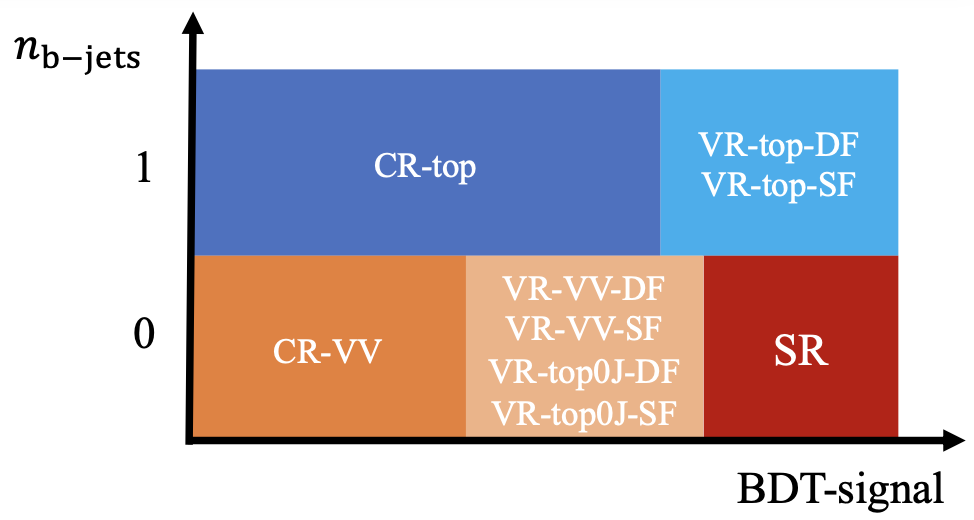}
\caption{A schematic illustration of the CR and VR strategy. Cuts on the BDT-signal score and number of $b$-jets are used to ensure orthogonality among regions. The VRs lie close in phase space to the SR. The BDT-VV and BDT-top scores are used to further increase the purity of each desired background.}
\label{fig:CRVRstrategy}
\end{figure}

In order to define the CRs and VRs, the BDT-signal cuts adopted in the DF and SF SRs are reversed and events in the CRs for top processes are selected with 1 $b$-jet, ensuring orthogonality with the SR and low signal contamination. As already described, for each event, the classifier outputs four scores corresponding to the likelihood of the event being signal, $VV$, top or other backgrounds. The sum of these four scores sum to one and the signal regions are defined with cuts on the BDT-signal score. Additional BDT-VV, BDT-top and BDT-others score cuts are then applied to achieve a good background purity. A $E_{\mathrm{T}}^{\mathrm{miss}}$ significance > 8 cut is also applied to ensure a selection close to the SRs one. The CRs and VRs analysis strategy is pictured in Fig.~\ref{fig:CRVRstrategy}.

\subsection{Control Regions}
The $VV$ background is selected by inverting the BDT-signal in the DF0J SR, thus requiring $\leq$ 0.81 and splitting this phase space into a CR and a VR. The CR-VV in the DF channel is defined by requiring 0.2 $<$ BDT-signal $\leq$ 0.65. Similarly to the DF0J channel, the $VV$ background is selected in the SF0J channel by requiring BDT-signal $\leq$ 0.77 and splitting this phase space into a CR and a VR. The CR-VV in the SF channel is defined by requiring 0.2 $<$ BDT-signal $\leq$ 0.65.
The purity of the $VV$ background is increased with a BDT-VV cut of $>$ 0.2 and BDT-top cut $<$ 0.1. These selections are determined by finding a balance between having sufficient statistics and maximizing the $VV$ purity. A BDT-others cut $<$ 0.01 is additionally applied in the SF channel to suppress $Z$+jets background, as done in the SF0J signal regions.\\

CRs for top background ($t\bar{t}$ and $Wt$) are based on the selection of events with one $b$-jet in the final state. This assures very good top purity and also ensures the orthogonality to the SRs, which have a $b$-veto applied. Then, dedicated CRs and VRs for DF and SF channels are defined. A 0.5 $<$ BDT-signal $\leq$ 0.7 cut is used to define the CR-top in the DF channel while a 0.7 $<$ BDT-signal $\leq$ 0.75 cut is used to define the CR-top in the SF channel. No cut on the BDT-top is applied in these regions, since the top backgrounds are very pure with a $b$-jet requirement, and so this does not provide any purity gain. Table~\ref{tab:table_CR_charg} illustrates the selections for these regions. \\

\begin{table}[!htb]
\begin{center}
\footnotesize 
\begin{tabular*}{\textwidth}{@{\extracolsep{\fill}}l|rrrr}
\noalign{\smallskip}\hline\noalign{\smallskip}
Control region (CR) 			&	\multicolumn{2}{c}{CR-VV} 		& \multicolumn{2}{c}{CR-top} \\
\noalign{\smallskip}\hline\noalign{\smallskip}
$E_{\mathrm{T}}^{\mathrm{miss}} \mathrm{significance}$  & \multicolumn{4}{c}{$>$ 8} \\ 
$m_{\mathrm{T2}}\,[\mathrm{GeV}]  $       		  & \multicolumn{4}{c}{$>$ 50} \\ 
$n_{\mathrm{non-} b \mathrm{-tagged\,jets}}$ 		&\multicolumn{4}{c}{$=$ 0}\\
\noalign{\smallskip}\hline\noalign{\smallskip}
Leptons flavour                                                      &	DF & SF	& DF	&SF \\    
$n_{b \mathrm{-tagged\,jets}}$                           & = 0 		& = 0		& = 1		& = 1	\\ 
BDT-other							& -			& $<$ 0.01 	& -			& $<$ 0.01\\
BDT-signal				&$\in(0.2,0.65]$    &$\in(0.2,0.65]$		&$\in(0.5,0.7]$	&$\in(0.7,0.75]$\\
BDT-VV							&$>$ 0.2		&$>$ 0.2		&-			&-\\
BDT-top							&$<$ 0.1		&$<$ 0.1		&-			&-\\
\noalign{\smallskip}\hline\noalign{\smallskip}
\end{tabular*}
\caption{Control region definitions for extracting the normalisation factors for the dominant background processes in the chargino search. The cuts are applied on top of the preselection. `DF' or `SF' refer to control regions with different lepton flavour or same lepton flavour pair combinations, respectively.}
\label{tab:table_CR_charg}
\end{center}
\end{table}

\begin{figure}[!htb]
\centering
\includegraphics[width=0.45\linewidth]{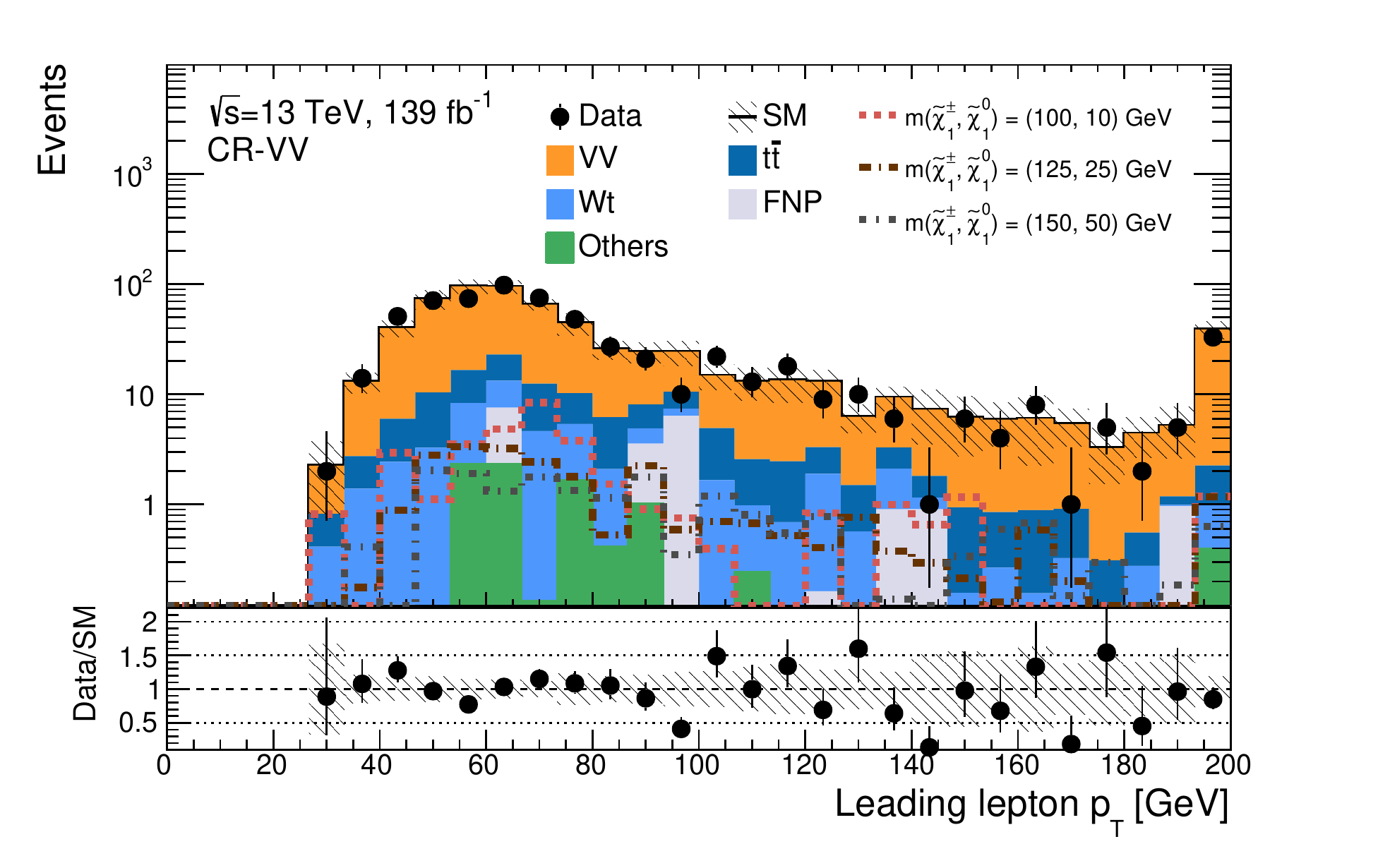}
\includegraphics[width=0.45\linewidth]{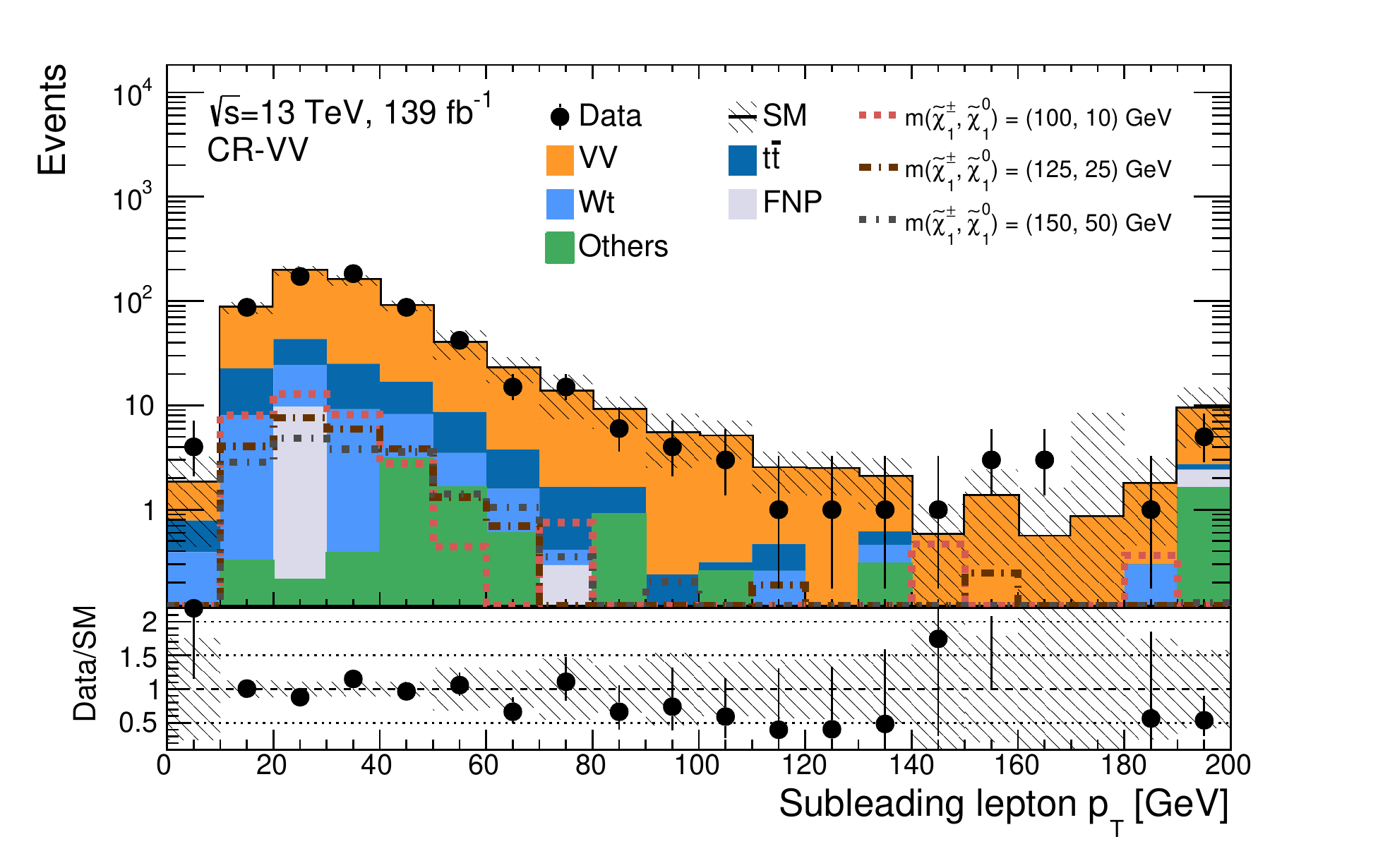}
\includegraphics[width=0.45\linewidth]{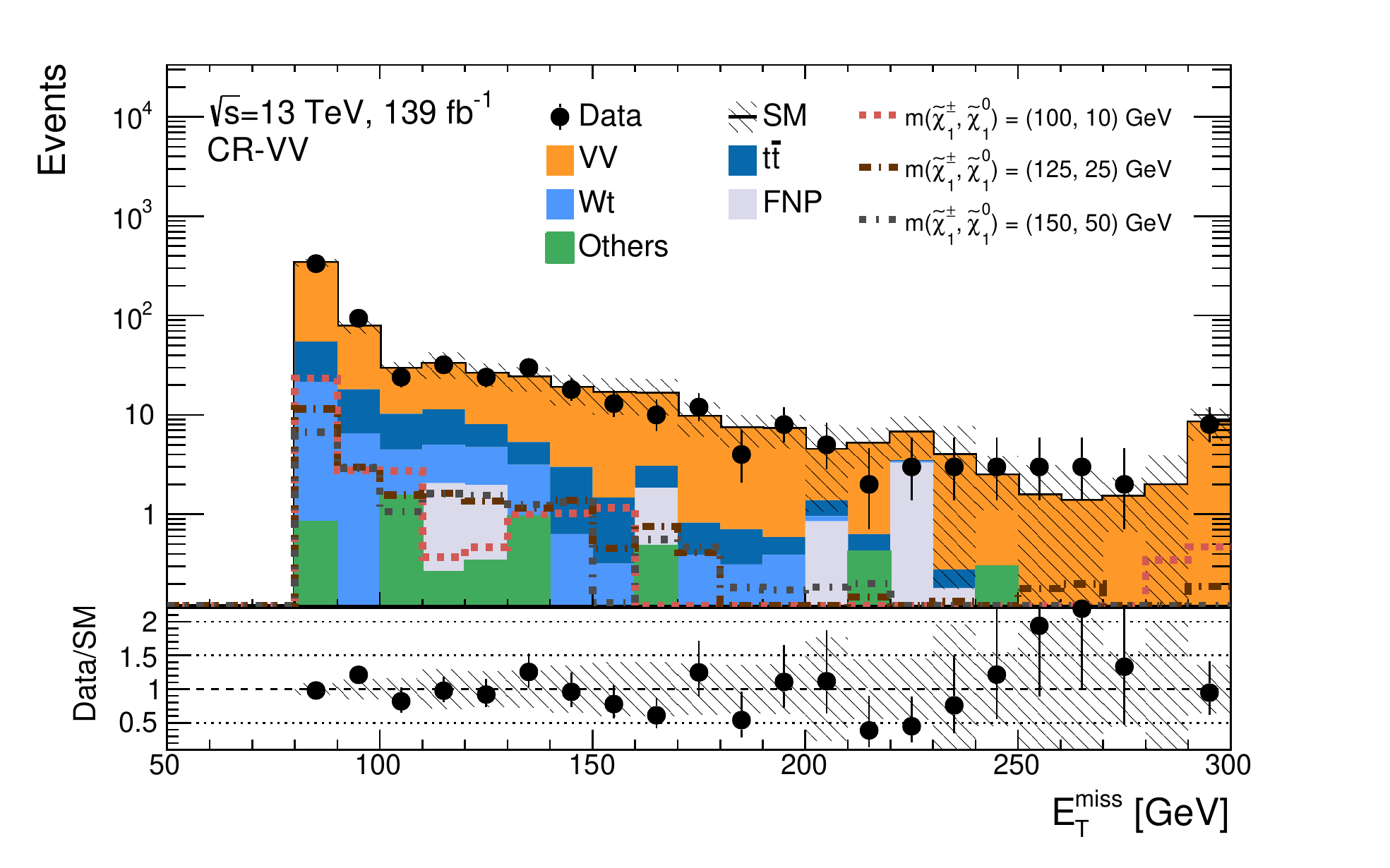}
\includegraphics[width=0.45\linewidth]{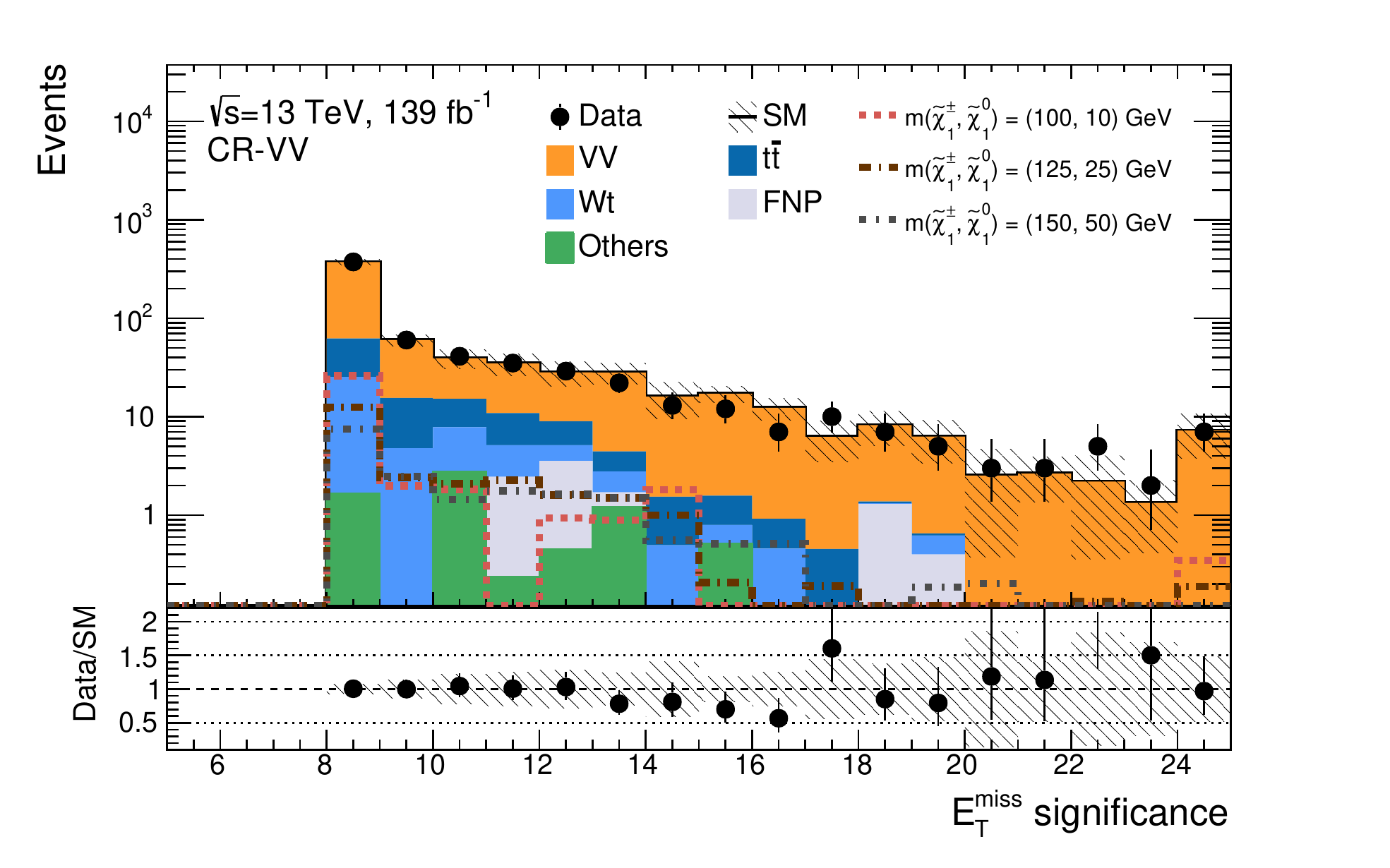}
\includegraphics[width=0.45\linewidth]{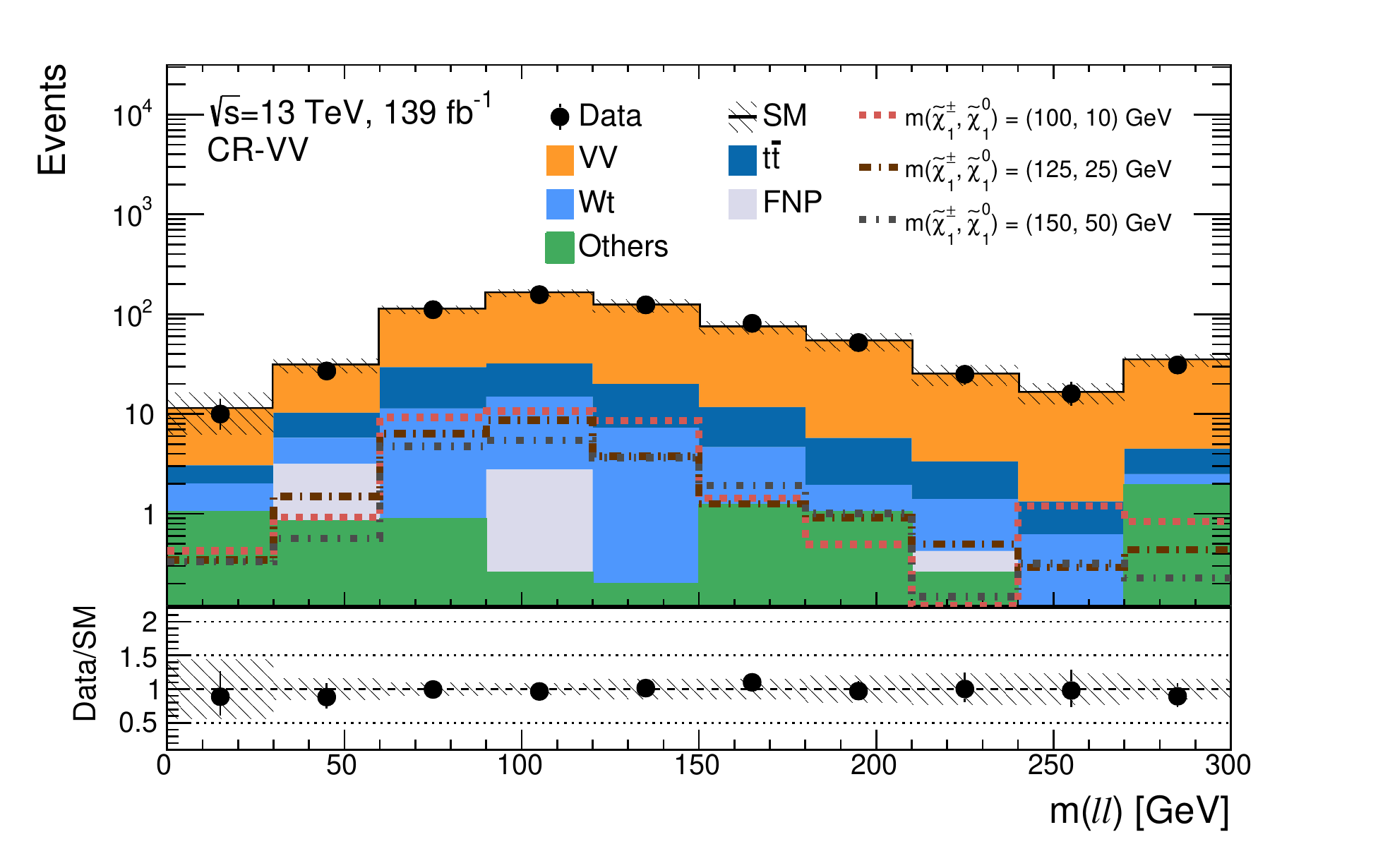}
\includegraphics[width=0.45\linewidth]{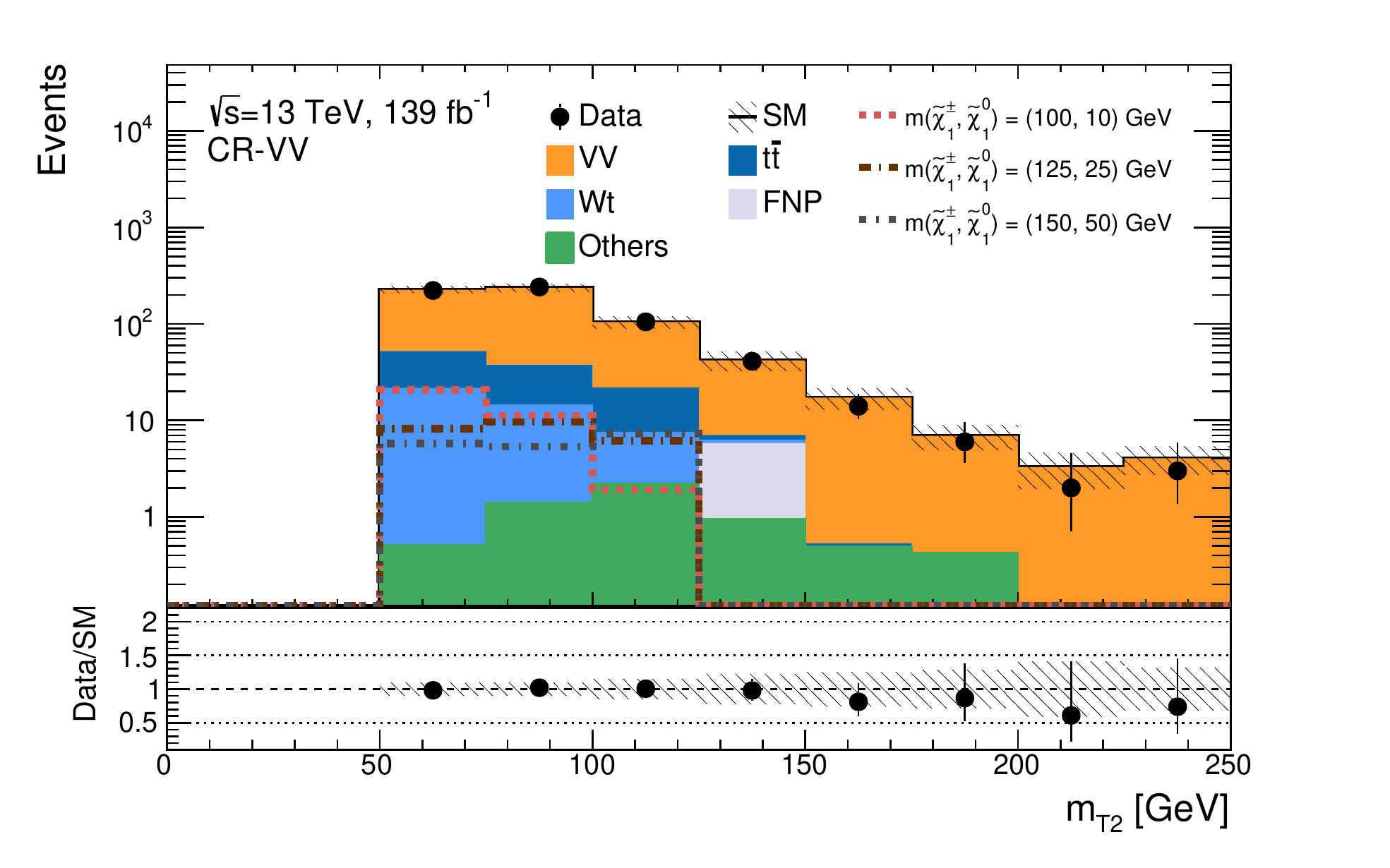}
\includegraphics[width=0.45\linewidth]{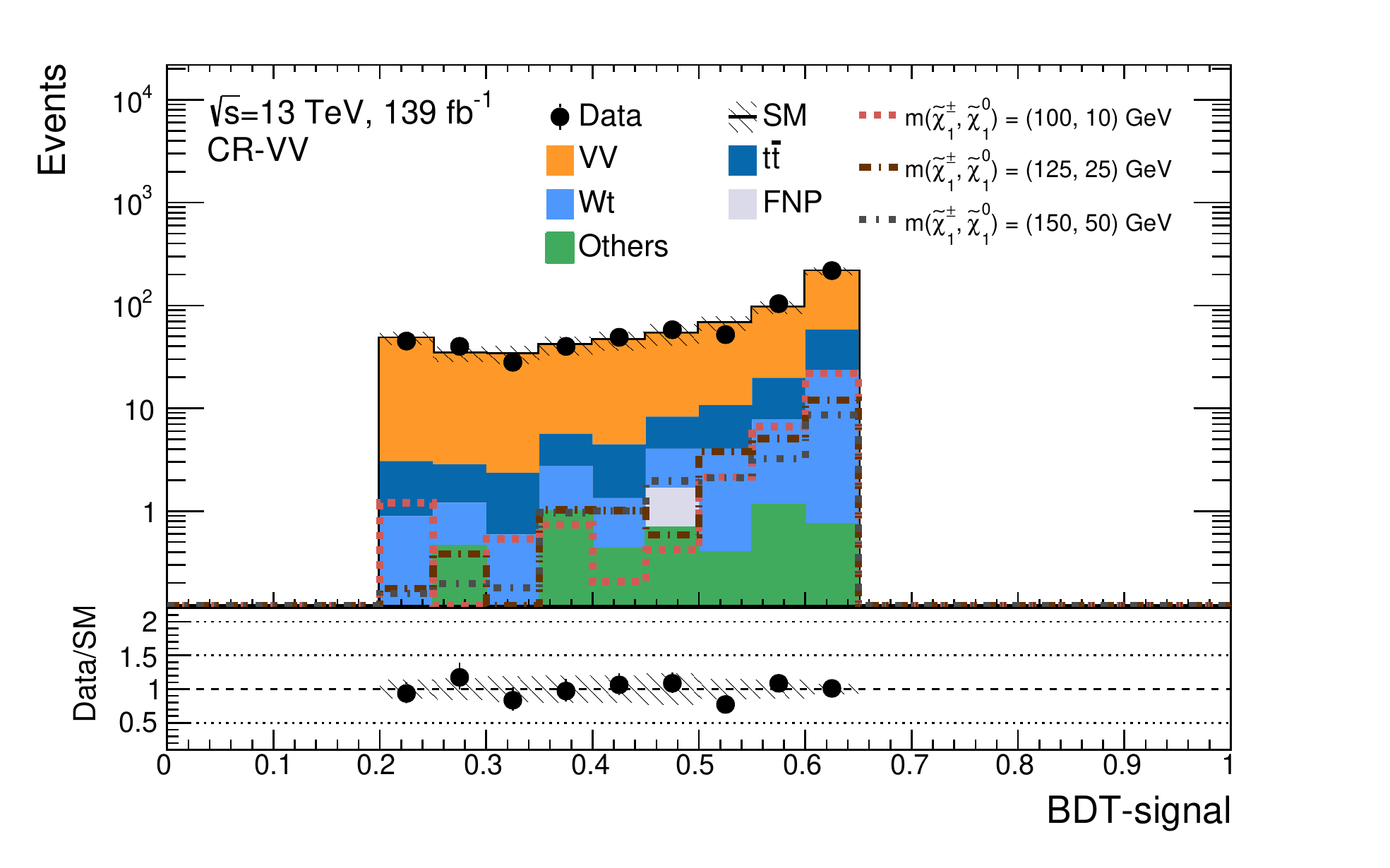}
\includegraphics[width=0.45\linewidth]{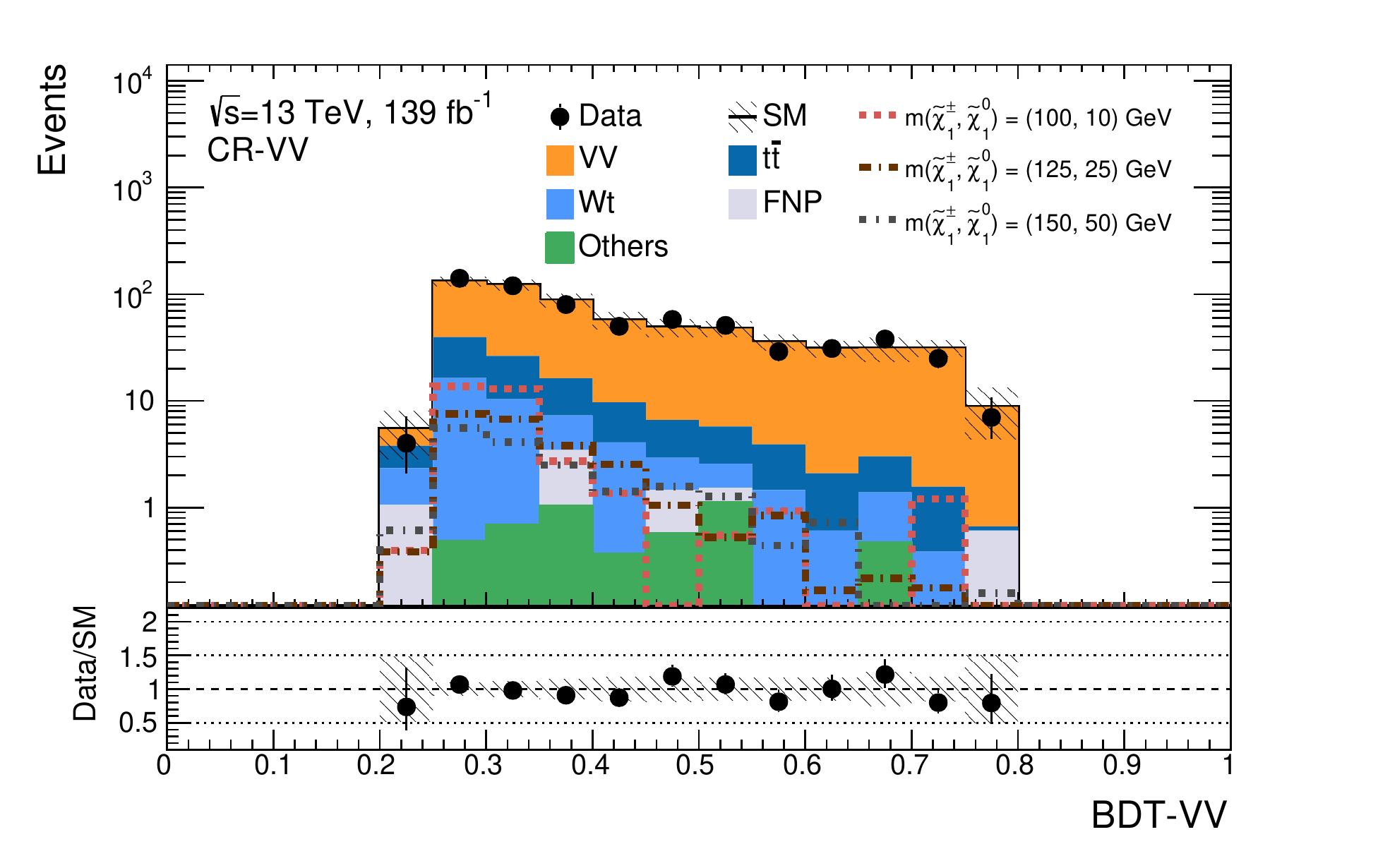}
\includegraphics[width=0.45\linewidth]{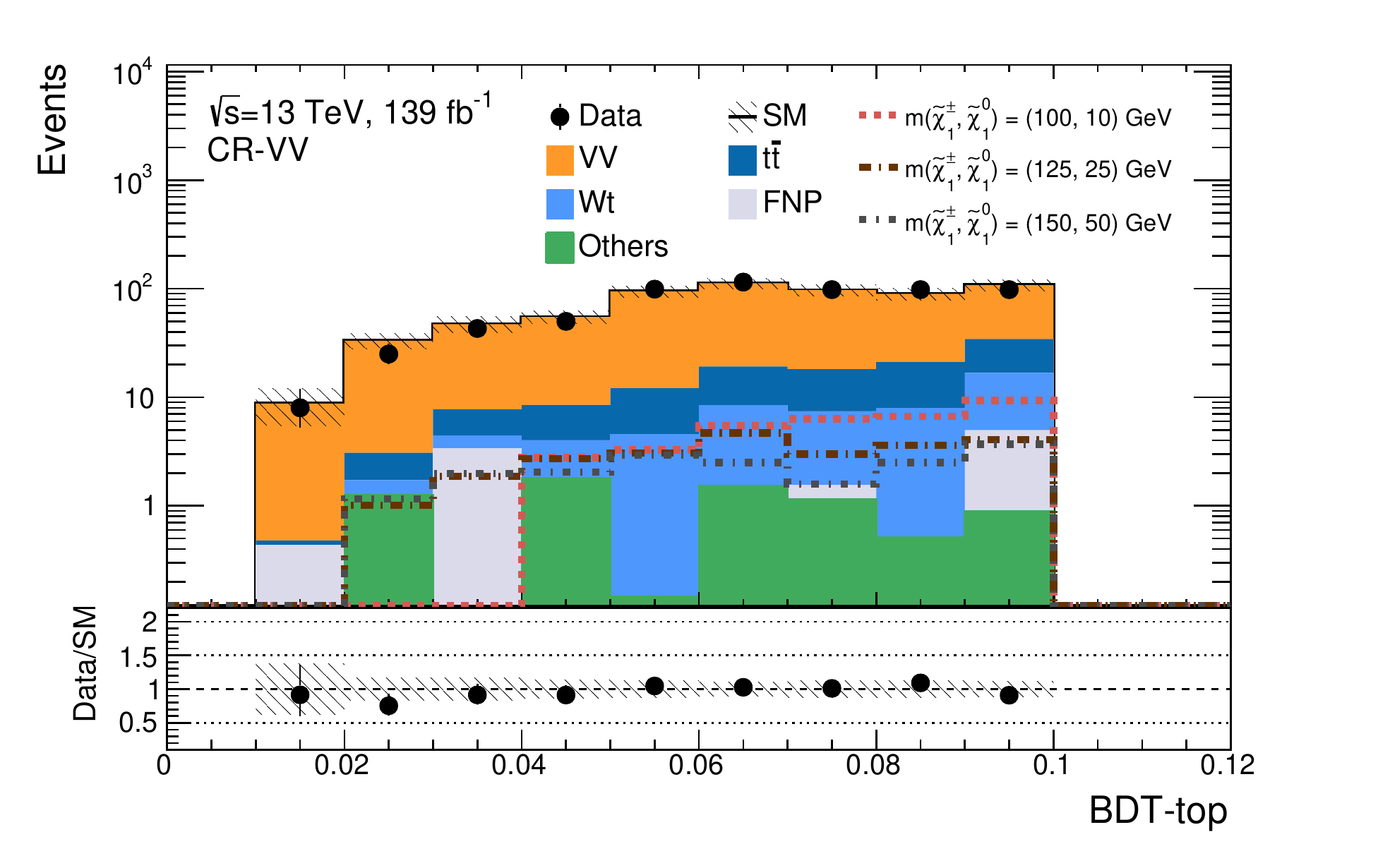}
\includegraphics[width=0.45\linewidth]{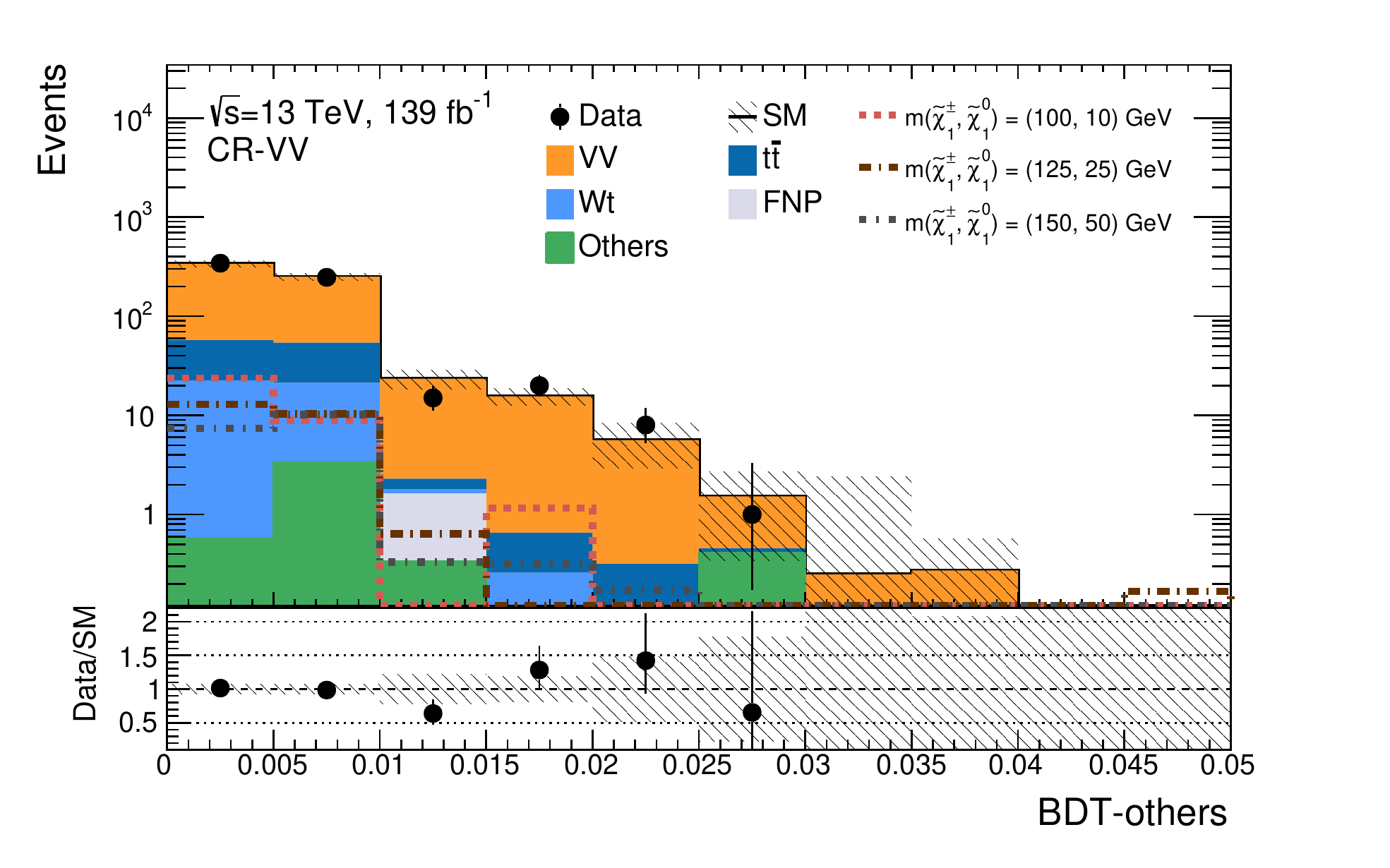}
\caption{The post-fit distributions of the relevant variables for this analysis in the CR-VV region. Both statistical and systematic uncertainties are shown.}
\label{fig:CR_VV_ML}
\end{figure}

\begin{figure}[!htb]
\centering
\includegraphics[width=0.45\linewidth]{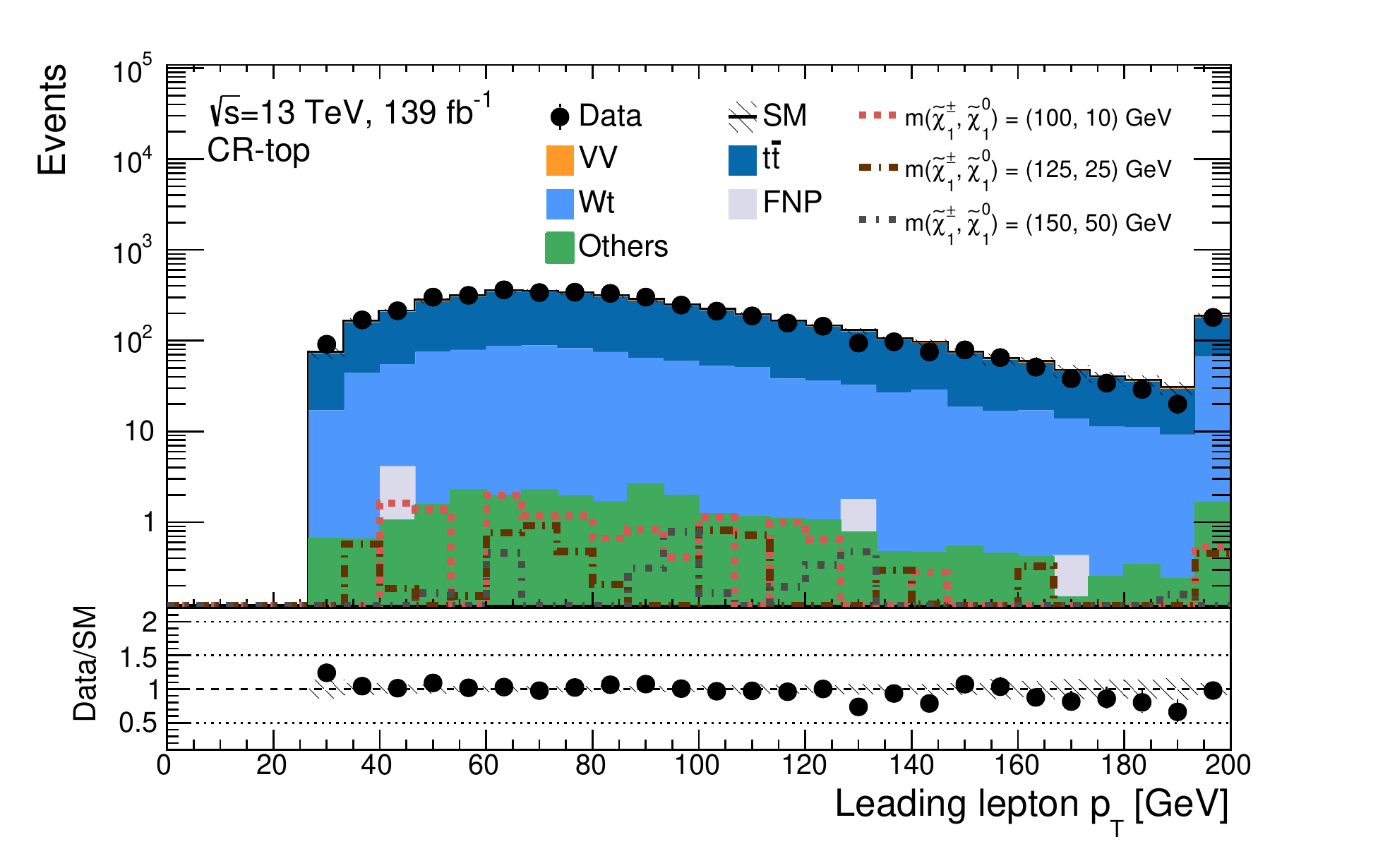}
\includegraphics[width=0.45\linewidth]{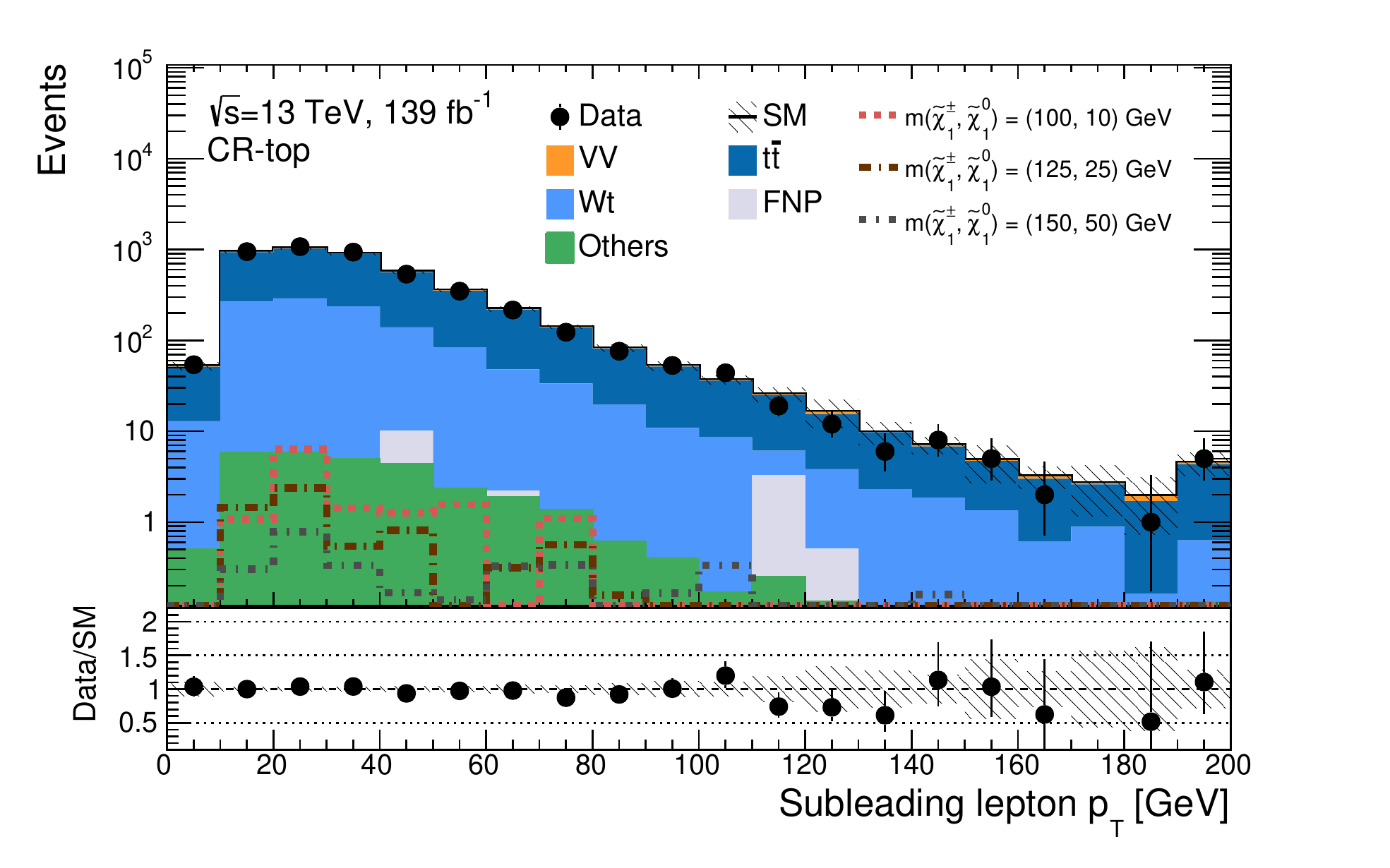}
\includegraphics[width=0.45\linewidth]{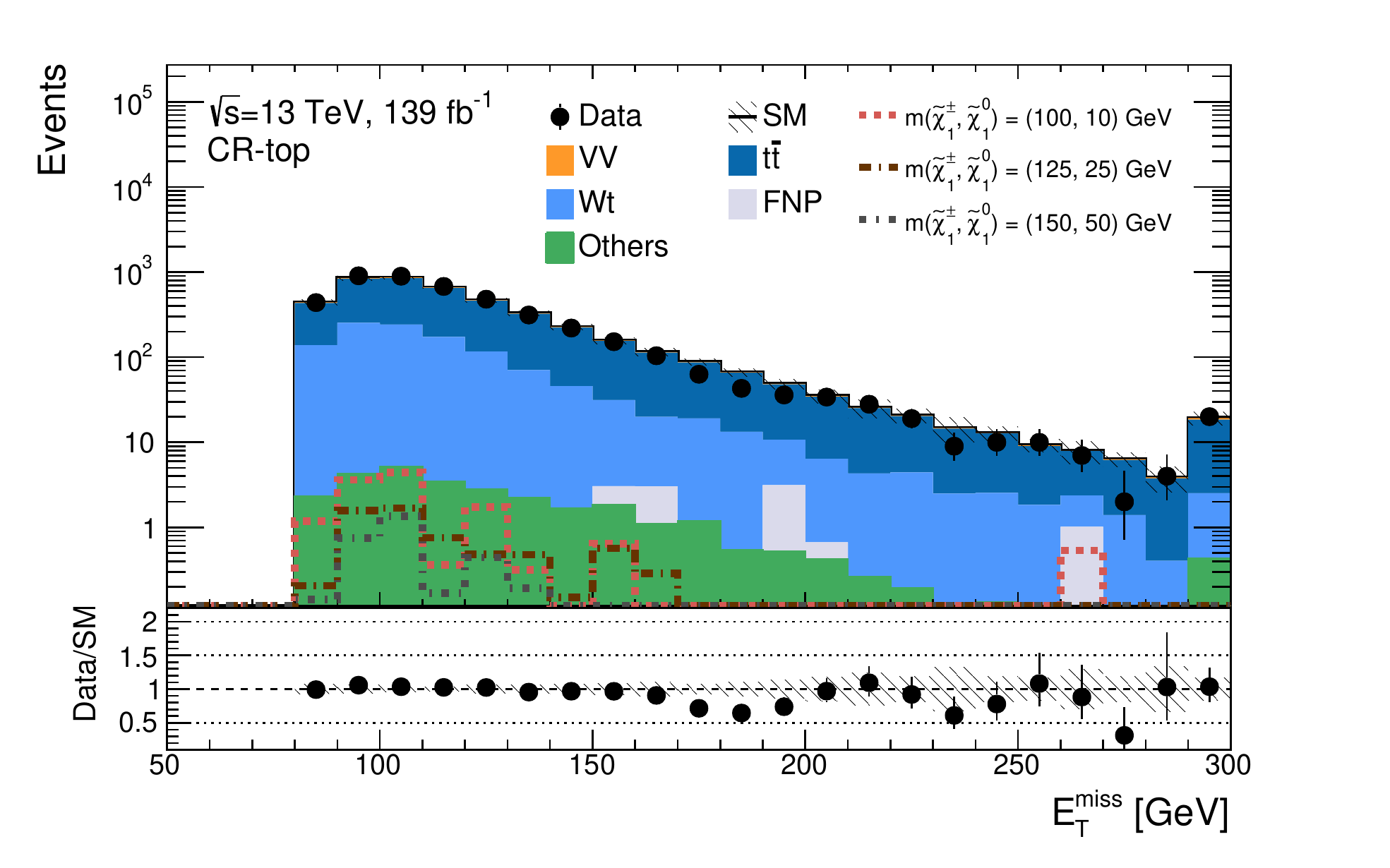}
\includegraphics[width=0.45\linewidth]{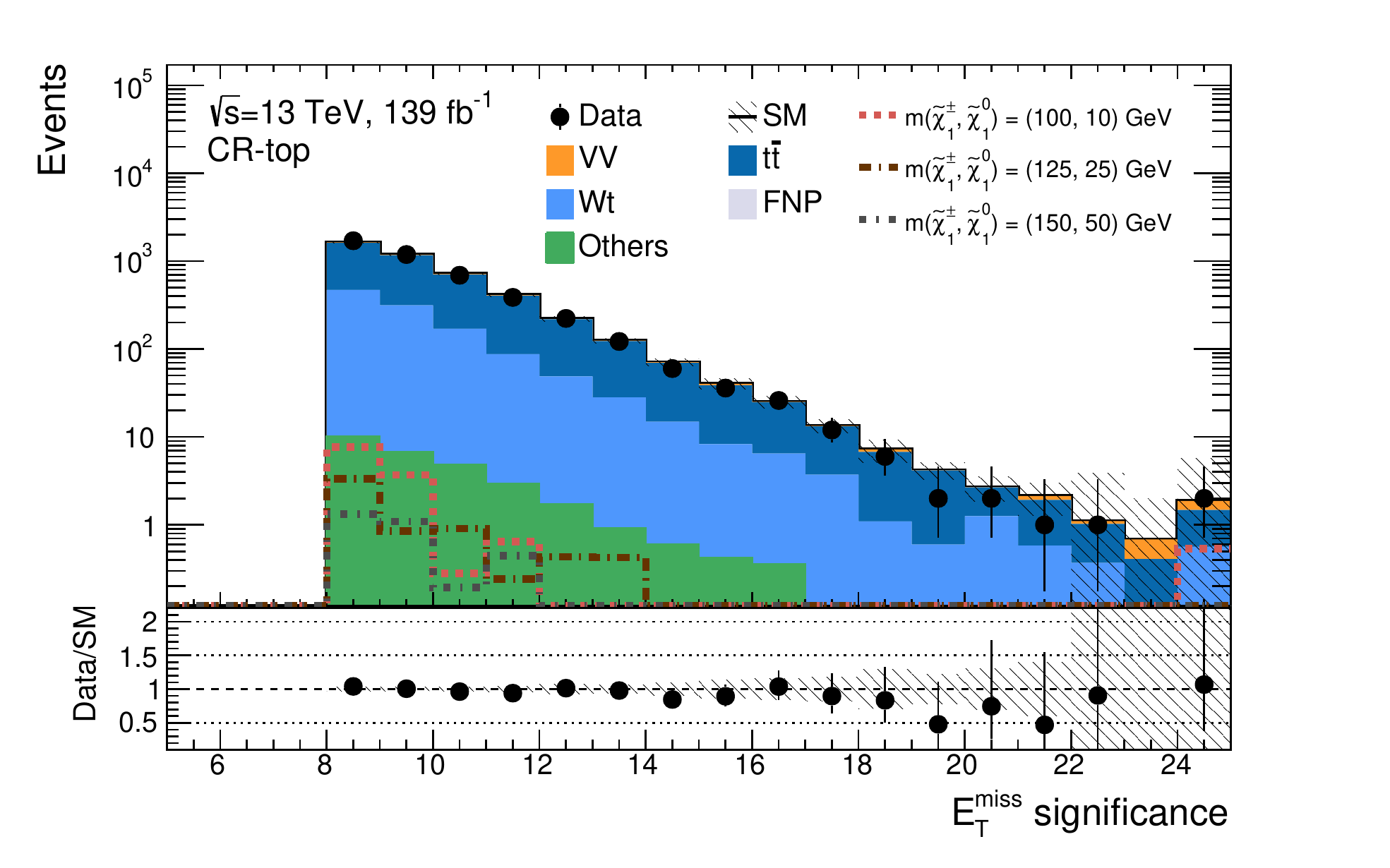}
\includegraphics[width=0.45\linewidth]{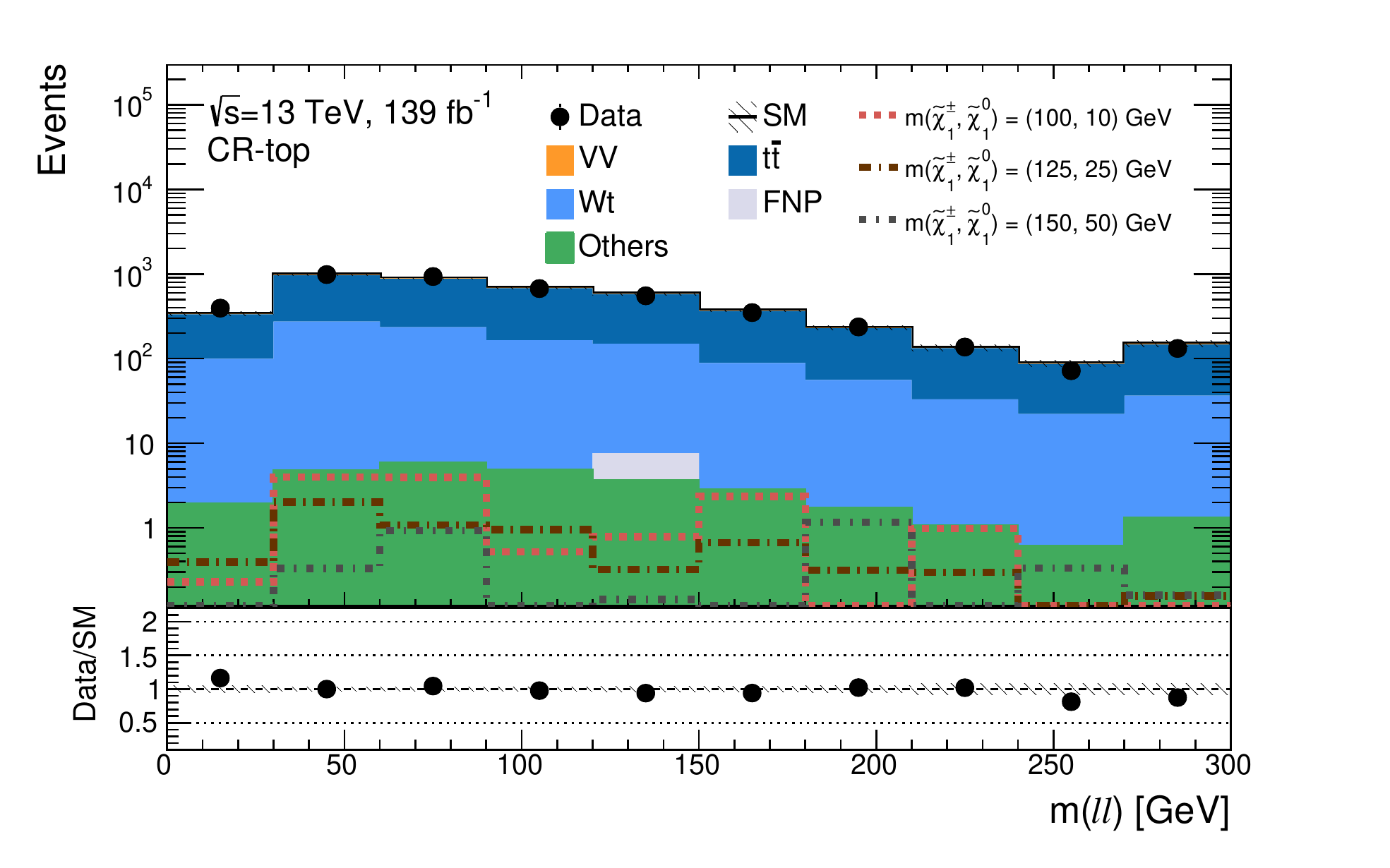}
\includegraphics[width=0.45\linewidth]{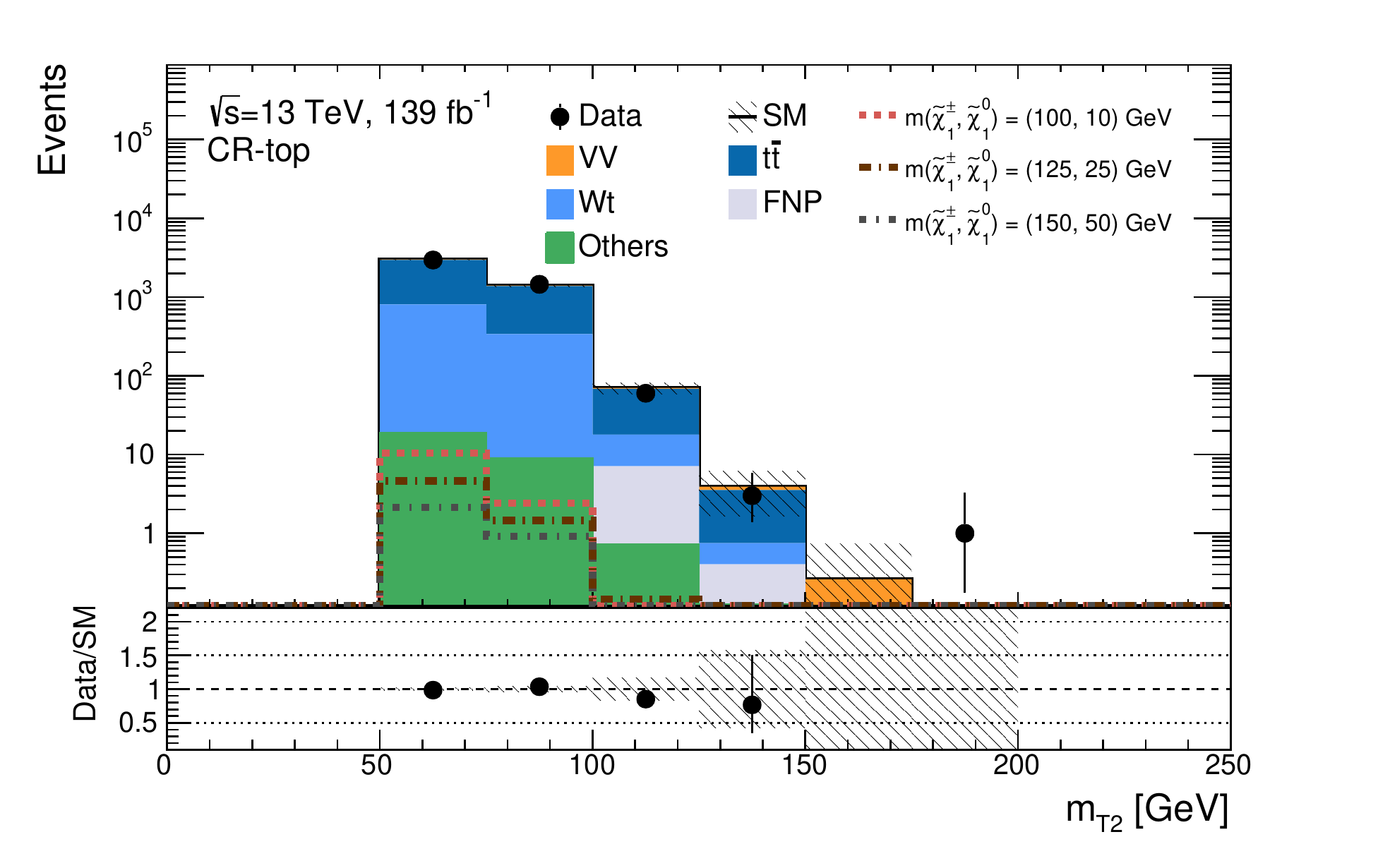}
\includegraphics[width=0.45\linewidth]{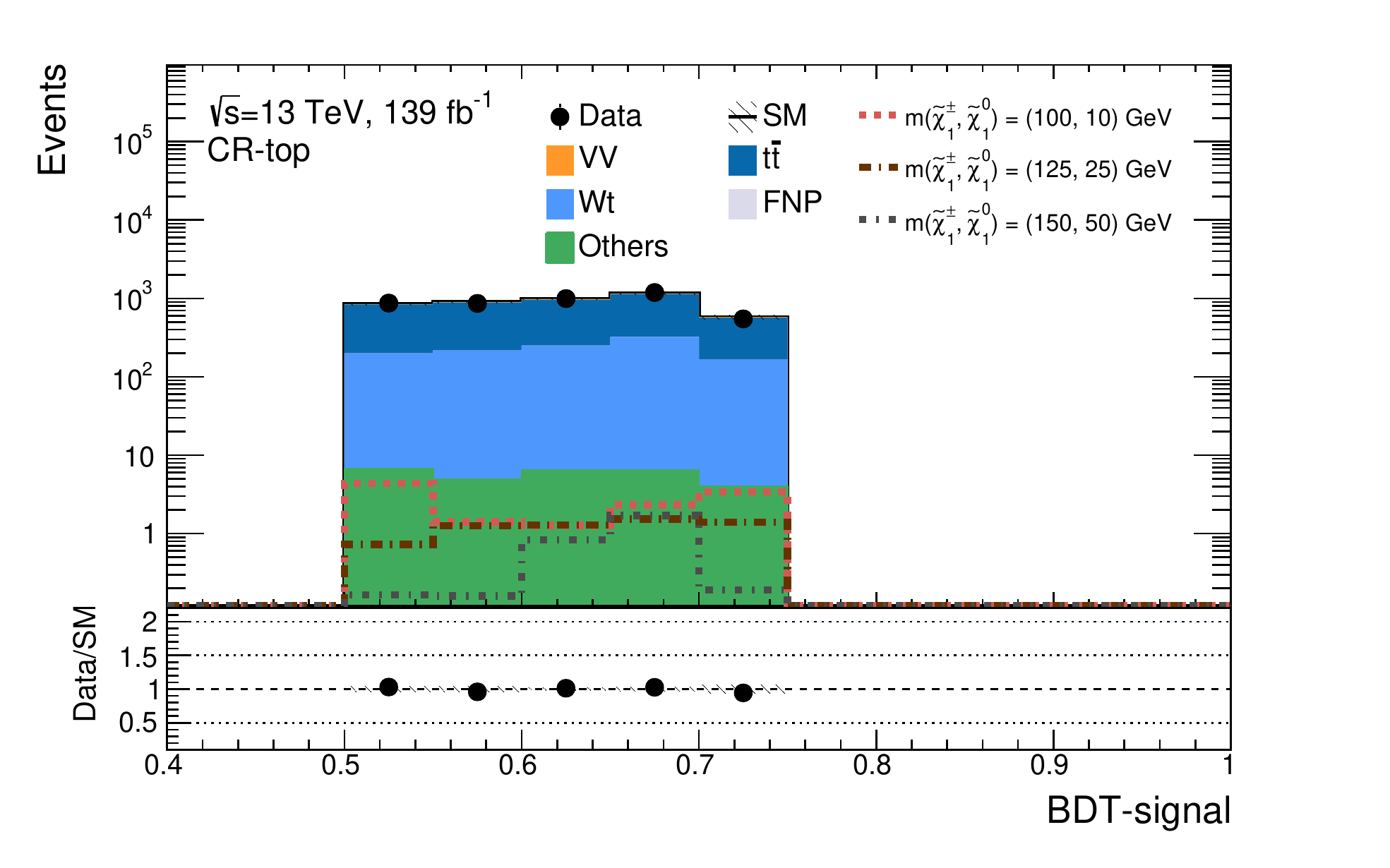}
\includegraphics[width=0.45\linewidth]{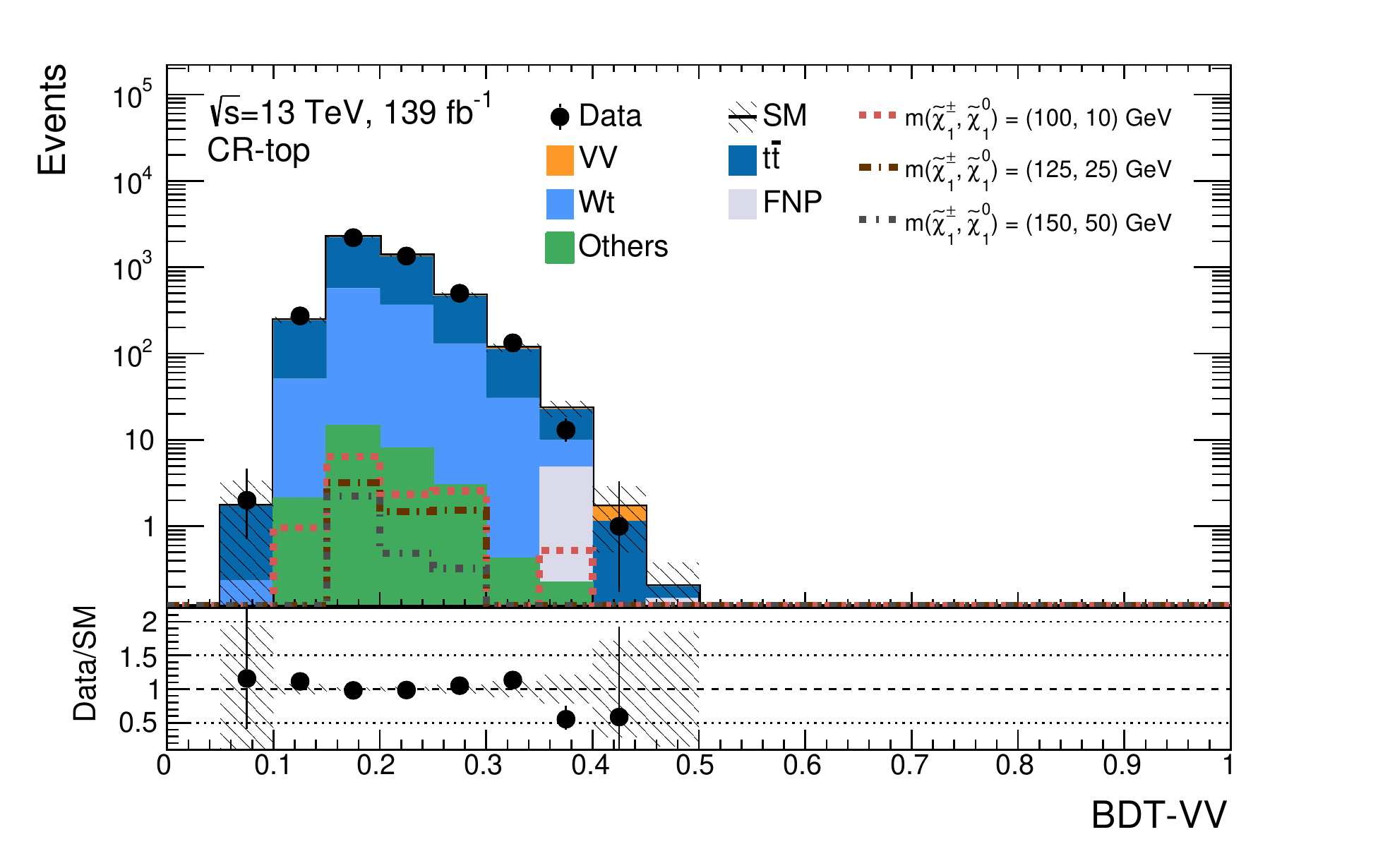}
\includegraphics[width=0.45\linewidth]{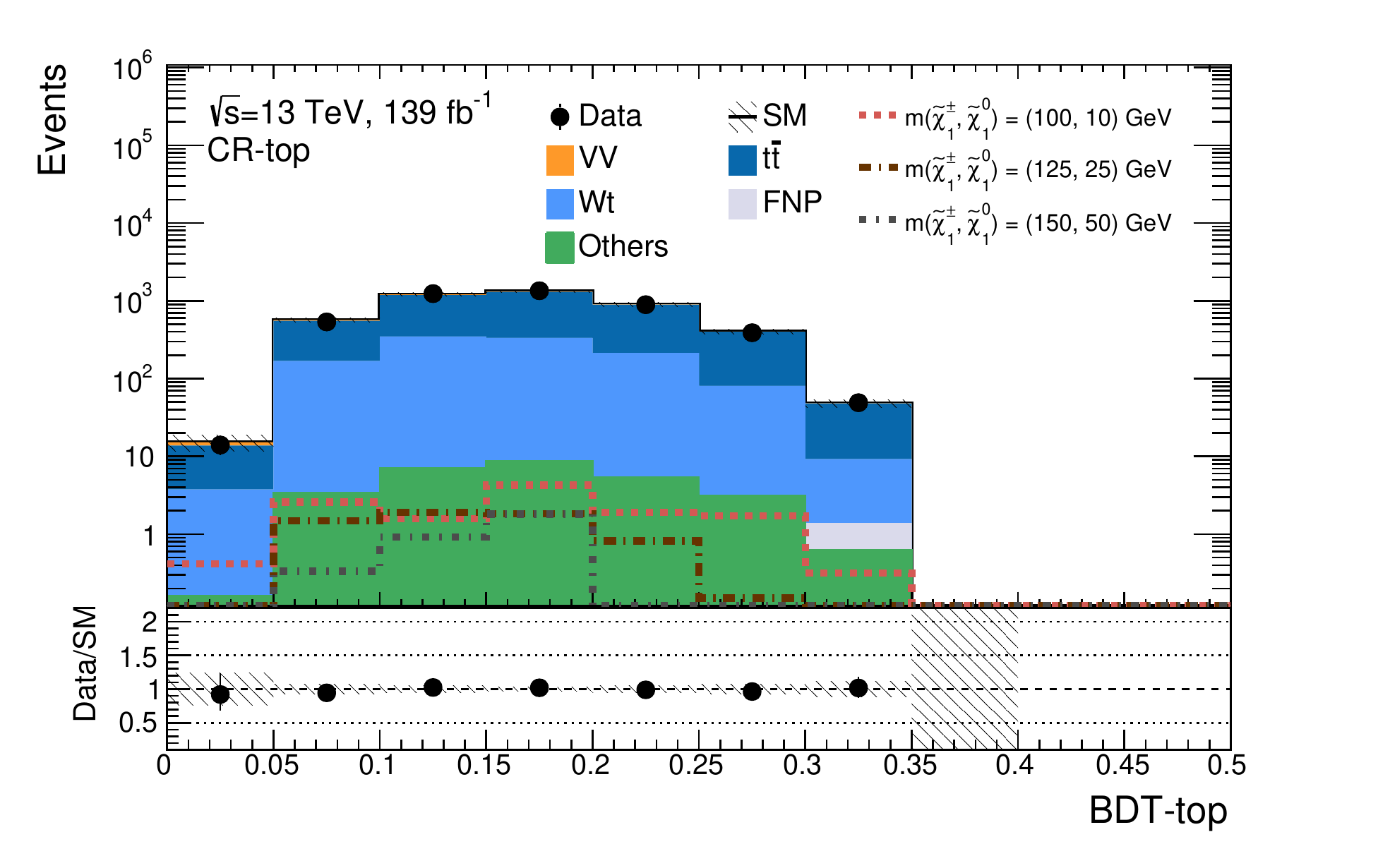}
\includegraphics[width=0.45\linewidth]{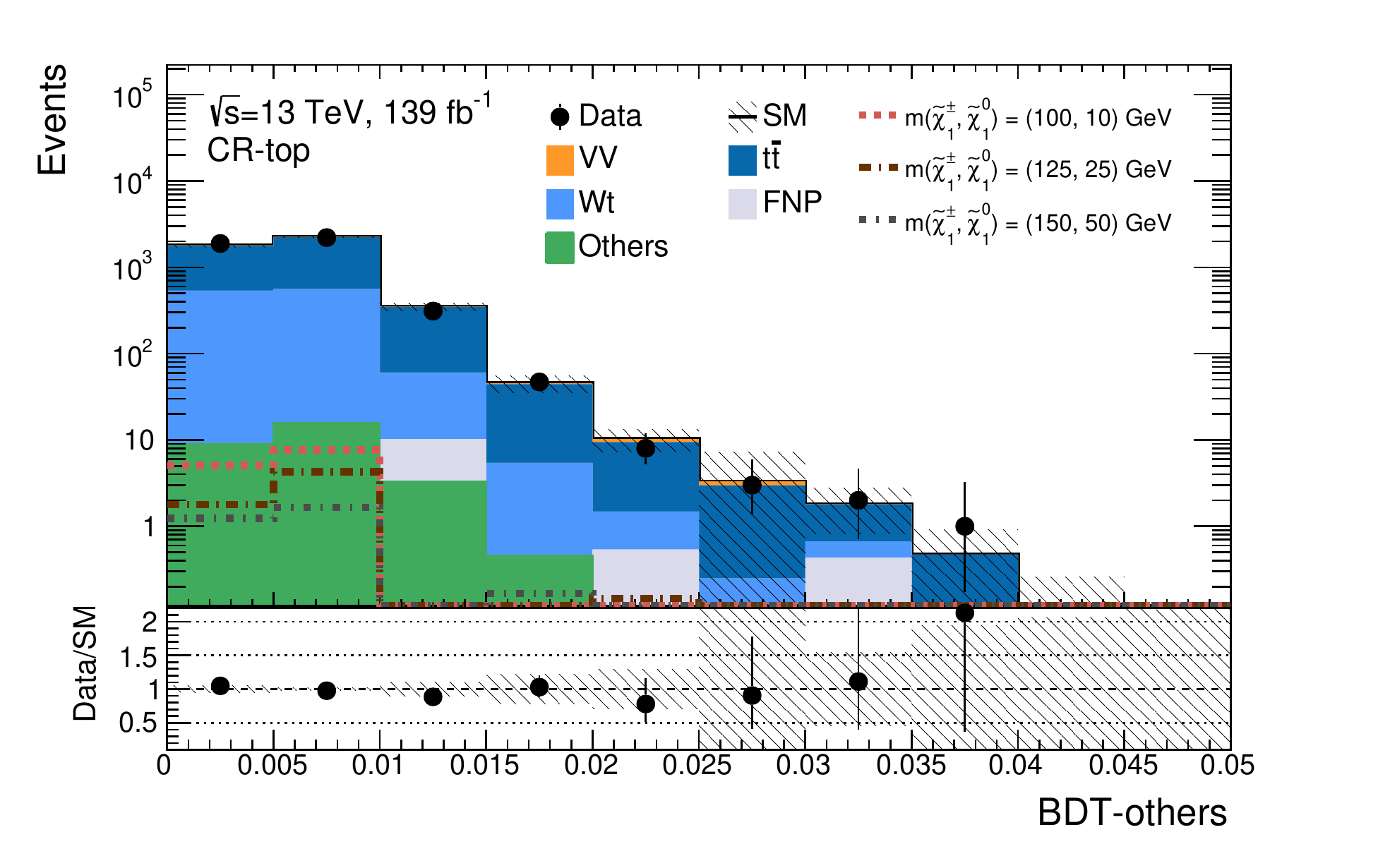}
\caption{The post-fit distributions of the relevant variables for this analysis in the CR-top  region. Both statistical and systematic uncertainties are shown.}
\label{fig:CR_top_ML}
\end{figure}

Diboson and top-quark backgrounds are normalised to the data observed in CR-VV and CR-top in a simultaneous likelihood fit, using a normalization factor for each background ($\mu_{VV}$ and $\mu_{top}$). 
The number of observed events in each CR, as well as the predicted yield of each SM process, is shown in Table~\ref{tab:CRresults}. 
For backgrounds whose normalisation is extracted from the likelihood fit, the yield expected from the MC simulation is also reported.
The normalisation factors applied to the VV and top-quark backgrounds are found to be $\mu_{VV} = 1.38 \pm 0.08$ and $\mu_{top}= 1.09 \pm 0.03$ respectively, where the errors include all uncertainties. These normalisation factors are applied in all of the chargino VRs and SRs. The shapes of kinematic distributions are well reproduced by the simulation in each CR, as shown in the post-fit data and MC distributions for variables relevant for the analysis, in Fig.~\ref{fig:CR_VV_ML} for CR-VV and in Fig.~\ref{fig:CR_top_ML} for CR-top.

\begin{table}[!htb]
\centering
\begin{tabular*}{\textwidth}{@{\extracolsep{\fill}}lrr}
\noalign{\smallskip}\hline\noalign{\smallskip}
Region           & CR-VV            & CR-top              \\
\noalign{\smallskip}\hline\noalign{\smallskip}
Observed events          & $634$              & $4468$                    \\
\noalign{\smallskip}\hline\noalign{\smallskip}
Fitted backgrounds  & $634 \pm 25$          & $4470 \pm 70$   \\
\noalign{\smallskip}\hline\noalign{\smallskip}
Fitted $VV$         & $520 \pm 27$          & $68 \pm 12$              \\
Fitted $t\bar{t}$         & $69 \pm 7$          & $3240 \pm 100$    \\
Fitted single top         & $40 \pm 6$          & $1130 \pm 90$ \\
Other backgrounds          & $4.8_{-4.8}^{+5.1}$          & $29 \pm 5$\\
FNP leptons         & $0.02_{-0.02}^{+1.4}$     & $0.06_{-0.06}^{+12}$\\
\noalign{\smallskip}\hline\noalign{\smallskip}
Simulated $VV$              & $376 $         & $49 $              \\
Simulated $t\bar{t}$        & $63 $          & $2974 $              \\
Simulated single top        & $37 $          & $1040 $             \\
\noalign{\smallskip}\hline\noalign{\smallskip}
\end{tabular*}
\caption{Observed event yields and predicted background yields from the likelihood fit in the CRs for the chargino search. For backgrounds with a normalisation extracted from the likelihood fit, the yield expected from the simulation before the likelihood fit is also shown. The FNP lepton background is calculated using the data-driven matrix method. `Other backgrounds' include the non-dominant background sources, e.g.\ $t\bar{t}$+$V$, Higgs boson and Drell--Yan events. The uncertainties include both statistical and systematic contributions.}
\label{tab:CRresults}
\end{table}

\clearpage

\clearpage

\FloatBarrier

\subsection{Validation Regions}
A set of six validation regions is used to verify the agreement of data and SM predictions within uncertainties in regions with a phase space kinematically close to the SRs, after performing the likelihood fit. The definitions are reported in Table~\ref{tab:table_VR_charg}. \\

\begin{table}[!htb]
\begin{center}
\scriptsize 
\begin{tabular*}{\textwidth}{@{\extracolsep{\fill}}l|rrrrrr}
\noalign{\smallskip}\hline\noalign{\smallskip}
Validation region (VR) 		&VR-VV-DF		&VR-VV-SF		&VR-top-DF		&VR-top-SF   &VR-top0J-DF		&VR-top0J-SF                \\
\noalign{\smallskip}\hline\noalign{\smallskip}
$E_{\mathrm{T}}^{\mathrm{miss}} \mathrm{significance}   $          				& \multicolumn{6}{c}{$>$ 8} \\ 
$m_{\mathrm{T2}} [\mathrm{GeV}]$            				& \multicolumn{6}{c}{$>$ 50} \\ 
$n_{\mathrm{non-} b \mathrm{-tagged\,jets}}$ 	&\multicolumn{6}{c}{$=$ 0}\\
\noalign{\smallskip}\hline\noalign{\smallskip}
$n_{ b \mathrm{-tagged\,jets}}$				& $= 0$		& $= 0$             & $= 1$             &$= 1$ 			&$= 0$ 			&$= 0$ \\
BDT-other							& -			& $<$ 0.01 	& -			& $<$ 0.01		& -				& $<$ 0.01\\
BDT-signal						&$\in(0.65,0.81]$    &$\in(0.65,0.77]$		&$\in(0.7,1]$	&$\in(0.75,1]$ &$\in(0.5,0.81]$	&$\in(0.5,0.77]$\\    BDT-VV							&$>$ 0.2		&$>$ 0.2		&-			&-				&$<$ 0.15			&$<$ 0.15\\
BDT-top							&$<$ 0.1		&$<$ 0.1		&-			&-				&-				&-\\			
\noalign{\smallskip}\hline\noalign{\smallskip}
\end{tabular*}
\caption{Validation region definitions for the dominant background processes in the chargino search used to study the modelling of the SM backgrounds. The cuts are applied on top of the preselection. `DF' or `SF' refer to validation regions with different lepton flavour or same lepton flavour pair combinations, respectively.}
\label{tab:table_VR_charg}
\end{center}
\end{table}

The regions VR-VV-DF, VR-VV-SF, VR-top-DF and VR-top-SF are designed to be in an intermediate BDT-signal selection compared to the corresponding CRs and SRs. The VR-VV-DF is defined by requiring 0.65 $<$ BDT-signal $\leq$ 0.81 while the VR-VV-SF is defined by requiring 0.65 $<$ BDT-signal $\leq$ 0.77. Similarly to the corresponding CRs, the purity of the $VV$ background is increased in the VRs with a BDT-VV $>0.2$ cut and a BDT-top $<$ 0.1 cut. \\
A BDT-signal $>$ 0.7 selection is used for defining the VR-top-DF and a BDT-signal $>$ 0.75 selection is used for defining the VR-top-SF. No cut on the BDT-top is applied in these regions since the top backgrounds are very pure with a $b$-jet requirement, and so this does not provide any purity gain.\\

The VR-top0J-DF and VR-top0J-SF regions are used to validate the extrapolation of the top normalization factor from the region with $n_{ b \mathrm{-tagged\,jets}}=1$ (CR-top) to regions with $n_{ b \mathrm{-tagged\,jets}}=0$ (SRs). These VRs are also used to check that the ratio of $Wt$/$t\bar{t}$ is consistent with the one of the SRs of 0.65, as $Wt$ and $t\bar{t}$ backgrounds are normalized with a common scale factor in the fit. These regions are defined with a selection on BDT-signal score close to the SRs, applying 0.5 $<$ BDT-signal $\leq$ 0.81 for VR-top0J-DF and 0.5 $<$ BDT-signal $\leq$ 0.77 for VR-top0J-SF. Also, a BDT-VV $<$ 0.15 cut is applied to both to reduce the VV yields and to ensure orthogonality with the other VRs. \\

\begin{figure}[!htb]
\centering
\includegraphics[width=1.\linewidth]{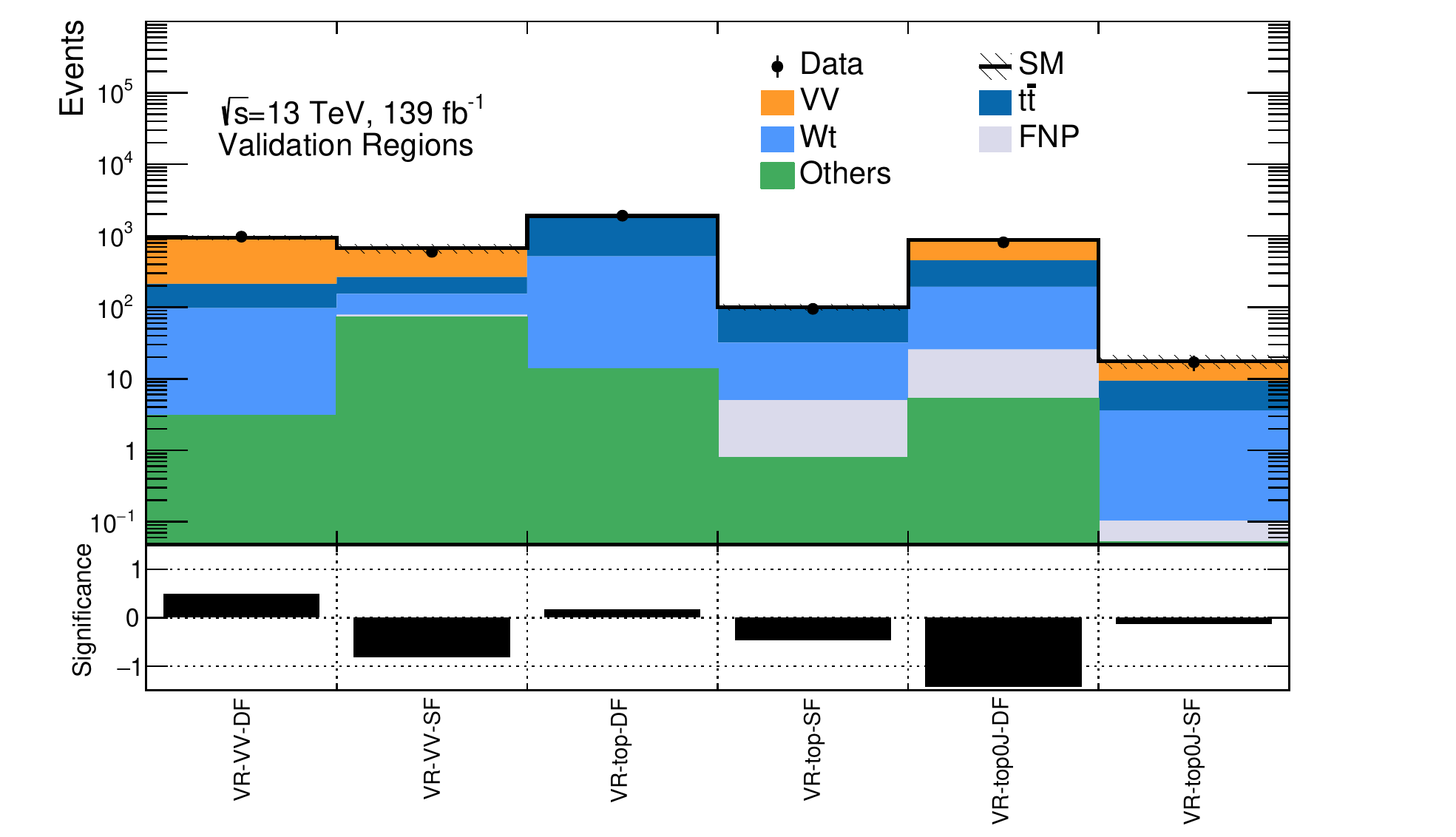}        
\caption{The upper panel shows the observed number of events in the VRs defined in Table~\ref{tab:table_VR_charg}, together with the expected SM backgrounds obtained after the background fit in the CRs. `Others' include the non-dominant background sources, e.g.\ $t \bar{t}$+$V$, Higgs boson and Drell-Yan events. The uncertainty band includes systematic and statistical errors from all sources. The lower panel shows the significance as defined in Section~\ref{sec:StatisticalSignificance}.}
\label{fig:VRpull_charg}
\end{figure}

The number of observed events and the predicted yields of each SM process in each VR is reported in Table~\ref{tab:VRresults} and shown in Fig.~\ref{fig:VRpull_charg}. 

\begin{table}[!htb]
\centering
{\scriptsize
\begin{tabular*}{\textwidth}{@{\extracolsep{\fill}}lrrrrrr}
\noalign{\smallskip}\hline\noalign{\smallskip}
Regions           & VR-VV-DF          & VR-VV-SF          & VR-top-DF                 & VR-top-SF            & VR-top0J-DF           & VR-top0J-SF                 \\[-0.05cm]
\noalign{\smallskip}\hline\noalign{\smallskip}
Observed events          & $972$              & $596$              & $1910$              & $95$              & $810$              & $17$                    \\
\noalign{\smallskip}\hline\noalign{\smallskip}
Fitted backgrounds           & $940 \pm 60$          & $670 \pm 90$          & $1900 \pm 90$          & $101 \pm 10$          & $880 \pm 40$          & $18 \pm 4$              \\       
\noalign{\smallskip}\hline\noalign{\smallskip}
Fitted $VV$ 	& $730 \pm 50$          & $400 \pm 50$          & $32 \pm 13$          & $2.2 \pm 2.1$          & $427 \pm 30$          & $8.1 \pm 2.6$              \\
Fitted $t\bar{t}$       & $116\pm 12$          & $111 \pm 11$          & $1350 \pm 50$          & $67 \pm 7$          & $260 \pm 21$          & $5.8 \pm 1.8$              \\
Fitted single top   & $94 \pm 19$          & $75 \pm 11$          & $500 \pm 60$          & $27 \pm 7$          & $168 \pm 18$          & $4 \pm 1$              \\      
Other backgrounds  & $3.1 \pm 1.5$          & $70 \pm 70$          & $13.6 \pm 2.5$          & $0.8 \pm 0.4$          & $5.2 \pm 1.9$          & $0.05 \pm 0.05$              \\       
FNP leptons      & $0.02_{-0.02}^{+2.3}$          & $7 \pm 4$          & $0.03_{-0.03}^{+5}$          & $4.2 \pm 1.3$          & $21 \pm 8$          & $0.05_{-0.05}^{+0.15}$              \\   
\noalign{\smallskip}\hline\noalign{\smallskip}
Simulated $VV$  & $527 $          & $291 $          & $23 $          & $1.6 $          & $309$          & $5.9 $   \\        
Simulated $t\bar{t}$         & $106$          & $102 $          & $1240 $          & $61 $          & $239 $          & $5.3 $          \\ 
Simulated single top      & $87 $          & $69 $          & $460 $          & $25 $          & $154 $          & $3.2 $            \\   
\noalign{\smallskip}\hline\noalign{\smallskip}
\end{tabular*}
}
\caption{Observed event yields and predicted background yields in the VRs defined in Table~\ref{tab:table_VR_charg}. For backgrounds with a normalisation extracted from the likelihood fit in the CRs, the yield expected from the simulation before the likelihood fit is also shown. The FNP lepton background is calculated using the data-driven matrix method. `Other backgrounds' include the non-dominant background sources, e.g.\ $t \bar{t}$+$V$, Higgs boson and Drell--Yan events. The uncertainties include both statistical and systematic contributions.} 
\label{tab:VRresults}
\end{table}

\begin{figure}[!htb]
\centering
\includegraphics[width=0.45\linewidth]{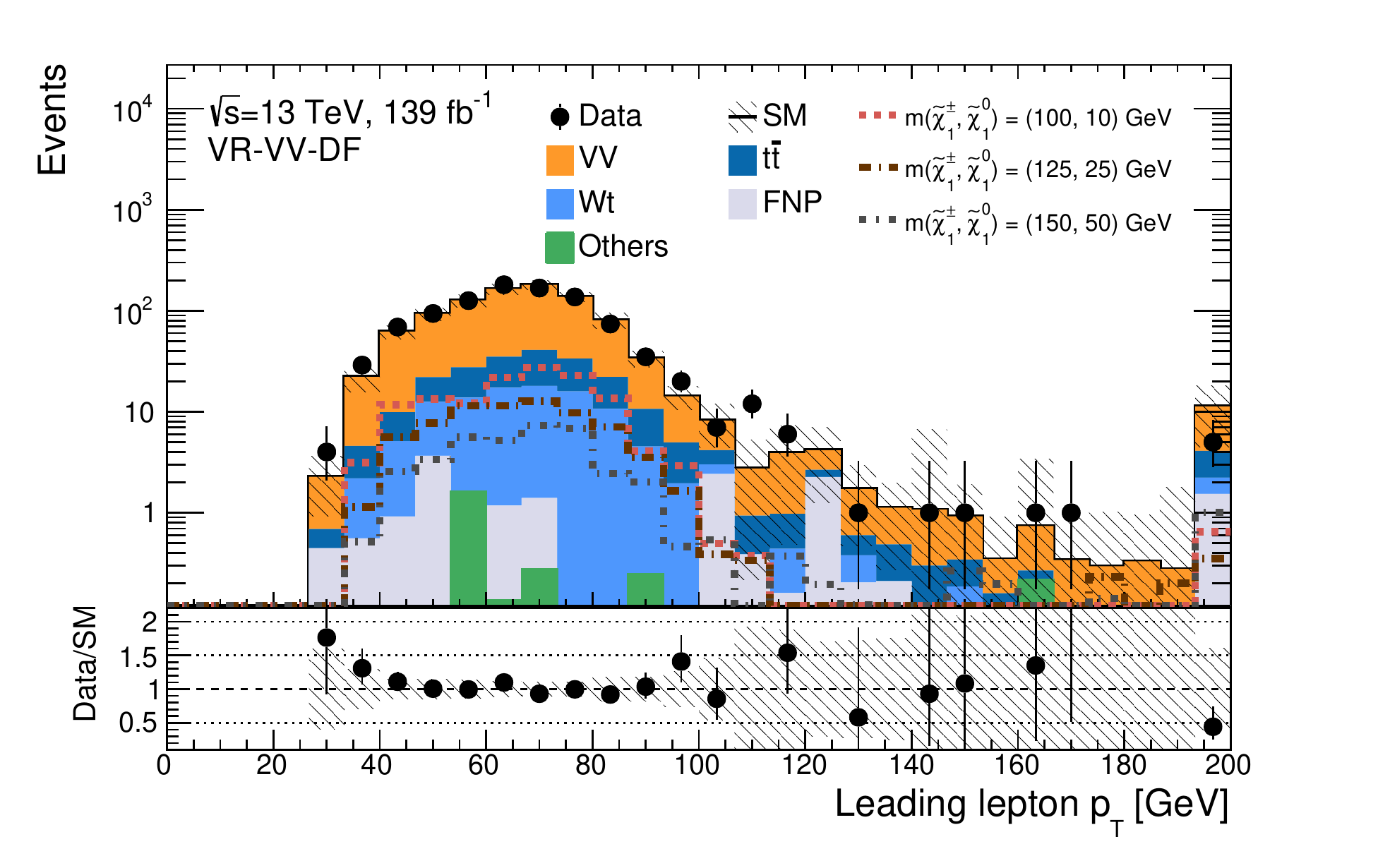}
\includegraphics[width=0.45\linewidth]{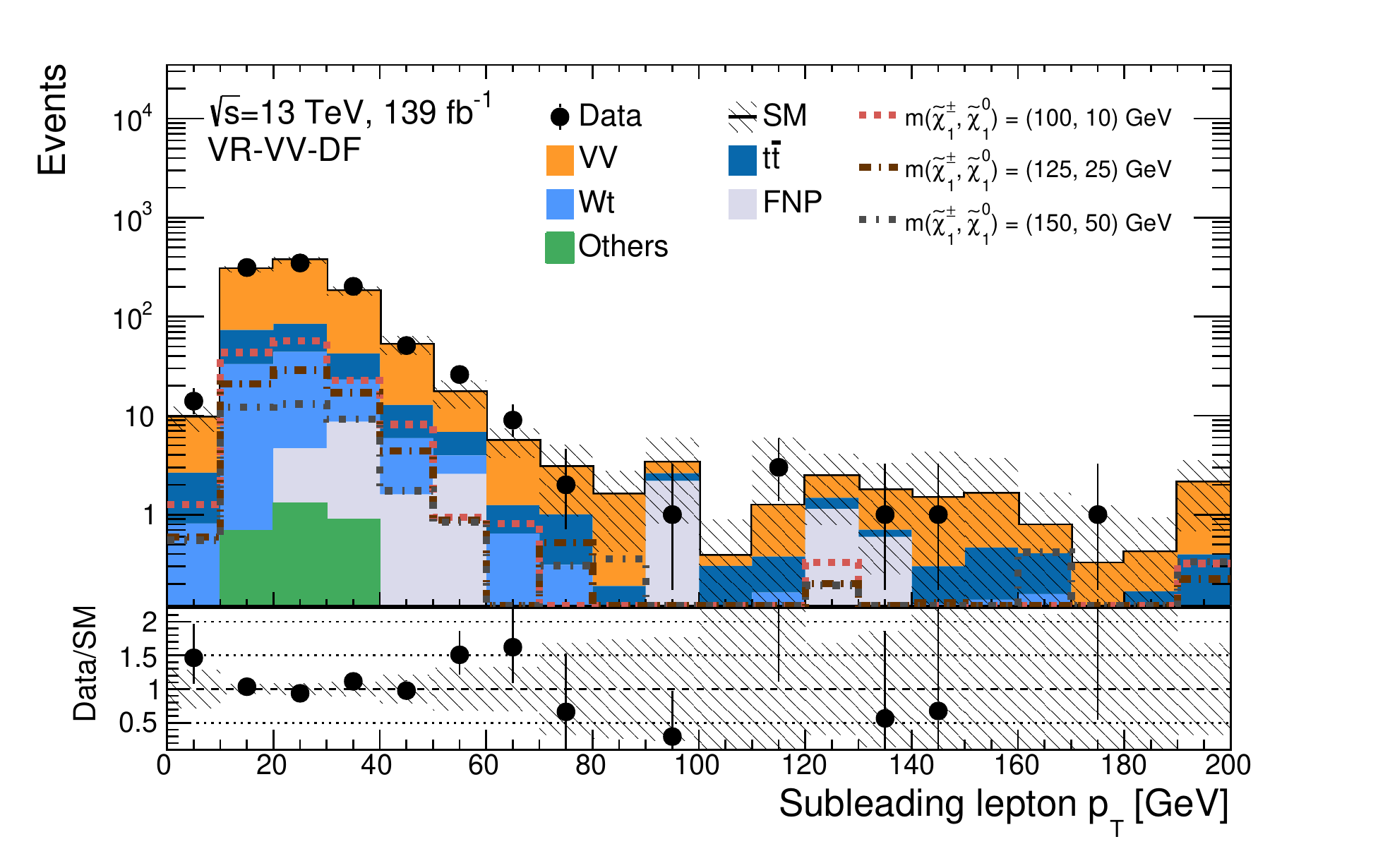}
\includegraphics[width=0.45\linewidth]{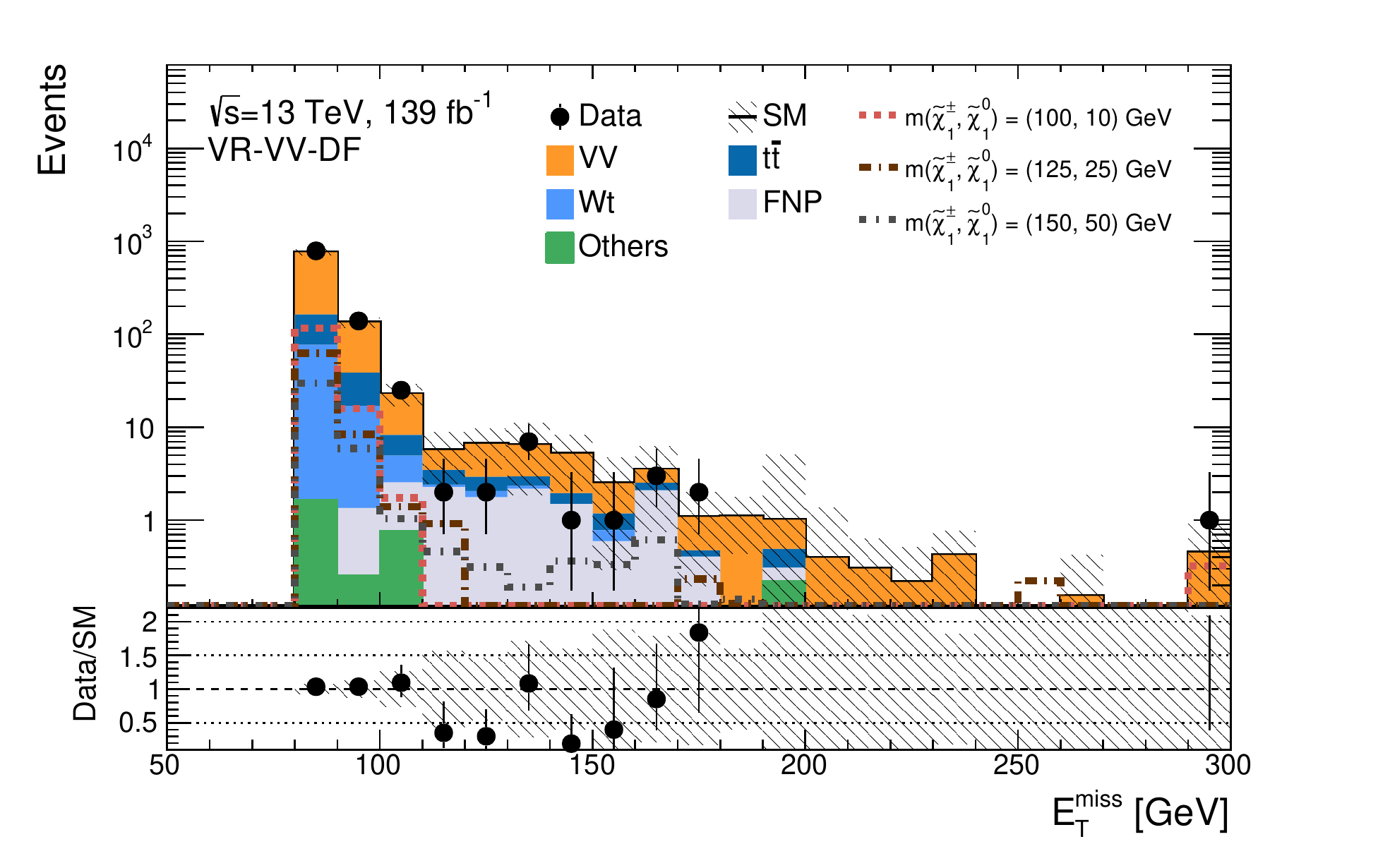}
\includegraphics[width=0.45\linewidth]{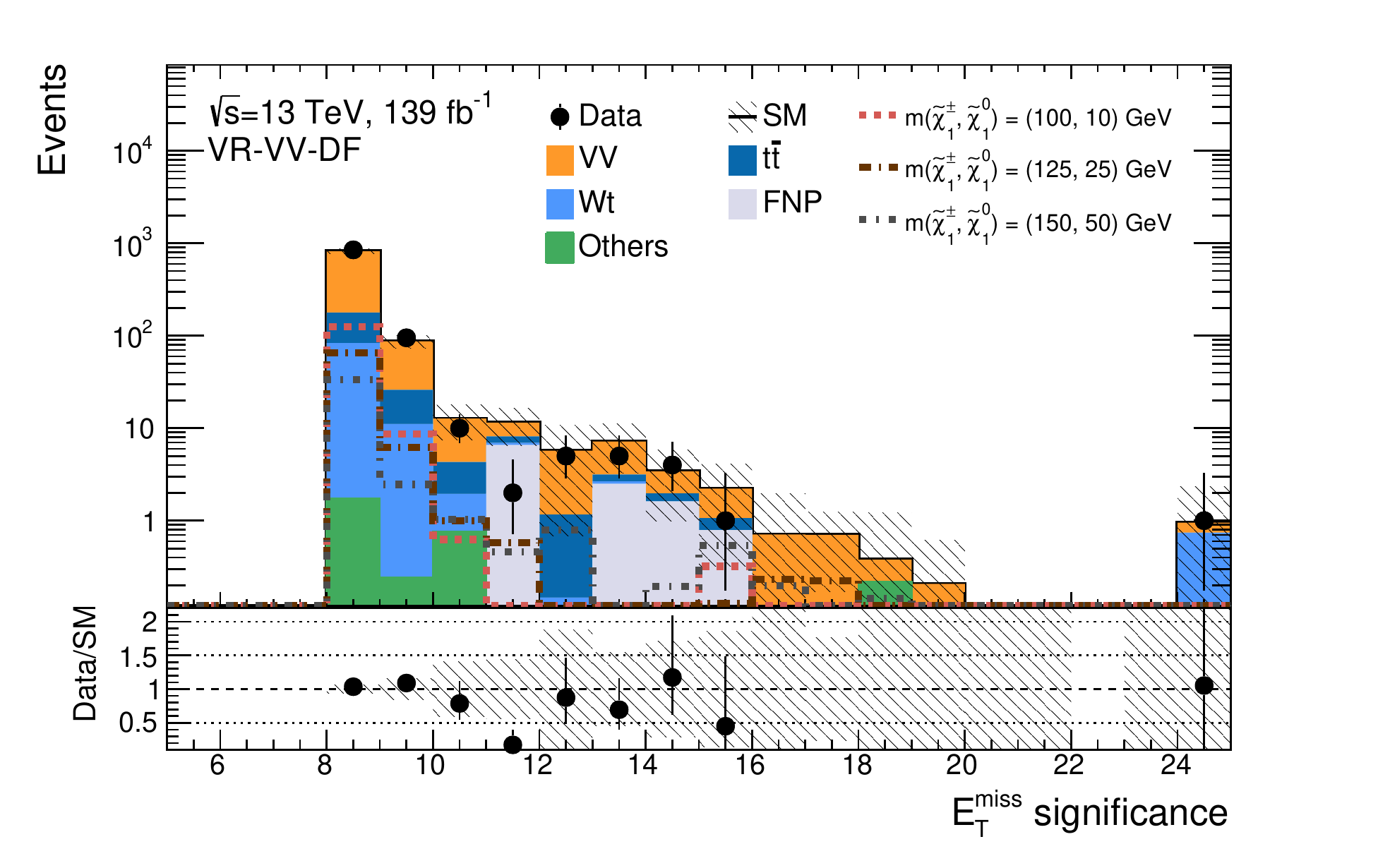}
\includegraphics[width=0.45\linewidth]{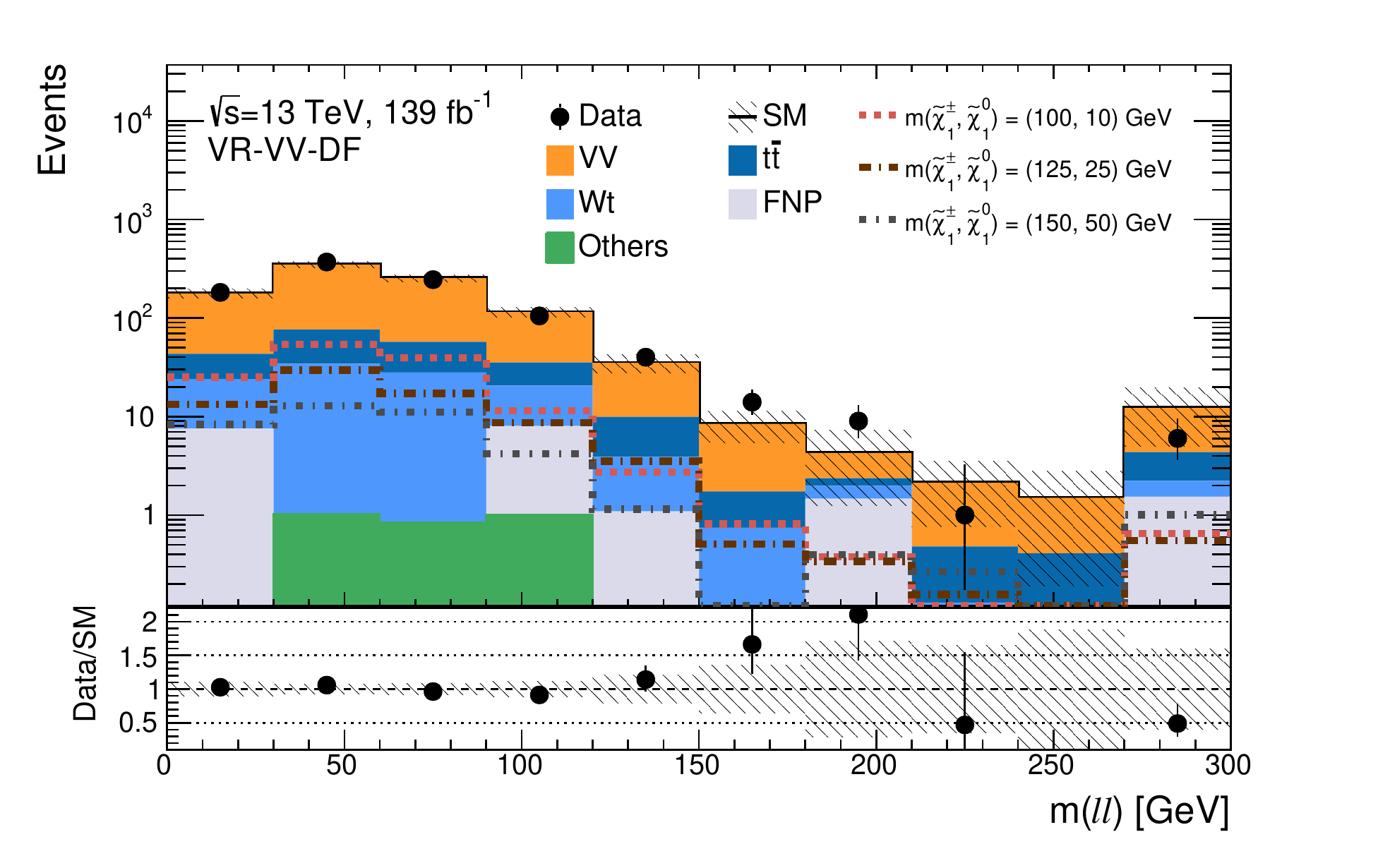}
\includegraphics[width=0.45\linewidth]{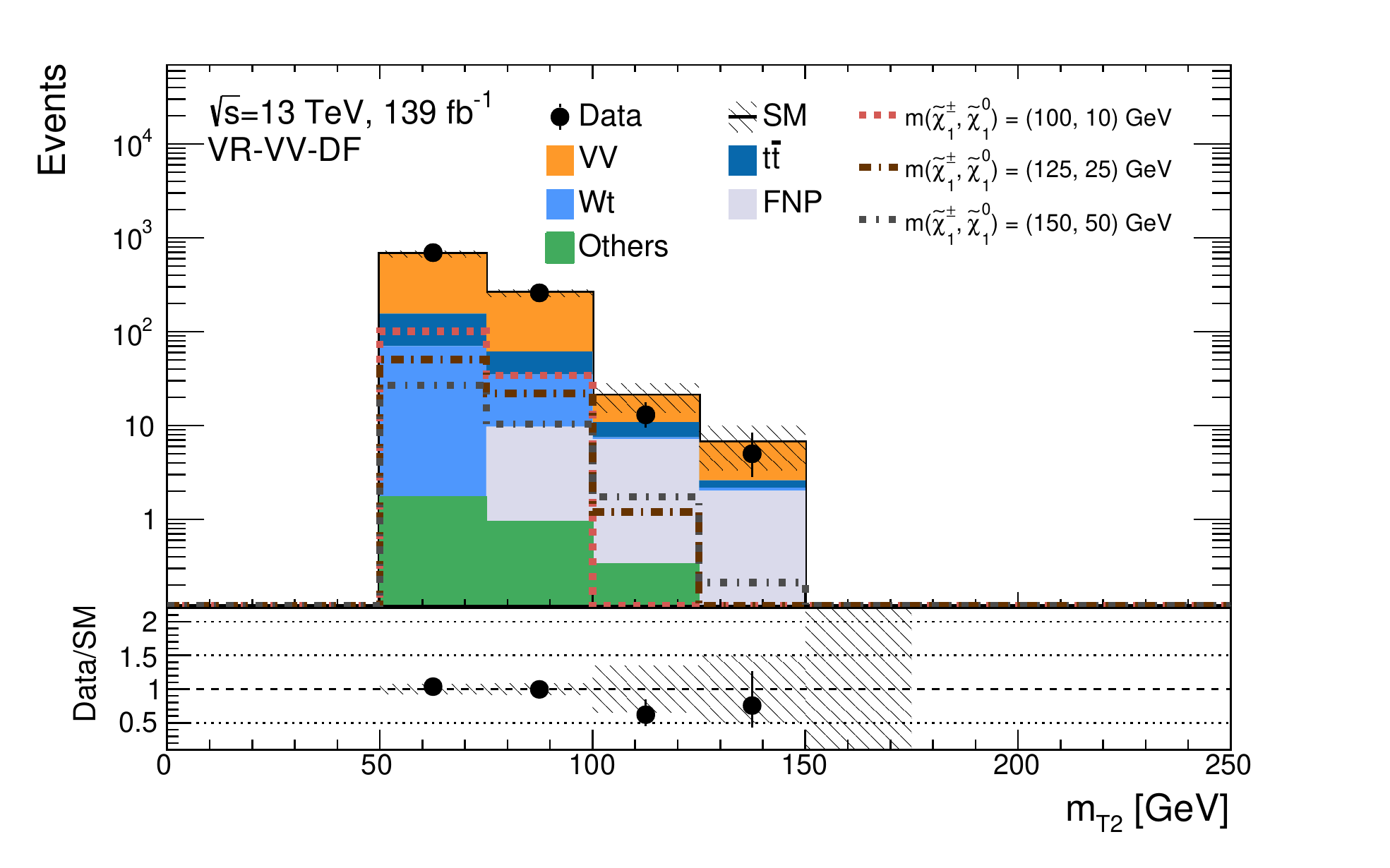}
\includegraphics[width=0.45\linewidth]{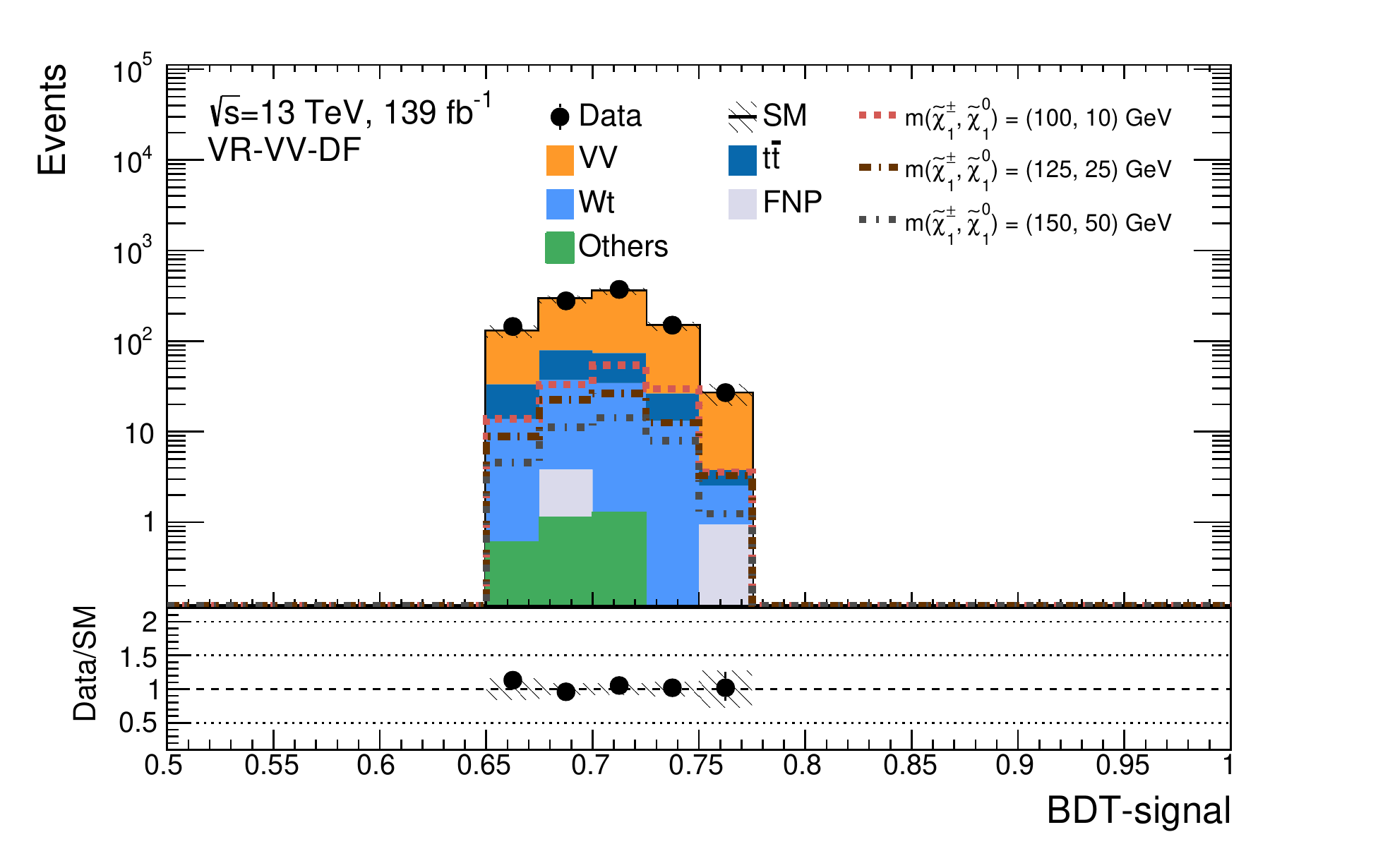}
\includegraphics[width=0.45\linewidth]{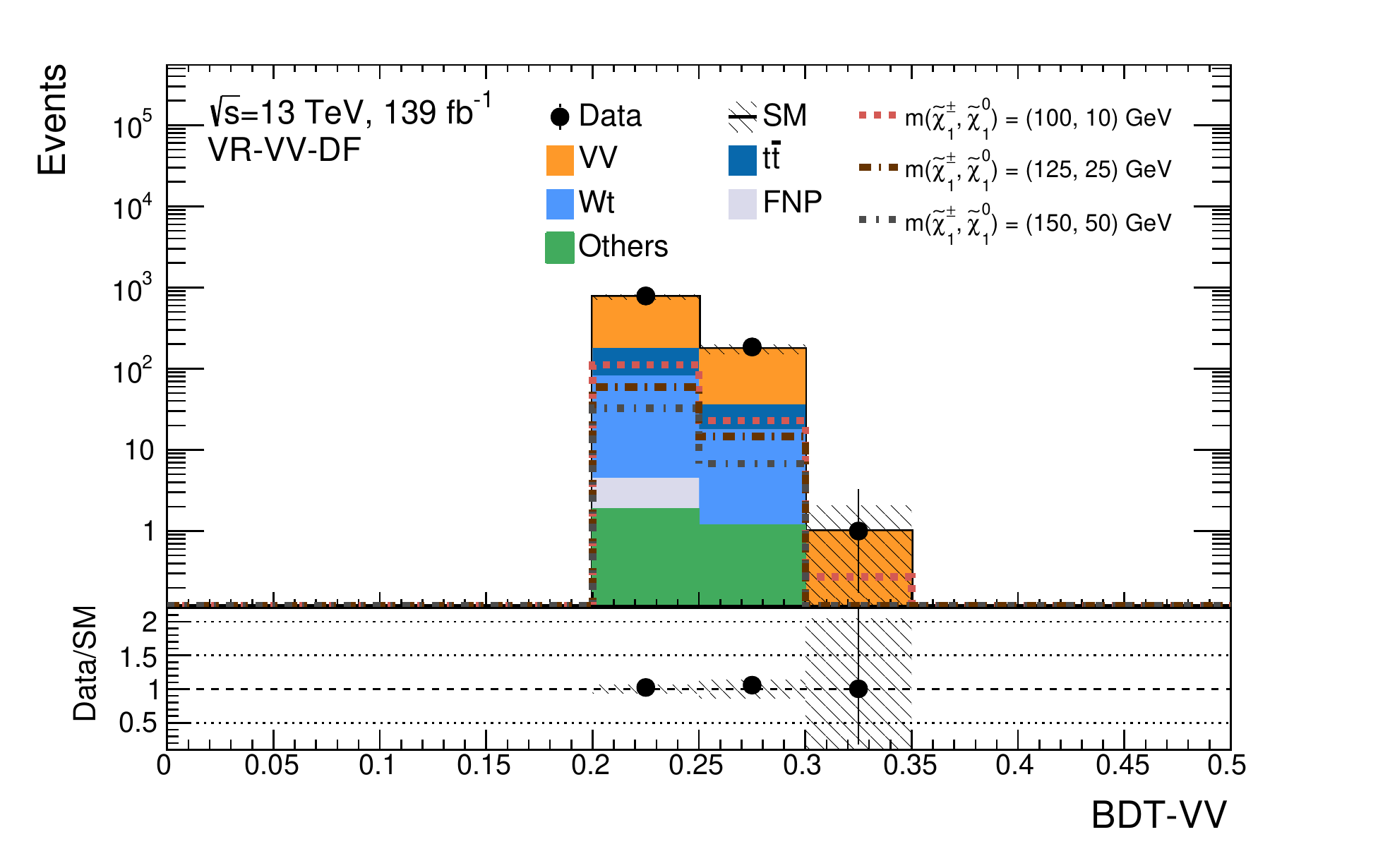}
\includegraphics[width=0.45\linewidth]{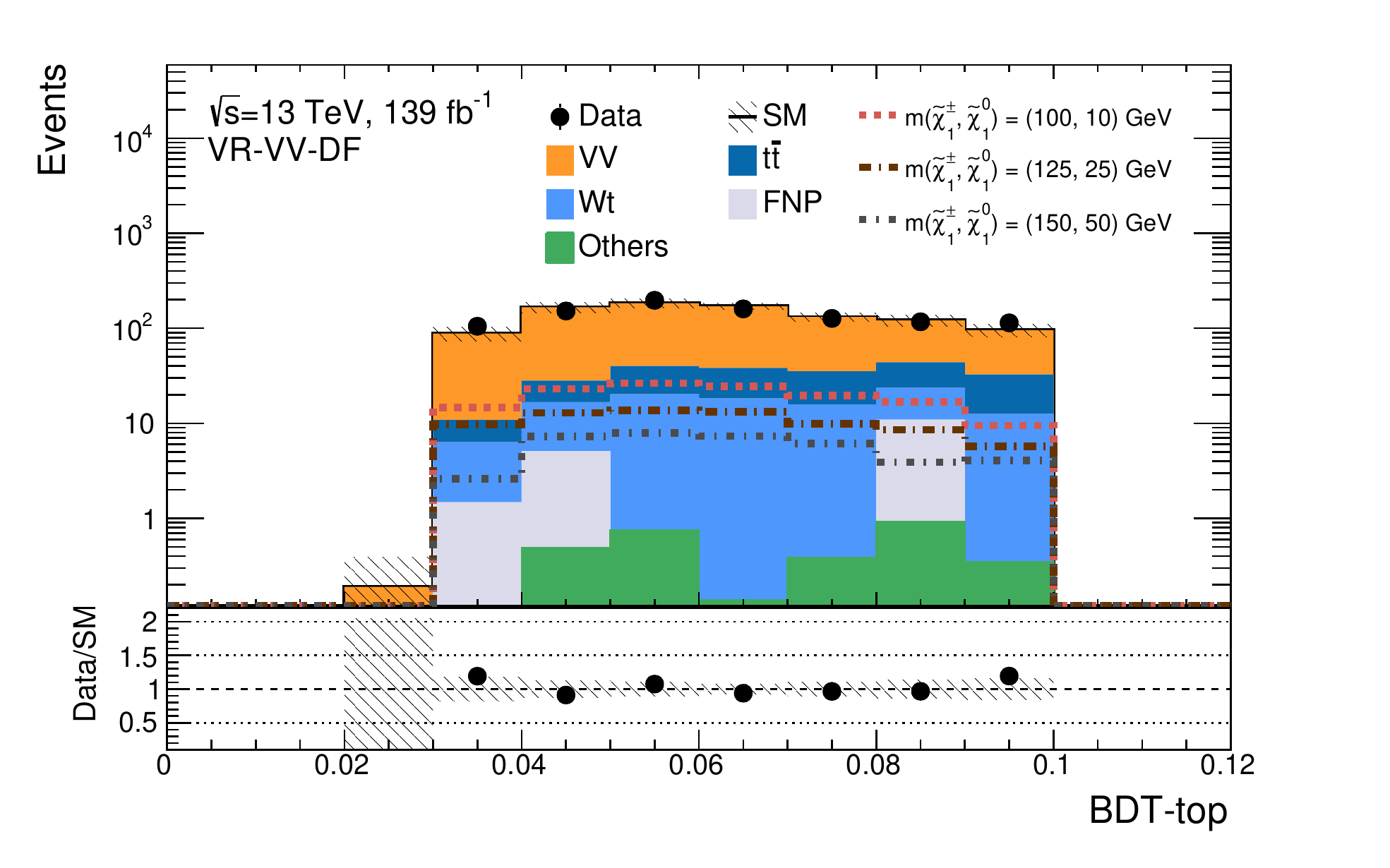}
\includegraphics[width=0.45\linewidth]{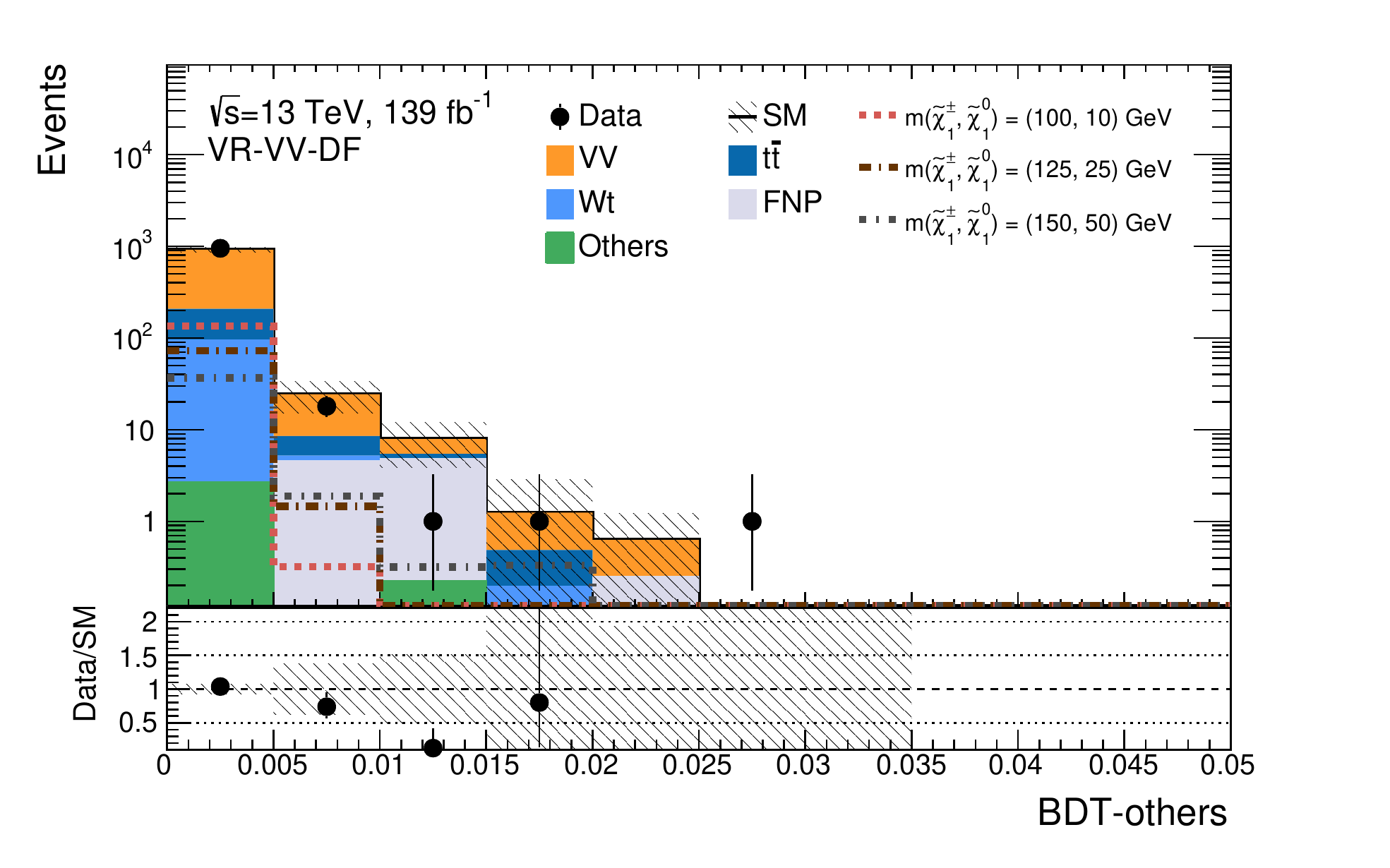}
\caption{The post-fit distributions in VR-VV-DF. Both statistical and systematic uncertainties are shown.}
\label{fig:VR_VVDF0J_ML}
\end{figure}

\begin{figure}[!htb]
\centering
\includegraphics[width=0.45\linewidth]{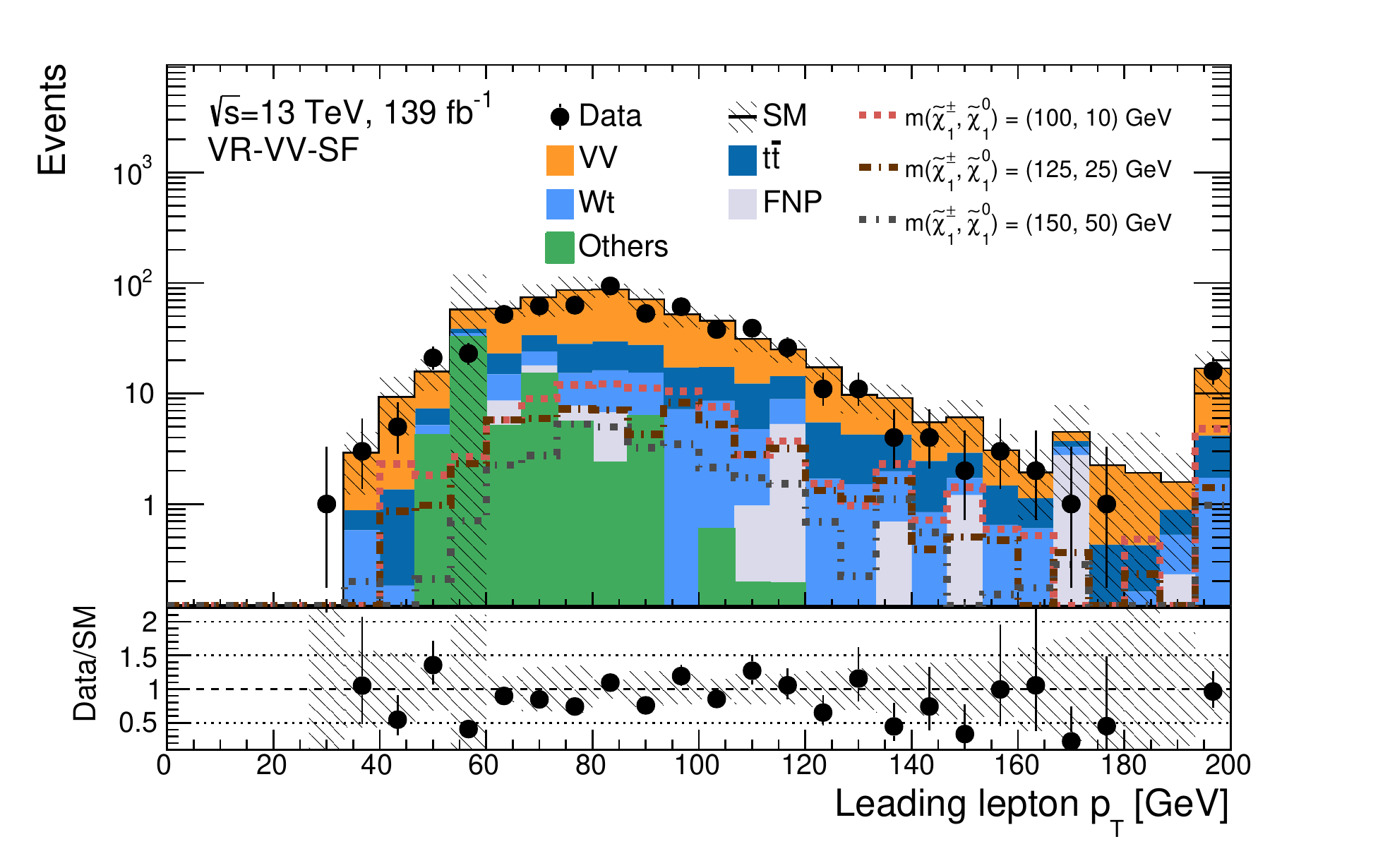}
\includegraphics[width=0.45\linewidth]{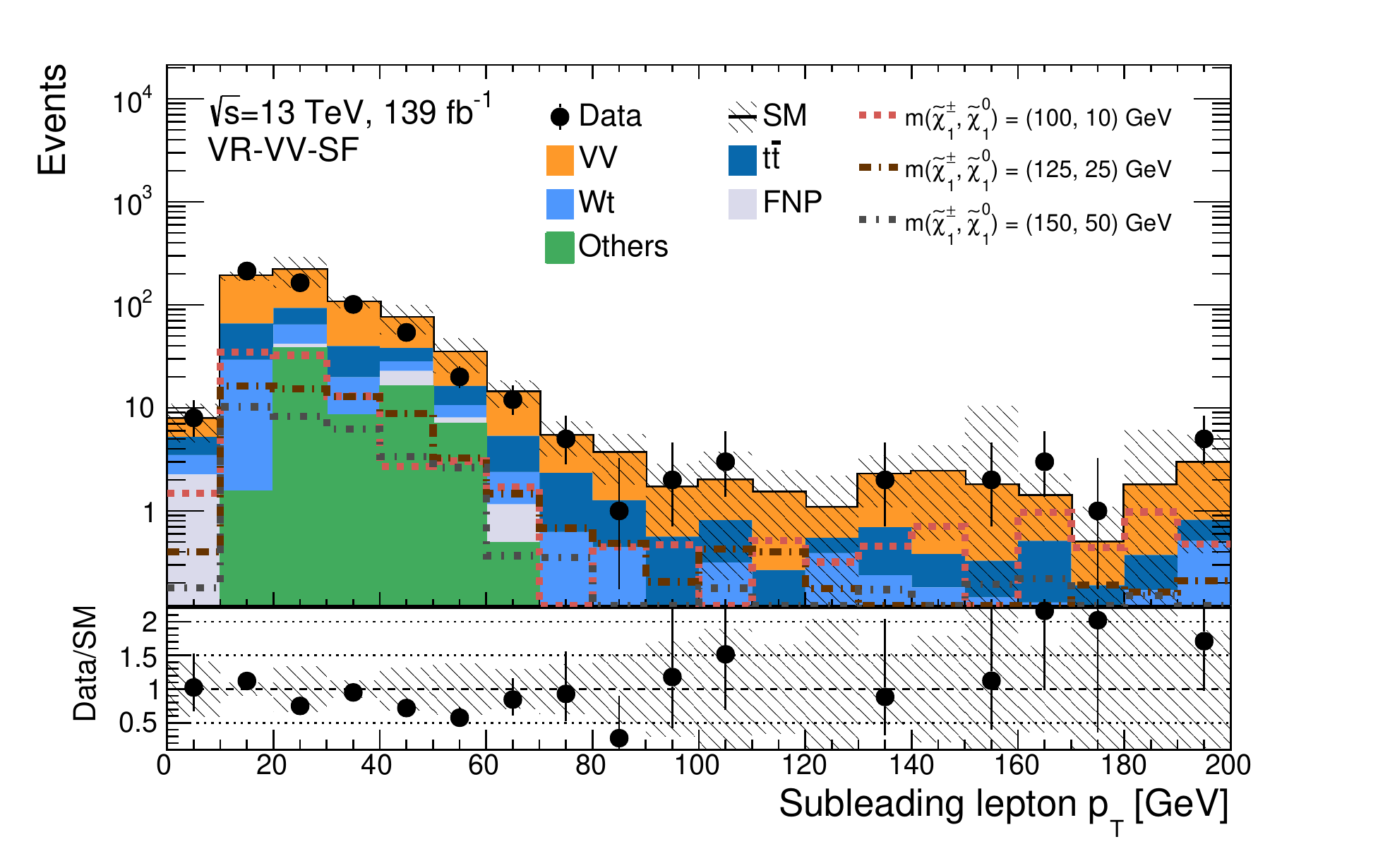}
\includegraphics[width=0.45\linewidth]{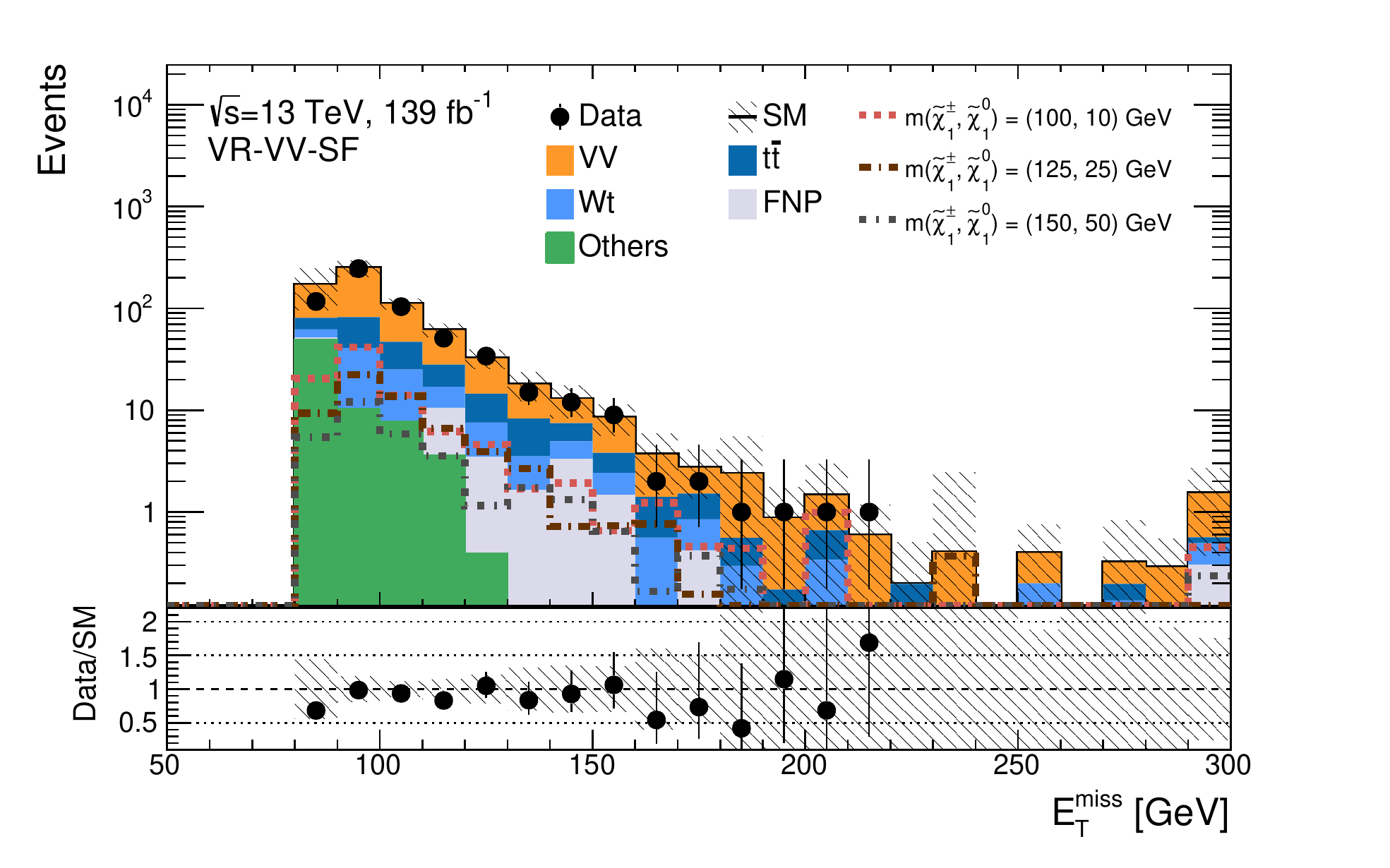}
\includegraphics[width=0.45\linewidth]{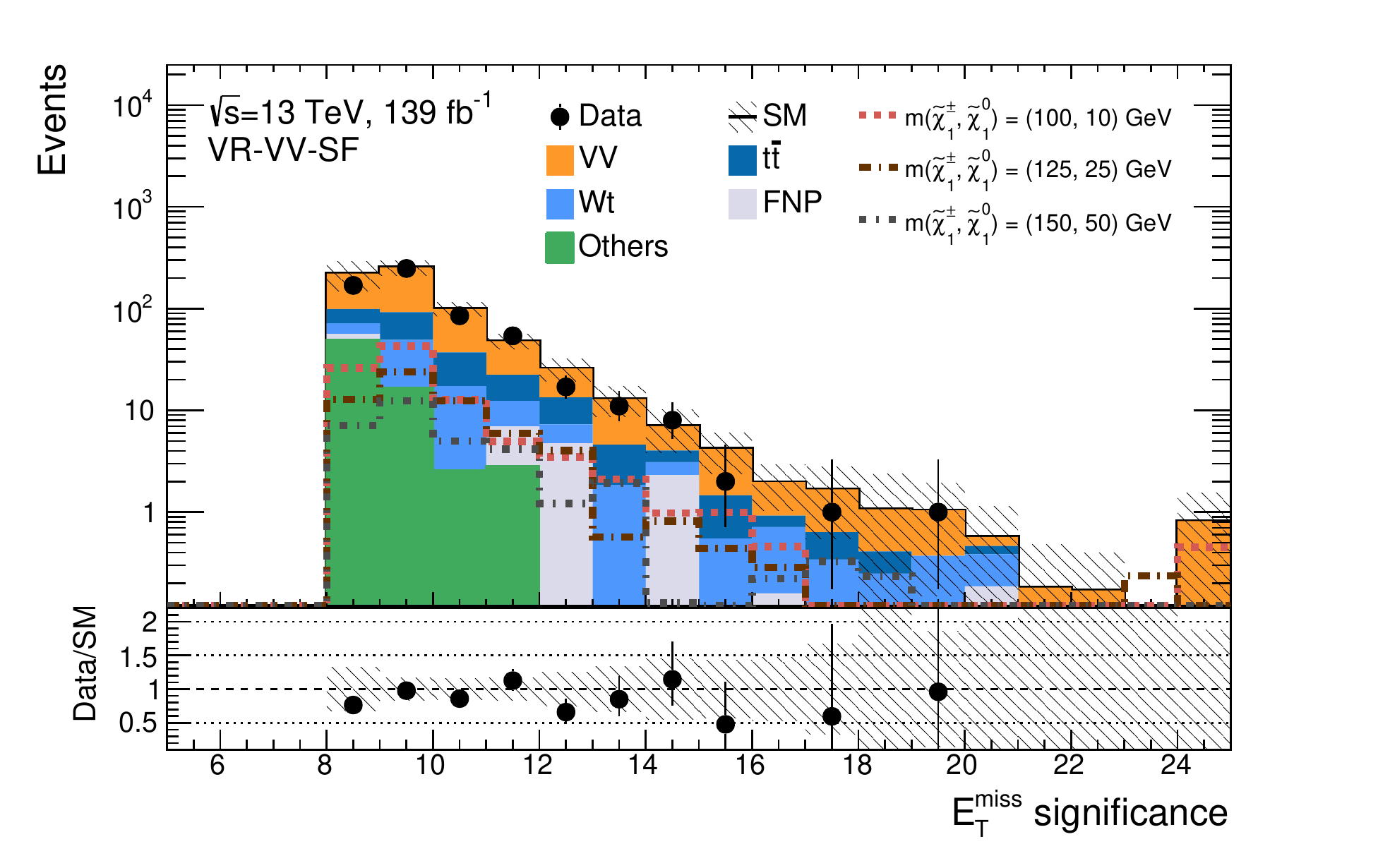}
\includegraphics[width=0.45\linewidth]{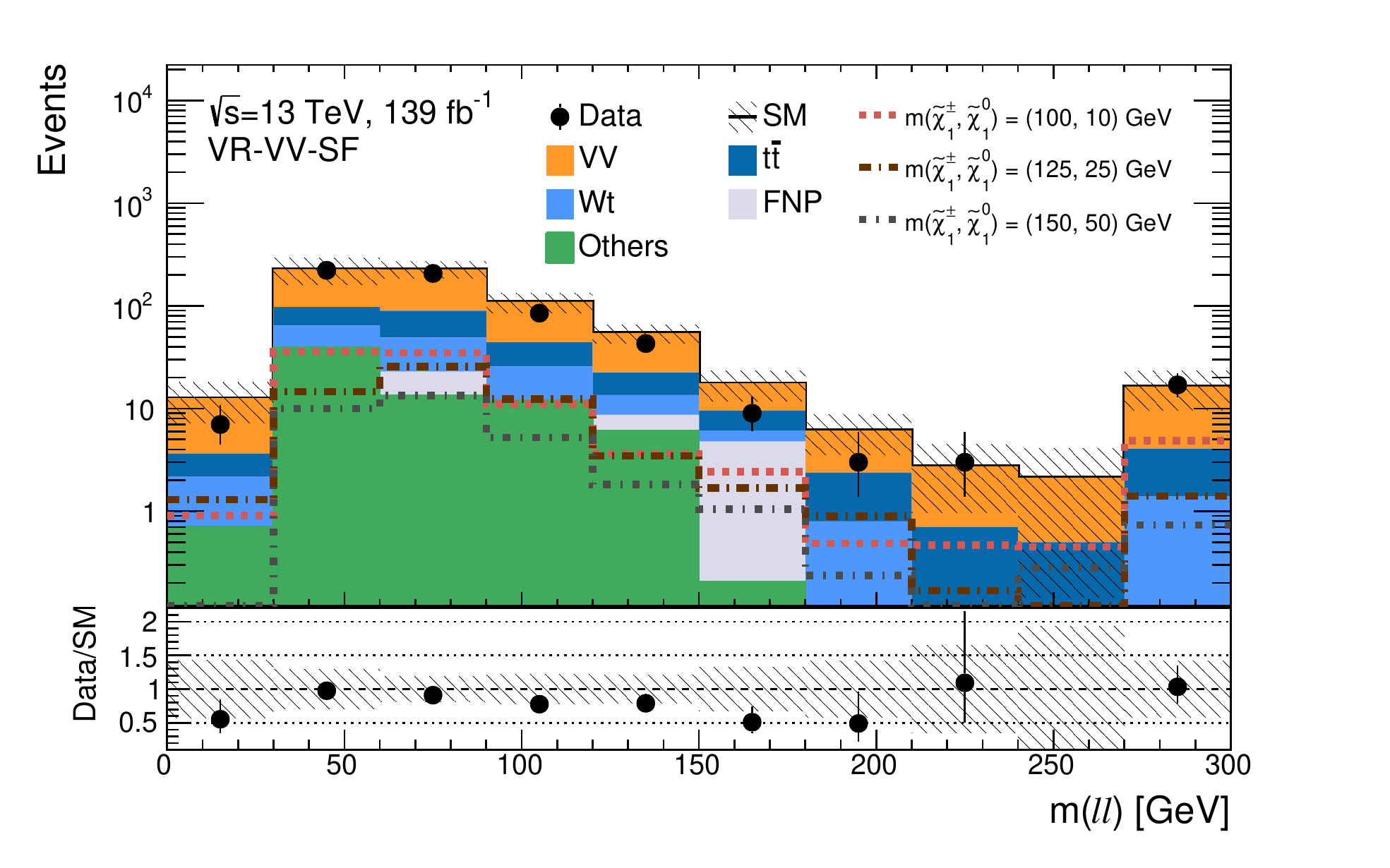}
\includegraphics[width=0.45\linewidth]{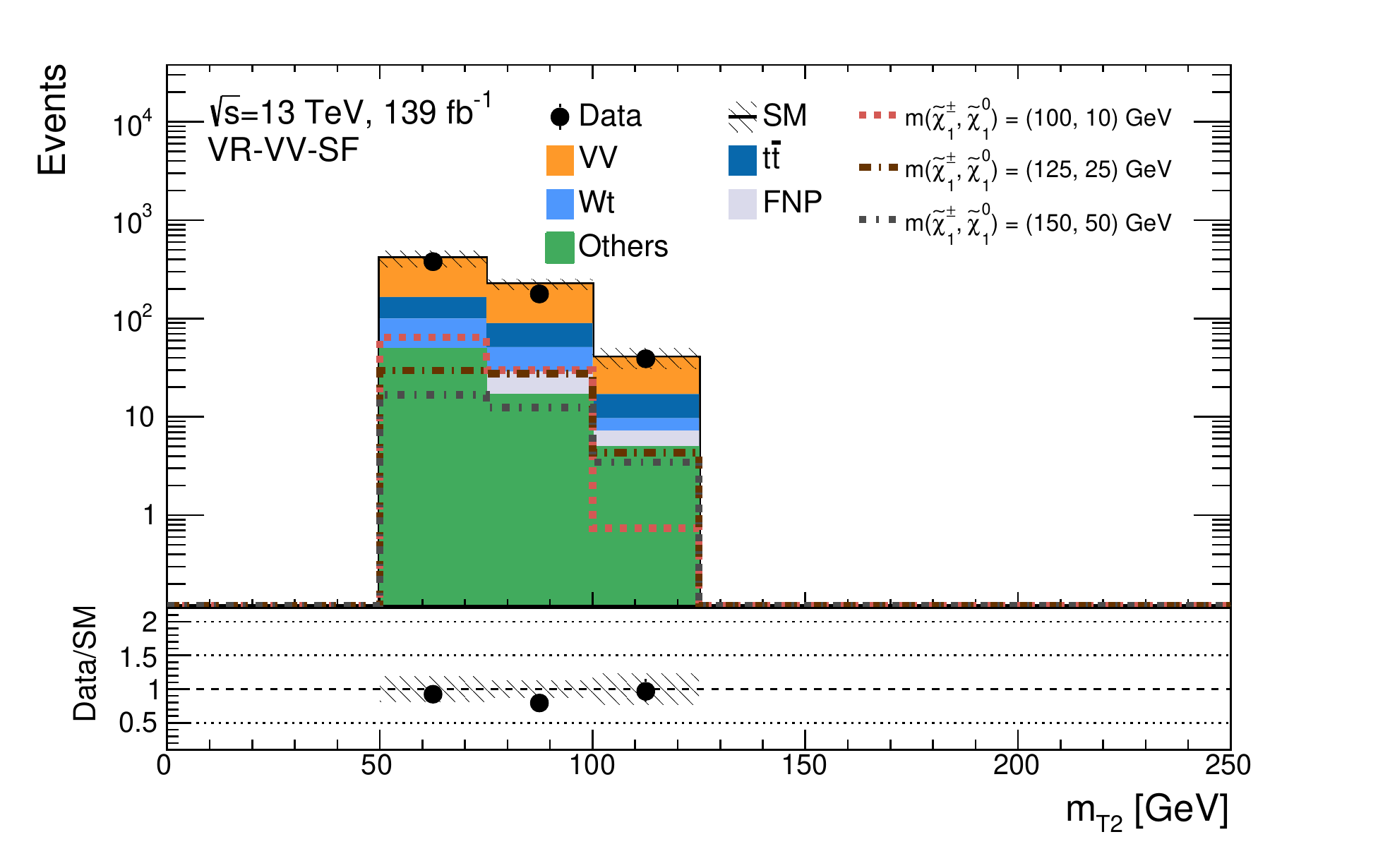}
\includegraphics[width=0.45\linewidth]{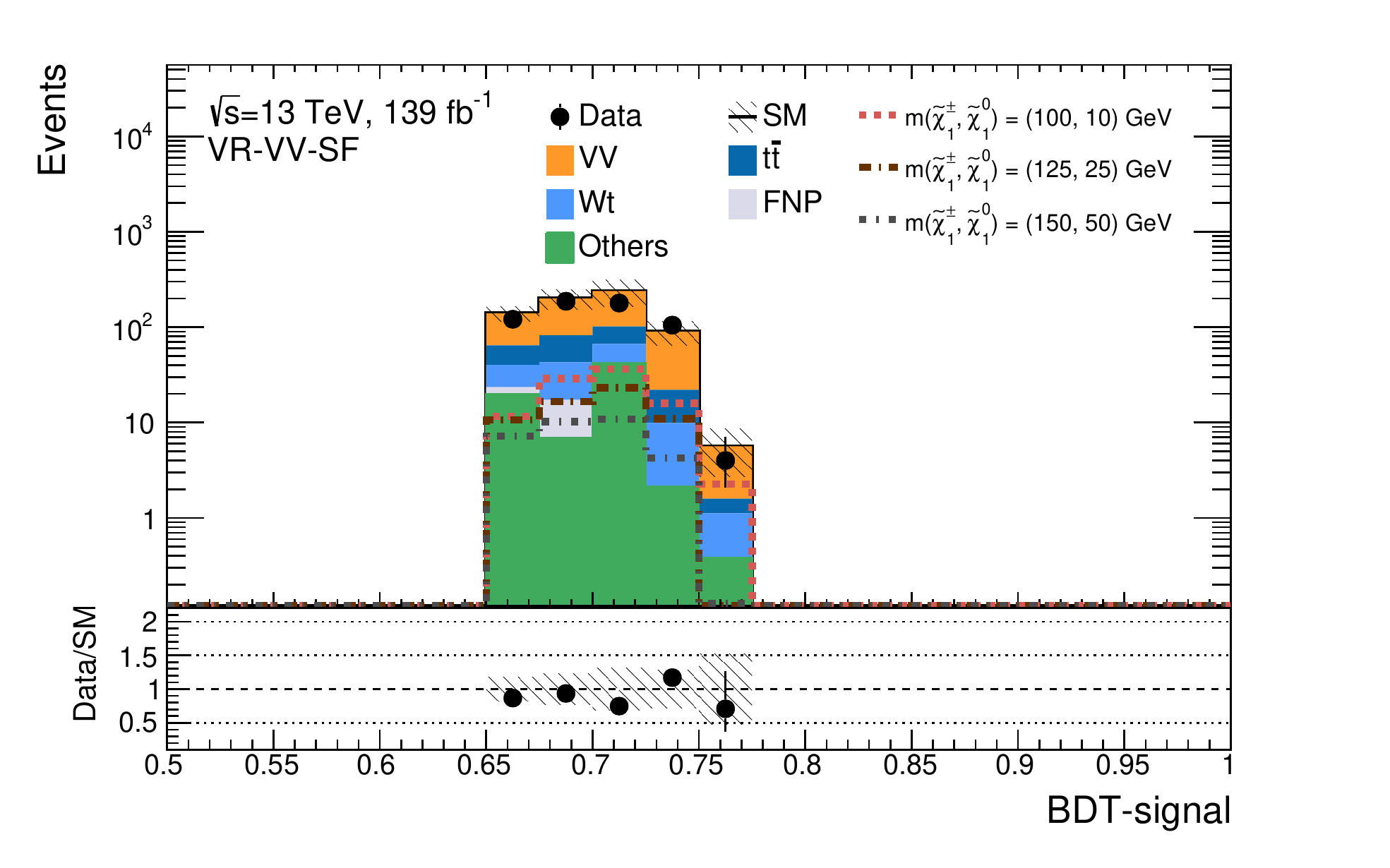}
\includegraphics[width=0.45\linewidth]{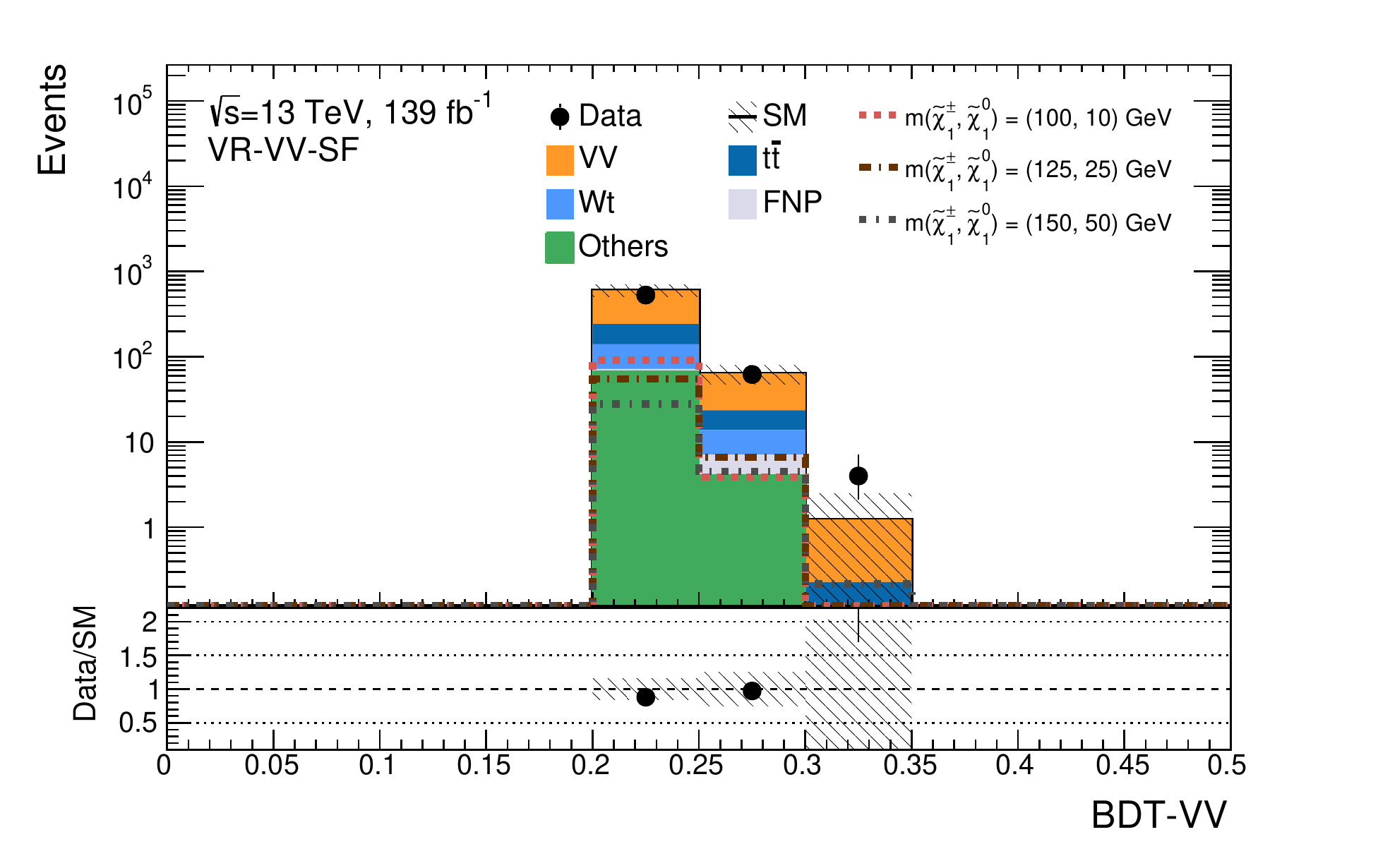}
\includegraphics[width=0.45\linewidth]{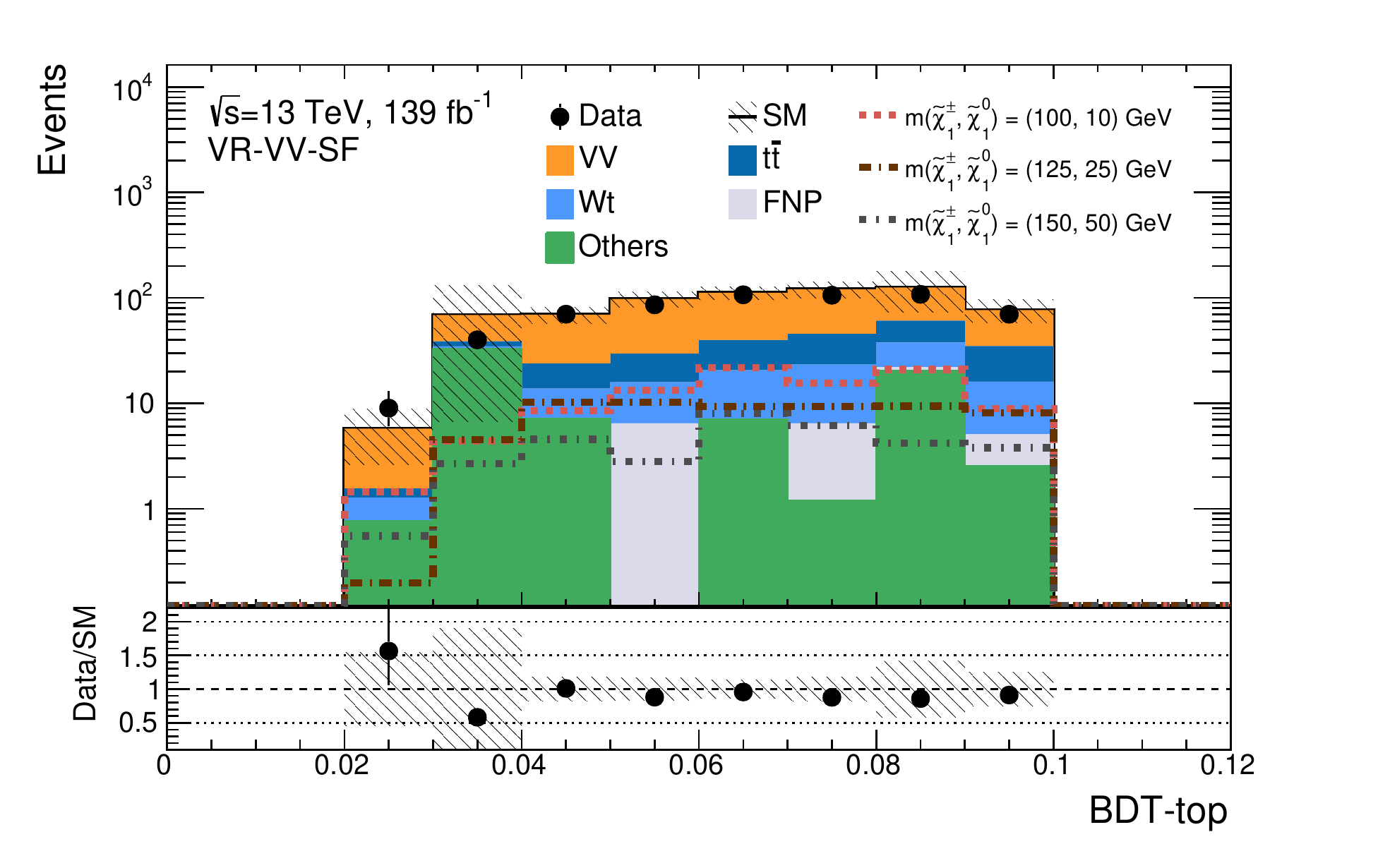}
\includegraphics[width=0.45\linewidth]{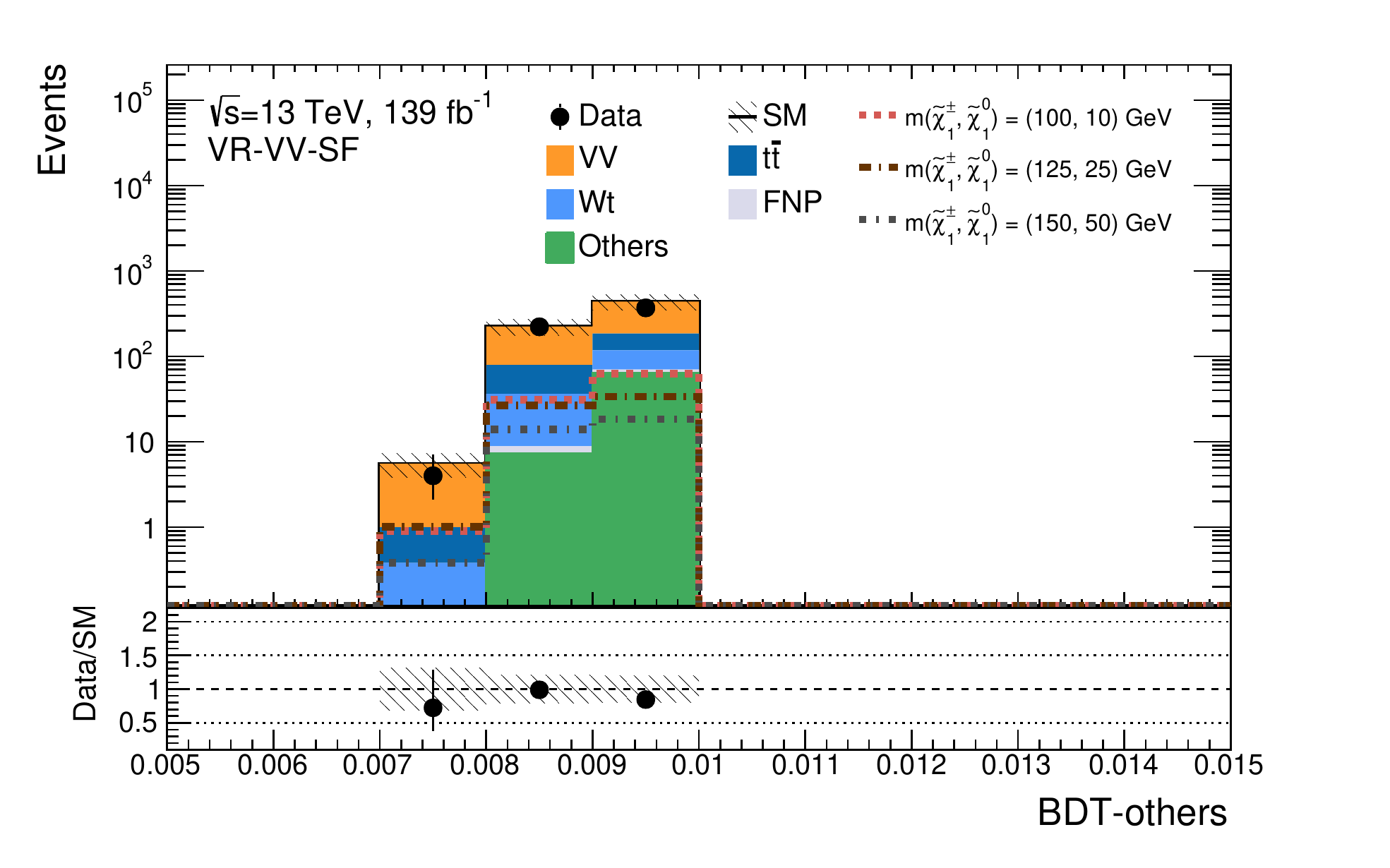}
\caption{The post-fit distributions in VR-VV-SF. Both statistical and systematic uncertainties are shown.}
\label{fig:VR_VVSF0J_ML}
\end{figure}

\begin{figure}[!htb]
\centering
\includegraphics[width=0.45\linewidth]{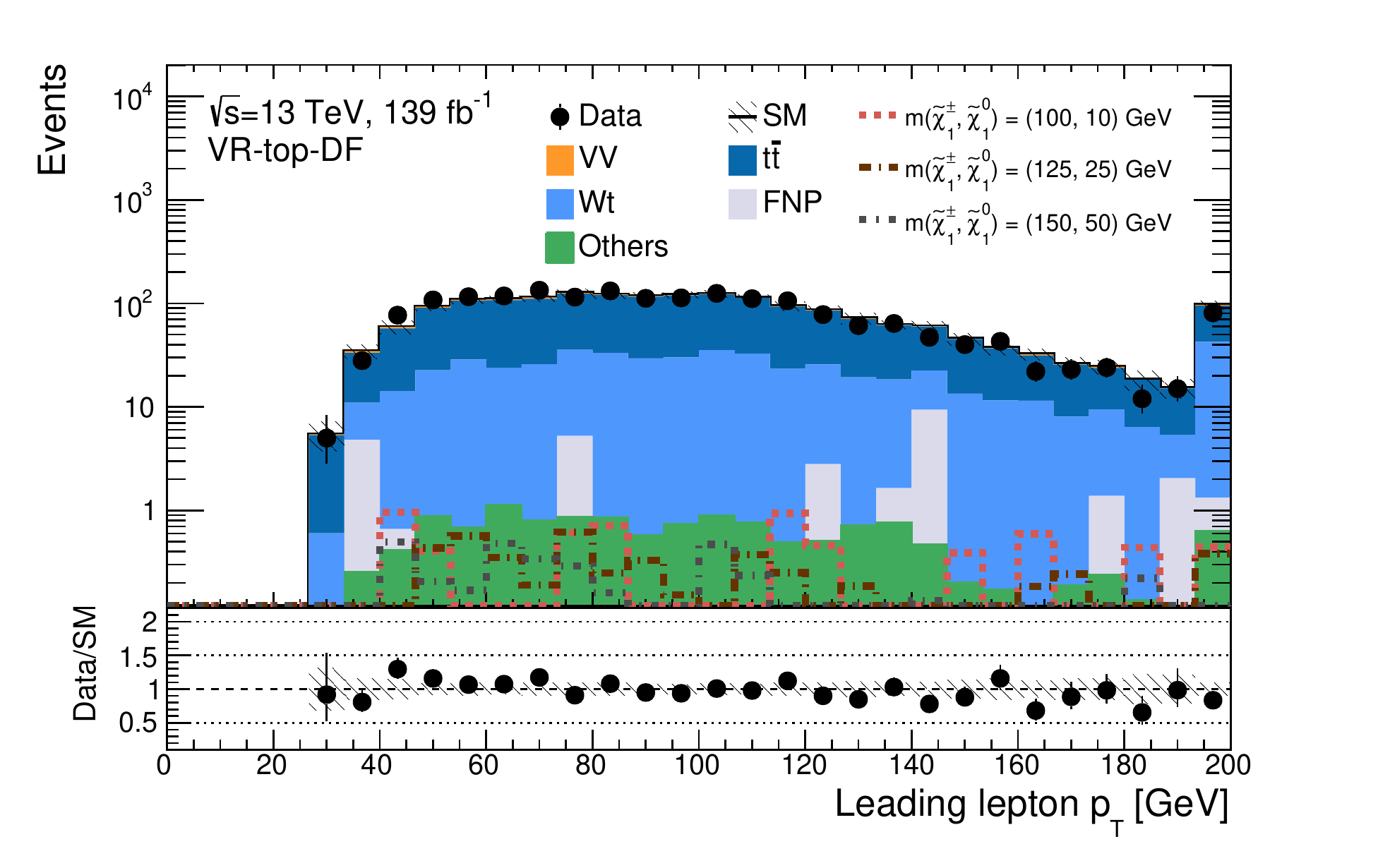}
\includegraphics[width=0.45\linewidth]{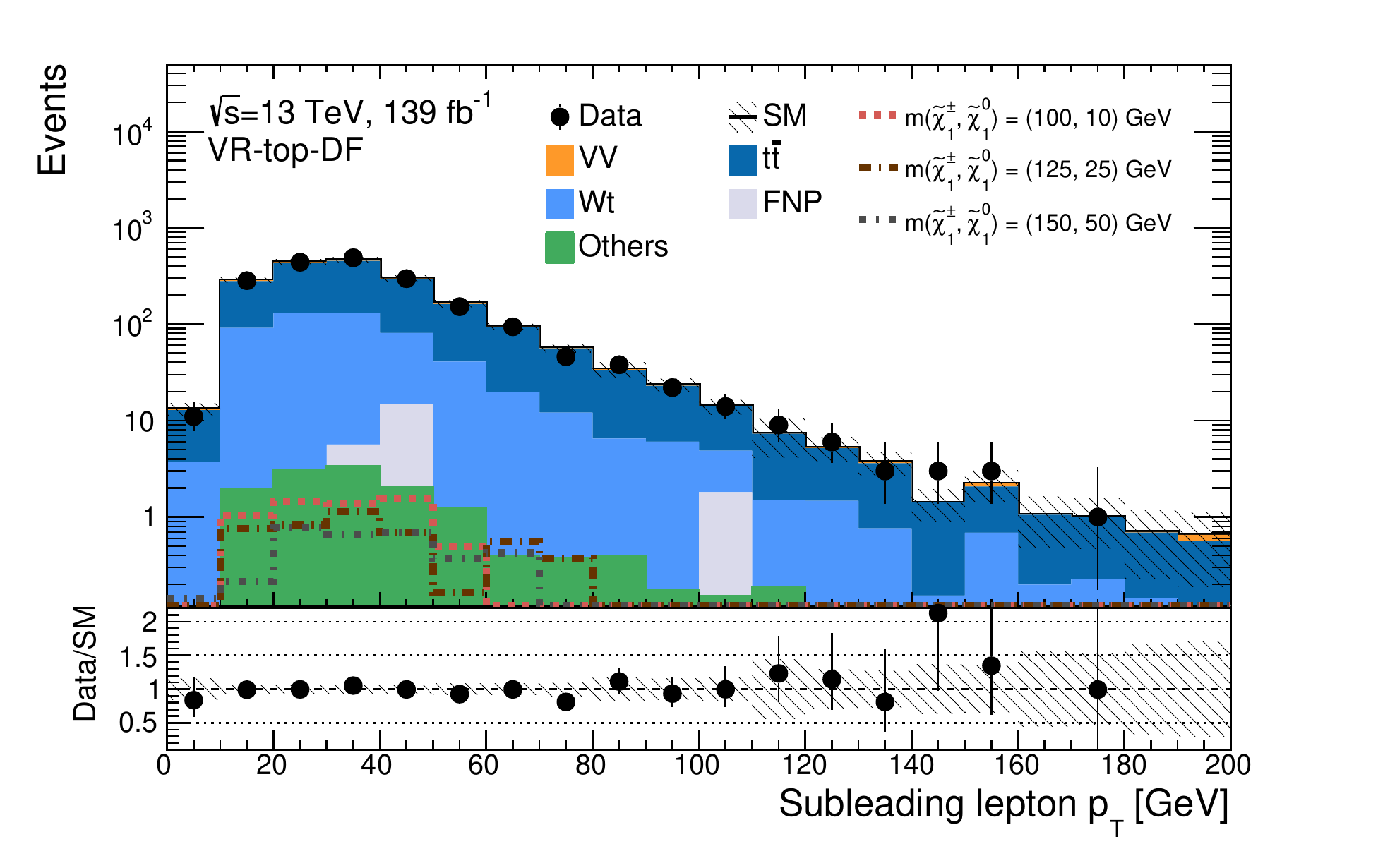}
\includegraphics[width=0.45\linewidth]{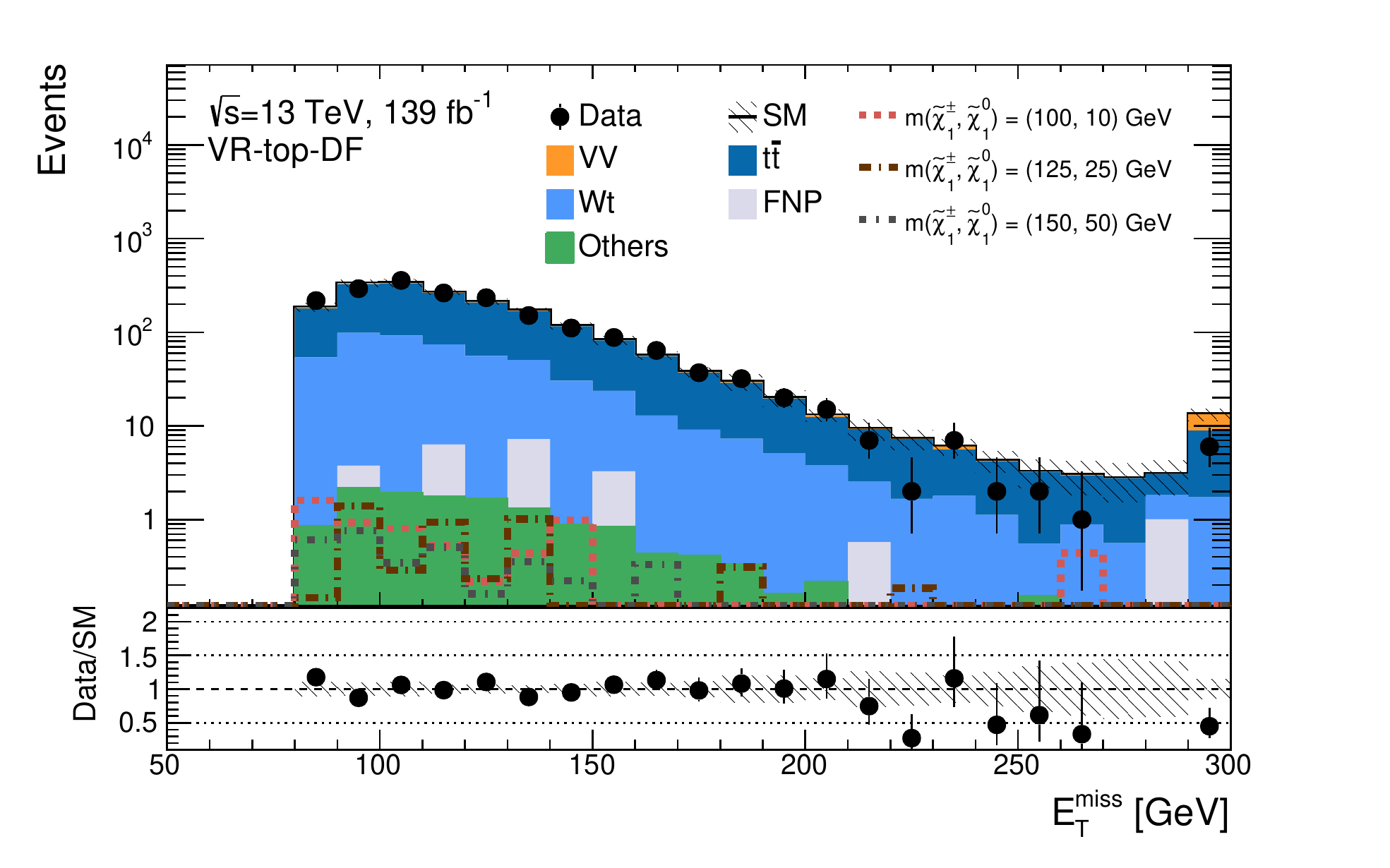}
\includegraphics[width=0.45\linewidth]{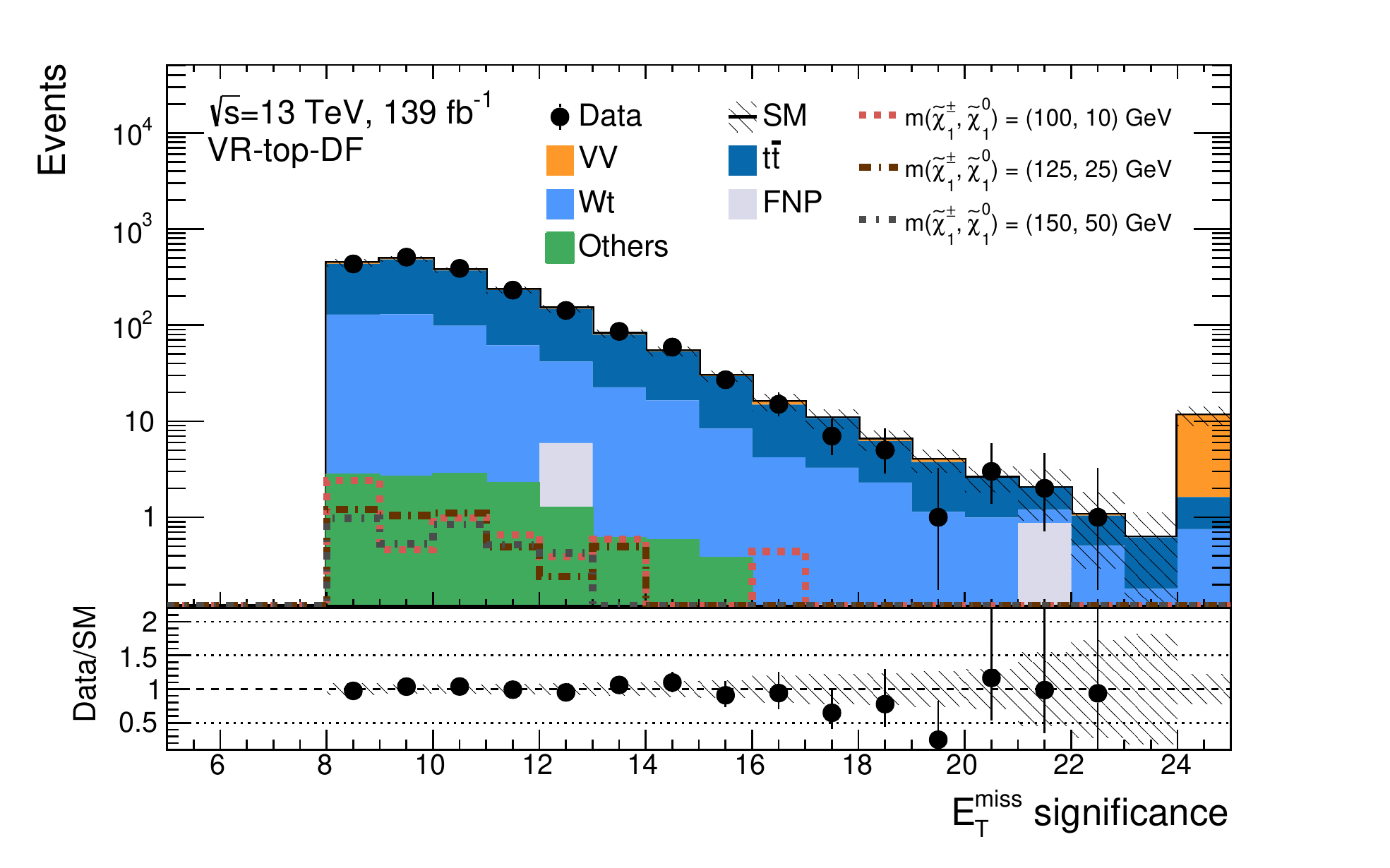}
\includegraphics[width=0.45\linewidth]{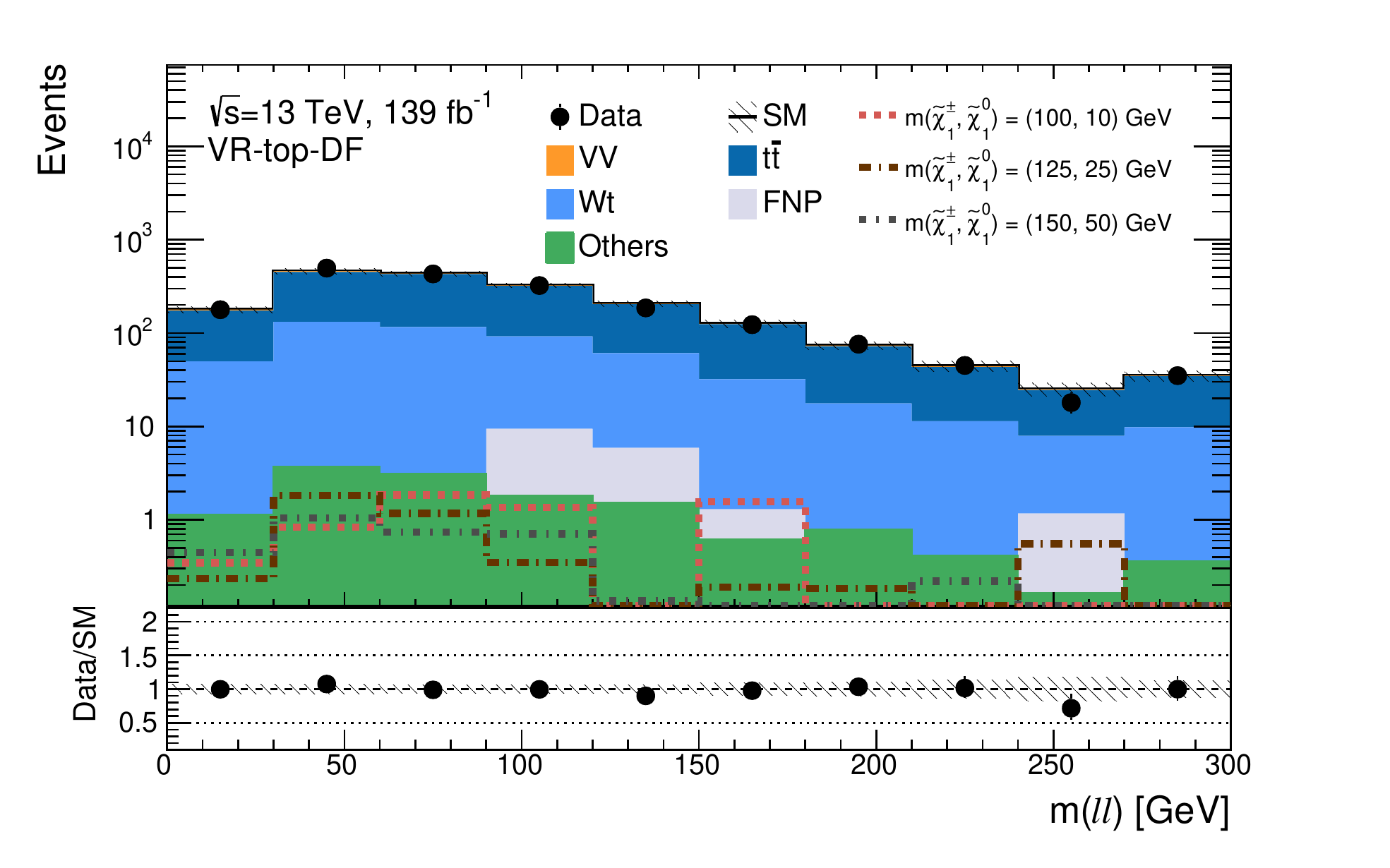}
\includegraphics[width=0.45\linewidth]{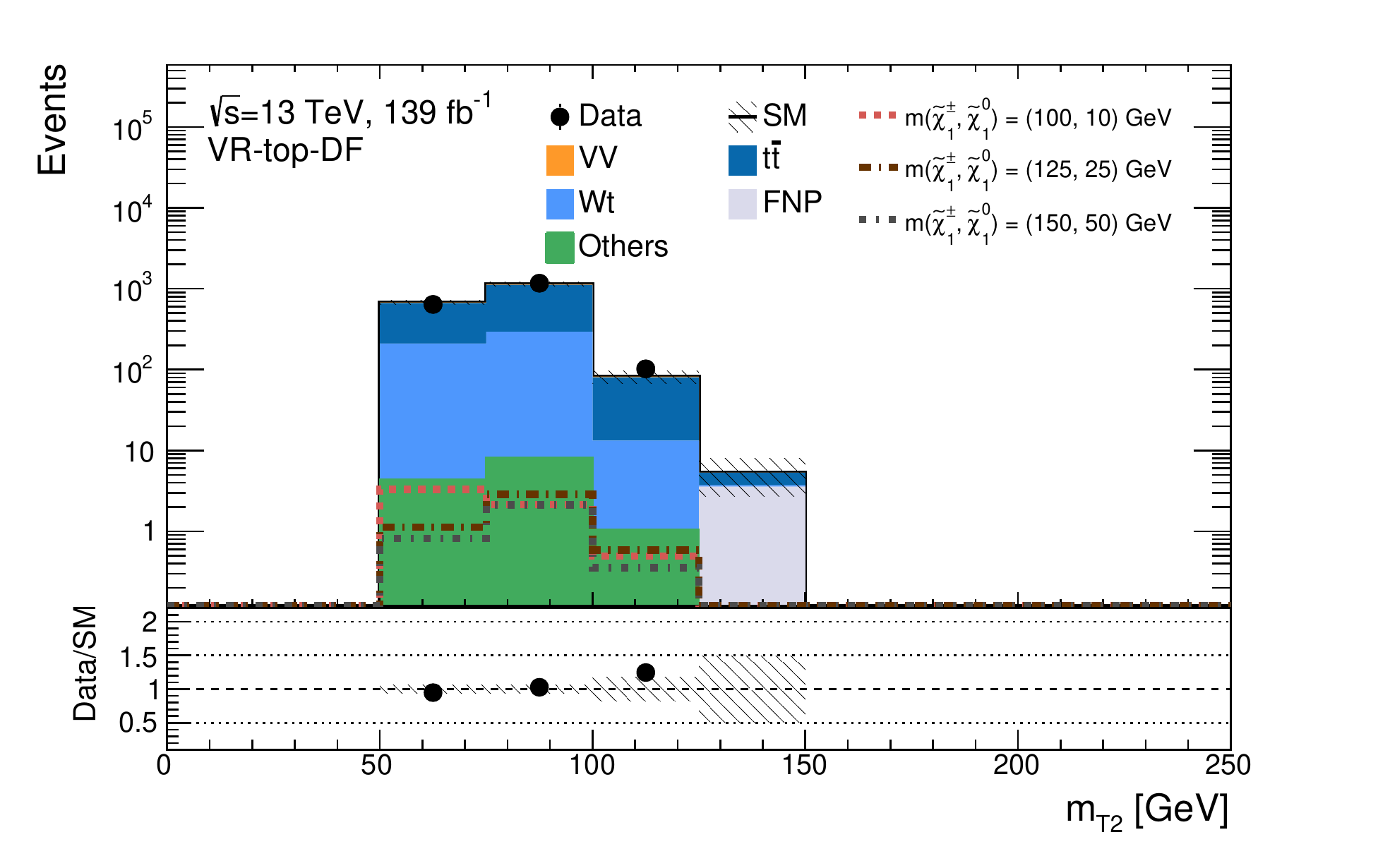}
\includegraphics[width=0.45\linewidth]{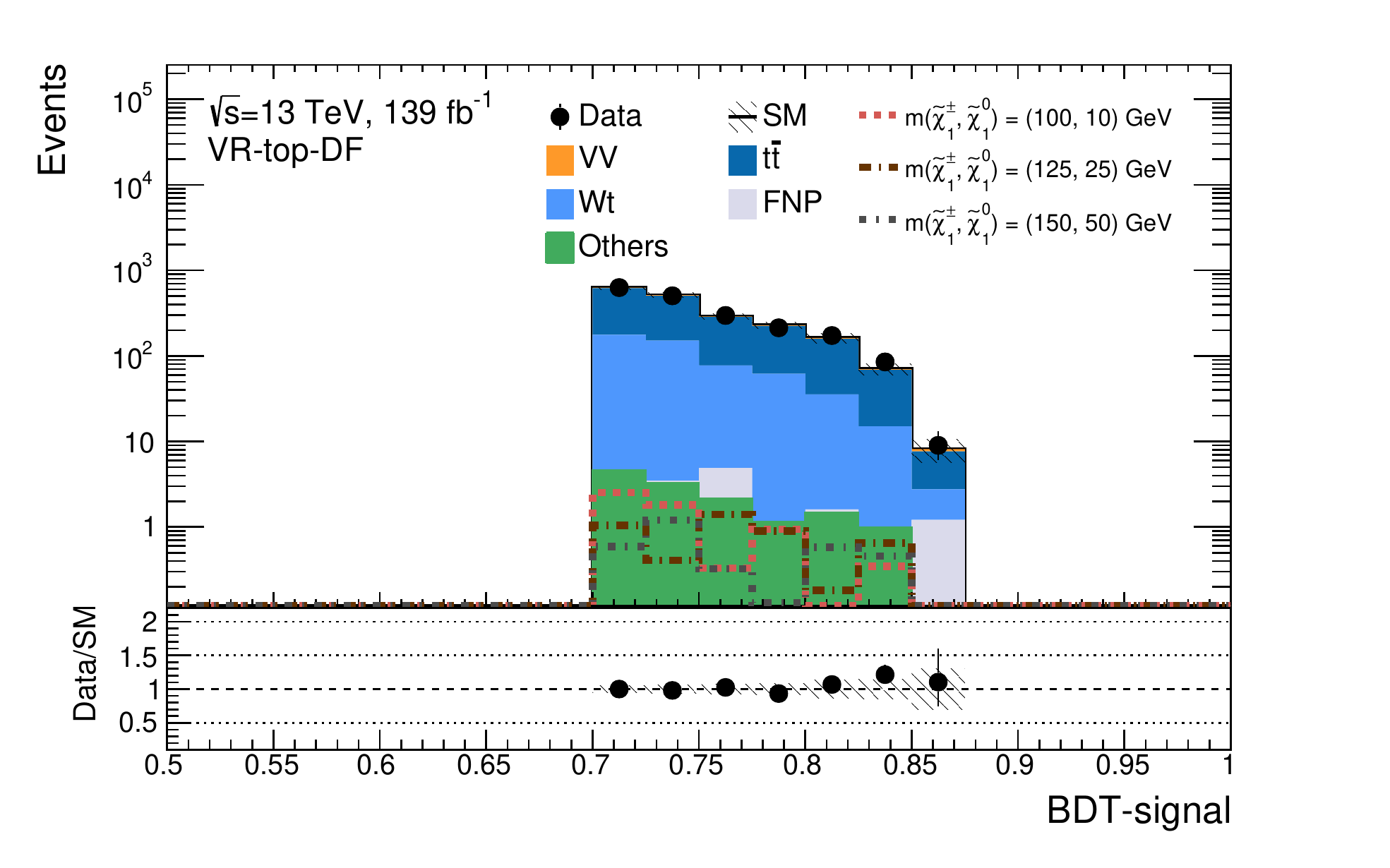}
\includegraphics[width=0.45\linewidth]{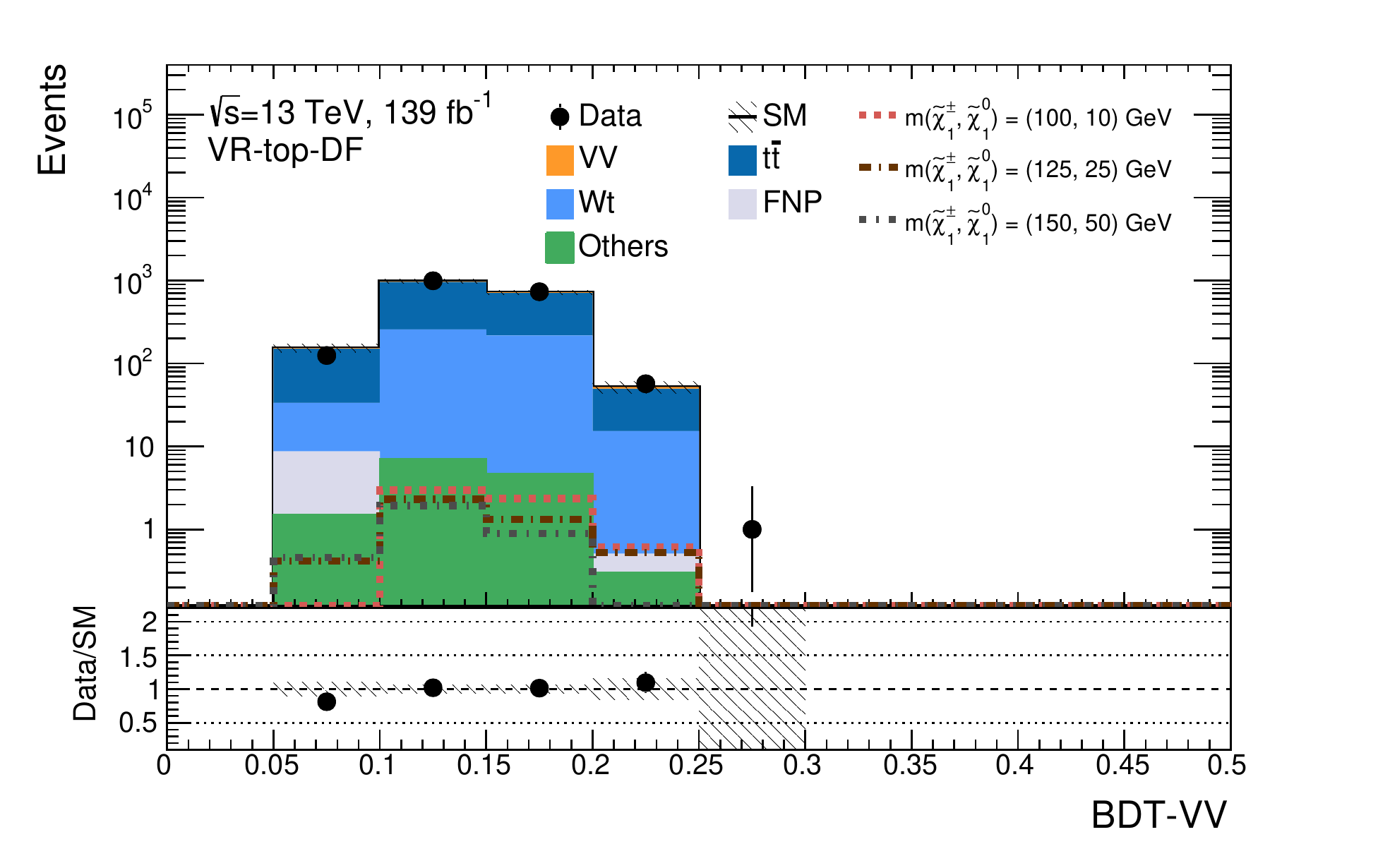}
\includegraphics[width=0.45\linewidth]{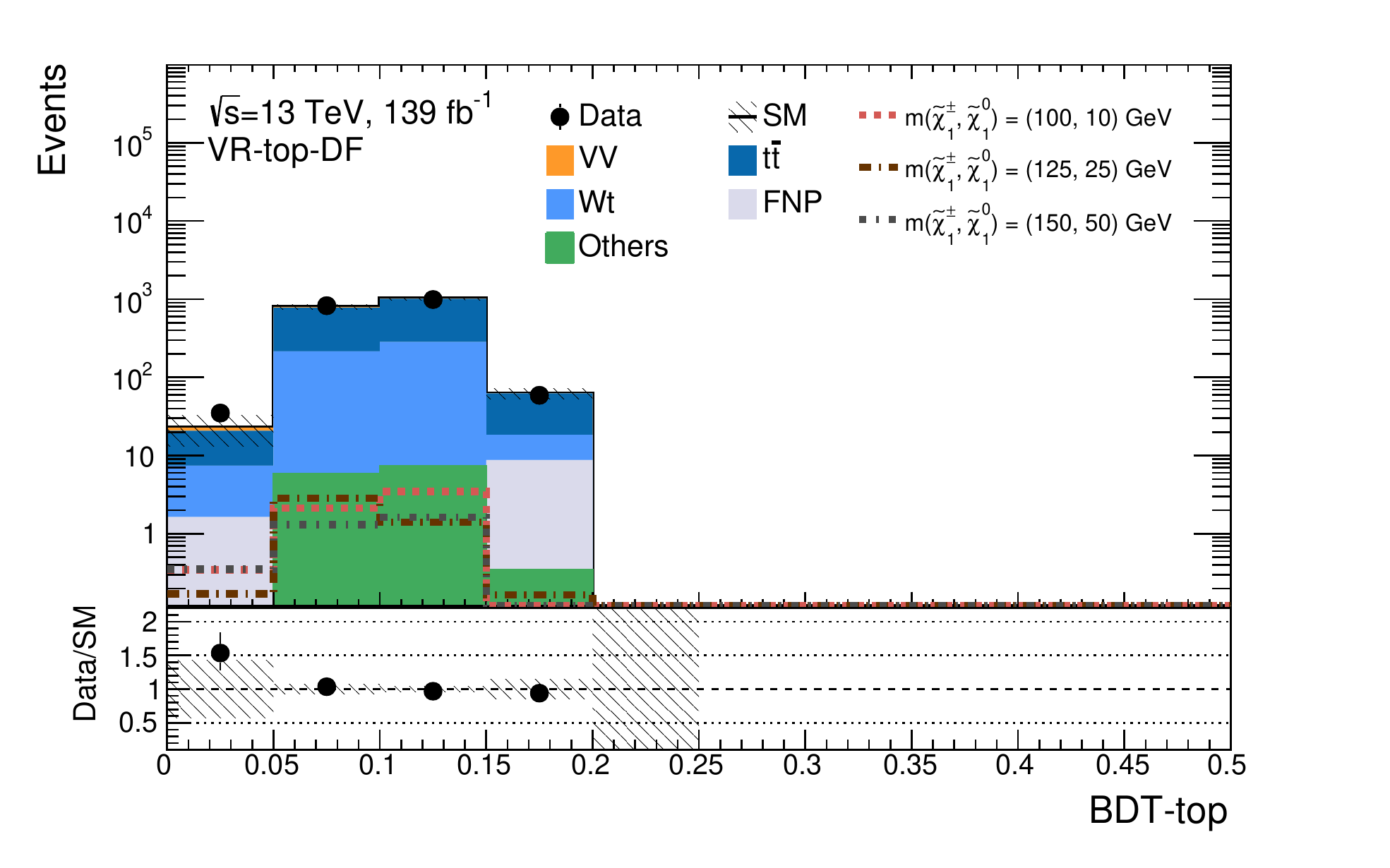}
\includegraphics[width=0.45\linewidth]{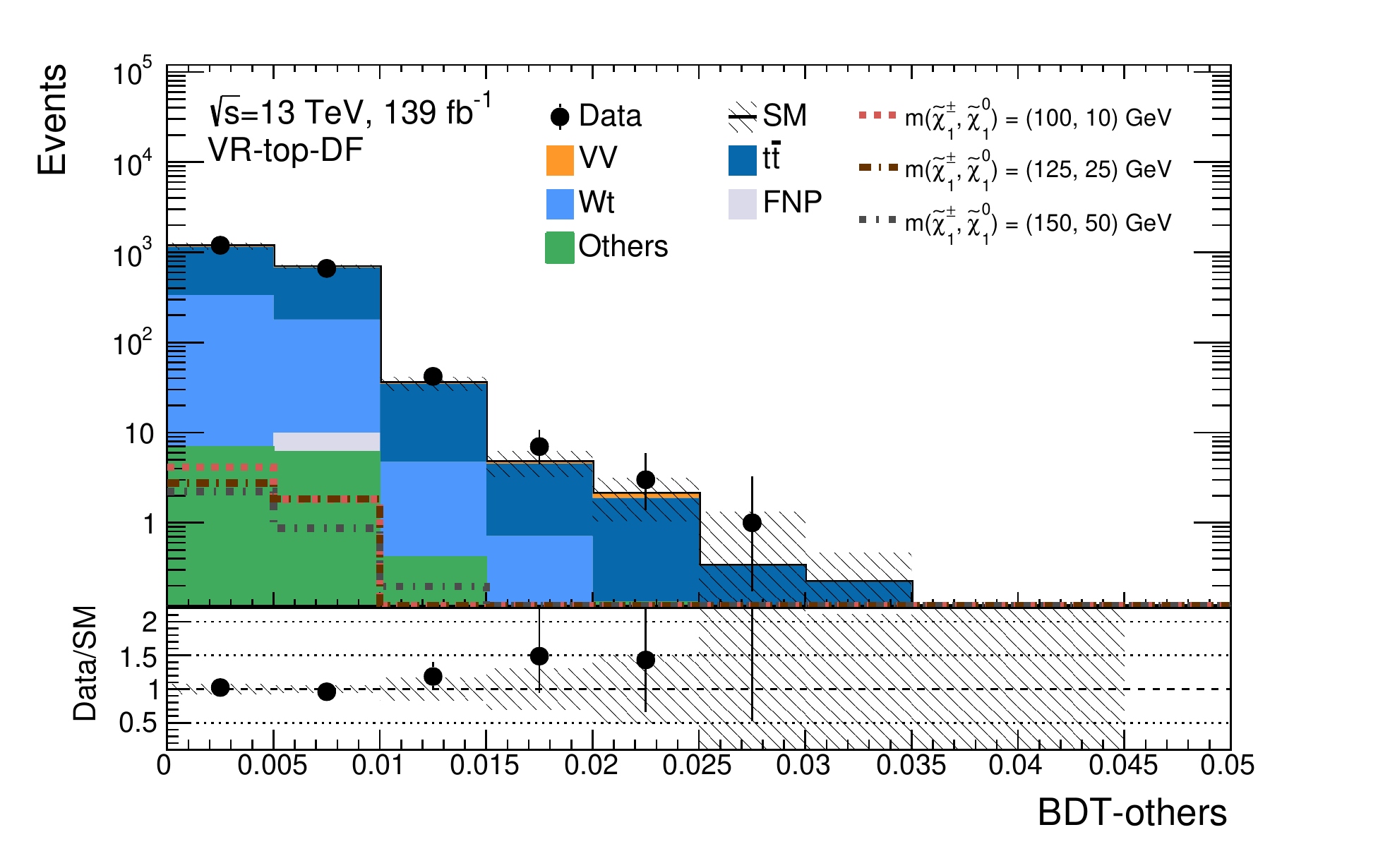}
\caption{The post-fit distributions in VR-top-DF. Both statistical and systematic uncertainties are shown.}
\label{fig:VR_topDF1J_ML}
\end{figure}

\begin{figure}[!htb]
\centering
\includegraphics[width=0.45\linewidth]{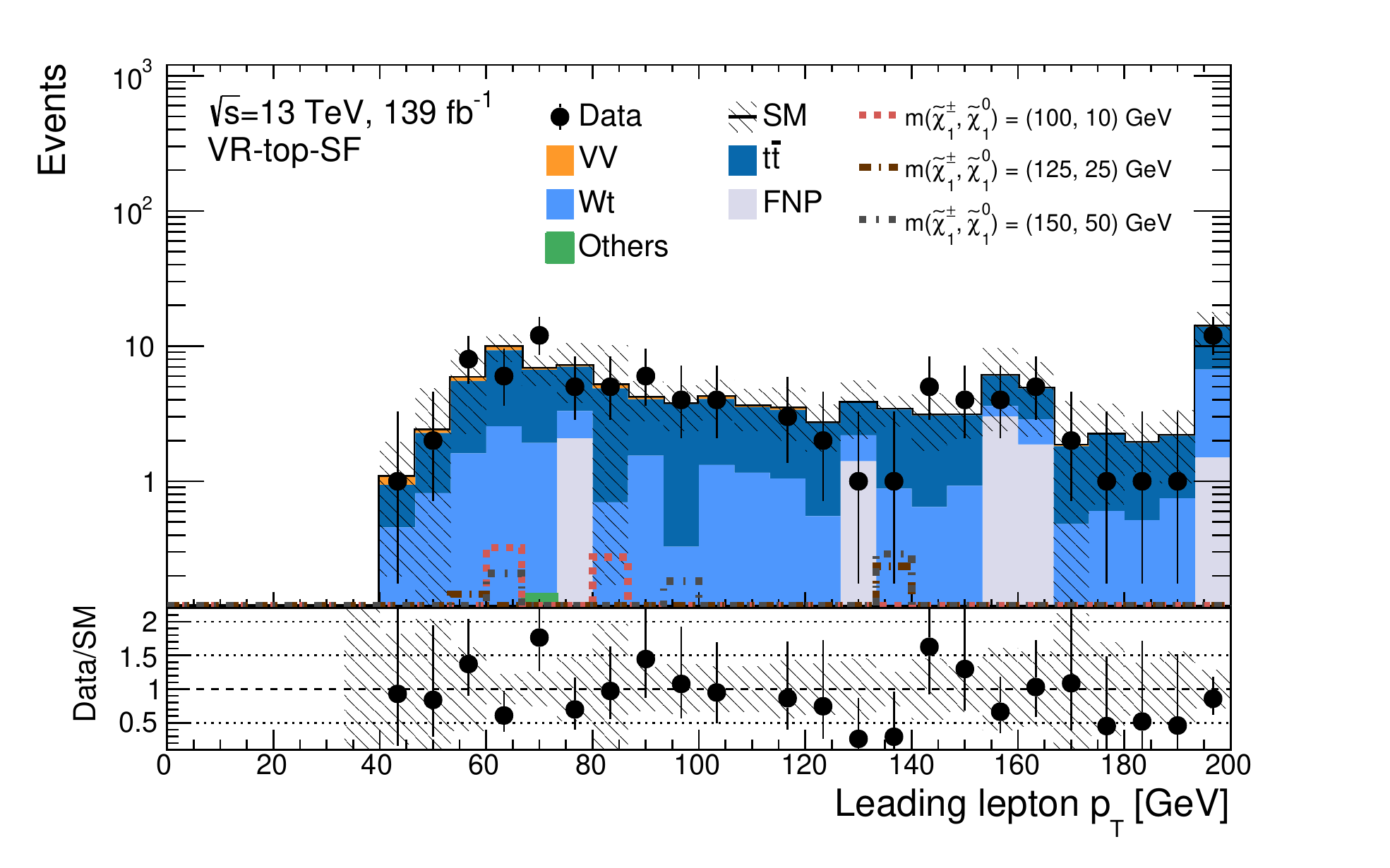}
\includegraphics[width=0.45\linewidth]{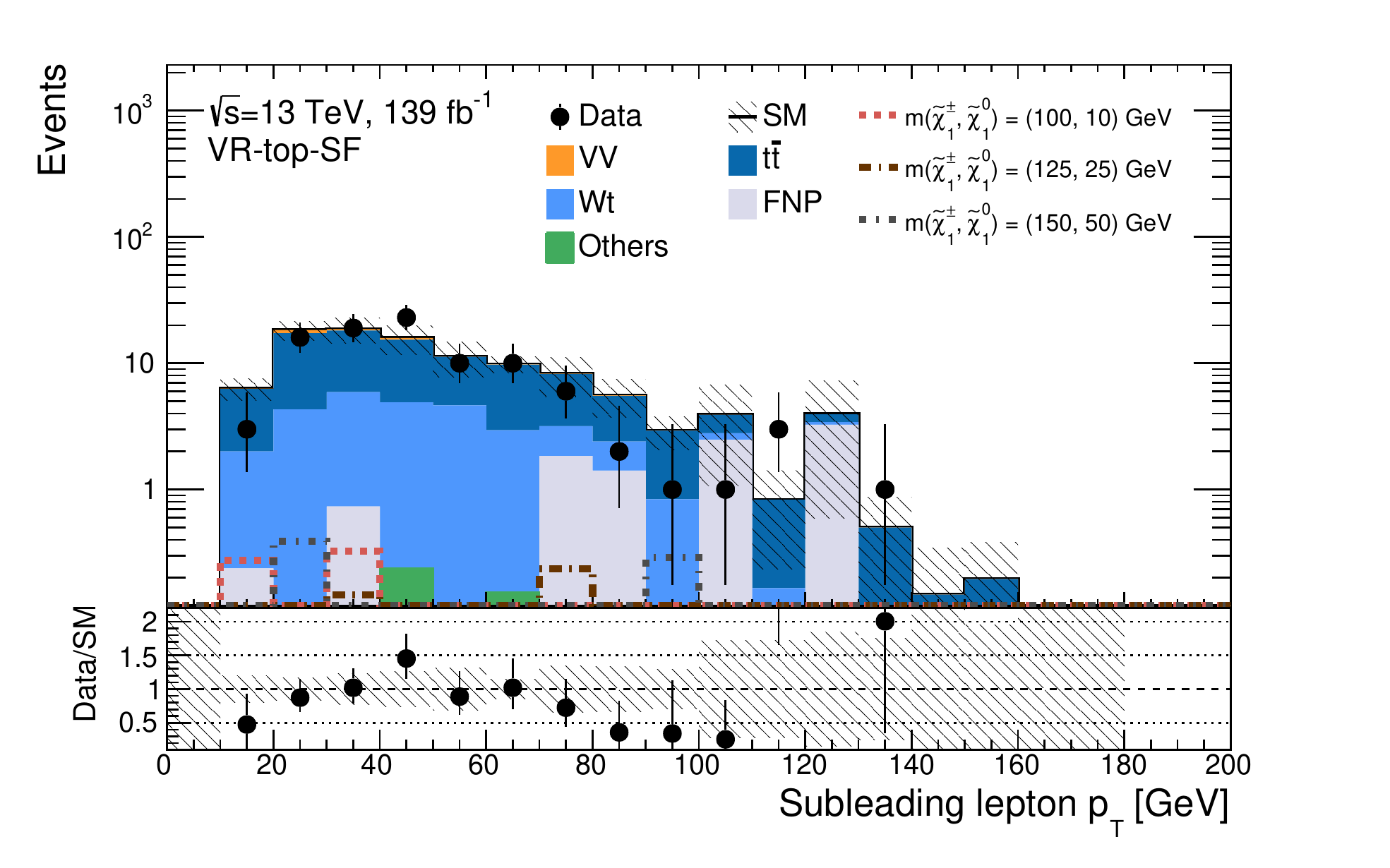}
\includegraphics[width=0.45\linewidth]{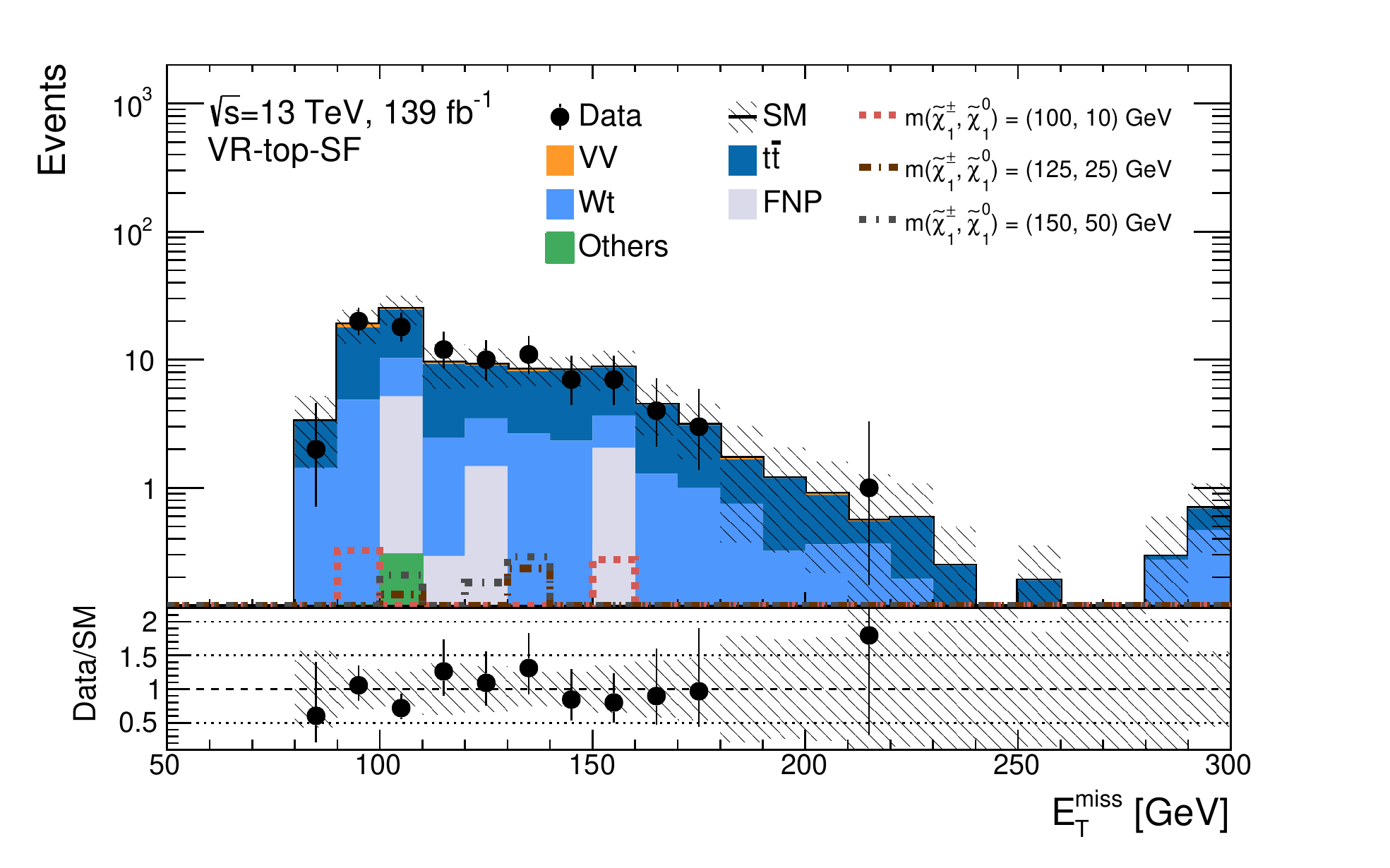}
\includegraphics[width=0.45\linewidth]{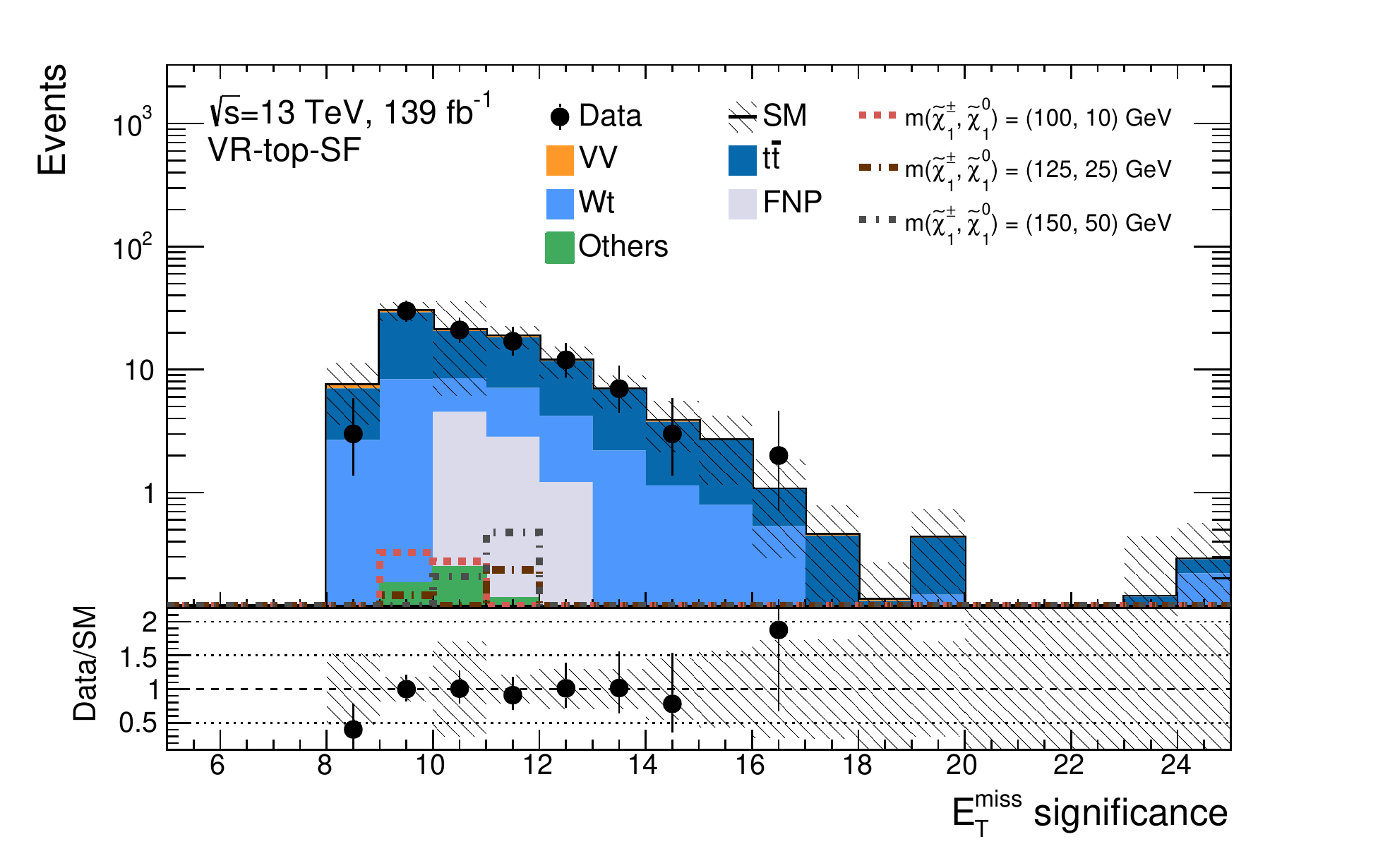}
\includegraphics[width=0.45\linewidth]{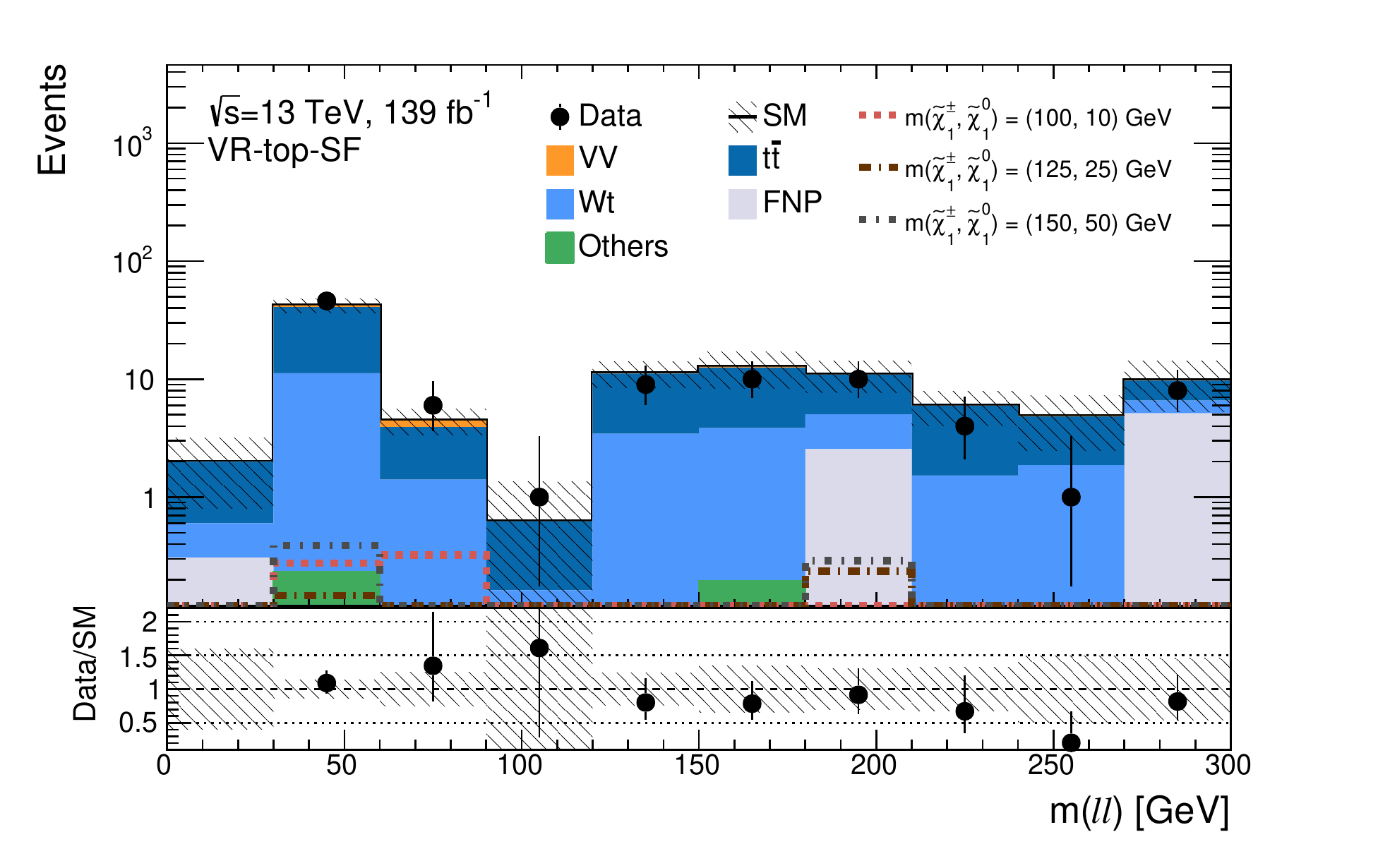}
\includegraphics[width=0.45\linewidth]{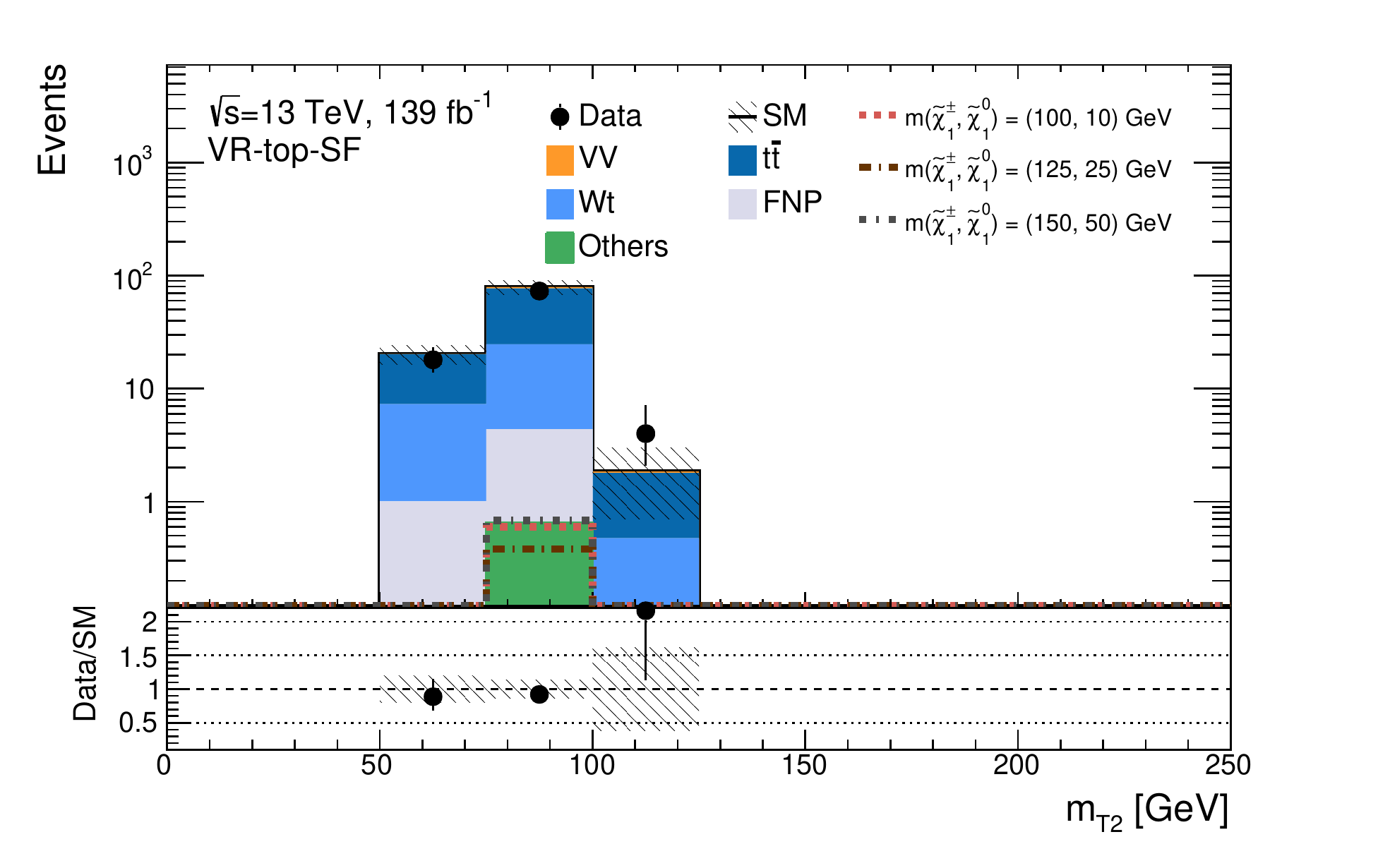}
\includegraphics[width=0.45\linewidth]{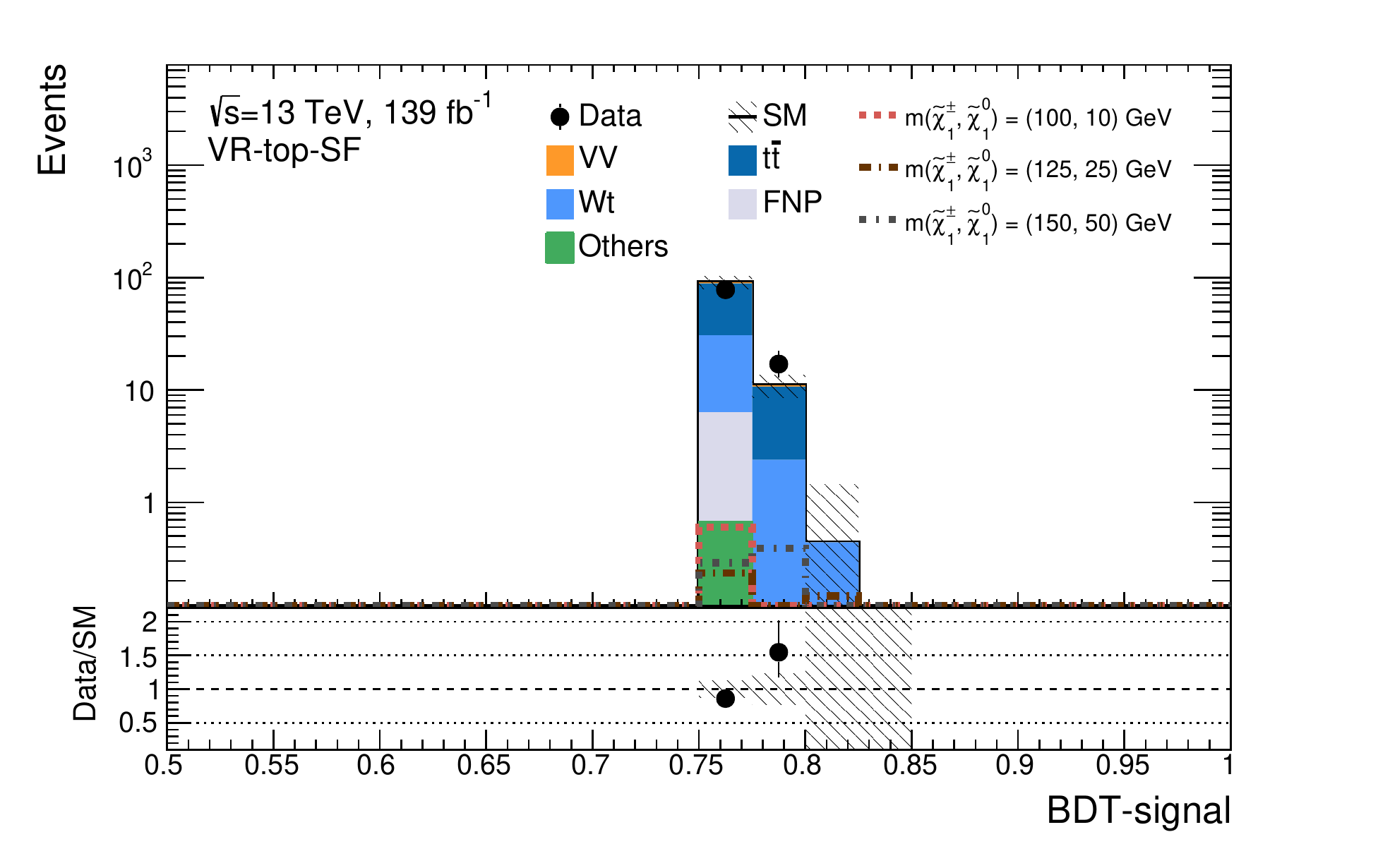}
\includegraphics[width=0.45\linewidth]{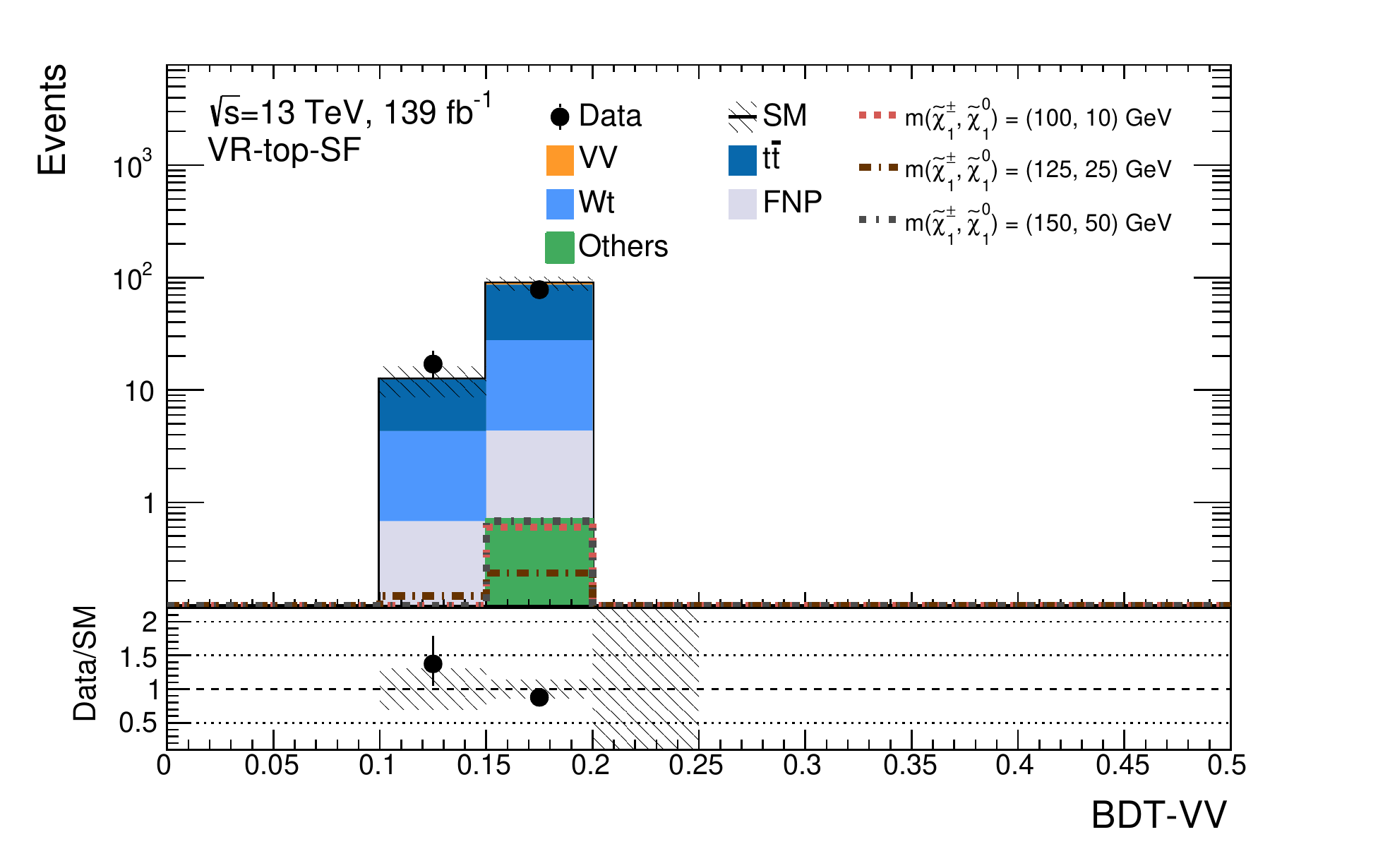}
\includegraphics[width=0.45\linewidth]{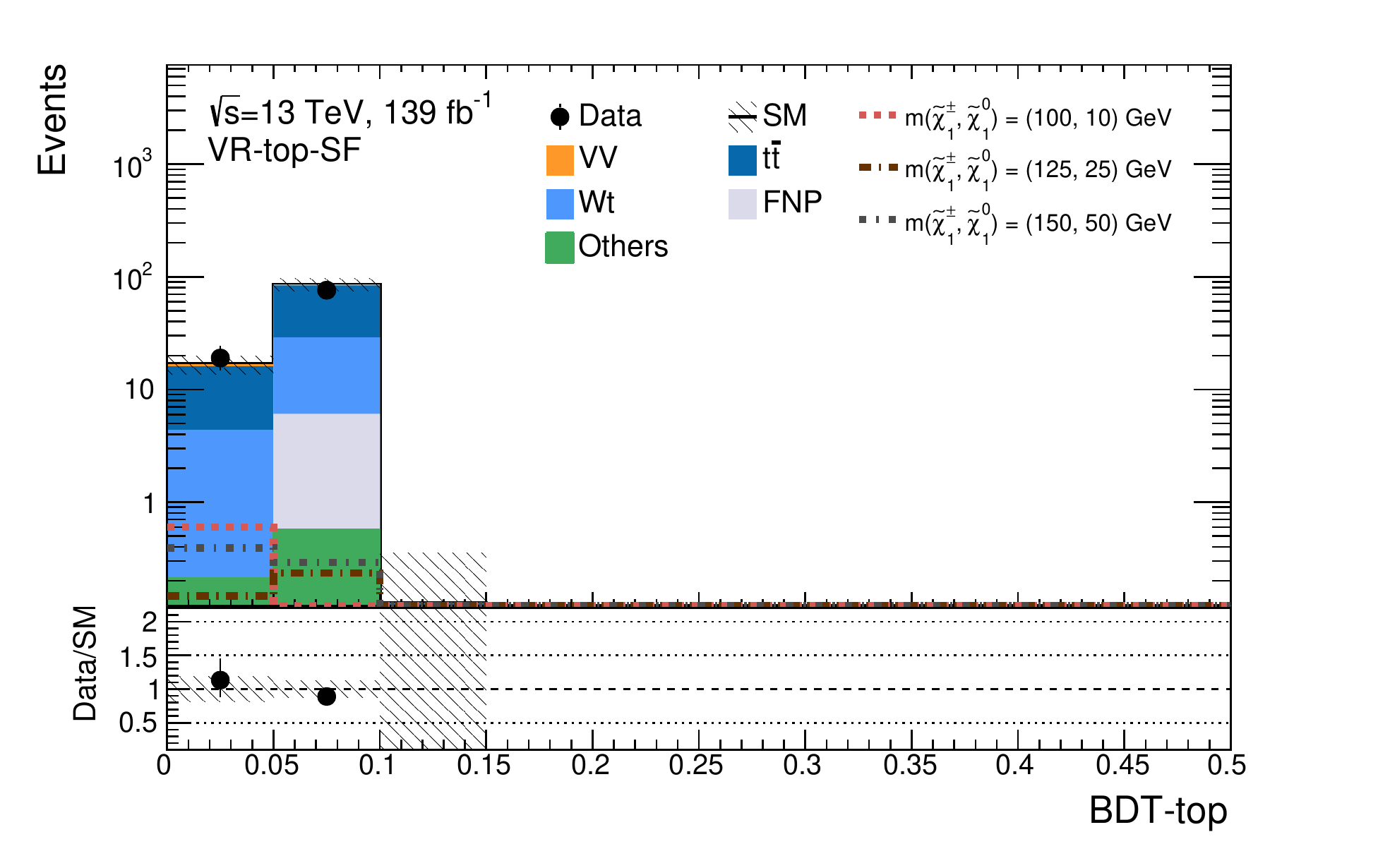}
\includegraphics[width=0.45\linewidth]{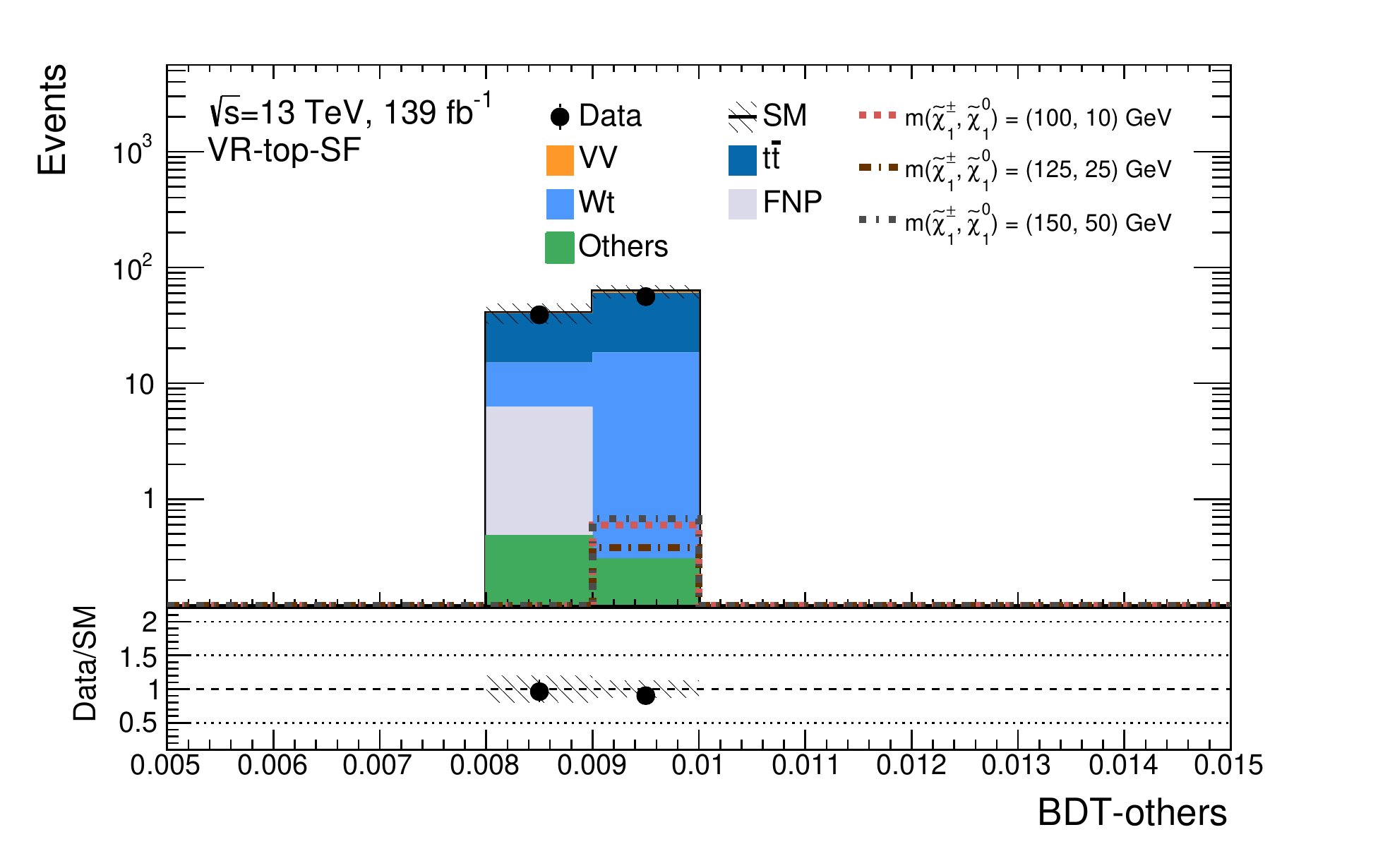}
\caption{The post-fit distributions in VR-top-SF. Both statistical and systematic uncertainties are shown.}
\label{fig:VR_topSF1J_ML}
\end{figure}

\begin{figure}[!htb]
\centering
\includegraphics[width=0.45\linewidth]{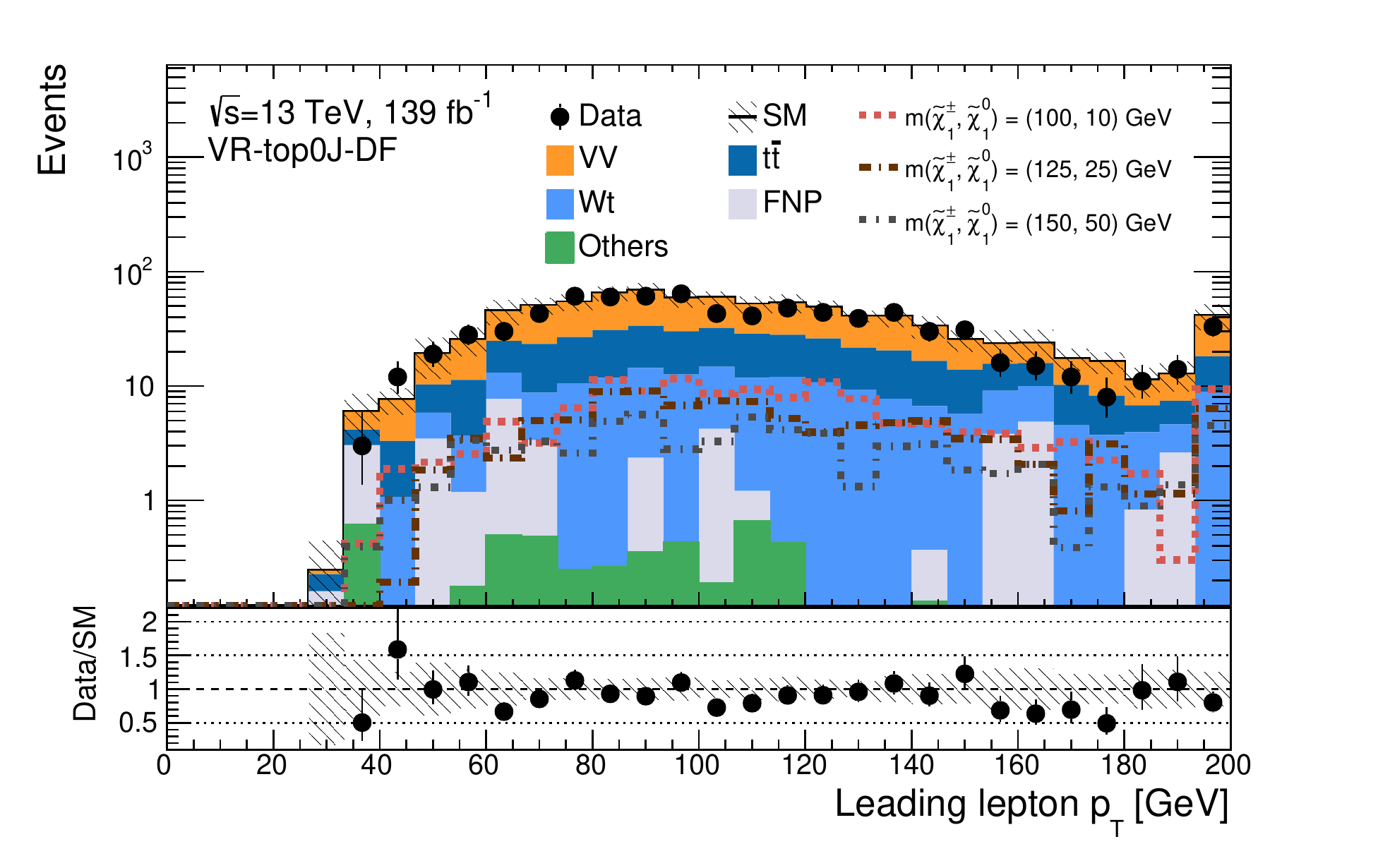}
\includegraphics[width=0.45\linewidth]{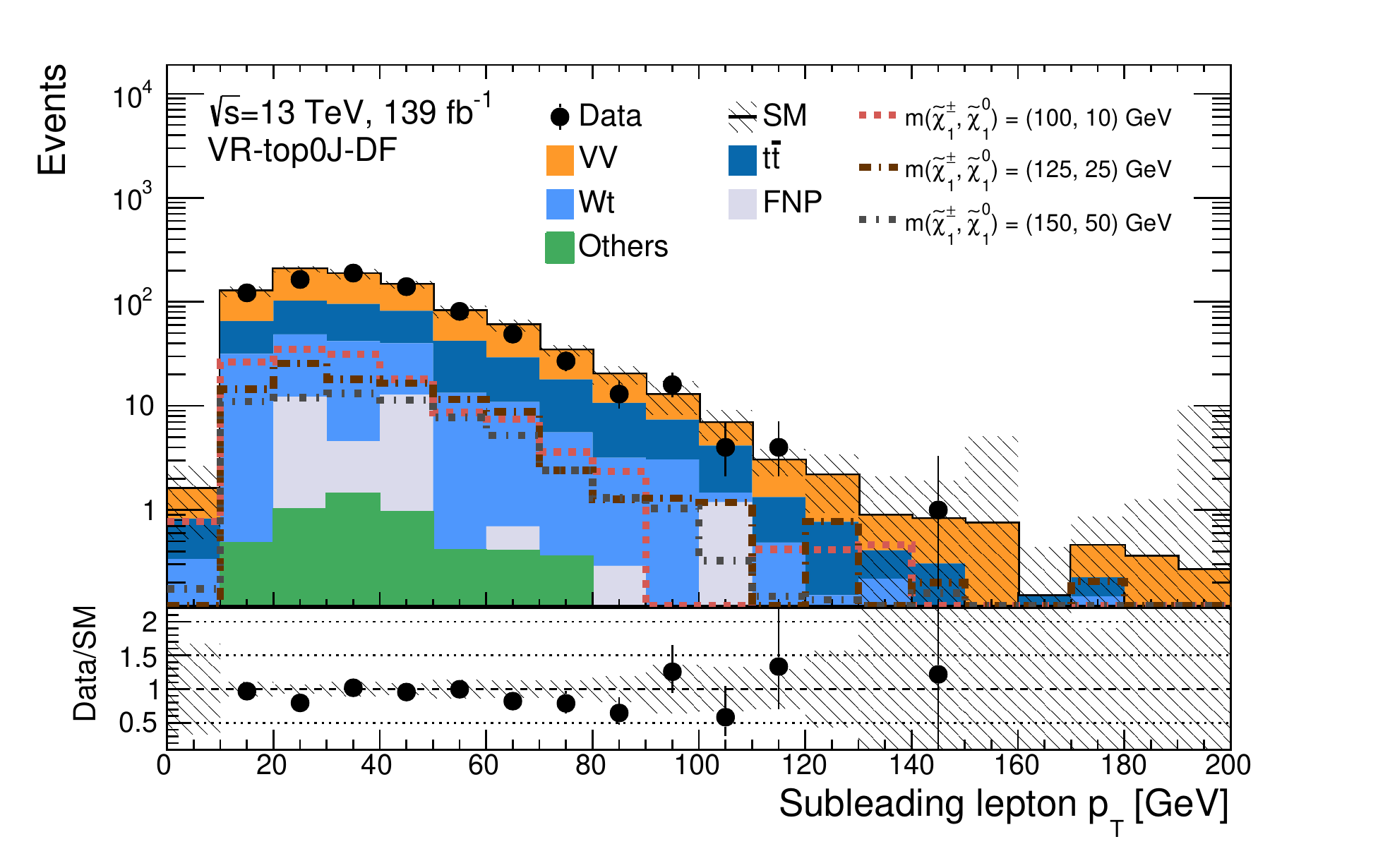}
\includegraphics[width=0.45\linewidth]{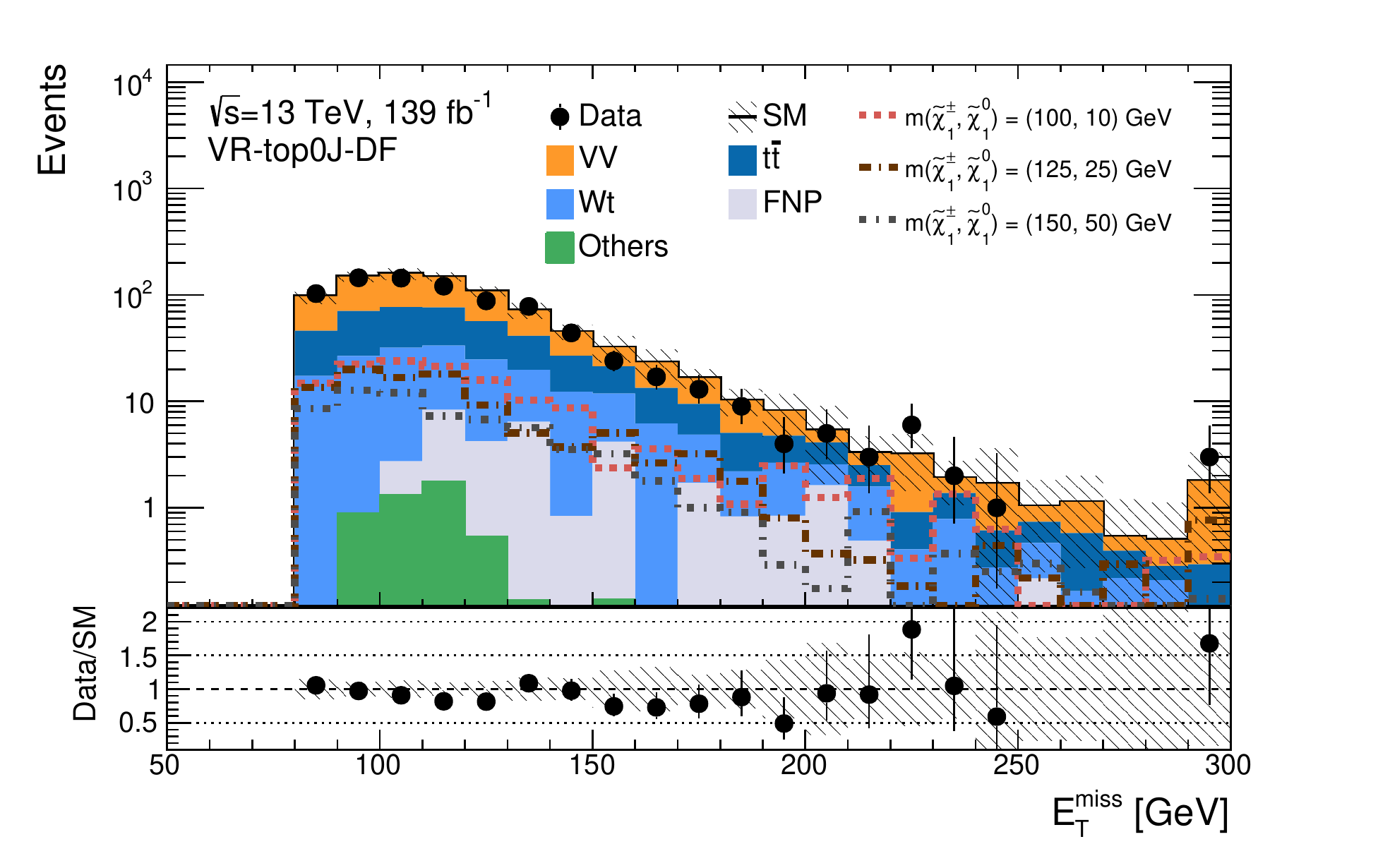}
\includegraphics[width=0.45\linewidth]{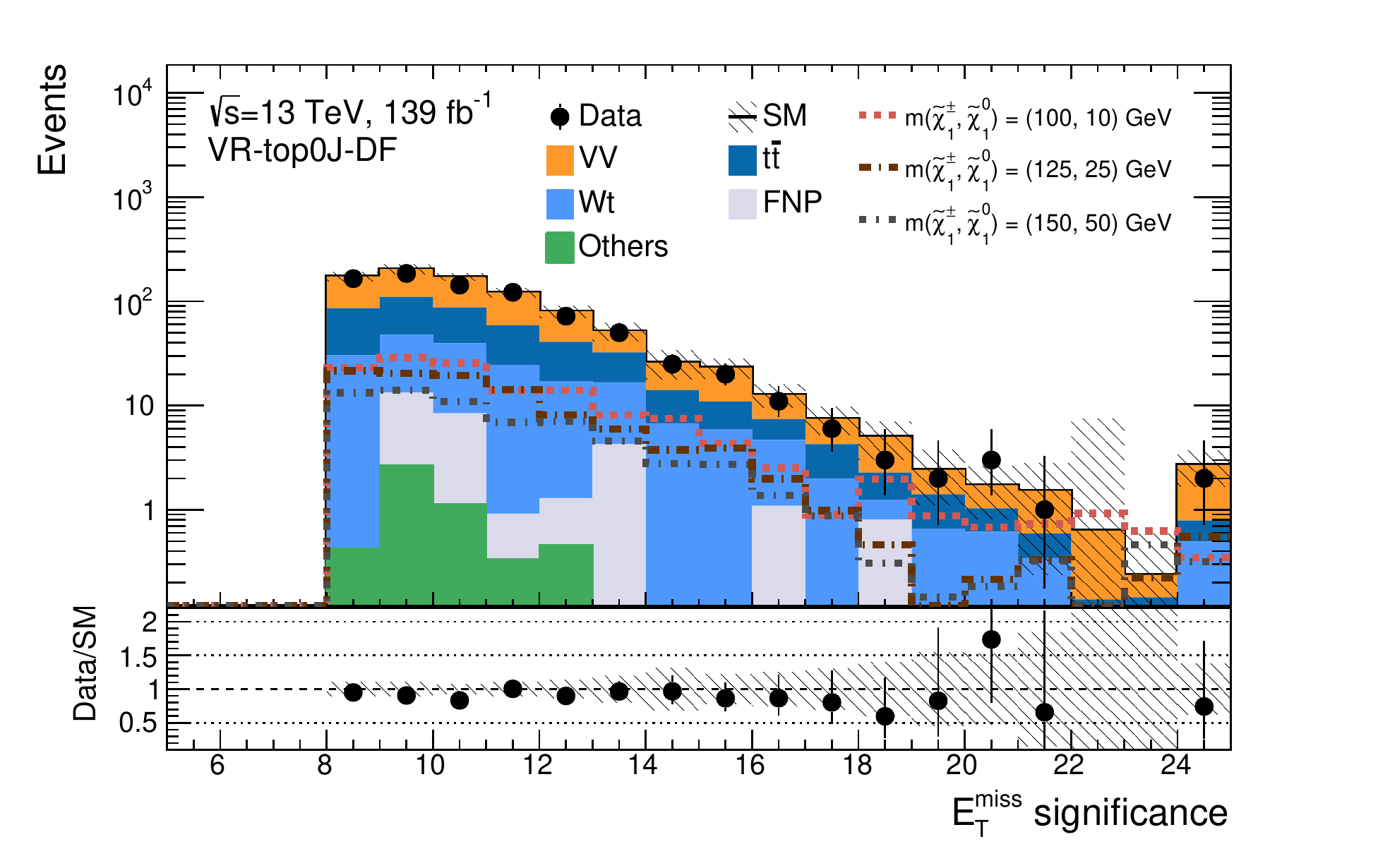}
\includegraphics[width=0.45\linewidth]{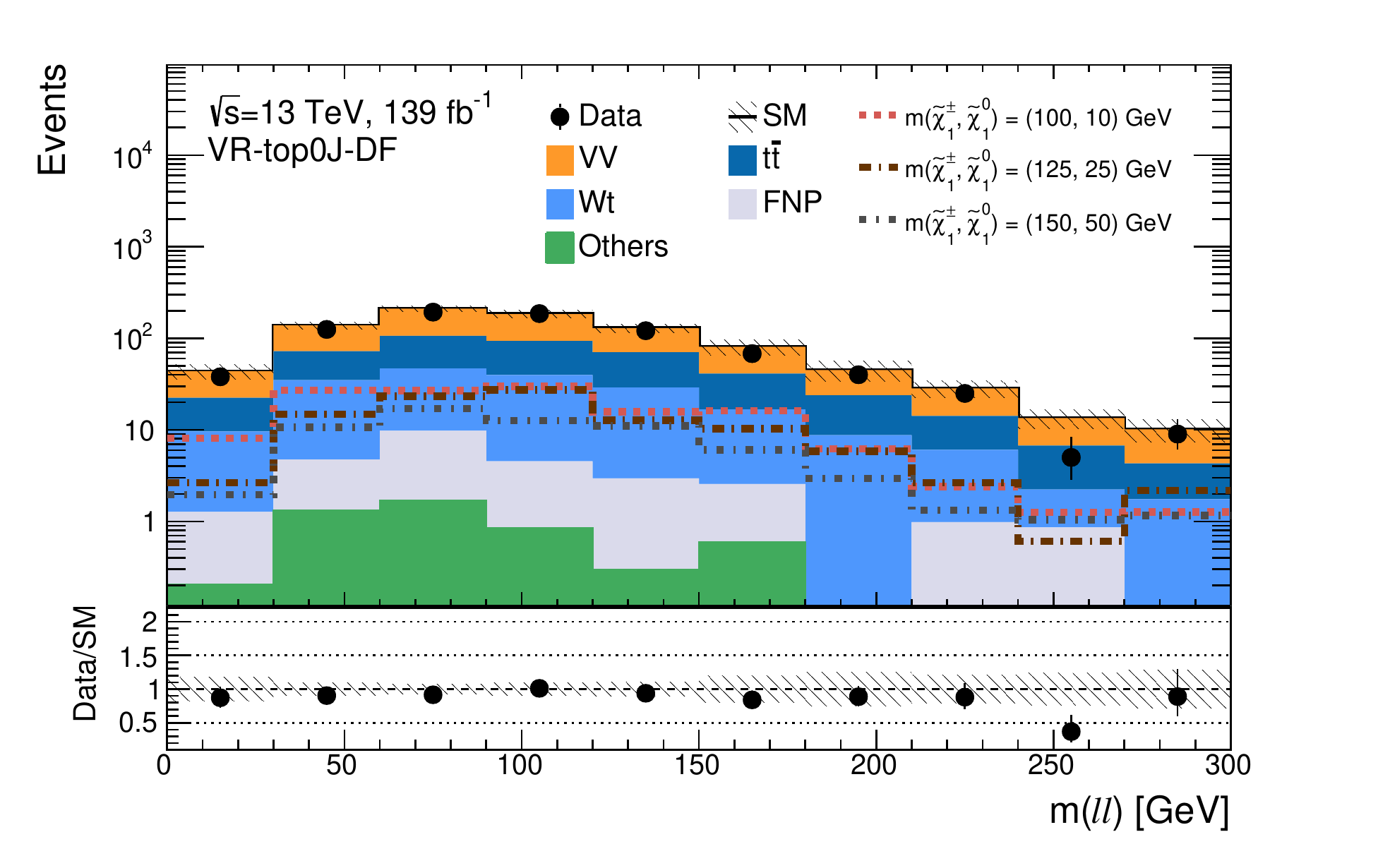}
\includegraphics[width=0.45\linewidth]{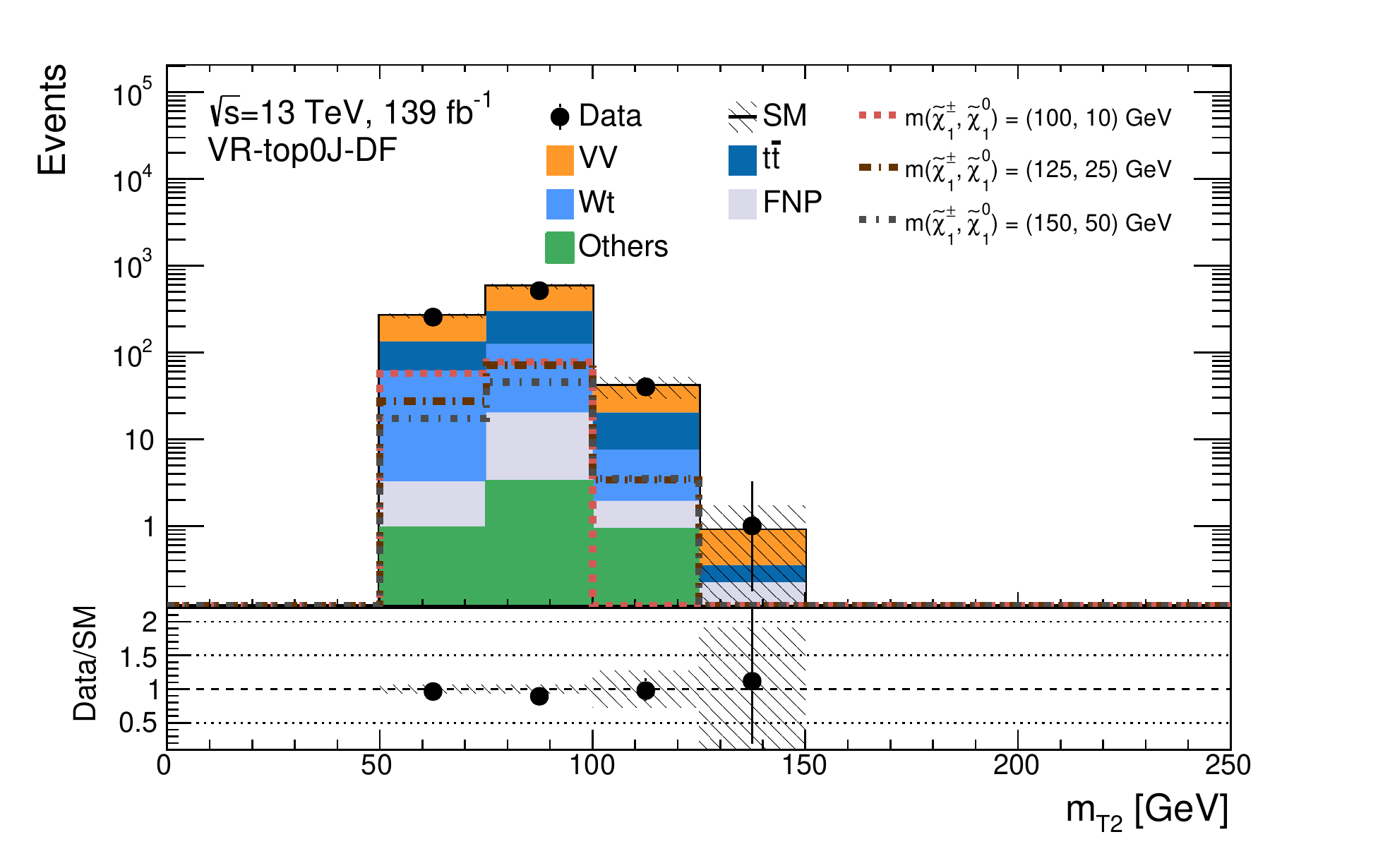}
\includegraphics[width=0.45\linewidth]{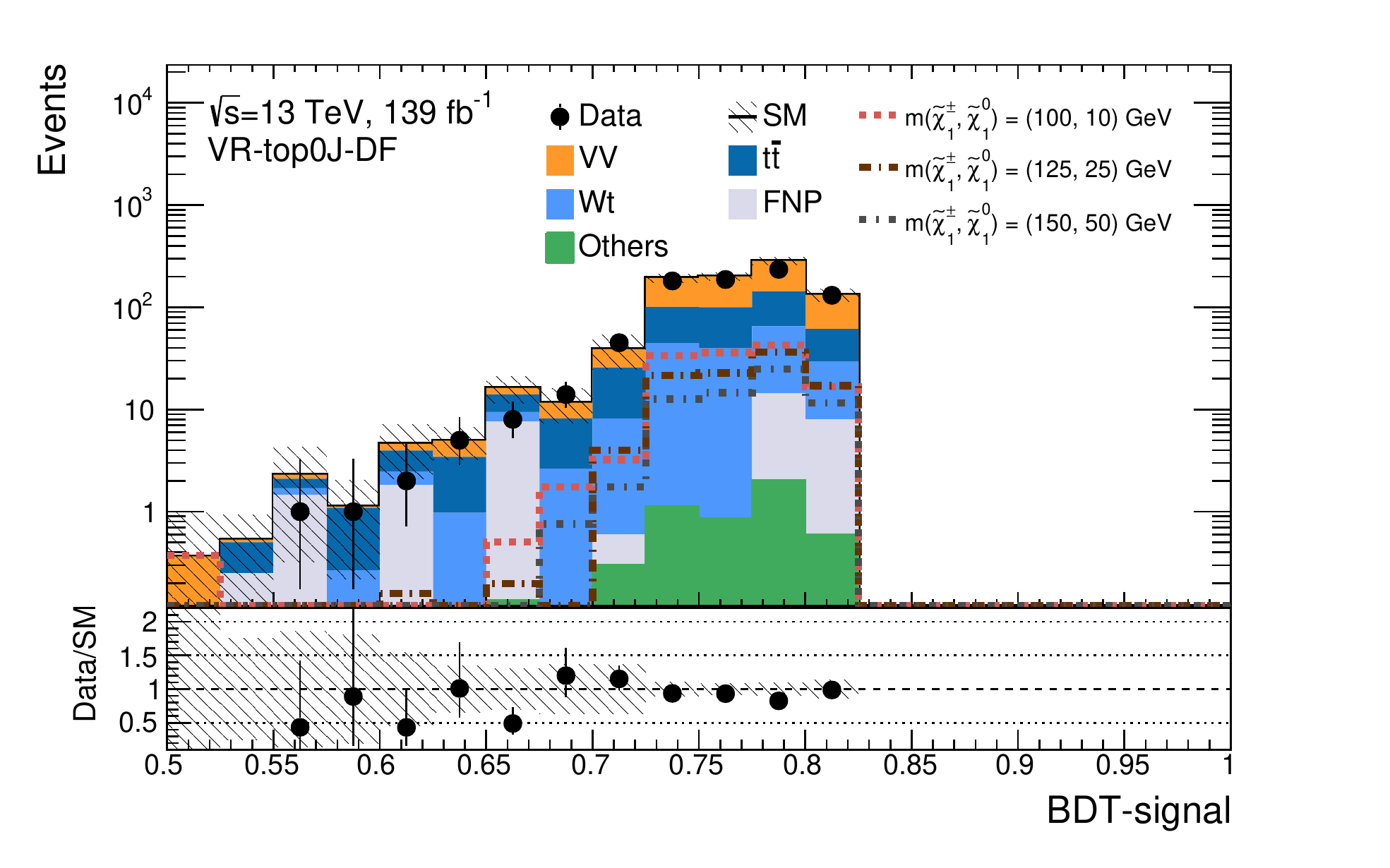}
\includegraphics[width=0.45\linewidth]{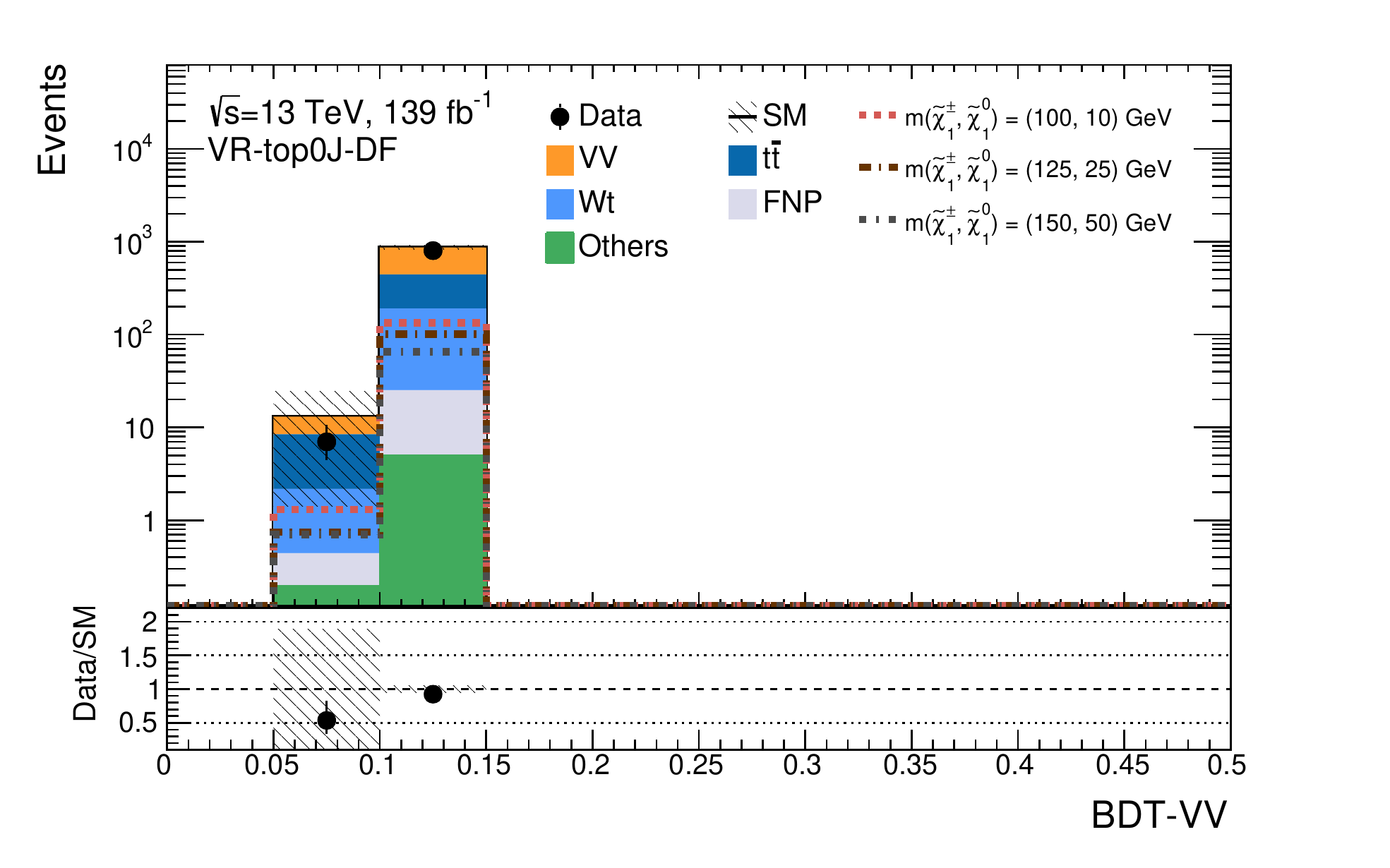}
\includegraphics[width=0.45\linewidth]{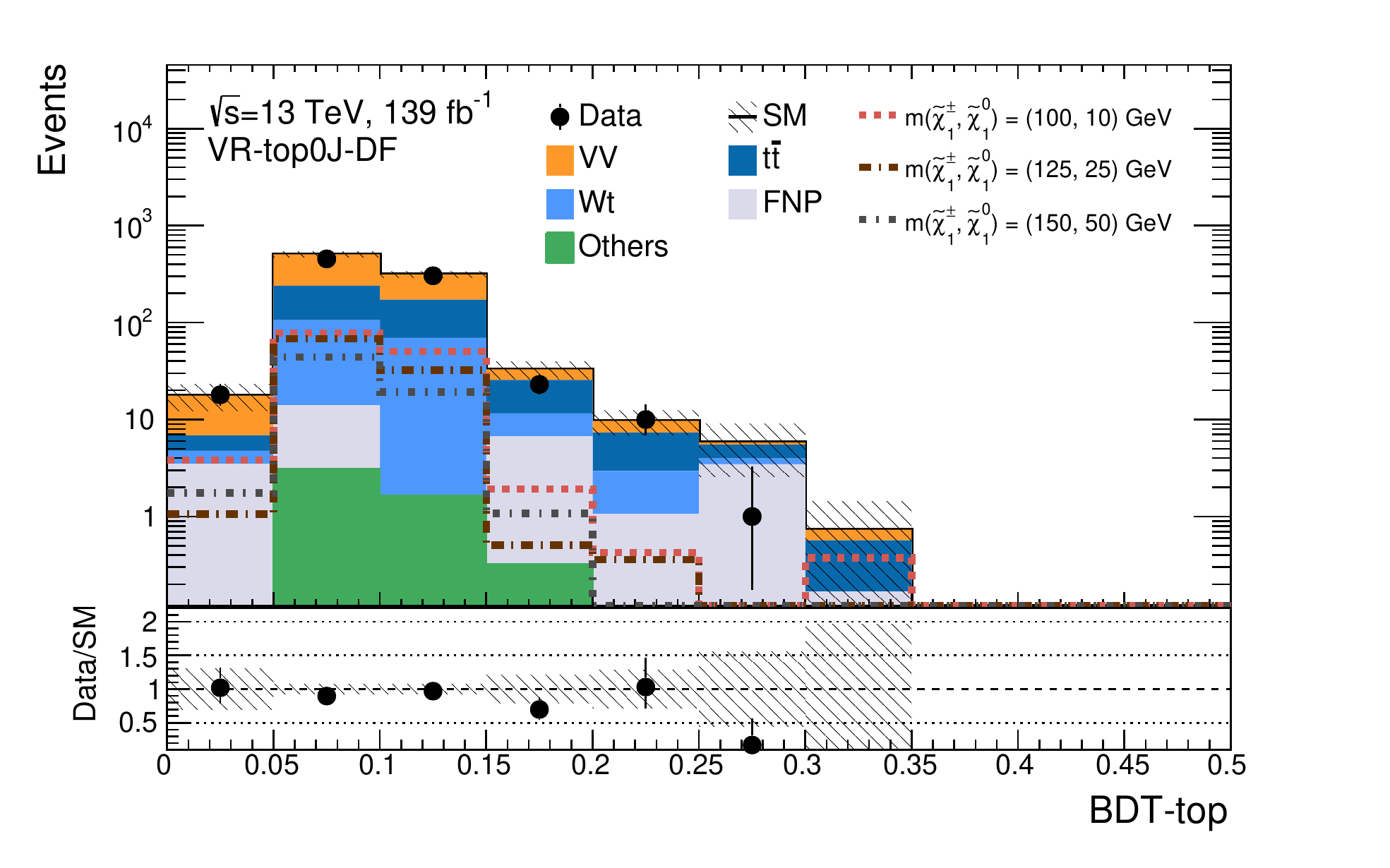}
\includegraphics[width=0.45\linewidth]{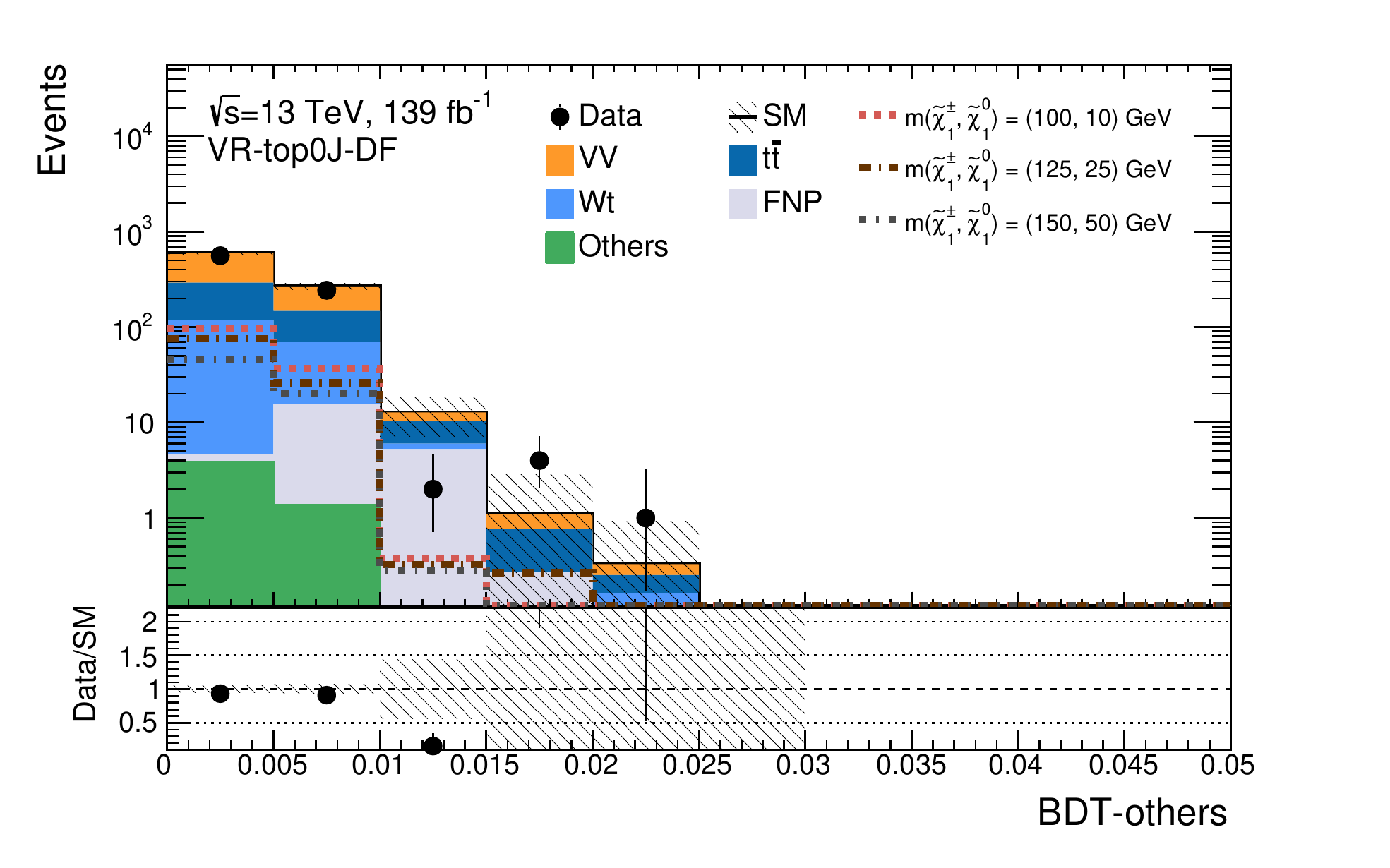}
\caption{The post-fit distributions in VR-top0J-DF. Both statistical and systematic uncertainties are shown.}
\label{fig:VR_top_DF0J_ML}
\end{figure}

\begin{figure}[!htb]
\centering
\includegraphics[width=0.45\linewidth]{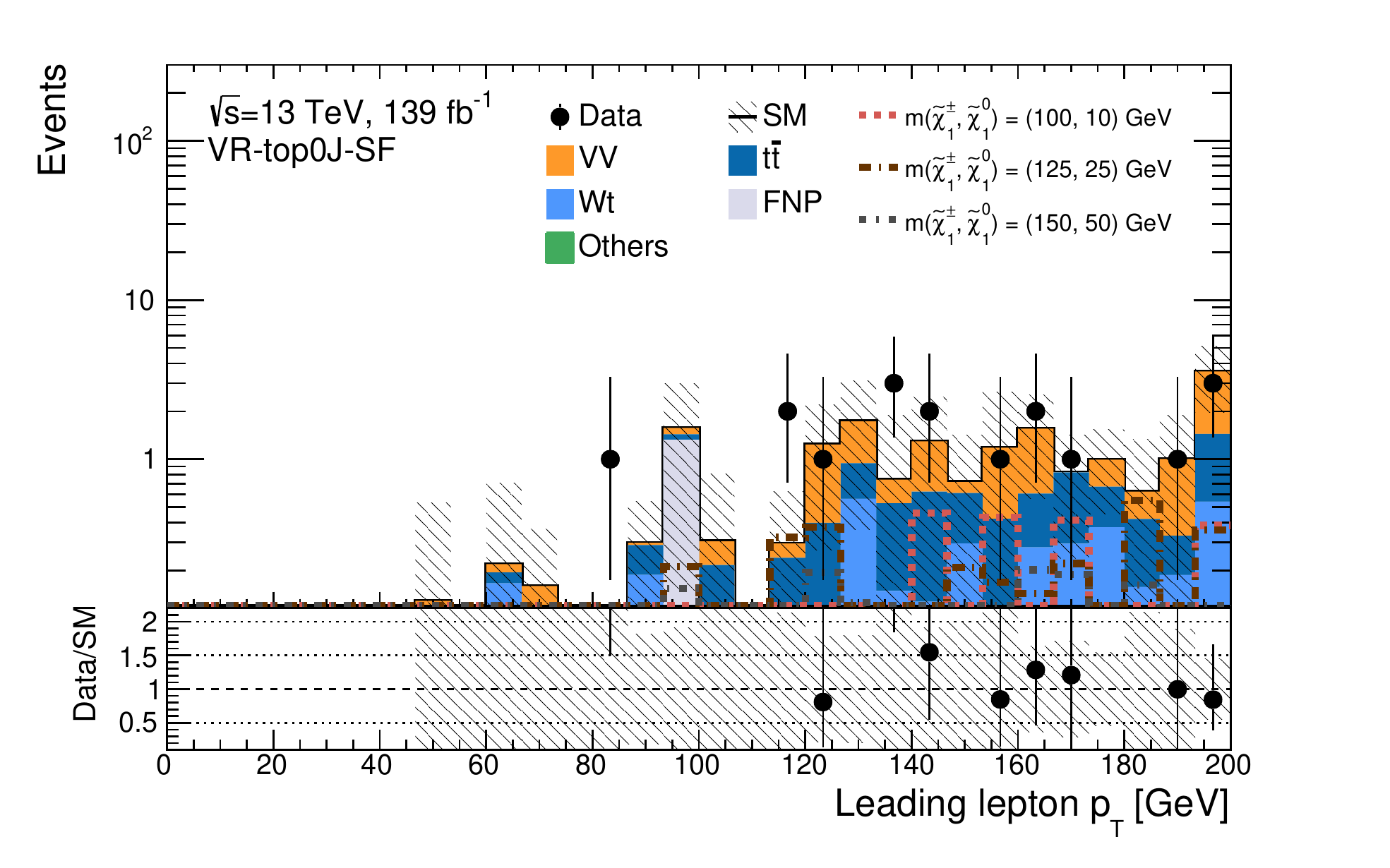}
\includegraphics[width=0.45\linewidth]{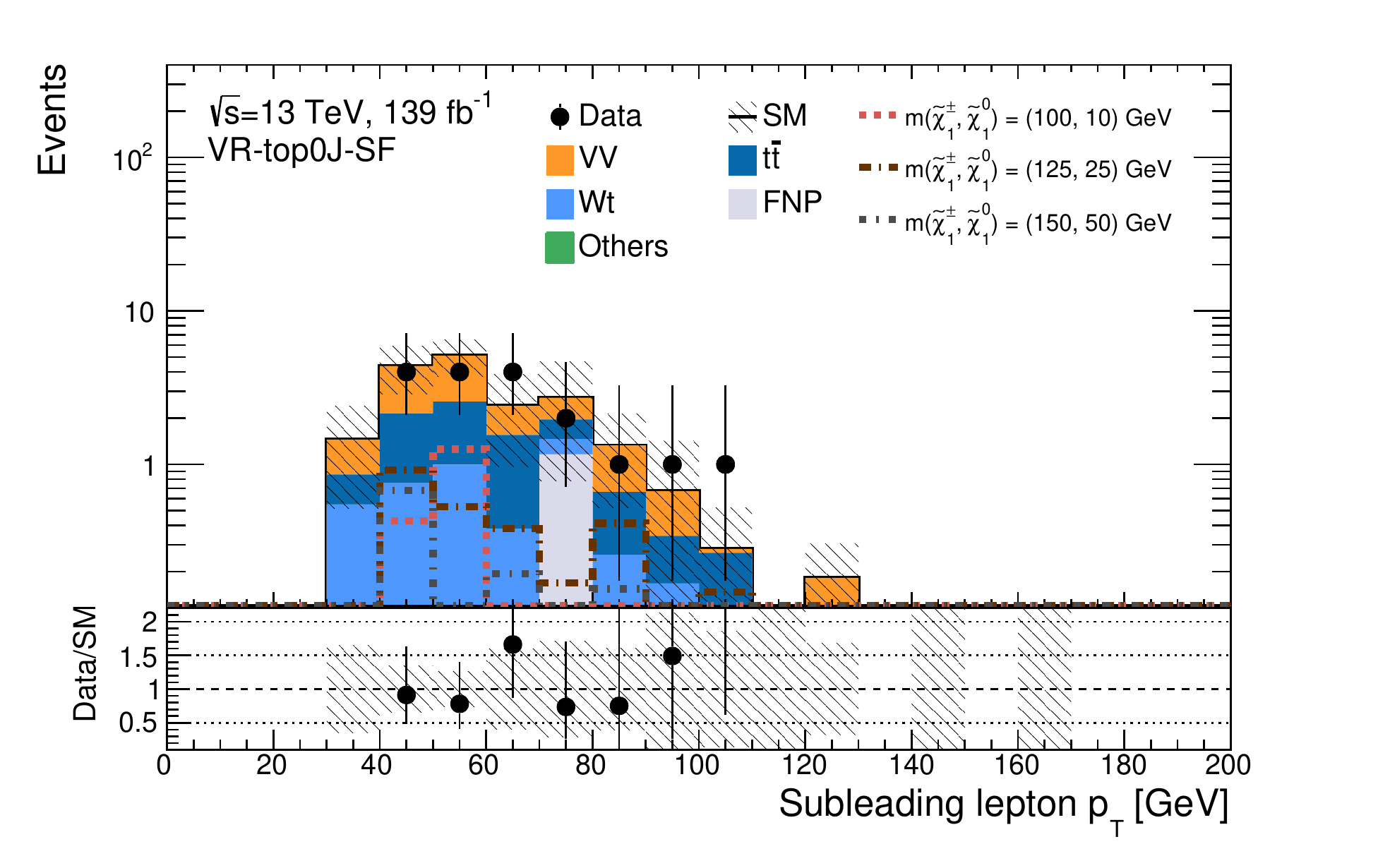}
\includegraphics[width=0.45\linewidth]{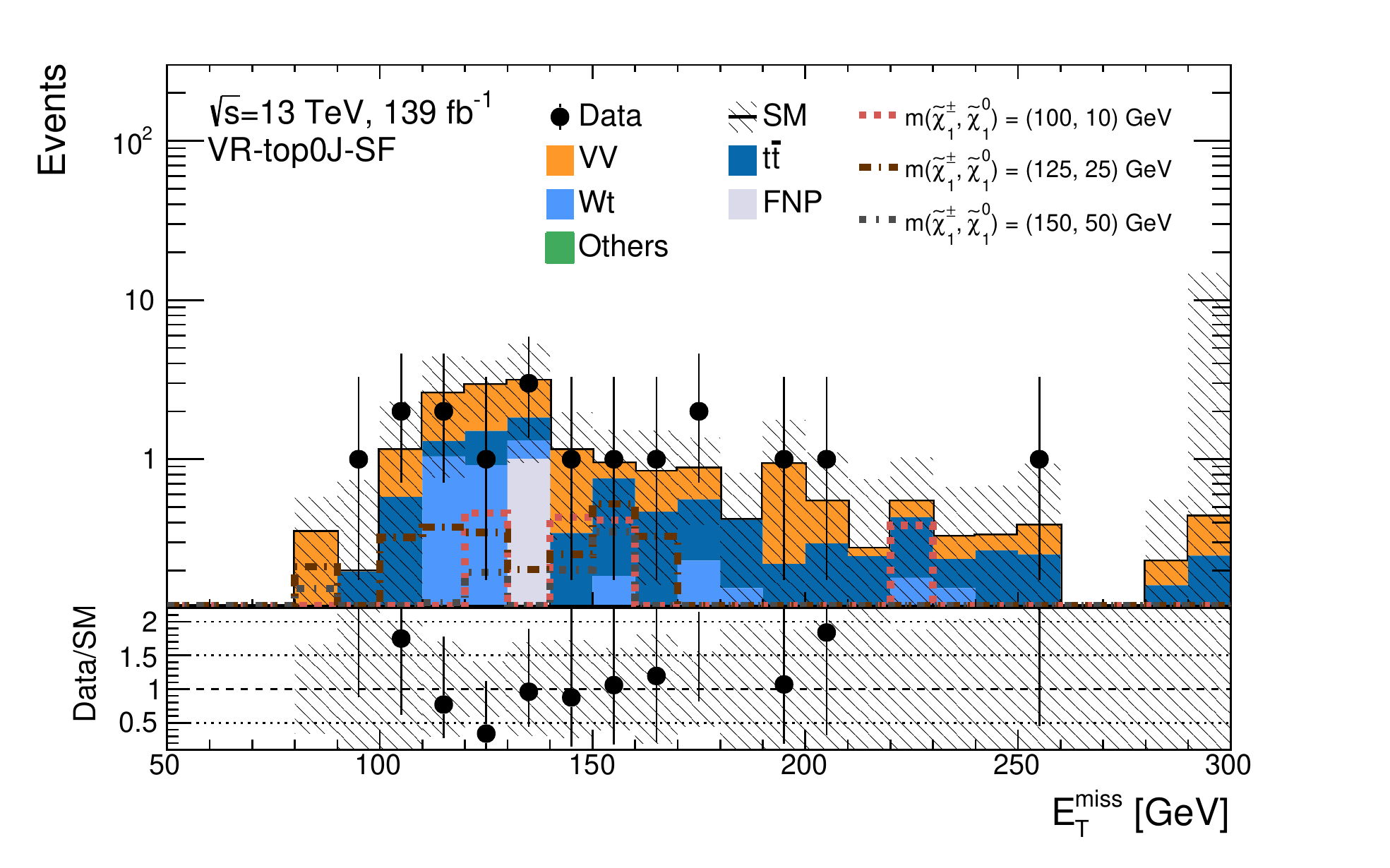}
\includegraphics[width=0.45\linewidth]{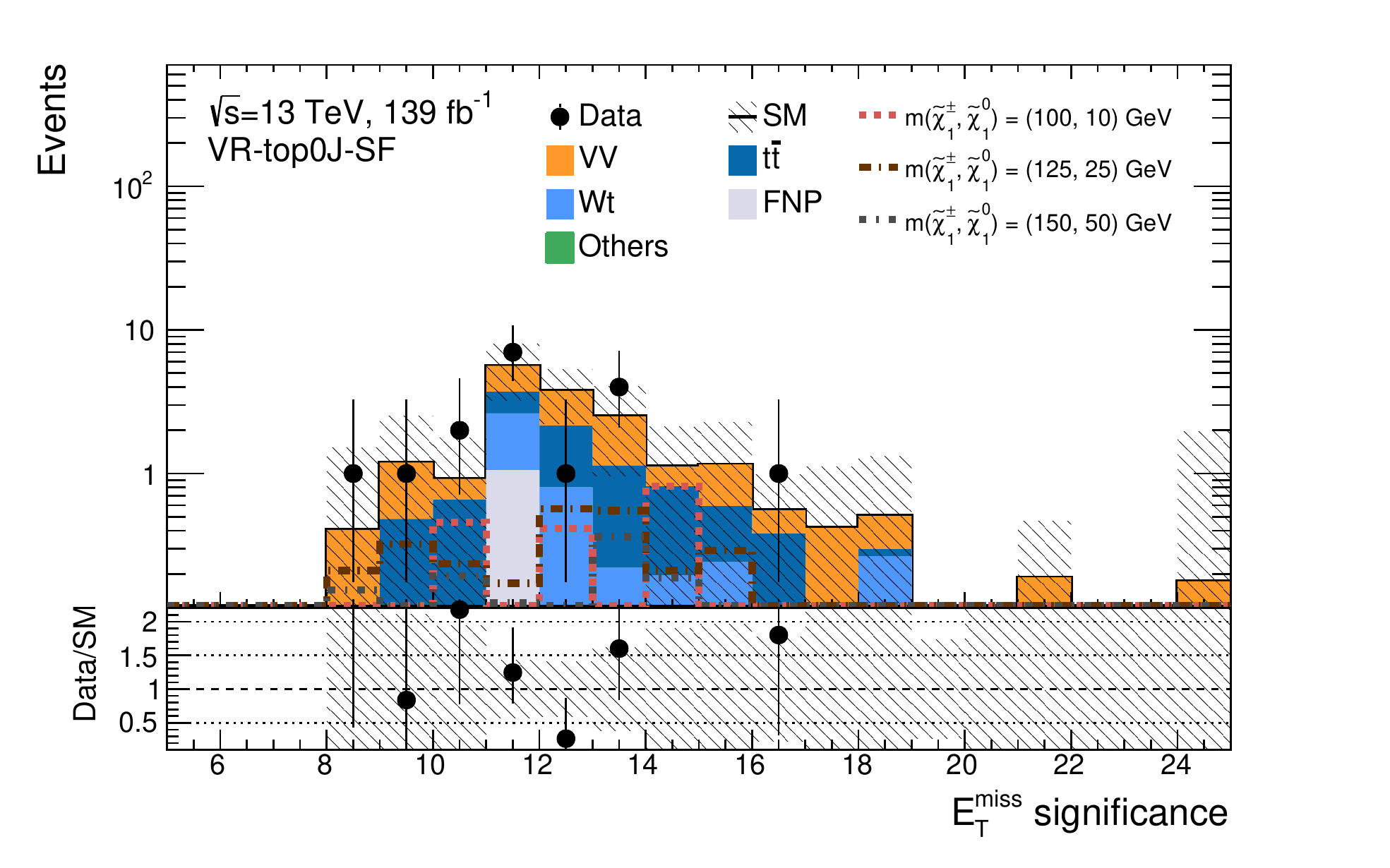}
\includegraphics[width=0.45\linewidth]{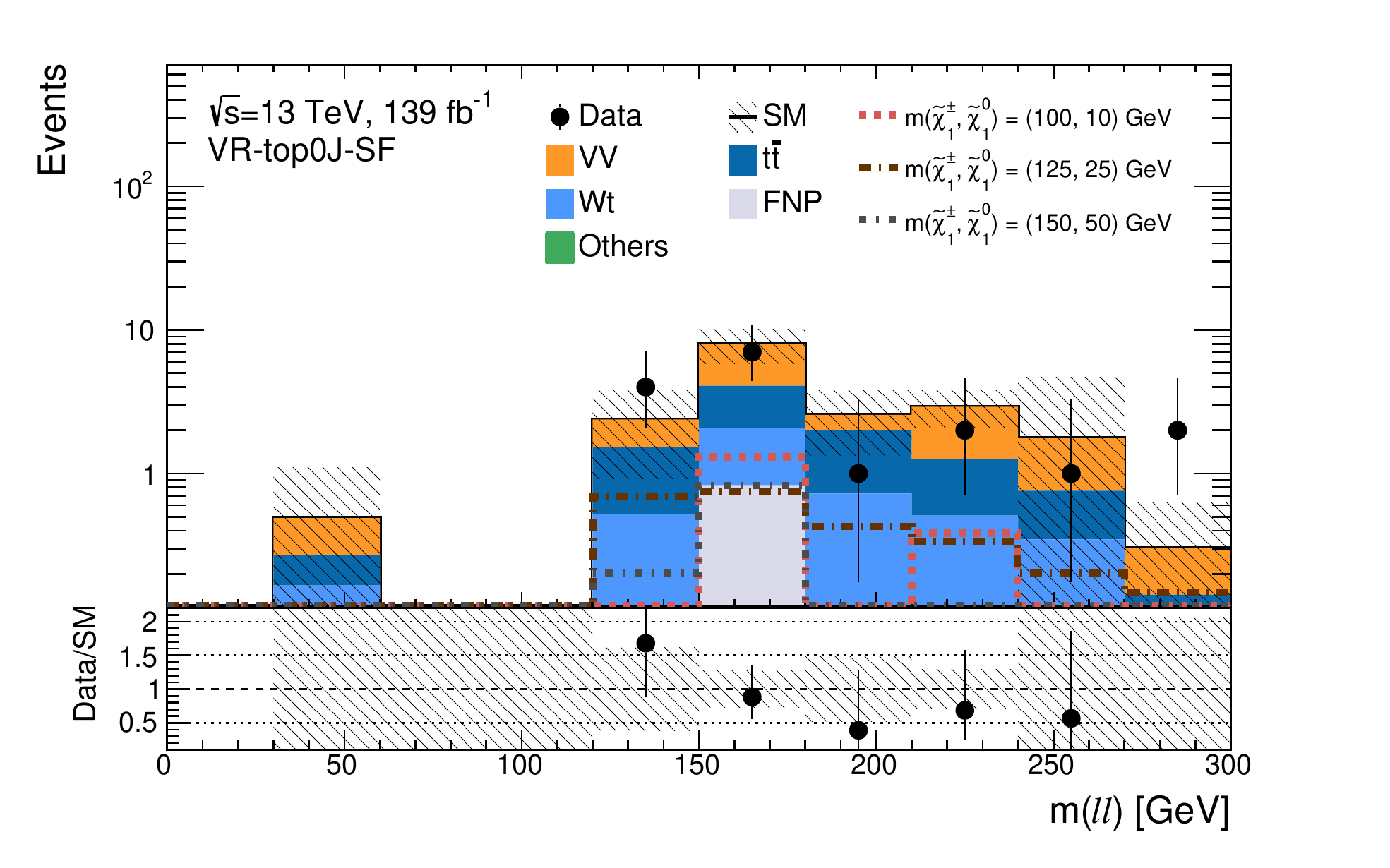}
\includegraphics[width=0.45\linewidth]{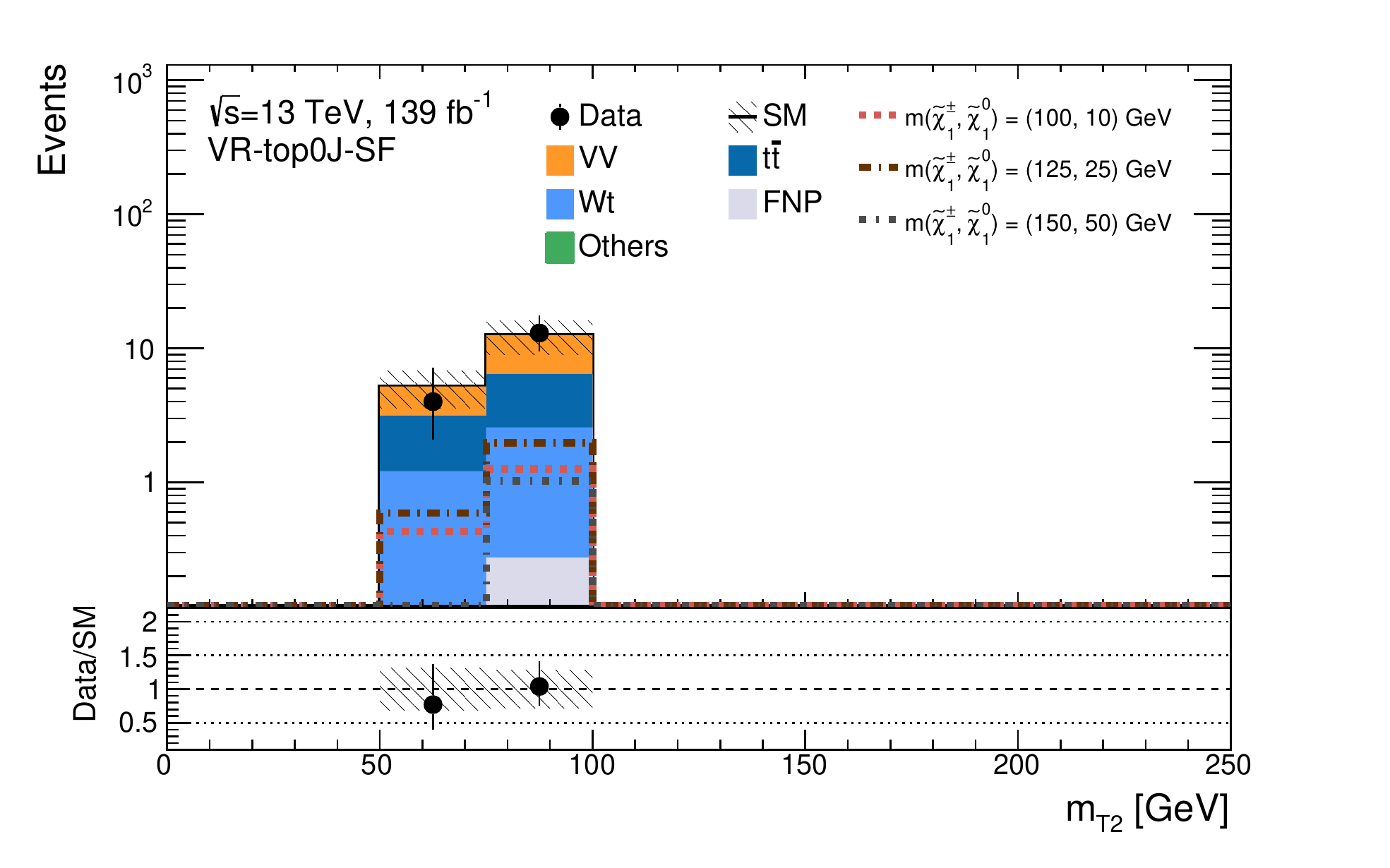}
\includegraphics[width=0.45\linewidth]{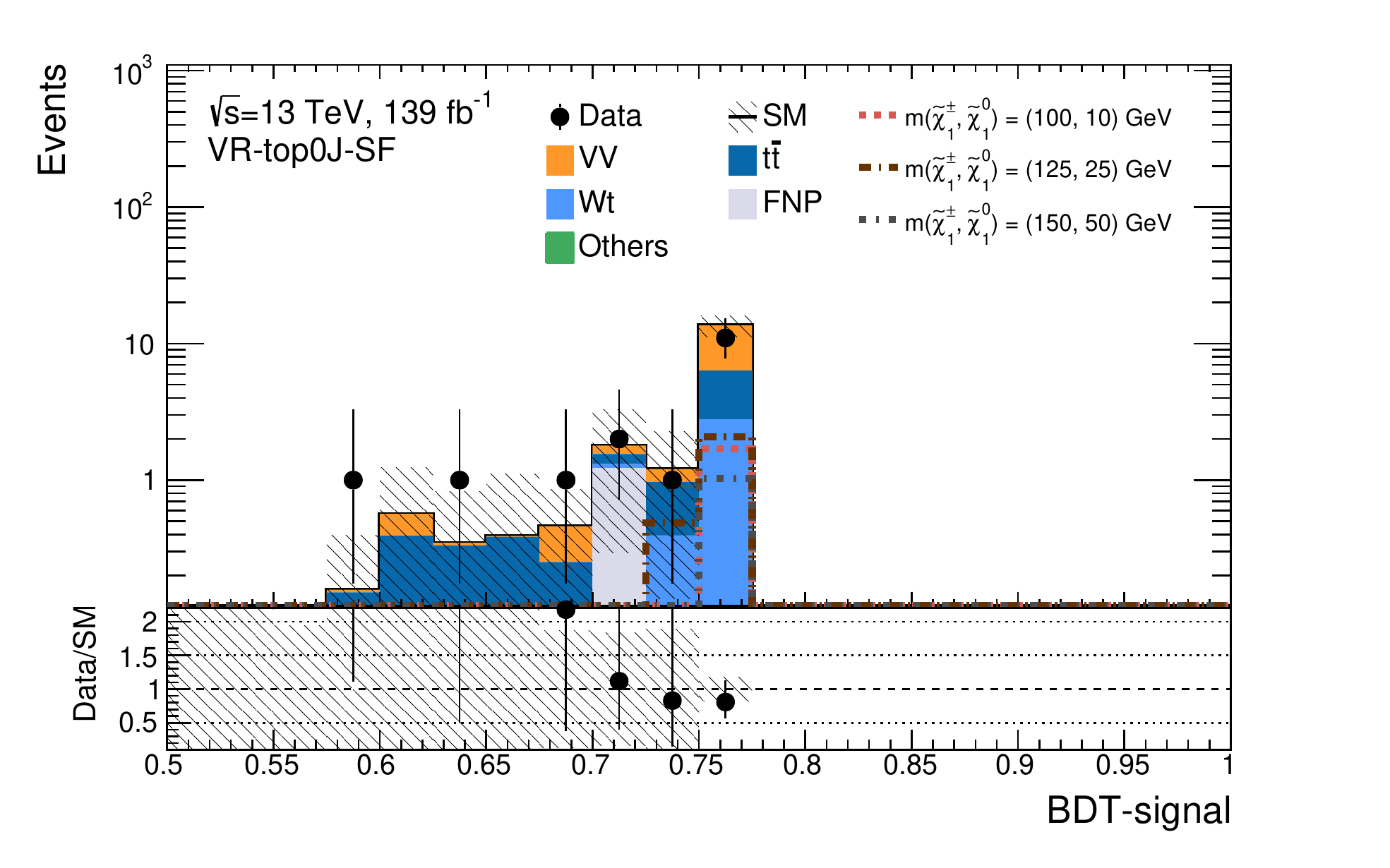}
\includegraphics[width=0.45\linewidth]{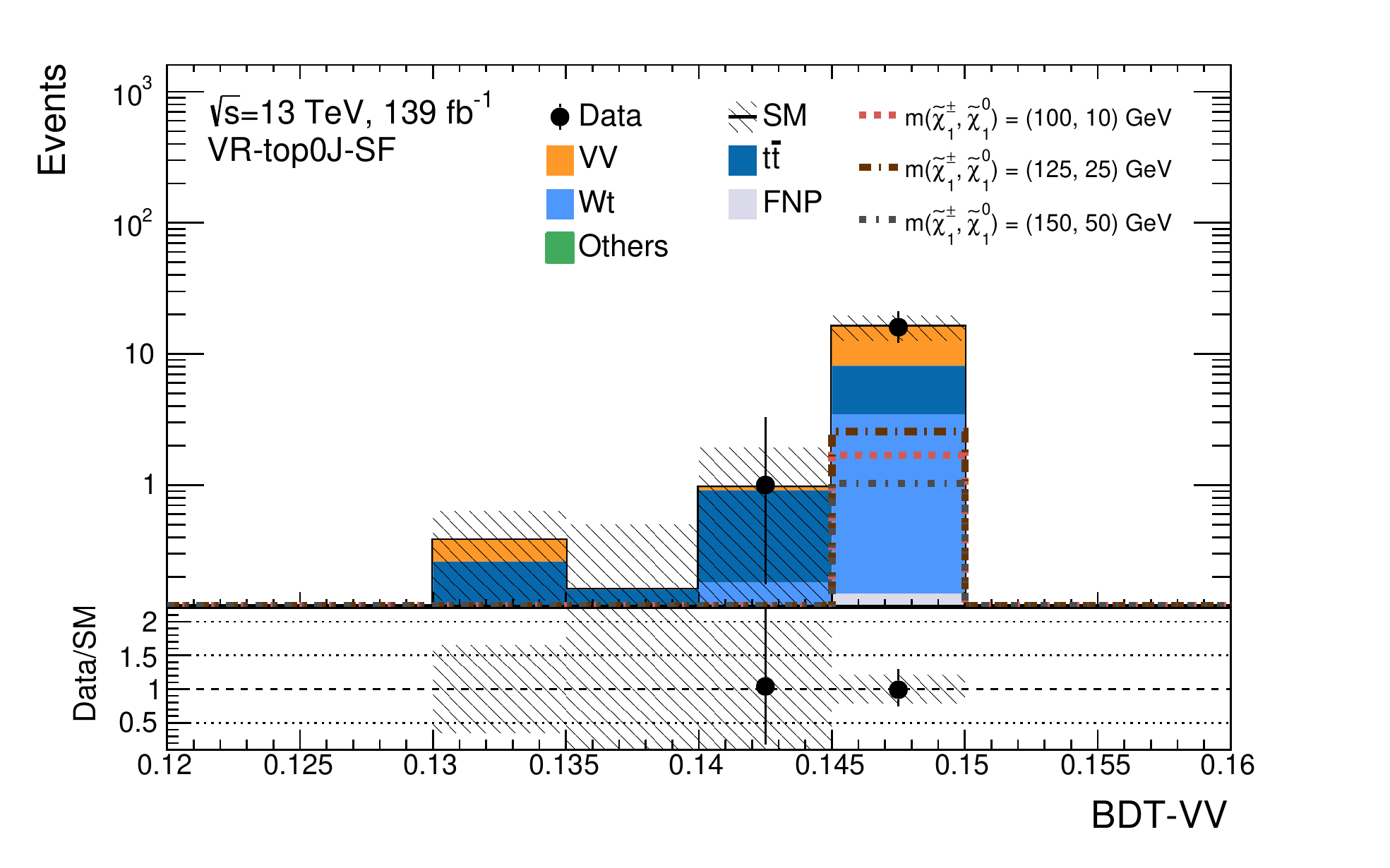}
\includegraphics[width=0.45\linewidth]{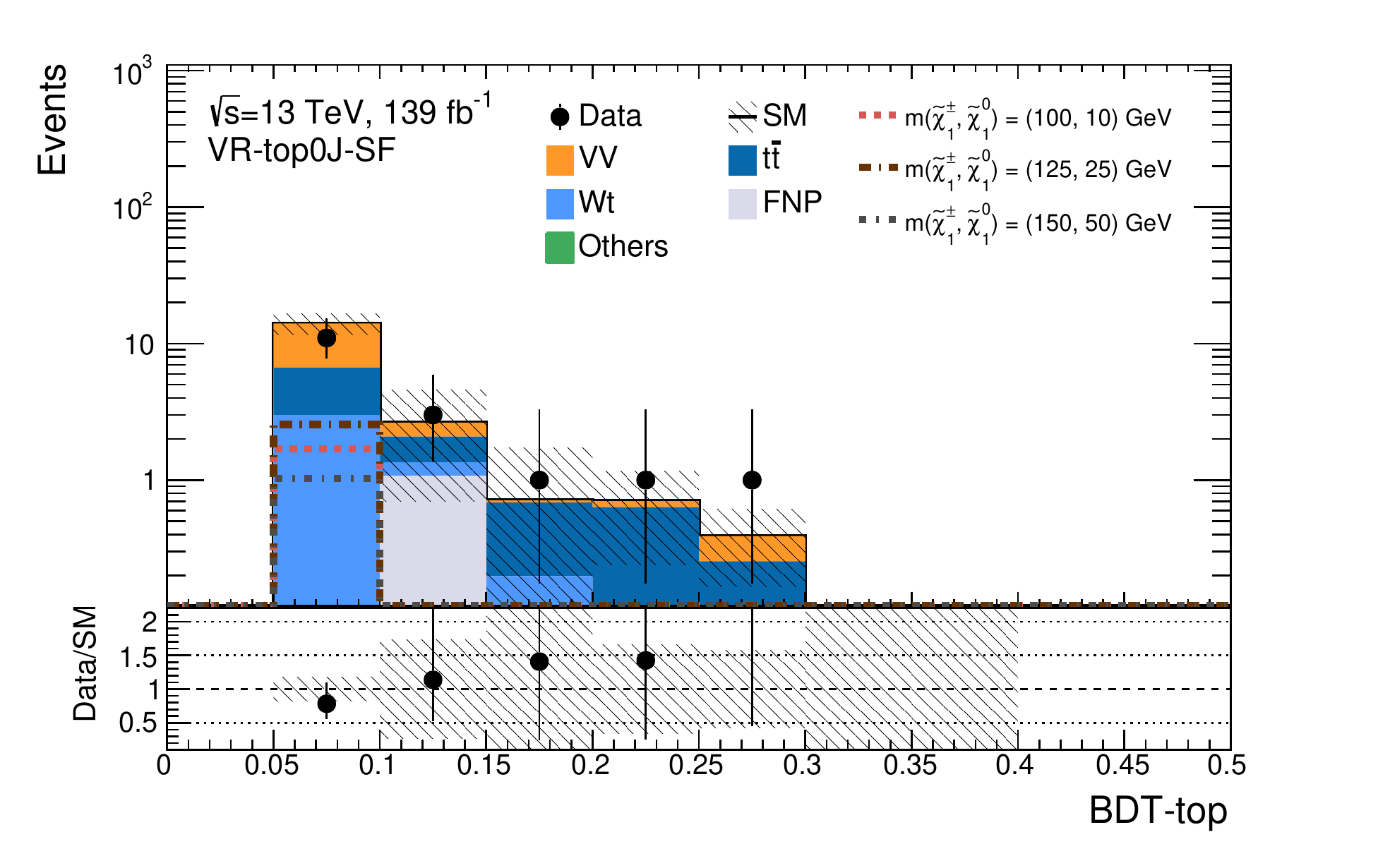}
\includegraphics[width=0.45\linewidth]{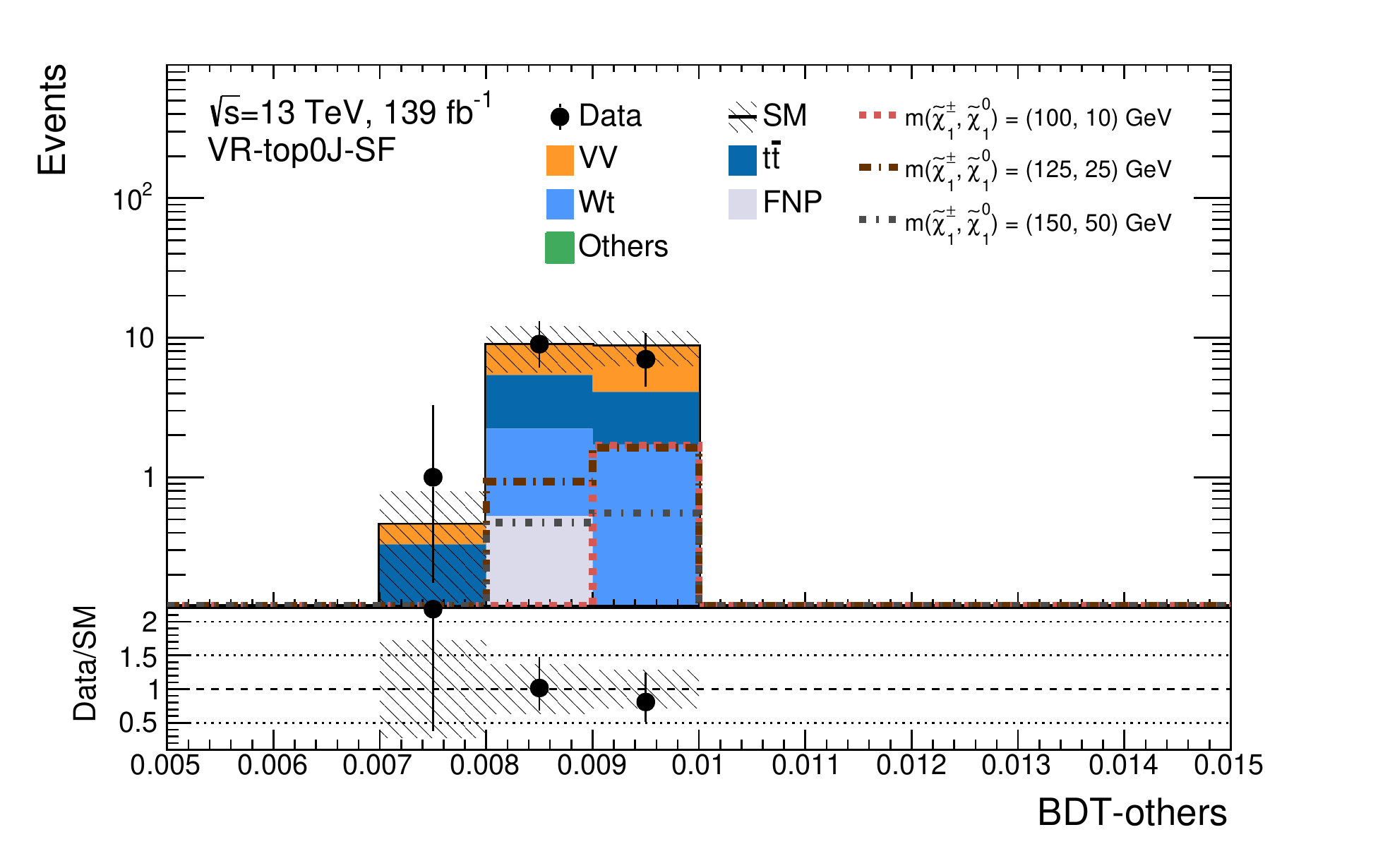}
\caption{The post-fit distributions in VR-top0J-SF. Both statistical and systematic uncertainties are shown.}
\label{fig:VR_top_SF0J_ML}
\end{figure}

Post-fit data and MC distributions of variables relevant for the analysis are shown in Fig.~\ref{fig:VR_VVDF0J_ML} for VR-VV-DF, in Fig. \ref{fig:VR_VVSF0J_ML} for VR-VV-SF, in Fig.~\ref{fig:VR_topDF1J_ML} for VR-VV-DF, in Fig. \ref{fig:VR_topSF1J_ML} for VR-VV-SF, in Fig.~\ref{fig:VR_top_DF0J_ML} for VR-top0J-DF, and in Fig.~\ref{fig:VR_top_SF0J_ML} for VR-top0J-SF.

\clearpage

\clearpage

\clearpage

\clearpage

\clearpage

\clearpage

\section{Systematic uncertainties}
\label{sec:compressedcharginos-systematicuncertainties}
The likelihood fits used for calculating the results of the analysis consider all relevant sources of experimental and theoretical systematic uncertainty affecting the SM background estimates and the signal predictions. 
The major sources of uncertainty come from the $VV$ theoretical uncertainty, normalisation of background processes and uncertainty associated to the jet energy scale and resolution and to the $p^\mathrm{miss}_\mathrm{T}$ soft-term scale and resolution.
Statistical uncertainties associated with the simulated MC samples are also accounted for. In the cases where the normalisation of background processes ($VV$ and top) are calculated using CRs, the systematic uncertainties only affect the extrapolation to the signal regions. 

The jet energy scale and resolution uncertainties are calculated as a function of the $p_{\mathrm{T}}$ and $\eta$ of the jet, and the pile-up conditions and flavour composition of the selected jet sample. They are derived using a combination of data and simulated samples, through studies including measurements of the transverse momentum balance between a jet and a reference object in dijet, $Z$+jets and $\gamma+$jets events~\cite{Aaboud:2017jcu}. 
An additional uncertainty in the modelling of $p^\mathrm{miss}_\mathrm{T}$ comes from the soft-term resolution and scale~\cite{Aaboud:2018tkc}. Experimental uncertainties on the scale factors to account for differences between the data and simulation in $b$-jet identification, lepton reconstruction efficiency and trigger efficiency are also included.
The remaining experimental uncertainties include lepton energy scale and resolution and are found to be negligible across all analysis regions.

Several sources of theoretical uncertainty in the modelling of the dominant backgrounds are considered. Modelling uncertainties on diboson, $t\bar{t}$, single-top and $Z$+jets backgrounds are considered.

The diboson modelling uncertainties are calculated by varying the PDF sets~\cite{Ball:2014uwa} as well as the QCD renormalisation and factorisation scales used to generate the samples. Uncertainties from missing higher orders are evaluated~\cite{Bothmann:2016nao} using six variations of the QCD factorisation and renormalisation scales in the matrix elements by factors of $0.5$ and $2$, avoiding variations in opposite directions.
Additional uncertainties on the resummation and matching scales between the matrix element generator and parton shower are considered.

The $t\bar{t}$ background is affected by modelling uncertainties associated with the parton shower modelling, the different approaches in the matching between the matrix element and the parton shower and the modelling of the initial and final-state radiation (ISR/FSR). Uncertainties in the parton shower simulation are estimated by comparing samples generated with \textsc{POWHEG-BOX} interfaced to either \textsc{PYTHIA 8.186} or \textsc{HERWIG 7.04}~\cite{Bahr:2008pv,Bellm:2015jjp}. The ISR/FSR uncertainties are calculated by comparing the predictions of the nominal sample with alternative scenarios with the relevant generator parameters varied~\cite{ATL-PHYS-PUB-2017-007}. The uncertainty associated with the choice of event generator is estimated by comparing the nominal samples with samples generated with {\textsc{MadGraph5}\_aMC@NLO} interfaced to \textsc{PYTHIA 8.186}~\cite{Sjostrand:2014zea}. Finally, an uncertainty is assigned for single-top-quark production to the treatment of the interference between the $Wt$ and $t\bar{t}$ samples. This is done by comparing the nominal sample generated using the diagram removal method with a sample generated using the diagram subtraction method~\cite{Frixione:2008yi,ATL-PHYS-PUB-2016-020}.

The $Z$+jets background is affected by QCD factorisation and renormalisation scales uncertainties.
Furthermore, uncertainties on the resummation and matching scales between the matrix element generator and parton shower are also considered.

There are several contributions to the uncertainty in the MM estimate of the FNP background. The real efficiencies and the electron light-flavoured fake rate (which are calculated using MC simulation) are affected by the experimental uncertainties on the scale factors applied to account for differences between data and simulation in the lepton trigger, identification, reconstruction and isolation efficiencies. For the heavy-flavour fake rate, an uncertainty is calculated to account for uncertainties in the subtraction of the prompt-lepton contamination in the CR, by varying this contamination and evaluating the effects on the resulting FNP background estimates. The uncertainties are evaluated by scaling up and down the real lepton contamination by 10\%. This means that one computes separate fake rates after subtraction of the nominal real-lepton contamination + 10\% and -10\%. These are then fed through the matrix inversion, and the resulting estimates are compared with the estimates using the nominal fake rate and differences are used to account for the systematic uncertainty on the subtraction of real lepton contamination in the CR. Finally, uncertainties in the expected composition of the FNP leptons in the signal regions are included, along with statistical uncertainties on all of the real efficiencies and fake rates used in the calculation.

Theoretical uncertainties related to the modelling of supersymmetric signals are considered. They include scale uncertainties computed as variations of the factorization, renormalization and merging scales, while the radiation uncertainty considers a combination of five different variations. Signal theoretical uncertainties are estimated from a subset of the signal mass hypotheses in samples with high statistics and are extended to the other signal samples. These signal theoretical uncertainties are then combined with the experimental ones. Their contribution is found to be negligible in the exclusion limits. The theoretical uncertainty related to the signal cross-section is not combined with the other uncertainties as it is used to build the additional contour around the observed exclusion limit, by scaling the nominal predicted value of the signal cross-section and recomputing the observed limits.

A summary of the impact of the systematic uncertainties on the background yields is shown in Table~\ref{tab:Signal_region_cuts_c1c1ww} after performing the likelihood fit. The systematic uncertainties are estimated in the inclusive region ${\mathrm{SR-DF}}_{\in(0.81,1]} {\mathrm{SF}}_{\in(0.77,1]}$, obtained as the integral of all the binned regions in Table~\ref{tab:Signal_region_cuts_c1c1ww} (thus requiring BDT-signal $\in (0.81,1]$ for DF events and $\in (0.77,1]$ for SF events).

\begin{table}[!htb]
    \begin{center}
    \setlength{\tabcolsep}{0.0pc}
        \begin{tabular}{p{0.52\linewidth}p{0.26\linewidth}}
            \noalign{\smallskip}\hline\noalign{\smallskip}
            {Region} & \hspace{-45pt} ${\mathrm{SR-DF}}_{\in(0.81,1]}{\mathrm{SF}}_{\in(0.77,1]}$   \\ \noalign{\smallskip}\hline\noalign{\smallskip}
        \end{tabular}
        \begin{tabular}{p{0.52\linewidth}p{0.26\linewidth}}
            Total background expectation      & \hspace{1.7pt}  630  \\
            \noalign{\smallskip}\hline\noalign{\smallskip}
            {$\!\begin{aligned}
            & E_{\mathrm{T}}^{\mathrm{miss}} \text{ modelling} \\[-2pt]
            & \text{$VV$ theoretical uncertainties}     \\[-2pt]
            & \text{Jet energy scale}                      \\[-2pt]
            & \text{$VV$ normalisation}                      \\[-2pt]
            & \text{Jet energy resolution}                 \\[-2pt]
            & \text{MC statistical uncertainties}          \\[-2pt]
            & \text{Lepton modelling}                      \\[-2pt]
            & \text{Top theoretical uncertainties}          \\[-2pt]
	    & t\bar{t} \text{ normalization}                \\[-2pt]
	    & \text{FNP leptons}                            \\[-2pt]
	    & \text{$b$-tagging}                              \\[-2pt]		  
	    & \text{$Z/\gamma^*(\rightarrow \ell\ell )$+jets theoretical uncertainties} \end{aligned} $}
            &   {$\!\begin{aligned}
                     6.6    &\%   \\[-2pt]
                     5.2    &\%   \\[-2pt]
                     5.1    &\%   \\[-2pt]
                     3.6    &\%   \\[-2pt]
                     1.8    &\%   \\[-2pt]
            	    1.7    &\%   \\[-2pt]
	             1.1    &\%   \\[-2pt]
 		        1    &\%  \\[-2pt]
		        1   &\%   \\[-2pt]
                     0.8    &\%   \\[-2pt]
                     0.7    &\%   \\[-2pt]
                     0.04  &\% 
                    \end{aligned}
            $} \\
            \noalign{\smallskip}\hline\noalign{\smallskip}
            {$\!\begin{aligned}
                & \text{Total systematic uncertainty}
                \end{aligned}
            $}
            &   {$\!\begin{aligned}
            \hspace{5pt} 8.7  &\%
                    \end{aligned}
            $} \\
            \noalign{\smallskip}\hline\noalign{\smallskip}
        \end{tabular}
    \end{center}
\caption{Breakdown of the dominant systematic uncertainties on background estimates in the inclusive region ${\mathrm{SR-DF}}_{\in(0.81,1]}{\mathrm{SF}}_{\in(0.77,1]}$ for the chargino search. The individual uncertainties can be correlated and do not necessarily add up quadratically to the total background uncertainty. The percentages show the size of the uncertainty relative to the total expected background.}
\label{tab:systematicSummary_charg}
\end{table}

\section{Results}
\label{sec:compressedcharginos-results}

The results are interpreted in the context of the chargino simplified model shown in Fig.~\ref{fig:c1c1viawwfeynman} and as general limits on new physics cross-sections.

The statistical interpretation of the results is performed using the {\textsc{HistFitter}} framework \cite{HistFitter}. The likelihood is a product of pdfs describing the observed number of events in each CR and SR as described in Section~\ref{sec:likelihoodfit}. Gaussian pdf distributions describe the nuisance parameters associated with each of the systematic uncertainties while Poisson distributions are used for MC statistical uncertainties. Systematic uncertainties that are correlated between different samples are accounted for in the fit configuration by using the same nuisance parameter. In particular, experimental systematic uncertainties are correlated between background and signal samples for all regions. The uncertainties are applied in each of the CRs and SRs and their effect is correlated for events across all regions in the fit.  

The search uses data in the CRs to constrain the nuisance parameters of the likelihood function, which include the background normalisation factors and parameters associated with the systematic uncertainties. The results of the background fit are used to test the compatibility of the observed data with the background estimates in the inclusive SRs.\\

\begin{figure}[!htb]
\centering
\includegraphics[width=0.85\linewidth]{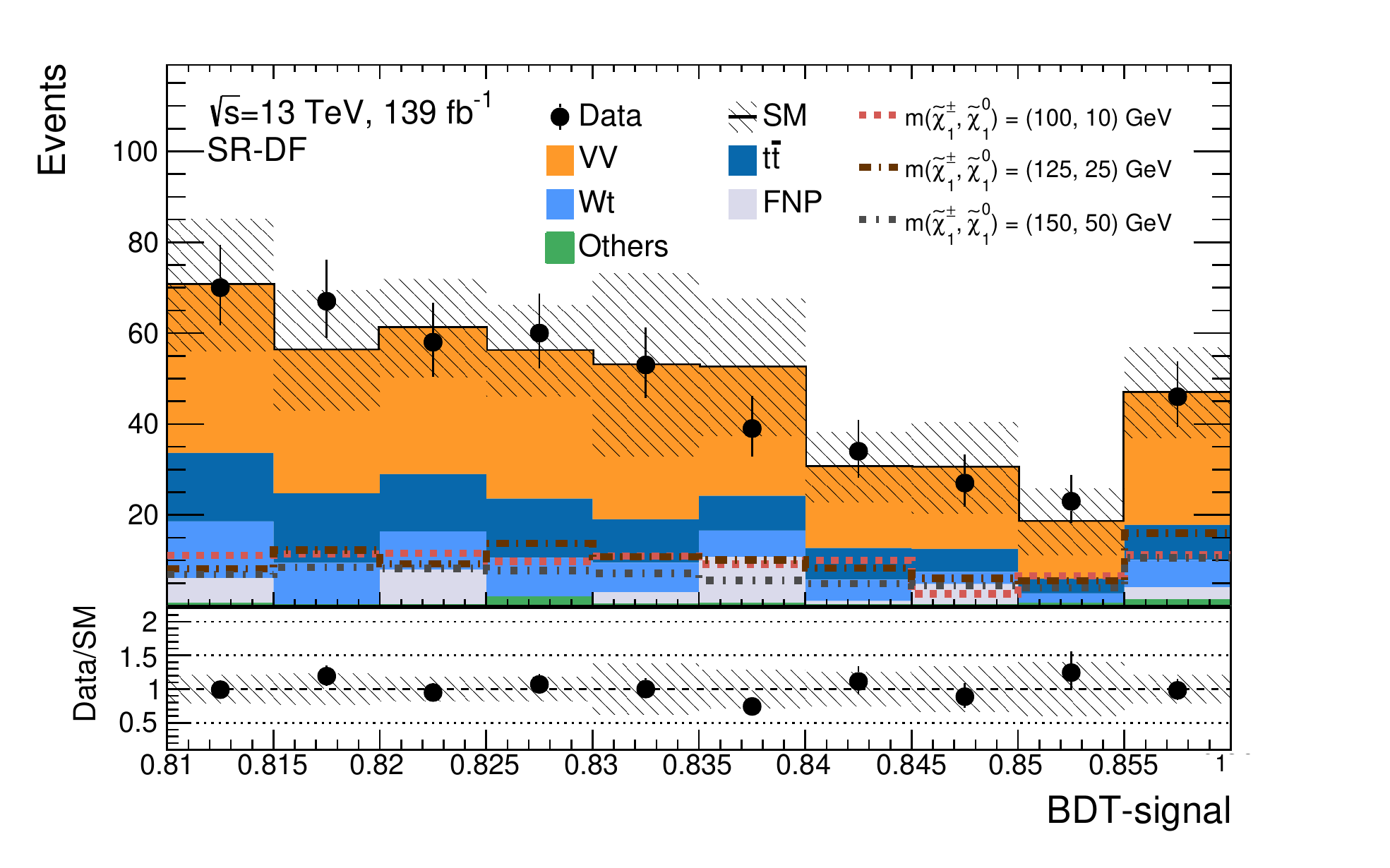}     
\includegraphics[width=0.85\linewidth]{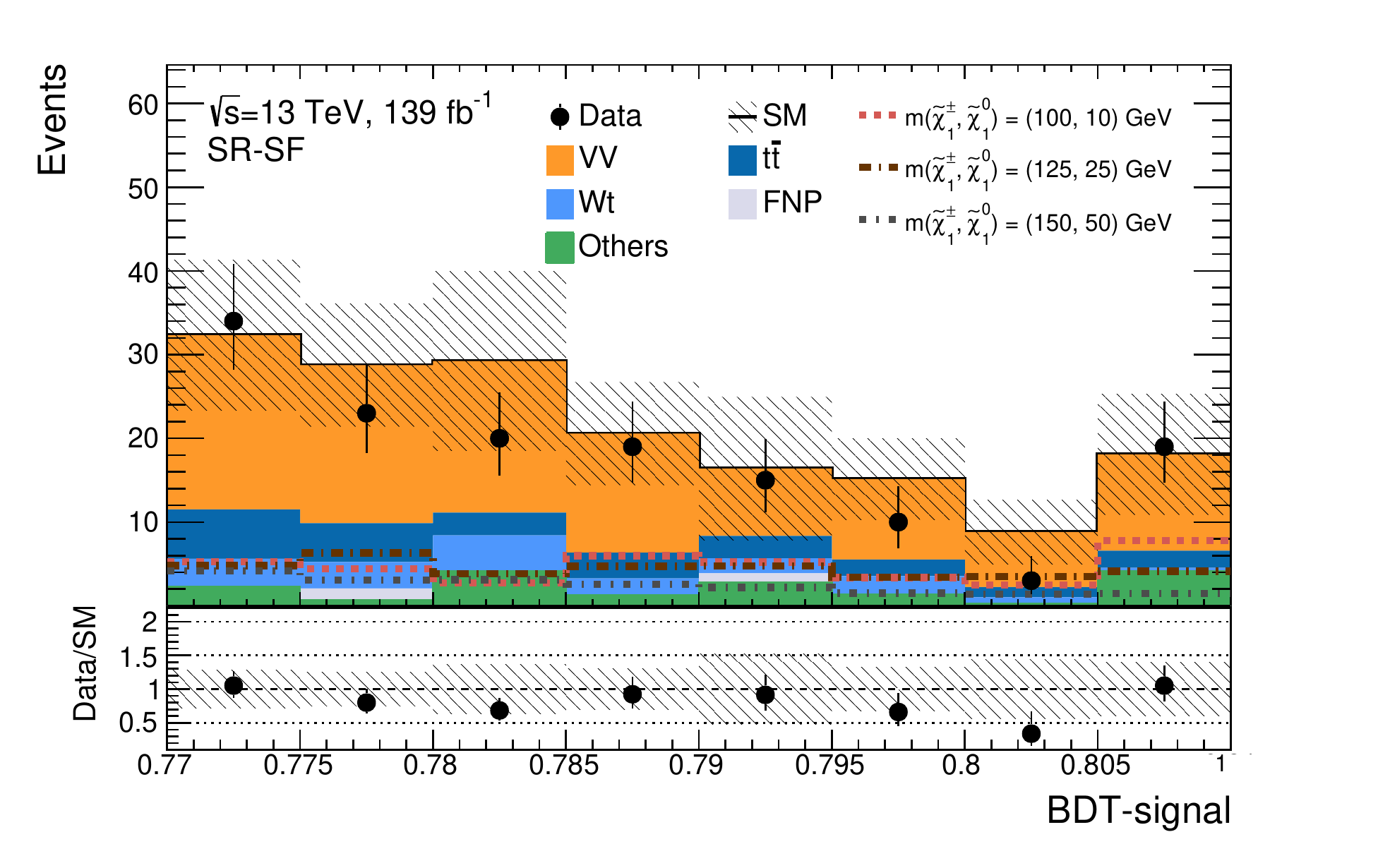}        
\caption{
The upper panel shows the observed number of events in SR-DF (top) and SR-SF (bottom) defined in Table~\ref{tab:Signal_region_cuts_c1c1ww}, together with the expected SM backgrounds obtained after the background fit in the CRs. `Others' include the non-dominant background sources, e.g.\ $t \bar{t}$+$V$, Higgs boson and Drell--Yan events. The uncertainty band includes systematic and statistical errors from all sources. Distributions for three benchmark signal samples are overlaid for comparison. The lower panel shows the data/SM ratio.}
\label{fig:SRBDT_charg}
\end{figure}

The post-fit background distributions of the BDT-signal scores, obtained applying the results of the background fit, are shown together with the observed data in Fig.~\ref{fig:SRBDT_charg}, for the SR-DF and for SR-SF defined in Table~\ref{tab:Signal_region_cuts_c1c1ww}. The shapes of the post-fit backgrounds and data agree well within uncertainties.

\begin{figure}[!htb]
\centering
\includegraphics[width=1.1\linewidth]{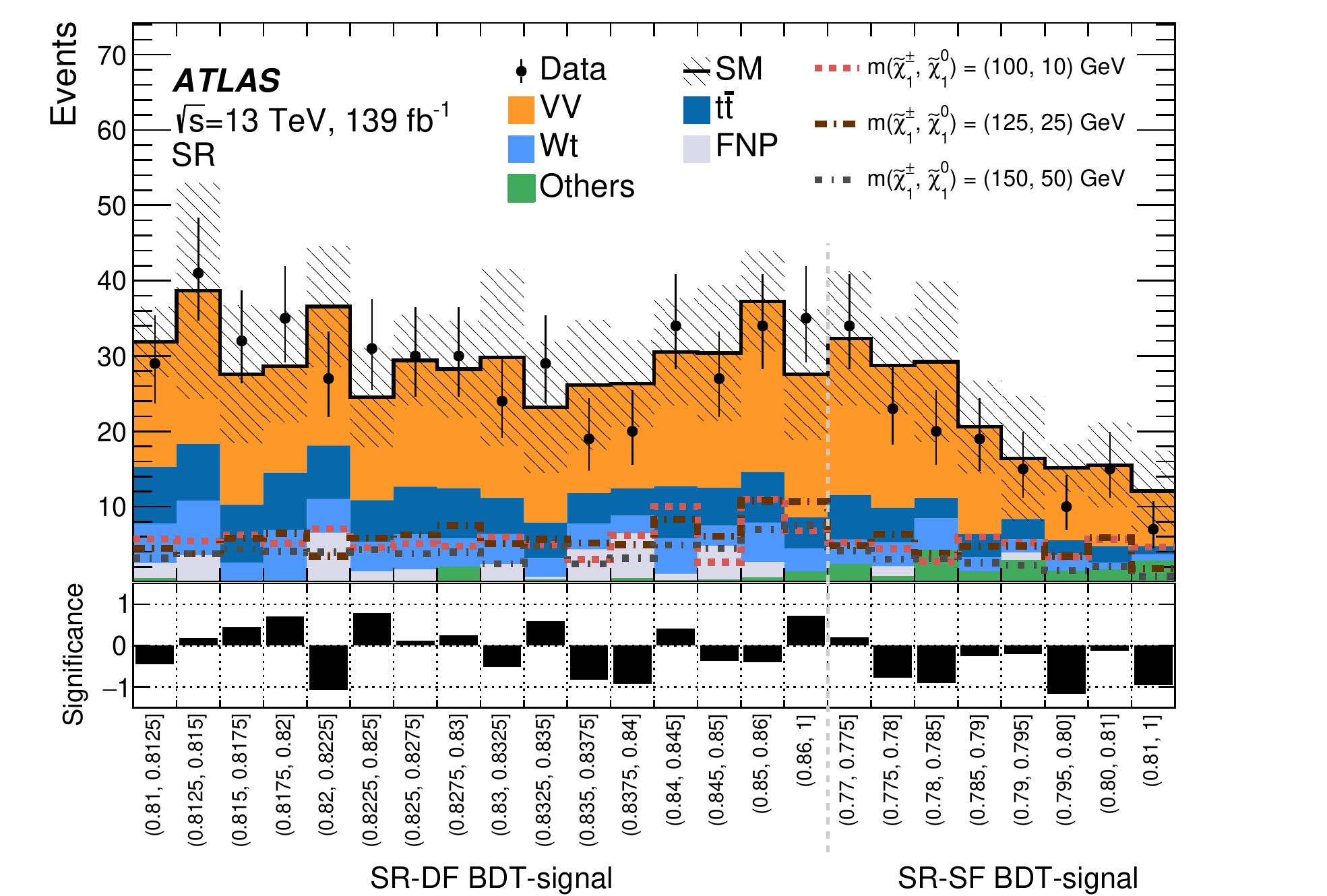}        
\caption{The upper panel shows the observed number of events in the SRs defined in Table~\ref{tab:Signal_region_cuts_c1c1ww}, together with the expected SM backgrounds obtained after the background fit in the CRs. `Others' include the non-dominant background sources, e.g.\ $t \bar{t}$+$V$, Higgs boson and Drell--Yan events. The uncertainty band includes systematic and statistical errors from all sources. Distributions for three benchmark signal samples are overlaid for comparison. The lower panel shows the significance as defined in Section~\ref{sec:StatisticalSignificance}. From Ref.~\cite{SUSY-2019-02}.}
\label{fig:SRpull_charg}
\end{figure}

The predicted numbers of background events obtained applying the results of the background fit in the binned SRs defined in Table~\ref{tab:Signal_region_cuts_c1c1ww} are shown together with the observed data in Fig.~\ref{fig:SRpull_charg}. No significant deviations from the SM expectations are observed in any of the SRs considered.\\

\begin{figure}[!htb]
\centering
\includegraphics[width=0.76\linewidth]{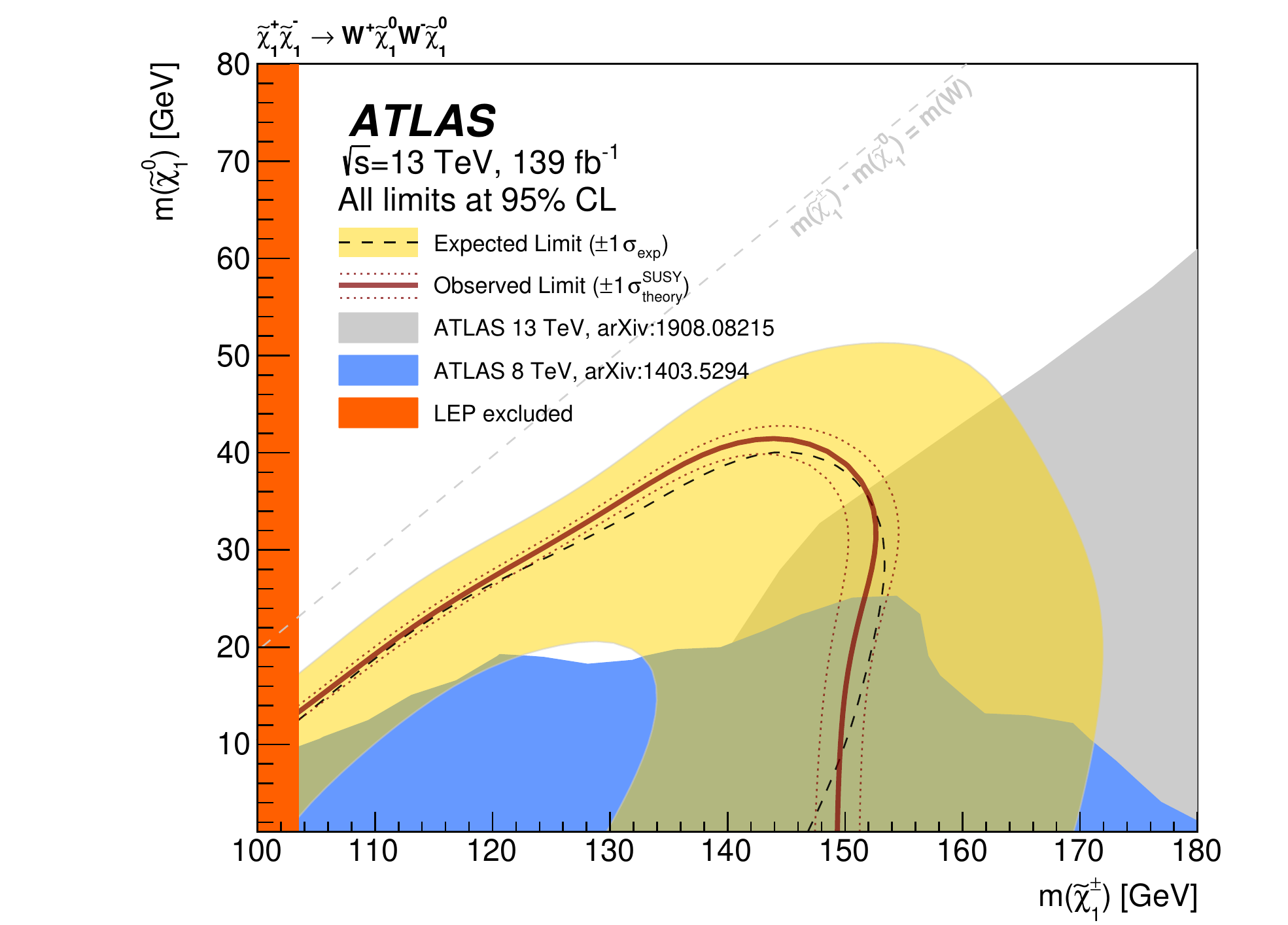}     
\includegraphics[width=0.76\linewidth]{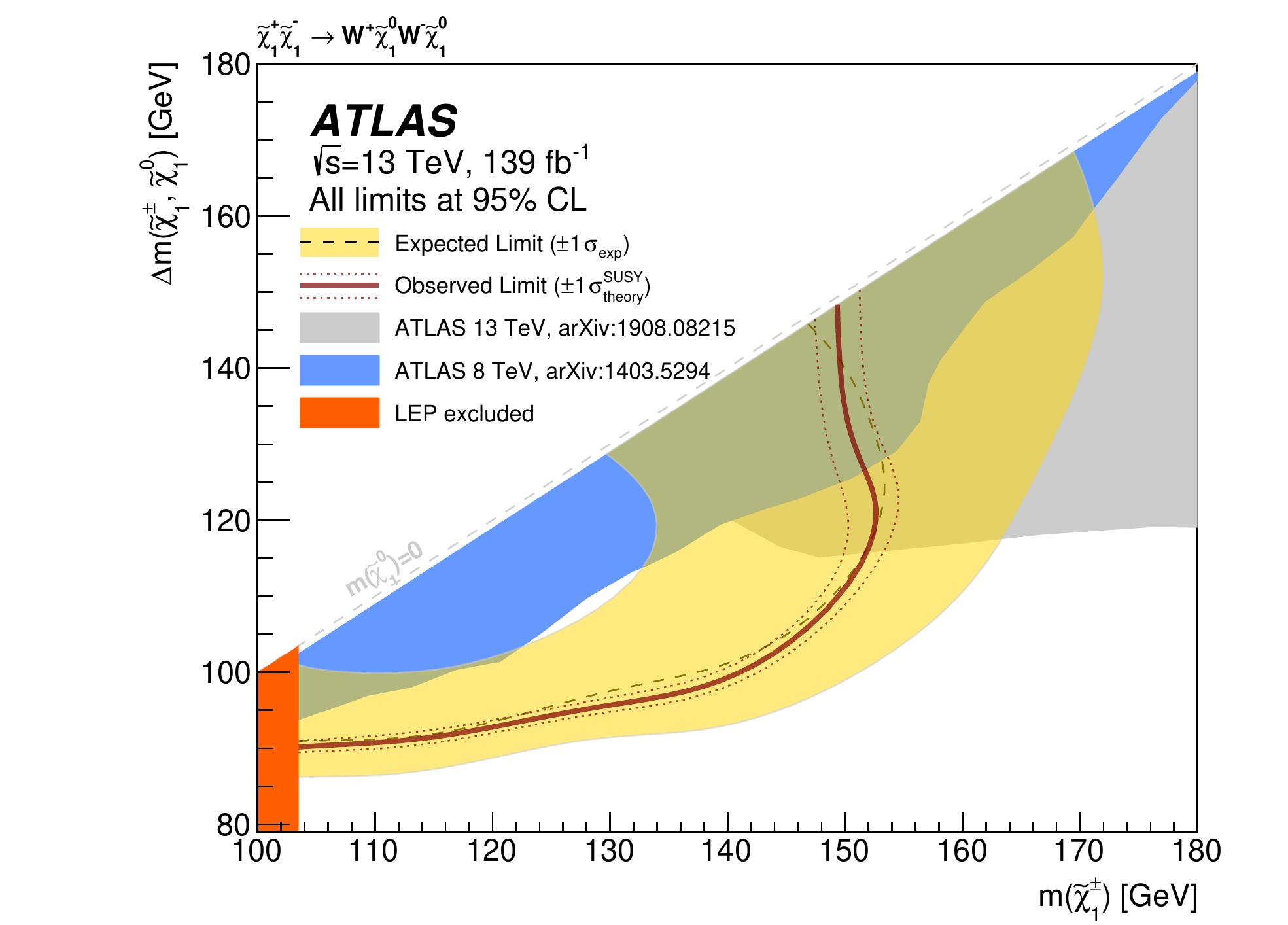}        
\caption{Observed and expected exclusion limits at 95\% CL on SUSY simplified models for chargino-pair production with $W$-boson-mediated decays in the $m(\tilde{\chi}_1^\pm)-m(\tilde{\chi}_1^0)$ (top) and $m(\tilde{\chi}_1^\pm)-\Delta m(\tilde{\chi}_1^\pm,\tilde{\chi}_1^0)$ (bottom) planes. The observed (solid thick line) and expected (thin dashed line) exclusion contours are indicated. The shaded band around the dashed line corresponds to the $\pm 1 \sigma $ variations in the expected limit, including all uncertainties except theoretical uncertainties in the signal cross-section. The dotted lines around the observed limit illustrate the change in the observed limit as the nominal signal cross-section is scaled up and down by the theoretical uncertainty. The observed limits obtained at LEP~\cite{LEP-chargino} and by the ATLAS experiment in previous searches are also shown~\cite{SUSY-2018-32,SUSY-2013-11}. In case of the search performed on ATLAS Run 1 data at $\sqrt{s} = 8$ TeV ~\cite{SUSY-2013-11} no sensitivity was expected for the exclusion in the mass plane. From Ref.~\cite{SUSY-2019-02}.}
\label{fig:exclusion_charg}
\end{figure}

Exclusion limits at 95\% CL are provided on the masses of the chargino and neutralino using the CL$_{\text{s}}$ prescription as described in Section~\ref{sec:CLtechnique}. The exclusion limits are shown in Fig.~\ref{fig:exclusion_charg}: chargino masses up to about 140 GeV are excluded at 95\% CL in the case of a mass splitting between chargino and neutralino down to about 100 GeV.

These results supersede the ATLAS 8 TeV results and extend the previous ATLAS 13 TeV results for small chargino masses in particularly interesting regions where the chargino pair production could have hidden behind the looking-alike $WW$ background.\\

\clearpage
The CL$_{\text{s}}$ method is also used to set model-independent upper limits at 95\% CL on the visible signal cross-section $\sigma^\mathrm{vis}$, defined as the cross-section times acceptance times efficiency, of processes beyond the SM. They are derived in each inclusive SR by performing a fit that includes the CRs, the observed yield in the SR as a constraint, and a signal yield in the SR as a free parameter of interest. The observed and predicted numbers of background events in the inclusive SRs are reported in Table~\ref{tab:SRinc_slep}, together with the model-independent upper limits on visible signal cross-section $\sigma^\mathrm{vis}$, the observed and expected limits at 95\% CL on the number of potential beyond the SM events $S^{0.95}_\mathrm{obs/exp}$, and the $p_0$-values. 
\begin{table}[!htb]
\centering
\setlength{\tabcolsep}{0.0pc}
\scalebox{0.95}{
\begin{tabular*}{\textwidth}{@{\extracolsep{\fill}}lcccccc}
\noalign{\smallskip}\hline\noalign{\smallskip}
{\textbf Signal region}        & Observed & Expected     & $\sigma^\mathrm{vis}$[fb]  &  $S_{\textrm{obs}}^{0.95}$  & $S_{\textrm{exp}}^{0.95}$  & $p_0$  \\ 
\noalign{\smallskip}\hline\noalign{\smallskip}
${\mathrm{SR-DF}}_{\in(0.81,1]}{\mathrm{SF}}_{\in(0.77,1]}$ & $620$ & $633 \pm 70$ & $1.20$ & $166.2$ & ${ 175.1 }^{ +44.9 }_{ -49.2 }$ & $ 0.50$ \\%
${\mathrm{SR-DF}}_{\in(0.81,1]}$        & $477$ & $470 \pm 50$ & $0.80$ & $111$ & ${ 108.9 }^{ +43.1 }_{ -31.1 }$ & $ 0.47$ \\%
${\mathrm{SR-DF}}_{\in(0.82,1]}$        & $340$ & $350 \pm 40$ & $0.55$ & $76.0$  & ${ 81.5 }^{ +32.7 }_{ -22.9 }$  & $ 0.50$ \\%
${\mathrm{SR-DF}}_{\in(0.83,1]}$        & $222$ & $231 \pm 26$ & $0.38$ & $52.3$  & ${ 57.8 }^{ +22.8 }_{ -16.1 }$  & $ 0.50$ \\%
${\mathrm{SR-DF}}_{\in(0.84,1]}$        & $130$ & $126 \pm 15$  & $0.29$ & $40.0$  & ${ 37.5 }^{ +15.0 }_{ -10.5 }$  & $ 0.41$ \\%
${\mathrm{SR-DF}}_{\in(0.85,1]}$        & $69$   & $65  \pm 10$  & $0.22$ & $30.9$  & ${ 28.0 }^{ +12.0 }_{ -8.3 }$   & $ 0.38$ \\%
${\mathrm{SR-SF}}_{\in(0.77,1]}$        & $143$ & $167 \pm 32$ & $0.47$ & $65.5$  & ${ 80.6 }^{ +19.4 }_{ -23.0 }$  & $ 0.50$ \\%
${\mathrm{SR-SF}}_{\in(0.78,1]}$        & $86$   & $108 \pm 23$   & $0.31$ & $42.8$  & ${ 53.9 }^{ +18.9 }_{ -13.6 }$  & $ 0.50$ \\%
${\mathrm{SR-SF}}_{\in(0.79,1]}$        & $47$   & $58 \pm 15$   & $0.21$ & $28.9$  & $ { 34.1 }^{ +10.8 }_{ -7.8 }$  & $ 0.50$ \\%
${\mathrm{SR-SF}}_{\in(0.80,1]}$        & $22$   & $28 \pm 8$   & $0.10$ & $14.3$  & $ { 16.8 }^{ +5.9 }_{ -4.5 }$   & $ 0.50$ \\%
\noalign{\smallskip}\hline\noalign{\smallskip}
\end{tabular*}}
\caption{Observed event yields and predicted background yields for the inclusive SRs defined in Table~\ref{tab:Signal_region_cuts_c1c1ww}. The model-independent upper limits at 95\% CL on the observed and expected numbers of beyond-the-SM events $S^{0.95}_\mathrm{obs/exp}$ and on the effective beyond-the-SM cross-section $\sigma^\mathrm{vis}$ ($\langle\mathrm{A}\epsilon\sigma\rangle^{0.95}_{\mathrm{obs}}$) are also shown. The $\pm 1 \sigma$ variations on $S^{0.95}_\mathrm{exp}$ are provided. The last column shows the $p_0$-value of the SM-only hypothesis. For SRs where the data yield is smaller than expected, the $p_0$-value is capped at 0.50.}
\label{tab:SRinc_slep}
\end{table}
The $p_0$-values, which represent the probability of the SM background alone to fluctuate to the observed number of events or higher, are capped at $p_0 = 0.50$. No significant deviations from the SM expectations are observed in any of the SRs considered.

\chapter{Compressed higgsino search}
\label{sec:compressedshiggsinos}

In this Chapter, the compressed higgsino search is presented. Supersymmetric higgsinos are triplets of states with $\tilde{\chi}_{2}^{0}$, $\tilde{\chi}_{1}^{\pm}$ and $\tilde{\chi}_{1}^{0}$ almost degenerate in mass. This search targets the direct production of higgsinos decaying into the LSP neutralinos $\tilde{\chi}_{1}^{0}$ with a mass splitting from 1 GeV down to 0.3 GeV. Charginos  $\tilde{\chi}_{1}^{\pm}$ and NLSP neutralinos $\tilde{\chi}_{2}^{0}$ produce soft pions whose tracks can be reconstructed a few millimetres away from the primary vertex. A signature with no leptons, $E_\mathrm{T}^{\mathrm{miss}}$ and no hadronic activity in the initial state is considered and a compressed mass spectrum is targeted.

\minitoc
\medskip

\section{Analysis overview}
\label{sec:compressedshiggsinos-analysisoverview}

\begin{figure}[!htb]
\begin{center}
\includegraphics[width=0.4\textwidth]{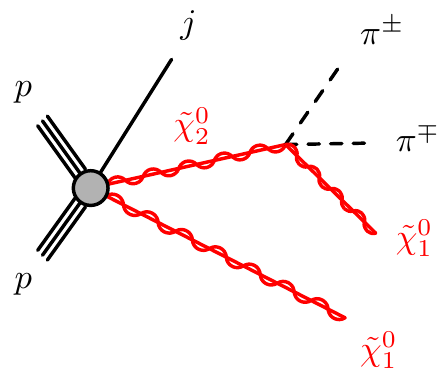}
\includegraphics[width=0.4\textwidth]{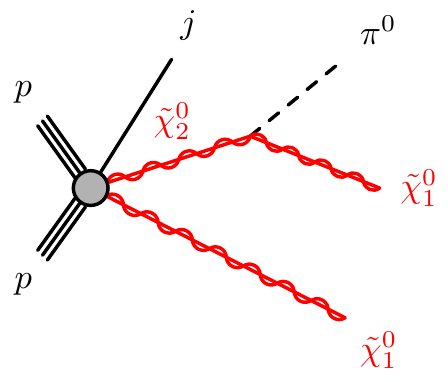}
\includegraphics[width=0.38\textwidth]{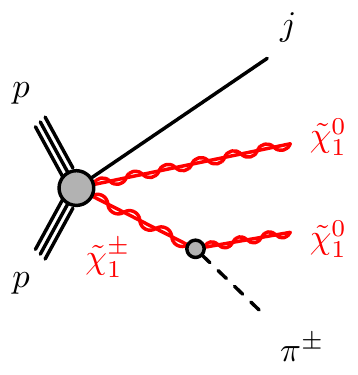}
\includegraphics[width=0.38\textwidth]{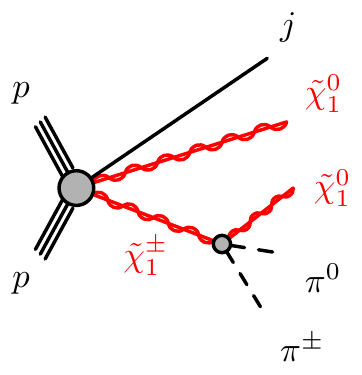}
\end{center}
\caption{Diagram of the supersymmetric production of charginos and neutralinos decaying into pions and with hadronic activity in the initial state.}
\label{fig:FeynmanHiggsinos}
\end{figure}

The analysis targets the direct production of higgsino pairs, $\tilde{\chi}_{2}^{0}\tilde{\chi}_{1}^{\pm}$, $\tilde{\chi}_{1}^{\pm}\tilde{\chi}_{1}^{0}$, $\tilde{\chi}_{1}^{\pm}\tilde{\chi}_{1}^{\mp}$, and $\tilde{\chi}_{2}^{0}\tilde{\chi}_{1}^{0}$, where each chargino $\tilde{\chi}_{1}^{\pm}$ or NLSP neutralino $\tilde{\chi}_{2}^{0}$ decays into the LSP neutrliano $\tilde{\chi}_{1}^{0}$ via the emission of one or more soft pions, leaving a discernible \textit{displaced track} in the detector a few millimetres away from the primary vertex. In order to boost the system, we also require a jet in the initial state. Fig.~\ref{fig:FeynmanHiggsinos} shows the production $\tilde{\chi}_{2}^{0}\tilde{\chi}_{1}^{\pm}$ and $\tilde{\chi}_{1}^{\pm}\tilde{\chi}_{1}^{0}$ decaying into pions, while the other production modes can be deducted from the ones shown. Only charged pions leave a track in the detector that can be reconstructed.\\

\begin{figure}[!htb]
\begin{center}
\includegraphics[width=0.9\textwidth]{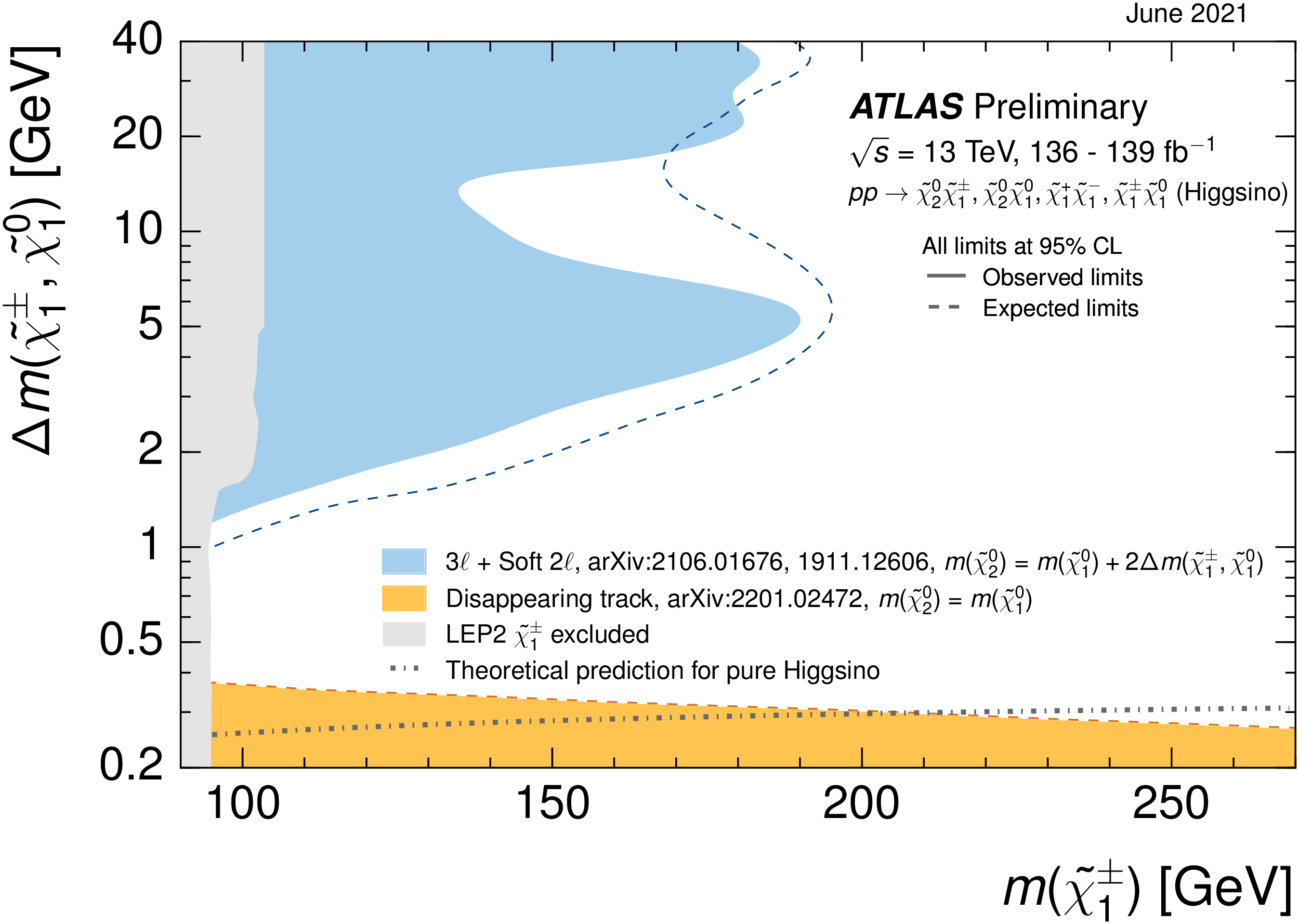}
\end{center}
\caption{Exclusion limits at 95\% CL for higgsino pair production $\tilde{\chi}_{2}^{0}\tilde{\chi}_{1}^{\pm}$, $\tilde{\chi}_{1}^{\pm}\tilde{\chi}_{1}^{0}$, $\tilde{\chi}_{1}^{\pm}\tilde{\chi}_{1}^{\mp}$, and $\tilde{\chi}_{2}^{0}\tilde{\chi}_{1}^{0}$ with off-shell SM-boson-mediated decays to the lightest neutralino, $\tilde{\chi}_{1}^{0}$, as a function of the $\Delta m(\tilde{\chi}_{1}^{\pm},\,\tilde{\chi}_{1}^{0})$. The production cross-section is for pure higgsinos.}
\label{fig:SusySummaryPlot}
\end{figure}

The search for higgsinos relies on different analysis strategies at the LHC according to the mass splitting between the produced higgsino and the particle into which it decays. If the mass splitting is greater than $\mathcal{O}$(1) GeV, the products of higgsino decays can be reconstructed as a multi-lepton signal, as shown in Fig.~\ref{fig:SusySummaryPlot}. 

On the other hand, if the mass splitting is $<300$ MeV, the charged higgsino has a lifetime long enough that it can be detected as a \textit{disappearing track} by the innermost tracking layers: this characteristic signature comes from the fact that, before it decays, the charged higgsino deposits energy in the innermost tracking layers and can be reconstructed as a short track if at least four pixel layers have been hit but its decay products, the LSP neutralino and the pion, are not reconstructed at all, the former because it escapes detection and leads to missing transverse momentum, the latter because it has too low momentum to be reconstructed as a track.\\
Both mass splittings have been ruled out by the ATLAS experiment up to chargino masses well beyond the LEP limits, as can be seen in Fig.~\ref{fig:SusySummaryPlot}. Meanwhile, the intermediate region, i.e. mass splitting of around $0.3 - 1$ GeV, has never been probed at the LHC. This analysis is inspired by a theoretical paper~\cite{CorneringHiggsino} which proposes a new strategy to explore this phase space region by using a soft pion from the decay of a charged higgsino.\\

In order to explore the compressed phase space, ML techniques are exploited. The analysis strategy relies on a Deep Neural Network (DNN) developed in the Keras framework~\cite{Keras} and uses track-level variables to discriminate the supersymmetric signal from the background. \\

The main SM backgrounds are the irreducible $Z(\nu\nu)$+jets and $W(e\nu)$+jets, $W(\mu\nu)$+jets, $W(\tau\nu)$+jets where the lepton is not reconstructed. The ABCD method is used to estimate the background composition in the SR. $E_{\mathrm{T}}^{\text{miss}}$ and the output score of the DNN training are used as the two uncorrelated variables to define the four regions in the ABCD plane.\\

The analysis is still in a preliminary phase, where the sensitivity to the targeted model is tried to be improved as much as possible and data are blinded in the SR. Sensitivity estimates and expected exclusion limits on the direct production of higgsinos are reported.

\section{Event samples}
\label{sec:compressedshiggsinos-eventsamples}

\subsection{Signal samples}
\label{sec:compressedshiggsinos-signalsamples}
The MC signal samples of 6 different production modes, $\tilde{\chi}_{2}^{0}\tilde{\chi}_{1}^{\pm}$, $\tilde{\chi}_{1}^{\pm}\tilde{\chi}_{1}^{0}$, $\tilde{\chi}_{1}^{\pm}\tilde{\chi}_{1}^{\mp}$, and $\tilde{\chi}_{2}^{0}\tilde{\chi}_{1}^{0}$ are generated from LO matrix elements using {\textsc{MadGraph5}\_aMC@NLO} v2.8.1 \cite{Alwall:2014hca} interfaced with \textsc{PYTHIA} v8.244 \cite{Sjostrand:2014zea} with the A14 tune \cite{ATL-PHYS-PUB-2014-021} and the NNPDF23LO set \cite{Ball:2012cx} of PDFs. EvtGen v1.7.0 \cite{Lange:2001uf} is also used. Higher order corrections up to NLO-NLL for $\sqrt{s}=13$ TeV to the cross-section are taken into account using \textsc{RESUMMINO} \cite{Fuks13040790}.
The \textsc{RESUMMINO} cross-sections are provided for events with $\geq 0$ jets in the final state, but we only generate events with $\geq 1$ jet with {\textsc{MadGraph5}\_aMC@NLO} because events with zero jets would not pass the analysis selections on jets and missing transverse momentum that are necessary to trigger the event. Thus, the \textsc{RESUMMINO} cross-section is multiplied by the \textit{MG ratio}, defined as the cross-section for events with $\geq 1$ jet divided for the inclusive cross-section for events with $\geq 0$ jets.\\
Four signal samples with different hypotheses of the masses of SUSY particles are produced:
\begin{itemize}
    \item $m(\tilde{\chi}_{2}^{0}, \, \tilde{\chi}_{1}^{\pm}, \,  \tilde{\chi}_{1}^{0}) = (151, 150.5, 150)$ GeV;
    \item $m(\tilde{\chi}_{2}^{0}, \, \tilde{\chi}_{1}^{\pm}, \,  \tilde{\chi}_{1}^{0}) = (150.5, 150.5, 150)$ GeV;
    \item $m(\tilde{\chi}_{2}^{0}, \, \tilde{\chi}_{1}^{\pm}, \,  \tilde{\chi}_{1}^{0}) = (150.7, 150.35, 150)$ GeV;
    \item $m(\tilde{\chi}_{2}^{0}, \, \tilde{\chi}_{1}^{\pm}, \,  \tilde{\chi}_{1}^{0}) = (151, 151, 150)$ GeV.
\end{itemize}
All four samples have higgsino masses around 150 GeV, a mass value that is expected to be excluded and that can be used as a benchmark of study to improve the sensitivity of the search. The mass splitting between NLSP neutralino or chargino and LSP neutralino is $0.3<\Delta m(\tilde{\chi}_{2}^{0}/\tilde{\chi}_{1}^{\pm},\,\tilde{\chi}_{1}^{0} ) < 1$ GeV, a range of supersymmetric higgsino masses which has not been excluded so far and which can be explored into by looking for displaced tracks in the ATLAS inner detector. The mass difference between $m(\tilde{\chi}_{2}^{0})$ and $m(\tilde{\chi}_{1}^{\pm})$ depends on the electroweak parameters $\tan\beta$, $M_1$ and $M_2$ and two theoretically well-motivated scenarios are considered:
\begin{itemize}
    \item $m(\tilde{\chi}_{2}^{0})= m(\tilde{\chi}_{1}^{\pm})$ for (150.5, 150.5, 150) GeV and (151, 151, 150) GeV signal samples;
    \item $m(\tilde{\chi}_{1}^{\pm}) = 1/2\, m(\tilde{\chi}_{2}^{0})$ for (151, 150.5, 150) GeV and (150.7, 150.35, 150) GeV signal samples.
\end{itemize}

\subsection{Background samples}
\label{sec:compressedshiggsinos-backgroundsamples}
The SM background processes considered in the search are $Z(\nu\nu)$+jets, $W(\ell\nu)$+jets (including $W(e\nu)$+jets, $W(\mu\nu)$+jets, $W(\tau\nu)$+jets), $t\bar{t}$ and $VV$.\\
\begin{itemize}
    \item $Z(\nu\nu)$+jets, $W(\ell\nu)$+jets and $Z(\tau\tau)$+jets processes. Samples of these processes are simulated with the \textsc{SHERPA} v2.2.11 \cite{Bothmann:2019yzt} at NLO accuracy using the NNPDF3.0NLO set of PDFs. \textsc{SHERPA} v2.2.11 also includes NLO virtual electroweak corrections. The matrix element calculations are matched and merged with the \textsc{SHERPA} parton shower based on Catani-Seymour dipole \cite{Gleisberg:2008fv,Schumann:2007mg} using the MEPS@NLO prescription \cite{Hoeche:2012yf,Catani:2001cc,Hoeche:2009rj}.
    \item $t\bar{t}$ processes. The production of $t\bar{t}$ is modelled using the \textsc{POWHEG-BOX-v2} \cite{Nason:2004rx,Frixione:2007vw,Alioli:2010xd} generator at NLO with the NNPDF3.0NLO \cite{Ball:2014uwa} set of PDFs. Events are interfaced to \textsc{PYTHIA8} ~\cite{Sjostrand:2014zea} to model the parton shower, hadronisation, and underlying event, with parameters set according to the A14 tune ~\cite{ATL-PHYS-PUB-2014-021} and using the NNPDF2.3LO \cite{Ball:2012cx} set of PDFs. The decays of bottom and charm hadrons are performed by \textsc{EvtGen~v1.6.0} \cite{Lange:2001uf}. 
    \item $VV$ processes. Samples of $VV$ processes are simulated with the \textsc{SHERPA} v2.2.1 or v2.2.2 \cite{Bothmann:2019yzt} at NLO accuracy in QCD for up to one additional parton and at LO accuracy for up to three additional parton emissions. The matrix element calculations are matched and merged with the \textsc{SHERPA} parton shower based on Catani-Seymour dipole \cite{Gleisberg:2008fv,Schumann:2007mg} using the MEPS@NLO prescription \cite{Hoeche:2012yf,Catani:2001cc,Hoeche:2009rj}. The virtual QCD correction are provided by the {\textsc{openloops}} library~\cite{Cascioli:2011va}. The set of PDFs used is NNPDF3.0NLO, along with the dedicated set of tuned parton shower parameters developed by the \textsc{SHERPA} authors.
\end{itemize}

\subsection{Data samples}
The full Run~2 ATLAS data set is exploited, corresponding to 3.2 fb$^{-1}$ of data collected in 2015, 33.0 fb$^{-1}$ of data collected in 2016, 44.3 fb$^{-1}$ of data collected in 2017 and 58.45 fb$^{-1}$ of data collected in 2018.
The data used in the analysis must satisfy the ATLAS GRL requirement. The dataset considered corresponds then to a combined total integrated luminosity of 138.95~fb$^{-1}$.

\section{Object definition}
\label{sec:compressedshiggsinos-objectdefinition}
This Section is dedicated to specifying the objects used for the analysis: tracks, electrons, muons, jets and $E_{\mathrm{T}}^{\text{miss}}$. Electrons and muons are not selected in our analysis but used in the veto. The set of quality cuts and trigger selection required for the events is first discussed. The object definition criteria for electrons, muons and jets can be found in Tables~\ref{tab:compressedshiggsinos-eledef},~\ref{tab:compressedshiggsinos-muondef} and~\ref{tab:compressedshiggsinos-jetsdef}, respectively.

\subsection{Event Quality Cuts}
\label{sec:compressedshiggsinos-techintro}
A set of requirements has been applied at the event level to reject non-collision background or to veto inactive regions of the detector:

\begin{itemize}
\item {GRL} (data only): events must satisfy the GRL;
\item {LAr/Tile error} (data only): events with noise bursts and data integrity errors in the LAr calorimeter/Tile corrupted are removed;
\item {Tile Trip} (data only): events with Tile trips are removed;
\item {SCT error} (data only): events affected by the recovery procedure for single event upsets in the SCT are removed;
\item {Bad jets} (data and MC): non-collision background is reduced by applying a \textit{\bf tight cleaning} on the jets. Tight cleaning is defined as a high-purity working point for analyses which aim to reduce as much as possible the non-collision backgrounds, it corresponds to the loose criteria plus an additional requirement on the ratio between the jet charged particle fraction and the jet energy fraction in the layer with maximum energy deposit \cite{ATLAS-tightcleaning}. In this analysis, it is found that it is sufficient to only apply this cleaning to the leading jet in the event within $|\eta|<2.4$;
\item {Primary Vertex} (data and MC): events must have a primary vertex (PV), selected as the one with the highest $\sum p_\mathrm{T}^{2}$ of associated tracks, with at least two tracks.
\end{itemize}

\subsubsection{Tracks}
Tracks are reconstructed by clustering the raw measurements from the pixel and SCT detectors, producing first track seeds and then track candidates, evaluating a quality score for each track and extending the reconstruction including TRT hits, as described in Section~\ref{Section:TrackReconstruction}. Two different sets of quality criteria are available in ATLAS for particle tracks, \texttt{Loose} and \texttt{Tight} \cite{ATLAS-trackreconstruction}. The selection of these quality criteria is reported in Table~\ref{tab:compressedshiggsinos-trackdef}. \texttt{Loose} selection is applied to all reconstructed tracks while \texttt{Tight} selection is applied at the preselection level in the analysis.

\begin{table}[!htb]
\begin{center}
\begin{tabular}{l|c}
\noalign{\smallskip}\hline\noalign{\smallskip}
\multicolumn{2}{c}{\texttt{Loose} track selection}\\
\noalign{\smallskip}\hline\noalign{\smallskip}
Acceptance       & Track $p_\mathrm{T}$ > 400 MeV, track $|\eta|$ < 2.5 \\
Number of silicon hits & $N_{\mathrm{Si}} \ge 7$ \\
Number of shared modules & $N^{\mathrm{sh}}_{\mathrm{Pix}} \le 1$ \\
Number of silicon holes & $N^{\mathrm{hole}}_{\mathrm{Si}} \le 2$ \\
Number of pixel holes & $N^{\mathrm{hole}}_{\mathrm{Pix}} \le 1$ \\
\noalign{\smallskip}\hline\noalign{\smallskip}
\noalign{\smallskip}\hline\noalign{\smallskip}
\multicolumn{2}{c}{\texttt{Tight} track selection}\\
\noalign{\smallskip}\hline\noalign{\smallskip}
Baseline       & \texttt{Loose} track selection \\
Number of silicon hits & $N_{\mathrm{Si}} \ge 9$ if $|\eta| \le 1.65$ \\
Number of silicon hits & $N_{\mathrm{Si}} \ge 11$ if $|\eta| > 1.65$ \\
IBL or B-Layer hit & $N_{\mathrm{Si}}^{\mathrm{IBL}} + N_{\mathrm{Si}}^{\mathrm{B-Layer}} > 0$ \\
Number of pixel holes & $N^{\mathrm{hole}}_{\mathrm{Pix}} = 0$ \\
\noalign{\smallskip}\hline\noalign{\smallskip}
\end{tabular}
\end{center}
\caption{Summary of the track selection criteria available for track reconstruction.}   
\label{tab:compressedshiggsinos-trackdef}
\end{table}

\subsubsection{Electrons}
Electrons are reconstructed as described in Section~\ref{ssec:reco-ele} and are required to reside within $|\eta|$ <2.47. At baseline level, electrons must have $p_\mathrm{T} > 4.5$ GeV, satisfy the \texttt{VeryLooseLLH} PID quality criteria and also satisfy the IP condition $|z_0 \sin \theta| < 0.5$~mm.
Signal electrons must have $p_\mathrm{T} > 4.5$ GeV and be isolated with respect to other high-$p_\mathrm{T}$ charged particles satisfying the \texttt{FCLoose} isolation criteria. Moreover signal electrons must pass \texttt{MediumLLH} quality criteria and also satisfy the IP condition $S(d_0)<5$. The electron selection is summarised in Table~\ref{tab:compressedshiggsinos-eledef}.

\begin{table}[!htb]
\begin{center}
\begin{tabular}{l|c}
\noalign{\smallskip}\hline\noalign{\smallskip}
\multicolumn{2}{c}{Baseline electron}\\
\noalign{\smallskip}\hline\noalign{\smallskip}
Acceptance       & $p_\mathrm{T}$ > 4.5 GeV, $|\eta^\mathrm{clust}|$ < 2.47 \\
PID Quality      & \texttt{VeryLooseLLH} \\
Impact parameter & $|z_0 \sin\theta|< 0.5$ mm \\
\noalign{\smallskip}\hline\noalign{\smallskip}
\noalign{\smallskip}\hline\noalign{\smallskip}
\multicolumn{2}{c}{Signal electron}\\
\noalign{\smallskip}\hline\noalign{\smallskip}
Acceptance       & $p_\mathrm{T}$ > 4.5 GeV, $|\eta^\mathrm{clust}|$ < 2.47 \\
PID Quality      & \texttt{MediumLLH}  \\
Isolation        & \texttt{FCLoose} \\ 
Impact parameter & $S(d_0)< 5$ \\ 
\noalign{\smallskip}\hline\noalign{\smallskip}
\end{tabular}
\end{center}
\caption{Summary of the electron selection criteria. The signal selection requirements are applied on top of the baseline selection and after Overlap Removal has been performed.}     
\label{tab:compressedshiggsinos-eledef}
\end{table}

\subsubsection{Muons}
Muons used in this analysis must have $p_\mathrm{T} > 3$ GeV and reside within $|\eta|<2.5$. Baseline muons must pass \texttt{Medium} quality requirements and also satisfy the IP condition $|z_0 \sin \theta|<0.5$~mm. Signal muons must have $p_\mathrm{T} > 3$ GeV, pass the \texttt{Medium} quality criteria, be isolated with respect to other high-$p_\mathrm{T}$ charged particles, satisfying the \texttt{Loose\_VarRad} isolation criteria and additionally having $S(d_0)<3$ constraint on the IP. The muon selection criteria are summarised in Table~\ref{tab:compressedshiggsinos-muondef}.

\begin{table}[!htb]
\begin{center}\renewcommand\arraystretch{1.2}
\begin{tabular}{l|c}
\noalign{\smallskip}\hline\noalign{\smallskip}
\multicolumn{2}{c}{Baseline muon}\\
\noalign{\smallskip}\hline\noalign{\smallskip}
Acceptance        & $p_\mathrm{T}$ > 3 GeV, $|\eta|$ < 2.5  \\
PID Quality       & \texttt{Medium}    \\
Impact parameter  & $|z_0 \sin\theta|$ < 0.5 mm \\
\noalign{\smallskip}\hline\noalign{\smallskip}
\noalign{\smallskip}\hline\noalign{\smallskip}
\multicolumn{2}{c}{Signal muon}\\
\noalign{\smallskip}\hline\noalign{\smallskip}
Acceptance        & $p_\mathrm{T}$ > 3 GeV, $|\eta|$ < 2.  \\
PID Quality       & \texttt{Medium}  \\
Isolation         & \texttt{Loose\_VarRad} \\
Impact parameter  & $S(d_0)$ < 3 \\
\noalign{\smallskip}\hline\noalign{\smallskip}             
\end{tabular}
\caption{Summary of the muon selection criteria. The signal selection requirements are applied on top of the baseline selection after Overlap Removal has been performed.} 
\label{tab:compressedshiggsinos-muondef}
\end{center}
\end{table}

\subsubsection{Jets}
This analysis uses \texttt{PFlow} jets reconstructed using the anti-$k_{\mathrm{T}}$ algorithm with distance parameter $D=0.4$. At baseline level these jets are required to have $p_\mathrm{T} > 20$ GeV and fulfill the pseudorapidity requirement of $|\eta|<4.5$.
To reduce the effects of pile-up, signal jets are further required to pass the \texttt{Tight} working point on the $JVT$ ~\cite{ATLAS-CONF-2014-018}, if their $p_\mathrm{T}$ are $> 20$ GeV range and they reside within $|\eta|<4.5$.

The DL1r algorithm \cite{DL1r} identifies $b$-jets. A selection that provides 85\% efficiency for tagging $b$-jets in simulated $t\bar{t}$ events is used. The choice of 85\% WP ensured a stronger $t\bar{t}$ and single top rejection, without a significant loss of signal statistics. The jet selection criteria are summarised in Table~\ref{tab:compressedshiggsinos-jetsdef}.

\begin{table}[!htb]
\begin{center}
\begin{tabular}{l|c}
\noalign{\smallskip}\hline\noalign{\smallskip}
\multicolumn{2}{c}{Baseline jet} \\
\noalign{\smallskip}\hline\noalign{\smallskip}
Collection     & \texttt{AntiKt4EMPFlowJets} \\
Acceptance     & $p_\mathrm{T}$ > 20 GeV, $|\eta|$ <4.5 \\
\noalign{\smallskip}\hline\noalign{\smallskip}
\noalign{\smallskip}\hline\noalign{\smallskip}
\multicolumn{2}{c}{Signal jet} \\
\noalign{\smallskip}\hline\noalign{\smallskip}
JVT         & \texttt{Tight} \\
Acceptance  & $p_\mathrm{T} > 20$ GeV, $|\eta | < 4.5$ \\ 
\noalign{\smallskip}\hline\noalign{\smallskip}
\noalign{\smallskip}\hline\noalign{\smallskip}
\multicolumn{2}{c}{Signal $b$-jet} \\
\noalign{\smallskip}\hline\noalign{\smallskip}
$b$-tagger Algorithm & DL1r \\
Efficiency & \texttt{FixedCutBEff\_85} \\
Acceptance & $p_\mathrm{T}$ > 20 GeV, $|\eta|$ < 4.5 \\ 
\noalign{\smallskip}\hline\noalign{\smallskip}
\end{tabular}
\end{center}
\caption{Summary of the jet and $b$-jet selection criteria. The signal selection requirements are applied on top of the baseline requirements after Overlap Removal has been performed.}
\label{tab:compressedshiggsinos-jetsdef}
\end{table}

\subsubsection{Missing transverse momentum}
The $E_\mathrm{T}^{\mathrm{miss}}$ is reconstructed using the \texttt{Tight} working point, where the jets with $|\eta|>$~2.4 are required to have $p_{\mathrm{T}}>$~30~GeV.

\subsection{Trigger selection}
$E_\mathrm{T}^{\mathrm{miss}}$ triggers requiring the presence of $E_\mathrm{T}^{\mathrm{miss}}$ in the final state are adopted. Events are required to satisfy a logical OR of the triggers listed in Table~\ref{tab:Higgsinotrigger}. The triggers are divided by year and ordered by $E_\mathrm{T}^{\mathrm{miss}}$ threshold at the HLT level, which is the number appearing in the name, and then ordered by $E_\mathrm{T}^{\mathrm{miss}}$ threshold at the L1 level, which is the number appearing second in some triggers, for each group having the same $E_\mathrm{T}^{\mathrm{miss}}$ threshold at the HLT level. Further details can be found in \cite{ATLAS-METTrigger}. The trigger selection is chosen to give high signal efficiency at the preselection level, where it is required $E_{\mathrm{T}}^{\mathrm{miss}} > 200$ GeV.

\begin{table}[!htb]
\begin{center}
\scalebox{0.68}{\begin{tabular}{llll}
\noalign{\smallskip}\hline\noalign{\smallskip}
2015      &  2016    &    2017    &    2018 \\
\noalign{\smallskip}\hline\noalign{\smallskip}
\texttt{HLT\_xe70\_mht}  & \texttt{HLT\_xe80\_tc\_lcw\_L1XE50}  & \texttt{HLT\_xe90\_pufit\_L1XE50}  & \texttt{HLT\_xe100\_pufit\_xe75\_L1XE60} \\
          & \texttt{HLT\_xe90\_mht\_L1XE50}                     & \texttt{HLT\_xe100\_pufit\_L1XE50} & \texttt{HLT\_xe110\_pufit\_L1XE55} \\
          & \texttt{HLT\_xe90\_tc\_lcw\_wEFMu\_L1XE50}          & \texttt{HLT\_xe100\_pufit\_L1XE55} & \texttt{HLT\_xe110\_pufit\_L1XE60} \\
          & \texttt{HLT\_xe90\_mht\_wEFMu\_L1XE50}              & \texttt{HLT\_xe110\_pufit\_L1XE50} & \texttt{HLT\_xe110\_pufit\_L1XE70} \\
          & \texttt{HLT\_xe100\_L1XE50}                         & \texttt{HLT\_xe110\_pufit\_L1XE55} & \texttt{HLT\_xe110\_pufit\_wEFMu\_L1XE55} \\
          & \texttt{HLT\_xe110\_pueta\_L1XE50}                  & \texttt{HLT\_xe110\_pufit\_L1XE60} & \texttt{HLT\_xe110\_pufit\_xe65\_L1XE50} \\
          & \texttt{HLT\_xe110\_mht\_L1XE50}                    & \texttt{HLT\_xe120\_pufit\_L1XE50} & \texttt{HLT\_xe110\_pufit\_xe65\_L1XE55} \\
          & \texttt{HLT\_xe120\_tc\_lcw\_L1XE50}                & \texttt{HLT\_xe120\_pufit\_L1XE55} & \texttt{HLT\_xe110\_pufit\_xe65\_L1XE60} \\
          & \texttt{HLT\_xe120\_pueta}                          & \texttt{HLT\_xe120\_pufit\_L1XE60} & \texttt{HLT\_xe110\_pufit\_xe70\_L1XE50} \\
          & \texttt{HLT\_xe120\_pufit}                          &                                    & \texttt{HLT\_xe120\_mht\_xe80\_L1XE55}  \\
          &                                                     &                                    & \texttt{HLT\_xe120\_mht\_xe80\_L1XE60}  \\
          &                                                     &                                    & \texttt{HLT\_xe120\_pufit\_L1XE50}  \\
          &                                                     &                                    & \texttt{HLT\_xe120\_pufit\_L1XE55}  \\
          &                                                     &                                    & \texttt{HLT\_xe120\_pufit\_L1XE60}  \\
          &                                                     &                                    & \texttt{HLT\_xe120\_pufit\_L1XE70}  \\
          &                                                     &                                    & \texttt{HLT\_xe120\_pufit\_wEFMu\_L1XE55}  \\
          &                                                     &                                    & \texttt{HLT\_xe120\_pufit\_wEFMu\_L1XE60}  \\
\noalign{\smallskip}\hline\noalign{\smallskip}
\end{tabular}}
\end{center}
\caption{Summary of the triggers used in the analysis. Further details on the triggers are provided in Ref.~\cite{ATLAS-METTrigger}.}  
\label{tab:Higgsinotrigger}
\end{table}

\subsection{Overlap Removal}\label{sec:compressedshiggsinos-overlapRemoval}
The OR procedure is performed with baseline objects (electrons, muons and jets) and follows the default prescription.

\section{Preselection}
Candidate events are firstly selected by applying a preselection, reported in Table \ref{tab:preselCuts_higgsinos} and following the one suggested in Ref.~\cite{CorneringHiggsino} with minor modifications that are described below. The preselection can be divided into an event level selection and a track level selection. The event level selection follows the selection of ATLAS mono-jet search \cite{ATLAS-monojet}, which employed this selection to search for dark matter particles that recoil against a high $p_{\mathrm{T}}$ jet. This scenario is also targeted by this search with the key difference that a displaced track selection is additionally performed on top of the mono-jet one to gain sensitivity for signals exhibiting a displaced track topology and having much smaller cross sections.\\

\begin{table}[!htb]
\begin{center}
\begin{tabular}{l | c }
\noalign{\smallskip}\noalign{\smallskip}\noalign{\smallskip}\hline
\noalign{\smallskip}\noalign{\smallskip}\noalign{\smallskip}
\multicolumn{2}{c}{Event level selection}\\
\noalign{\smallskip}\noalign{\smallskip}\noalign{\smallskip}\hline
\noalign{\smallskip}\noalign{\smallskip}\noalign{\smallskip}
Variable & Selection\\
\noalign{\smallskip}\noalign{\smallskip}\noalign{\smallskip}\hline
\noalign{\smallskip}\noalign{\smallskip}\noalign{\smallskip}
$n_{\mathrm{leptons}}$ &  $=0$ \\
$n_{\mathrm{jets}}$ & $\geq 1$ \\
Leading jet $p_\mathrm{{T}}$ & $> 250$ GeV \\
Leading jet $|\eta|$ & < 2.4 \\
$n_{\mathrm{jets}}$ with jet $p_{\mathrm{T}} > 30$ GeV  and $|\eta|<2.8$ & $\leq 4$ \\
$E_{\mathrm{T}}^{\mathrm{miss}}$ & $> 200$ GeV \\
$\min_{i}|\Delta\phi(\mathrm{jet}_{i}, {E}_{\mathrm{T}}^{\mathrm{miss}})|$ &   $> 0.4$  \\
\noalign{\smallskip}\noalign{\smallskip}\hline
\noalign{\smallskip}\noalign{\smallskip}
\multicolumn{2}{c}{Track level selection}\\
\noalign{\smallskip}\noalign{\smallskip}\noalign{\smallskip}\hline
\noalign{\smallskip}\noalign{\smallskip}\noalign{\smallskip}
\multirow{5}{*}{Baseline}&1 GeV < track $p_{\mathrm{T}}$ < 5 GeV\\&track $|\eta|$ < 1.5\\&\texttt{Tight} track\\&track $d_{0}$ < 10 mm\\&track $|\Delta z_{0} \sin \theta|$ < 3 mm \smallskip\\
\multirow{4}{*}{Isolation} &  $\min_{i} \Delta R(\mathrm{track}, \mathrm{track}_{i}) > 0.3$ \\& with $\mathrm{track}_{i}$ having $p_{\mathrm{T}}>1$ GeV, \\& $|d_{0}|<1.5$ mm, $|\Delta z_{0} \sin \theta| < 1.5$ mm \smallskip\\
Displacement & No selection on $S(d_{0})$ \smallskip\\ 
$E_{\mathrm{T}}^{\mathrm{miss}}$ alignment & $|\Delta\phi(\mathrm{track}, {E}_{\mathrm{T}}^{\mathrm{miss}})| < 1$ \\
\noalign{\smallskip}\noalign{\smallskip}\hline
\noalign{\smallskip}\noalign{\smallskip}
\end{tabular}
\end{center}
\caption{Preselection cuts on the higgsino search.}
\label{tab:preselCuts_higgsinos}
\end{table}

The event selection is performed by first imposing a veto to all electrons or muons with $p_{\mathrm{T}} > 7$ GeV. This veto is useful for reducing $W \rightarrow \tau \bar{\nu}$ events in which $\tau$ decays leptonically, while a veto on the hadronic decays of the $\tau$ has not been introduced because the identification of such decays in ATLAS has low efficiency. Also, the veto helps to reduce the $t\bar{t}$ background and all the other backgrounds in which a $W$ boson decays in a lepton (typically with high $p_{\mathrm{T}}$) and its antineutrino.
Events are required to have at least one jet, and the leading jet is required to have  $p_\mathrm{T} > 250$ GeV and $|\eta|<2.4$. Then, a list of \textit{good jets} which are likely to come from the hard scatter rather than from pile-up is created. These jets are chosen to have $p_{\mathrm{T}}>30$ GeV and $|\eta|<2.6$. A superior limit of 4 on high $p_{\mathrm{T}}$ jets is posed because events with many high $p_{\mathrm{T}}$ jets are not expected to come from the primary collisions. In contrast to the suggested $E_{\mathrm{T}}^{\mathrm{miss}} > 500$ GeV cut in Ref.~\cite{CorneringHiggsino}, a selection of $E_{\mathrm{T}}^{\mathrm{miss}} > 200$ GeV is performed here to allow the definition of CRs in the loose $E_{\mathrm{T}}^{\mathrm{miss}}$ phase space region and because the significance improves when training the DNN inclusively in the $E_{\mathrm{T}}^{\mathrm{miss}} > 200$ GeV region with respect to the $E_{\mathrm{T}}^{\mathrm{miss}} > 500$ GeV, even if the phase space with $200$ GeV $< E_{\mathrm{T}}^{\mathrm{miss}} <$ 500 GeV is more background-like. Finally, a cut regarding the separation of jets with the direction of $E_{\mathrm{T}}^{\mathrm{miss}}$ is applied: only events with $\min_{i}|\Delta\phi(\mathrm{jet}_{i}, {E}_{\mathrm{T}}^{\mathrm{miss}})| > 0.4$, where the index $i$ runs over all selected jets, are accepted. This cut is necessary to reduce the very high cross-section of the multi-jet background when the energy of a jet is mismeasured and generates a fake $E_{\mathrm{T}}^{\mathrm{miss}}$ aligned to the direction of the jet.\\

The track level selection is performed considering all the tracks associated to the selected events. The preselection cuts can be grouped into four groups: baseline selection, isolation selection, displacement selection and $E_{\mathrm{T}}^{\mathrm{miss}}$ alignment selection. 
The baseline selection is the first applied and aims to suppress the background tracks while keeping the number of signal tracks as high as possible. The selection 1 GeV < track $p_{\mathrm{T}}$ < 5 GeV is based on the fact that under 1 GeV the number of background tracks increases much faster than the number of signal tracks while the number of tracks above 5 GeV is negligible. The selection of track $|\eta|$ < 1.5 is performed because signal tracks correspond to heavy particles that are produced more centrally than objects from soft interactions, such as background tracks. The \texttt{Tight} track selection ensures a reliable track reconstruction, as described in \cite{ATLAS-trackreconstruction}. Baseline tracks are also required to have $d_{0}$ < 10 mm and $|\Delta z_{0} \sin \theta|$ < 3 mm. The cut on $d_{0}$ is loose enough to accept most signal events for the average decay lengths considered in the search and eliminates poorly measured background tracks or tracks produced by interactions with the detector material. The goal of the cut on $|\Delta z_{0} \sin \theta|$ is analogous with the only difference being that it mostly rejects tracks from pile-up vertices. However, the track displacement is not visible in the $|\Delta z_{0} \sin \theta|$ distribution as it is in the $d_{0}$ distribution because the resolution in $z_{0}$ is much worse than the one in $d_{0}$.
The isolation selection requires a minimum $\Delta R$ among the selected track and any other track that has $p_{\mathrm{T}}>1$ GeV, $|d_{0}|<1.5$ mm and $|\Delta z_{0} \sin \theta| < 1.5$ mm. 
In contrast to the suggested cut $S(d_{0}) > 6$ in Ref.~\cite{CorneringHiggsino}, no displacement selection on track $S(d_{0})$ is performed here because the introduction of post-training selections on this variable improves the significance more than performing a pre-training selection and training over a subset of pre-training selected events. In fact, the absence of a displacement selection helps the DNN to discriminate better the backgrounds from the signals. 
Finally, the $E_{\mathrm{T}}^{\mathrm{miss}}$ alignment selection, $|\Delta\phi(\mathrm{track}, {E}_{\mathrm{T}}^{\mathrm{miss}})| < 1$, is designed to select signal tracks which are close to the direction of the ${E}_{\mathrm{T}}^{\mathrm{miss}}$ and thus, being a boosted topology, in opposite direction of the leading jet.\\

\begin{figure}[!p]
\centering
\includegraphics[width=0.49\linewidth]{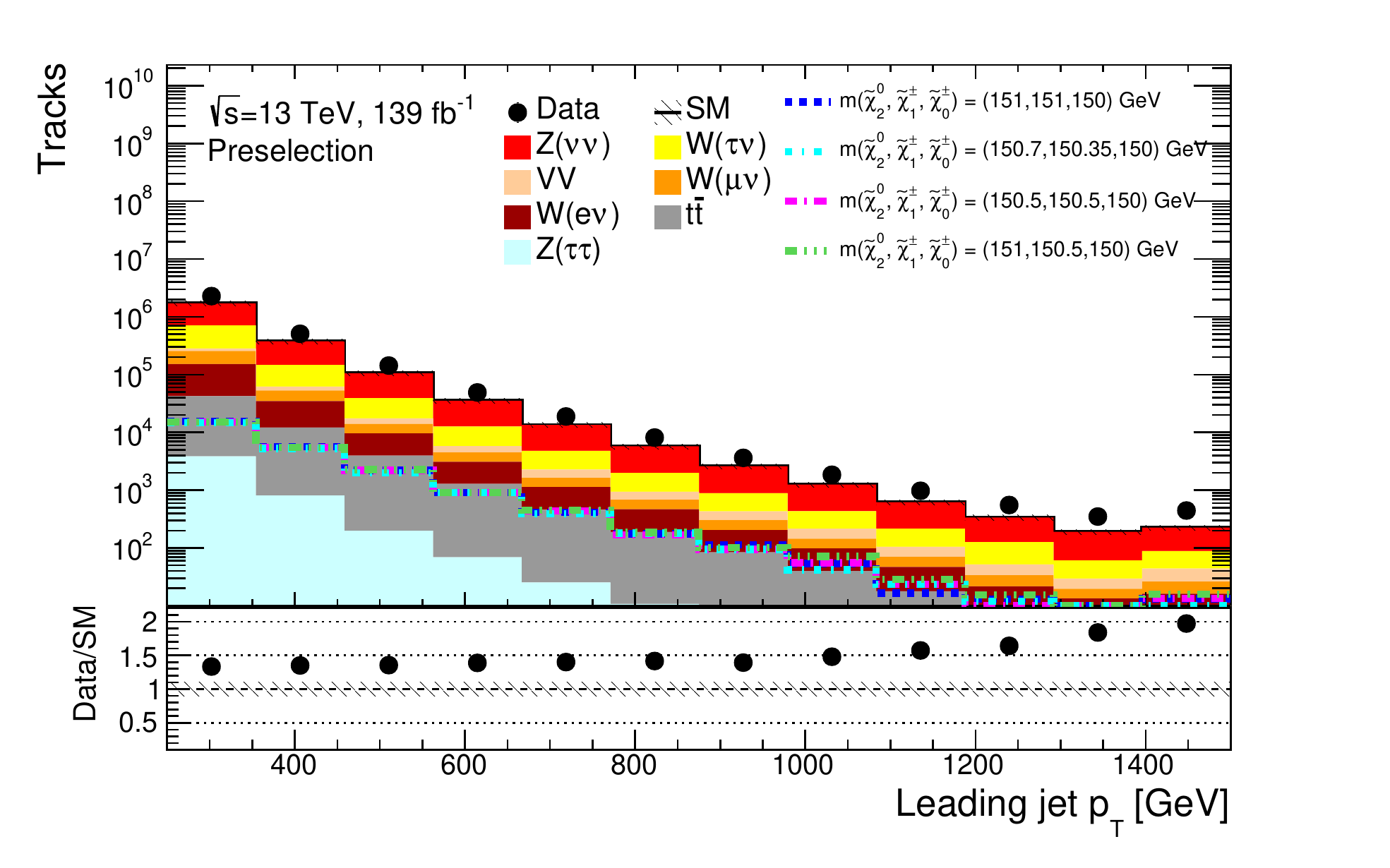}
\includegraphics[width=0.49\linewidth]{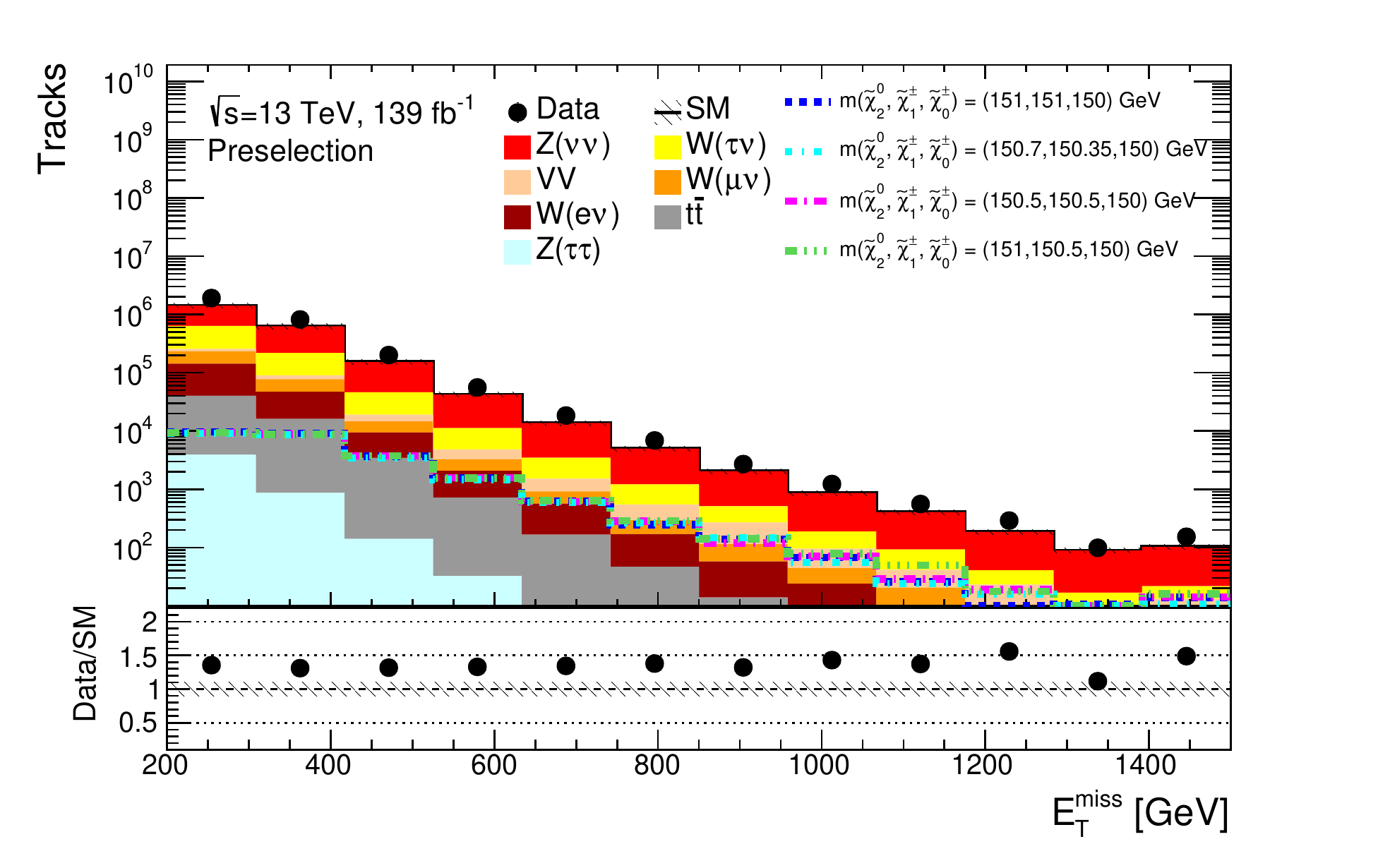}
\includegraphics[width=0.49\linewidth]{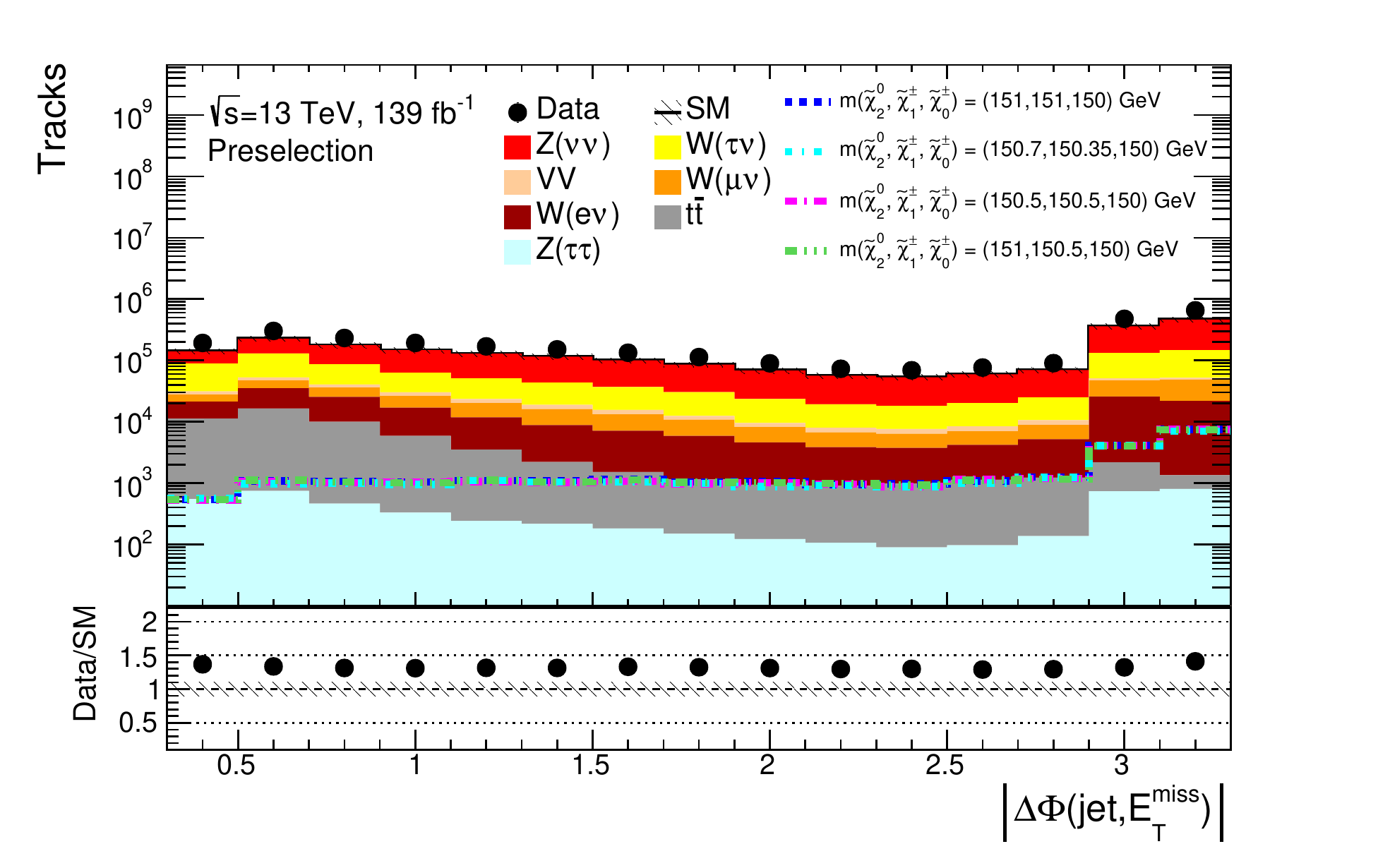}
\includegraphics[width=0.49\linewidth]{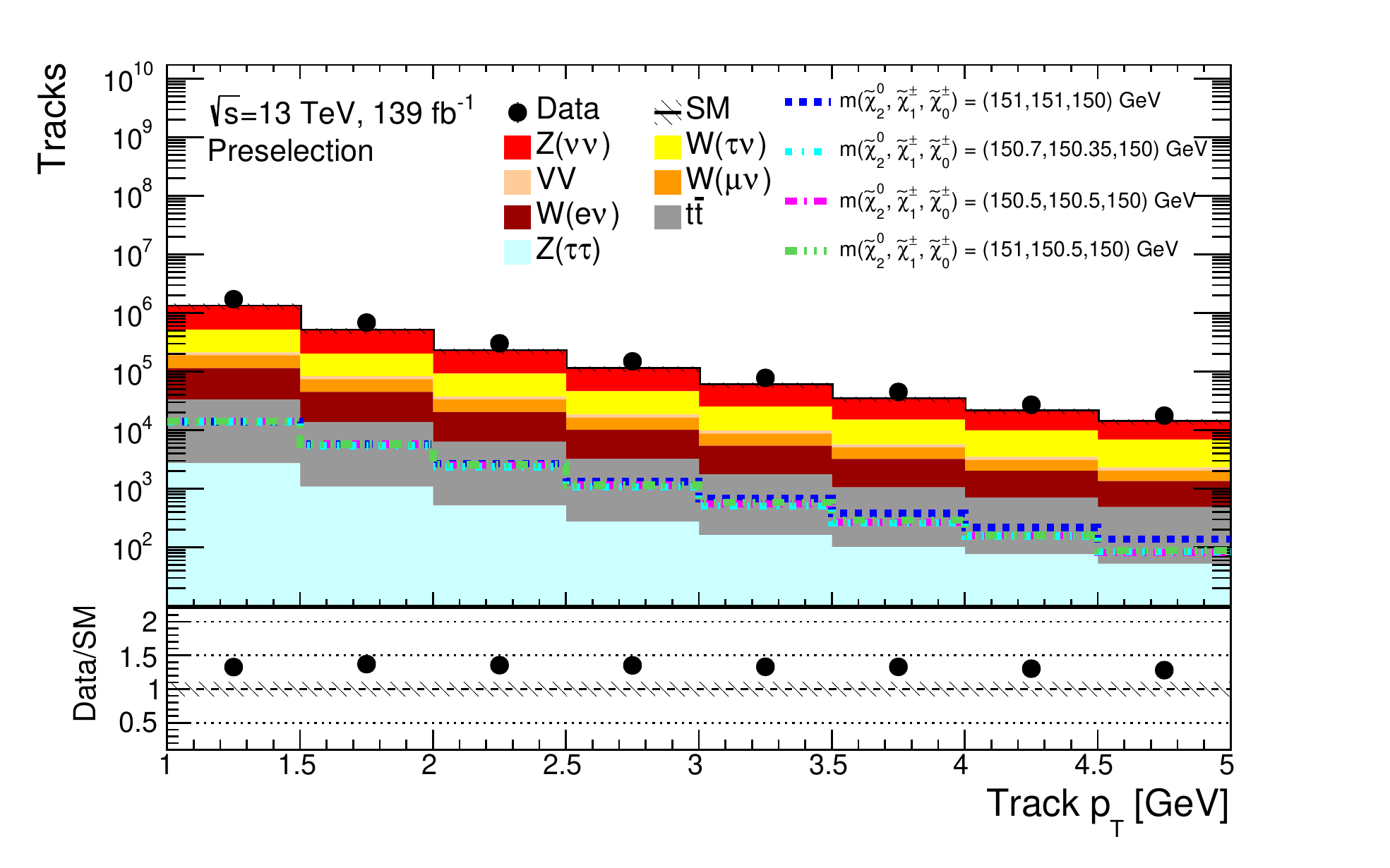}
\includegraphics[width=0.49\linewidth]{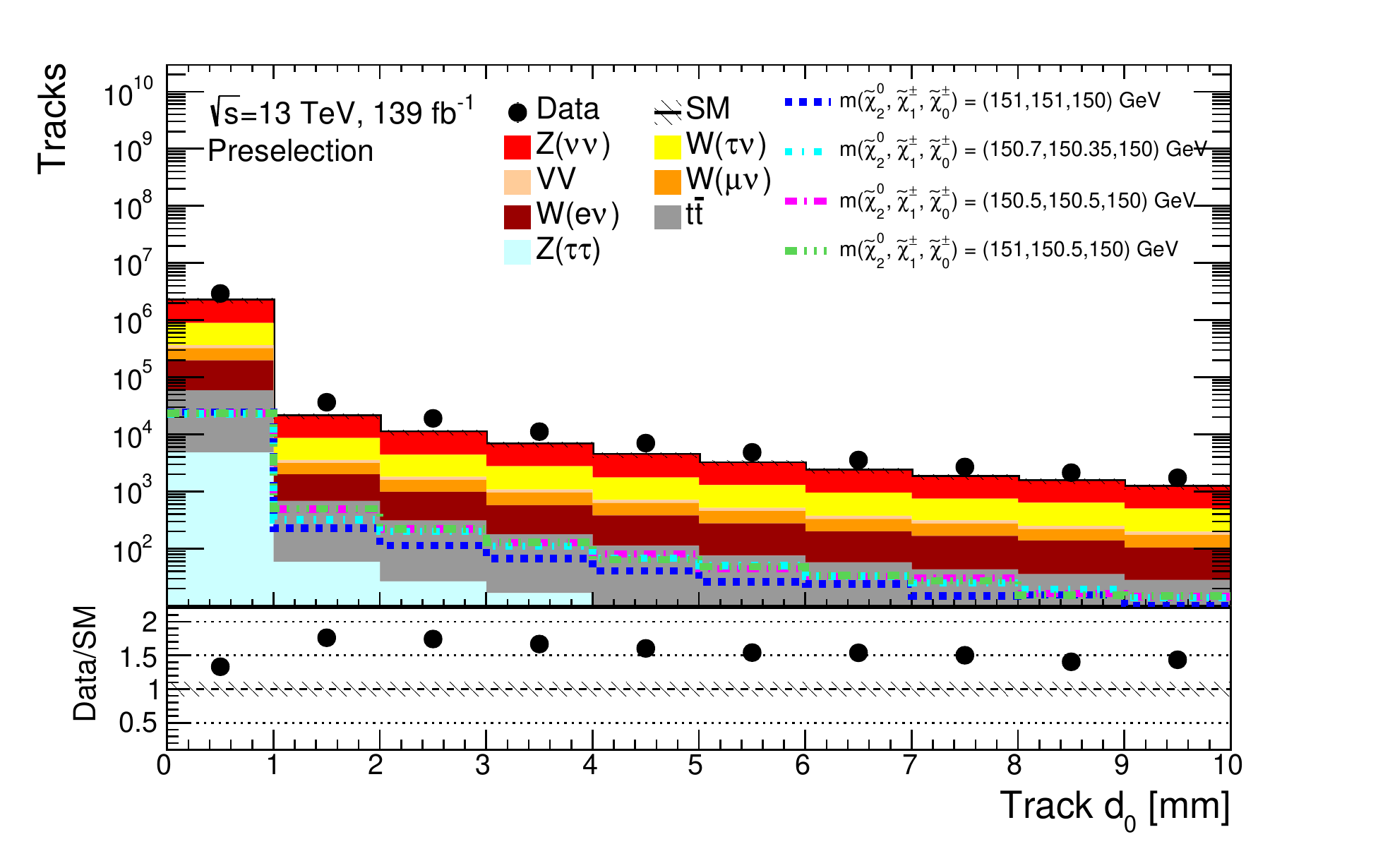}
\includegraphics[width=0.49\linewidth]{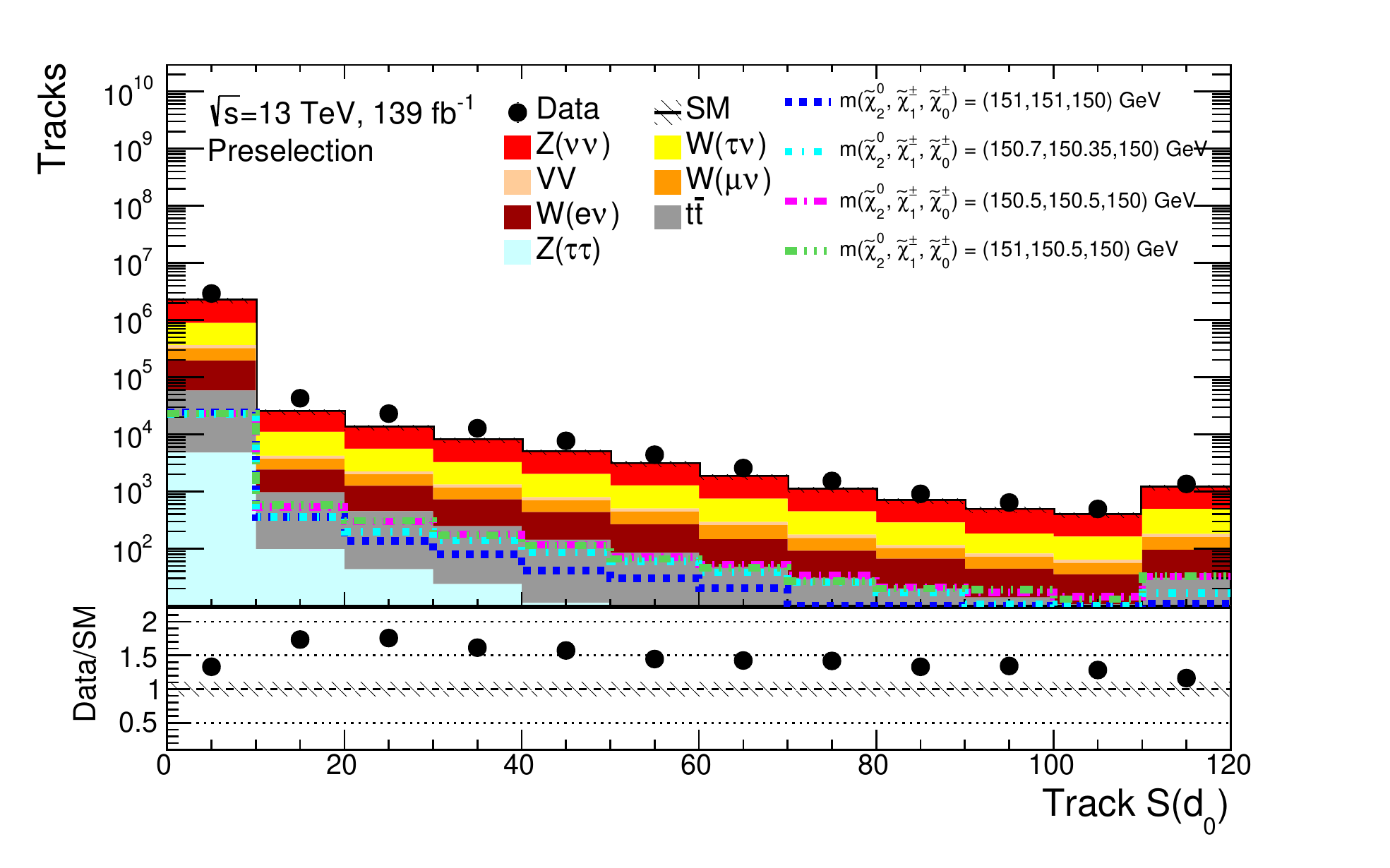}
\includegraphics[width=0.49\linewidth]{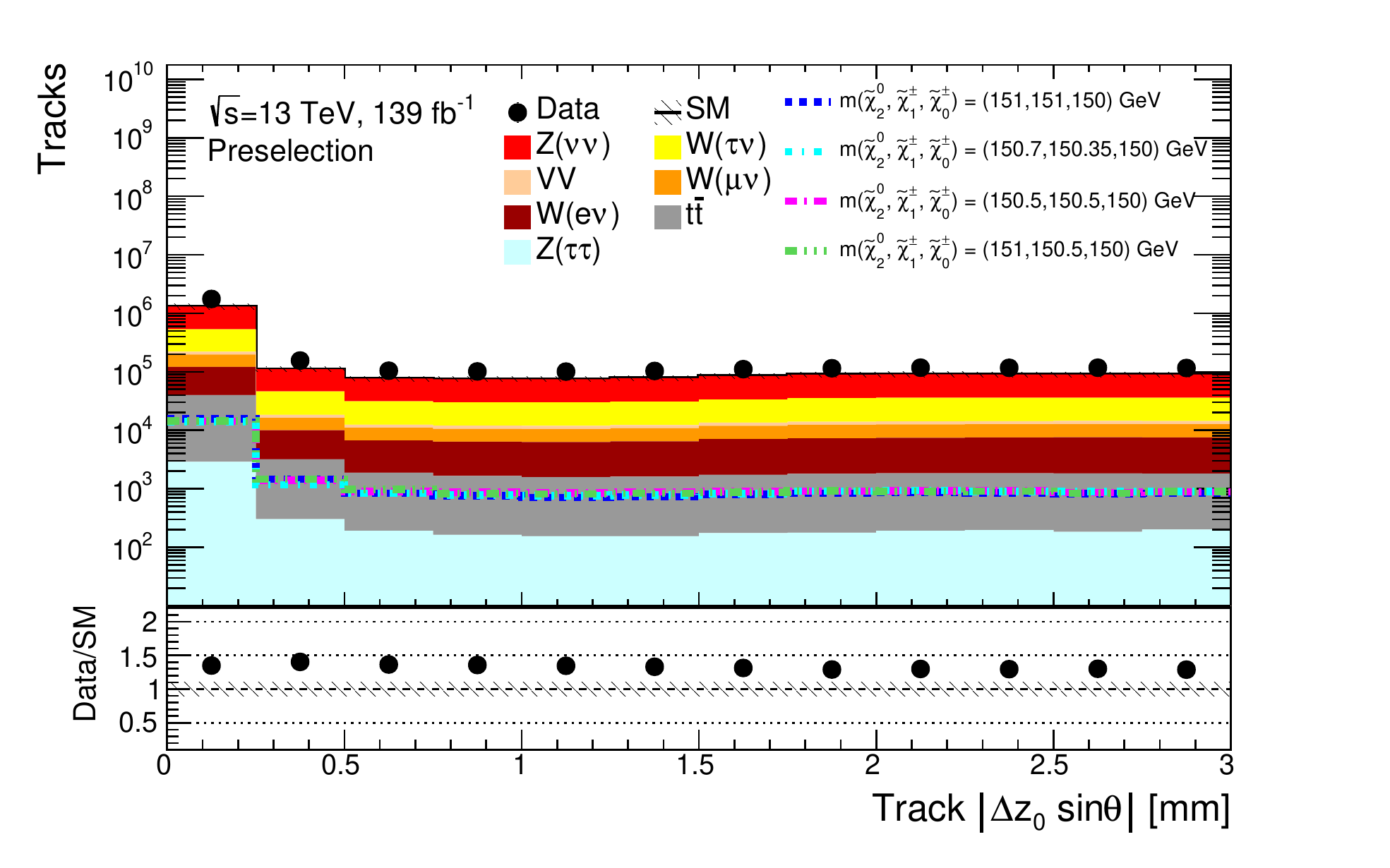}
\includegraphics[width=0.49\linewidth]{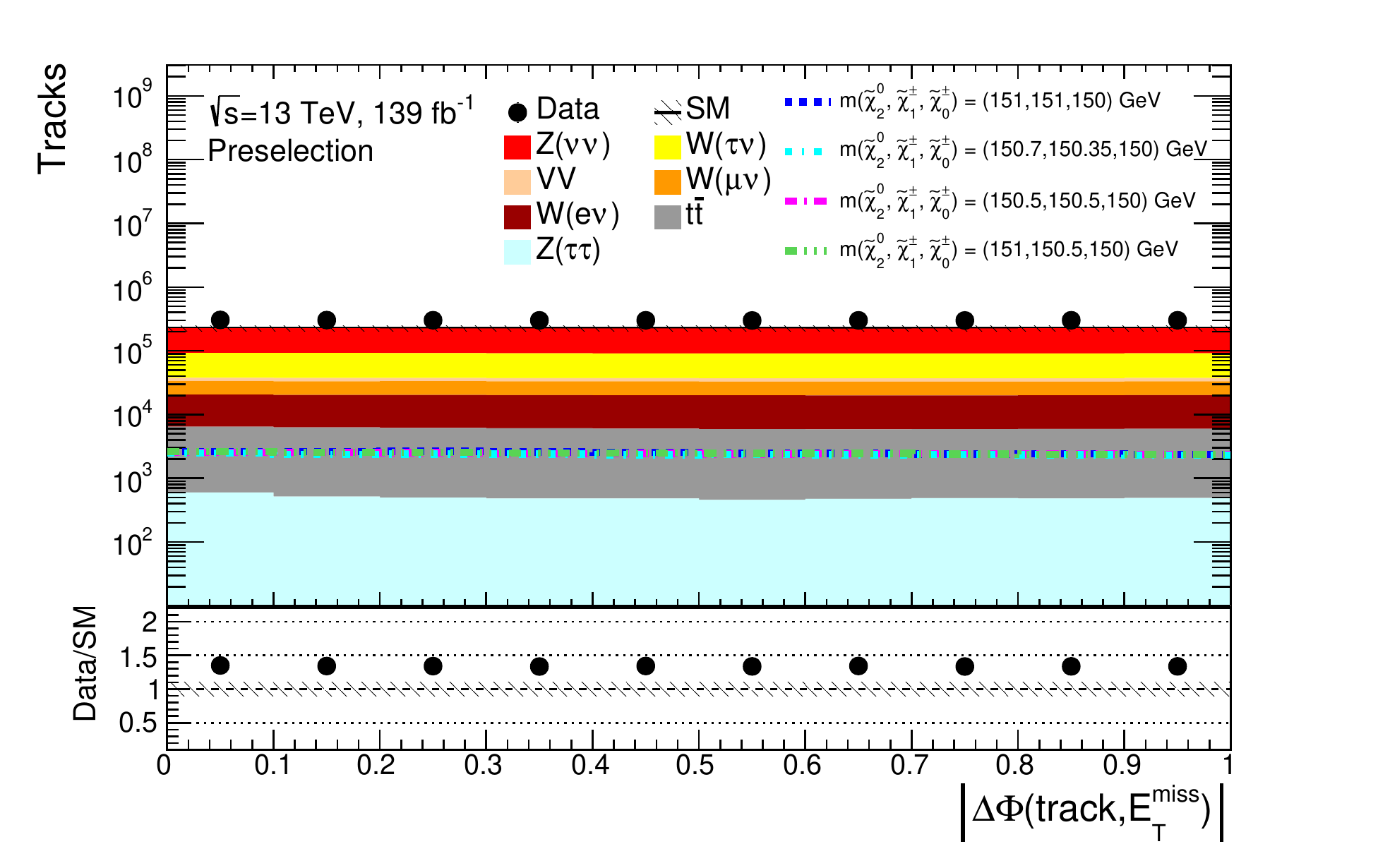}
\caption{The data and MC distributions of the most important variables at the preselection level. Uncertainties include the statistical contributions and a 10\% flat uncertainty as a preliminary estimate of the systematic uncertainty.}
\label{fig:Higgsino_presel}
\end{figure}

Fig.~\ref{fig:Higgsino_presel} shows the data and MC distributions of the most important variables of the search at the preselection level.
It is important to notice that the $d_{0}$ and $S(d_{0})$ distributions are similar for the  
$m(\tilde{\chi}_{2}^{0}, \, \tilde{\chi}_{1}^{\pm}, \,  \tilde{\chi}_{1}^{0}) = (151, 150.5, 150)$ GeV, $ (150.5, 150.5, 150)$ GeV, $ (150.7, 150.35, 150)$ GeV signal samples which have $\Delta m(\tilde{\chi}_{1}^{\pm}, \tilde{\chi}_{1}^{0}) \leq 0.5$ GeV, but differ for the $m(\tilde{\chi}_{2}^{0}, \, \tilde{\chi}_{1}^{\pm}, \,  \tilde{\chi}_{1}^{0}) = (151, 151, 150)$ GeV signal sample which has $\Delta m(\tilde{\chi}_{1}^{\pm}, \tilde{\chi}_{1}^{0}) = 1$ GeV. This behaviour is predicted by the following equation \cite{DecayWidth1,DecayWidth2}, 

\begin{equation}
    \Gamma^{-1}_{\tilde{\chi}^{\pm}_{1}\rightarrow\tilde{\chi}^{0}_{1}\pi^{\pm}} \simeq \frac{\mathrm{14\,mm}}{\hbar c} \times \Big[ \Big( \frac{\Delta m(\tilde{\chi}_{1}^{\pm}, \tilde{\chi}_{1}^{0})}{\mathrm{340\,MeV}} \Big)^{3} \sqrt{1-\frac{m^{2}_{\pi^{\pm}}}{\Delta m(\tilde{\chi}_{1}^{\pm}, \tilde{\chi}_{1}^{0})}}\Big]^{-1},
\end{equation}

which relates  to the $\Delta m(\tilde{\chi}_{1}^{\pm}, \tilde{\chi}_{1}^{0})$ mass splitting and essentially states that the shorter the mass difference between $\tilde{\chi}_{1}^{\pm}$ and $\tilde{\chi}_{1}^{0}$, the longer the decay length associated to the $\tilde{\chi}_{1}^{\pm}$ decay. Table~\ref{Table:DecayLengthsHiggsinos} reports the decay lengths $\delta=\Gamma^{-1}\hbar c=\tau c$, with $\tau$ representing the mean lifetime, for $\tilde{\chi}_{2}^{0}$ and $\tilde{\chi}_{1}^{\pm}$ and corresponding to our signal samples.

\begin{table}[!htb]
\centering
\begin{tabular}{lcc}
\noalign{\smallskip}\hline\noalign{\smallskip}
$m(\tilde{\chi}_{2}^{0}, \, \tilde{\chi}_{1}^{\pm}, \,  \tilde{\chi}_{1}^{0})$ [GeV]     & $\delta$ for $\tilde{\chi}_{2}^{0}$ [mm] & $\delta$ for $\tilde{\chi}_{1}^{\pm}$ [mm] \\
\noalign{\smallskip}\hline\noalign{\smallskip}
(151, 150.5, 150)    & 0.60  & 3.88 \\
(150.5, 150.5, 150)  & 8.32  & 3.88 \\
(150.7, 150.35, 150) & 2.71  & 12.96\\
(151, 151, 150)      & 0.60  & 2.50  \\
\noalign{\smallskip}\hline\noalign{\smallskip}
\end{tabular}
\caption{Decay length $\delta=\Gamma^{-1}\hbar c=\tau c$ corresponding to the signal samples listed in Sec~\ref{sec:compressedshiggsinos-signalsamples}.}
\label{Table:DecayLengthsHiggsinos}
\end{table}

\section{DNN classification}
\label{sec:compressedhiggsinos-DNN}

\subsection{Training setup}
The signal and background MC samples are split into two sets, the training and test sets. A classifier is trained over the training set, and the classifier tests the testing set. The two sets are then inverted, and the test set is used for training a second classifier which tests the original training set. The training and test sets are split such that they both contain half of the total MC samples, according to the \say{EventNumber} being odd or even. In this way, the entire signal and background samples get tested and we can use the entire MC statistics for the fit, whilst always testing over a statistically independent data sample.

The training is performed in the region with no lepton at the preselection level, as defined in Sec.~\ref{tab:preselCuts_higgsinos}. For each event, all the tracks satisfying the preselection cuts are saved and passed to the training algorithm. The training performs a binary classification between tracks from signal samples that are labelled as \textit{signal} with a score value of 1 and tracks from background samples that are labelled as \textit{background} with a score value of 0.

Due to the statistical imbalance between signal tracks and background tracks at the preselection level, the background tracks overwhelm the signal tracks. A subset of randomly chosen background tracks is extracted from the initial set of all background tracks coming from the various background processes to have the same size as the signal track set. Hence, the training is performed between these two balanced sets. This process improves the range of the DNN output score of the test set, which otherwise would be too compressed towards 0 values (background tracks), without reducing the signal statistical significance.

\subsection{Signal samples}
Concerning the signal samples used for the training, different training strategies were tested. Depending on the mass splitting $\Delta m(\tilde{\chi}_{2}^{0}/\tilde{\chi}_{1}^{\pm}, \tilde{\chi}_{1}^{0})$, the lifetime of the produced particles is different and consequently the decay length reconstructed in the detector. Differences in the decay length translate into differences in track $d_{0}$ and $S(d_{0})$ input features, and so dedicated trainings are adopted to tackle such differences. Two different training strategies are used:
\begin{itemize}
\item A training combining $m(\tilde{\chi}_{2}^{0}, \, \tilde{\chi}_{1}^{\pm}, \,  \tilde{\chi}_{1}^{0}) = (151, 150.5, 150)$ GeV, $(150.5, 150.5, 150)$ GeV, $(150.7, 150.35, 150)$ GeV signal samples which have longer decay lengths.
\item A training using $m(\tilde{\chi}_{2}^{0}, \, \tilde{\chi}_{1}^{\pm}, \,  \tilde{\chi}_{1}^{0}) = (151, 151, 150)$ GeV signal sample which has a shorter decay length.
\end{itemize}

\subsection{Input features}
\label{sec:compressedshiggsino-inputfeature}

Different variables are considered as input features, both at the event level and at the track level. 

A key consideration is that the DNN output score has to be not correlated with the $E_{\mathrm{T}}^{\mathrm{miss}}$ to use these two variables as the independent variables to estimate the main backgrounds via the ABCD method. For this reason, some variables cannot be included in the training because adding them would correlate the DNN output score to the $E_{\mathrm{T}}^{\mathrm{miss}}$. This explains the choice of our variables, which are almost all track variables, apart from $|\Delta\phi(\mathrm{jet}, {E}_{\mathrm{T}}^{\mathrm{miss}})|$ which is an angular variable and does not provide correlation. Variables such as $E_{\mathrm{T}}^{\mathrm{miss}}$, $E_{\mathrm{T}}^{\mathrm{miss}}$ significance and the leading jet $p_{\mathrm{T}}$, which were used in the chargino analysis, are absent in the training as they would have correlated the DNN output score to the $E_{\mathrm{T}}^{\mathrm{miss}}$.
After training, the correlation between these two variables has been studied and the DNN output score has been confirmed to be independent from the $E_{\mathrm{T}}^{\mathrm{miss}}$.

The performance when each variable is removed and the DNN retrained is assessed. A variable that, when removed, resulted in an increase of performance is removed and this process is repeated until no gain in performance is gained by removing any of the variables.

\begin{figure}[!htb]
\centering
\includegraphics[width=0.8\linewidth]{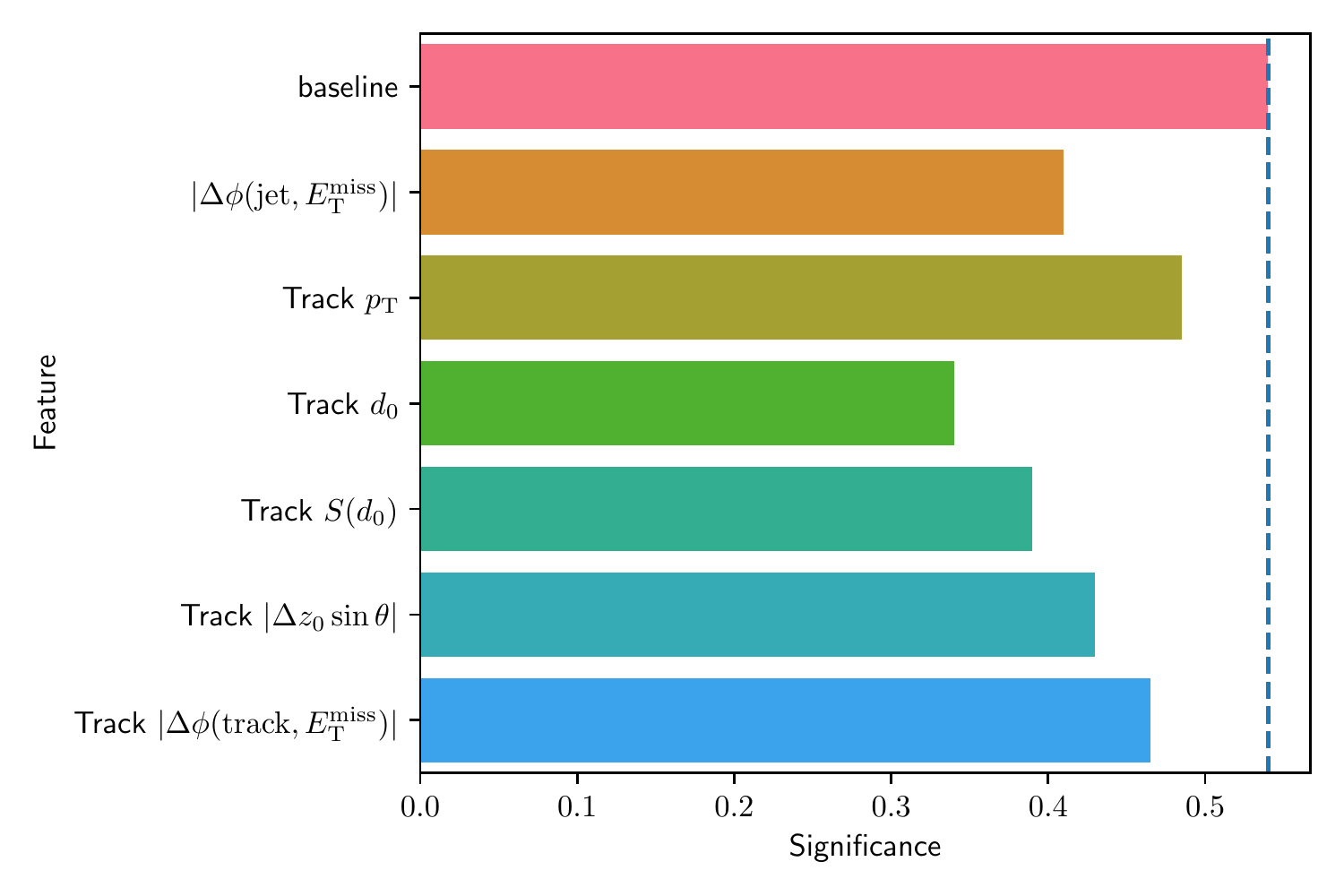}
\caption{The significance estimated as an average of the significance values of the four signal samples when removing each variable and retraining the classifier. The blue vertical line indicates our current significance.}
\label{fig:Higgsino_FeatureImportance}
\end{figure}

The best variable set is found to be: $|\Delta\phi(\mathrm{jet}, {E}_{\mathrm{T}}^{\mathrm{miss}})|$, track $p_{\mathrm{T}}$, track $d_{0}$, track $S(d_{0})$, track $|\Delta z_{0}\sin\theta|$ and $|\Delta\phi(\mathrm{track}, {E}_{\mathrm{T}}^{\mathrm{miss}})|$. Fig.~\ref{fig:Higgsino_FeatureImportance} shows the significance when each feature is removed and the DNN retrained, with the significance estimated as an average of the significance values of the four signals. 

Track $d_{0}$ and track $S(d_{0})$ are the most important ones, as we would expect from the fact that our signals tend to statistically have larger displacements than the backgrounds. $|\Delta\phi(\mathrm{jet}, {E}_{\mathrm{T}}^{\mathrm{miss}})|$ is also found to have a sizeable impact if removed, signalling that it is one of the most beneficial for the training. Furthermore, we cannot gain any significant sensitivity by reducing this set of variables trained over while training with the extra variables showed no increase in sensitivity.

\subsection{Network structure}
\label{sec:compressedshiggsino-networkstructure}
\begin{figure}[!htb]
\centering
\includegraphics[width=0.6\linewidth]{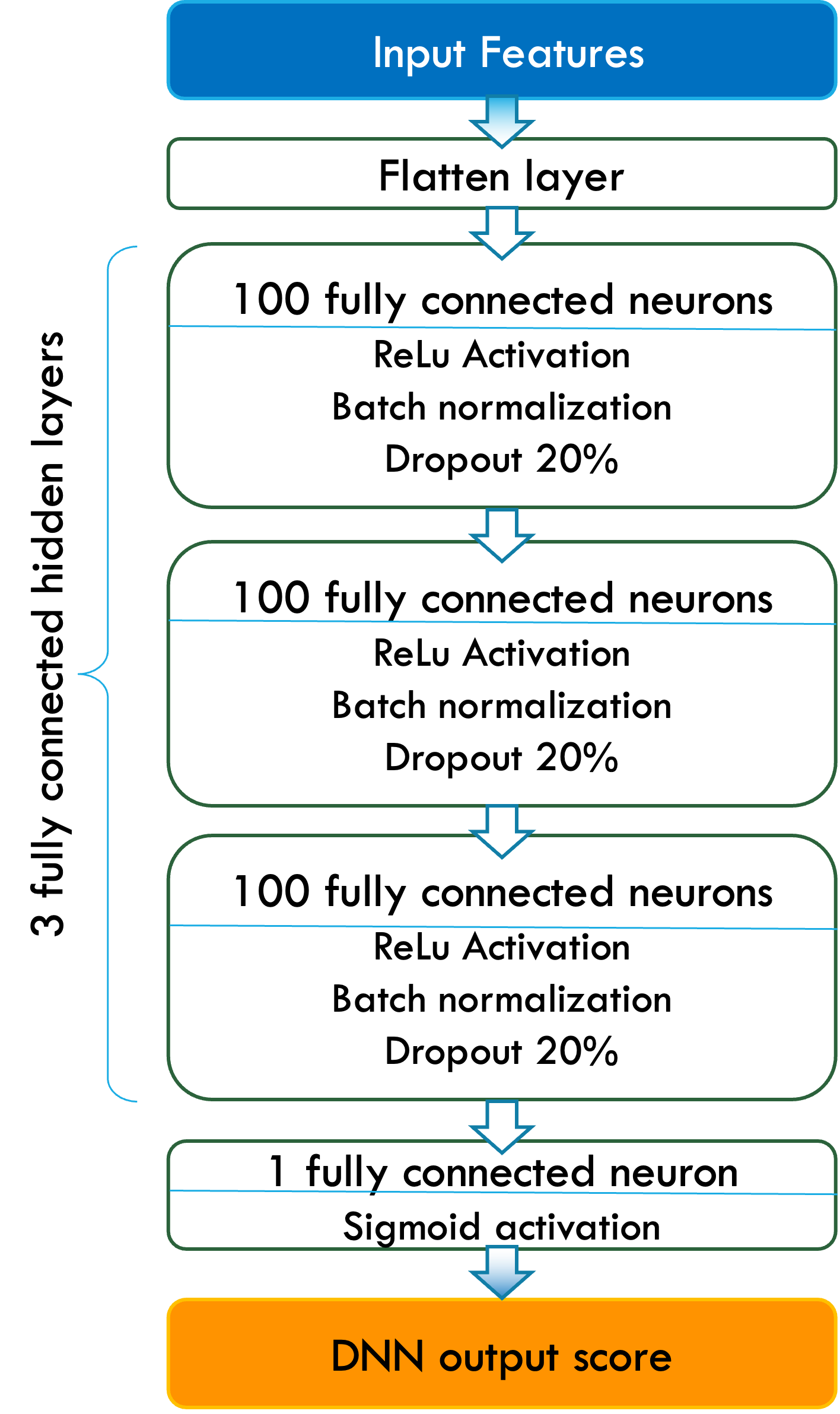}
\caption{Internal structure of the DNN, with a flatten layer receiving input features, 3 fully connected hidden layers and a final output neuron which provides the DNN output score.}
\label{fig:Higgsino_NetworkStructure}
\end{figure}

The structure of the DNN is shown in Fig.~\ref{fig:Higgsino_NetworkStructure}.

It is a Sequential algorithm, a kind of feedforward neural network with fully connected layers, developed using the Tensorflow~\cite{Tensorflow} and Keras~\cite{Keras} packages.
The input features are passed to a flatten layer which is connected to the subsequent fully connected layers (or \textit{dense layers}). A DNN has more than one dense layer by definition and, in our case, 3 hidden layers are used, a choice that guarantees good performance after training without overtraining. Each hidden layer is composed of 100 fully connected neurons, activated by the Rectified Linear Unit (ReLU) function and using a 20\% of dropout. For a given input $z$, ReLu is defined as $\mathrm{ReLu}=\max(0,z)$. Being continuous and a simple function to compute, ReLu has become the default choice when using the gradient descent algorithm, but unfortunately not differentiable at $z = 0$.
Dropout, instead, efficiently avoids overtraining: for every training step, every neuron (including the input neurons and excluding the output one) has a probability of being temporarily \textit{dropped out}, i.e. completely ignored by the other neurons. This probability, referred to as \textit{dropout rate}, is a hyperparameter that is usually selected between 10\% and 50\%.

Batch Normalization is added to each layer to reduce the risk of vanishing/exploding gradients in the gradient descent method and simplify the success of the training. Batch Normalization is usually applied near the activation function of each hidden layer, centering to zero and normalizing with a gaussian function for each input function. Then, it scales and shifts the results using two hyperparameters per layer, imposed by the user. In other words, Batch Normalization lets the model learn the optimal scale for each of the layer inputs using their mean values. Even if this method adds some complexity to the model, because the DNN makes slower predictions due to extra computations required at each layer, it can be considered as a fundamental element of the inner structure because is very efficient and reduce the need for other regularization techniques.
The Adam optimizer \cite{Adam} is implemented to obtain better performance of the DNN. Adam stands for \textit{adaptive moment estimation}, and can be thought as an improved gradient descent method: it keeps track of an exponentially decaying average of past gradients and combines them according to a learning rate fixed by the user.
A fully connected neuron activated by the sigmoid function provides the prediction of the DNN output score for each track.

\subsection{ROC curve}
\begin{figure}[!htb]
\centering
\includegraphics[width=0.8\linewidth]{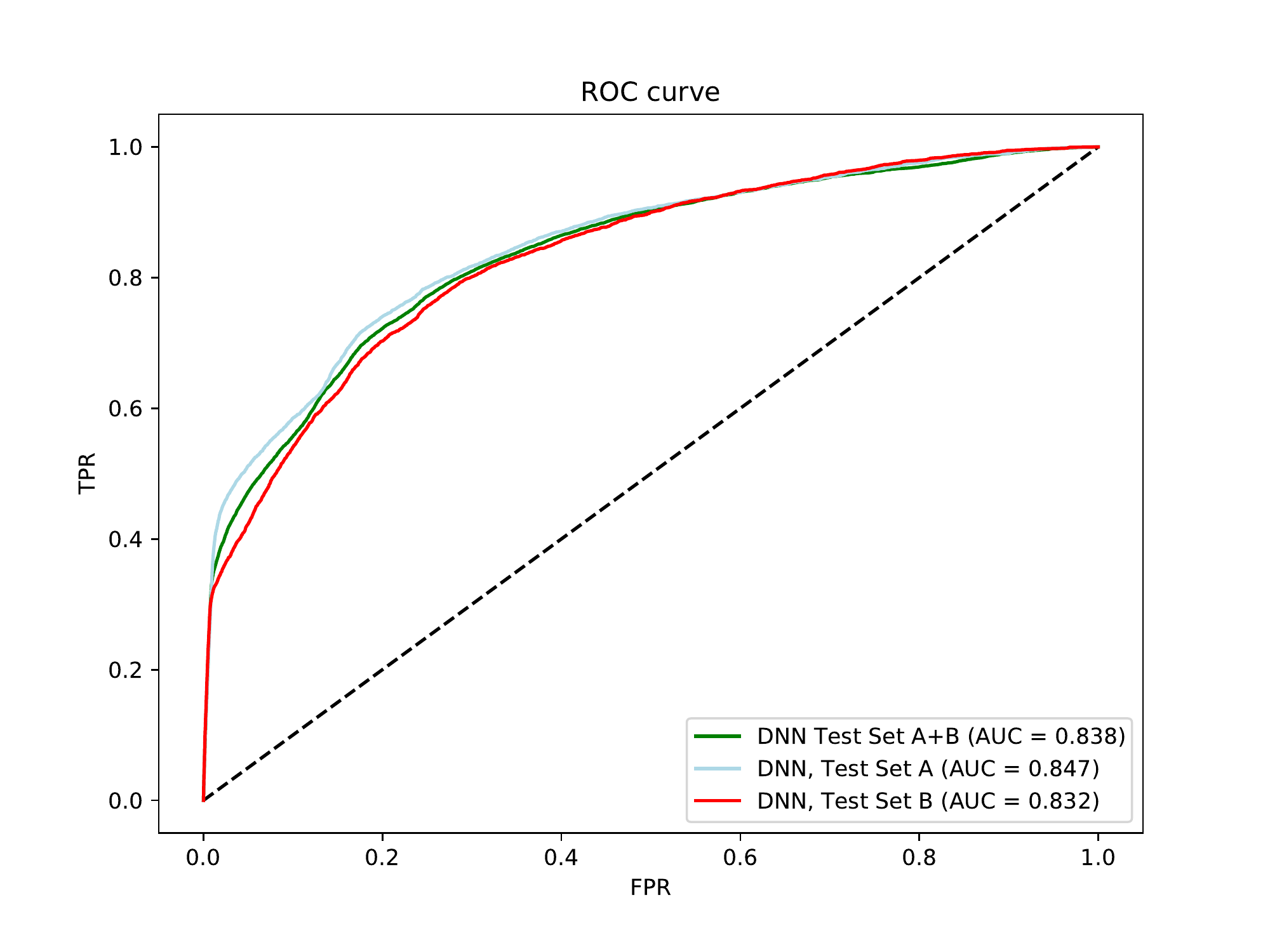}
\caption{The ROC curve for the DNN with the AUC values included in the legend.}
\label{fig:Higgsino_ROCcurve}
\end{figure}

The ROC curve for the DNN training is shown in Fig.~\ref{fig:Higgsino_ROCcurve}. 

The different curves show the performance as evaluated for the test set A (selected with even EventNumbers), for the test set B (selected with odd EventNumbers) and for the whole test set (A+B). The three curves agree well as there are no significant differences between the three trainings, which might be instead caused by limited statistics of the training set after being split into two halves. The combined AUC value is 0.838.

\subsection{SHAP values}
\begin{figure}[!htb]
\centering
\includegraphics[width=1.\linewidth]{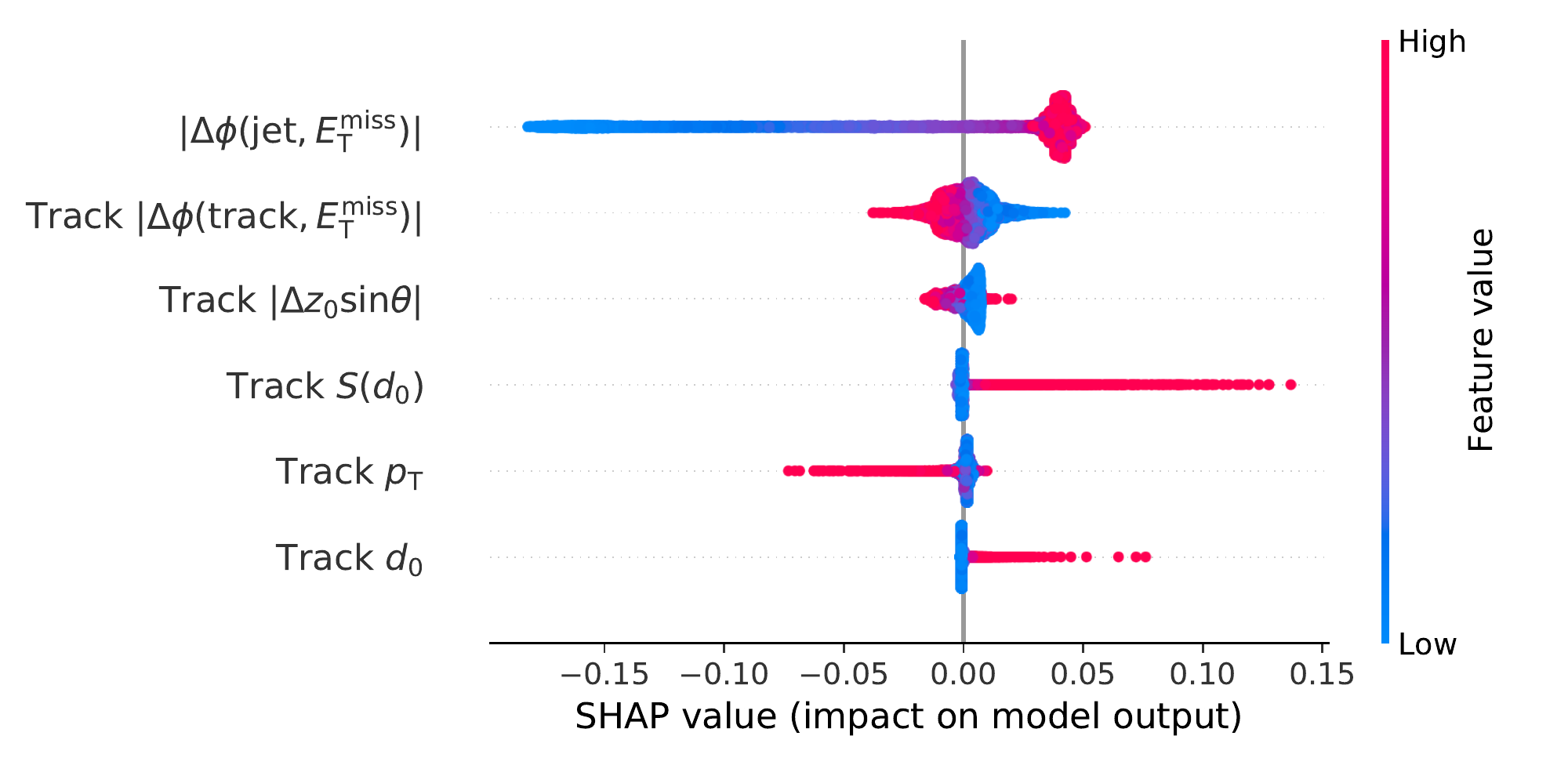}
\caption{The SHAP values for the DNN output score. Points to the right are more signal-like and to the left are more background-like. The colour of the point indicated the value of the corresponding variable.}
\label{fig:HiggsinoSHAP}
\end{figure}

The SHAP values for the DNN training are shown in Fig.~\ref{fig:HiggsinoSHAP}.

The $x$-axis indicates the SHAP value, e.g. the impact on the model output. Higher SHAP values mean more signal-like for the event while lower SHAP values mean more background-like for the event. Each dot on the plot corresponds to one track, and its colour (on the scale blue to red) indicates the value of the corresponding variable the SHAP value is calculated for, labelled on the $y$-axis. As we can see, low $|\Delta\phi(\mathrm{jet}, {E}_{\mathrm{T}}^{\mathrm{miss}})|$ values are only associated to background-like events, while signal-like events have higher angular separation between the $\mathrm{jet}$ and ${E}_{\mathrm{T}}^{\mathrm{miss}}$. Conversely, low $|\Delta\phi(\mathrm{track}, {E}_{\mathrm{T}}^{\mathrm{miss}})|$ values are associated to signal-like tracks, while background-like tracks have higher angular separation with respect to ${E}_{\mathrm{T}}^{\mathrm{miss}}$ direction. Displaced tracks having large track $d_{0}$ and $S(d_{0})$ are classified as signal ones, as we would expect. Instead, both low and high values of track $p_{\mathrm{T}}$ and $|\Delta z_{0}\sin\theta|$ features is attributed to signal-like tracks by the classifier. We can also notice that $|\Delta\phi(\mathrm{jet}, {E}_{\mathrm{T}}^{\mathrm{miss}})|$ can have a large impact in recognizing tracks as associated to background processes, as well as track $d_{0}$ and $S(d_{0})$ have a sizeable impact in recognizing tracks as associated to supersymmetric signal processes.

\subsection{Correlations}
\begin{figure}[!htb]
\centering
 \hspace*{-0.5cm}
 \includegraphics[width=1.2\linewidth]{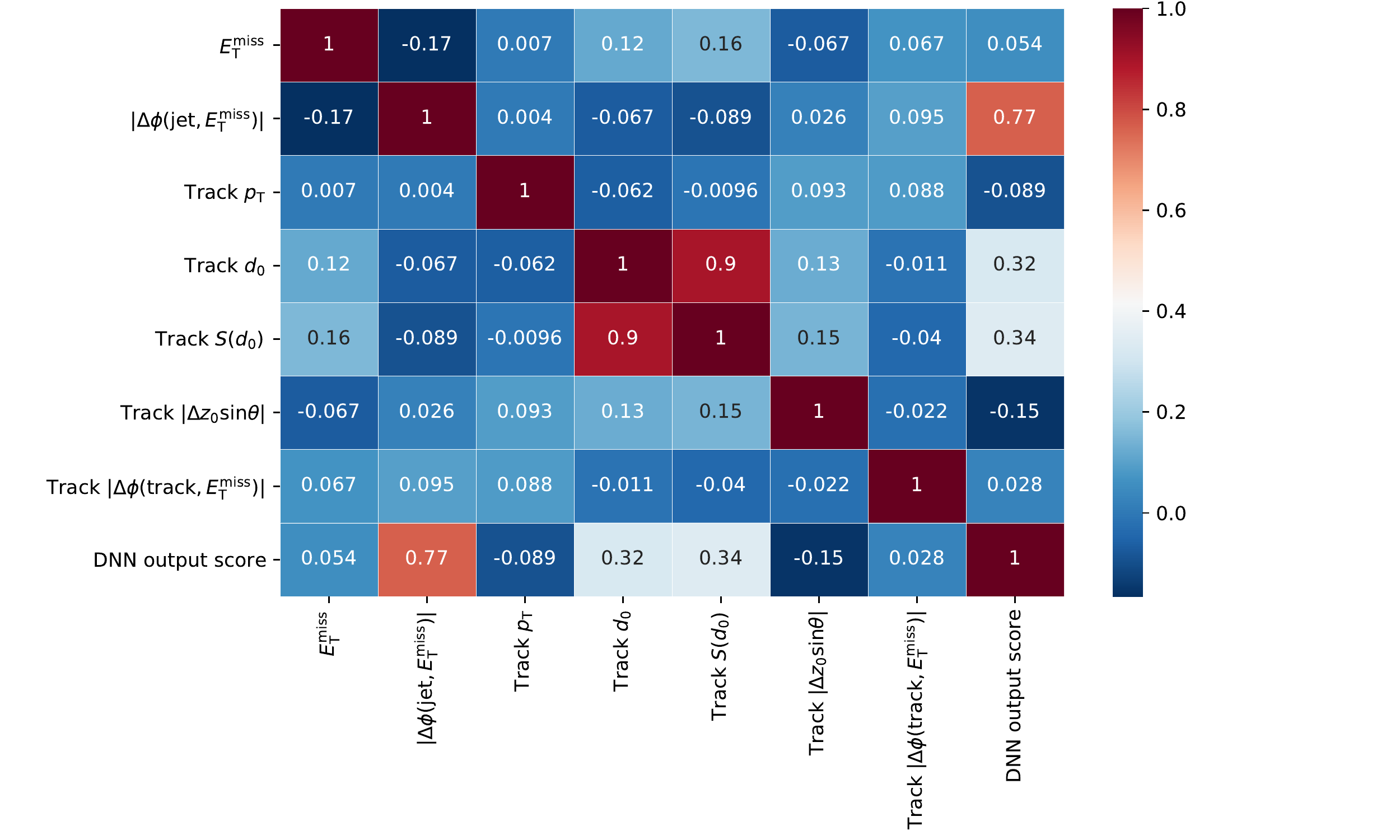}
\caption{The correlation values computed as Pearson correlation coefficients for different variables. The colour of each box corresponds to the axis on the right.}
\label{fig:HiggsinoCorrelation}
\end{figure}

The correlation values computed as Pearson correlation coefficients for the ${E}_{\mathrm{T}}^{\mathrm{miss}}$, for the set of input features and for the DNN output score are shown in Fig.~\ref{fig:HiggsinoCorrelation}.  

The Pearson correlation coefficient is defined as the ratio between the covariance of two variables and the product of their standard deviations. It can assume a value between $-1$ and 1, with $-1$ indicating anti-correlation between the two variables, 1 indicating correlation between the two variables and 0 no correlation or anti-correlation. As we can see, track $d_{0}$ and $S(d_{0})$ are highly correlated, as we would expect from the fact that track $S(d_{0})$ is obtained from track $d_{0}$ by additionally considering its associated uncertainties. No other relevant correlations can be found among the set of input features. Interestingly, the DNN output score is highly correlated to the $|\Delta\phi(\mathrm{jet}, {E}_{\mathrm{T}}^{\mathrm{miss}})|$. Coherently with what we have learned from the plot of the SHAP values, high DNN output score values tend to be associated to events with high $|\Delta\phi(\mathrm{jet}, {E}_{\mathrm{T}}^{\mathrm{miss}})|$ values. A moderate level of correlation is also found between the DNN output score and the $d_{0}$ and $S(d_{0})$ variables. Remarkably, ${E}_{\mathrm{T}}^{\mathrm{miss}}$ is uncorrelated with the DNN output score or any other input features.

\subsection{DNN output score}
\begin{figure}[!htb]
\centering
\includegraphics[width=0.8\linewidth]{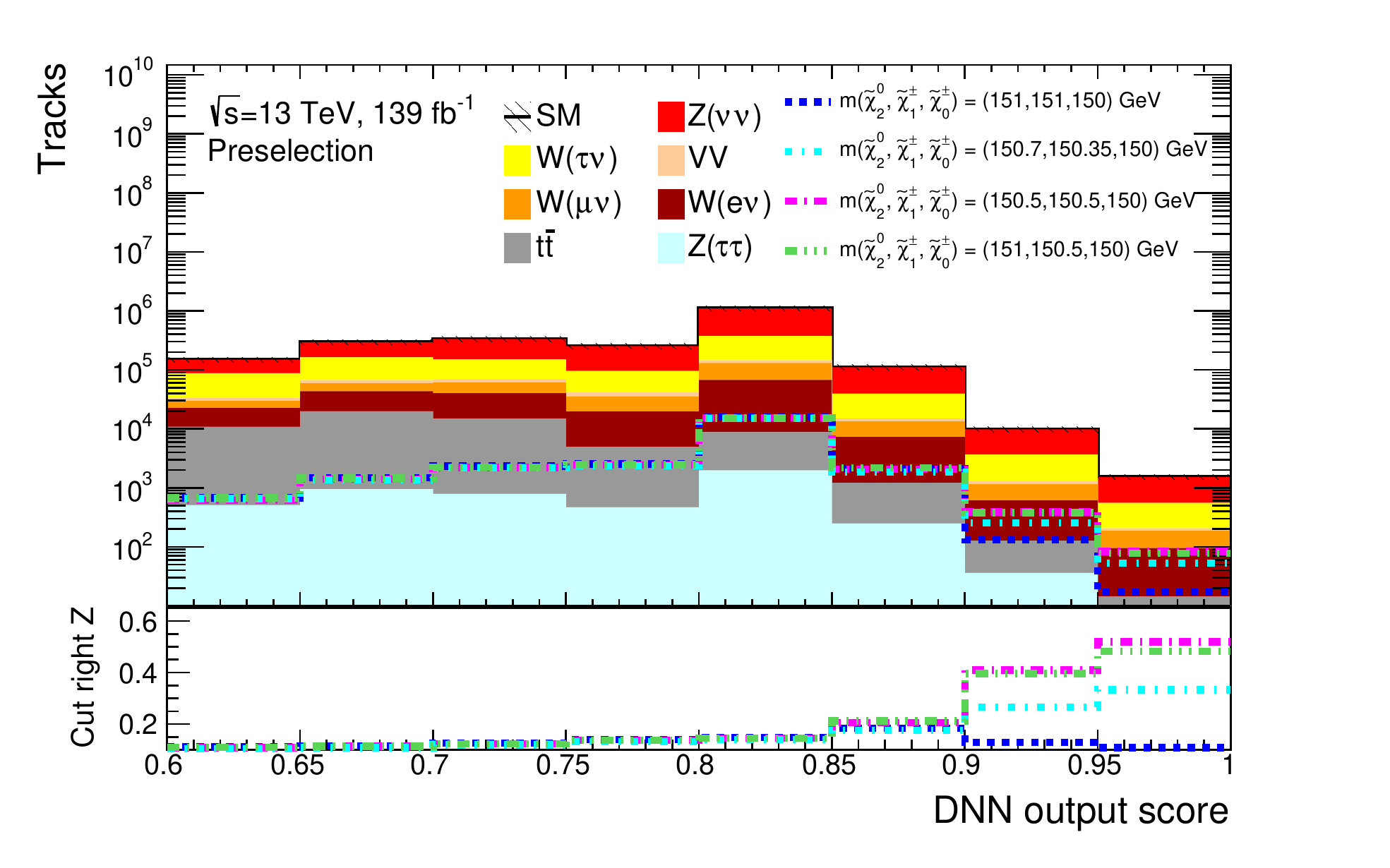}
\includegraphics[width=0.8\linewidth]{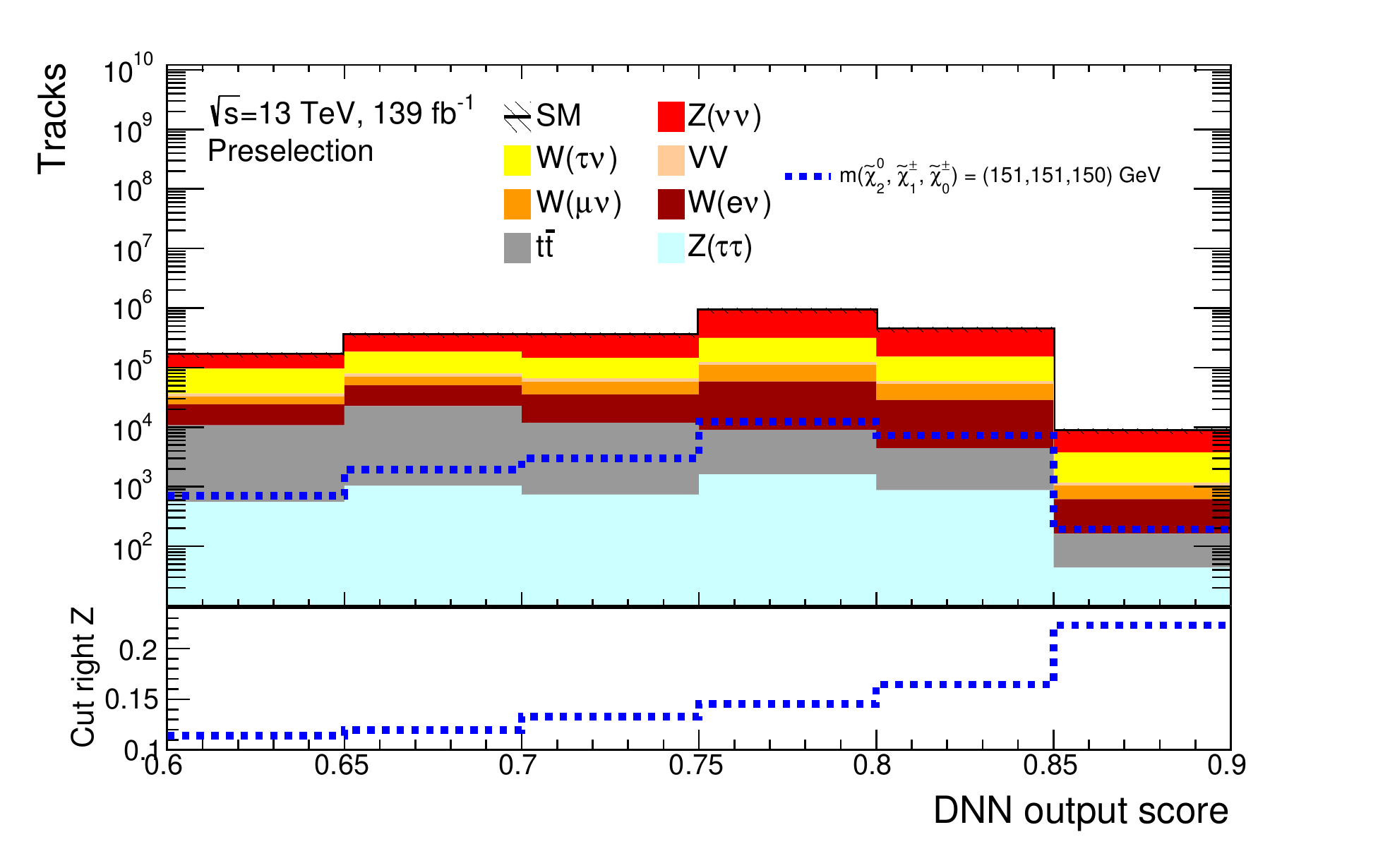}
\caption{The DNN output score for the signal and the background samples, for the training combining $m(\tilde{\chi}_{2}^{0}, \, \tilde{\chi}_{1}^{\pm}, \,  \tilde{\chi}_{1}^{0}) = (151, 150.5, 150)$ GeV, $ (150.5, 150.5, 150)$ GeV, $ (150.7, 150.35, 150)$ GeV signal samples (top) and for the training using $m(\tilde{\chi}_{2}^{0}, \, \tilde{\chi}_{1}^{\pm}, \,  \tilde{\chi}_{1}^{0}) = (151, 151, 150)$ GeV signal sample (bottom). Uncertainties include the statistical contributions and a 10\% flat uncertainty as a preliminary estimate of the systematic uncertainty.}
\label{fig:DNNscore}
\end{figure}

The DNN output score at the preselection level for the signal and the background samples is shown in Fig.~\ref{fig:DNNscore}. The plot on the top refers to the training combining $m(\tilde{\chi}_{2}^{0}, \, \tilde{\chi}_{1}^{\pm}, \,  \tilde{\chi}_{1}^{0}) = (151, 150.5, 150)$ GeV, $(150.5, 150.5, 150)$ GeV, $(150.7, 150.35, 150)$ GeV signal samples that have longer decay lengths. The $m(\tilde{\chi}_{2}^{0}, \, \tilde{\chi}_{1}^{\pm}, \,  \tilde{\chi}_{1}^{0}) = (151, 151, 150)$ GeV signal sample is only tested here, but not trained over. The plot on the bottom refers to the training using $m(\tilde{\chi}_{2}^{0}, \, \tilde{\chi}_{1}^{\pm}, \,  \tilde{\chi}_{1}^{0}) = (151, 151, 150)$ GeV signal sample that has a shorter decay length. 
The dedicated training using only $m(\tilde{\chi}_{2}^{0}, \, \tilde{\chi}_{1}^{\pm}, \,  \tilde{\chi}_{1}^{0}) = (151, 151, 150)$ GeV is found to give better performance than a training combining all of the four signal samples, or using the three signal samples that have longer decay lengths for training and using this signal sample for evaluating the performance. In general, it is very hard to separate the signals from the backgrounds, especially for the $m(\tilde{\chi}_{2}^{0}, \, \tilde{\chi}_{1}^{\pm}, \,  \tilde{\chi}_{1}^{0}) = (151, 151, 150)$ signal sample which has a $S(d_{0})$ distribution more background-like than the other signal samples.

As it is possible to see, we gain signal sensitivity at higher DNN output scores, where the DNN has identified the events to be more signal-like.

\section{SR definition}
The definition of the SR proceeds by first applying two baseline cuts, then cuts on $E_{\mathrm{T}}^{\mathrm{miss}}$ GeV and DNN output score.

\subsection{Baseline cuts}
Considering the significance as a function of the track $p_{\mathrm{T}}$ distribution, we can apply a more stringent cut of track $p_{\mathrm{T}} > 1.5$ GeV, this way reducing the tracks as coming from pile-up and decays of secondary particles. Similarly, a more stringent cut of $|\Delta\phi(\mathrm{track}, {E}_{\mathrm{T}}^{\mathrm{miss}})| < 0.5$ is applied, making the signal tracks more likely to come from the decays of supersymmetric particles.

\subsection{MET and DNN output score cuts}
The SR is defined by applying two additional cuts with respect to the baseline cuts. The two cuts are $E_{\mathrm{T}}^{\mathrm{miss}} > 600$ GeV and DNN output score $>0.9$. These two cuts are designed to maximise the signal significance of the search for the three benchmark signals with a longer decay length. 
For the benchmark signal with a smaller decay length, the SR is selected with $E_{\mathrm{T}}^{\mathrm{miss}} > 600$ GeV and DNN output score $>0.85$, where the DNN output score refers to its dedicated training.\\

The significance in the SR is compared to the significance we could obtain by using the track $S(d_{0})$ instead of the DNN output score following a cut\&count approach. The usage of the $E_{\mathrm{T}}^{\mathrm{miss}} > 600$ and track $S(d_{0})$ variables was proposed by \cite{CorneringHiggsino}.
An alternative SR following this approach, SR-cut\&count, is defined by applying $E_{\mathrm{T}}^{\mathrm{miss}} > 600$ GeV and track $S(d_{0}) > 30$ for the three benchmark signals with longer decay length, and $E_{\mathrm{T}}^{\mathrm{miss}} > 600$ GeV and track $6 < S(d_{0}) < 30$ for the benchmark signal with a smaller decay length. A comparison between the significance values when using the DNN instead of the cut\&count approach is reported in Table~\ref{tab:compressedshiggsinos-ZforSR}. The cut on the DNN output score defining the SR is found to give a higher significance than the cut on track $S(d_{0})$. This is expected from the fact that the DNN output score contains more information than the track $S(d_{0})$ alone, therefore, the DNN can better discriminate the signal from the backgrounds.

\begin{table}[!htb]
\begin{center}
\begin{tabular}{l|cc}
\noalign{\smallskip}\hline\noalign{\smallskip}
$m(\tilde{\chi}_{2}^{0}, \, \tilde{\chi}_{1}^{\pm}, \,  \tilde{\chi}_{1}^{0})$ [GeV] & Z for SR-DNN & Z for SR-cut\&count \\
\noalign{\smallskip}\hline\noalign{\smallskip}
$(151, 150.5, 150)$    & $3.92$ & $3.57$ \\ 
$(150.5, 150.5, 150)$  & $3.94$ & $3.75$ \\ 
$(150.7, 150.35, 150)$ & $2.01$ & $1.87$ \\ 
$(151, 151, 150)$ & $1.76$ & $1.73$ \\ 
\noalign{\smallskip}\hline\noalign{\smallskip}
\end{tabular}
\end{center}
\caption{Comparison of the estimated significance values for the SR, selected using the DNN output score or track $S(d_{0})$. Uncertainties include the statistical contributions and a 10\% flat uncertainty as a preliminary estimate of the systematic uncertainty.}
\label{tab:compressedshiggsinos-ZforSR}
\end{table}

\section{Background estimation}
The main backgrounds of the search, $Z(\nu\nu)$+jets and $W(\ell\nu)$+jets, are generally hard to reconstruct at the LHC due to the presence of invisible particles in the final states. Instead of relying on a single CR, one can estimate the background contributions by comparing the MC predictions to data in a set of specific CRs, all in a phase space close to the SR and normalize to data the MC predictions in the SR. This strategy, called the ABCD method, consists in the definition of a two-dimensional plane generated by two uncorrelated variables and divided into four regions: a SR to collect most of the signal events, and three CRs which, ideally, shall be rich in events produced from the background processes that we are trying to estimate with the method. The goal of the ABCD method is to produce a prediction for the number of non-signal events in the SRs, starting from the measurements of the background rates in CRs.

\subsection{Control Regions}
\label{sec:compressedshiggsinos-CRs}
\begin{figure}[!htb]
\centering
\includegraphics[width=0.6\linewidth]{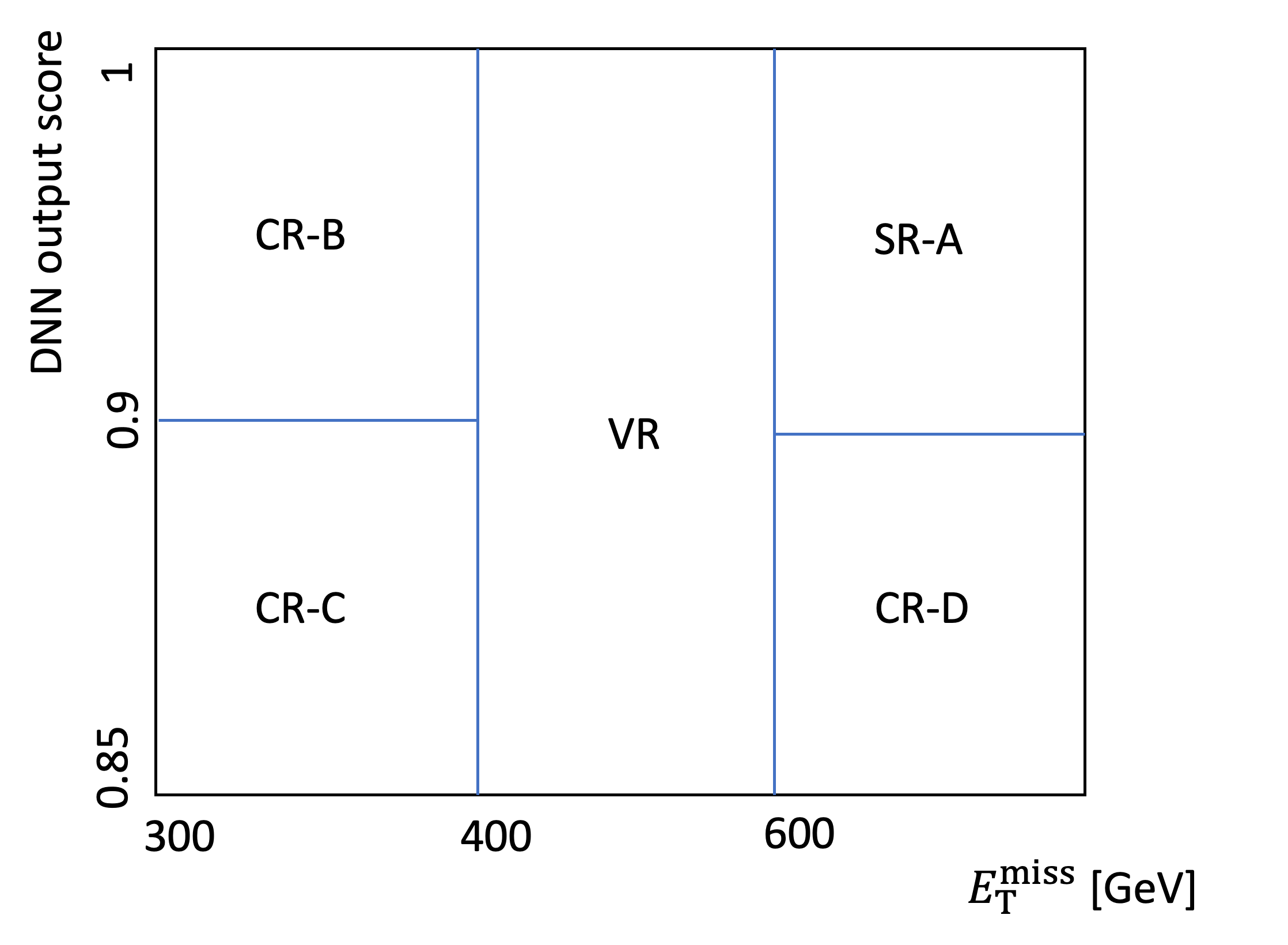}
\caption{The ABCD background estimation strategy. The $E_{\mathrm{T}}^{\mathrm{miss}}$ and DNN output score cuts refer to the selections for the signals having longer decay lengths.}
\label{fig:Higgsino_ABCD}
\end{figure}

The ABCD plane is constructed considering the uncorrelated $E_{\mathrm{T}}^{\mathrm{miss}}$ and DNN output score variables, as shown Fig.~\ref{fig:Higgsino_ABCD}. The cuts on $E_{\mathrm{T}}^{\mathrm{miss}}$ and DNN output score variables define a SR, called SR-A, and three CRs, called CR-B, CR-C and CR-D. In Fig.~\ref{fig:Higgsino_ABCD}, the cuts refer to the strategy for the three signal benchmark samples having longer decay lengths.

Due to the different shapes of the DNN output score between the training using signal samples with longer decay lengths and the training using a signal sample with a shorter decay length, two different analysis selections are adopted. The cuts defining the two strategies are similar, the difference is in the DNN output selection which has been modified for the training using a signal sample with a shorter decay length to improve the significance and the statistics in SR-A. The cuts are reported in Table~\ref{tab:compressedshiggsinos-AnalysisRegions}.

\begin{table}[!htb]
\begin{center}
\begin{tabular}{l|cc}
\noalign{\smallskip}\hline\noalign{\smallskip}
\multicolumn{3}{c}{Longer decay length}\\
\noalign{\smallskip}\hline\noalign{\smallskip}
Region    &  $E_{\mathrm{T}}^{\mathrm{miss}}$ [GeV] & DNN output score \\ 
\noalign{\smallskip}\hline\noalign{\smallskip}
SR-A    & > 600               & > 0.9 \\ 
CR-B    &  $\in$ [300, 400]   & > 0.9 \\
CR-C    &  $\in$ [300, 400]   & $\in$ [0.85, 0.9] \\
CR-D    &  > 600              & $\in$ [0.85, 0.9] \\
\noalign{\smallskip}\hline\noalign{\smallskip}
\multicolumn{3}{c}{Shorter decay length}\\
\noalign{\smallskip}\hline\noalign{\smallskip}
Region    &  $E_{\mathrm{T}}^{\mathrm{miss}}$ [GeV] & DNN output score \\ 
\noalign{\smallskip}\hline\noalign{\smallskip}
SR-A*    & > 600               & > 0.85 \\ 
CR-B*    & $\in$ [300, 400]    & > 0.85 \\
CR-C*    & $\in$ [300, 400]    & $\in$ [0.8, 0.85]\\
CR-D*    & > 600               & $\in$ [0.8, 0.85]\\
\noalign{\smallskip}\hline\noalign{\smallskip}
\end{tabular}
\end{center}
\caption{Definition of the analysis regions for the two different strategies, the one training over signal samples with longer decay lengths and the one training over a signal sample with a shorter decay length.}
\label{tab:compressedshiggsinos-AnalysisRegions}
\end{table}

These cuts have been chosen in order to minimise the signal contamination in the three CRs while keeping them as close as possible to the SRs. The estimated backgrounds via ABCD are the $Z(\nu\nu)$+jets and $W(e\nu)$+jets, $W(\mu\nu)$+jets, $W(\tau\nu)$+jets simultaneously. It is possible to estimate the background composition in SR-A, $N_{\mathrm{bkg}}^{\mathrm{A}}$, as
\begin{equation}
    N_{\mathrm{bkg}}^{\mathrm{A}} = \frac{N_{\mathrm{bkg}}^{\mathrm{B}} \cdot N_{\mathrm{bkg}}^{\mathrm{D}}}{N_{\mathrm{bkg}}^{\mathrm{C}}},
\label{Eq:ABCD-method}
\end{equation}
assuming that the two variables used for defining the analysis regions are not correlated.
The fit evolves by normalising MC predicted events in CRs to data, so that for each CR there is a scale factor $\mu$ applied to the backgrounds we want to estimate, with $\mu^{\mathrm{CR}} = N_{\mathrm{bkg-postfit}}^{\mathrm{CR}} / N_{\mathrm{bkg-prefit}}^{\mathrm{CR}}$. Therefore, Eq.~(\ref{Eq:ABCD-method}) can be rewritten in terms of the scale factors, with the scale factor to be applied in SR-A, $\mu^{\mathrm{A}}$, estimated as
\begin{equation}
    \mu^{\mathrm{A}} = \frac{\mu^{\mathrm{B}} \cdot \mu^{\mathrm{D}}}{\mu^{\mathrm{C}}}.
\end{equation}

\begin{figure}[!p]
\centering
\includegraphics[width=0.49\linewidth]{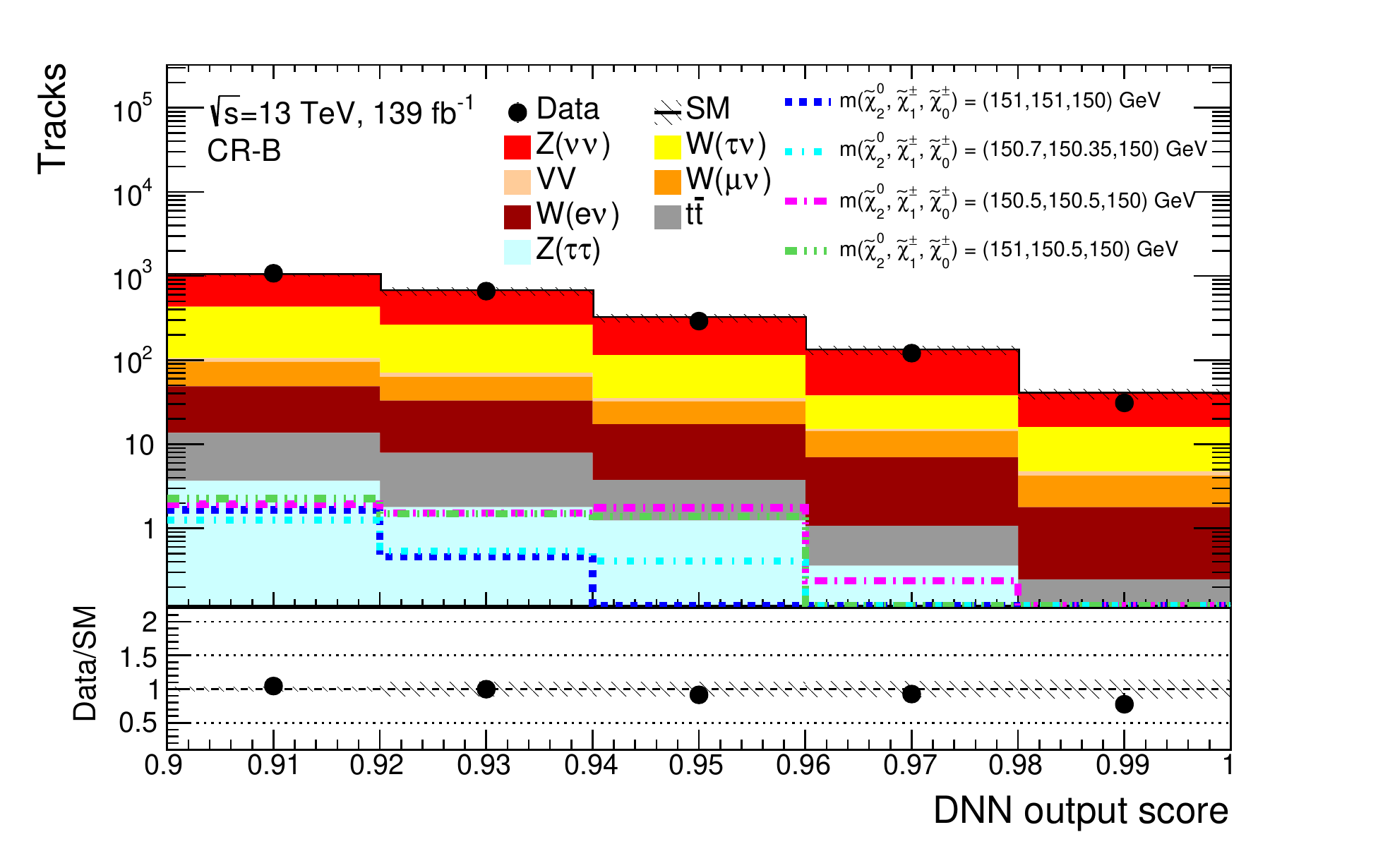}
\includegraphics[width=0.49\linewidth]{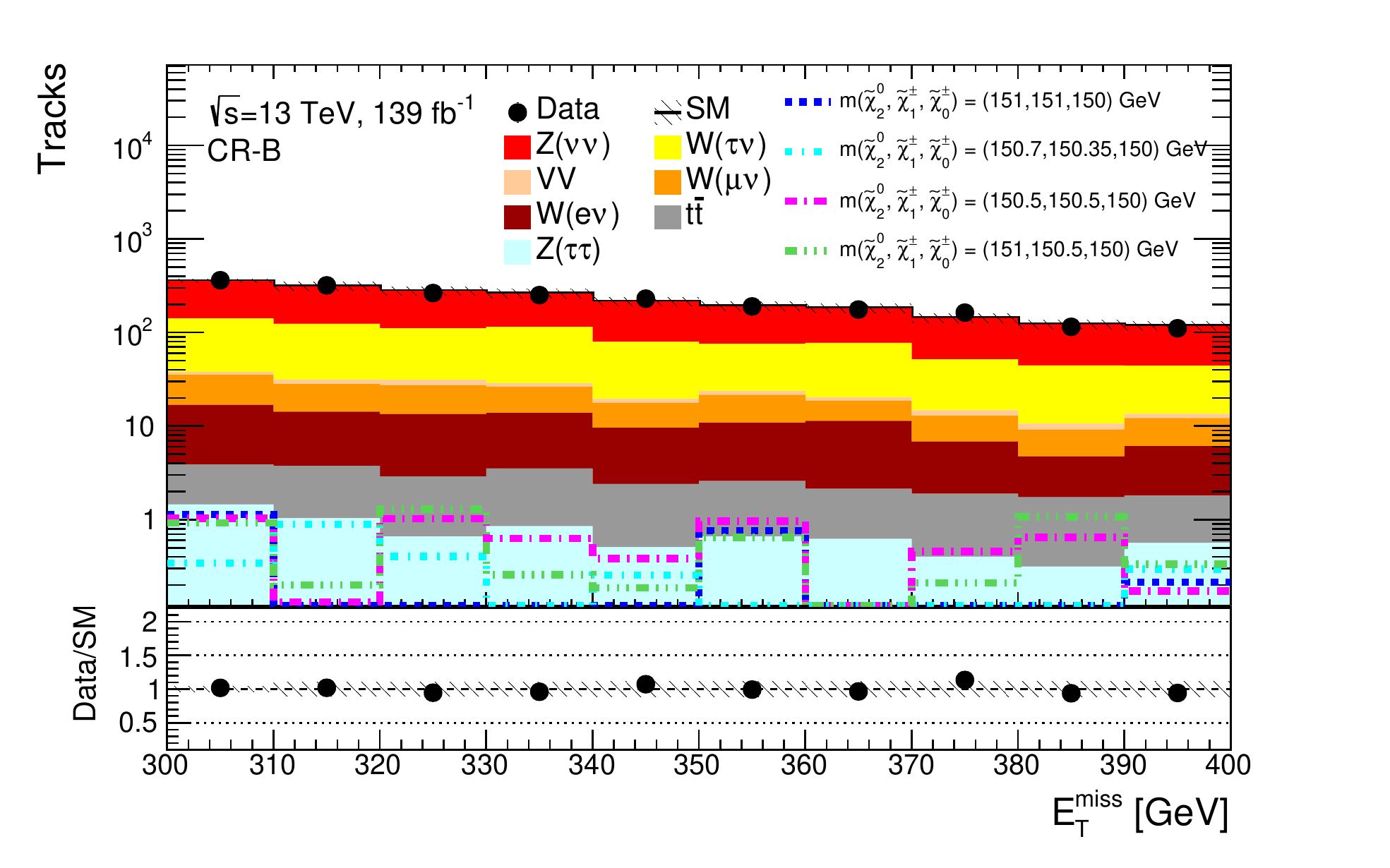}
\includegraphics[width=0.49\linewidth]{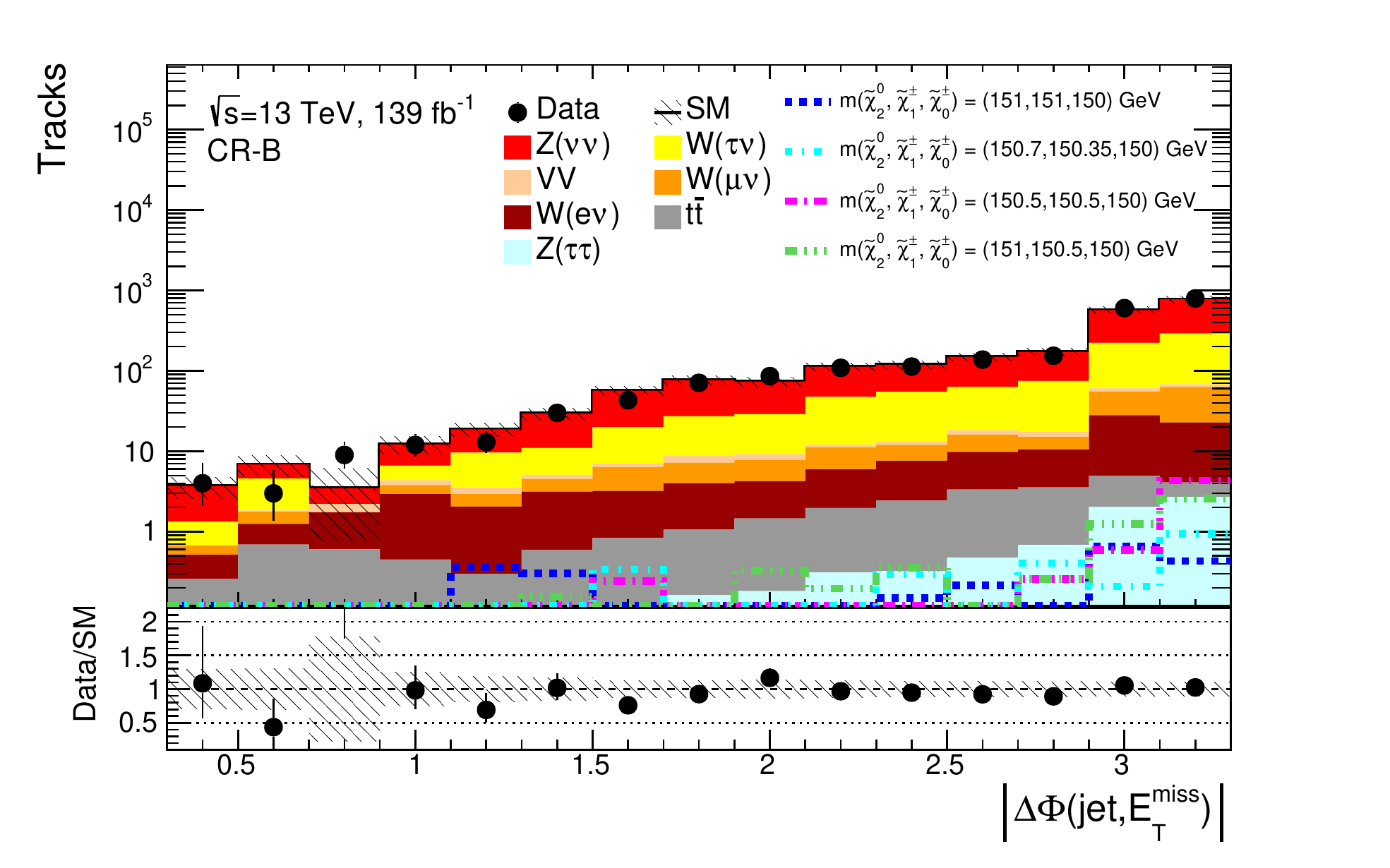}
\includegraphics[width=0.49\linewidth]{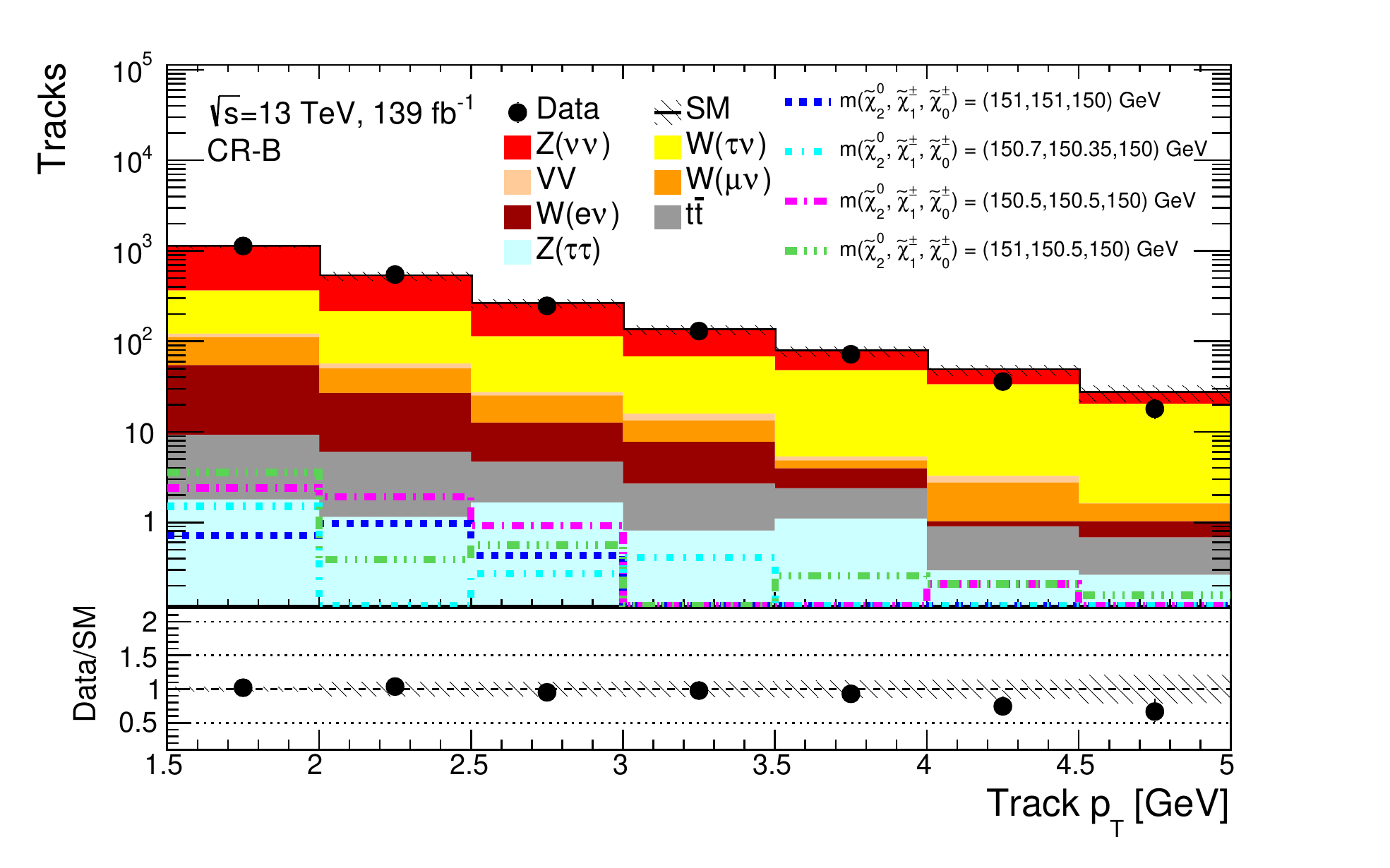}
\includegraphics[width=0.49\linewidth]{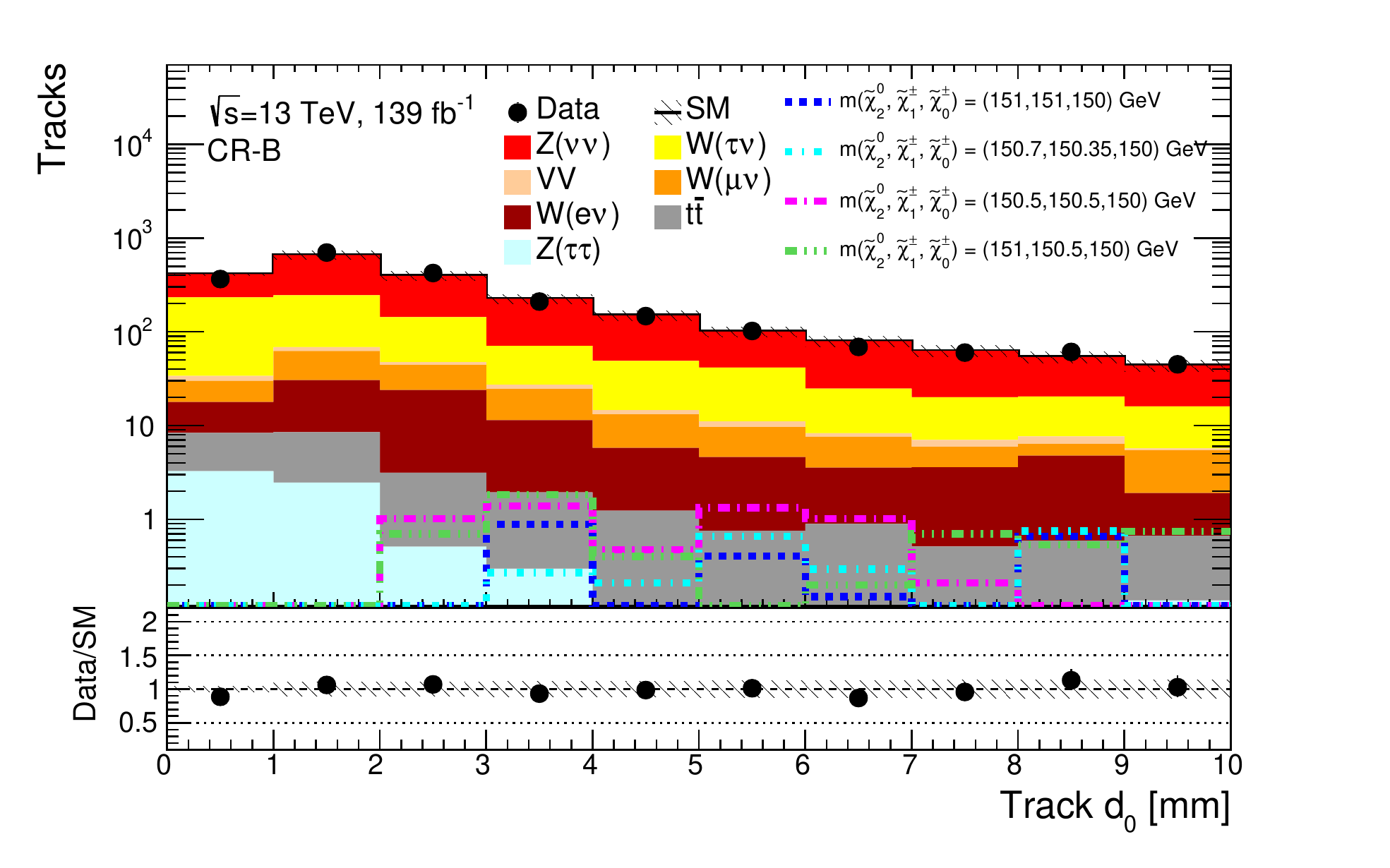}
\includegraphics[width=0.49\linewidth]{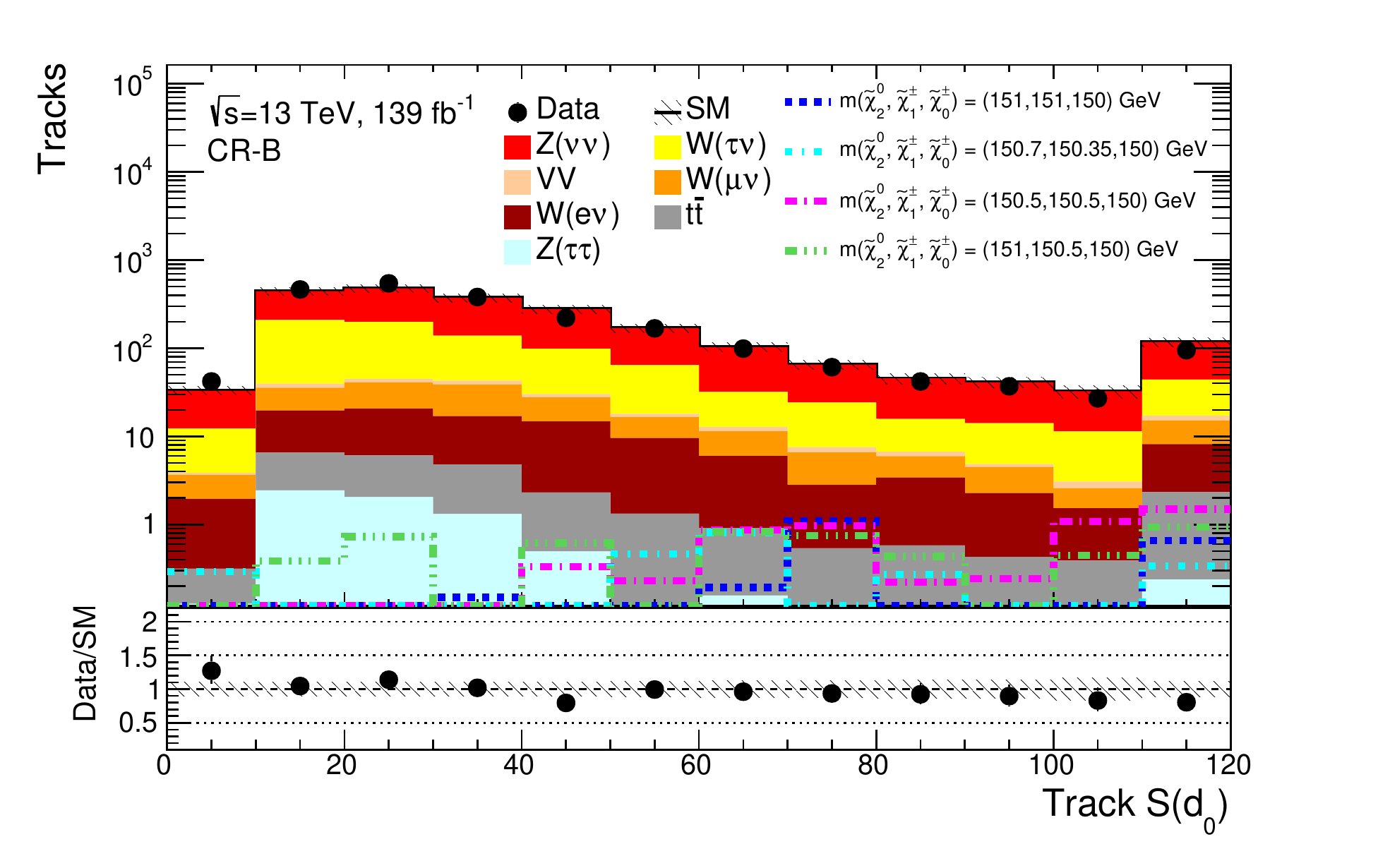}
\includegraphics[width=0.49\linewidth]{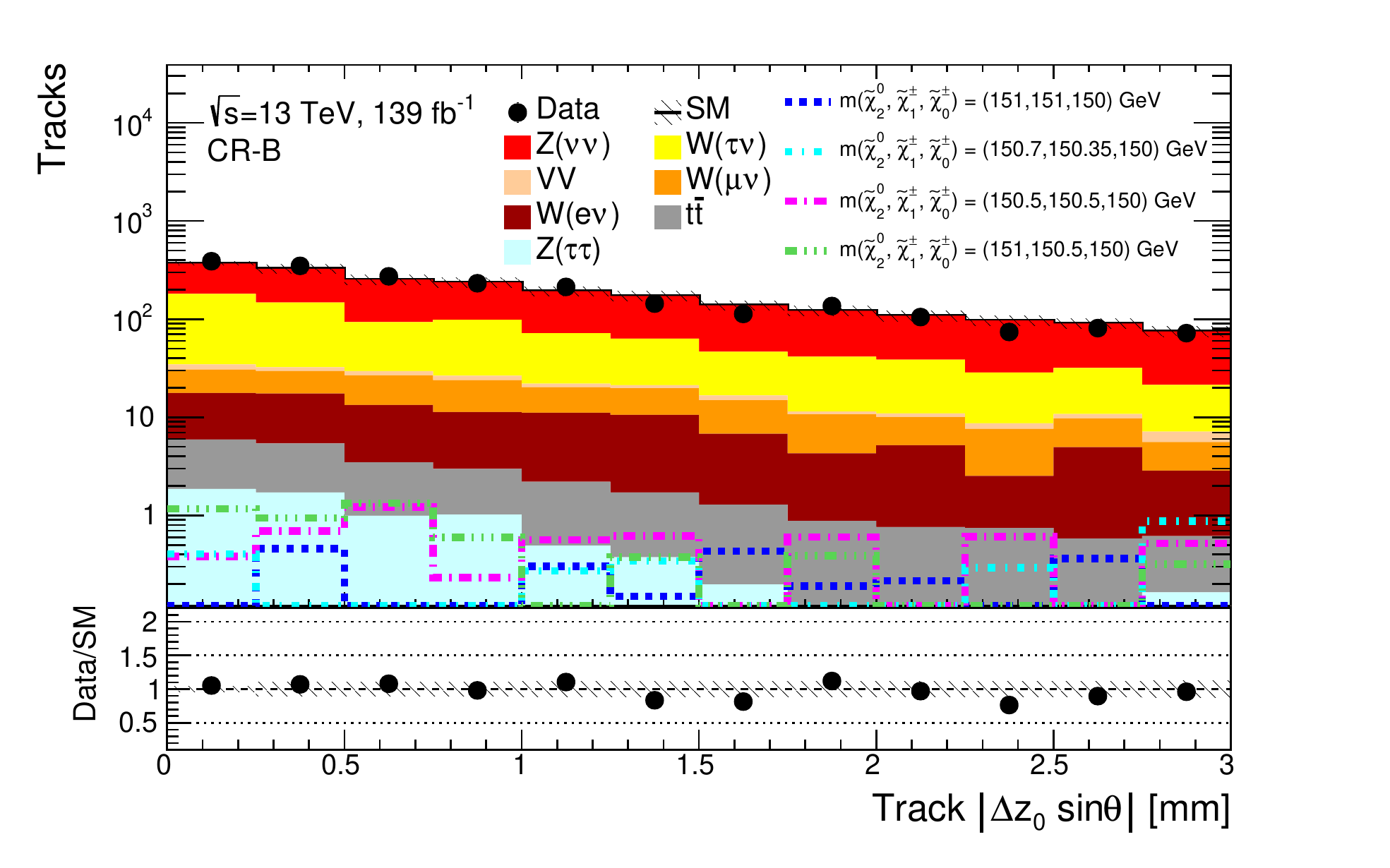}
\includegraphics[width=0.49\linewidth]{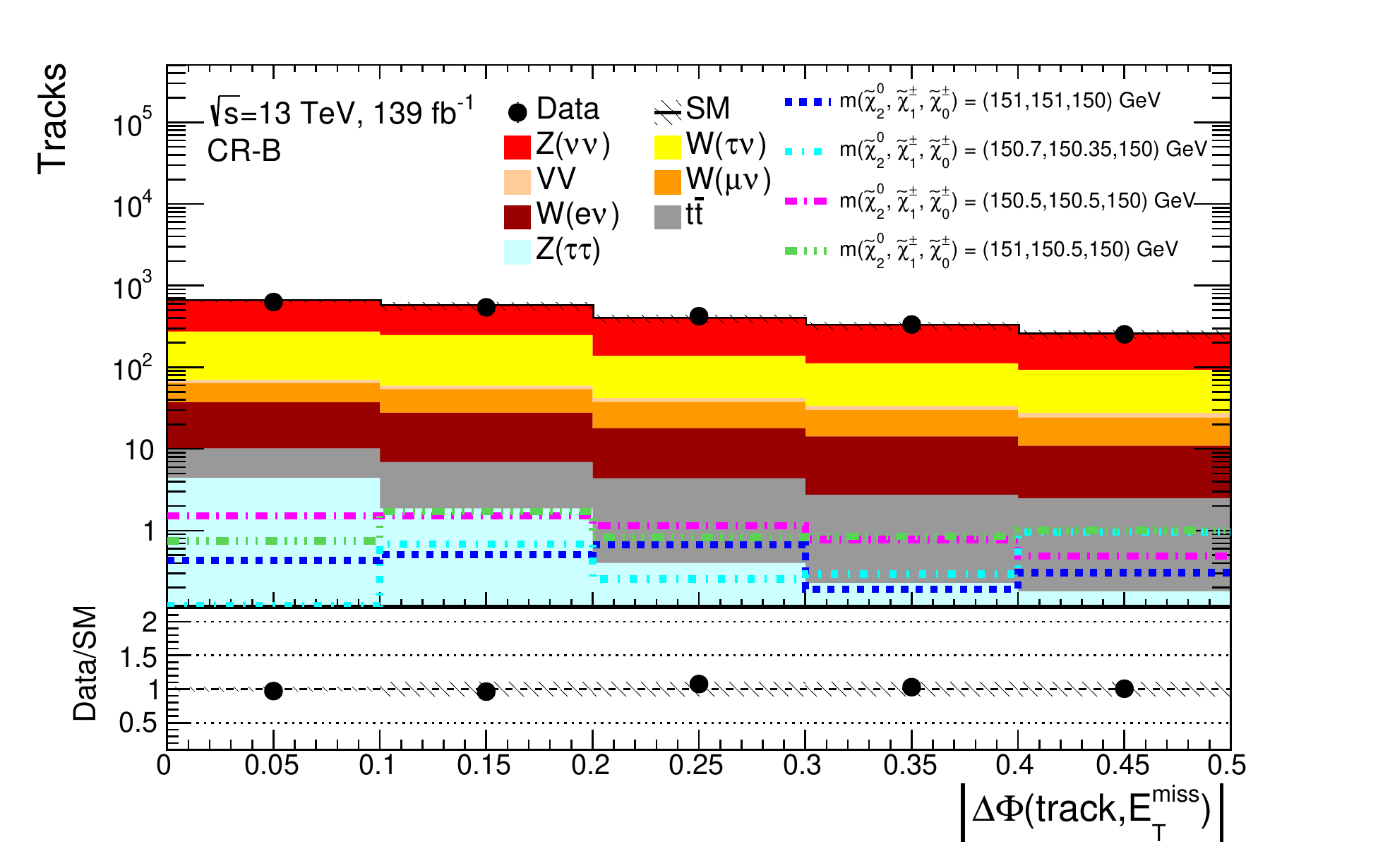}
\caption{The data and MC distributions of the variables used in the search in CR-B. Uncertainties include the statistical contributions and a 10\% flat uncertainty as a preliminary estimate of the systematic uncertainty.}
\label{fig:Higgsino_CRB}
\end{figure}

\begin{figure}[!p]
\centering
\includegraphics[width=0.49\linewidth]{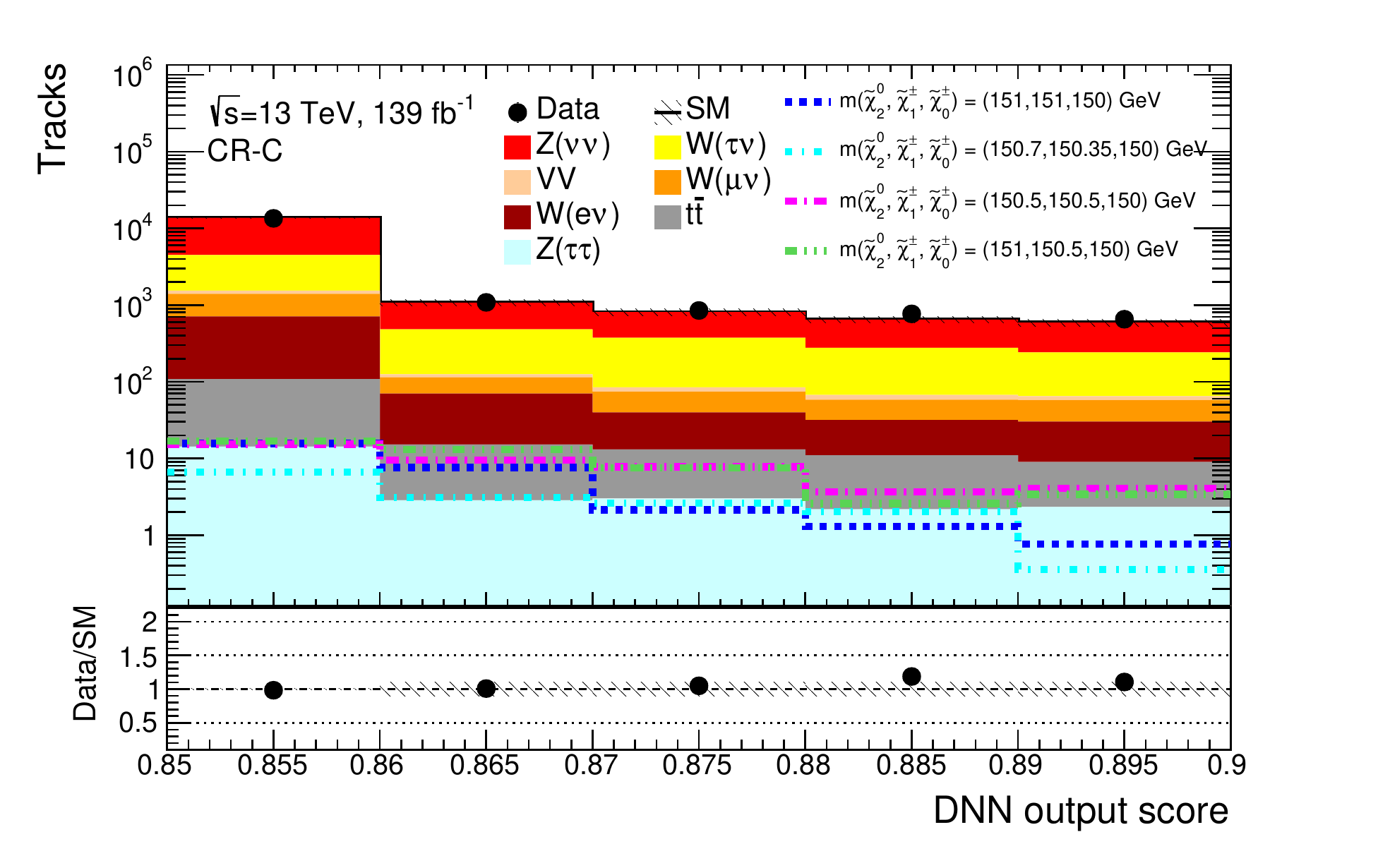}
\includegraphics[width=0.49\linewidth]{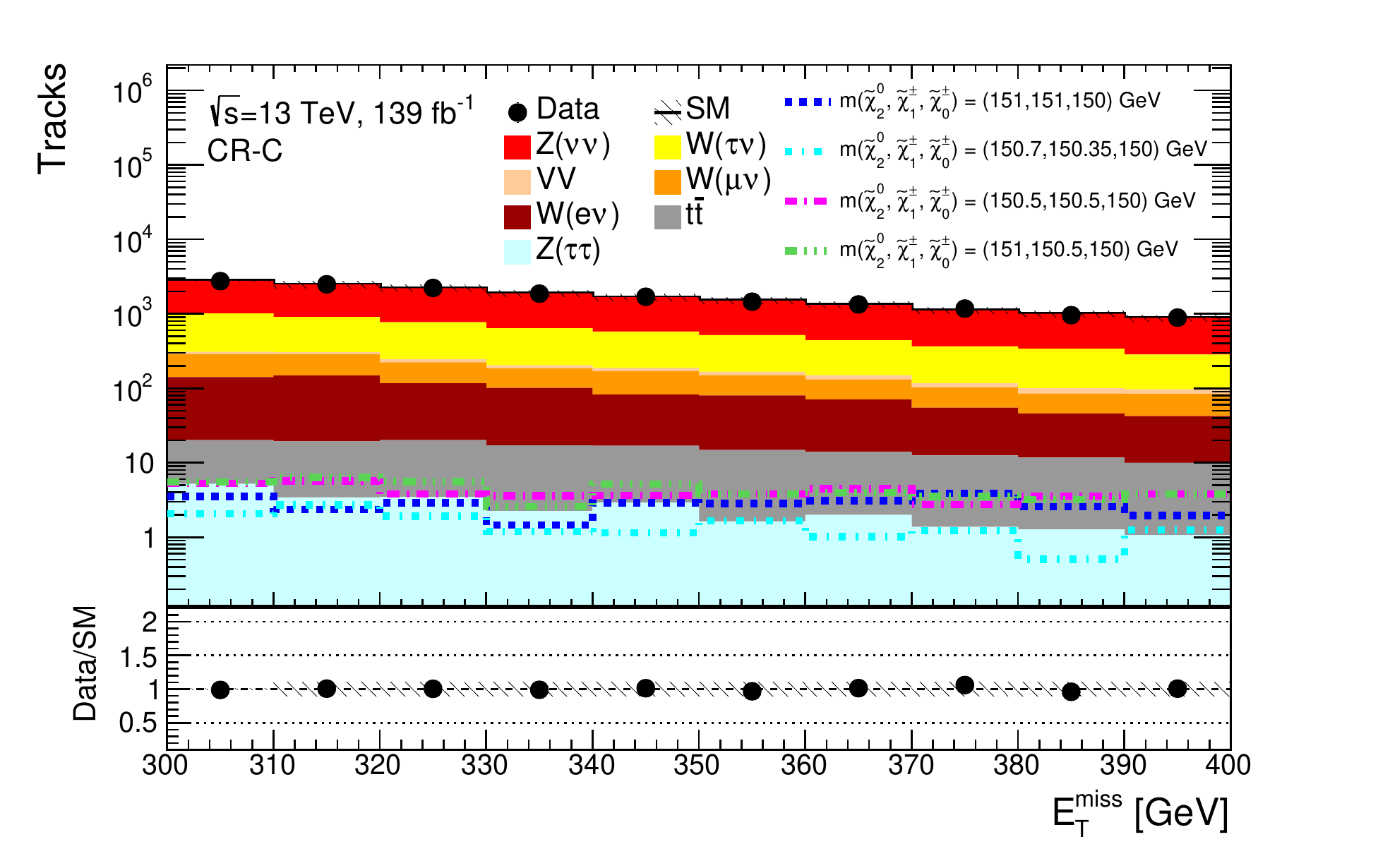}
\includegraphics[width=0.49\linewidth]{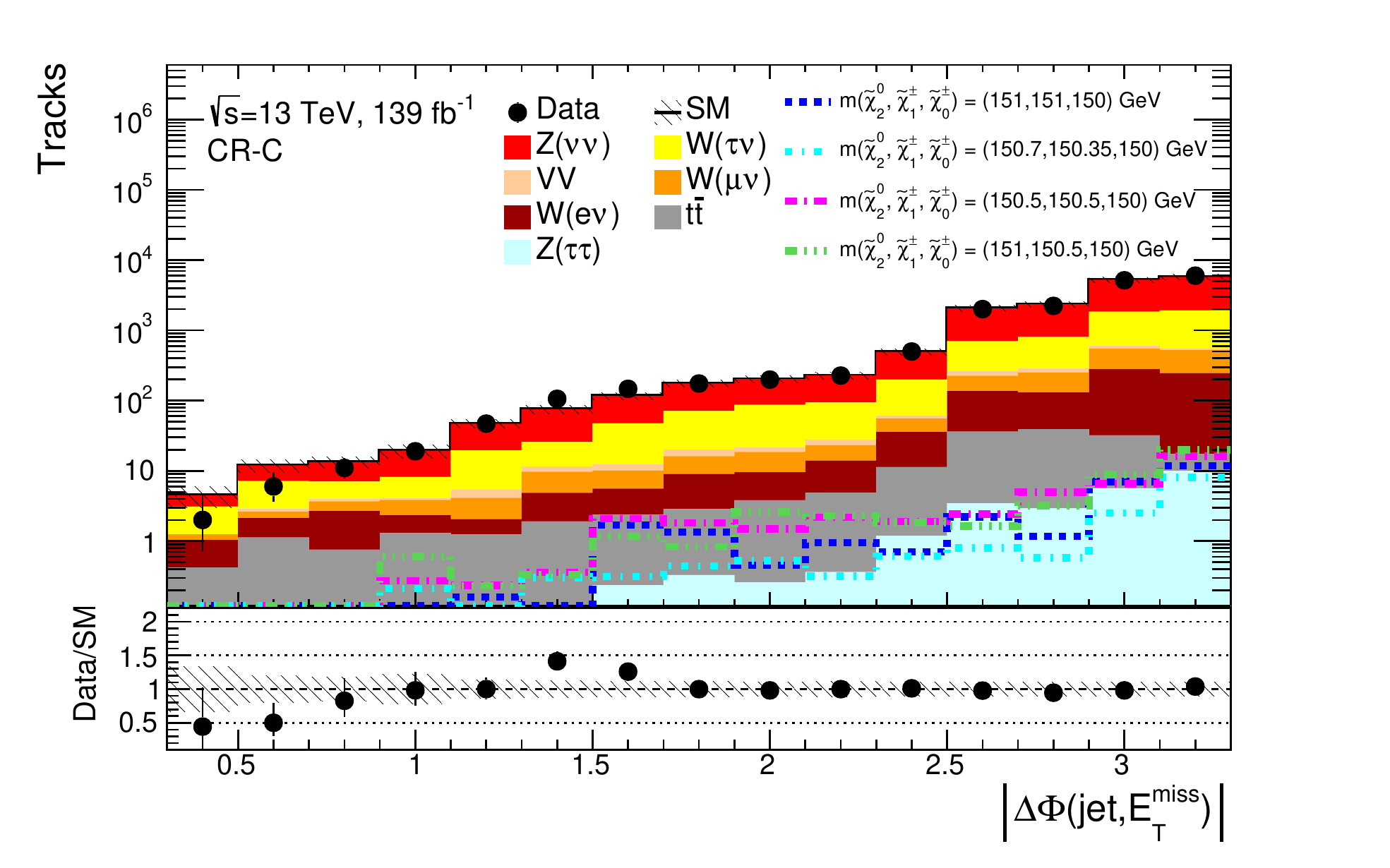}
\includegraphics[width=0.49\linewidth]{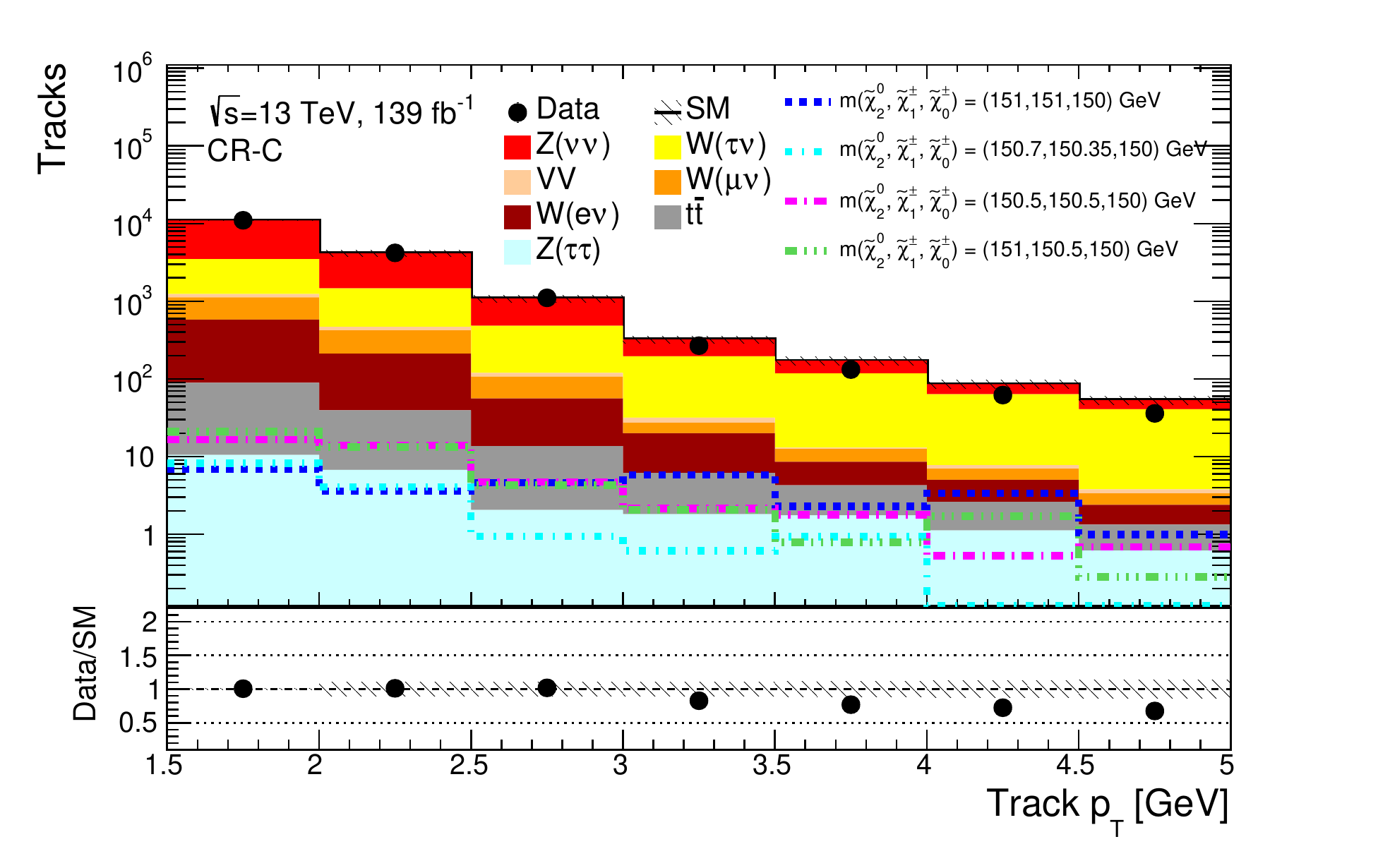}
\includegraphics[width=0.49\linewidth]{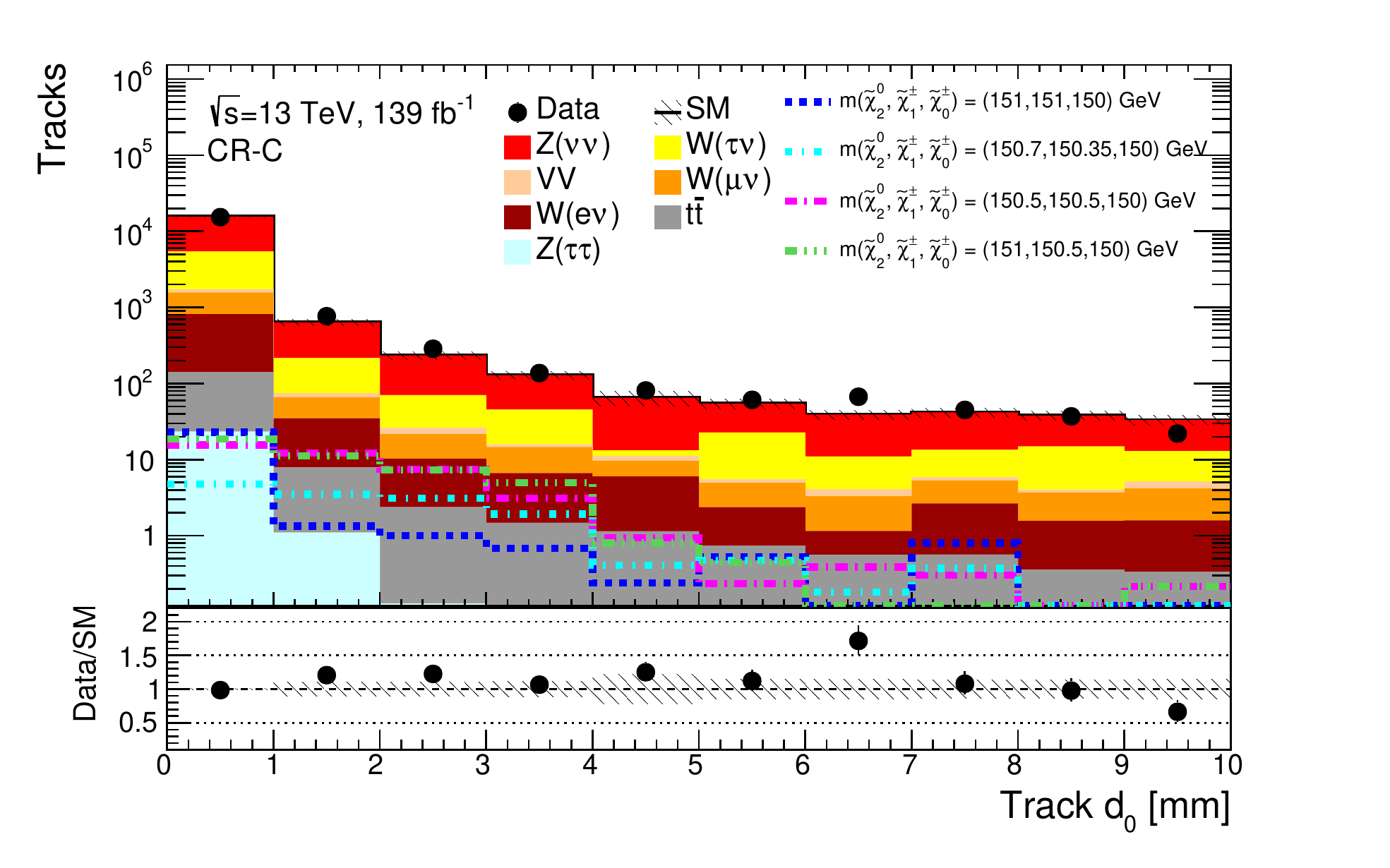}
\includegraphics[width=0.49\linewidth]{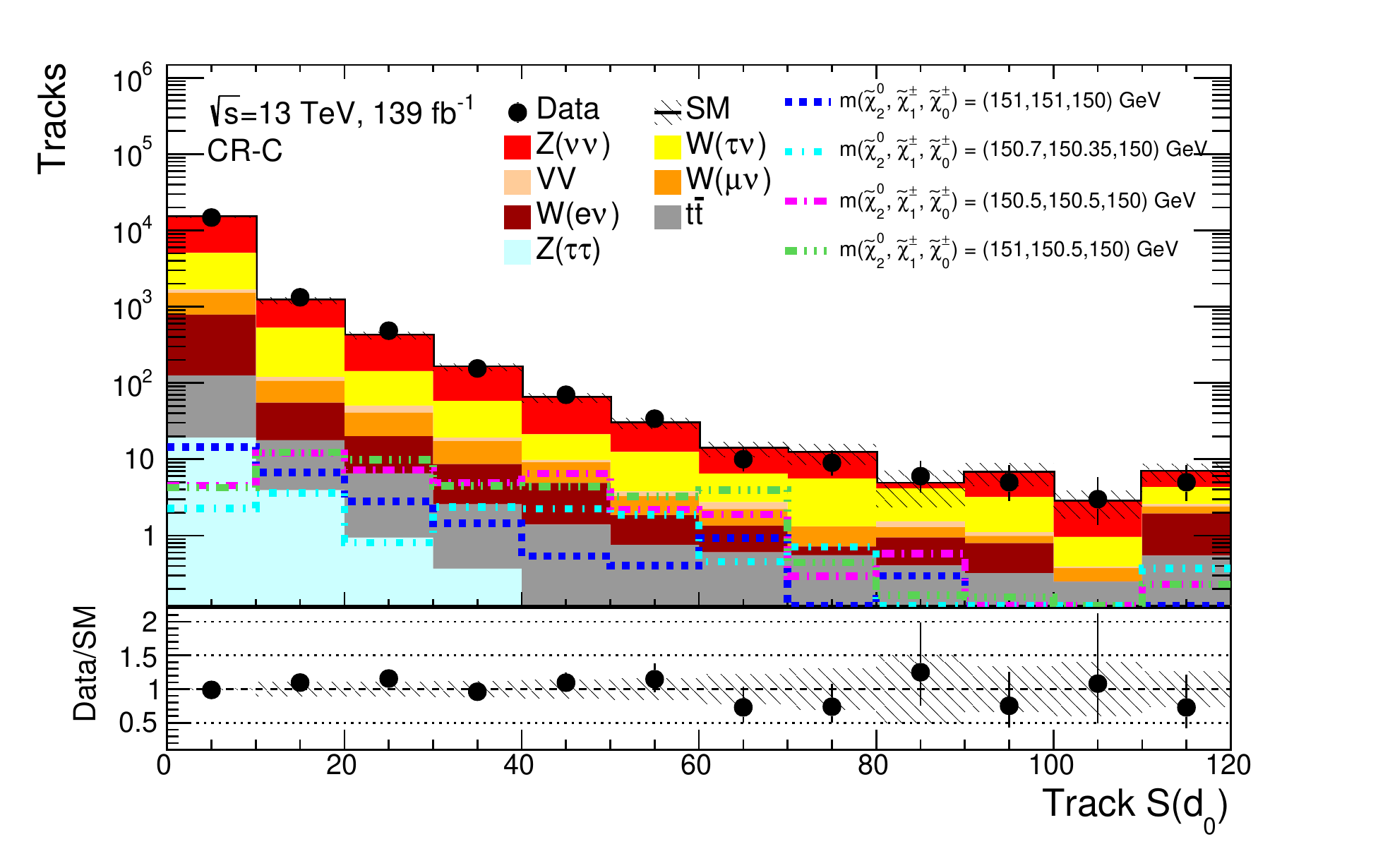}
\includegraphics[width=0.49\linewidth]{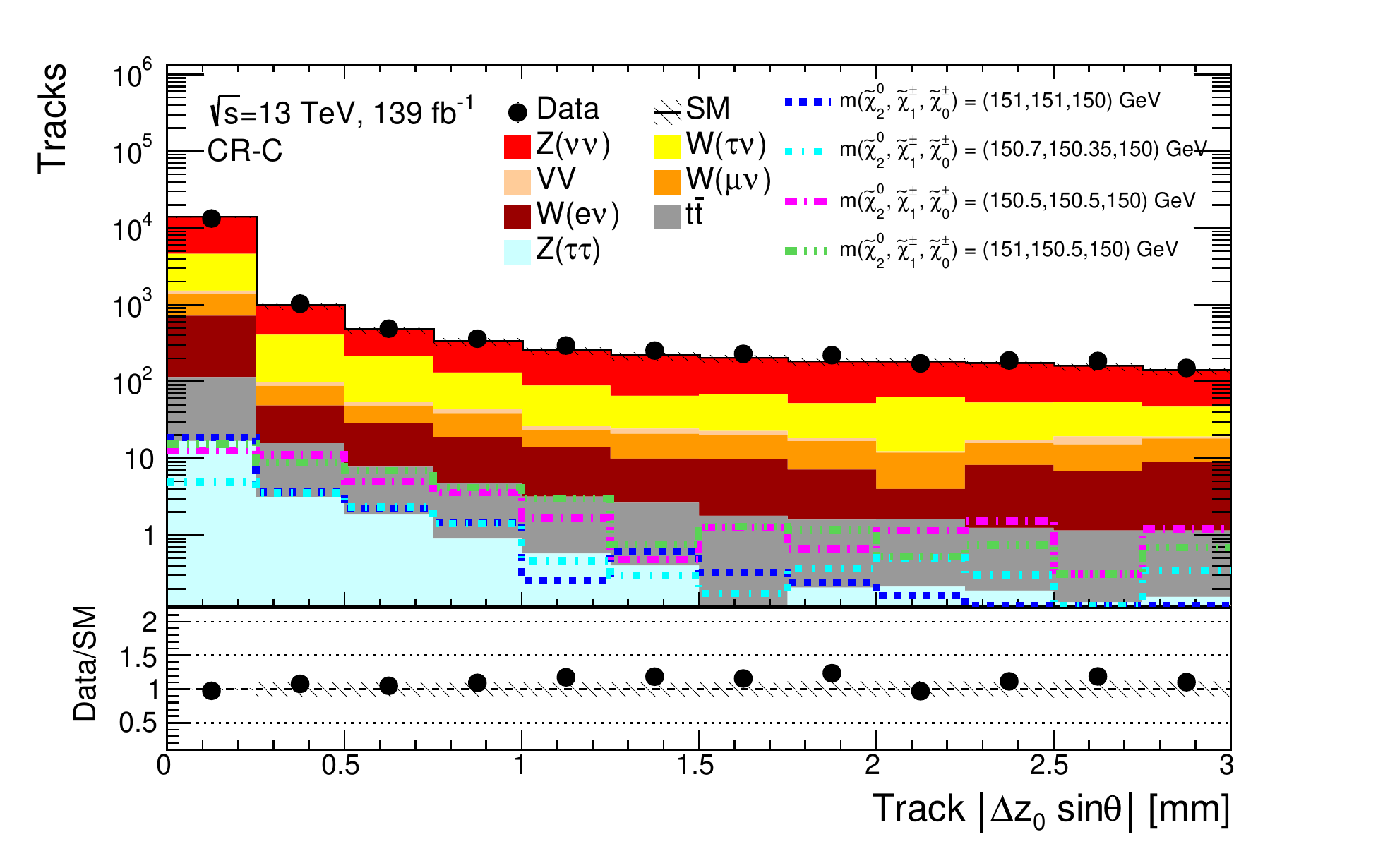}
\includegraphics[width=0.49\linewidth]{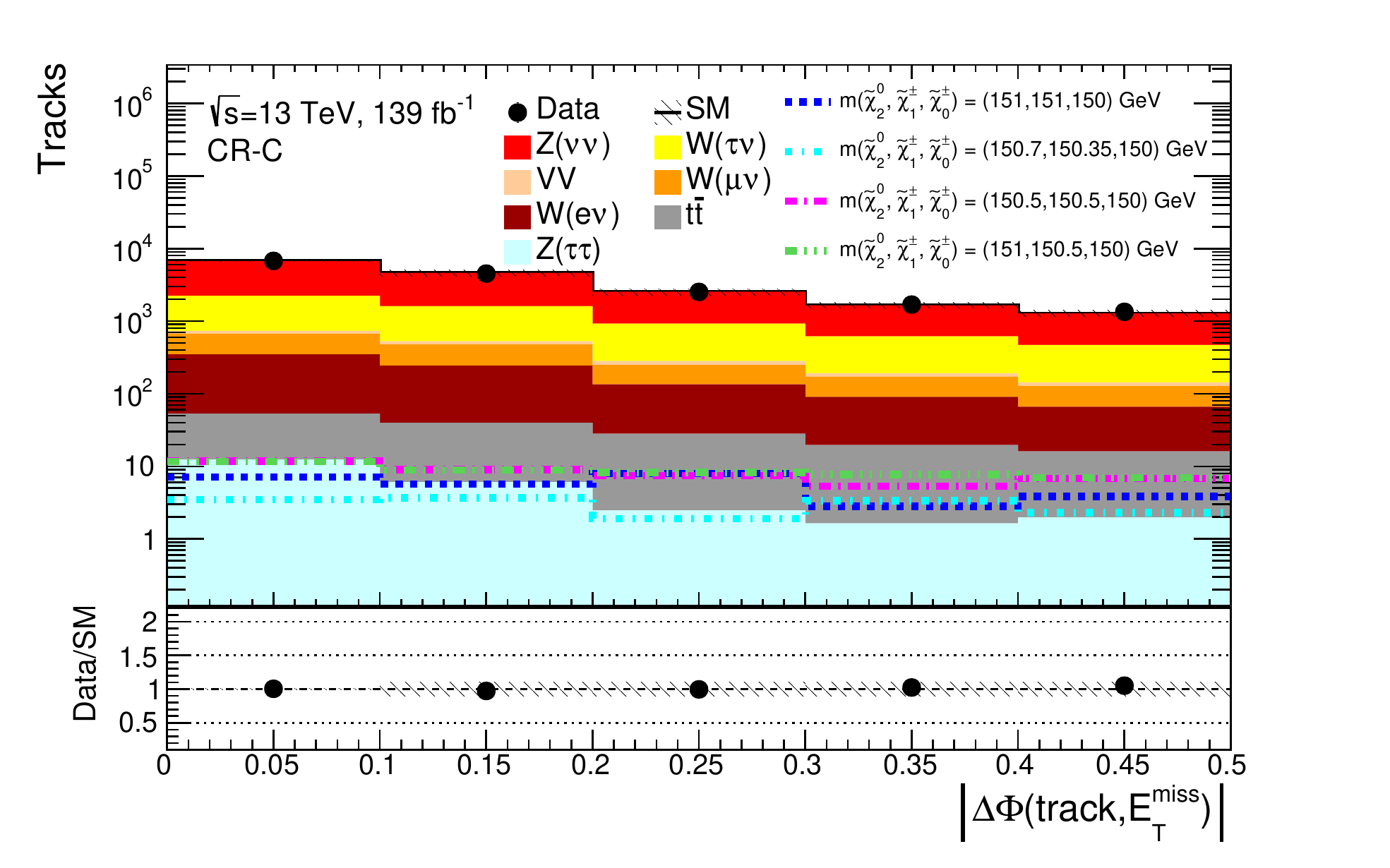}
\caption{The data and MC distributions of the variables used in the search in CR-C. Uncertainties include the statistical contributions and a 10\% flat uncertainty as a preliminary estimate of the systematic uncertainty.}
\label{fig:Higgsino_CRC}
\end{figure}

\begin{figure}[!p]
\centering
\includegraphics[width=0.49\linewidth]{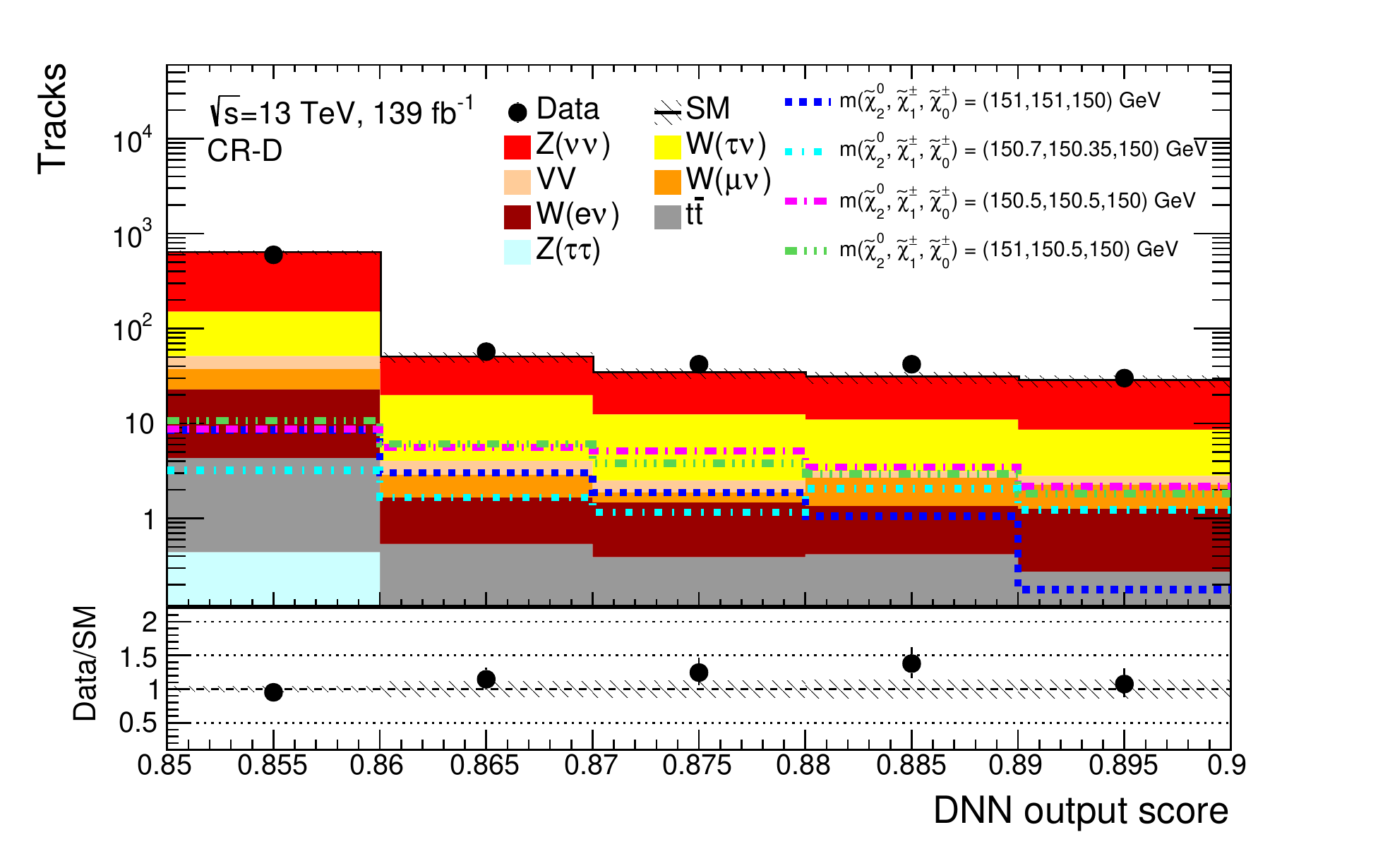}
\includegraphics[width=0.49\linewidth]{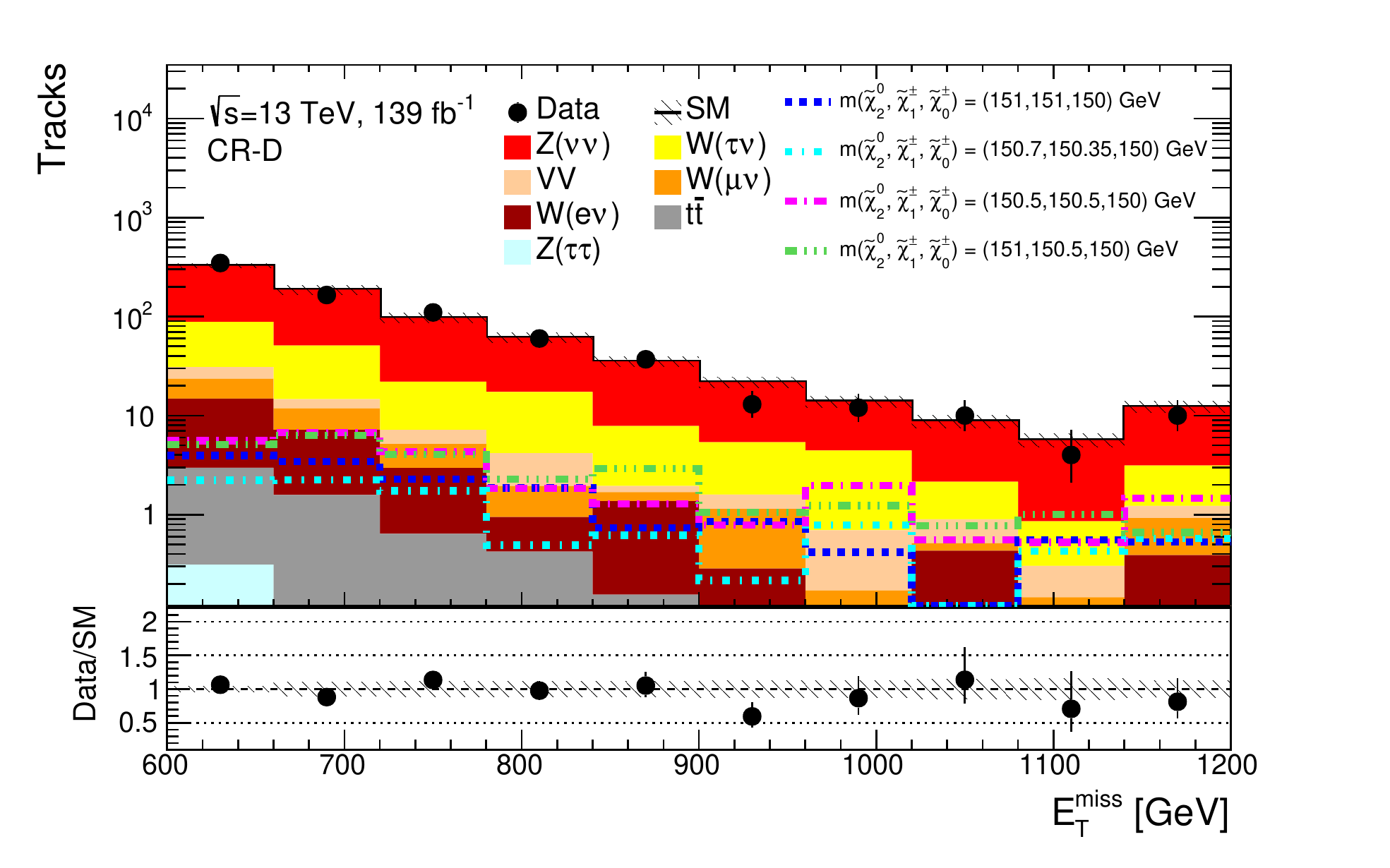}
\includegraphics[width=0.49\linewidth]{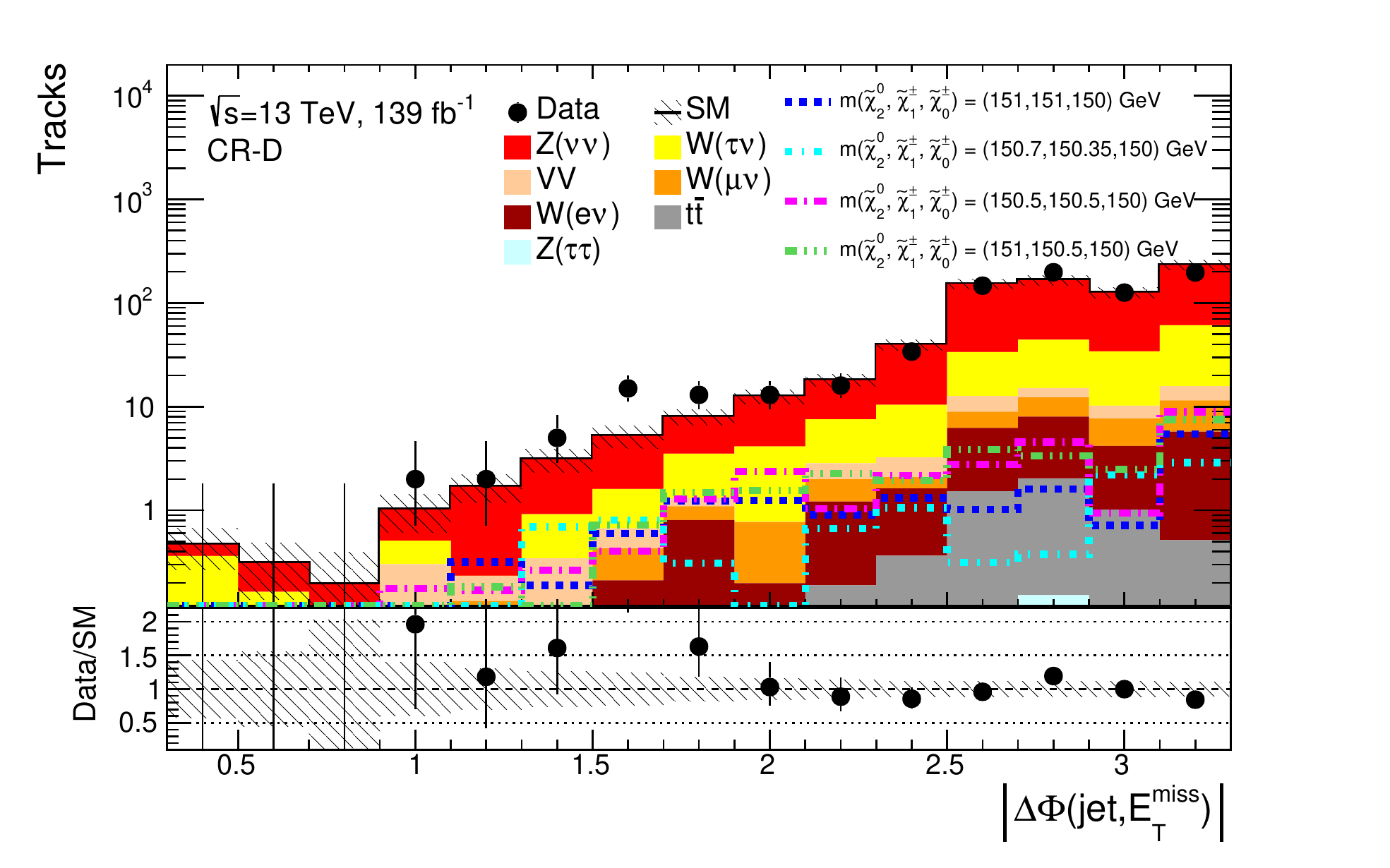}
\includegraphics[width=0.49\linewidth]{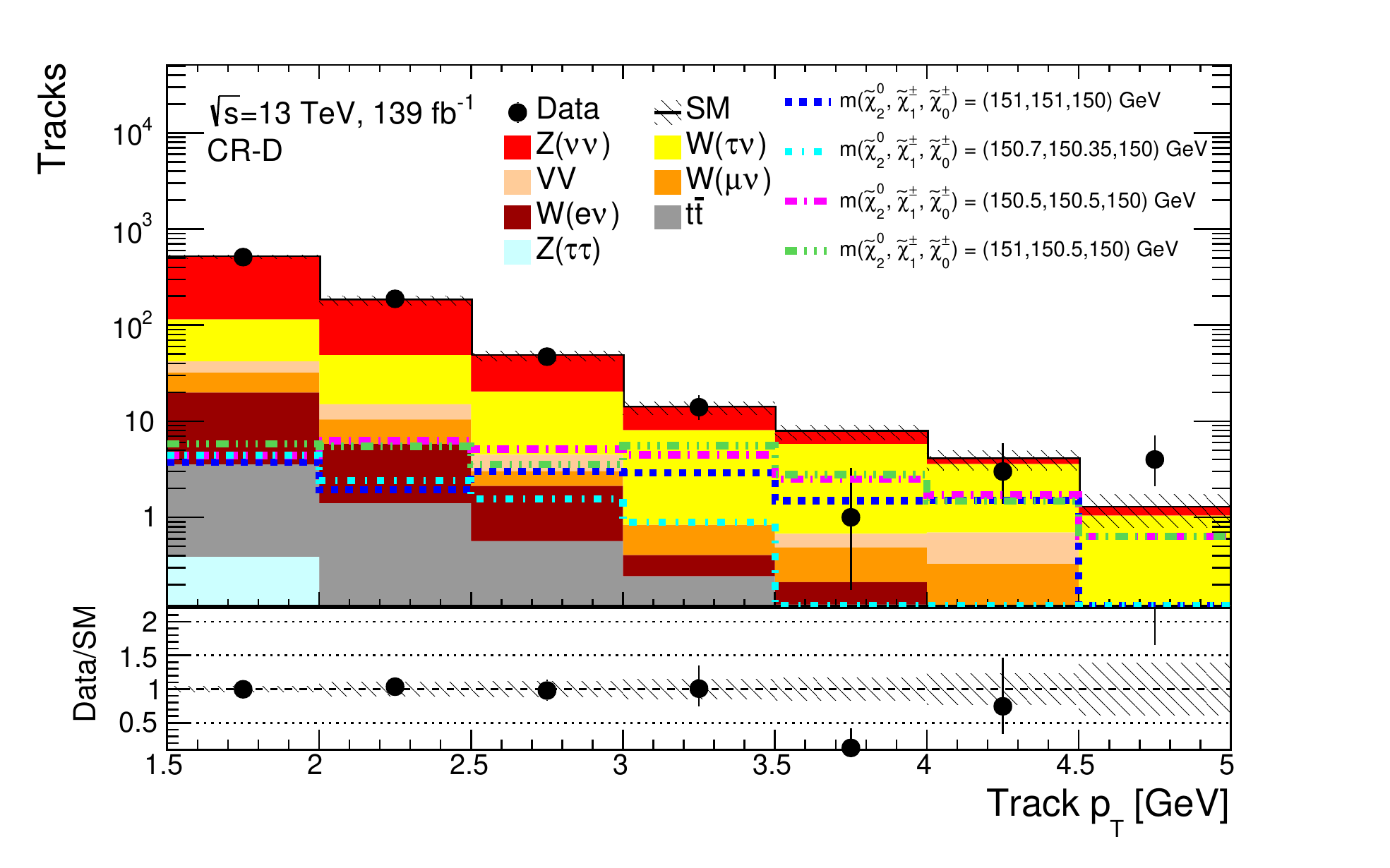}
\includegraphics[width=0.49\linewidth]{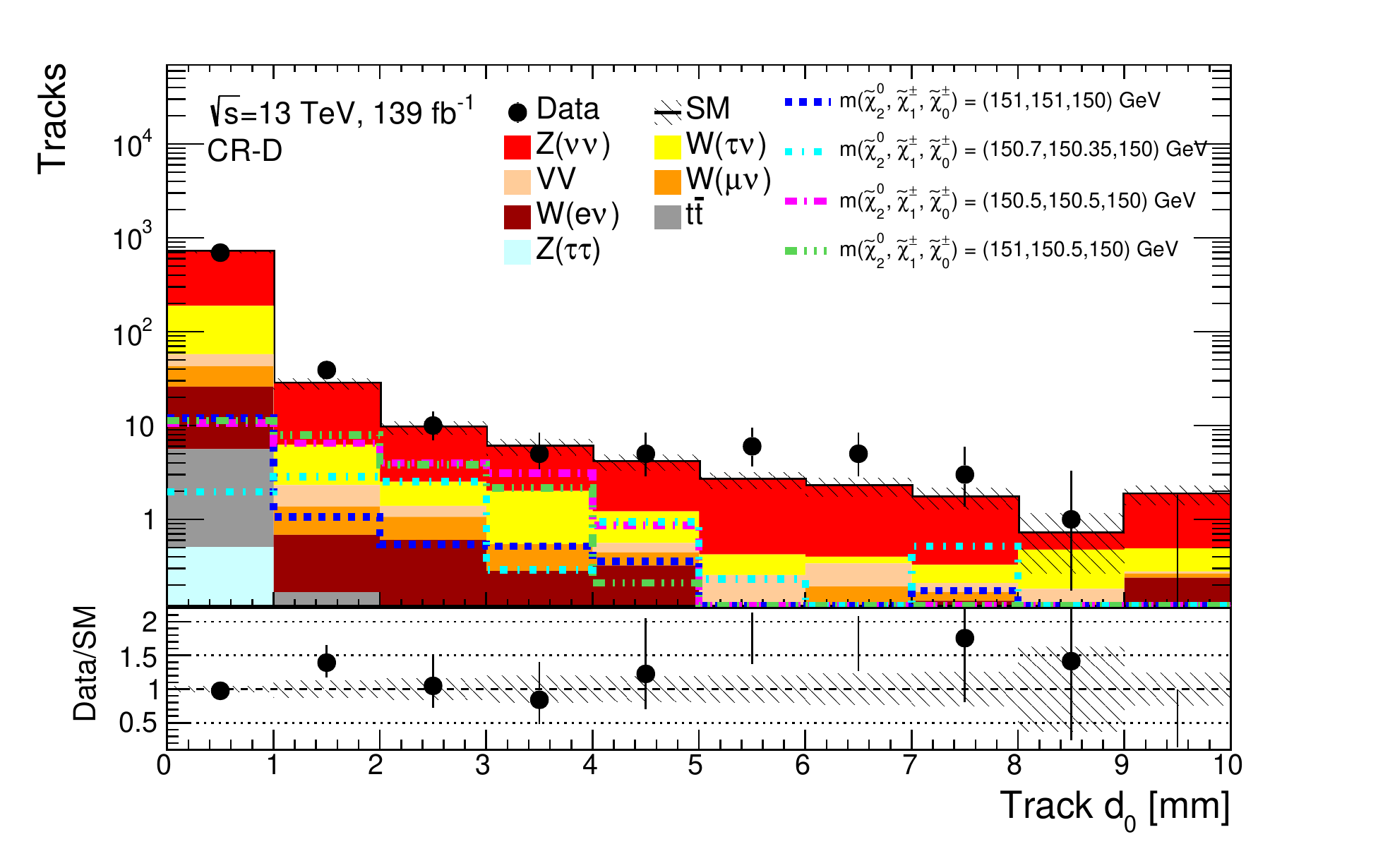}
\includegraphics[width=0.49\linewidth]{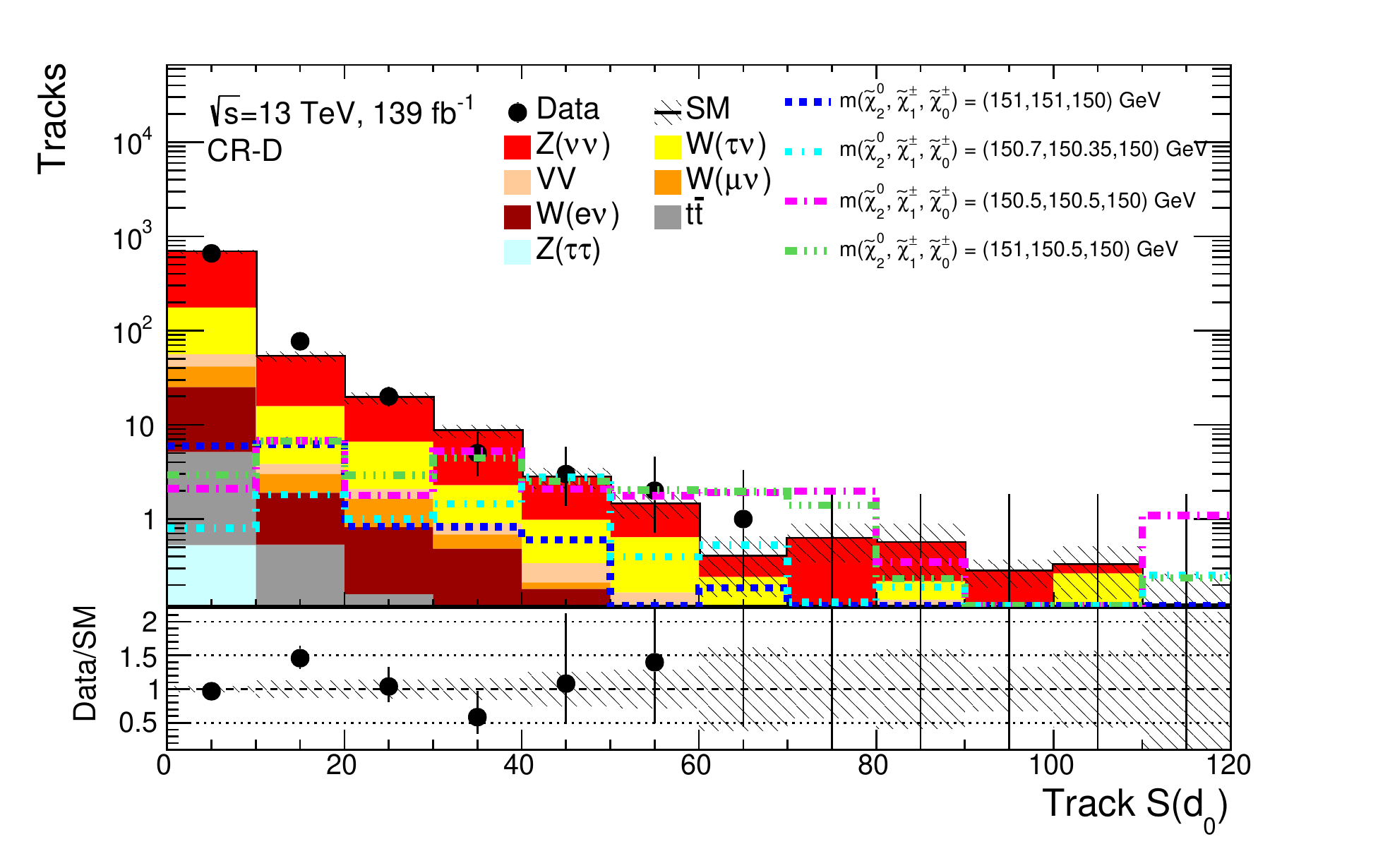}
\includegraphics[width=0.49\linewidth]{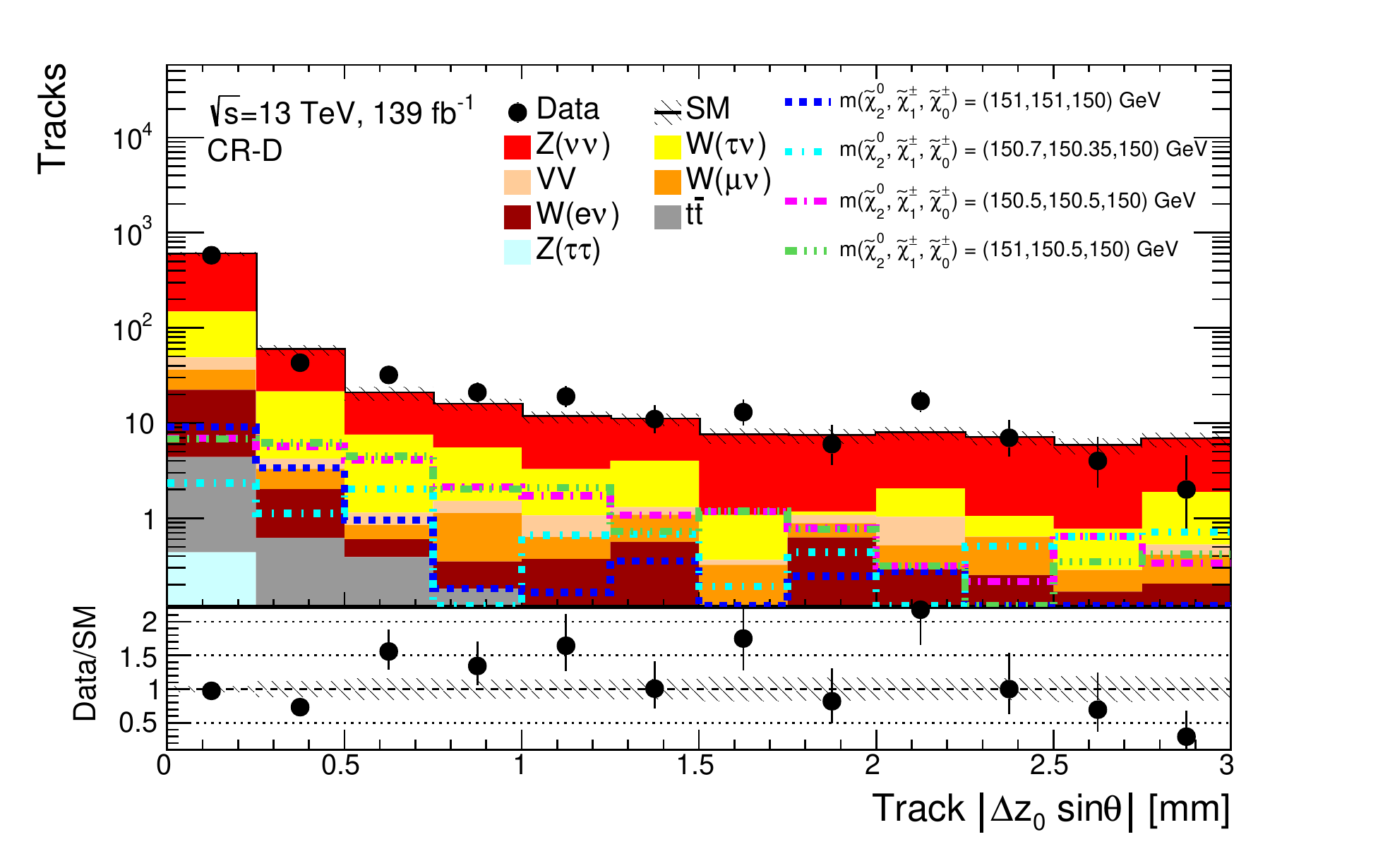}
\includegraphics[width=0.49\linewidth]{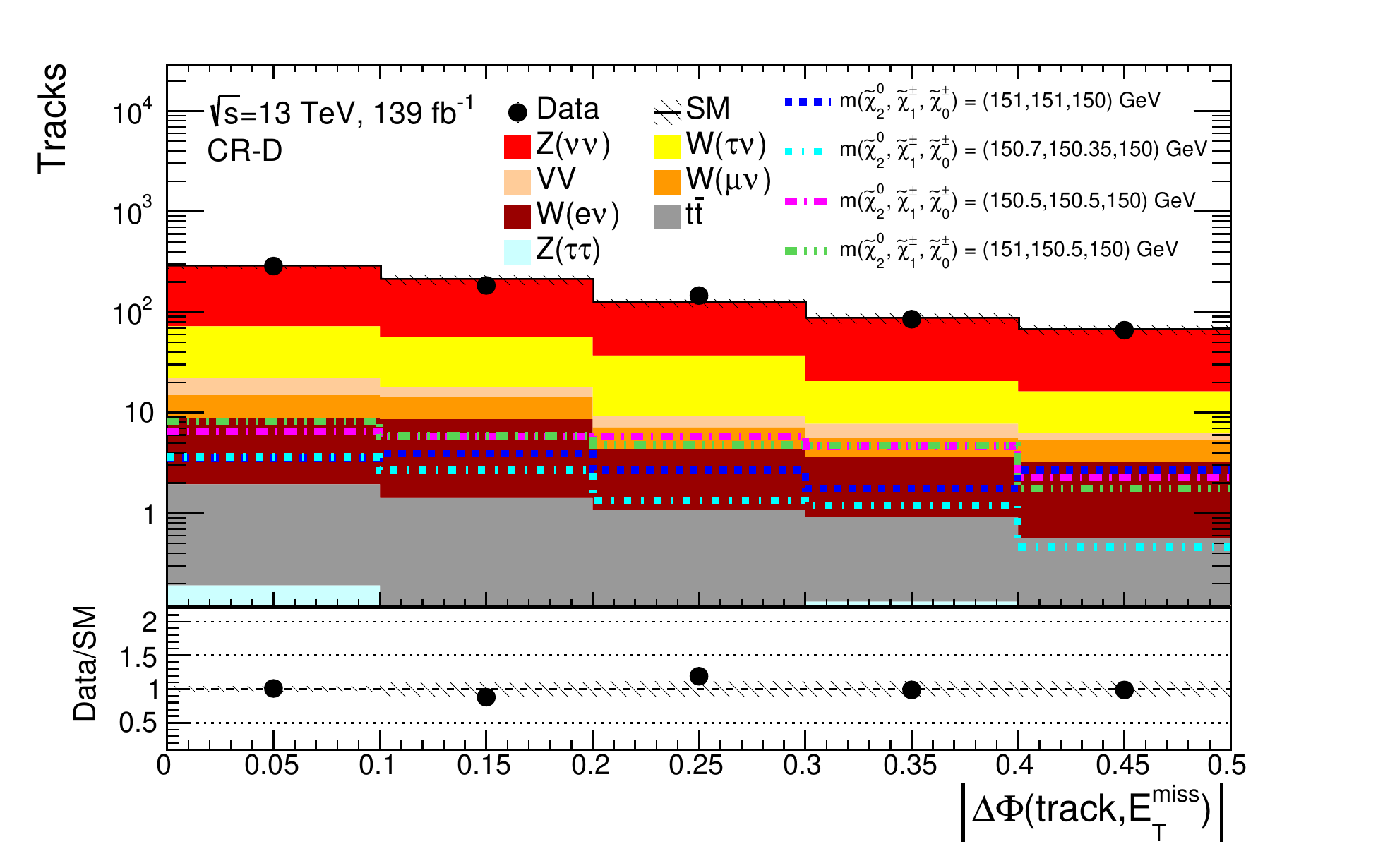}
\caption{The data and MC distributions of the variables used in the search in CR-D. Uncertainties include the statistical contributions and a 10\% flat uncertainty as a preliminary estimate of the systematic uncertainty.}
\label{fig:Higgsino_CRD}
\end{figure}

The expected track yields in the CRs and the SR for the strategy based on signals with longer decay lengths are shown in Table~\ref{Table:Higgsino-ABCD-yields}. The number of observed tracks in SR-A is not reported as the analysis is currently blinded. The scale factors estimated for the CRs and for the SR are reported in Table~\ref{tab:compressedshiggsinos-scalefactors}. Uncertainties include the statistical contributions and a 10\% flat uncertainty as a preliminary estimate of the systematic uncertainty. To emulate the effects of real systematic uncertainties, the flat systematic uncertainty is implemented as two 7\% systematic variations summed in quadrature, one correlated and one uncorrelated across the analysis regions (of type histoSysy and overallSys as described in Sec.~\ref{Sec:nuisanceparameters}). Fig.~\ref{fig:Higgsino_CRB}, Fig.~\ref{fig:Higgsino_CRC} and Fig.~\ref{fig:Higgsino_CRD} show the post-fit data and MC distributions of the variables used in the search in CR-B, CR-C and CR-D, respectively. The shapes of the kinematic distributions are well reproduced by the simulation in each CR. 

\begin{table}[!htb]
\centering
\scriptsize
\begin{tabular*}{\textwidth}{@{\extracolsep{\fill}}lrrrr}
\noalign{\smallskip}\hline\noalign{\smallskip}
Regions         & SR-A  & CR-B      & CR-C      & CR-D \\
\noalign{\smallskip}\hline\noalign{\smallskip}
Observed tracks & --    & $2181$    & $16838$   & $768$     \\
\noalign{\smallskip}\hline\noalign{\smallskip}
Fitted backgrounds      & $92 \pm 6$      & $2181 \pm 47$   & $16838 \pm 130$ & $768 \pm 28$    \\  \noalign{\smallskip}\hline\noalign{\smallskip}
Fitted $t\bar{t}$        & $0.85 \pm 0.10$      & $19 \pm 2$      & $132 \pm 14$    & $5.4 \pm 0.6$ \\
Fitted $W(e\nu)$+jets    & $2.43 \pm 0.15$       & $81.5 \pm 1.8$  & $727 \pm 6$     & $22.6 \pm 0.8$ \\
Fitted $W(\mu\nu)$+jets  & $2.61 \pm 0.17$       & $102 \pm 2$     & $821 \pm 7$     & $18.6 \pm 0.7$ \\
Fitted $W(\tau\nu)$+jets & $20.0 \pm 1.3$     & $633 \pm 14$    & $3975 \pm 33$   & $138 \pm 5$     \\
Fitted $Z(\nu\nu)$+jets  & $64 \pm 4$      & $1316 \pm 29$   & $10982 \pm 91$          & $566 \pm 21$   \\
Fitted $Z(\tau\tau)$+jets   & $0.11 \pm 0.01$   & $7.1 \pm 0.8$     & $24 \pm 2$    & $0.54 \pm 0.06$   \\
Fitted $VV$ events         & $1.56 \pm 0.18$    & $22 \pm 2$    & $177 \pm 18$  & $16.4 \pm 1.8$        \\
\noalign{\smallskip}\hline\noalign{\smallskip}
Simulated $W(e\nu)$+jets   & $1.51 \pm 0.16$    & $51 \pm 5$    & $494 \pm 49$  & $15.2 \pm 1.5$    \\
Simulated $W(\mu\nu)$+jets  & $1.62 \pm 0.17$   & $64 \pm 6$    & $558 \pm 55$  & $12.5 \pm 1.2$            \\
Simulated $W(\tau\nu)$+jets        & $12.4 \pm 1.3$          & $396 \pm 40$     & $2701 \pm 268$    & $94 \pm 9$     \\
Simulated $Z(\nu\nu)$+jets      & $40 \pm 4$          & $823 \pm 82$          & $7464 \pm 740$  & $381 \pm 38$  \\
\noalign{\smallskip}\hline\noalign{\smallskip}
\end{tabular*}
\caption{Observed track yields and predicted background track yields in the regions defined in the ABCD method for the strategy based on signals with longer decay lengths. For backgrounds with a normalisation extracted from the likelihood fit in the CRs, the yield expected from the simulation before the likelihood fit is also shown. The uncertainties include both statistical and preliminary systematic contributions.}
\label{Table:Higgsino-ABCD-yields}
\end{table}

\begin{table}[!htb]
\begin{center}
\begin{tabular}{l|c}
\noalign{\smallskip}\hline\noalign{\smallskip}
Scale factor & Estimated value \\
\noalign{\smallskip}\hline\noalign{\smallskip}
$\mu^{\mathrm{B}}$ & $1.59 \pm 0.16$ \\ 
$\mu^{\mathrm{C}}$ & $1.47 \pm 0.15$ \\ 
$\mu^{\mathrm{D}}$ & $1.48 \pm 0.16$ \\ \hline
$\mu^{\mathrm{A}}$ & $1.60 \pm 0.29$ \\
\noalign{\smallskip}\hline\noalign{\smallskip}
\end{tabular}
\end{center}
\caption{Scale factors estimated from the ABCD method in a background-only fit for the strategy based on signals with longer decay lengths.}
\label{tab:compressedshiggsinos-scalefactors}
\end{table}

The expected track yields in the CRs and in the SR for the strategy based on the signal with a smaller decay length are shown in Table~\ref{Table:Higgsino-ABCD-yields-smallDL}. 
The resulting scale factors are reported in Table~\ref{tab:compressedshiggsinos-scalefactors-smallDL}.  The post-fit data and MC track yields in the CRs agree well.

\begin{table}[!htb]
\centering
\scriptsize
\begin{tabular*}{\textwidth}{@{\extracolsep{\fill}}lrrrr} 
\noalign{\smallskip}\hline\noalign{\smallskip}
Regions         & SR-A*   & CR-B*         & CR-C*          & CR-D* \\ 
\noalign{\smallskip}\hline\noalign{\smallskip}
Observed tracks & --     & $1766$       & $50593$       & $2319$    \\ 
\noalign{\smallskip}\hline\noalign{\smallskip}
Fitted backgrounds      & $59 \pm 4$    & $1766 \pm 42$ & $50593 \pm 225$   & $2319 \pm 48$   \\  
\noalign{\smallskip}\hline\noalign{\smallskip}
Fitted $t\bar{t}$       & $0.88 \pm 0.10$  & $24.8 \pm 2.7$  & $399 \pm 42$ & $13.2 \pm 1.4$ \\
Fitted $W(e\nu)$+jets   & $2.03 \pm 0.14$   & $66.9 \pm 1.6$  & $2251 \pm 11$  & $64.1 \pm 1.4$              \\
Fitted $W(\mu\nu)$+jets & $1.18 \pm 0.08$   & $59.7 \pm 1.5$  & $2603 \pm 14$  & $54.5 \pm 1.2$              \\
Fitted $W(\tau\nu)$+jets & $21.5 \pm 1.4$   & $721 \pm 18$  & $10762 \pm 57$   & $393 \pm 9$             \\
Fitted $Z(\nu\nu)$+jets & $31.7 \pm 2.1$    & $867 \pm 21$  & $33930 \pm 178$  & $1714 \pm 38$              \\
Fitted $Z(\tau\tau)$+jets   & $0.16 \pm 0.02$   & $8.5 \pm 0.9$  & $60 \pm 6$   & $1.38 \pm 0.15$              \\
Fitted $VV$ events  & $1.47 \pm 0.17$   & $18.4 \pm 2.0$  & $588 \pm 63$ & $79 \pm 8$        \\ 
\noalign{\smallskip}\hline\noalign{\smallskip}
Simulated $W(e\nu)$+jets    & $1.41 \pm 0.16$   & $46 \pm 5$    & $1601 \pm 159$    & $46 \pm 4$ \\
Simulated $W(\mu\nu)$+jets  & $0.82 \pm 0.09$   & $41 \pm 4$    & $1851 \pm 183$    & $39 \pm 4$           \\
Simulated $W(\tau\nu)$+jets & $14.9 \pm 1.7$    & $491 \pm 49$  & $7655 \pm 759$    & $284 \pm 28$     \\
Simulated $Z(\nu\nu)$+jets  & $21.9 \pm 2.5$    & $590 \pm 59$  & $24134 \pm 2392$  & $1241 \pm 124$             \\ 
\noalign{\smallskip}\hline\noalign{\smallskip}
\end{tabular*}
\caption{Observed track yields and predicted background track yields in the regions defined in the ABCD method for the strategy based on the signal with a smaller decay length. For backgrounds with a normalisation extracted from the likelihood fit in the CRs, the yield expected from the simulation before the likelihood fit is also shown. The uncertainties include both statistical and preliminary systematic contributions.}
\label{Table:Higgsino-ABCD-yields-smallDL}
\end{table}

\begin{table}[!htb]
\begin{center}
\begin{tabular}{l|c}
\noalign{\smallskip}\hline\noalign{\smallskip}
Scale factor & Estimated value \\
\noalign{\smallskip}\hline\noalign{\smallskip}
$\mu^{\mathrm{B*}}$ & $1.46 \pm 0.16$ \\ 
$\mu^{\mathrm{C*}}$ & $1.40 \pm 0.14$ \\ 
$\mu^{\mathrm{D*}}$ & $1.38 \pm 0.15$ \\ \hline
$\mu^{\mathrm{A*}}$ & $1.44 \pm 0.26$ \\
\noalign{\smallskip}\hline\noalign{\smallskip}
\end{tabular}
\end{center}
\caption{Scale factors estimated from the ABCD method in a background-only fit for the strategy based on the signal with a smaller decay length.}
\label{tab:compressedshiggsinos-scalefactors-smallDL}
\end{table}

\FloatBarrier

\subsection{Validation Regions}
\label{sec:compressedshiggsinos-VRs}
\begin{figure}[!p]
\centering
\includegraphics[width=0.49\linewidth]{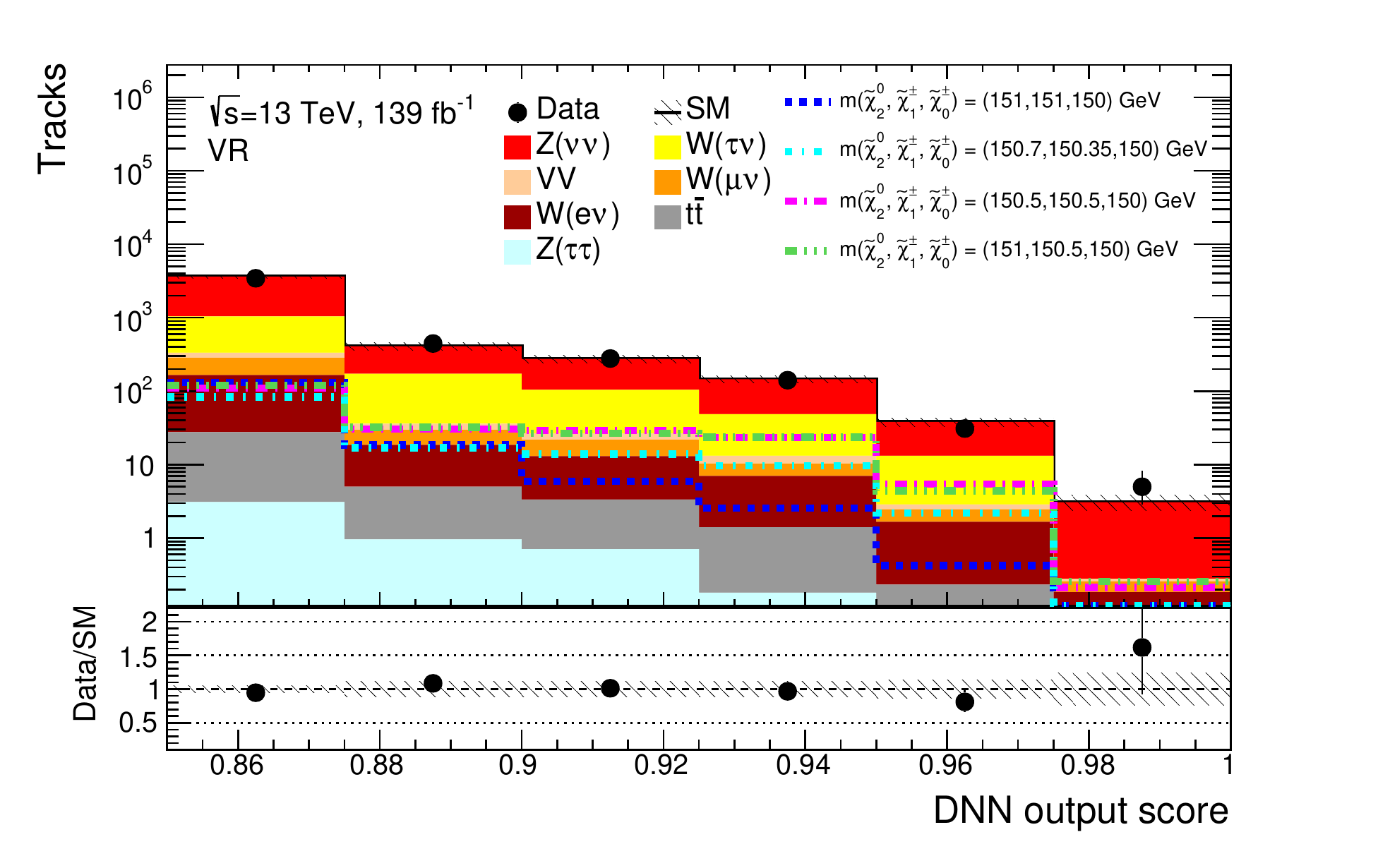}
\includegraphics[width=0.49\linewidth]{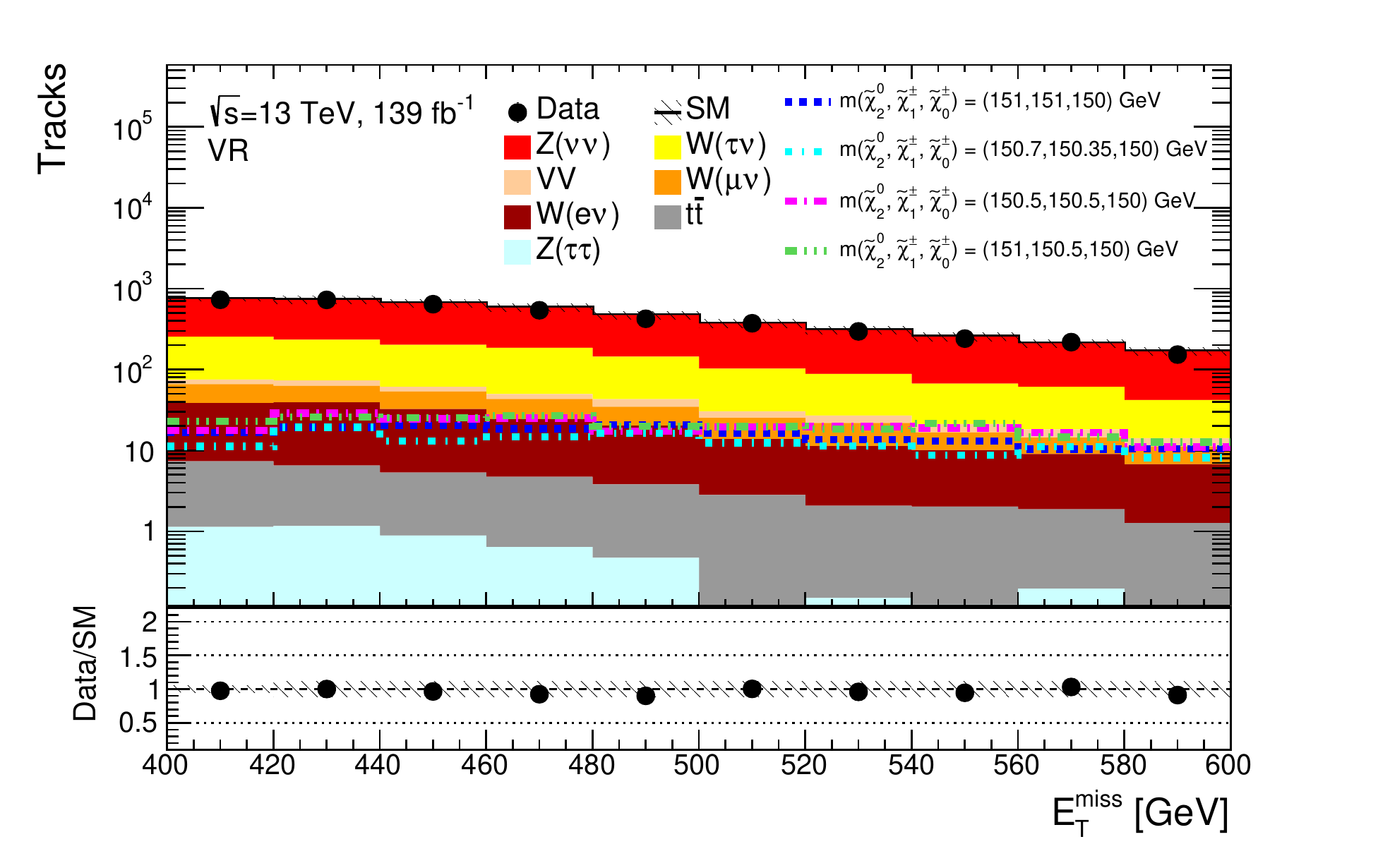}
\includegraphics[width=0.49\linewidth]{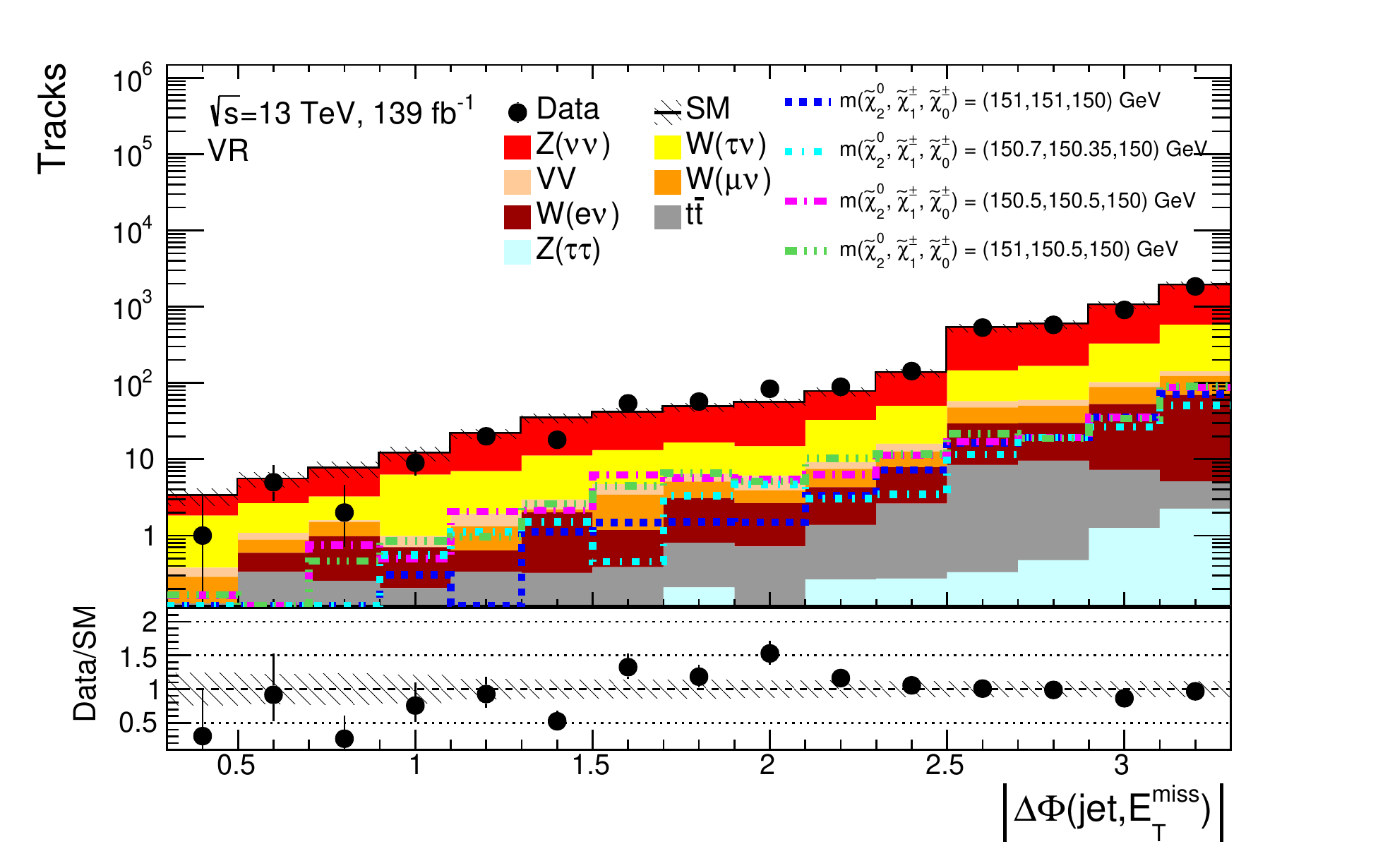}
\includegraphics[width=0.49\linewidth]{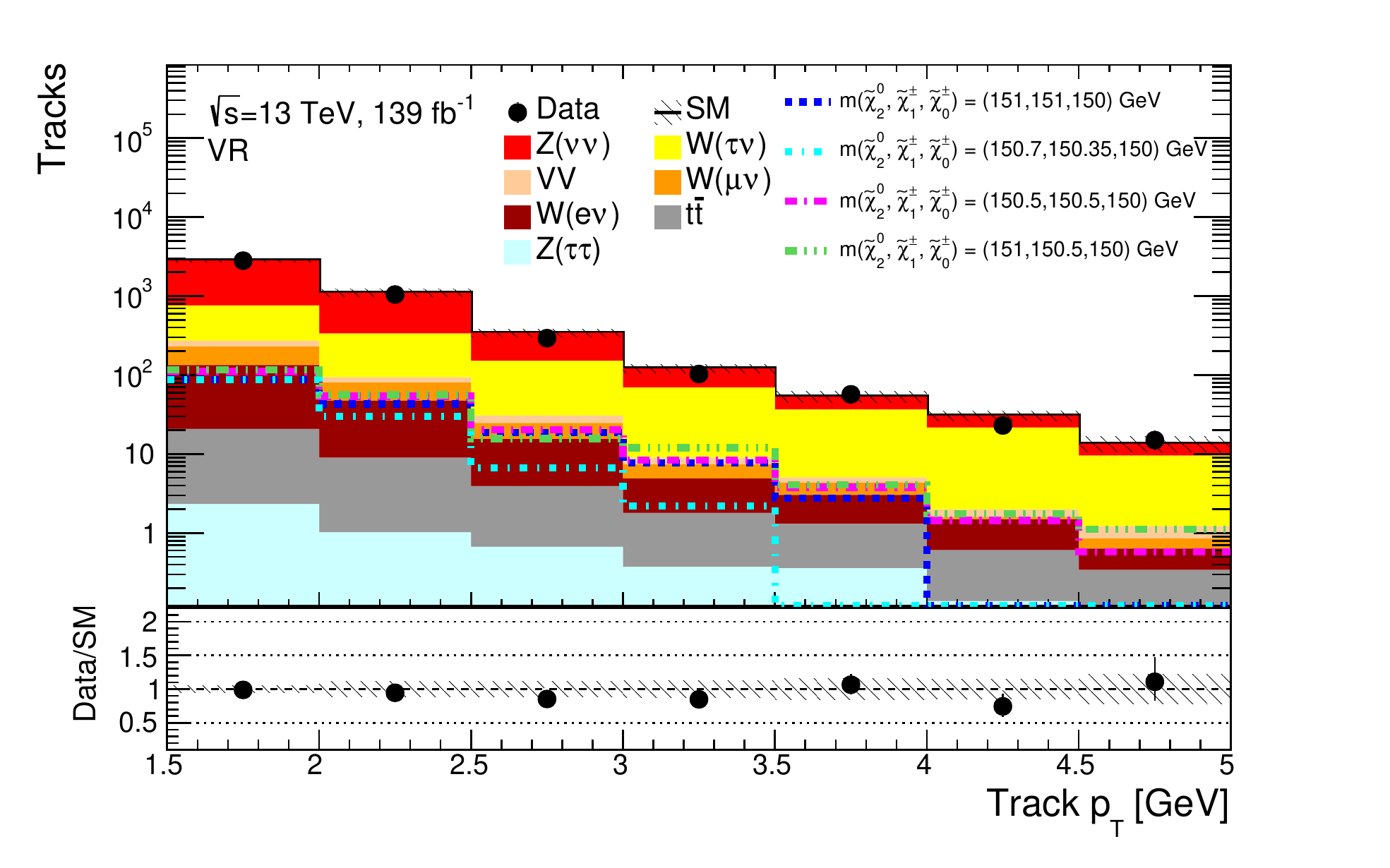}
\includegraphics[width=0.49\linewidth]{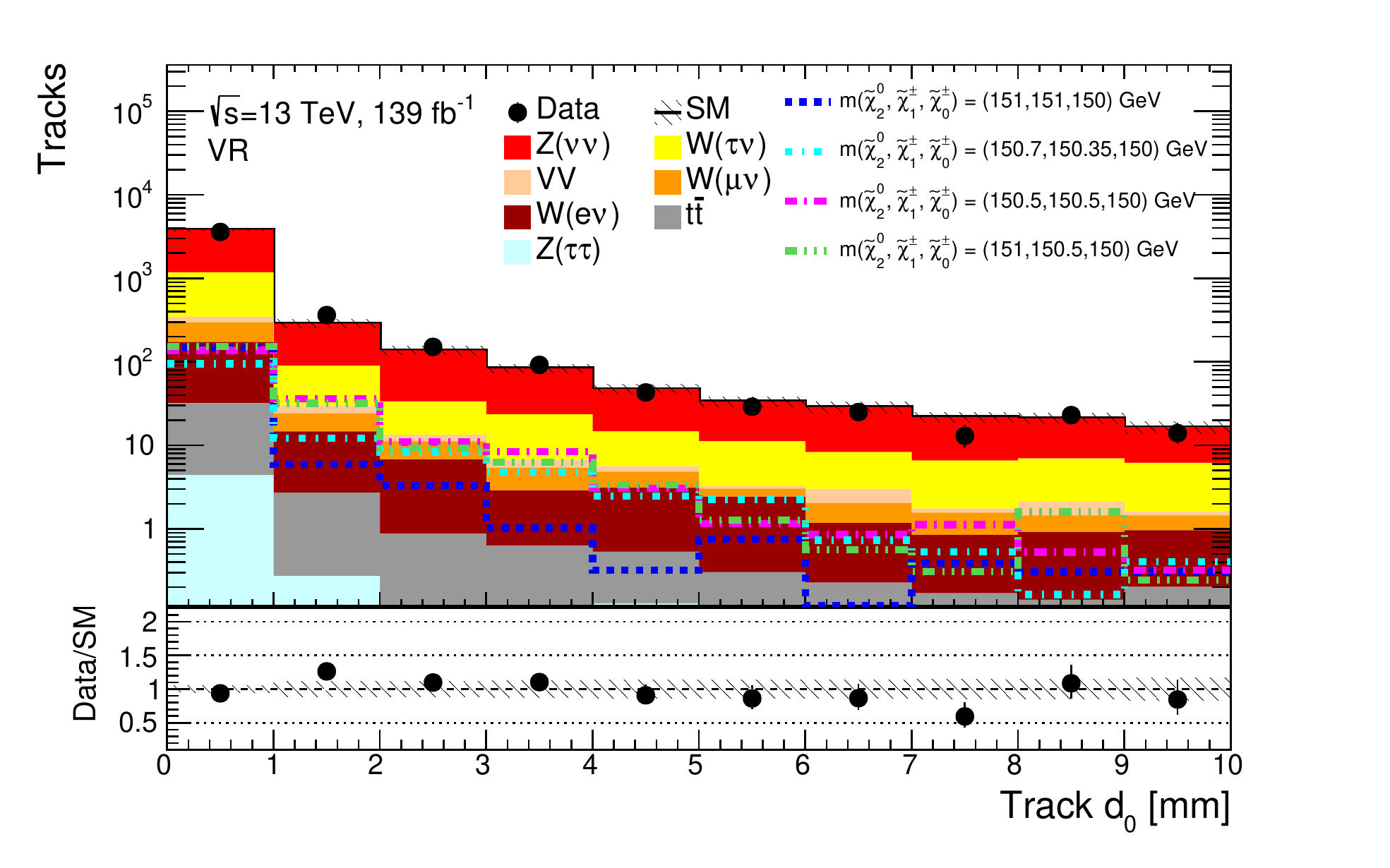}
\includegraphics[width=0.49\linewidth]{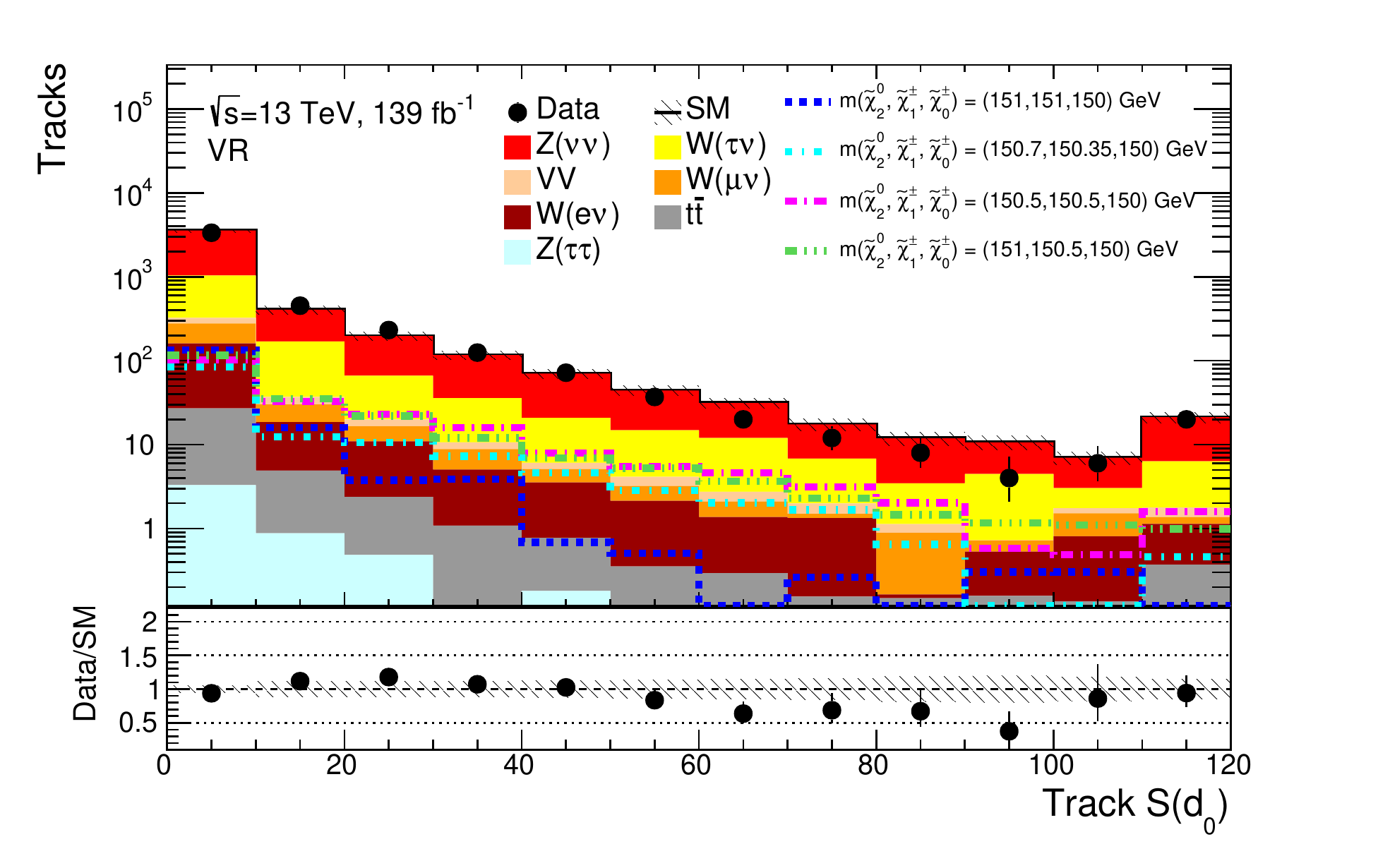}
\includegraphics[width=0.49\linewidth]{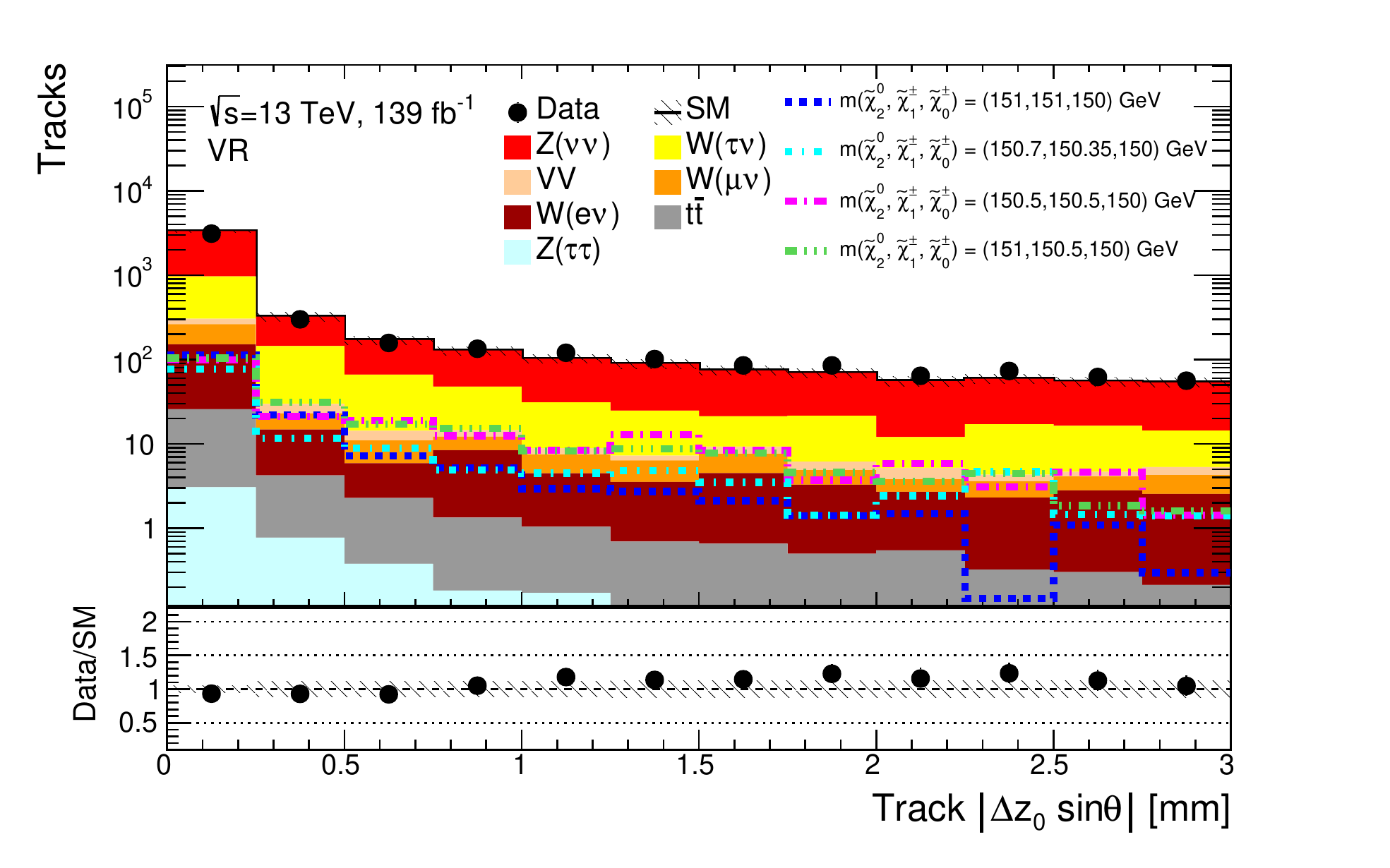}
\includegraphics[width=0.49\linewidth]{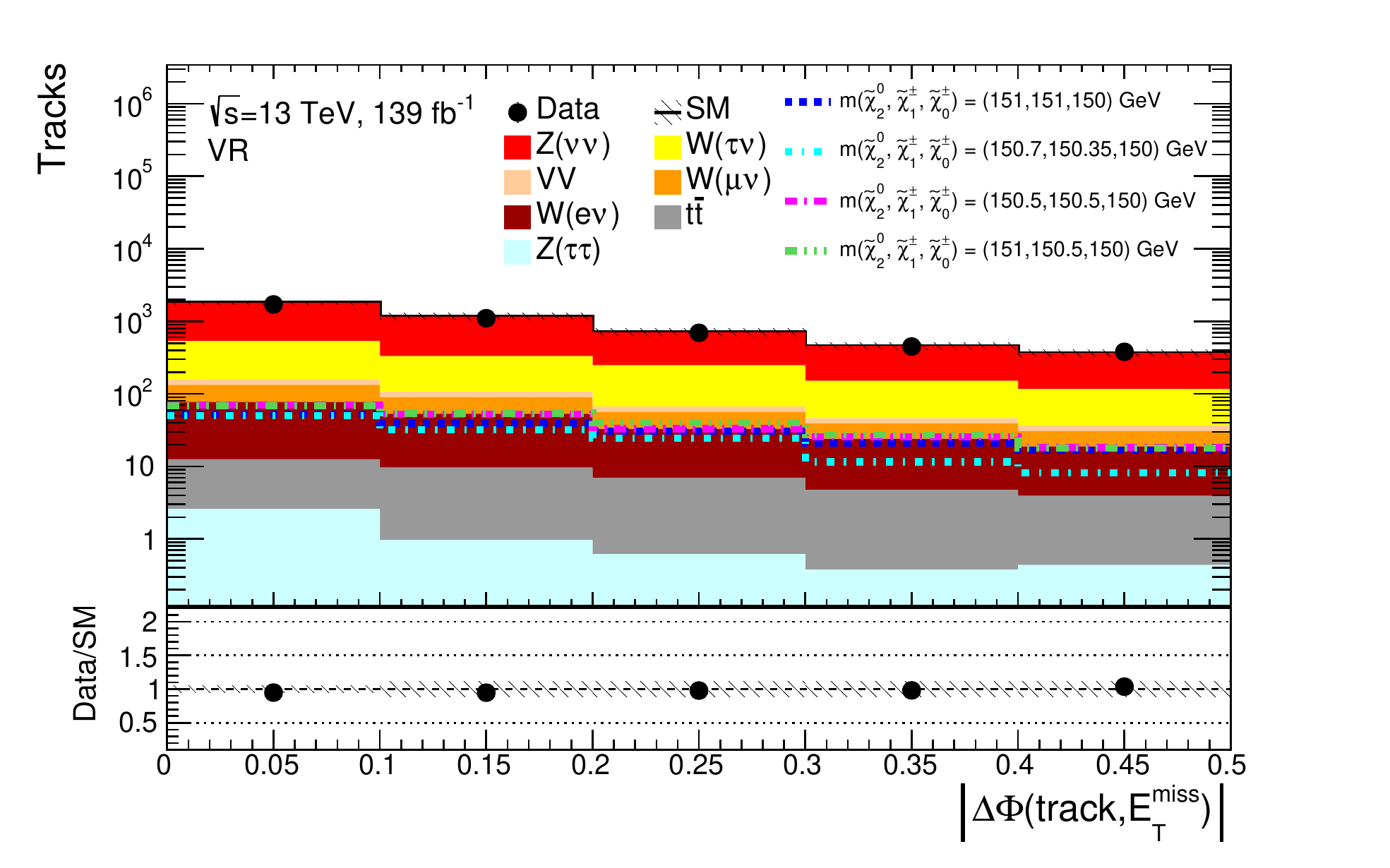}
\caption{The data and MC distributions of the variables used in the search in VR. Uncertainties include the statistical contributions and a 10\% flat uncertainty as a preliminary estimate of the systematic uncertainty.}
\label{fig:Higgsino_VR}
\end{figure}

\begin{figure}[!p]
\centering
\includegraphics[width=0.49\linewidth]{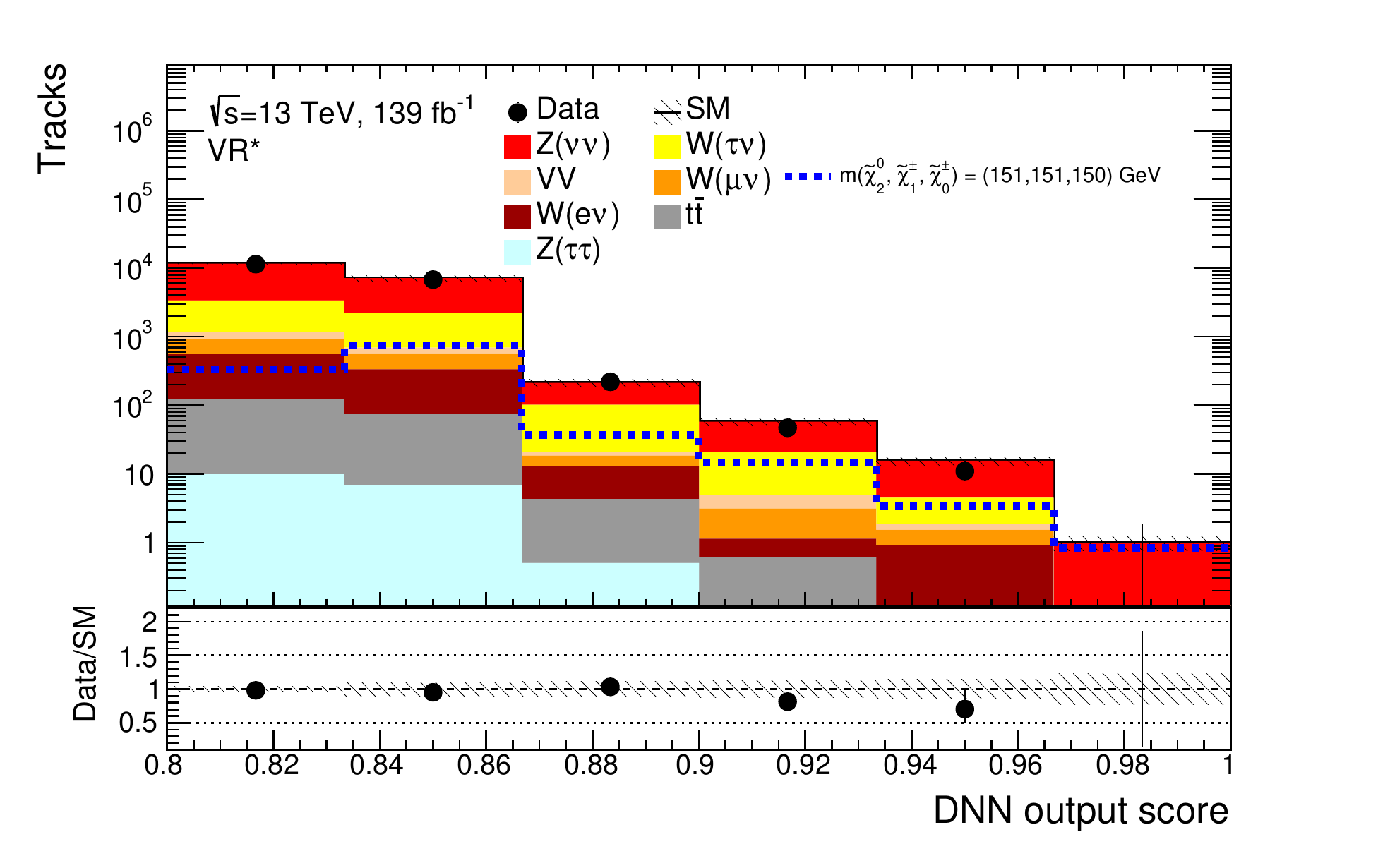}
\includegraphics[width=0.49\linewidth]{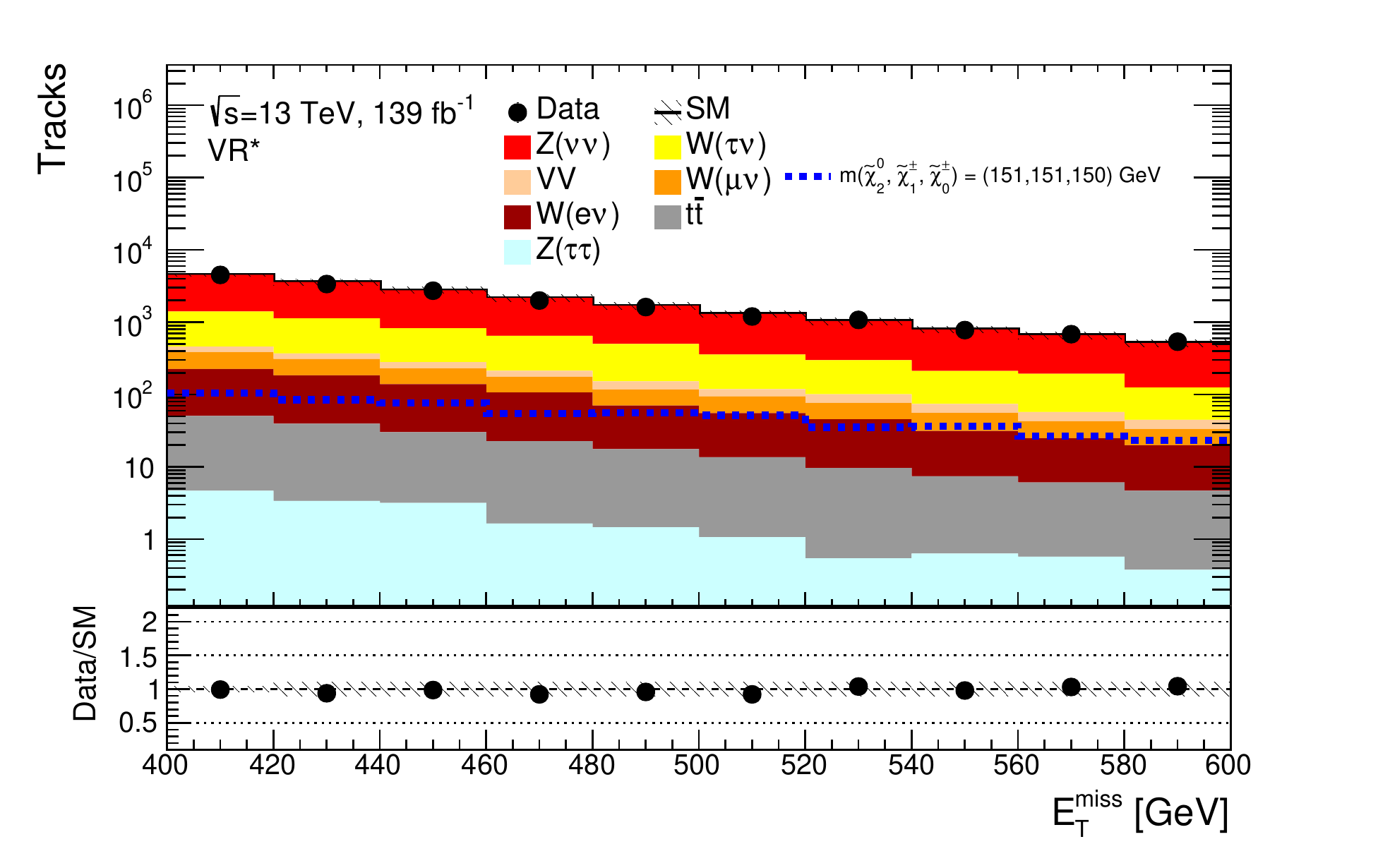}
\includegraphics[width=0.49\linewidth]{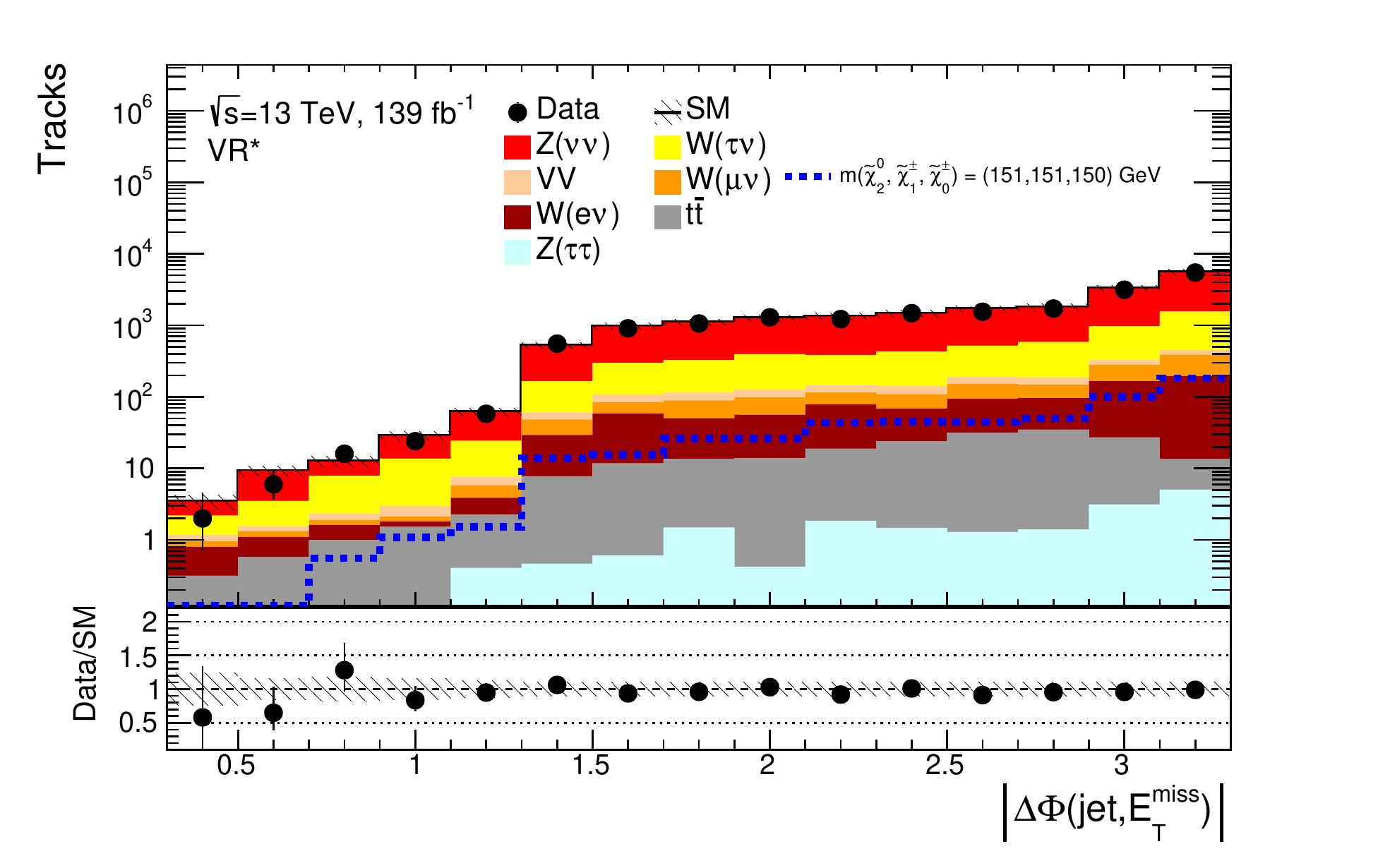}
\includegraphics[width=0.49\linewidth]{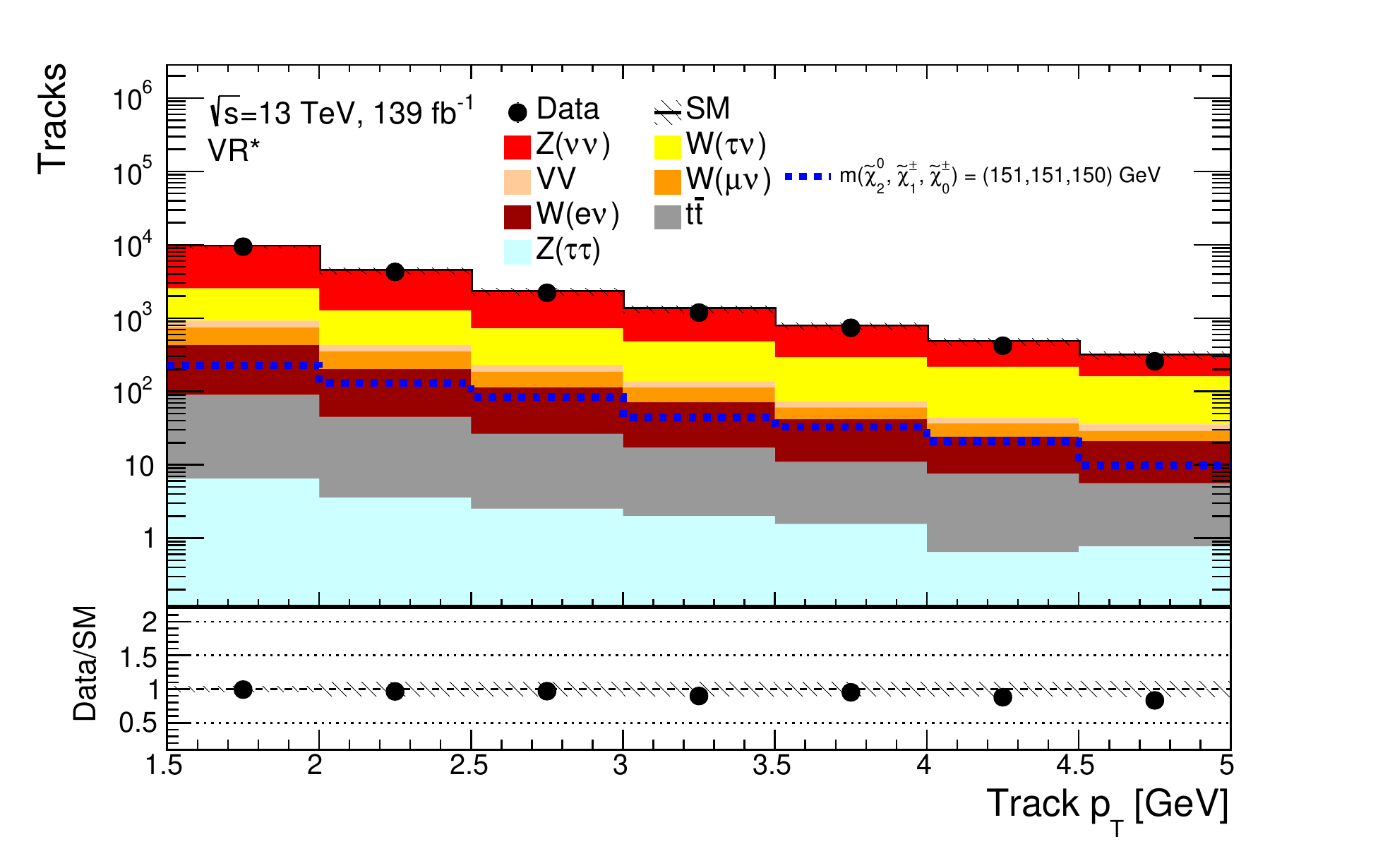}
\includegraphics[width=0.49\linewidth]{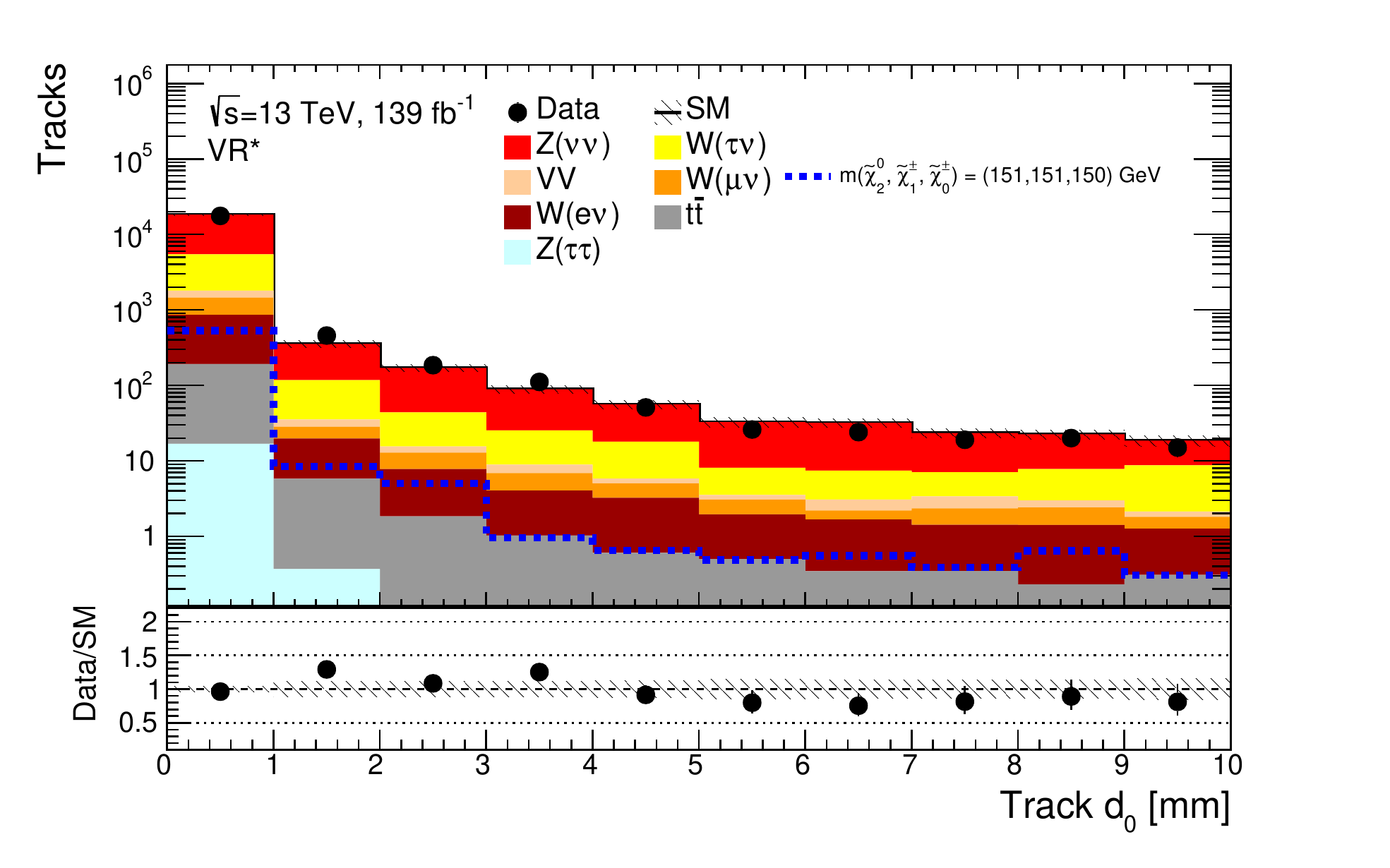}
\includegraphics[width=0.49\linewidth]{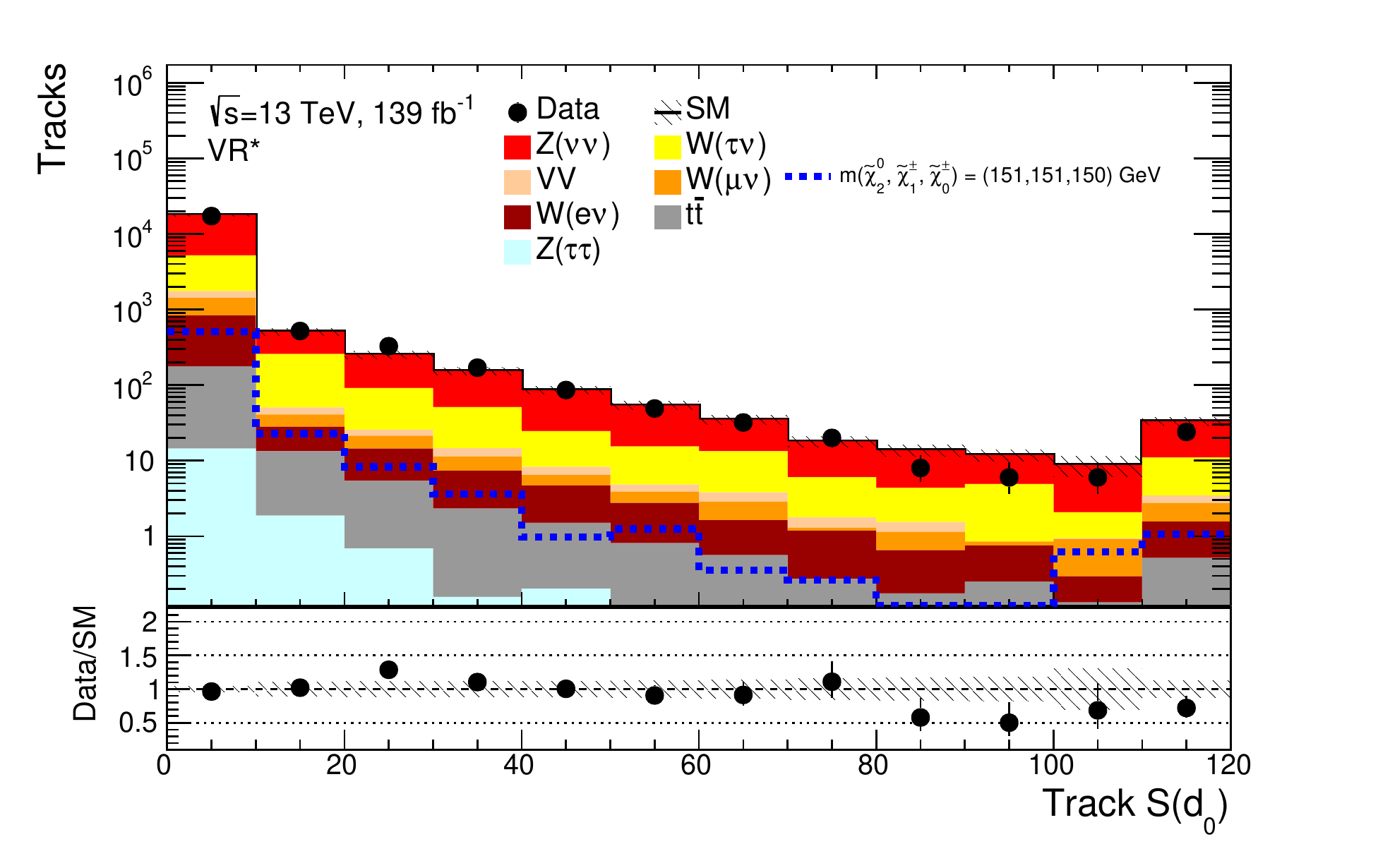}
\includegraphics[width=0.49\linewidth]{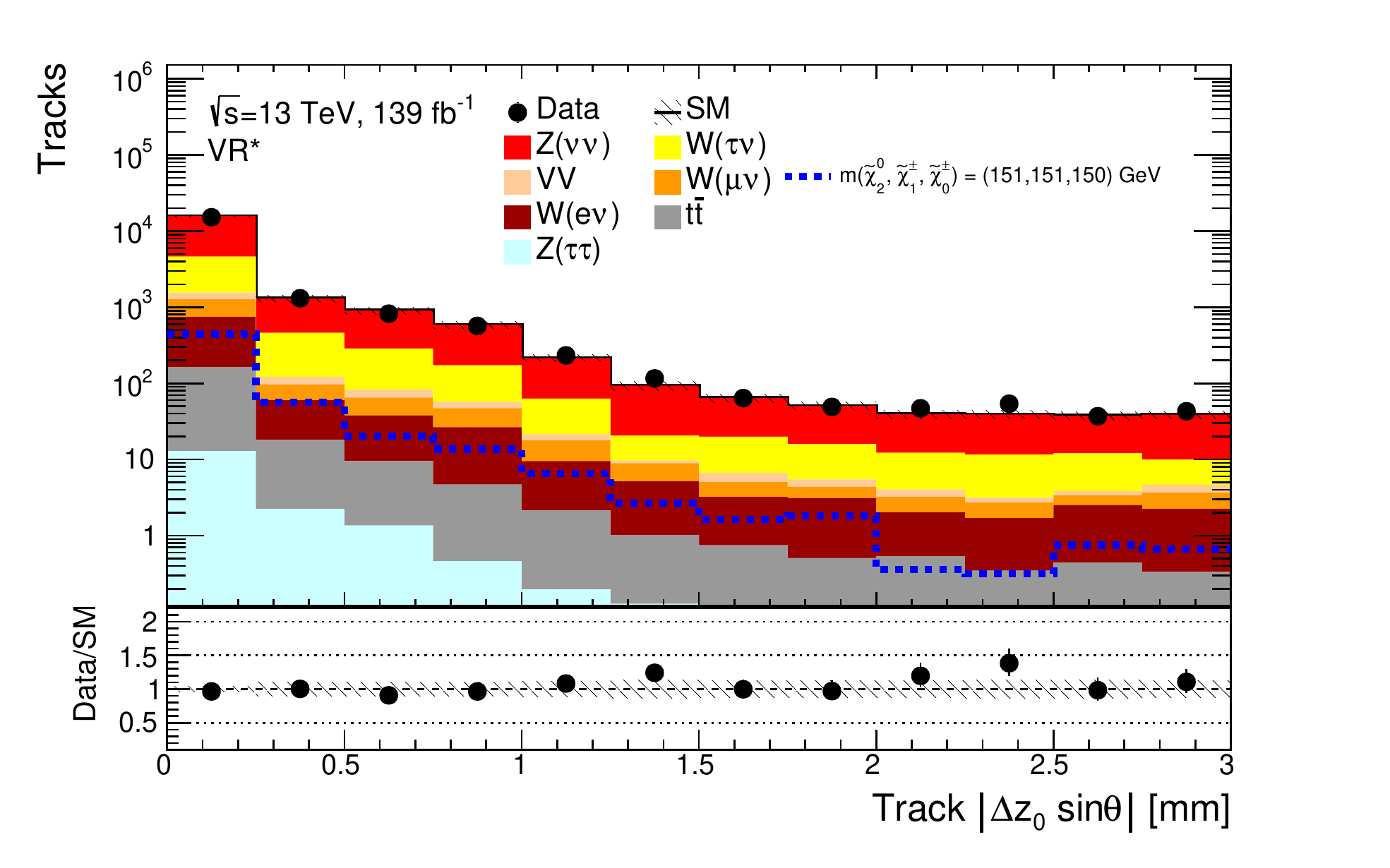}
\includegraphics[width=0.49\linewidth]{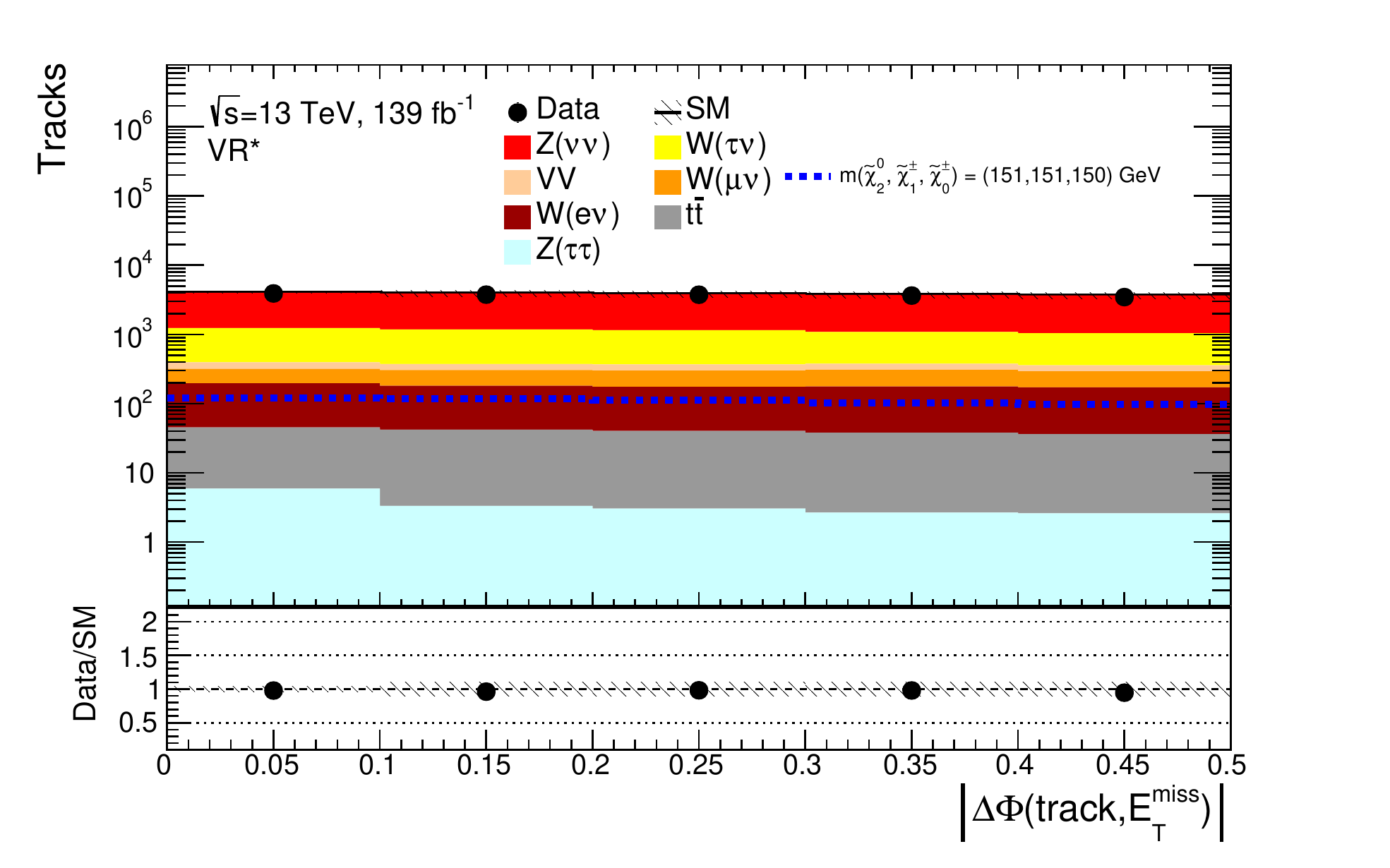}
\caption{The data and MC distributions of the variables used in the search in VR*. Uncertainties include the statistical contributions and a 10\% flat uncertainty as a preliminary estimate of the systematic uncertainty.}
\label{fig:Higgsino_VR2}
\end{figure}

VRs are introduced to check the modelling of the most relevant distributions in regions between the CRs and the SR.
Due to the different shapes of the DNN output score between the two trainings, 2 different VRs are defined. They are denoted as VR and VR*, targeting signals with longer and shorter decay lengths, respectively. They both are defined in the $400 < E_\mathrm{T}^{\mathrm{miss}} < 600$ GeV range, as shown in Fig.~\ref{fig:Higgsino_ABCD}, with the difference that VR is defined using DNN output score > 0.85 while VR* is defined using DNN output score > 0.8.
The post-fit data and MC distributions of the variables used in the search in VR and VR* are shown in Fig.~\ref{fig:Higgsino_VR} and Fig.~\ref{fig:Higgsino_VR2}, respectively. The number of observed tracks and the predicted track yields of each SM process in each VR are reported in Table~\ref{Table:Higgsino-ABCD-VR}. The scale factor $\mu_{\mathrm{A}}$ reported in Table~\ref{tab:compressedshiggsinos-scalefactors} is applied in VR while the scale factor $\mu^{\mathrm{A*}}$ reported in Table~\ref{tab:compressedshiggsinos-scalefactors-smallDL} is applied in VR*. The shapes of the data and MC distributions agree well within the uncertainties.

\begin{table}[!htb]
\centering
\scriptsize
\begin{tabular*}{\textwidth}{@{\extracolsep{\fill}}lrr} 
\noalign{\smallskip}\hline\noalign{\smallskip}
Regions         & VR  & VR*   \\ 
\noalign{\smallskip}\hline\noalign{\smallskip}
Observed tracks & 4341 & $18498$        \\ 
\noalign{\smallskip}\hline\noalign{\smallskip}
Fitted backgrounds      & $4506 \pm 214$    & $19074 \pm 728$  \\  
\noalign{\smallskip}\hline\noalign{\smallskip}
Fitted $t\bar{t}$       & $33 \pm 4$  & $184 \pm 20$  \\
Fitted $W(e\nu)$+jets   & $168 \pm 8$   & $701 \pm 28$   \\
Fitted $W(\mu\nu)$+jets & $144 \pm 7$   & $623 \pm 25$    \\
Fitted $W(\tau\nu)$+jets & $968 \pm 47$   & $3822 \pm 152$      \\
Fitted $Z(\nu\nu)$+jets & $3125 \pm 152$    & $13377 \pm 530$     \\
Fitted $Z(\tau\tau)$+jets   & $5.0 \pm 0.5$   & $17.5 \pm 1.8$     \\
Fitted $VV$ events  & $63 \pm 7$   & $349 \pm 37$     \\ 
\noalign{\smallskip}\hline\noalign{\smallskip}
Simulated $W(e\nu)$+jets    & $105 \pm 10$   & $485 \pm 41$   \\
Simulated $W(\mu\nu)$+jets  & $89 \pm 9$   & $431 \pm 37$          \\
Simulated $W(\tau\nu)$+jets & $601 \pm 60$    & $2646 \pm 224$     \\
Simulated $Z(\nu\nu)$+jets  & $1939 \pm 193$    & $9263 \pm 698$            \\ 
\noalign{\smallskip}\hline\noalign{\smallskip}
\end{tabular*}
\caption{Observed track yields and predicted background track in the VRs defined in the ABCD method. For backgrounds with a normalisation extracted from the likelihood fit in the CRs, the yield expected from the simulation before the likelihood fit is also shown. The uncertainties include both statistical and preliminary systematic contributions.}
\label{Table:Higgsino-ABCD-VR}
\end{table}

\FloatBarrier

\section{Results}
\label{sec:compressedshiggsinos-results}

\begin{figure}[!htb]
\centering
\includegraphics[width=0.8\linewidth]{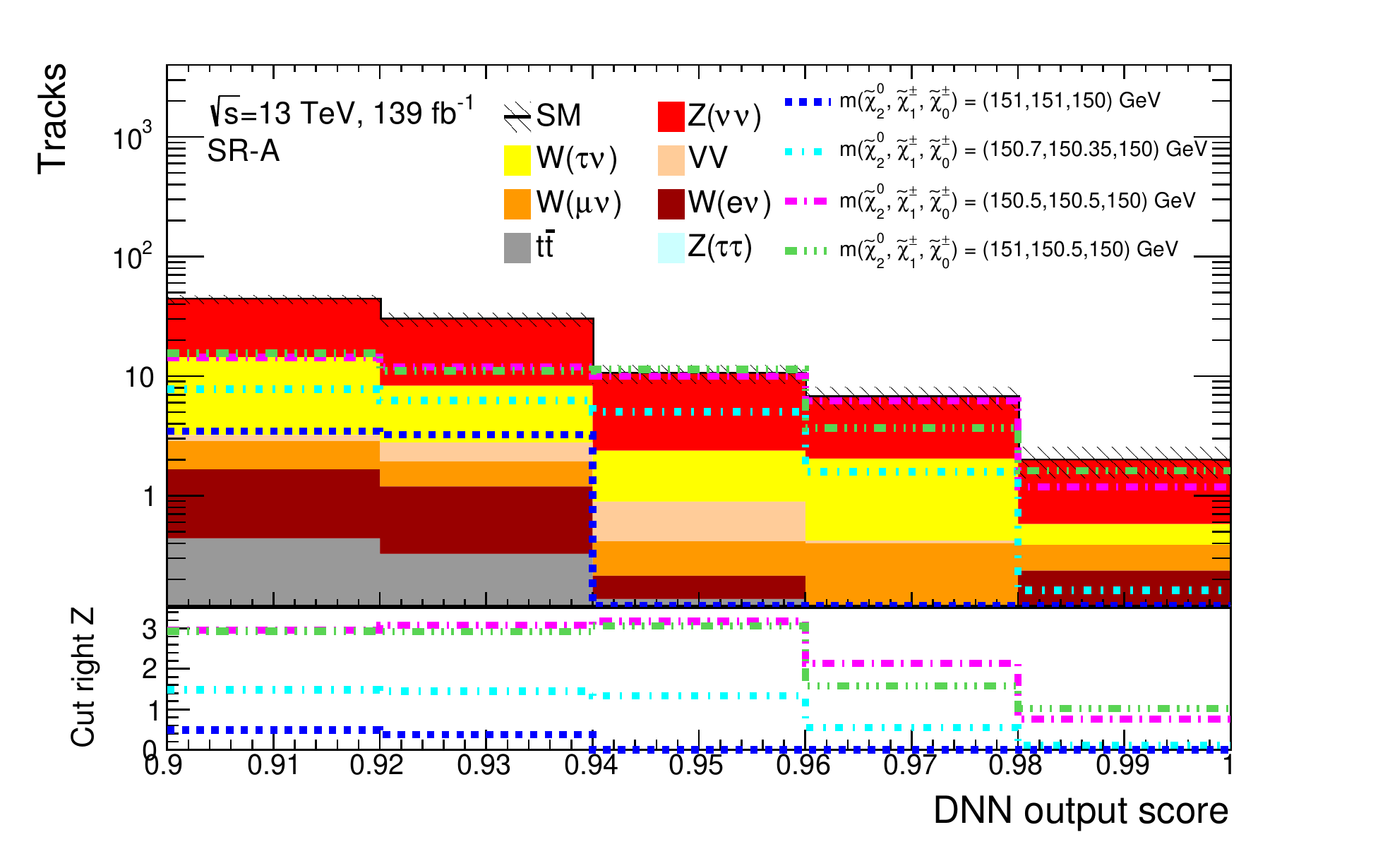}
\caption{The signal and background distributions of the DNN output score in SR-A. A preliminary estimate of the systematic uncertainty is shown.}
\label{fig:SR}
\end{figure}

\begin{figure}[!htb]
\centering
\includegraphics[width=0.8\linewidth]{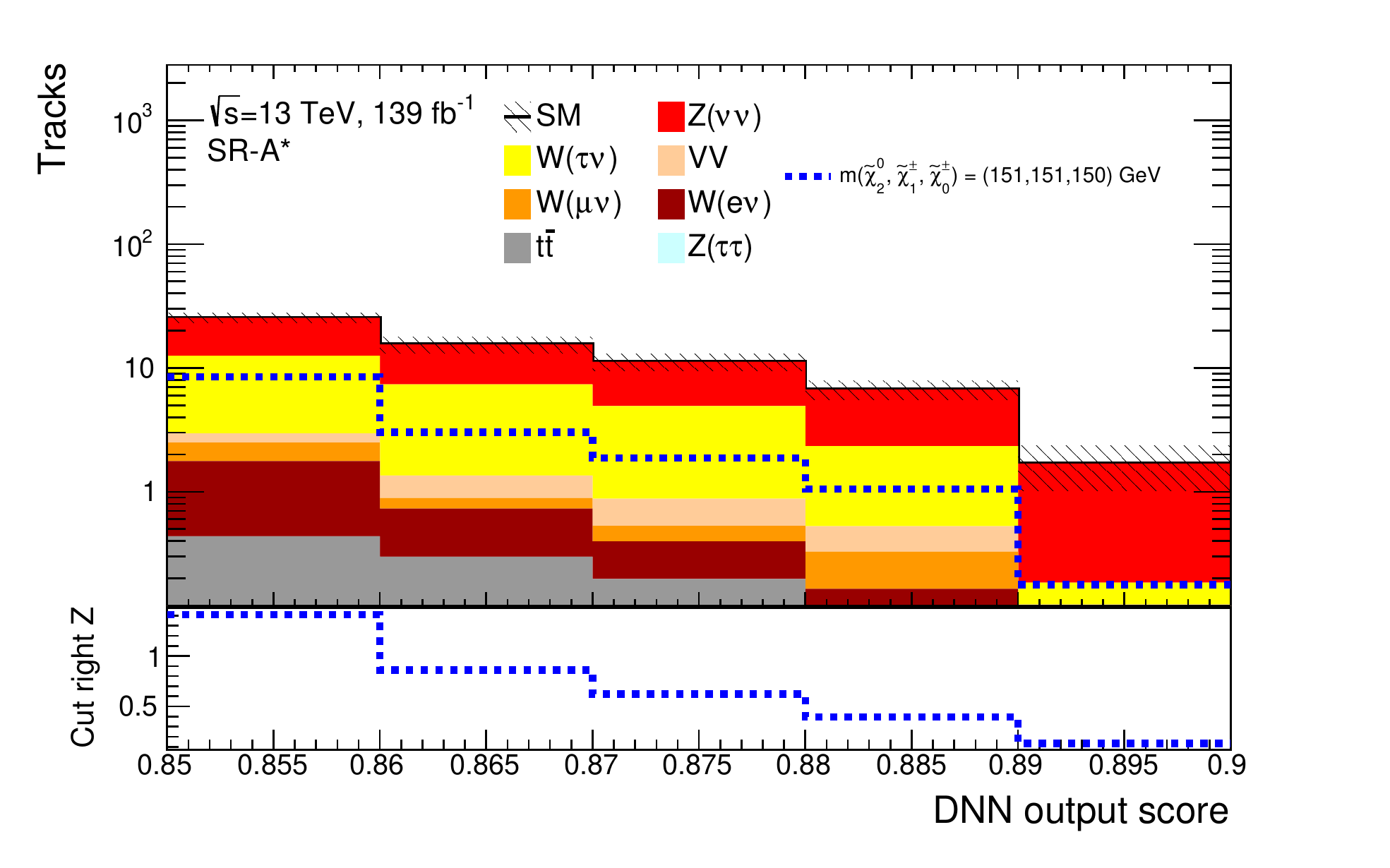}
\caption{The signal and background distributions of the DNN output score in SR-A*. A preliminary estimate of the systematic uncertainty is shown.}
\label{fig:SR2}
\end{figure}

The results are interpreted in the context of the higgsino simplified model shown in Fig.~\ref{fig:FeynmanHiggsinos}. The DNN distributions at the post-fit level in the blinded SRs defined in Table~\ref{tab:compressedshiggsinos-AnalysisRegions} are shown in Fig.~\ref{fig:SR} and in Fig.~\ref{fig:SR2} for SR-A and SR-A*, respectively. 

The SRs are defined by considering the maximum of the DNN output score significance and allow to have good sensitivity to test different hypotheses of higgsino masses around 150 GeV. A tighter selection on the DNN output score to define the SRs would only reduce the signal statistics without yielding a substantial increment of the significance because, as can be seen on the bottom panel of these plots in the SRs, the cut right significances do not significantly increase as a function of the DNN output score. Moreover, the advantage of using a dedicated training for the $m(\tilde{\chi}_{2}^{0}, \, \tilde{\chi}_{1}^{\pm}, \,  \tilde{\chi}_{1}^{0}) = (151, 151, 150)$ GeV signal sample can be understood by comparing the cut right significance for this signal sample in~Fig.\ref{fig:SR2}, where it is trained over alone and tested, to~Fig.\ref{fig:SR}, where it only gets tested with a classifier trained over the other signal samples.\\

Preliminary results for the higgsino search are reported as exclusion limits at 95\% CL on the masses of the higgsinos using the CL$_{\text{s}}$ prescription as described in Section~\ref{sec:CLtechnique}. $\mathrm{CL_{s}}$ values obtained from a blinded exclusion fit in the SR are used to determine the expected sensitivity of the search. The regions are setup in the fit following the definitions in Table~\ref{tab:compressedshiggsinos-AnalysisRegions} and the two scenarios are considered according to the masses of the signal samples. A comparison between $\mathrm{CL_{s}}$ values obtained from the strategy described using the DNN and the cut\&count strategy is provided in Table~\ref{tab:compressedshiggsinos-CLspreliminary}. 

\begin{table}[!htb]
\begin{center}
\begin{tabular}{l|cc}
\noalign{\smallskip}\hline\noalign{\smallskip}
$m(\tilde{\chi}_{2}^{0}, \, \tilde{\chi}_{1}^{\pm}, \,  \tilde{\chi}_{1}^{0})$ [GeV] & $\mathrm{CL_{s}}$ using the DNN & $\mathrm{CL_{s}}$ with cut\&count \\
\noalign{\smallskip}\hline\noalign{\smallskip}
$(151, 150.5, 150)$    & $0.006$ & $0.016$ \\ 
$(150.5, 150.5, 150)$  & $0.005$ & $0.007$ \\ 
$(150.7, 150.35, 150)$ & $0.19$ & $0.21$ \\ 
$(151, 151, 150)$      & $0.24$ & $0.44$ \\ 
\noalign{\smallskip}\hline\noalign{\smallskip}
\end{tabular}
\end{center}
\caption{Comparison of the estimated $\mathrm{CL_{s}}$ values obtained from a blinded exclusion fit in the SR-A defined using the DNN output score or the cut\&count procedure. Uncertainties include the statistical contributions and a 10\% flat uncertainty as a preliminary estimate of the systematic uncertainty.}
\label{tab:compressedshiggsinos-CLspreliminary}
\end{table}

Again, we can see the advantages of using a DNN instead of the cut\&count approach as the $\mathrm{CL_{s}}$ values are smaller using the DNN. From the table, only $m(\tilde{\chi}_{2}^{0}, \, \tilde{\chi}_{1}^{\pm}, \,  \tilde{\chi}_{1}^{0}) = (151, 150.5, 150)$ GeV and $ (150.5, 150.5, 150)$ GeV signal samples are expected to be excluded. However, introducing a shape fit should allow to easily exclude also the $m(\tilde{\chi}_{2}^{0}, \, \tilde{\chi}_{1}^{\pm}, \,  \tilde{\chi}_{1}^{0}) = (150.7, 150.35, 150)$ GeV signal sample. The $m(\tilde{\chi}_{2}^{0}, \, \tilde{\chi}_{1}^{\pm}, \,  \tilde{\chi}_{1}^{0}) = (151, 151, 150)$ GeV signal sample is not expected to be excluded, due to its low significance and the difficulty of discriminating it from the backgrounds.\\

\begin{figure}[!htb]
\centering
\includegraphics[width=0.99\linewidth]{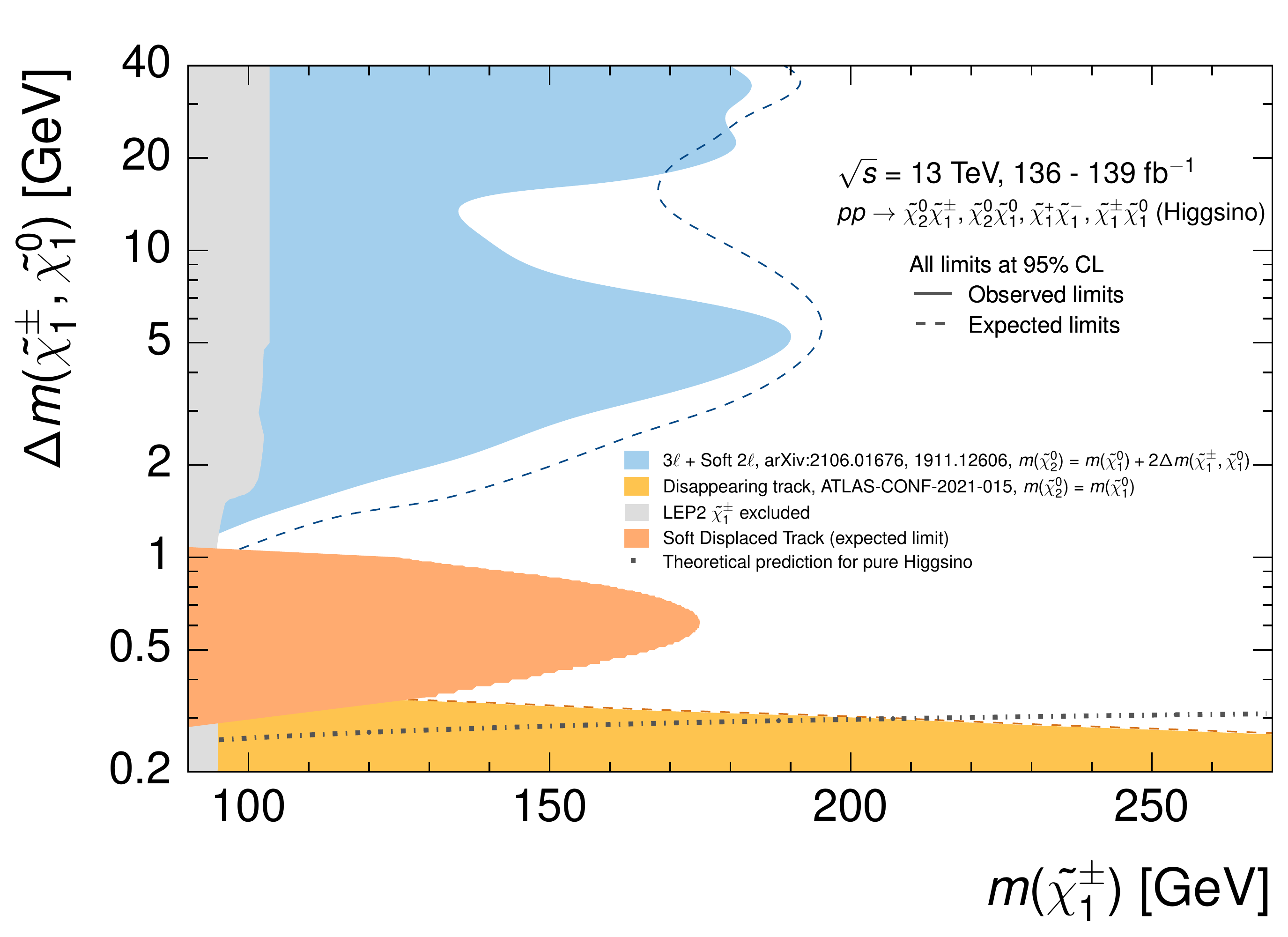}
\caption{Exclusion limits on SUSY simplified models for higgsino production with the addition of the orange shaded band corresponding to the expected exclusion limit achieved by this search. All limits are computed at 95\% CL.}
\label{fig:Higgsino_Contour}
\end{figure}

The expected exclusion limits at 95\% CL on the masses of the higgsinos achieved by this search are shown in Fig.~\ref{fig:Higgsino_Contour}.

They are obtained by considering different mass hypotheses for the higgsino candidates, all having different cross-sections according to the $\tilde{\chi}_{2}^{0}, \, \tilde{\chi}_{1}^{\pm}$ masses but the same kinematics of the benchmark points for the corresponding mass splitting $\Delta m(\tilde{\chi}_{1}^{\pm},\,\tilde{\chi}_{1}^{0})$. \\

Higgsino masses up to 175 GeV are expected to be excluded at 95\% CL in the case of a mass splitting between chargino and neutralino of about 0.5 GeV, for both $m(\tilde{\chi}_{2}^{0})= m(\tilde{\chi}_{1}^{\pm})$ and $m(\tilde{\chi}_{1}^{\pm})= m(\tilde{\chi}_{2}^{0})$ scenarios. The expected limits supersede the previous ATLAS observed limits and cover the gap between the multi-lepton search and the disappearing track search, in particularly interesting regions where the higgsino model with masses around the electroweak scale is theoretically well-motivated.\\

These results show that the analysis has the sensitivity to exclude different signal hypotheses for higgsino masses around 150 GeV if no excess is observed in data. At the lower mass of 100 GeV, the larger signal cross-section allows to achieve significance values larger than 5, corresponding to the conventional discovery sensitivity, for mass splittings of 0.35 and 0.5 GeV. All these signal hypotheses have not been probed by any existing analysis of LHC data.

\chapter*{Conclusions}
\addcontentsline{toc}{chapter}{Conclusion}
\label{sec:Conclusions}

In this thesis, two supersymmetric analyses were presented.\\

The chargino analysis searched for the production of a pair of charginos, each decaying into a neutralino via the emission of a $W$ boson, and targeted a phase space with compressed mass splittings between the chargino and the neutralino. Such a signal, having a relatively low production cross section and being very similar to the $WW$ background, is very challenging to search for and has evaded all searches at the LHC. The new search presented here considered a final state with two leptons from the $W$ boson decays, missing transverse momentum and no hadronic activity. Advanced machine learning techniques were adopted to enhance the supersymmetric signal from the backgrounds and to achieve an unprecedented sensitivity. Results obtained with 139 $\mathrm{fb^{-1}}$ of proton-proton collision data using the ATLAS detector at a centre-of-mass energy of 13 TeV were reported. Chargino masses up to about 140 GeV are excluded at 95\% CL in the case of a mass splitting between chargino and neutralino down to about 100 GeV. The results supersede the previous ATLAS results in particularly interesting regions where the difference in mass between the chargino and neutralino is close to the mass of the $W$ boson and the chargino pair production could have hidden behind the looking-alike $WW$ background.\\

The higgsino analysis searched for the production of pairs of charginos decaying into pions and neutralinos. The charginos and neutralinos have small mass splittings in the targeted scenario, thus pions leave mildly displaced tracks in the detector a few millimetres away from the production vertex. The experimental signature is one jet, missing transverse momentum, and a low momentum charged track with an origin displaced from the collision point, the last element being the first time it is used in a search of this kind at a hadron collider. Sensitivity estimates relative to 139 $\mathrm{fb^{-1}}$ of proton-proton collision data using the ATLAS detector at a centre-of-mass energy of 13 TeV were reported. The results show that the analysis has the sensitivity to exclude different signal hypotheses for higgsino masses around 150 GeV if no excess is observed in data. For lower masses, the larger signal cross-section allows to achieve higher significance values for different mass splitting scenarios. All these signal hypotheses have not been probed by any existing analysis of LHC data. \\

In addition to the two analyses, the work carried out at CERN as a Doctoral Student during the last year of my PhD was presented. This work involved the assembly and quality control of pixel modules for the ATLAS Phase-II upgrade during which time the experiment aims to accumulate a total dataset of 4000 $\mathrm{{fb}^{-1}}$ with a peak instantaneous luminosity of $7.5 \times 10^{34} \, \mathrm{cm^{-2}s^{-1}}$. CERN is one of the assembly centres of pixel modules for the barrel layers of the Phase-II Inner Tracker, ITk, with several key activities conducted to ensure sustained and reliable production rate of hybrid pixel modules. As part of the ATLAS pixel module team at CERN, I contributed to the activities carried out in the laboratory, taking part in the assembly of the pixel modules and in their quality control through visual inspection, metrology and electrical tests, following and developing new guidelines.

\printbibliography[heading=bibintoc]

\end{document}